\documentclass[a4paper,12pt,twoside,openany]{book}
\usepackage{amssymb}
\usepackage{amsmath}
\usepackage[utf8]{inputenc}
\usepackage{color}
\usepackage[margin=1in]{geometry}
\usepackage{fancyhdr}
\usepackage{feynmp-auto} 
\usepackage{caption}
\usepackage{subcaption}  
\usepackage{graphicx}
\usepackage{inputenc}
\usepackage{xspace}
\usepackage{url}
\usepackage{epstopdf}
\usepackage[nottoc,numbib]{tocbibind}
\graphicspath{{./figures/}}
\usepackage{ulem}
\usepackage{multirow}
\usepackage{float}
\newcommand{\be}{\begin{equation}}
\newcommand{\ee}{\end{equation}}
\newcommand{\beq} {\begin{equation}}
\newcommand{\eeq} {\end{equation}}
\newcommand{\ba}{\begin{eqnarray}}
\newcommand{\ea}{\end{eqnarray}}

\usepackage{amsmath}
\usepackage{graphicx}
\usepackage{amsfonts}
\usepackage{latexsym}
\usepackage{tikz}
\usepackage{bbold}
\usepackage{wasysym}
\usepackage{calligra}
\usepackage{float}
\usepackage{ulem}
\usepackage{inputenc}
\usepackage{xspace}
\usepackage{url}
\usepackage{epstopdf}

\newcommand\Lie{\pounds}

\begin{document}
\frontmatter
%titlepage
\thispagestyle{empty}
\begin{center}
\begin{minipage}{0.75\linewidth}
    \centering

    \includegraphics[scale=1.3]{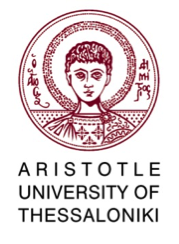}
    \rule{0\linewidth}{0\linewidth}\par
   \vspace{2cm}
%Thesis title
    {\uppercase{\textbf{\large Metric-Affine Gravity and Cosmology/ Aspects of Torsion and Non-metricity in Gravity Theories}\par}}
    \vspace{1.5cm}
%Author's name
    {\large by \textbf{Iosifidis Damianos}\footnote{email:diosifid@auth.gr}\\
     Institute of Theoretical Physics \\
   Physics Department of Aristotle University of Thessaloniki \par}
    \vspace{2.0cm}
%Degree
    {\Large A thesis submitted for the degree of Doctor of Philosophy\par}
    \vspace{2.2cm}
    
%Date
{ Supervised by Anastasios Petkou and Christos Tsagas \par}
 \vspace{0.5cm}
    {\Large January 2019}

\end{minipage}
\end{center}
\clearpage

\begin{flushright}
''Don't wish it was easier, wish you were better''

Jim Rohn

\end{flushright}

\chapter{Preface}

General Relativity is, in its own right, a very elegant and mathematically well established Theory of Gravity. Even though it has passed many tests, it has some serious drawbacks and therefore cannot be regarded as the fundamental  Theory to describe Gravitation. As it is well known the basic  drawbacks of GR are, the inability to explain the late time Cosmological acceleration, the dark matter problem, the early Universe description and the fact that it is not renormalizable. Along with the aforementioned problems, there is also the need for the Grand Unification, that certainly demands that GR should be modified, at least in the microscopic world. Therefore, the past few years, there was a tremendous interest in Modified Gravity. In general, there are many ways to modify General Relativity. To name a few, one can consider extra fields (scalar, vector,... etc), extend the spacetime dimensions ($n>4$), generalize the geometry to include torsion and so on (see for instance \cite{charmousis2015self} for a more extensive discussion). Among the many approaches to modify gravity, in this Thesis we will consider modifications that arise due to the generalization of the spacetime geometry. In particular, we will study what is broadly known as Metric-Affine Theories of Gravity(MAG).  

 The main feature of MAGs is that the underlying geometry is no longer Riemannian and possesses both torsion and non-metricity. In this generalized non-Riemannian geometry vectors rotate (torsion) and undergo a length change (non-metricity) when transported on the manifold (see detailed discussion in $1^{st}$ Chapter). The geometry on the manifold can be fully described once a metric $g_{\mu\nu}$ and an independent affine connection $\Gamma^{\lambda}_{\;\;\;\mu\nu}$ are given. In this framework $\left\lbrace \mathcal{M},g,\Gamma\right\rbrace$, the metric and the connection are not  a priori related and a relation among them may be found only after solving the field equations. This general procedure for solving for the affine connection for generic actions is also presented in this Thesis. The main advantage of MAG that distinguishes it from the rest of Modified Gravity Theories is that the modifications in this case are naturally produced by the generalization of the geometry and have a well established geometrical meaning. The new degrees of freedom that come from torsion and non-metricity are beautifully encoded in the affine connection. In addition, in the general Metric-Affine formalism, the connection couples with the matter fields and the inclusion of particle's spin is  easily formulated into the Theory. This is what happens for instance in the so-called Einstein-Cartan Theory where the space apart from curvature, also possesses torsion (but zero non-metricity). There also exists the Weyl space which is torsionless but with curvature and a specific part of non-metricity. Therefore we see that many theories can be obtained as special cases of the general MAG. Another great advantage of Metric-Affine Gravity is that, when written in the language of differential forms, it can be seen as a gauge theory of gravity. Furthermore, Metric-Affine Gravity is an excellent tool that allows microscopic properties of matter to act as sources for the gravitational field since it takes into account the intrinsic characteristics of particles such as spin , dilation current, hypercharge etc.(as mentioned above). For more information on the advantages and  motivation for MAG see \cite{hehl1995metric}. 
 
 Even though the extensive study of MAG has started a few decades ago there are many things that need to be addressed. To name a few, what is the role of projective invariance and its relevance to physics(if there is any) and how can we break it?. In addition, since both torsion and non-metricity can be determined once a connection is given, how can one solve for the affine connection for general theories? Can we classify Theories with dynamical/non-dynamical connections? Under which circumstances a connection becomes dynamical? How does the presence of torsion and non-metricity affects the Raychaudhuri equation and which are the most general modified Friedmann equations with torsion and non-metricity? Can we formulate actions that are conformally and/or projective invariant? The above questions are some of the many that are addressed in this thesis. To be more specific let us sketch the most important novel features about MAG that are presented in this thesis
\begin{itemize}
	\item A very detailed and extensive introduction into the generalized geometry is presented. The geometrical role of torsion and non-metricity is discussed in depth and many illustrative examples (most of which were absent from the literature) are given.
	\item A new way to break the projective invariance in Metric-Affine $f(R)$ theories of gravity is proposed, that treats the torsion and non-metricity vectors on equal footing.
	\item The general proof on how to solve for the affine connection in MAG is presented (for the first time in the literature) and the results are collected in 3 subsequent Theorems.
	\item The classification of a broad class of Theories that yield Einstein's Gravity in vacuum is presented and proved.
	\item A method to excite torsion by coupling surface terms with scalar is extended to include also non-metricity.
    \item The Kinematics of of torsion and non-metricity in FRW Universes is presented. The most general expression for non-metricity\footnote{For torsion the results were, for  long time, known in the literature.} in such spaces is derived along with the modified Friedmann equations with non-metricity.
    \item The peculiar Metric-Affine $f(R)\propto R^{2}$ is extensively studied and the duality between torsion and non-metricity is derived for such Theories. The key point that allows to map vectorial torsion to Weyl non-metricity is found and a proof of the map is given. The cosmological solutions are also found for this peculiar case.
    \item The most general form of the Raychaudhuri equation is derived (for the first time in the literature) for spaces of arbitrary dimension that have both torsion and non-metricity. The result is applied to Cosmology and Cosmological solution in the presence of torsion and non-metricity are found respectively. For completeness we also derive the vorticity evolution equation.
    \item Scale Transformations in Metric-Affine Geometry are considered and scale invariant Theories are constructed with respect to the three possible scale transformations in the Metric-Affine Geometry. The identities that come along with the Invariances are obtained and the parameter space, of Theories that respect each transformation, is found.
\end{itemize}

Of course, the above list is but a very small contribution to the fruitful field called Metric-Affine Gravity and many other questions need to be answered. We will touch upon future projects and possible extensions of this study in the last chapter.

\newpage

	\section{Publications/Collaborations}
The content of this PhD Thesis is mainly based (but not restricted to) on the following publications (newest first)
\newline
\newline
\begin{enumerate}
	\item ''Exactly Solvable Connections in Metric-Affine Gravity'' \cite{iosifidis2018exactly}
	\newline
	\newline
	\item ''Scale Transformations in Metric-Affine Geometry'' \cite{iosifidis2018scale}
	\newline
	\newline
	\item ''Torsion/non-metricity Duality in f (R) Gravity'' \cite{iosifidis2018torsion}
	\newline
	\newline
	\item ''Friedmann-like Universes with Torsion'' \cite{kranas2018friedmann}
	\newline
	\newline
	\item ''Raychaudhuri Equation in Spacetimes with Torsion and Non-metricity''\cite{iosifidis2018raychaudhuri}
	\newline
	\newline
	\item ''Self tuning scalar tensor black holes'' \cite{charmousis2015self}
\end{enumerate}
where (most of) the publications were done in a collaboration with the respective Professors and Colleagues (see References and also Acknowledgments for more details).

\newpage
\section{Conventions/Notations}

\begin{tabular}{|l|l|}
	\hline
	\multicolumn{2}{|c|}{Conventions/Notations} \\
	$\delta_{\mu}^{\nu}$& Kronecker's delta \\
	$\varepsilon_{\mu\nu\rho\sigma}$ & Totally antisymmetric Levi-Civita symbol \\
	$\epsilon_{\mu\nu\rho\sigma}=\sqrt{-g}\varepsilon_{\mu\nu\rho\sigma}$ & Levi-Civita tensor \\
	$\delta^{\mu_{1}\mu_{2}...\mu_{k}}_{\nu_{1}\nu_{2}...\nu_{k}}$ & Generalized Kronecker delta \\
	$g_{\mu\nu }=g_{\mu\nu}(x^{\alpha})$ & Metric tensor of general spacetimes  \\
	$g=det(g_{\mu\nu})$ & Determinant of the metric tensor \\
	$Tr(A_{\mu\nu})=g^{\mu\nu}A_{\mu\nu}$ & Generalization of the trace of a matrix $A_{\mu\nu}$ in curved spacetimes \\
	$\eta_{\mu\nu}=diag(-1,1,1,1)$ & Minkowski metric tensor \\
	$\Gamma^{\alpha}_{\;\;\;\mu\nu}$ & Linear  Affine Connection \\
	$\tilde{\Gamma}^{\alpha}_{\;\;\;\mu\nu}$ & Levi-Civita connection (or Christoffel symbols) \\
	$\nabla_{\mu}$ & Covariant derivative with respect to the Affine Connection  \\
	$R^{\mu}_{\;\;\;\nu\rho\sigma}$ & Riemann (or curvature) tensor \\
	$S_{\mu\nu}^{\;\;\;\;\lambda}\equiv  \Gamma^{\lambda}_{\;\;\;\;[\mu\nu]}$ & Torsion tensor\\
	$Q_{\alpha\mu\nu}\equiv -\nabla_{\alpha}g_{\mu\nu}$ & Non-metricity tensor \\
	$\Psi$ & Collectively denotes matter fields \\
	$\mathcal{L}$ & Lagrangian density \\
	$T_{\mu\nu} \equiv -\frac{2}{\sqrt{-g}}\frac{\delta S_{M}[g,\Gamma,\Psi]}{\delta g^{\mu\nu}}$ & Energy-momentum (or stress-energy) tensor \\ 
	$\Delta_{\lambda}^{\;\;\;\;\mu\nu} \equiv -\frac{2}{\sqrt{-g}}\frac{\delta S_{M}[g,\Gamma, \Psi]}{\delta \Gamma^{\lambda}_{\;\;\;\mu\nu}}$ & Hyper-momentum Tensor \\
	$a(t)$ & Scale factor \\
	$H(t)\equiv \frac{\dot{a}}{a}$ & Hubble parameter \\
	$p(t)$ & Pressure of a perfect fluid \\
	$\rho(t)$ & Density of a perfect fluid \\
	$\psi,\phi$ & Scalar fields \\
	$\kappa=8\pi G$ & Gravitational constant \\
	$G$ & Newton's Gravitational constant \\
	$m_{pl}^{2}=\kappa^{-1}$ & Squared Plank mass \\
	\hline
\end{tabular}
\newline \\
\\
A list of some of the conventions/notations that are used throughout the thesis is given above. We try to be as standard and self-consistent  as possible with the conventions and there is an indication whenever a different notation is used.\\\\

We furthermore adopt natural units, that is, $c=1=\hbar$.
Notice also, that we have chosen the convention $\eta_{\mu\nu}=diag(-1,1,1,1)$ for the Minkowski metric tensor as this is the most common choice made in GR textbooks. As a result, for the Robertson-Walker metric we have\footnote{One usually adopts the gauge $N(t)=1$.}
\begin{equation}
ds^{2}=-N^{2}(t)dt^{2}+a^{2}(t)\left( \frac{dr^{2}}{1-Kr^{2}}+r^{2}(d\theta^{2}+\sin^{2}\theta d\phi^{2} )\right)
\end{equation}
We denote the symmetric and antisymmetric parts of a tensor with round and square brackets respectively. For instance, given a rank-$2$ tensor $T_{\mu\nu}$, its symmetric part is expressed as
\begin{equation}
T_{(\mu\nu)}=\frac{1}{2}(T_{\mu\nu}+T_{\nu\mu})
\end{equation}
while the antisymmetric reads
\begin{equation}
T_{[\mu\nu]}=\frac{1}{2}(T_{\mu\nu}-T_{\nu\mu})
\end{equation}

As a final remark, we use the standard notation for the indices. Namely, the Greek indices $\mu,\nu,\rho,...$ etc. run from $0$ to $3$ (or from $0$ to $n-1$ for general dimensions) while Latin indices run over the spatial part of spacetime, that is $i,j,k,...=1,2,3$. Tangent space indices will be denoted by the first letters of the Latin alphabet, namely $a,b,c,...$ and will run from $0$ to (n-1).

\pagenumbering{roman}
\tableofcontents

\mainmatter
\pagenumbering{arabic}

\include{chapters}

	%$\Box$
	
	\newpage

	%\chapter*{Preface}
	%\addcontentsline{toc}{chapter}{Preface}

	\chapter*{Introduction}

	Geometrical modifications of Gravity by generalizing the affine connection have a long history and date back to the works of Weyl \cite{Weyl:1918ib} and Cartan \cite{cartan1922equations}. In Weyl's theory the connection was symmetric but not metric compatible while Cartan's was a metric one but with an antisymmetric part (torsion). A general space that has an affine connection that is neither metric compatible nor symmetric constitutes what is broadly known as non-Riemannian Geometry. The underlying  Gravity theory in such a geometry is called Metric-Affine Gravity\cite{hehl1995metric}. In the Metric-Affine formulation, the metric tensor $g_{\mu\nu}$ and the affine connection $\Gamma^{\lambda}_{\;\;\;\mu\nu}$ are treated as independent variables and a relation among them may be found only after using the field equations. In the general formulation, both the gravity and matter sectors can depend on the affine connection. The additional contributions in the Metric-Affine theories come from torsion and non-metricity. Torsion is the antisymmetric part of the connection and the non-metricity measures the failure of the connection to be metric compatible (see definitions in next chapter). Both of these features can be computed once an affine connection $\Gamma^{\lambda}_{\;\;\;\mu\nu}$ is given\footnote{To be more specific, this is true only for torsion. In order to compute the non-metricity tensor one also needs to have a metric (along with the affine connection).}.

	Metric-Affine Theories of Gravitation are particularly interesting for studying modifications of Gravity (beyond General Relativity) because the modifications, in this case, are introduced naturally by extending the geometry to be non-Riemannian. In view of this, along with the need to modify General Relativity, the latter have attracted some attention during the past few years \cite{vitagliano2011dynamics,olmo2011palatini,sotiriou2007metric,vitagliano2010dynamics,olmo2009dynamical,sotiriou2010f}, especially when it comes to Palatini $f(R)$ Gravity \cite{olmo2011palatini,sotiriou2007metric}. The Palatini approach is based on the assumption that the matter part of the action does not depend on the connection. With such a simplifying assumption, it can be shown (see for instance  \cite{sotiriou2009f}) that the connection in Palatini $f(R)$ lacks dynamics and can be expressed in terms of the metric, its derivatives and the matter fields. The situation changes radically when one allows matter to couple to the connection. In this case (Metric-Affine $f(R)$) the connection becomes dynamical in general \cite{vitagliano2011dynamics}. Staying in the realm of Palatini Gravity it was shown in   \cite{allemandi2004accelerated}   (and also in \cite{olmo2009dynamical}) that for Ricci squared families of the type  $f(R,R_{(\mu\nu)}R^{(\mu\nu)})$ the affine connection can still be algebraically eliminated and carries no dynamics. The way to solve for the affine connection was also presented there \cite{allemandi2004accelerated,olmo2009dynamical}. This is not the case however when one generalizes to families of the type $f(R,R_{\mu\nu}R^{\mu\nu})$ and in this case the connection becomes dynamical, as shown in \cite{vitagliano2010dynamics}, even for the simplifying case of vanishing torsion. From an effective field theory perspective, theories containing second order invariants of torsion and non-metricity were studied\footnote{The renormalizability of theories containg quadratic torsion and non-metricity scalars was studied in \cite{pagani2015quantum}.} in \cite{vitagliano2014role} and \cite{vitagliano2011dynamics} where it was found that to this order the connection lacks dynamics, but of course will become dynamical once higher order terms are added.

	Therefore, from the above discussion we see that it is important to have a tool for obtaining the form of the affine connection for a given theory and see whether the latter becomes dynamical or not.
	This is one of the  subjects that we investigate throughout this thesis. To be more specific, we present a systematic way to do so for specific  Metric-Affine theories, for the first time in the literature. With this at hand, one can study in depth, Metric-Affine Theories. We will be using the above result throughout the thesis in order to study the various MAG theories. In addition, there are many questions that arise when one is working in a generalized geometry such as, what is the form of the Raychaudhuri equation in non-Riemannian spaces, how do  torsion and non-metricity look like in a highly symmetric spacetime such as an FLRW Universe, etc. The aforementioned questions along with many others are answered in the various chapters of the thesis. To be more specific, the thesis is organized as follows. 
	
	In the first Chapter we define the various geometrical quantities that characterize a non-Riemannian geometry. We also present many examples in order to illustrate the geometric meaning of torsion and non-metricity since we believe that such examples are absent from the literature and will help one gain a deeper understanding of the generalized geometry. In addition, we carefully compute all the tricky parts that arise in computations due to the presence of torsion and non-metricity such as non-trivial surfaces terms, generalized Bianchi identities etc. Finally we introduce    and develop all the necessary machinery needed to study MAG model building. 
	
	In the second Chapter we explore the MAG model building. We start by the Einstein Hilbert action, solve the equations of motion and discuss the projective freedom. We then proceed with $f(R)$ theories and
	 we also touch upon projective invariance breaking in Metric-Affine $f(R)$ theories. In particular, we review the two methods that have been suggested in the literature (\cite{1981GReGr..13.1037H,sotiriou2007metric})  in order to break this invariance and also present another possibility. Our approach on breaking the projective invariance treats the torsion and non-metricity vectors on equal footing and can therefore be considered as the more appropriate one. We then continue by deriving the field equation of more general Metric-Affine theories.

	In Chapter $3$ we use a well known procedure to excite torsional degrees of freedom by coupling surface terms to scalars. We extend this procedure in order to excite non-metric degrees of freedom. We then apply these methods to excite torsion and non-metric degrees of freedom simultaneously.
	
	Then, in Chapter $4$ which seems to be the most important Chapter of the thesis, at least with regards to its use in applications, we present a step by step way to solve for the affine connection in non-Riemannian geometries, for the first time in the literature. We start with certain assumptions which we relax later on. We collect our results and present them as $3$ subsequent Theorems. We then consider three simple examples  and apply  our results  in order to illustrate the procedure and discuss the cases of dynamical/non-dynamical connections.
	
	A peculiar $f(R)$ case is studied in Chapter $5$. This is the conformally (as well as projective invariant) invariant theory $f(R)=\alpha R^{2}$ which contains an undetermined scalar degree of freedom. For this case we study separately cosmological solutions\footnote{Of course this model cannot be regarded as a viable cosmological model since there is an undetermined scalar degree of freedom.} for the pure non-metricity case and observe that the solution is identical to the pure torsion case presented in \cite{capozziello2008f}. We can map one theory to another by making a simple exchange between the torsion and non-metricity vectors. We then go on and study the same model but now allowing  both torsion and non-metricity to be present. We then discuss the torsion-nonmetricity duality for this simple case and show how one can treat vectorial torsion for Weyl non-metricity in projective invariant theories.
	
	We then turn our attention to Cosmology with torsion and non-metricity (Chapter $6$). After studying to some degree the kinematics with torsion and non-metricity we derive the allowed forms of torsion and non-metricity that can live in such spacetimes. In addition we obtain the form of fixed length non-metricity that is allowed in FLRW spacetimes. We also find cosmological solutions with torsion and also derive the modified Friedmann equations in the presence of non-metricity.
	
	In Chapter $7$,  we formulate the necessary setup for the $1+3$ splitting of the generalized spacetime.  Having clarified the subtle points (that generally stem from non-metricity) in the aforementioned formulation  we carefully derive the generalized Raychaudhuri equation in the presence of both torsion and non-metricity (along with curvature). This, as it stands, is the most general form of the Raychaudhuri equation that exists in the literature. It is the Raychaudhuri equation in generic non-Riemannian spaces. We then specialize to the pure torsion and pure non-metricity cases and discuss similarities and differences. In addition,  considering first vectorial torsion (and vanishing non-metricity) and then Weyl non-metricity (and vanishing torsion) we find find cosmological solutions for each case. As it turns out the two cosmological solutions are identical by again exchanging (with the appropriate factors) the torsion and Weyl vectors. This, as we have discussed in previous chapter, is a consequence of the simplified forms of torsion and non-metricity we have chosen and we will not expect it to hold true for more general geometries since non-metricity caries more degrees of torsion in general. Closing this Chapter we also find some solutions for fixed length vector non-metricity and also derive the evolution equation for vorticity, in non-Riemannian spaces, for completeness.

	We close this Thesis by considering three possible scale transformations that one can consider in Metric-Affine Geometry. These are,  conformal transformations of the metric (keeping the connection fixed), projective transformations of the connection (keeping the metric fixed) and frame rescalings which are composed by a combination of conformal transformation accompanied by a special projective transformation. After computing how  the torsion and non-metricity tensors change under these transformations, we prove the identities that have to be obeyed by actions that are invariant with respect to these transformations. We then construct theories quadratic in torsion and non-metricity, derive the general field equations and impose conditions on the parameters.  Continuing, we also consider parity violating terms and write down the most general quadratic action with torsion and non-metricity including all  possible parity-even and parity-odd scalars. For this case we also restrict the parameter space and find the conditions on the parameters in order for the given theory to be invariant under conformal, projective and frame rescaling transformations respectively.
	
	Let us now review the vielbein and coordinate formalisms for Gravity, and discuss the variational approaches before starting with our introduction for the non-Riemannian geometry. We start by defining the notions of frame fields and spin connection and discuss the exterior forms approach to MAG.

	\section{Frames and Co-frames on a differential manifold $\mathcal{M}_{n}$}
	We shall start with a very brief discussion of Metric-Affine Gravity written in the language of exterior differential forms. This formalism, certainly has many advantages with regards to computations as well as the gauge nature of MAG. The independent fields used in the variational principle  are now the co-frame $e^{a}$ and the linear (or oftentimes called spin) connection $\omega^{a}_{\;\; b}$(see definitions in what follows). These variables ($e,\omega$) are related to the basic fields of the coordinate formalism ($g,\Gamma$). After giving the basic definitions of this approach we will almost exclusively work on the coordinate formalism having as our variables the metric tensor and the independent affine connection. Let us for te time being review the MAG setup using forms.
	Consider an $n-dim$ differentiable manifold $\mathcal{M}_{n}$. Then, at each point $P$ we can define the so-called $tangent$    vector space (at $P$) $T_{P}(\mathcal{M}_{n})$ which is of the same dimensionality $n$ with the manifold. Having defined the tangent space we can consequently introduce a local vector basis on it. We call this local vector basis -$e_{a}$ local frame\footnote{It is also very common to refer to it as vielbein (the German word for many-legs$;$ as many as the dimensionality of the manifold). In $4-dim$ it is called $vierbein$(four legs) or a tetrad which is the Greek word for a group of four.}, $a=0,1,...,n-1$. Our conventions for the indices are the following. Latin ones $a,b,c,...$ are frame indices (anholonomic)\footnote{Holonomic indices are those which can be expanded in a local coordinate basis $\partial_{\mu}$. On the contrary, anholonomic indices cannot be written as partials, that is we cannot write $\partial_{a}$.} while the Greek ones $\mu,\nu,\rho,...$ are coordinate indices. Both of them run over the dimensionality of the manifold, that is from $0$ to $n-1$. Given a local coordinate basis $\partial_{\mu}=\frac{\partial}{\partial x^{\mu}}$ we can expand the frame $e_{a}$ as
	\begin{equation}
	e_{a}=e^{\mu}_{\;\;a}\partial_{\mu}
	\end{equation}
	Note that $e_{a}$ can be regarded as an anholonomic basis so long as, $\det(e^{\mu}_{\;\;a})\neq 0$. Given a tangent space $T_{P}(\mathcal{M}_{n})$ we can define its dual space $T_{P}^{\ast }(\mathcal{M}_{n})$ which is referred to as the $co$-$tangent$ space. On the co-tangent space $T_{P}^{\ast }(\mathcal{M}_{n})$ we can define the co-frame $\vartheta^{a}$ which is, by definition, given by
	\begin{equation}
	e_{a} \rfloor \vartheta^{b}=\delta_{a}^{b}
	\end{equation}
	where $\rfloor$ denotes interior product. In addition, in the co-tangent space there exists an $1$-form basis $dx^{\mu}$ and the co-frame may be expanded as\footnote{We use both symbols $	\vartheta^{b}$ and $e^{b}$ for the co-frame since both of them appear in the literature depending on ones preference in notations.}
	\begin{equation}
	\vartheta^{b}=e^{b}=e_{\mu}^{\;\; b}dx^{\mu}
	\end{equation}
	From the above definitions we also conclude that
	\begin{equation}
	e^{\mu}_{\;\;a}e_{\mu}^{\;\;b}=\delta_{a}^{b}
	\end{equation}
	that is $e_{\mu}^{\;\; b}$ is the inverse of $e^{\mu}_{\;\;a}$. Note that the requirement $\det(e^{\mu}_{\;\;a})\neq 0$ also forces  $\det{(e_{\mu}^{\;\;b})}\neq 0$. Indeed, taking the determinant of the above equation it follows that
	\begin{equation}
	\det{(e^{\mu}_{\;\;a})} \det{(e_{\mu}^{\;\;b})}=\det{(\delta_{a}^{b})}=1\Rightarrow \det{(e_{\mu}^{\;\;b})}=\frac{1}{\det{(e^{\mu}_{\;\;a})}} \neq 0
	\end{equation}
	given that $\det(e^{\mu}_{\;\;a})\neq 0$. With the help of veirbeins one can write the metric tensor (in $4-dim$ for instance) as 
	\beq
	g_{\mu\nu}=e_{\mu}^{\;\;a}e_{\nu}^{\;\;b}\eta_{ab}
	\eeq
	where $\mu,\nu$ are coordinate indices while $a,b$ are tangent space indices and $n_{ab}=(-1,1,1,1)$ is the tangent-space Minkowski metric. Then, one also has
	\beq
	e \equiv \det{(e^{\mu}_{\;\;a})}=\sqrt{-g}
	\eeq
	as can be easily checked by taking the determinant of the equation above. In addition, the tetrad field may be used to relate the internal and external index tensors, according to
	\beq
	A^{\mu}= e^{\mu}_{\;\; a}A^{a} \;,\;\;\; A^{b}= e_{\mu}^{\;\; b}A^{\mu}
	\eeq
	Notice now that given a metric tensor, the orthonormal frame is not unique. Indeed, we can always perform a general  linear transformation parametrized by the matrices $\Lambda^{a}_{\;\;b}$ , on the frame 
	\beq
	e_{\mu}^{\;\; a}\rightarrow \tilde{e}_{\mu}^{\;\; a} =\Lambda^{a}_{\;\;b}e_{\mu}^{\;\; b}
	\eeq
	\beq
	e^{\mu}_{\;\; a}\rightarrow \tilde{e}^{\mu}_{\;\; a} =(\Lambda^{-1})^{b}_{\;\;a}e^{\mu}_{\;\; b}
	\eeq
	with $\Lambda^{a}_{\;\;c}(\Lambda^{-1})^{c}_{\;\;b}=\delta_{b}^{a}$ and we then get the same metric
	\beq
	g_{\mu\nu}=\tilde{e}_{\mu}^{\;\;a}\tilde{e}_{\nu}^{\;\;b}\eta_{ab}=e_{\mu}^{\;\;a}e_{\nu}^{\;\;b}\eta_{ab}
	\eeq
	where we have used the fact that $\eta_{ab}\Lambda^{a}_{\;\;c}\Lambda^{b}_{\;\;d}=\eta_{cd}$. Let us now define the spin connection $\omega^{a}_{\;\; b}$. Consider a vector valued form $v^{a}$. The exterior covariant derivative is defined to be
	\beq
	Dv^{a}:= dv^{a}+\omega^{a}_{\;\; b}\wedge v^{b}
	\eeq
	The above definition may of course be defined for tensor valued forms. With this definition at hand one defines the torsion $2-form$ 
	\beq
	\mathcal{T}^{a}:=De^{a}=de^{a}+\omega^{a}_{\;\; b}\wedge e^{b}
	\eeq
	from which we see that torsion is the field strength of local translations. Continuing we may define the curvature $2-form$
	\beq
	\mathcal{R}^{a}_{\;\; b}:=d\omega^{a}_{\;\; b}+\omega^{a}_{\;\; c}\wedge \omega^{c}_{\;\; b}
	\eeq
	and the non-metricity $1-form$
	\beq
	\mathcal{Q}_{ab}:=-D\eta_{ab}=2\omega_{(ab)}
	\eeq
	The first two of the above are sometimes called Cartan's first and second structural equations. With these definitions the Bianchi identities may be easily derived\cite{hehl1995metric}
    \beq
    D\mathcal{T}^{a}=\mathcal{R}^{a}_{\;\; b}\wedge e^{b}
    \eeq
	\beq
	D\mathcal{R}^{a}_{\;\; b}=0
	\eeq
	\beq
	D\mathcal{Q}_{ab}=2 \mathcal{R}_{(ab)}
	\eeq
	Now note that any   $p-form$  in an $n-dim$ space (with $p \leq n$) can be expanded in a coordinate basis according to
	\beq
	\alpha =\frac{1}{p!}\alpha_{\mu_{1},...,\mu_{p}} dx^{\mu_{1}}\wedge...\wedge dx^{\mu_{p}}
	\eeq
	The same of course holds true for vector valued forms too. 
	For instance, for torsion we have\footnote{Note that our definition when we expand forms in a coordinate basis is the following. The coordinate indices come first (on the very left)  and then the Lorentz ones follow (right).  }
	\beq
	\mathcal{T}^{a}=\frac{1}{2}\mathcal{T}_{\mu\nu}^{\;\;\;a}dx^{\mu}\wedge dx^{\nu}
	\eeq
	Then multiplying by the vielbeins $e_{\mu}^{\;\; a}$ we can switch from Lorentz indices to coordinate (or world) indices. So for the above example with torsion, we have
	\beq
	\mathcal{T}_{\mu\nu}^{\;\;\;\;\lambda}=e_{\lambda}^{\;\; a}\mathcal{T}_{\mu\nu}^{\;\;\;a}
	\eeq
	The above tensor, as we will show later, is equal (up to numerical factors) to the torsion defined in the coordinate formalism as the anti-symmetric part of the affine connection $\Gamma^{\lambda}_{\;\;\;\mu\nu}$. The link between frame and coordinate formalism is provided by the so-called "tetrad postulate"\footnote{It is common in the literature to call it postulate, however this linking equation for the two formalism need not be postulated.} which we will discuss briefly later on.

	We will return to the orthonormal frames at some  point and also define the spin connection (or tangent space connection) but for the most part of this thesis we will almost exclusively work on the coordinate formalism. For more on frames and the veilbein formalism the reader is referred to (\cite{hehl1995metric,aldrovandi2010introduction}).

	\section{Einstein-Hilbert action in metric and vielbein formulations}
	As it is known, Einstein's field equations can be derived by the variation of the so-called Einstein-Hilbert action
	\begin{equation}
	S_{EH}=\int d^{n}x \sqrt{|g|}R \label{mn}
	\end{equation}
	with respect to the metric tensor and subsequent application of the Principle of Least Action $\delta_{g} S=0$. In the above $R=g^{\mu\nu}R_{\mu\nu}$ is the Ricci scalar, $g$ is the determinant of the metric tensor and $\sqrt{|g|}d^{n}x$ the $n-dim$ invariant volume element. Action (\ref{mn}) as it stands, yields Einstein's equations in vacuum and if one wants to derive the full field equations, one has to add a matter action to it. Then, upon applying the Principle of least action  to
	\begin{equation}
	S[g_{\mu\nu}]=S_{EH}[g_{\mu\nu}]+S_{M}[g_{\mu\nu}]=\int d^{n}x \sqrt{|g|}R+S_{M}[g_{\mu\nu}]
	\end{equation}
	and making the identification 
	\begin{equation}
	T_{\mu\nu}:=-\frac{2}{\sqrt{|g|}}\frac{\delta S_{M}}{\delta g^{\mu\nu}}
	\end{equation}
	qs the energy-momentum tensor of matter, one arrives at the full Einstein equations in the presence of matter. We should point out something that was extremely crucial in our discussion so-far. That the actions considered above depend only on the metric tensor and not on the connection. To arrive at such a result two assumption have to be made, firstly that the connection is metric compatible and secondly that is torsionless. These two assumptions together force the connection to be uniquely given by the expression
	\begin{equation}
	\Gamma^{\alpha}_{\;\;\;\mu\nu}=\frac{1}{2}g^{\alpha\beta}(\partial_{\mu}g_{\beta\nu}+\partial_{\nu}g_{\beta\mu}-\partial_{\beta}g_{\mu\nu})
	\end{equation}
	which is called the Levi-Civita connection. The crucial point is that in such a formalism the connection carries no dynamics as  it is uniquely specified in terms of the metric and its first derivatives as seen above. This formalism (where both $\nabla_{\alpha}g_{\mu\nu}=0$ and $\Gamma^{\lambda}_{\;\;\;[\mu\nu]}=0$) is called the Metric formalism of Gravity. These two constraints form essentially what we call a Riemannian Geometry\footnote{When the assumptions of vanishing non-metricity and torsion are abandoned one is dealing with a non-Riemannian Geometry. This is exactly what we have in a Metric-Affine Gravity framework as we will see below.}. There also exist the so-called Palatini and Metric Gravity. We define what exactly do we mean by that in the next subsection.
	\
	We now go on by writing the Einstein-Hilbert action in the language of differential forms. This is exactly the same as the one we have given above, just written in a different language. However, one should be able (given the appropriate tools) to jump from one formalism to another. This is what we do here. Firstly, we give the Einstein-Hilbert action in the language of differential forms and then translate it to the one in the metric formalism.
	
	\section{Gravity in the Language of differential forms}
	An equivalent formulation of gravity an be achieved my writing the action in terms of exterior differential forms and consider as basic independent fields the co-frame $e^{a}$ and the so-called linear connection (or spin connection) $\omega^{a}_{\;\; b}$ instead of $g_{\mu\nu}$ and $\Gamma^{\lambda}_{\;\;\;\;\mu\nu}$. For intsance, the Einstein-Hilbert action, in this formalism, is given by
	\beq
	S_{EH}[e,\omega]=\frac{1}{4 \kappa}\int _{\mathcal{M}}\epsilon_{abcd}e^{a}\wedge e^{b}\wedge R^{cd}
	\eeq
	where $R^{ab}$ is the curvature $2-form$ constructed out of the spin connection $1-form$ as
	\beq
	R^{a}_{\;\;b}=d\omega^{a}_{\;\; b} + \omega^{a}_{\;\; c}\wedge  \omega^{c}_{\;\; b}
	\eeq
	Variation of the above action with respect to the two independent fields $e,\omega$ gives Einstein equations in vacuum but with an undetermined vectorial projective mode. We will discuss thoroughly about this projective freedom later on, but we should point out that this mode exists because the above action is invariant under projective transformations of the linear connection
	\beq
	\omega^{a}_{\;\; b} \rightarrow \omega^{a}_{\;\; b}+\delta^{a}_{b}\xi
	\eeq
	where $\xi$ is an arbitrary one-form. The same holds true when one is working in the coordinate formalism (with the fields $g,\Gamma$) where the Ricci scalar is also invariant under projective transformations of the affine connection $\Gamma^{\lambda}_{\;\;\;\;\mu\nu}$(see discussion in Chapter-2). As a last note we remark that the above action can be shown to be equivalent to the one in the coordinate formalism. More specifically, it holds that
	\beq
	\epsilon_{abcd}e^{a}\wedge e^{b}\wedge R^{cd}=2 R\sqrt{-g} d^{4}x
	\eeq 
	as can be easily seen by expanding $R^{ab}=\frac{1}{2}R^{ab}_{\;\;\;\;\;\;\mu\nu}dx^{\mu}\wedge dx^{\nu}$, $e^{a}=e_{\mu}^{\;\; a}dx^{\mu}$ and using the fact that $dx^{0}\wedge dx^{1}\wedge dx^{2} \wedge dx^{3}=d^{4}x$. We will come back to the frame formalism at some point but from the most part we will be working in the coordinate formalism. Let us now discuss the three most common (among the many) variational approaches to gravity.

	\subsection{Most common Variational Approaches for Gravity}
	There are $3$ basic variational approaches to Gravity\footnote{In fact there is also the purely affine theory of Gravity where the basic field is only the affine connection but we will not discuss it here. For a review see for instance \cite{poplawski2006nonsymmetric} and references therein.}. These are the $Metric$, the $Palatini$ and the $Metric-Affine$ Gravities. Each one of them predicts different dynamics, in general, but in some particular cases they coincide, as we discuss in what follows. Before introducing any of them, we should point out that throughout the thesis we are going to be focusing mostly on the $Metric-Affine$ approach which is the most general among them. Let us now explore the aforementioned approaches.
	\subsubsection{$Metric-Theories$ of Gravity}
	In $Metric-Theories$ of Gravity one makes two assumptions. Firstly, that the connection is metric compatible, that is
	\begin{equation}
	\nabla_{\alpha}g_{\mu\nu}=0
	\end{equation}
	along with the assumption of a torsion-free connection
	\begin{equation}
	\Gamma^{\alpha}_{\;\;\;[\mu\nu]}=S_{\mu\nu}^{\;\;\;\;\alpha}=0 \Rightarrow \Gamma^{\alpha}_{\;\;\;\mu\nu}=\Gamma^{\alpha}_{\;\;\;\nu\mu}
	\end{equation}
	The above two conditions completely fix the connection to be the $Levi-Civita$ connection
	\begin{equation}
	\Gamma^{\alpha}_{\;\;\;\mu\nu}=\frac{1}{2}g^{\alpha\beta}(\partial_{\mu}g_{\beta\nu}+\partial_{\nu}g_{\beta\mu}-\partial_{\beta}g_{\mu\nu})
	\end{equation}
	Then, the only independent quantity on the manifold is the metric tensor. This defines a Riemannian geometry (pseudo-Riemannian in our case) and the space is fully described by the metric tensor. We should point out that this need not be the case in general. Indeed, the metric tensor and the connection define, in general, different notions on the manifold. For the former defines distances and angles between vectors, while the latter defines parallel transfer of vector and tensor fields on the manifold. Thus, in the case of $Metric$ Gravity there is not much of a choice, if one were to write a Gravity action this can only depend on the metric tensor. The mathematical expression of such an action is
	\begin{equation}
	S=S_{G}[g_{\mu\nu}]+S_{M}[g_{\mu\nu},\Psi]
	\end{equation}
	with both gravity and matter actions metric-dependent only. Here $S_{G}$ and $S_{M}$ stand for the Gravity and matter parts of the action respectively. Relaxing the assumptions of vanishing non-metricity and torsion we have the Palatini and Metric-Affine approaches which we give below.

	\subsubsection{$Palatini$ Gravity}
	In Palatini Gravity no a-priori assumptions about the compatibility of the metric or the torsionlessness of the connection are made. Thus, metric and connection are independent fields, both fundamental each with their own geometrical significance. However, one assumption is made; that the matter part of the action does not depend on the connection\footnote{The covariant conservation of the energy-momentum tensor in Palatini Gravity was shown in \cite{koivisto2006note}.}. A general action in the Palatini formulation reads
	\begin{equation}
	S=S_{G}[g_{\mu\nu},\Gamma^{\kappa}_{\;\;\;\alpha\beta}]+S_{M}[g_{\mu\nu},\Psi]
	\end{equation} 
	Then, in order to derive the field equations one has to independently vary with respect to the metric tensor-$\delta_{g}$ as well as with respect to the connection-$\delta_{\Gamma}$. Note that  the connection here is not symmetric in general and also the metric compatibility condition does not hold true. As a result the connection-$\Gamma^{\kappa}_{\;\;\;\alpha\beta}$ is not (in general) the Levi-Civita connection. It is said in the literature that if one chooses the Einstein-Hilbert Lagrangian density, that is $\mathcal{L}_{G}=R$ , then the Palatini procedure coincides with Einstein's theory formulated in $Metric$-approach. However, this is not true. To achieve so, an additional assumption for the vanishing of either the torsion vector or the Weyl vector must also be made. As we will see in what follows, for non vanishing torsion and Weyl vectors, the theory obtained is Einstein's Gravity with an additional vector degree of freedom which is left unspecified. We will also show how this vectorial degree of freedom can be canceled through a projective transformation of the connection.

	\subsubsection{$Metric-Affine$ Gravity}
	A generalization of the $Palatini$-Gravity is the so-called $Metric-Affine$ Gravity in which the matter action does depend on the independent connection as well. The general action is then written as
	\begin{equation}
	S=S_{G}[g_{\mu\nu},\Gamma^{\kappa}_{\;\;\;\alpha\beta}]+S_{M}[g_{\mu\nu},\Gamma^{\kappa}_{\;\;\;\alpha\beta},\Psi]
	\end{equation}
	Exactly this dependence of the matter action on the connection, defines a new tensor
	\begin{equation}
	\Delta_{\alpha}^{\;\; \mu\nu}\equiv -\frac{2}{\sqrt{-g}} \frac{\delta S_{M}[g_{\mu\nu}, \Gamma^{\alpha}_{\;\;\; \mu\nu}]}{\delta \Gamma^{\alpha}_{\;\; \mu\nu}}
	\end{equation}
	which is called the $Hypermomentum-tensor$ \cite{hehl1976hypermomentum}. Note that the above quantity is indeed a tensor. This is so because even though $\Gamma^{\mu}_{\alpha\beta}$ is not a tensor, the variation $\delta \Gamma^{\mu}_{\alpha\beta}$ being a difference of connections, is a tensor. It is worth noting that there seems to exist a relation between the spin of a particle (intrinsic property) and the non-vanishing of the Hypermomentum-tensor which results in different gravitational effects. In particular it can be shown that the antisymmetric part of the Hypermomentum tensor identically vanishes for spinless particles but has a non-zero value for particles that do have spin\cite{hehl1977hadron}. Under certain assumptions, Metric-Affine Gravity can reproduce Einstein's theory and therefore be compatible with observational analysis. We should also point out that among these three approaches, $Metric-Affine$ Gravity is the least studied, mostly because of its complexity. The dynamics of the latter has  been studied to some extend in works like \cite{hehl1995metric,1981GReGr..13.1037H,hehl1989progress,vitagliano2011dynamics}, and some solutions have found in \cite{hehl1999metric,obukhov1996exact,tresguerres1995exact,hehl1998gauge,puetzfeld2001cosmological}  however its effects are not completely understood\footnote{For some recent inflationary scenarios in Metric-Affine Gravity see \cite{Shimada:2018lnm}. }. It is our purpose to analyze it here as thoroughly as possible and study its consequences both as a pure Gravitational theory as well as with regard to its cosmological consequences.

	\chapter{Introduction to Non-Riemannian Geometry}
	In this chapter we review, define and develop the necessary notions that constitute the generalized geometry. We pay special attention to the geometrical meaning of torsion and non-metricity by giving many illustrative examples. We also deal with subtle points that arise in computations and develop the necessary machinery needed in order to study MAG theories.

	\section{Introduction to Non-Riemannian geometry}
	Let us introduce here the basic mathematical quantities that constitute a generalized non-Riemannian geometry. The most general Gravity Theory that is based on a non-Riemannian geometry is the so called Metric-Affine Gravity\cite{hehl1995metric}. First of all note that the term non-Riemannian refers to a generalized geometry where apart from the curvature the space is also endowed with torsion (i.e. vectors rotate upon parallel transport and as a result infinitesimal parallelograms do not exist) and non-metricity (dot products and lengths of vectors are not preserved while moving on the manifold). It is important to stress out that curvature, torsion and non-metricity are different geometrical entities and one can have the one without necessarily the others. For example, we may have a space that is metric and flat but has a non-vanishing torsion. This is the case in what is known as the teleparallel formulation of Gravity. In this formulation curvature and non-metricity are zero and gravity is due to torsion (see \cite{aldrovandi2010introduction,cai2016f} for instance). There also exists the symmetric teleparallel formulation \cite{nester1999symmetric,jimenez2018teleparallel}  where one has zero curvature and torsion but a non-vanishing non-metricity. A space with  zero torsion and non-metricity but non-vanishing  curvature is our familiar Riemannian space of General Relativity. A space that has all three vanishing will be a Euclidean  (or Minkowski) space. \\
	Note that the three aforementioned geometrical quantities can all be calculated when the two fundamental objects of a manifold are given, a metric $g_{\mu\nu}$ and a connection $\Gamma^{\lambda}_{\;\;\;\mu\nu}$. The former defines distances and angles between vectors and the latter defines parallel transfer of vectors (or tensor fields in general) on the manifold. In a general non-Riemannian space, these two quantities (metric and connection) are independent\footnote{To quote Albert Einstein himself: ''The essential achievement of GR,
		namely to overcome rigid space, is
		only indirectly connected with the
		introduction of a Riemannian metric.
		The directly relevant conceptual
		element is the displacement field $\Gamma^{\lambda}_{\;\;\;\mu\nu}$
		which expresses the infinitesimal
		displacement of vectors''.} and only become  interrelated when further assumptions are made. For instance, when one assumes  a torsion-free and  metric-compatible connection, the resulting connection is uniquely defined in terms of the metric tensor and its derivatives and is the familiar Levi-Civita connection (see subsequent discussion). We now proceed by giving the basic definitions that built a non-Riemannian geometry.

	\subsection{Connection and Riemann tensor}
	We will give now the general definitions of the connection and the Riemann tensor. We should point out that these definitions do not need the existence of a metric. Let us firstly introduce a general connection $\Gamma^{\alpha}_{\;\;\;\;\mu\nu}$ which is used in order to define parallel transport (through covariant differentiation) of tensorial fields. For a general tensorial field of rank (n,m) one has
	\begin{gather}
	\nabla_{\mu}T^{\alpha_{1}\alpha_{2}...\alpha_{n}}_{\;\;\;\;\beta_{1}\beta_{2}...\beta_{m}}=\partial_{\mu}T^{\alpha_{1}\alpha_{2}...\alpha_{n}}_{\;\;\;\;\beta_{1}\beta_{2}...\beta_{m}}+\Gamma^{\alpha_{1}}_{\;\;\;\rho\mu}T^{\rho\alpha_{2}...\alpha_{n}}_{\;\;\;\;\beta_{1}\beta_{2}...\beta_{m}}+...+\Gamma^{\alpha_{2}}_{\;\;\;\rho\mu}T^{\alpha_{a}\alpha_{2}...\alpha_{n-1}\rho}_{\;\;\;\;\beta_{1}\beta_{2}...\beta_{m}}  \\ 
	-\Gamma^{\rho}_{\;\;\;\beta_{1}\mu}T^{\alpha_{1}\alpha_{2}...\alpha_{n}}_{\;\;\;\;\rho\beta_{2}...\beta_{m}}-...-\Gamma^{\rho}_{\;\;\;\beta_{m}\mu}T^{\alpha_{1}\alpha_{2}...\alpha_{n}}_{\;\;\;\;\rho\beta_{2}...\beta_{m-1}\rho} \nonumber
	\end{gather}
	Notice that according to our definition the index $\mu$ that appears in the covariant derivative is placed at the very right of the connection.\footnote{Some authors define it the other way around. It is important to strictly stick to whichever definition one adopts, since this will have an impact on the definition of the Riemann tensor.} 
	In particular, for a rank-$2$ tensorial field,\footnote{That is, all possible combinations of rank-$2$, namely $(2,0)$,$(1,1)$ and $(0,2)$.}
	the following hold true
	\begin{equation}
	\nabla_{\mu}T^{\alpha\beta}=\partial_{\mu}T^{\alpha\beta}+\Gamma^{\alpha}_{\;\;\;\rho\mu}T^{\rho\beta}+\Gamma^{\beta}_{\;\;\;\rho\mu}T^{\alpha\rho}
	\end{equation}
	\begin{equation}
	\nabla_{\mu}T^{\alpha}_{\;\;\beta}=\partial_{\mu}T^{\alpha}_{\;\;\beta}+\Gamma^{\alpha}_{\;\;\;\rho\mu}T^{\rho}_{\;\;\beta}-\Gamma^{\rho}_{\;\;\;\beta\mu}T^{\alpha}_{\;\;\rho}
	\end{equation}
	\begin{equation}
	\nabla_{\mu}T_{\alpha\beta}=\partial_{\mu}T_{\alpha\beta}-\Gamma^{\rho}_{\;\;\;\alpha\mu}T_{\rho\beta}-\Gamma^{\rho}_{\;\;\;\beta\mu}T_{\alpha\rho}
	\end{equation}
	Contracting in $\alpha$, $\beta$ the second equation above ($i.e.$ forming the scalar quantity $T\equiv T^{\alpha}_{\;\;\alpha}$) we immediately conclude that
	\begin{gather}
	\nabla_{\mu}T=\partial_{\mu}T+\Gamma^{\alpha}_{\;\;\;\rho\mu}T^{\rho}_{\;\;\alpha}-\Gamma^{\rho}_{\;\;\;\alpha\mu}T^{\alpha}_{\;\;\rho}= \\ \nonumber
	=\partial_{\mu}T+\Gamma^{\alpha}_{\;\;\;\rho\mu}T^{\rho}_{\;\;\alpha}-\Gamma^{\alpha}_{\;\;\;\rho\mu}T^{\rho}_{\;\;\alpha}=\partial_{\mu}T \Rightarrow  \nonumber
	\end{gather}
	\begin{equation}
	\nabla_{\mu}T=\partial_{\mu}T
	\end{equation}
	confirming that on scalars covariant differentiation reduces to partial one. Now regarding scalar densities\footnote{Recall that a scalar density-$\mathcal{P}$ of weight $w$ transforms as $\mathcal{P} \rightarrow \mathcal{P}^{'}=J^{w}\mathcal{P}$ under a general coordinate transformation $x \rightarrow x^{'}=f(x)$. Notice that  the Jacobian of the transformation reads $J\equiv \Big|\frac{\partial x}{\partial x'}\Big|$ according to our definition. As a result the determinant of the metric tensor and the square root of it, are scalar densities of weights $+2$ and $+1$ respectively! If one defines the Jacobian $J\equiv \Big|\frac{\partial x'}{\partial x} \Big|$ then the above weights are $-2$ and $-1$ respectively.}, it holds that
	\beq
	\nabla_{\mu}\mathcal{P}=\partial_{\mu}\mathcal{P}-w\Gamma^{\lambda}_{\;\;\;\lambda\mu}\mathcal{P}
	\eeq
	for a scalar density $\mathcal{P}$ of weight $w$. Also, for a  tensor density $\mathcal{T}^{\alpha_{1}...\alpha_{n}}_{\;\;\;\;\beta_{1}...\beta_{m}}$ of weight $w$ one has
	\begin{gather}
	\nabla_{\mu}\mathcal{T}^{\alpha_{1}...\alpha_{n}}_{\;\;\;\;\beta_{1}...\beta_{m}}=\partial_{\mu}\mathcal{T}^{\alpha_{1}...\alpha_{n}}_{\;\;\;\;\beta_{1}...\beta_{m}}+\Gamma^{\alpha_{1}}_{\;\;\;\lambda\mu}\mathcal{T}^{\lambda...\alpha_{n}}_{\;\;\;\;\beta_{1}...\beta_{m}}+...+\Gamma^{\alpha_{n}}_{\;\;\;\lambda\mu}\mathcal{T}^{\alpha_{1}...\lambda}_{\;\;\;\;\beta_{1}...\beta_{m}} \nonumber \\
	-\Gamma^{\lambda}_{\;\;\;\beta_{1}\mu}\mathcal{T}^{\alpha_{1}...\alpha_{n}}_{\;\;\;\;\lambda...\beta_{m}}-...-\Gamma^{\lambda}_{\;\;\;\beta_{m}\mu}\mathcal{T}^{\alpha_{1}...\alpha_{n}}_{\;\;\;\;\beta_{1}...\lambda}-w\Gamma^{\lambda}_{\;\;\;\lambda\mu}\mathcal{T}^{\alpha_{1}...\alpha_{n}}_{\;\;\;\;\beta_{1}...\beta_{m}}
	\end{gather}
	Notice the appearance of the term $-w\Gamma^{\lambda}_{\;\;\;\lambda\mu}\mathcal{T}^{\alpha_{1}...\alpha_{n}}_{\;\;\;\;\beta_{1}...\beta_{m}}$ with regards to the definition of the covariant derivative of a tensor field $(n,m)$.

	Let us proceed now by giving the Riemann tensor. Forming the commutator of two covariant derivatives and acting it on a vector $u^{\mu}$ we arrive at
	\begin{equation}
	[\nabla_{\alpha} ,\nabla_{\beta}]u^{\mu}=2\nabla_{[\alpha} \nabla_{\beta]}u^{\mu}=R^{\mu}_{\;\;\;\nu\alpha\beta} u^{\nu}+2 S_{\alpha\beta}^{\;\;\;\;\;\nu}\nabla_{\nu}u^{\mu}
	\end{equation}
	where
	\begin{equation}
	R^{\mu}_{\;\;\;\nu\alpha\beta}:=2\partial_{[\alpha}\Gamma^{\mu}_{\;\;\;|\nu|\beta]}+2\Gamma^{\mu}_{\;\;\;\rho[\alpha}\Gamma^{\rho}_{\;\;\;|\nu|\beta]}
	\end{equation}
	is the so-called Riemann tensor and the horizontal bars around an index denote that this index is left out of the (anti)-symmetrization. In addition, it appears the torsion tensor $ S_{\alpha\beta}^{\;\;\;\;\;\nu}$ which is given by the antisymmetric part of the connection\footnote{Note that even though the connection is not a tensor the difference between two connections does behave as a tensor.}
	\begin{equation}
	S_{\alpha\beta}^{\;\;\;\;\;\nu}:=\Gamma^{\nu}_{\;\;[\alpha\beta]}=\frac{1}{2}(\Gamma^{\nu}_{\;\;\alpha\beta}-\Gamma^{\nu}_{\;\;\beta\alpha})
	\end{equation}
	Alternatively, one may also define the torsion tensor by acting the anti-symmetrized double covariant derivative to  a scalar, namely
	\beq
	\nabla_{[\mu} \nabla_{\nu]}\phi = S_{\mu\nu}^{\;\;\;\;\lambda}\nabla_{\lambda}\phi
	\eeq
	for any scalar $\phi$. We should point out that by the above definition of the Riemann tensor alone, the only symmetry that the latter possesses is antisymmetry in its last two indices. Further symmetries appear only after imposing a torsionless connection ($ S_{\alpha\beta}^{\;\;\;\;\;\nu}=0$) and a metric compatible metric ($\nabla_{\alpha}g_{\mu\nu}=0$). This allows one to form the following contractions
	\begin{equation}
	R^{\mu}_{\;\;\;\mu\alpha\beta} \nonumber
	\end{equation}
	\begin{equation}
	R^{\mu}_{\;\;\;\nu\mu\beta} \nonumber
	\end{equation}
	\begin{equation}
	R^{\mu}_{\;\;\;\nu\alpha\mu} \nonumber
	\end{equation}
	Note that the last contraction above is up to a minus sign equal to the second one and need not be considered separately. This defines the Ricci tensor 
	\begin{equation}
	R_{\nu\beta}:=  R^{\mu}_{\;\;\;\nu\mu\beta} = 2\partial_{[\mu}\Gamma^{\mu}_{\;\;\;|\nu|\beta]}+2\Gamma^{\mu}_{\;\;\;\rho[\mu}\Gamma^{\rho}_{\;\;\;|\nu|\beta]}
	\end{equation}
	which, is not symmetric in $\nu,\beta$ in general. In addition, the very first contraction above defines a new tensor which is non-vanishing only when non-metricity is present ($\nabla_{\mu}g_{\alpha\beta}\neq 0$), and goes by the name homothetic curvature
	\begin{equation}
	\hat{R}_{\alpha\beta}:=R^{\mu}_{\;\;\;\mu\alpha\beta}=2\partial_{[\alpha}\Gamma^{\mu}_{\;\;\;|\mu|\beta]}=\partial_{\alpha}\Gamma^{\mu}_{\;\;\;\mu\beta}-\partial_{\beta}\Gamma^{\mu}_{\;\;\;\mu\alpha}
	\end{equation}
	Note now that for the above considerations no metric is required. When the space is also endowed with a metric tensor there is a third independent contraction that can be formed
	\begin{equation}
	\check{R}^{\mu}_{\;\;\beta} =g^{\nu\alpha}R^{\mu}_{\;\;\;\nu\alpha\beta}:=2 g^{\nu\alpha}\partial_{[\alpha}\Gamma^{\mu}_{\;\;\;|\nu|\beta]}+2 g^{\nu\alpha} \Gamma^{\mu}_{\;\;\;\rho[\alpha}\Gamma^{\rho}_{\;\;\;|\nu|\beta]}
	\end{equation}
	However, the Ricci scalar is still uniquely defined since\footnote{Of course the other scalar that we can form by contracting the homothetic curvature with the metric is automatically zero since the former is antisymmetric and the latter symmetric in their indices.}
	\begin{equation}
	\check{R}=\check{R}^{\alpha}_{\;\;\alpha}=R^{\alpha}_{\;\;\;\beta\mu\alpha}g^{\beta\mu}=-R^{\alpha}_{\;\;\;\beta\alpha\mu}g^{\beta\mu}=-R_{\beta\mu}g^{\beta\nu}=-R
	\end{equation}
	
	\subsection{Torsion tensor and related vectors}
	As we have already seen, the torsion tensor is defined as
	\begin{equation}
	S_{\mu\nu}^{\;\;\;\;\;\lambda}=\Gamma^{\lambda}_{\;\;[\mu\nu]}
	\end{equation}
	with this at hand we can define two new quantities. The first one is obtained by contracting in $(\mu=\lambda)$,
	\begin{equation}
	S_{\mu} \equiv S_{\mu\lambda}^{\;\;\;\;\;\lambda}
	\end{equation}
	which we shall call the torsion vector. The second is a pseudo-vector that comes about when contracting with the Levi-Civita tensor, and in $4$-dim reads
	\begin{equation}
	\tilde{S}^{\mu} \equiv \epsilon^{\mu\nu\rho\sigma}S_{\nu\rho\sigma}
	\end{equation}

	\subsubsection{Torsion decomposition}
	Having defined the two torsion vectors $S_{\mu}$ and $\tilde{S}_{\mu}$ and recalling that torsion is antisymmetric in its first two indices, we may write
	\begin{equation}
	S_{\mu\nu\lambda}=a g_{\lambda[\mu}S_{\nu]}+b \epsilon_{\mu\nu\lambda\rho}\tilde{S}^{\rho}+Z_{\mu\nu\lambda}
	\end{equation}
	where $a$,$b$ constants to be determined and $Z_{\mu\nu\lambda}$ is the remaining piece of torsion when we subtract $S_{\mu}$ and $\tilde{S}_{\mu}$ out. Contracting the above with $\epsilon^{\mu\nu\lambda\alpha}$ we get
	\begin{equation}
	-\tilde{S}^{\alpha}=0-3!\;b\tilde{S}^{\alpha}+\epsilon^{\mu\nu\lambda\alpha}Z_{\mu\nu\lambda}
	\end{equation}
	from which we conclude that
	\begin{equation}
	b=\frac{1}{6}\;,\;\; and \;\; \tilde{Z}^{\alpha}\equiv-\epsilon^{\mu\nu\lambda\alpha}Z_{\mu\nu\lambda}=0
	\end{equation}
	Now, contracting with $g^{\nu\lambda}$ it follows that
	\begin{equation}
	S_{\mu}=\frac{a}{2}(1-n)S_{\mu}+0+Z_{\mu\nu\lambda}g^{\nu\lambda}
	\end{equation}
	that is
	\begin{equation}
	a=\frac{2}{1-n}\;,\;\; and \;\; Z_{\mu}\equiv Z_{\mu\nu\lambda}g^{\nu\lambda}=0
	\end{equation}
	We have therefore fully decomposed the torsion tensor
	\begin{equation}
	S_{\mu\nu\lambda}=\frac{2}{1-n} g_{\lambda[\mu}S_{\nu]}+\frac{1}{6} \epsilon_{\mu\nu\lambda\rho}\tilde{S}^{\rho}+Z_{\mu\nu\lambda}
	\end{equation}
	where the $Z$-tensor satisfies
	\begin{equation}
	\tilde{Z}^{\alpha}\equiv-\epsilon^{\mu\nu\lambda\alpha}Z_{\mu\nu\lambda}=0 \;,\; and \;\; Z_{\mu}\equiv Z_{\mu\nu\lambda}g^{\nu\lambda}=0
	\end{equation}
	For the torsion decomposition in the language of differential forms, the reader is refereed to \cite{mccrea1992irreducible,obukhov1997irreducible,hehl1995metric}.

	\subsubsection{Geometrical Meaning of torsion}
	
	The effect of torsion on geometrical grounds is that infinitesimal parallelograms do not exist due to it. In others words we cannot form small parallelograms by parallel transportation of one vector to the direction of the other and vice versa. The end result is a pentagon.  To see this consider two curves $\mathcal{C}:x^{\mu}=x^{\mu}(\lambda)$ and $\mathcal{\tilde{C}}:\tilde{x}^{\mu}=\tilde{x}^{\mu}(\lambda)$ with tangent vectors
	\begin{equation}
	u^{\mu}=\frac{dx^{\mu}}{d\lambda}
	\end{equation}
	and
	\begin{equation}
	\tilde{u}^{\mu}=\frac{d\tilde{x}^{\mu}}{d\lambda}
	\end{equation}
	respectively. Now, let us $d\tilde{x}^{\mu}$-displace $u^{\alpha}$ along $\mathcal{\tilde{C}}$ to obtain $u^{'\alpha}$ which in first order is given by
	\begin{equation}
	u^{'\alpha}=u^{\alpha}+(\partial_{\mu}u^{\alpha})d\tilde{x}^{\mu} \label{toru}
	\end{equation}
	but since $u^{\alpha}$ is parallely transported along $\mathcal{\tilde{C}}$, it holds that
	\begin{equation}
	\frac{d\tilde{x}^{\mu}}{d\lambda}\nabla_{\mu}u^{\alpha}=0= \frac{d\tilde{x}^{\mu}}{d\lambda}\partial_{\mu}u^{\alpha}+\Gamma^{\alpha}_{\;\;\;\nu\mu}\frac{d\tilde{x}^{\mu}}{d\lambda}u^{\nu} \Rightarrow  \nonumber
	\end{equation}
	\begin{equation}
	(\partial_{\mu}u^{\alpha})d\tilde{x}^{\mu}=-\Gamma^{\alpha}_{\;\;\;\nu\mu}u^{\nu}\tilde{u}^{\mu}d\lambda
	\end{equation}
	which when substituted back in $(\ref{toru})$ results in
	\beq
	u^{'\alpha}=u^{\alpha}-\Gamma^{\alpha}_{\;\;\;\nu\mu}u^{\nu}\tilde{u}^{\mu}d\lambda
	\eeq
	Doing the same job but now for a $dx^{\mu}$-displacement of $\tilde{u}^{a}$ along $\mathcal{C}$, we get
	\beq
	\tilde{u}^{'\alpha}=\tilde{u}^{\alpha}-\Gamma^{\alpha}_{\;\;\;\nu\mu}\tilde{u}^{\nu}u^{\mu}d\lambda= \tilde{u}^{\alpha}-\Gamma^{\alpha}_{\;\;\;\mu\nu}\tilde{u}^{\mu}u^{\nu}d\lambda
	\eeq
	Subtracting the latter two, it follows that
	\beq
	(\tilde{u}^{\alpha}+u^{'\alpha})-(u^{\alpha}+\tilde{u}^{'\alpha})=2 S_{\mu\nu}^{\;\;\;\;\;\alpha}\tilde{u}^{\mu}u^{\nu}d\lambda
	\eeq
	Notice now that for the infinitesimal parallelogram to exist, the vectors $(\tilde{u}^{\alpha}+u^{'\alpha})$ and $(u^{\alpha}+\tilde{u}^{'\alpha})$ should be equal and as it is clear from the above, this is not true in the presence of torsion. Defining the vector that shows this deviation as $V^{\alpha}d\lambda=(\tilde{u}^{\alpha}+u^{'\alpha})-(u^{\alpha}+\tilde{u}^{'\alpha})$ the latter can also be written as\footnote{This only holds true for small displacements in the directions of $\tilde{u}^{\mu}$ and $u^{\nu}$ which themselves are computed at the starting point of the path.}
	\beq
	V^{\alpha}=2 S_{\mu\nu}^{\;\;\;\;\;\alpha}\tilde{u}^{\mu}u^{\nu}
	\eeq
	which is the vector that shows how much the parallelogram has been deformed.

	\subsubsection{Illustrative Example}
	
	Let us examine now the role of torsion, with a simple two dimensional example. Consider a $2-dim$ Euclidean (i.e flat) space with vanishing non-metricity but with a non-vanishing torsion. Take the  familiar orthonormal vector  basis $\{\bold{e}_{i}\}$ , $\;i=1,2$ on the $xy$-plane. Next, consider the lines $C$ :$y=0$ and $\tilde{C}$ :$x=0$ with tangent vectors $\bold{u}=\bold{e}_{1}$ and $\bold{\tilde{u}}=\bold{e}_{2}$ respectively. Now, take the vector $\bold{\tilde{u}}=\bold{e}_{2}$ and parallel transport it along the line $C$  a parameter distance $\lambda_{1}=1$ to obtain $\bold{\tilde{u}'}$. Also,  parallel transport  $\bold{u}=\bold{e}_{1}$ along $\tilde{C}$ a parameter distance $\lambda_{2}=1$ to obtain $\bold{u'}$. The connecting vector between the two is
	\begin{equation}
	V^{\alpha}=2 S_{\mu\nu}^{\;\;\;\;\;\alpha}\tilde{u}^{\mu}u^{\nu} \label{vv}
	\end{equation}
	as we have already seen,and depends solely on torsion. To see now how is torsion related to rotations, denote as $\theta$ the angle between $\bold{\tilde{u}'}$ and the $x$-axis and as $\phi$ the angle between the vector $\bold{u'}$ and the $y$-axis\footnote{Bear in mind that the resulting vectors $\bold{\tilde{u}'}$,$\bold{u'}$ retain the length of the initial vectors $\bold{u}$,$\bold{\tilde{u}}$  which lengths in our case are both equal to one. If non-metricity was present their lengths would also change under parallel transport. In this example, however, we consider only torsion in order explore its geometrical meaning. }. Then, by means of elementary vector analysis we find
	\begin{equation}
	\bold{\tilde{u}'}=\cos{\theta}\bold{e}_{1}+\sin{\theta}\bold{e}_{2}
	\end{equation}
	and
	\begin{equation}
	\bold{u'}=\sin{\phi}\bold{e}_{1}+\cos{\phi}\bold{e}_{2}
	\end{equation}
	Also, it holds that
	\begin{equation}
	\bold{\tilde{u}}+\bold{u'}+\bold{V}=\bold{u}+\bold{\tilde{u}'}
	\end{equation}
	so that
	\begin{equation}
	\bold{V}= (1+\cos{\theta}-\sin{\phi})  \bold{e}_{1}   +(\sin{\theta}-1-\cos{\phi})\bold{e}_{2} \label{tt}
	\end{equation}

	\begin{tikzpicture}
	\draw[thin,gray!40] (0,0) grid (5,5);
	\draw[->] (0,0)--(5,0) node[right]{$x$};
	\draw[->] (0,0)--(0,5) node[above]{$y$};
	\draw[line width=1pt,blue,-stealth](0,0)--(0,1) node[anchor=north east]{$\tilde{u}$};
	\draw[line width=1pt,red,-stealth](0,0)--(1,0) node[anchor=north east]{$u$};
	\draw[line width=1pt,blue,-stealth](1,0)--(1.4,0.9) node[anchor=north east]{$\bold{\tilde{u}'}$};
	\draw[line width=1pt,red,-stealth](0,1)--(0.9,1.4) node[anchor=north east]{$u'$};
	\draw[line width=1pt,orange,-stealth](0.9,1.4)--(1.4,0.9) node[anchor=south west]{V};
	
	\end{tikzpicture}

	Furthermore, using the fact that $u^{\mu}=\delta^{\mu}_{1}$ and $\tilde{u}^{\mu}=\delta^{\mu}_{2}$ equation ($\ref{vv}$) becomes
	\begin{equation}
	V^{\alpha}=2 S_{21}^{\;\;\;\;\alpha}
	\end{equation}
	or in components
	\begin{equation}
	V^{1}=2 S_{21}^{\;\;\;\;1}\;,\; \;V^{2}=2 S_{21}^{\;\;\;\;2}
	\end{equation}
	and by writing out $\bold{V}$ in the $\{\bold{e}_{i}\} $ basis
	\begin{equation}
	\bold{V}=V^{1}\bold{e}_{1} +V^{2}\bold{e}_{2}=2 S_{21}^{\;\;\;\;1}\bold{e}_{1} +2 S_{21}^{\;\;\;\;2}\bold{e}_{2}=S_{x}\bold{e}_{1}+S_{y}\bold{e}_{2}
	\end{equation}
	where we have defined $S_{x}\equiv 2 S_{21}^{\;\;\;\;1} $\;, $S_{y}\equiv 2 S_{21}^{\;\;\;\;2} $ the 2 only components of torsion in $2-dim$\footnote{Recall that in general $n-dim$ spaces the torsion tensor has $n^{2}(n-1)/2$ components.}. Comparing the above equation with ($\ref{tt}$) we find the relation between the components of torsion and the angles of rotation of the transported vectors
	\begin{equation}
	S_{x}=1+\cos{\theta}-\sin{\phi}
	\end{equation}
	\begin{equation}
	S_{y}=\sin{\theta}-1-\cos{\phi}
	\end{equation}
	From these it is now pretty apparent how is torsion related to the rotation of vectors. Let us go one step further and compute the actual area of the pentagon that is formed due to torsion. Notice that if no torsion was present we would have the formation of a square (since we have picked $\lambda_{1}=\lambda_{2}=1$) with area $\sigma_{0}=1$ but now we have a pentagon and we would like to compute its area. One way to do this is by a specific application of Green's theorem which gives the area enclosed by a closed curve in terms of a closed line integral. As it is well known, it holds that 
	\begin{equation}
	\sigma =\oint_{C_{0}} x dy
	\end{equation}
	Breaking up the integral into its five individual line integrals that constitute the pentagon we finally arrive at
	\begin{gather}
	\sigma ( \theta,\phi) =\frac{1}{2}\Big[ 2\cos{\theta} +\sin{\theta}\cos{\theta}-\sin{\phi}\cos{\phi}+\nonumber \\
	(1+\sin{\phi}-\cos{\theta})(1+\sin{\theta}+\cos{\phi})\Big]
	\end{gather}
	After some rearranging, it can also be brought to the more symmetric form
	\begin{gather}
	\sigma ( \theta,\phi) =\frac{1}{2}\Big[ 1+\cos{\theta}+\cos{\phi}+\sin{\theta}+\sin{\phi}-\cos{(\theta +\phi)} \Big]
	\end{gather}
	and this is the area of the pentagon that did not close to square due to torsion. Notice that when there is no rotation (i.e torsion is zero) $\theta=0=\phi$ and $\sigma(0,0)=1$ the area of the square. Now, in the case where the effect of torsion is small, one can approximate $\sin{x}\simeq x$ and $\cos{x}\simeq 1$ where $x\ll 1$ stands for both $\theta,phi$ such that $S_{x}\simeq\theta$\;, \; $S_{y}\simeq-\phi$ and the pentagon area is given by
	\begin{gather}
	\sigma ( \theta,\phi) \simeq 1+\frac{\theta+\phi}{2}
	\end{gather}
	or
	\begin{gather}
	\sigma ( \theta,\phi) \simeq 1+\frac{S_{x}-S_{y}}{2}=1+S_{21}^{\;\;\;\;1}+S_{12}^{\;\;\;\;2}
	\end{gather}
	in terms of the torsion components. Again, the unity on the right hand side is the area of the square that is formed when there is no torsion, and the rest is the modification of the original area due to torsion effects.

	\subsection{The non-metricity Tensor}
	In a general metric affine space, as we have already pointed out, the connection is not metric compatible. This failure of the connection to covariantly conserve the metric is called the non-metricity tensor and is defined as
	\begin{equation}
	Q_{\alpha\mu\nu} :=-\nabla_{\alpha}g_{\mu\nu} \label{p}
	\end{equation}
	We should also mention that the non-metricity is a quantity that depends both on the metric tensor and the connection. Indeed, expanding (\ref{p}) we obtain
	\begin{equation}
	Q_{\alpha\mu\nu} :=-\nabla_{\alpha}g_{\mu\nu} =-\partial_{\alpha}g_{\mu\nu}+\Gamma^{\rho}_{\;\;\;\mu\alpha}g_{\rho\nu}+\Gamma^{\rho}_{\;\;\;\nu\alpha}g_{\mu\rho}
	\end{equation}
	from which, the dependence on $\Gamma^{\lambda}_{\;\;\;\mu\nu}$ and $g_{\mu\nu}$ is apparent. Notice that the corresponding expression of the non-metricity with upper indices is given by
	\begin{gather}
	Q_{\rho}^{\;\;\;\alpha\beta}:= g^{\mu\alpha}g^{\nu\beta}Q_{\rho\mu\nu}=-g^{\mu\alpha}g^{\nu\beta}\nabla_{\rho}g_{\mu\nu}= \nonumber \\
	=-\nabla_{\rho}(g^{\mu\alpha}g^{\nu\beta}g_{\mu\nu})+g_{\mu\nu}g^{\mu\alpha}\nabla_{\rho}g^{\nu\beta}+g_{\mu\nu}g^{\nu\beta}\nabla_{\rho}g^{\mu\alpha}= \nonumber \\
	=-\nabla_{\rho}g^{\alpha\beta}+\delta_{\nu}^{\alpha}\nabla_{\rho}g^{\nu\beta}+\delta_{\mu}^{\beta}\nabla_{\rho}g^{\mu\alpha}= \nonumber \\
	=-\nabla_{\rho}g^{\alpha\beta}+\nabla_{\rho}g^{\alpha\beta}+\nabla_{\rho}g^{\alpha\beta}=+\nabla_{\rho}g^{\alpha\beta} \Rightarrow  \nonumber
	\end{gather}
	\begin{equation}
	Q_{\rho}^{\;\;\;\alpha\beta}=+\nabla_{\rho}g^{\alpha\beta} \label{ol}
	\end{equation}
	where on going from the first to the second line we have employed Leibniz's rule. Notice also the sign difference compared to the expression $(\ref{p})$. Having defined the non-metricity tensor there exist two independent vectors that one can form out of it. The first one is formed by contracting the second and third indices of the latter with the metric tensor \footnote{In the literature it is common to also divide this vector by the spacetime dimensionality. That is $Q_{\mu}\rightarrow  Q_{\mu}/n$. However, our definition here does not include this factor.} and goes by the name Weyl vector
	\begin{equation}
	Q_{\alpha}:= g^{\mu\nu}Q_{\alpha\mu\nu}=Q_{\alpha\mu}^{\;\;\;\;\;\mu}=Q_{\alpha\;\;\;\;\mu}^{\;\;\;\mu}
	\end{equation}
	The second vector is formed by contracting the first and second indices with the metric\footnote{Note that the possibility to contract first and third index also exists. However, since non-metricity is symmetric in the second and third indices this vector would be the same with the one formed here.}, namely
	\begin{equation}
	\tilde{Q}_{\nu}:= g^{\mu\alpha}Q_{\alpha\mu\nu}=Q^{\mu}_{\;\;\;\mu\nu}=-g^{\mu\alpha}\nabla_{\alpha}g_{\mu\nu}
	\end{equation}
	and does not seem to go with any particular name in the literature. We shall call it $2^{nd}$ non-metricity vector and write it in the suppressed notation $2$nmv in what follows. We should point out that this is the same vector that one can form by contracting ($\ref{ol}$) in $\rho$ and $\alpha$ (or $\rho$ and $\beta$). Indeed, one has
	\begin{equation}
	\tilde{Q}^{\beta}:= Q_{\alpha}^{\;\;\;\alpha\beta}=\nabla_{\alpha}g^{\alpha\beta}=g^{\nu\beta}g^{\mu\alpha}Q_{\alpha\mu\nu}=g^{\nu\beta}\tilde{Q}_{\nu}
	\end{equation} 
	Thus, two independent vectors can be formed out of non-metricity and metric tensor alone. 
	
	\subsubsection{Non-metricity decomposition}
	As we mentioned we have two independent vectors (a priori) of non-metricity, $Q_{\mu}$ and $\tilde{Q}_{\mu}$ so we may decompose the non-metricity tensor as
	\beq
	Q_{\alpha\mu\nu}=a Q_{\alpha}g_{\mu\nu}+b Q_{\mu}g_{\nu\alpha}+c Q_{\nu}g_{\mu\alpha}+ d \tilde{Q}_{\alpha}g_{\mu\nu}+e\tilde{Q}_{\mu}g_{\nu\alpha}+f \tilde{Q}_{\nu}g_{\mu\alpha}+\Omega_{\alpha\mu\nu}
	\eeq
	where the coefficients $a,b,c,d,e,f$ are to be found and $\Omega_{\alpha\mu\nu}$ is the traceless part of non-metricity. To find the coefficients we simply impose upon the above decomposition the definitions 
	\beq
	Q_{\alpha}=Q_{\alpha\mu\nu}g^{\mu\nu}
	\eeq
	\beq
	\tilde{Q}_{\nu}=Q_{\alpha\mu\nu}g^{\alpha\mu} \;,\;\; \tilde{Q}_{\mu}=Q_{\alpha\mu\nu}g^{\alpha\nu} 
	\eeq
	to obtain the system of equation
	\begin{gather}
	a n+b+c=1 \nonumber \\
	d n+e+f=0 \nonumber \\
	a+b+ c n=0 \nonumber \\
	d+e+f n=1 \nonumber \\
	a+b n+c=0 \nonumber \\
	d+e n+f=1 \nonumber
	\end{gather}
	along with the conditions\footnote{Note that $\Omega_{\alpha\mu\nu}$ is of course symmetric in $\mu,\nu$.} $\Omega_{\alpha\mu\nu}g^{\mu\nu}=0$ and $\Omega_{\alpha\mu\nu}g^{\alpha\mu}=0$. The latter is easily solved and we obtain
	\beq
	a=\frac{n+1}{(n+2)(n-1)}
	\eeq
	\beq
	b=c=-\frac{1}{(n+2)(n-1)}
	\eeq
	\beq
	d=-\frac{2}{(n+2)(n-1)}
	\eeq
	\beq
	e=f=\frac{n}{(n+2)(n-1)}
	\eeq
	which when substituted back in the non-metricity tensor give us its decomposition
	\begin{gather}
	Q_{\alpha\mu\nu}=\frac{n+1}{(n+2)(n-1)} Q_{\alpha}g_{\mu\nu}-\frac{2}{(n+2)(n-1)} Q_{(\mu}g_{\nu )\alpha} \nonumber \\
	-\frac{2}{(n+2)(n-1)}\tilde{Q}_{\alpha}g_{\mu\nu}+ \frac{2 n}{(n+2)(n-1)} \tilde{Q}_{(\mu}g_{\nu )\alpha}+\Omega_{\alpha\mu\nu}
	\end{gather}
	or
	\beq
	Q_{\alpha\mu\nu}=\frac{\Big( (n+1)Q_{\alpha}-2 \tilde{Q}_{\alpha} \Big)}{(n+2)(n-1)}g_{\mu\nu}+\frac{2\Big(  n \tilde{Q}_{(\mu}g_{\nu )\alpha}-Q_{(\mu}g_{\nu )\alpha}   \Big) }{(n+2)(n-1)}+\Omega_{\alpha\mu\nu}
	\eeq
	where $\Omega_{\alpha\mu\nu}$ is the traceless part, satisfying $\Omega_{\alpha\mu\nu}g^{\mu\nu}=0$ and $\Omega_{\alpha\mu\nu}g^{\alpha\mu}=0$. Again, for the same decomposition but in the language of differential forms, the reader is refereed to \cite{mccrea1992irreducible,obukhov1997irreducible,hehl1995metric}.

	\subsubsection{Geometrical meaning of Non-Metricity}
	
	To see the effect on non-metricity in the space let us consider two vectors $a^{\mu}$ and $b^{\mu}$ and form their inner product $a\cdot b=a^{\mu}b^{\nu}g_{\mu\nu}$. Now, let us parallel transport both vectors along a given curve $\mathcal{C}: x^{\mu}=x^{\mu}(\lambda)$. For a Riemannian space (both torsion and non-metricity vanish) we know that upon such a transportation their inner product does not change, that is 
	\beq
	\frac{D}{d\lambda}(a\cdot b)=0
	\eeq
	When non-metricity is present a computation now reveals
	\beq
	\frac{D}{d\lambda}(a\cdot b)=\frac{dx^{\alpha}}{d\lambda}(\nabla_{\alpha} a^{\mu})b_{\mu}+\frac{dx^{\alpha}}{d\lambda}(\nabla_{\alpha} b^{\nu})a_{\nu}+\frac{dx^{\alpha}}{d\lambda}(\nabla_{\alpha}g_{\mu\nu})a^{\mu}b^{\nu}
	\eeq
	Now, since $a^{\mu}$ and $b^{\mu}$ are parallel transported along the curve, it holds that
	\beq
	\frac{dx^{\alpha}}{d\lambda}(\nabla_{\alpha} a^{\mu})=0\;,\;\;\frac{dx^{\alpha}}{d\lambda}(\nabla_{\alpha} b^{\nu})=0
	\eeq
	so we are left with
	\beq
	\frac{D}{d\lambda}(a\cdot b)=-Q_{\alpha\mu\nu}\frac{dx^{\alpha}}{d\lambda}a^{\mu}b^{\nu}
	\eeq
	from which we conclude that, when non-metricity is present, the inner product of two vectors does change when we parallel transport them along a curve. Note that for $b^{\mu}=a^{\mu}$ the above becomes
	\beq
	\frac{D}{d\lambda}(\|a \|^{2} )=-Q_{\alpha\mu\nu}\frac{dx^{\alpha}}{d\lambda}a^{\mu}a^{\nu} \label{fixedlvq}
	\eeq
	which means that the magnitude of a vector changes when we parallel transport it along a given curve! Therefore non-metricity has to do with vectors non-preserving their magnitudes and inner products.
	
	\subsubsection{An illustrative example}
	Let us find how does the length of a vector change in the case where the non-metricity is Weyl non-metricity. Recall that for Weyl geometry, we have
	\beq
	Q_{\alpha\mu\nu}=\frac{1}{n}Q_{\alpha}g_{\mu\nu}
	\eeq 
	and the length of a vector $a^{\mu}$, when transfered along a given curve $C:$ $x^{\alpha}=x^{\alpha}(\lambda)$, satisfies
	\beq
	\frac{D}{d\lambda}(\|a \|^{2} )=-\frac{1}{n}Q_{\alpha}\frac{dx^{\alpha}}{d\lambda}g_{\mu\nu}a^{\mu}a^{\nu}=-\frac{1}{n}Q_{\alpha}\frac{dx^{\alpha}}{d\lambda}\|a \|^{2}
	\eeq
	Setting $l^{2}=\|a \|^{2}$ and integrating that last one, it follows that
	\beq
	l(x)=l_{0}e^{-\frac{1}{2n}\int_{c} Q_{\alpha}dx^{\alpha}}
	\eeq
	from which we see that the change of the length is generally path dependent. In the case where the Weyl vector is exact, that is $Q_{\mu}=\partial_{\mu}\phi$ , we have what is known as a Weyl integrable geometry (WIG) for which the change on the vector's length depends only on the endpoints of the curve $C$, and for a closed loop the vector retains its initial length.

	\subsubsection{Geometric Meaning of Homothetic Curvature}
	Recall, that in a previous section we defined the homothetic curvature tensor  $\hat{R}_{\mu\nu}$ as the first contraction of the Riemann tensor $\hat{R}_{\mu\nu}:=R^{\alpha}_{\;\;\;\alpha\mu\nu}$. This tensor has  a purely non-metric nature and is in fact related to the non-metricity vector through
	\beq
	\hat{R}_{\mu\nu}=\frac{1}{2}(\partial_{\mu}Q_{\nu}-\partial_{\nu}Q_{\mu})=\partial_{[\mu}Q_{\nu]}
	\eeq 
	as we prove in later section. That is, the homothetic curvature is the curl of the non-metricity vector. To see its geometrical meaning, let us go back to the length change of a vector when transfered along a curve $C$. If  $C$ is taken to be a closed curve (loop) then the length varies as
	\beq
	l(x)=l_{0}e^{-\frac{1}{2n}\oint_{c} Q_{\alpha}dx^{\alpha}} \label{leng}
	\eeq
	but, using Stoke's theorem
	\beq
	\oint_{c} Q_{\alpha}dx^{\alpha}=\iint_{S}\partial_{[\mu}Q_{\nu]}d S^{\mu\nu}=\iint_{S}\hat{R}_{\mu\nu} d S^{\mu\nu}
	\eeq
	where $S$ is a surface that is enclosed by $C$ and $d S^{\mu\nu}$ the differential area element. Using this ($\ref{leng}$) becomes
	\beq
	l(x)=l_{0}e^{-\frac{1}{2 n}\iint_{S}\hat{R}_{\mu\nu} d S^{\mu\nu}} \label{hom}
	\eeq
	and from this we see that homothetic curvature is related with the length change that a vector experiences when transported along a closed loop. If non-metricity is weak, or the loop is small enough, by Taylor expanding we see that the total length change is given by
	\beq
	\delta l \simeq  -\frac{l_{0}}{2 n}\iint_{S}\hat{R}_{\mu\nu} d S^{\mu\nu}
	\eeq
	from which we see that the homothetic curvature serves as a generator of length changes of vector fields along closed paths.

	\subsubsection{Toy Model}
	Having established ($\ref{hom}$) let us play a little bit with the form of non-metricity to arrive at an interesting formula. To be more specific, consider a flat Euclidean $3-dim$ space that may posses non-vanishing non-metricity as well as torsion\footnote{The presence of torsion does not modify anything here, it simply rotates the vector when it is parallely transported along the curve. So, torsion rotates the vectors and non-metricity changes their lengths!}. Furthermore, assume we have a non-metric configuration with a non-metricity vector such that
	\beq
	\bold{Q}=\frac{3}{\alpha}(y \bold{e}_{1}-x \bold{e}_{2})  
	\eeq
	where $\alpha$ is a constant with area dimensions and $\bold{e}_{i}$, $\;i=1,2$ the usual orthonormal basis on the $xy$-plane. Take now the closed curve to lie  on the $xy$-plane, then
	\begin{gather}
	\oint_{c} Q_{\alpha}dx^{\alpha}=\oint_{c} \bold{Q} \cdot d\bold{r}=\iint_{S} (\bold{\nabla}\times \bold{Q})\cdot d \bold{S} = \nonumber \\
	=-\frac{6}{\alpha}\iint _{S} d\sigma =-\frac{6}{\alpha} \sigma
	\end{gather}
	where $\sigma$ is the area enclosed by $C$. Substituting this back to ($\ref{leng}$) and setting $n=3$, we get
	\beq
	l(x)=l_{0}e^{-\frac{1}{6}\oint_{c} Q_{\alpha}dx^{\alpha}}=l_{0}e^{\frac{\sigma}{\alpha}} \Rightarrow  \nonumber
	\eeq
	\beq
	l(x)=l_{0}e^{\frac{\sigma}{\alpha}} 
	\eeq
	Thus, for such an arrangement of non-metricity the change in length of a vector transported along a closed curve $C$ depends on the surface area that $C$ encloses! In addition, if the ratio $\sigma/\alpha$ is small enough, the total change in length is exactly proportional to that surface, namely
	\beq
	\delta l \simeq \frac{l_{0}}{\alpha}\sigma
	\eeq

	\subsubsection{Fixed Length Vectors}
	
	Now as we have seen, one consequence of non-metricity is that it changes the length of the vectors\footnote{The other consequence is the change of the dot product of two vectors.} when we transport them in space. So, one may ask are their any vectors, that retain their length in the presence of non-metricity? For generic non-metricity the answer is no. However, there exists a type of non-metricity for which we have vectors that remain unchanged. These are called $\bold{fixed}$ $\bold{length}$ $\bold{vectors}$. To see what kind of non-metricity allows for the existence of such vectors let us take a careful look at $(\ref{fixedlvq})$,
	\beq
	\frac{D}{d\lambda}(\|a \|^{2} )=-Q_{\alpha\mu\nu}\frac{dx^{\alpha}}{d\lambda}a^{\mu}a^{\nu}
	\eeq
	Taking $a^{\mu}$ to be proportional to $dx^{\mu}/d\lambda$ we obtain
	\beq
	\frac{D}{d\lambda}(\|a \|^{2} )\propto -Q_{\alpha\mu\nu}a^{\alpha}a^{\mu}a^{\nu}=-Q_{(\alpha\mu\nu)}a^{\alpha}a^{\mu}a^{\nu}
	\eeq
	Form the above we see that in order to have fixed lengths the right hand side must be zero, and given that $a^{\mu}$ is random we must have
	\beq
	Q_{(\alpha\mu\nu)}=0 \label{flnmt}
	\eeq
	in order for the theory to possess fixed length vectors. Any non-metricity that has vanishing totally symmetric part will admit fixed length vectors. This condition is also presented in the classic Schroendinger's $Spacetime-Structure$ \cite{schrodinger1985space}.  Let us go one step further and actually compute the simplest form of such non-metricity. The most straightforward decomposition of such a tensor would be in terms of a vector field, say $A_{\mu}$ and the metric $g_{\mu\nu}$, so that
	\beq
	Q_{\alpha\mu\nu}= a A_{\alpha}g_{\mu\nu}+b g_{\alpha(\mu}A_{\nu)}+c A_{\mu}A_{\nu}A_{\alpha}
	\eeq
	where $a,b,c$ are parameters to be computed and we demanded that the combinations are symmetric in $\mu,\nu$. Now since $Q_{\alpha\mu\nu}$ cannot have a totally symmetric part the last term on the right hand side of the above must be absent, and hence $c=0$. Now, demanding $Q_{(\alpha\mu\nu)}a^{\alpha}a^{\mu}a^{\nu}=0$ for random $a^{\mu}$ we get the relation
	\beq
	a+b=0\Rightarrow a= -b
	\eeq
	Notice also that we may set $b=1$ since this $b$ can be absorbed in a redefinition of $A_{\mu}$. Taking all the above into consideration, we finally arrive at
	\beq
	Q_{\alpha\mu\nu}=  A_{\alpha}g_{\mu\nu}-g_{\alpha(\mu}A_{\nu)}
	\eeq
	we can check that this form of non-metricity indeed satisfies $Q_{(\alpha\mu\nu)}=0$. We have
	\begin{gather}
	Q_{(\alpha\mu\nu)}=\frac{1}{3!}(Q_{\alpha\mu\nu}+Q_{\alpha\nu\mu}+Q_{\mu\nu\alpha}+Q_{\mu\alpha\nu}+Q_{\nu\alpha\mu}+Q_{\nu\mu\alpha}= \nonumber \\
	=\frac{1}{3!}(2 Q_{\alpha\mu\nu}+2 Q_{\mu\nu\alpha}+2 Q_{\nu\alpha\mu} ) =\nonumber \\
	=\frac{1}{3!}\Big( 2 A_{\alpha}g_{\mu\nu}-g_{\alpha\mu}A_{\nu}-g_{\alpha\nu}A_{\alpha} +2 A_{\mu}g_{\nu\alpha}-g_{\mu\nu}A_{\alpha}  \nonumber \\
	-g_{\mu\alpha}A_{\nu}+2 A_{\nu}g_{\alpha\mu}-g_{\nu\alpha}A_{\mu}  -g_{\nu\mu}A_{\alpha}\Big)=0
	\end{gather}
	Now, as can be easily checked by contracting with the metric tensor, the Weyl and second non-metricity vectors, are related to this $A_{\mu}$ through
	\beq
	Q_{\mu}=(n-1)A_{\mu}\;,\;\; \tilde{Q}_{\mu}=-\frac{(n-1)}{2}A_{\mu}
	\eeq
	From which we establish the relation between the two non-metricity vectors
	\beq
	Q_{\mu}=-2 \tilde{Q}_{\mu}
	\eeq
	Interestingly, this kind of non-metricity (that preserves lengths) overcomes Einstein's objection to the Weyl theory of unification\footnote{In Weyl's theory the non-metric tensor was given by $Q_{\alpha\mu\nu}=\frac{1}{n}Q_{\alpha}g_{\mu\nu}$ which definitely does not satisfy $Q_{(\alpha\mu\nu)}=0$ and therefore does not preserve the lengths of vectors.} . To recap, If the non-metricity is of of the form $(\ref{flnmt})$ the theory possesses fixed length vectors.

	\subsection{Connection decomposition}
	Having defined torsion and non-metricity we are now in a position to decompose the general connection in terms of the latter plus the Levi-Civita connection. To do so, we start by writing out the definition of the non-metricity 
	\begin{equation}
	Q_{\alpha\mu\nu}=-\nabla_{\alpha}g_{\mu\nu} =-\partial_{\alpha}g_{\mu\nu}+\Gamma^{\rho}_{\;\;\;\mu\alpha}g_{\rho\nu}+\Gamma^{\rho}_{\;\;\;\nu\alpha}g_{\mu\rho}
	\end{equation}
	and upon successive  permutations $\alpha\rightarrow \mu$, $\mu\rightarrow \nu$, $\nu\rightarrow \alpha$ on the above we may also write\footnote{We do assume that the metric tensor is symmetric since any antisymmetric part of it lacks a geometrical interpretation.}
	\begin{equation}
	Q_{\mu\nu\alpha}=-\nabla_{\mu}g_{\nu\alpha} =-\partial_{\mu}g_{\nu\alpha}+\Gamma^{\rho}_{\;\;\;\nu\mu}g_{\alpha\rho}+\Gamma^{\rho}_{\;\;\;\alpha\mu}g_{\nu\rho}
	\end{equation}
	and
	\begin{equation}
	Q_{\nu\alpha\mu}=-\nabla_{\nu}g_{\alpha\mu} =-\partial_{\nu}g_{\alpha\mu}+\Gamma^{\rho}_{\;\;\;\mu\nu}g_{\alpha\rho}+\Gamma^{\rho}_{\;\;\;\alpha\nu}g_{\mu\rho}
	\end{equation}
	Now, upon subtracting the very first equation above from the last two, we derive
	\begin{gather}
	-Q_{\alpha\mu\nu}+Q_{\mu\nu\alpha}+Q_{\nu\alpha\mu}=-(\partial_{\mu}g_{\nu\alpha}+\partial_{\nu}g_{\alpha\mu}-\partial_{\alpha}g_{\mu\nu}) \nonumber \\
	+2\Gamma^{\rho}_{\;\;\;(\mu\nu)}g_{\alpha\rho}+2\Gamma^{\rho}_{\;\;\;[\alpha\nu]}g_{\mu\rho}+2\Gamma^{\rho}_{\;\;\;[\alpha\mu]}g_{\nu\rho}
	\end{gather}
	In addition, substituting
	\begin{equation}
	\Gamma^{\alpha}_{\;\;\;[\beta\gamma]}=S_{\beta\gamma}^{\;\;\;\;\alpha}
	\end{equation}
	and using
	\begin{equation}
	\Gamma^{\rho}_{\;\;\;(\mu\nu)}=\Gamma^{\rho}_{\;\;\;\mu\nu}-S_{\mu\nu}^{\;\;\;\;\rho}
	\end{equation}
	the last one recasts to
	\begin{gather}
	-Q_{\alpha\mu\nu}+Q_{\mu\nu\alpha}+Q_{\nu\alpha\mu}=-(\partial_{\mu}g_{\nu\alpha}+\partial_{\nu}g_{\alpha\mu}-\partial_{\alpha}g_{\mu\nu}) \nonumber \\
	+2\Gamma^{\rho}_{\;\;\;\mu\nu}g_{\alpha\rho}-2 S_{\mu\nu\alpha}+2 S_{\alpha\nu\mu} +2 S_{\alpha\mu\nu}
	\end{gather}
	where $S_{\mu\nu\alpha}:=S_{\mu\nu}^{\;\;\;\;\rho}g_{\alpha\rho}$. Finally, multiplying (and contracting) through $g^{\alpha\lambda}$ we can bring the last equation to the form
	\begin{gather}
	\Gamma^{\lambda}_{\;\;\;\mu\nu}=\frac{1}{2}g^{\alpha\lambda}(\partial_{\mu}g_{\nu\alpha}+\partial_{\nu}g_{\alpha\mu}-\partial_{\alpha}g_{\mu\nu}) \nonumber \\
	+\frac{1}{2}g^{\alpha\lambda}(Q_{\mu\nu\alpha}+Q_{\nu\alpha\mu}-Q_{\alpha\mu\nu}) -g^{\alpha\lambda}(S_{\alpha\mu\nu}+S_{\alpha\nu\mu}-S_{\mu\nu\alpha}) \label{affcondec}
	\end{gather}
	We recognize the first part on the right-hand side as the Levi-Civita connection for which we use the tilde notation to distinguish it from the general connection, namely
	\begin{equation}
	\tilde{\Gamma}^{\lambda}_{\;\;\;\mu\nu}:=\frac{1}{2}g^{\alpha\lambda}(\partial_{\mu}g_{\nu\alpha}+\partial_{\nu}g_{\alpha\mu}-\partial_{\alpha}g_{\mu\nu})
	\end{equation}
	Thus,
	\begin{equation}
	\Gamma^{\lambda}_{\;\;\;\mu\nu}=\tilde{\Gamma}^{\lambda}_{\;\;\;\mu\nu}+\frac{1}{2}g^{\alpha\lambda}(Q_{\mu\nu\alpha}+Q_{\nu\alpha\mu}-Q_{\alpha\mu\nu}) -g^{\alpha\lambda}(S_{\alpha\mu\nu}+S_{\alpha\nu\mu}-S_{\mu\nu\alpha}) \label{affconection}
	\end{equation}
	we have fully decomposed the connection into a Riemannian-part (Levi-Civita connection), a contribution coming from non-metricity and another one due to torsion. It is common to introduce, at this point, a tensor which measures the deviation of the general connection with respect to the Levi-Civita one. This is the so-called $distortion$ tensor\footnote{Again, even though connections are not tensors,  the difference between connections defines $'legal'$ tensors.}
	\begin{gather}
	N^{\lambda}_{\;\;\;\;\mu\nu}:=\Gamma^{\lambda}_{\;\;\;\mu\nu}-\tilde{\Gamma}^{\lambda}_{\;\;\;\mu\nu}= \nonumber \\
	\frac{1}{2}g^{\alpha\lambda}(Q_{\mu\nu\alpha}+Q_{\nu\alpha\mu}-Q_{\alpha\mu\nu}) -g^{\alpha\lambda}(S_{\alpha\mu\nu}+S_{\alpha\nu\mu}-S_{\mu\nu\alpha})
	\end{gather} 
	or
	\beq
	N_{\alpha\mu\nu}=\frac{1}{2}(Q_{\mu\nu\alpha}+Q_{\nu\alpha\mu}-Q_{\alpha\mu\nu}) -(S_{\alpha\mu\nu}+S_{\alpha\nu\mu}-S_{\mu\nu\alpha}) \label{l}
	\eeq
	In addition, the combination 
	\begin{equation}
	K_{\mu\nu}^{\;\;\;\;\lambda} =g^{\alpha\lambda}(S_{\alpha\mu\nu}+S_{\alpha\nu\mu}-S_{\mu\nu\alpha})
	\end{equation}
	appearing above is oftentimes referred to as the $contorsion$. Note that we can split the distortion tensor into some symmetric and antisymmetric parts. Indeed, taking the symmetric part of (\ref{l}) in $\alpha, \mu$ and using the symmetries of $Q_{\alpha\mu\nu}$ and $S_{\alpha\mu\nu}$ we arrive at\footnote{Another way to derive this is by starting from the definition of non-metricity, (covariant derivative of the metric tensor )decompose the connection into the Levi-Civita and its non-Riemannian parts and use the fact that the non-metricity of the Levi-Civita connection is zero. }
	\beq
	Q_{\nu\alpha\mu}=2 N_{(\alpha\mu)\nu}
	\eeq
	While, when one takes the antisymmetric part in $\mu,\nu$ arrives at
	\beq
	S_{\mu\nu\alpha}=N_{\alpha[\mu\nu]}
	\eeq
	In addition, its totally antisymmetric part is given by
	\beq
	N_{[\alpha\mu\nu]}=S_{[\mu\nu\alpha]}=S_{[\alpha\mu\nu]}
	\eeq
	as can be easily checked. Note also that when we are looking at the autoparallels only the symmetric part   $N^{\lambda}_{\;\;\;\;(\mu\nu)}$ contributes to the equation, which is equal to
	\beq
	N^{\lambda}_{\;\;\;\;(\mu\nu)}=\frac{1}{2}g^{\alpha\lambda}(2 Q_{(\mu\nu)\alpha}-Q_{\alpha\mu\nu})-g^{\alpha\lambda}2 S_{\alpha(\mu\nu)}
	\eeq
	and from this, it is apparent that a completely antisymmetric torsion $(S_{\alpha\mu\nu}=S_{[\alpha\mu\nu]})$ has no effect on autoparallels. In addition, we see that as far as the motion of a test particle is concerned, a particle that $'feels'$ torsion but does not $'feel'$ non-metricity, will follow the same trajectory\footnote{Maybe this is too strong a statement, since whether the particle would follow an autoparallel or a geodesic is an open subject in the literature. } with a particle that $'experiences'$ only non-metricity but not torsion, when the two are related via
	\beq
	2 Q_{(\mu\nu)\alpha}-Q_{\alpha\mu\nu}=4 S_{\alpha(\mu\nu)}
	\eeq
	Thus, for such configurations torsion and non-metricity are indistinguishable as long as autoparallel motion is concerned, and there is a duality of a sub-space of torsion with a sub-space of non-metricity. We will see this duality clearly later on when we study specific theories of gravity. Now, having decomposed the affine connection (see eq. ($\ref{affcondec}$)) ), we may compute the contractions $\Gamma^{\lambda}_{\;\;\;\lambda\mu}$, $\Gamma^{\lambda}_{\;\;\;\mu\lambda}$  which will prove to be useful in later calculations. A straightforward computations yields
	\beq
	\Gamma^{\lambda}_{\;\;\;\lambda\mu}=\tilde{\Gamma}^{\lambda}_{\;\;\;\lambda\mu}+\frac{1}{2}Q_{\mu}
	\eeq
	\beq
	\Gamma^{\lambda}_{\;\;\;\mu\lambda}=\tilde{\Gamma}^{\lambda}_{\;\;\;\mu\lambda}+\frac{1}{2}Q_{\mu}+2 S_{\mu}
	\eeq
	By subtracting them and using the fact that the Levi-Civita connection $\tilde{\Gamma}^{\lambda}_{\;\;\;\mu\nu}$ is symmetric, we obtain
	\beq
	\Gamma^{\lambda}_{\;\;\;\lambda\mu}-\Gamma^{\lambda}_{\;\;\;\mu\lambda}=-2 S_{\mu}
	\eeq

	Let us now proceed with the decomposition of the Riemann tensor.
	\subsection{Riemann Tensor Decomposition}
	The general Riemann tensor (calculated with respect to the affine connection) can be fully decomposed in terms of a Riemannian part plus contributions from torsion and non-metricity. This decomposition is very helpful since it also allows one to fully decompose the Ricci tensor and scalar as well. Let us derive this decomposition here. By the definition of the Riemann tensor, one has
	\begin{equation}
	R^{\mu}_{\;\;\;\nu\alpha\beta}:=2\partial_{[\alpha}\Gamma^{\mu}_{\;\;\;|\nu|\beta]}+2\Gamma^{\mu}_{\;\;\;\rho[\alpha}\Gamma^{\rho}_{\;\;\;|\nu|\beta]}
	\end{equation}
	Then, substituting 
	\begin{equation}
	\Gamma^{\lambda}_{\;\;\;\mu\nu}=\tilde{\Gamma}^{\lambda}_{\;\;\;\mu\nu}+N^{\lambda}_{\;\;\;\;\mu\nu}
	\end{equation}
	it follows that
	\begin{equation}
	R^{\mu}_{\;\;\;\nu\alpha\beta}=\tilde{R}^{\mu}_{\;\;\;\nu\alpha\beta}+ \tilde{\nabla}_{\alpha}N^{\mu}_{\;\;\;\nu\beta}-\tilde{\nabla}_{\beta}N^{\mu}_{\;\;\;\nu\alpha}+N^{\mu}_{\;\;\;\rho\alpha}N^{\rho}_{\;\;\;\nu\beta}-N^{\mu}_{\;\;\;\rho\beta}N^{\rho}_{\;\;\;\nu\alpha} \label{rtdec}
	\end{equation}
	where we have added and subtracted the term $\tilde{\Gamma}^{\rho}_{\;\;\;\beta\alpha}N^{\mu}_{\;\;\;\nu\rho}$ in order to form the covariant derivative constructed out of the Levi-Civita connection. Note that quantities that appear with a tilde are computed with respect to the Levi-Civita connection and are therefore the Riemannian parts. Having this we  can immediately decompose the Ricci tensor, to find
	\begin{equation}
	R_{\nu\beta}=\tilde{R}_{\nu\beta}+ \tilde{\nabla}_{\mu}N^{\mu}_{\;\;\;\nu\beta}-\tilde{\nabla}_{\beta}N^{\mu}_{\;\;\;\nu\mu}+N^{\mu}_{\;\;\;\rho\mu}N^{\rho}_{\;\;\;\nu\beta}-N^{\mu}_{\;\;\;\rho\beta}N^{\rho}_{\;\;\;\nu\mu}
	\end{equation}
	As long as the Ricci tensor is concerned, a further contraction of the above with the metric tensor, reveals
	\begin{equation}
	R=\tilde{R}+ \tilde{\nabla}_{\mu}( A^{\mu}-B^{\mu})+ B_{\mu}A^{\mu}-N_{\alpha\mu\nu}N^{\mu\nu\alpha} \label{Recomp}
	\end{equation}
	where we have defined $A^{\mu}\equiv g^{\nu\beta}N^{\mu}_{\;\;\;\nu\beta}$ and $B^{\mu}\equiv N^{\alpha\mu}_{\;\;\;\;\alpha}$. Note now that when taken into an integral the second term appearing above is a surface term due to the fact that the metric tensor is compatible with the Levi-Civita-formed covariant derivative. Note also that the vectors $A_{\mu}$ and $B_{\mu}$ can be expressed in terms of the non-metricity and torsion vectors as follows
	\beq
	A_{\mu}=\tilde{Q}_{\mu}-\frac{1}{2}Q_{\mu}-2 S_{\mu}
	\eeq
	\beq
	B_{\mu}=\frac{1}{2}Q_{\mu}+2 S_{\mu}
	\eeq
	Note now that these two vectors also appear when decomposing the contracted covariant derivative of a vector field\footnote{In a flat space and in the absence of torsion and non-metricity this is of course the divergence of the vector field.} in  Riemannian and non-Riemannian parts. Indeed, for any vector $w^{\mu}$ it holds that
	\beq
	\nabla_{\mu}w^{\mu}=\tilde{\nabla}_{\mu}w^{\mu}+N^{\mu}_{\;\;\;\;\nu\mu}w^{\nu}=\tilde{\nabla}_{\mu}w^{\mu}+B_{\mu}w^{\mu}
	\eeq
	The other combination $(the\; A^{\mu})$ appears when we contract the covariant derivative of a covariant vector $w_{\nu}$ with the metric tensor. More specifically, we have
	\begin{gather}
	g^{\mu\nu}\nabla_{\mu}w_{\nu}=  \nabla_{\mu}(w_{\nu} g^{\mu\nu})-w_{\nu}\nabla_{\mu} g^{\mu\nu}    =         \nabla_{\mu}w^{\mu}-\tilde{Q}^{\mu}w_{\mu}= \nonumber \\ 
	\tilde{\nabla}_{\mu}w^{\mu}  +(B_{\mu}-\tilde{Q}_{\mu})w^{\mu}= \tilde{\nabla}_{\mu}w^{\mu}- A_{\mu}w^{\mu}
	\end{gather}
	where we have used the obvious fact  $B_{\mu}-\tilde{Q}_{\mu}=-A_{\mu}$. To conclude, we have
	\beq
	\nabla_{\mu}w^{\mu}=\tilde{\nabla}_{\mu}w^{\mu}+B_{\mu}w^{\mu}=\tilde{\nabla}_{\mu}w^{\mu}+\left( \frac{1}{2}Q_{\mu}+2 S_{\mu}  \right)w^{\mu}
	\eeq
	as well as
	\beq
	g^{\mu\nu}\nabla_{\mu}w_{\nu}=\tilde{\nabla}_{\mu}w^{\mu}-A_{\mu}w^{\mu}=\tilde{\nabla}_{\mu}w^{\mu}+\left( -\tilde{Q}_{\mu}+\frac{1}{2}Q_{\mu}+2 S_{\mu}  \right) w^{\mu}
	\eeq
	Looking  at ($\ref{Recomp}$) we see that for zero non-metricity $Q_{\alpha\mu\nu}=0$ but non-zero torsion, one finds that\footnote{We easily arrive at this result by substituting $N_{\alpha\mu\nu}$ in terms of torsion and execute the calculations.}
	\beq
	R=\tilde{R}+S_{\mu\nu\alpha}S^{\mu\nu\alpha}-2S_{\mu\nu\alpha}S^{\alpha\mu\nu}-4S_{\mu}S^{\mu}-4\tilde{\nabla}_{\mu}S^{\mu}
	\eeq
	then defining the torsion $\mathcal{T}$-scalar by
	\beq
	\mathcal{T} \equiv S_{\mu\nu\alpha}S^{\mu\nu\alpha}-2S_{\mu\nu\alpha}S^{\alpha\mu\nu}-4S_{\mu}S^{\mu}
	\eeq
	we have
	\beq
	R=\tilde{R}+\mathcal{T}-4\tilde{\nabla}_{\mu}S^{\mu}
	\eeq
	and for a flat geometry $R=0$ it follows that the Riemannian Ricci scalar differs from the $\mathcal{T}$-scalar by a total derivative. Therefore their variation is the same and considering the action built up from the $\mathcal{T}$-scalar one obtains the teleparallel equivalent of $GR$ (see \cite{aldrovandi2010introduction} for instance). Similarly, considering only non-metricity and vanishing torsion $S_{\mu\nu\lambda}=0$ we find
	\beq
	R=\tilde{R}+\frac{1}{4}Q_{\alpha\mu\nu}Q^{\alpha\mu\nu}-\frac{1}{2}Q_{\alpha\mu\nu}Q^{\mu\nu\alpha}    -\frac{1}{4}Q_{\mu}Q^{\mu}+\frac{1}{2}Q_{\mu}\tilde{Q}^{\mu}+\tilde{\nabla}_{\mu}(\tilde{Q}^{\mu}-Q^{\mu})
	\eeq
	and defining the non-metricity $\mathcal{Q}$-scalar as
	\beq
	\mathcal{Q}\equiv \frac{1}{4}Q_{\alpha\mu\nu}Q^{\alpha\mu\nu}-\frac{1}{2}Q_{\alpha\mu\nu}Q^{\mu\nu\alpha}    -\frac{1}{4}Q_{\mu}Q^{\mu}+\frac{1}{2}Q_{\mu}\tilde{Q}^{\mu}
	\eeq
	it follows that
	\beq
	R=\tilde{R}+\mathcal{Q}+\frac{1}{2}Q_{\mu}\tilde{Q}^{\mu}
	\eeq
	which again for a flat geometry $R=0$ gives
	\beq
	\tilde{R}=-\mathcal{Q}-\frac{1}{2}Q_{\mu}\tilde{Q}^{\mu}+\tilde{\nabla}_{\mu}(\tilde{Q}^{\mu}-Q^{\mu})
	\eeq
	then the variation of $\tilde{R}$ is the same with the variation of $\mathcal{Q}$ and considering the action built from the $\mathcal{Q}$-scalar one obtains the symmetric teleparallel equivalent of General Relativity (\cite{nester1999symmetric},\cite{jimenez2018coincident}). Now, if we allow for both torsion and non-metricity then it follows that
	\beq
	R=\tilde{R}+\mathcal{T}+\mathcal{Q} +\mathcal{Q\ast T}  +\tilde{\nabla}_{\mu}(\tilde{Q}^{\mu}-Q^{\mu}-4S^{\mu})
	\eeq
	where we have defined the mixed $\mathcal{Q\ast T}$-scalar 
	\beq
	\mathcal{Q\ast T} \equiv 2 Q_{\alpha\mu\nu}S^{\alpha\mu\nu}+2 S_{\mu}(\tilde{Q}^{\mu}-Q^{\mu})
	\eeq
	Further defining 
	\beq
	\mathcal{Z} \equiv \mathcal{T}+\mathcal{Q} +\mathcal{Q\ast T}
	\eeq
	we have that
	\beq
	R=\tilde{R}+\mathcal{Z}+\tilde{\nabla}_{\mu}(\tilde{Q}^{\mu}-Q^{\mu}-4S^{\mu})
	\eeq
	and we see that for a flat geometry
	\beq
	\tilde{R}=-\mathcal{Z}-\tilde{\nabla}_{\mu}(\tilde{Q}^{\mu}-Q^{\mu}-4S^{\mu})
	\eeq
	and the variations of $\tilde{R}$ and $\mathcal{Z}$ are the same since they only differ by a total derivative. We should mention that the general teleparallel equivalent with both torsion and non-metricity (but vanishing curvature), as given from the above equation,  has not been studied extensively in the literature. Therefore, it seems to be a very interesting subject worth further studying in the future.

	\section{Autoparallels and Geodesics}
	There are two distinctively different notions on a manifold.  The autoparallel and the geodesic curves. In general relativity where both torsion and non-metricity are assumed to vanish these two notions coincide. When dealing with a Metric-Affine geometry though, these need not be the same and separately define two different geometrical concepts. In what follows we give their definitions and study some cases where the two give entirely different results.

	\subsection{Autoparallels}
	Consider a smooth  manifold on an n-dim space and a curve $C$ parametrized as $x^{\mu}=x^{\mu}(\lambda)$ where $\mu=0,1,...,n-1$ and $\lambda$ is the curve parameter. The curve $C$ is said to be autoparallel if and only if the tangent vector of the curve $u^{\mu}=\frac{dx^{\mu}}{d\lambda}$ is parallel transported along $C$, namely
	\begin{equation}
	\frac{D}{d\lambda}u^{\mu}=\frac{d x^{\nu}}{d\lambda}\nabla_{\nu}u^{\mu}=u^{\nu}\nabla_{\nu}u^{\mu}=0
	\end{equation} 
	or, expanding the covariant derivative
	\begin{equation}
	\frac{d^{2} x^{\mu}}{d\lambda^{2}}+\Gamma^{\mu}_{\;\;\;\alpha\beta}\frac{dx^{\alpha}}{d\lambda}\frac{dx^{\beta}}{d\lambda}=0 \label{auto}
	\end{equation}
	where $\Gamma^{\mu}_{\;\;\;\alpha\beta}$ is the general affine connection which is, in general, different from the Levi-Civita connection as we have already seen. The solution $C:\;x^{\mu}=x^{\mu}(\lambda;x^{\alpha}_{0};\dot{x}^{\alpha}_{0})$ where $(x^{\alpha}_{0}, \dot{x}^{\alpha}_{0})$ \footnote{The dot indicates differentiation with respect to $\lambda$, that is $\dot{f}=df/d\lambda$.} are initial conditions, define the autoparallel curve. More generally, for every tensorial field $T_{\mu\nu...\kappa}$ we can define the parallel transport of the latter by
	\begin{equation}
	\frac{D}{d\lambda}T_{\mu\nu...\kappa}=\frac{d x^{\alpha}}{d\lambda}\nabla_{\alpha}T_{\mu\nu...\kappa}=0
	\end{equation}
	The above is known as the equation of parallel transport.

	\subsection{Geodesics}
	As it is well known from Euclidean geometry (flat plane geometry) the shortest distance between two given points is the straight line joining the points. Given now a space that has curvature, the natural question as to what the shortest curve joining two points would look like rises. The way to find the differential equations of the shortest curve in a curved space is to minimize the path length 
	\begin{equation}
	S=\int ds=\int \sqrt{\mid g_{\mu\nu}dx^{\mu}dx^{\nu} \mid }=\int \sqrt{\Big| g_{\mu\nu}\frac{dx^{\mu}}{d\lambda}\frac{dx^{\nu}}{d\lambda} \Big| }d\lambda
	\end{equation}
	where $ds^{2}=g_{\mu\nu}dx^{\mu}dx^{\nu}$ is the line element also known as the first fundamental or just the metric. Note that when the space is Lorentzian the above is written as
	\begin{equation}
	S=\int \sqrt{-g_{\mu\nu}\frac{dx^{\mu}}{d\lambda}\frac{dx^{\nu}}{d\lambda} }d\lambda
	\end{equation}
	By either directly varying the path $x^{\mu}\rightarrow x^{\mu}+\delta x^{\mu}$ in the latter and using the principle of least action or by using Lagrange equations 
	\begin{equation}
	\frac{d}{d\lambda}\left(\frac{\partial L}{\partial \dot{x}^{\mu}} \right)-\frac{\partial L}{\partial {x}^{\mu}}=0
	\end{equation} 
	with $L=\sqrt{g_{\mu\nu}\dot{x}^{\mu}\dot{x}^{\nu}}$\footnote{Again dot indicates differentiation with respect to $\lambda$.}, we derive the equations giving the geodesic curves
	\begin{equation}
	\frac{d^{2} x^{\mu}}{d\lambda^{2}}+\frac{1}{2}g^{\mu\rho}({\partial_{\alpha}g_{\beta\rho}+\partial_{\beta}g_{\alpha\rho}-\partial_{\rho}g_{\alpha\beta}})\frac{dx^{\alpha}}{d\lambda}\frac{dx^{\beta}}{d\lambda}=0
	\end{equation}
	It is worth noting now that the above combination of the metric and its first derivatives is exactly the Levi-Civita connection. Therefore, the geodesic equations are written as
	\begin{equation}
	\frac{d^{2} x^{\mu}}{d\lambda^{2}}+\tilde{\Gamma}^{\mu}_{\;\;\;\alpha\beta}\frac{dx^{\alpha}}{d\lambda}\frac{dx^{\beta}}{d\lambda}=0
	\end{equation}
	The solution $\gamma:\;x^{\mu}=x^{\mu}(\lambda;x^{\alpha}_{0};\dot{x}^{\alpha}_{0})$\footnote{Note that we chose to use the letter $\gamma$ for the geodesic curve here in order to distinguish it from the autoparallel curve $C$.}. We should emphasize now the difference of the latter with equation ($\ref{auto}$) that we derived for autoparallels. The connection here is the Levi-Civita connection, but for autoparallels the connection used is the general affine connection of the space. Although the two coincide in general relativity, in a general Metric-Affine framework these are not the same. Indeed, using the connection decomposition ($\ref{affconection}$) that we derived in previous chapter, the autoparallel equation becomes
	\begin{gather}
	\frac{d^{2} x^{\lambda}}{d\lambda^{2}}+\tilde{\Gamma}^{\lambda}_{\;\;\;\mu\nu}\frac{dx^{\mu}}{d\lambda}\frac{dx^{\nu}}{d\lambda}= \nonumber \\
	=-g^{\alpha\lambda}\frac{dx^{\mu}}{d\lambda}\frac{dx^{\nu}}{d\lambda}\left[ \frac{1}{2}Q_{\mu\nu\alpha}+\frac{1}{2}Q_{\nu\alpha\mu}-\frac{1}{2}Q_{\alpha\mu\nu}-S_{\alpha\mu\nu}-S_{\alpha\nu\mu}+S_{\mu\nu\alpha}\right]
	\end{gather}
	Now, recalling that $Q_{\alpha\mu\nu}=Q_{\alpha\nu\mu}$ along with $S_{\alpha\mu\nu}=-S_{\mu\alpha\nu}$, and using the fact that $\dot x^{\mu}\dot x^{\nu}$ is symmetric,  we have
	\begin{gather}
	\frac{d^{2} x^{\lambda}}{d\lambda^{2}}+\tilde{\Gamma}^{\lambda}_{\;\;\;\mu\nu}\frac{dx^{\mu}}{d\lambda}\frac{dx^{\nu}}{d\lambda}= \nonumber \\
	=-g^{\alpha\lambda}\frac{dx^{\mu}}{d\lambda}\frac{dx^{\nu}}{d\lambda}\left[ \frac{1}{2}Q_{\mu\nu\alpha}+\frac{1}{2}Q_{\mu\alpha\nu}-\frac{1}{2}Q_{\alpha\mu\nu}-2S_{\alpha\mu\nu}\right]
	\end{gather}
	From the above equation we see that because of torsion and non-metricity, geodesics and autoparallels are, in general, different curves.
	We illustrate this difference between geodesics and autoparallels with some examples and prove the Theorems regarding projectively equivalent connections in the Appendix.  We may now proceed by giving the definitions of the energy-momentum and hyper-momentum tensors.

	\subsection{Energy-momentum and Hyper-momentum Tensors}
	Having defined and explored the generalized geometry let us continue by introducing the physical content that gives rise to such a geometry. Following the literature we define the
	Energy-Momentum Tensor as the variation of the matter sector (of the action) with respect to the metric, namely
	\beq
	T_{\alpha\beta}:= -\frac{2}{\sqrt{-g}}\frac{\delta S_{M}}{\delta g^{\alpha\beta}}=-\frac{2}{\sqrt{-g}}\frac{\partial(\sqrt{-g} \mathcal{L}_{M})}{\partial g^{\alpha\beta}}
	\eeq
	Now, since matter can also depend on the affine connection, its variation with respect to it defines the  Hyper-momentum tensor \cite{hehl1976hypermomentum}
	\beq
	\Delta_{\lambda}^{\;\;\;\mu\nu}:= -\frac{2}{\sqrt{-g}}\frac{\delta S_{M}}{\delta \Gamma^{\lambda}_{\;\;\;\mu\nu}}=-\frac{2}{\sqrt{-g}}\frac{\partial ( \sqrt{-g} \mathcal{L}_{M})}{\partial \Gamma^{\lambda}_{\;\;\;\mu\nu}}
	\eeq
	An important thing that is almost never mentioned in the literature is that the above two tensors are not completely independent. Indeed, since $g_{\alpha\beta}$ and $\Gamma^{\lambda}_{\;\;\;\mu\nu}$ are independent variables, it holds that
	\beq
	\frac{\partial^{2}(\sqrt{-g} \mathcal{L}_{M})}{\partial g^{\alpha\beta}\partial \Gamma^{\lambda}_{\;\;\;\mu\nu}}=\frac{\partial^{2} (\sqrt{-g}\mathcal{L}_{M})}{\partial \Gamma^{\lambda}_{\;\;\;\mu\nu} \partial g^{\alpha\beta} }
	\eeq
	and as a result
	\beq
	\frac{1}{\sqrt{-g}}\frac{\partial}{\partial g^{\alpha\beta}}\Big( \sqrt{-g} \Delta_{\lambda}^{\;\;\;\mu\nu} \Big)= \frac{\partial T_{\alpha\beta}}{\partial \label{emhpt} \Gamma^{\lambda}_{\;\;\;\mu\nu}}
	\eeq
	Therefore we see that the energy-momentum and hyper-momentum tensors are not independent. If the latter is applied for a perfect fluid for instance, where $T_{\mu\nu}$ is independent of the connection, the hyper-momentum tensor has to satisfy
	\beq
	\frac{\partial}{\partial g^{\alpha\beta}}\Big( \sqrt{-g} \Delta_{\lambda}^{\;\;\;\mu\nu} \Big)=0
	\eeq
	So for a perfect fluid\footnote{Assuming that its form remains the same as in GR.}
	\beq
	\sqrt{-g} \Delta_{\lambda}^{\;\;\;\mu\nu}=independent \;of\;\; g_{\mu\nu}
	\eeq
	In addition, in the so-called Palatini Theories the matter action $S_{M}$ is assumed to be independent of the connection and therefore $\Rightarrow \Delta_{\lambda}^{\;\;\;\mu\nu} =0$. The latter means that in this case (Palatini Gravity) the energy momentum tensor is independent of the connection, as seen from $(\ref{emhpt})$. This result is crucial when studying the dynamical content of a connection and we will use it latter on when we touch upon the subject of dynamical/non-dynamical connections.

	\subsection{Curvature Identities}
	
	As we have already seen, the anti-symmetrized covariant derivative on a vector field, yields
	\begin{equation}
	[\nabla_{\alpha} ,\nabla_{\beta}]u^{\mu}=2\nabla_{[\alpha} \nabla_{\beta]}u^{\mu}=R^{\mu}_{\;\;\;\nu\alpha\beta} u^{\nu}+2 S_{\alpha\beta}^{\;\;\;\;\;\nu}\nabla_{\nu}u^{\mu}
	\end{equation}
	When acting on a co-vector gives
	\begin{equation}
	[\nabla_{\alpha} ,\nabla_{\beta}]u_{\mu}=2\nabla_{[\alpha} \nabla_{\beta]}u_{\mu}=-R^{\lambda}_{\;\;\;\mu\alpha\beta} u_{\lambda}+2 S_{\alpha\beta}^{\;\;\;\;\;\nu}\nabla_{\nu}u_{\mu}
	\end{equation}
	And in the trivial case of a scalar, one has
	\beq
	\nabla_{[\alpha} \nabla_{\beta]}\phi= S_{\alpha\beta}^{\;\;\;\;\;\nu}\nabla_{\nu}\phi
	\eeq
	Of course we can generalize the above considerations for higher rank tensors, for instance for a rank-$2$ tensor we have
	\beq
	[\nabla_{\alpha} ,\nabla_{\beta}]T_{\mu\nu} =-R^{\lambda}_{\;\;\;\mu\alpha\beta}T_{\lambda\nu}-R^{\lambda}_{\;\;\;\nu\alpha\beta}T_{\lambda\mu}+2 S_{\alpha\beta}^{\;\;\;\;\;\lambda}\nabla_{\lambda}T_{\mu\nu}
	\eeq
	which when applied to the metric tensor  yields
	\beq
	[\nabla_{\alpha} ,\nabla_{\beta}]g_{\mu\nu} =-R^{\lambda}_{\;\;\;\mu\alpha\beta}g_{\lambda\nu}-R^{\lambda}_{\;\;\;\nu\alpha\beta}g_{\lambda\mu}+2 S_{\alpha\beta}^{\;\;\;\;\;\lambda}\nabla_{\lambda}g_{\mu\nu}
	\eeq
	Recalling now the definition of non-metricity $Q_{\alpha\mu\nu}\equiv -\nabla_{\alpha}g_{\mu\nu}$, we get the identity
	\beq
	R_{(\mu\nu)\alpha\beta}=\nabla_{[\alpha}Q_{\beta]\mu\nu}-S_{\alpha\beta}^{\;\;\;\;\;\lambda}Q_{\lambda\mu\nu}
	\eeq
	Notice from the above that the Riemann tensor is antisymmetric in its first two indices only for a metric connection ($Q_{\alpha\mu\nu}=0$). Another identity comes about when we fully anti-symmetrize the Riemann tensor in its three lower indices. In words
	\beq
	R^{\alpha}_{\;\;\;[\beta\mu\nu]}=\frac{2}{3!}\Big( R^{\alpha}_{\;\;\;\beta[\mu\nu]}+R^{\alpha}_{\;\;\;\nu[\beta\mu]}+R^{\alpha}_{\;\;\;\mu[\nu\beta]} \Big)=\frac{2}{3!}\Big( R^{\alpha}_{\;\;\;\beta\mu\nu}+R^{\alpha}_{\;\;\;\nu\beta\mu}+R^{\alpha}_{\;\;\;\mu\nu\beta} \Big)
	\eeq
	where we used the fact that the Riemann tensor is already antisymmetric in its last two indices. Carrying out the calculations, we can easily arrive at
	\beq
	R^{\alpha}_{\;\;\;[\beta\mu\nu]}=-2 \nabla_{[\beta}S_{\mu\nu]}^{\;\;\;\;\;\alpha}-4 S_{[\beta\mu}^{\;\;\;\;\;\lambda}S_{\nu]\lambda}^{\;\;\;\;\;\alpha}
	\eeq
	Notice that the latter vanishes for a torsion-free space ($S_{\alpha\beta}^{\;\;\;\;\;\lambda}=0$) even when non-metricity is present. Contracting the above in $\alpha=\beta$ we obtain a further identity
	\beq
	R_{[\mu\nu]}=\frac{1}{2}\hat{R}_{\mu\nu}+\nabla_{\alpha}S_{\mu\nu}^{\;\;\;\;\;\alpha}+2\nabla_{[\mu}S_{\nu]} +2(S_{\alpha\mu}^{\;\;\;\;\;\lambda}S_{\nu\lambda}^{\;\;\;\;\;\alpha}+S_{\nu\alpha}^{\;\;\;\;\;\lambda} S_{\mu\lambda}^{\;\;\;\;\;\alpha}  - S_{\mu\nu}^{\;\;\;\;\;\alpha}S_{\alpha})
	\eeq
	or
	\begin{gather}
	R_{[\mu\nu]}=\frac{1}{2}\hat{R}_{\mu\nu}+\nabla_{\alpha}S_{\mu\nu}^{\;\;\;\;\;\alpha}+2\partial_{[\mu}S_{\nu]} +4 S_{\alpha[\mu}^{\;\;\;\;\;\lambda}S_{\nu]\lambda}^{\;\;\;\;\;\alpha}= \nonumber \\
	=\frac{1}{2}\hat{R}_{\mu\nu}+\partial_{\alpha}S_{\mu\nu}^{\;\;\;\;\;\alpha}+2\partial_{[\mu}S_{\nu]} +2\Gamma^{\lambda}_{\;\;\;\alpha[\mu}S_{\nu]\lambda}^{\;\;\;\;\alpha}+\Gamma^{\alpha}_{\;\;\;\lambda\alpha}S_{\mu\nu}^{\;\;\;\;\;\lambda}
	\end{gather}
	from which we conclude that for the Ricci tensor to be symmetric, both torsion and non-metricity have to vanish. Another identity that is so scary even Schouten does not bother writing it down is the one that involves interchanging the first two indices of the Riemann tensor with its last two. The starting point here is the identity
	\begin{gather}
	A_{\mu\lambda\kappa\nu}=A_{\kappa\nu\mu\lambda}-\frac{3}{2}\Big( A_{\mu[\lambda\nu\kappa]}+A_{\lambda[\mu\nu\kappa]}+A_{\kappa[\nu\mu\lambda]}+A_{\nu[\kappa\mu\lambda]}  \Big) \nonumber\\
	+A_{(\lambda\mu)\nu\kappa}+A_{(\kappa\mu)\lambda\nu}+A_{(\nu\lambda)\kappa\mu}+A_{(\nu\kappa)\mu\lambda}+A_{(\lambda\kappa)\mu\nu}+A_{(\nu\mu)\lambda\kappa}
	\end{gather}
	which holds for any rank-$4$ tensor. Applying this to the Riemann tensor, and using the identities we derived above, it follows that
	\begin{gather}
	R_{\mu\nu\kappa\lambda}-R_{\kappa\lambda\mu\nu}=3\Big( g_{\mu\alpha}\nabla_{[\nu}S_{\lambda\kappa]}^{\;\;\;\;\;\alpha} + g_{\nu\alpha}\nabla_{[\mu}S_{\lambda\kappa]}^{\;\;\;\;\;\alpha}+ g_{\kappa\alpha}\nabla_{[\lambda}S_{\mu\nu]}^{\;\;\;\;\;\alpha}+g_{\lambda\alpha}\nabla_{[\kappa}S_{\mu\nu]}^{\;\;\;\;\;\alpha}
	\Big) \nonumber \\
	+6\Big( g_{\mu\alpha}S_{[\nu\lambda}^{\;\;\;\;\;\beta}S_{\kappa]\beta}^{\;\;\;\;\;\alpha}+g_{\nu\alpha}S_{[\mu\lambda}^{\;\;\;\;\;\beta}S_{\kappa]\beta}^{\;\;\;\;\;\alpha} +g_{\kappa\alpha}S_{[\lambda\mu}^{\;\;\;\;\;\beta}S_{\nu]\beta}^{\;\;\;\;\;\alpha}+g_{\lambda\alpha}S_{[\kappa\mu}^{\;\;\;\;\;\beta}S_{\nu]\beta}^{\;\;\;\;\;\alpha}   \Big) \nonumber \\
	+\nabla_{[\lambda}Q_{\kappa]\nu\mu}+\nabla_{[\nu}Q_{\lambda]\kappa\mu}+\nabla_{[\kappa}Q_{\mu]\lambda\nu}+\nabla_{[\mu}Q_{\nu]\lambda\kappa}+\nabla_{[\mu}Q_{\lambda]\nu\kappa}+\nabla_{[\nu}Q_{\kappa]\lambda\mu} \nonumber \\
	-\Big( S_{\lambda\kappa}^{\;\;\;\;\;\alpha}Q_{\alpha\nu\mu}+ S_{\nu\lambda}^{\;\;\;\;\;\alpha}Q_{\alpha\kappa\mu}+ S_{\kappa\mu}^{\;\;\;\;\;\alpha}Q_{\alpha\lambda\nu} \nonumber \\
	+ S_{\mu\nu}^{\;\;\;\;\;\alpha}Q_{\alpha\lambda\kappa}+ S_{\mu\lambda}^{\;\;\;\;\;\alpha}Q_{\alpha\nu\kappa}+ S_{\nu\kappa}^{\;\;\;\;\;\alpha}Q_{\alpha\lambda\mu}\Big) \label{xamoulhs}
	\end{gather}
	and we see that the symmetry $R_{\mu\nu\kappa\lambda}=R_{\kappa\lambda\mu\nu}$ only holds when both torsion and non-metricity vanish. The moral of the above identity is crystal clear: If you were to interchange the two first with the two last indices in  the Riemann tensor, when the space is not Riemannian  ( i.e has  non vanishing torsion and non-metricity) you'd better not do it! Another identity can be obtained by contracting the Riemann tensor with the Levi-Civita tensor to form the (parity violating) scalar
	\beq
	\varepsilon^{\mu\nu\alpha\beta}R_{\mu\nu\alpha\beta}=2 \tilde{\nabla}_{\alpha}(\varepsilon^{\alpha\mu\nu\beta}N_{\mu\nu\beta})+2 \varepsilon^{\mu\nu\alpha\beta}S_{\alpha\beta}^{\;\;\;\;\lambda}( Q_{[\mu\nu]\lambda}-S_{\mu\nu\lambda})
	\eeq
	where we have used the decomposition ($\ref{rtdec}$) for the Riemann tensor, the fact that $\varepsilon^{\mu\nu\alpha\beta}\tilde{R}_{\mu\nu\alpha\beta}=0$ for the Riemannian part and the fact that the Levi-Civita tensor is covariantly conserved with respect to the Levi-Civita connection $(\tilde{\nabla}_{\lambda}\varepsilon_{\mu\nu\alpha\beta}=0)$. In a Riemannian space the right hand term is zero. Notice also that if torsion is zero, the right hand side of the above equation vanishes even if the space has non-metricity. This may become more apparent by observing that
	\beq
	N_{[\alpha\mu\nu]}=S_{[\mu\nu\alpha]}=S_{[\alpha\mu\nu]}
	\eeq
	such that
	\beq
	\varepsilon^{\mu\nu\alpha\beta}R_{\mu\nu\alpha\beta}=2 \tilde{\nabla}_{\alpha}(\varepsilon^{\alpha\mu\nu\beta}S_{[\mu\nu\beta]})+2 \varepsilon^{\mu\nu\alpha\beta}S_{\alpha\beta}^{\;\;\;\;\lambda}( Q_{[\mu\nu]\lambda}-S_{\mu\nu\lambda})
	\eeq
	from which we conclude that
	\beq
	\varepsilon^{\mu\nu\alpha\beta}R_{\mu\nu\alpha\beta}=0
	\eeq
	if torsion is zero. In addition, using the definition of the torsion pseudo-vector $\tilde{S}^{\mu}\equiv \varepsilon^{\mu\rho\nu\beta}S_{[\rho\nu\beta]}$, for a non-Riemannian space we have the above identity written in a more compact form
	\beq
	\varepsilon^{\mu\nu\alpha\beta}R_{\mu\nu\alpha\beta}=2 \tilde{\nabla}_{\alpha}\tilde{S}^{\alpha}+2 \varepsilon^{\mu\nu\alpha\beta}S_{\alpha\beta}^{\;\;\;\;\lambda}( Q_{[\mu\nu]\lambda}-S_{\mu\nu\lambda})
	\eeq

	\subsection{Weitzenbock identities
		(generalized Bianchi identities)}
	Let us now give the generalized Bianchi identities for a torsionfull, non-metric connection. These identities are also known as Weitzenbock identities. The first one we obtain by taking the covariant derivative of the Riemann tensor and antisymmetrize in three indices, which results in
	\beq
	\nabla_{[\rho}R^{\alpha}_{\;\;\;|\beta|\mu\nu]}=2 R^{\alpha}_{\;\;\;\beta\lambda[\rho}S_{\mu\nu]}^{\;\;\;\;\;\lambda} \label{bian}
	\eeq
	where the vertical bars around an index indicate that this index is left out of the (anti-)symmetrization. Contracting the above in $\alpha=\beta$ we also obtain another identity
	\beq
	\nabla_{[\rho}\hat{R}_{\mu\nu]}=2 \hat{R}_{\lambda[\rho}S_{\mu\nu]}^{\;\;\;\;\;\lambda}
	\eeq
	We can also obtain an identity by contracting ($\ref{bian}$) in $\rho=\alpha$ to obtain
	\beq
	\nabla_{\alpha}R^{\alpha}_{\;\;\;\beta\mu\nu}-2\nabla_{[\mu}R_{|\beta|\nu]}=-2 R_{\beta\lambda}S_{\mu\nu}^{\;\;\;\;\;\lambda}+4 R^{\alpha}_{\;\;\;\beta\lambda[\nu}S_{\mu]\alpha}^{\;\;\;\;\;\lambda} 
	\eeq

	\section{The Lie Derivative}
	As can be found in any GR textbook (see \cite{carroll1997lecture} for instance) when one is dealing with a Riemannian geometry, the partial derivatives that appear in a Lie derivative, can be replaced with the covariant ones. For instance, the Lie derivative of the contravariant vector $u^{\mu}$ in the direction of $\xi^{\mu}$, reads
	\beq
	\Lie_{\xi}u^{\alpha}=\xi^{\mu}\partial_{\mu}u^{\alpha}-u^{\mu}\partial_{\mu}\xi^{\alpha}=\xi^{\mu}\nabla_{\mu}u^{\alpha}-u^{\mu}\nabla_{\mu}\xi^{\alpha}
	\eeq
	in a Riemannian space (both torsion and non-metricity are zero and the covariant derivative is computed with respect to the Levi-Civita connection). However, in an affine space where both torsion and non-metricity are not zero, the above is written as
	\beq
	\Lie_{\xi}u^{\alpha}=\xi^{\mu}\partial_{\mu}u^{\alpha}-u^{\mu}\partial_{\mu}\xi^{\alpha}=\xi^{\mu}\nabla_{\mu}u^{\alpha}-u^{\mu}\nabla_{\mu}\xi^{\alpha}-2 S_{\mu\nu}^{\;\;\;\;\alpha}u^{\mu}\xi^{\nu}
	\eeq
	We should stress out the the Lie derivative on any tensor field, in its original form, contains only partial derivatives of the associated quantities. If we insist upon expressing the final result in terms of the full covariant derivative we must also include the additional terms that appear due to torsion and non-metricity. The way to do this is to expand the covariant derivative, solve the partial derivative in terms of it and substitute it back in the Lie derivative. For example, let us prove the above result for the contravariant vector $u^{\alpha}$. We start by expanding
	\beq
	\nabla_{\mu}u^{\alpha}=\partial_{\mu}u^{\alpha}-\Gamma^{\alpha}_{\;\;\;\lambda\mu}u^{\lambda}
	\eeq
	\beq
	\nabla_{\mu}\xi^{\alpha}=\partial_{\mu}\xi^{\alpha}-\Gamma^{\alpha}_{\;\;\;\lambda\mu}\xi^{\lambda}
	\eeq
	multiplying the former by $\xi^{\mu}$ and the latter by $u^{\mu}$ and subtracting them, it follows that
	\beq
	\xi^{\mu}\nabla_{\mu}u^{\alpha}-u^{\mu}\nabla_{\mu}\xi^{\alpha}=\xi^{\mu}\partial_{\mu}u^{\alpha}-u^{\mu}\partial_{\mu}\xi^{\alpha}+2 S_{\mu\nu}^{\;\;\;\;\alpha}u^{\mu}\xi^{\nu} \Rightarrow \nonumber
	\eeq
	\beq
	\xi^{\mu}\partial_{\mu}u^{\alpha}-u^{\mu}\partial_{\mu}\xi^{\alpha}=\xi^{\mu}\nabla_{\mu}u^{\alpha}-u^{\mu}\nabla_{\mu}\xi^{\alpha}-2 S_{\mu\nu}^{\;\;\;\;\alpha}u^{\mu}\xi^{\nu} \Rightarrow  \nonumber
	\eeq
	\beq
	\Lie_{\xi}u^{\alpha}=\xi^{\mu}\nabla_{\mu}u^{\alpha}-u^{\mu}\nabla_{\mu}\xi^{\alpha}-2 S_{\mu\nu}^{\;\;\;\;\alpha}u^{\mu}\xi^{\nu}
	\eeq
	In a similar manner we find the Lie derivative of a covariant vector $v_{\mu}$,
	\beq
	\Lie_{\xi}v_{\mu}=\xi^{\alpha}\partial_{\alpha}v_{\mu}-v_{\alpha}\partial_{\mu}\xi^{\alpha}=\xi^{\alpha}\nabla_{\alpha}v_{\mu}-v_{\alpha}\nabla_{\mu}\xi^{\alpha}-2 S_{\alpha\mu}^{\;\;\;\;\lambda}\xi^{\alpha}v_{\lambda}
	\eeq
	Now, for a rank $2$ covariant tensor field-$T_{\mu\nu}$ one computes
	\beq
	\Lie_{\xi}T_{\mu\nu}=\xi^{\lambda}\partial_{\lambda}T_{\mu\nu}+T_{\lambda\nu}\partial_{\mu}\xi^{\lambda}+T_{\mu\lambda}\partial_{\nu}\xi^{\lambda}
	\eeq
	If $T_{\mu\nu}$ is taken to be the metric tensor $g_{\mu\nu}$, the above gives
	\begin{gather}
	\Lie_{\xi}g_{\mu\nu}=\xi^{\lambda}\partial_{\lambda}g_{\mu\nu}+g_{\lambda\nu}\partial_{\mu}\xi^{\lambda}+g_{\mu\lambda}\partial_{\nu}\xi^{\lambda}=\nonumber \\
	=\xi^{\lambda}\partial_{\lambda}g_{\mu\nu}+\partial_{\mu}\xi_{\nu}-\xi^{\lambda}\partial_{\mu}g_{\lambda\nu}+\partial_{\nu}\xi_{\mu}-\xi^{\lambda}\partial_{\nu}g_{\mu\lambda}= \nonumber \\
	=\partial_{\mu}\xi_{\nu}+\partial_{\nu}\xi_{\mu}-\xi^{\lambda}( \partial_{\mu}g_{\lambda\nu}+\partial_{\nu}g_{\lambda\mu}-\partial_{\lambda}g_{\mu\nu})= \nonumber \\
	=\partial_{\mu}\xi_{\nu}+\partial_{\nu}\xi_{\mu}-2 \tilde{\Gamma}^{\lambda}_{\;\;\;\mu\nu}\xi_{\lambda}=\tilde{\nabla}_{\mu}\xi_{\nu}+\tilde{\nabla}_{\nu}\xi_{\mu} \label{lieg}
	\end{gather}
	where on going from the first to the second line we employed Leibniz's rule and on going  third to forth we used the definition of the Levi-Civita connection. Note that in the last line the covariant derivative computed with respect to the Levi-Civita connection, appears. If we want to express the Lie derivative in terms of the general covariant derivative, we may use the connection decomposition
	\beq
	\Gamma^{\lambda}_{\;\;\;\mu\nu}=\tilde{\Gamma}^{\lambda}_{\;\;\;\mu\nu}+N^{\lambda}_{\;\;\;\mu\nu}
	\eeq
	and recast ($\ref{lieg}$) to
	\beq
	\Lie_{\xi}g_{\mu\nu}=\nabla_{\mu}\xi_{\nu}+\nabla_{\nu}\xi_{\mu}+2 N^{\lambda}_{\;\;\;(\mu\nu)}\xi_{\lambda}
	\eeq
	where we have used the relation
	\beq
	\nabla_{\mu}\xi_{\nu}=\partial_{\mu}\xi_{\nu}-\Gamma^{\lambda}_{\;\;\;\nu\mu}\xi_{\lambda}=\tilde{\nabla}_{\mu}\xi_{\nu}-N^{\lambda}_{\;\;\;\nu\mu}\xi_{\lambda}
	\eeq
	The above procedure can be generalized to derive the Lie derivative for tensor fields of arbitrary rank.

	\subsection{Non-trivial Surface terms}
	In Metric Gravity (GR is a special case of it) where we have neither torsion nor non-metricity, the following hold true
	\begin{equation}
	S_{\mu\nu}^{\;\;\;\;\alpha}=0
	\end{equation}
	\begin{equation}
	\nabla_{\mu}\sqrt{-g}=0
	\end{equation}
	which lead to trivial surface terms
	\begin{gather}
	\int d^{4}x\nabla_{\mu}(\sqrt{-g}u^{\mu})=\int d^{4}x\sqrt{-g}\nabla_{\mu}u^{\mu}= \nonumber \\
	=\int d^{4}x\partial_{\mu}(\sqrt{-g}u^{\mu})=surface\;term \label{b}
	\end{gather}
	for any vector field $u^{\mu}$. However, in Metric-Affine spaces both non-metricity and torsion are non-vanishing and one has\footnote{Recall that $\sqrt{-g}$ is a scalar density of weight $-1$.}
	\begin{equation}
	S_{\mu\nu}^{\;\;\;\;\alpha} \neq 0
	\end{equation}
	\begin{equation}
	\nabla_{\mu}\sqrt{-g}=\partial_{\mu}\sqrt{-g}-\Gamma^{\alpha}_{\;\;\;\alpha\mu}\sqrt{-g} \neq 0
	\end{equation}
	Furthermore, the covariant derivative on contravariant vectors yields
	\begin{equation}
	\nabla_{\mu}u^{\nu}=\partial_{\mu}u^{\nu}+\Gamma^{\nu}_{\;\;\;\alpha\mu}u^{\alpha}
	\end{equation}
	and contracting in $\mu$,$\nu$ we obtain
	\begin{equation}
	\nabla_{\mu}u^{\mu}=\partial_{\mu}u^{\mu}+\Gamma^{\mu}_{\;\;\;\alpha\mu}u^{\alpha}
	\end{equation}
	Thus, using the above, one has
	\begin{equation}
	\nabla_{\mu}(\sqrt{-g}u^{\mu}) =\partial_{\mu}(\sqrt{-g}u^{\mu})+(\Gamma^{\alpha}_{\;\;\;\mu\alpha}-\Gamma^{\alpha}_{\;\;\;\alpha\mu})\sqrt{-g}u^{\mu} \Rightarrow \nonumber
	\end{equation}
	\begin{equation}
	\nabla_{\mu}(\sqrt{-g}u^{\mu}) =\partial_{\mu}(\sqrt{-g}u^{\mu})+\sqrt{-g} 2 S_{\mu}u^{\mu}
	\end{equation}
	where $S_{\mu}:=S_{\mu\nu}^{\;\;\;\nu}$. As a result the integral in $(\ref{b})$ now takes the form
	\begin{gather}
	\int d^{4}x\nabla_{\mu}(\sqrt{-g}u^{\mu})=\int d^{4}x\partial_{\mu}(\sqrt{-g}u^{\mu})+\int d^{4}x \sqrt{-g}2 S_{\mu}u^{\mu}=  \nonumber \\
	=surface\;term+\int d^{4}x \sqrt{-g} 2 S_{\mu}u^{\mu}
	\end{gather}
	Therefore, when both torsion and non-metricity are present there are additional contributions to the surface terms that need to be taken into account. However, note from the last equation that this contribution is proportional to the torsion vector $S_{\mu}$ and as a result the additional term comes solely by the presence of torsion and not of that of non-metricity. That is, equation $(\ref{b})$ holds true in the absence of torsion even when the non-metricity is not zero. Now, consider an integral of the form
	\begin{equation}
	\int d^{4}x\sqrt{-g}\nabla_{\mu}u^{\mu}
	\end{equation}
	Using Leibniz's rule and the above result we may express it as
	\begin{gather}
	\int d^{4}x\sqrt{-g}\nabla_{\mu}u^{\mu}=\int d^{4}x\nabla_{\mu}(\sqrt{-g}u^{\mu})-\int d^{4}x u^{\mu}\nabla_{\mu}\sqrt{-g}= \nonumber \\
	=surface\;term+\int d^{4}x \sqrt{-g}2 S_{\mu}u^{\mu}-\int d^{4}x \sqrt{-g}u^{\mu}\frac{1}{2}g^{\alpha\beta}\nabla_{\mu}g_{\alpha\beta}= \nonumber \\
	=surface\;term+\int d^{4}x \sqrt{-g}u^{\mu}\left( 2 S_{\mu}+\frac{1}{2}g^{\alpha\beta}Q_{\mu\alpha\beta}\right)    \nonumber
	\end{gather}
	therefore we obtain
	\begin{equation}
	\int d^{4}x\sqrt{-g}\nabla_{\mu}u^{\mu}=\int d^{4}x \sqrt{-g}u^{\mu}\left( 2 S_{\mu}+\frac{1}{2}Q_{\mu}\right)+surface\;term
	\end{equation}
	where $Q_{\mu}:=Q_{\mu\nu}^{\;\;\;\;\nu}=g^{\nu\alpha}Q_{\mu\nu\alpha}$ is the Weyl vector. Notice that on going from the first to the second line we employed the identity
	\begin{equation}
	\nabla_{\mu}\sqrt{-g}=-\sqrt{-g}\frac{1}{2}g_{\alpha\beta}\nabla_{\mu}g^{\alpha\beta}=+\sqrt{-g}\frac{1}{2}g^{\alpha\beta}\nabla_{\mu}g_{\alpha\beta}
	\end{equation}
	
	\section{Useful Identities and Proofs}
	We begin here by giving some general identities for the Levi-Civita symbol firstly in arbitrary spacetime dimensions and then specializing to four. In general the following hold true
	\begin{equation}
	\epsilon_{a_{1}a_{2}...a_{k}a_{k+1}...a_{n}}\epsilon^{a_{1}a_{2}...a_{k}b_{k+1}...b_{k}...b_{n}}=(-1)^{t}(n-k)!k!\delta^{b_{k+1}}_{[a_{k+1}}...\delta^{b_{n}}_{a_{n}]}
	\end{equation}
	\begin{equation}
	\epsilon_{a_{1}a_{2}...a_{n-2}ed}\epsilon^{a_{1}a_{2}...a_{n-2}l m}=(-1)^{t}(n-2)!2\delta^{l}_{[e}\delta^{m}_{d]}
	\end{equation}
	\begin{equation}
	\epsilon_{a_{1}a_{2}...a_{n-1}d}\epsilon^{a_{1}a_{2}...a_{n-1}e}=(-1)^{t}(n-1)!\delta_{d}^{e}
	\end{equation}
	\begin{equation}
	\epsilon_{a_{1}a_{2}...a_{n}}\epsilon^{a_{1}a_{2}...a_{n}}=(-1)^{t}n!
	\end{equation}
	where $k\leq n$ and $t$ is the signature of the space. Having the Levi-Civita symbol we can write the determinant of the metric tensor as follows
	\begin{equation}
	g:=\det{(g_{ab})}=\frac{(-1)^{t}}{n!}\epsilon^{a_{1}a_{2}...a_{n}}\epsilon^{b_{1}b_{2}...b_{n}}g_{a_{1}b_{1}}...g_{a_{n}b_{n}}
	\end{equation}
	which is a scalar density of weight $-2$. Now, specializing in a $4-dim$ Lorentzian spacetime\footnote{Of course the same results hold true for any $n$-dim Lorentzian spacetime. We take $n=4$ here just for convenience.} (i.e. $n=4$, $t=1$) the above reduce to
	\begin{equation}
	\epsilon_{\mu\nu\rho\lambda}\epsilon^{\mu\alpha\beta\gamma}=-1! 3! \delta_{[\nu}^{\alpha}\delta_{\rho}^{\beta}\delta_{\lambda]}^{\gamma} \label{eiden1}
	\end{equation}
	\begin{equation}
	\epsilon_{\mu\nu\rho\lambda}\epsilon^{\mu\nu\kappa\sigma}=-2! 2! \delta_{[\rho}^{\kappa}\delta_{\lambda]}^{\sigma}
	\end{equation}
	\begin{equation}
	\epsilon_{\mu\nu\rho\lambda}\epsilon^{\mu\nu\rho\sigma}=-3!\delta_{\lambda}^{\sigma}
	\end{equation}
	\begin{equation}
	\epsilon_{\mu\nu\rho\sigma}\epsilon^{\mu\nu\rho\sigma}=-4!
	\end{equation}
	Also note that when no contraction among indices is involved, we have
	\begin{equation}
	\epsilon_{\mu\nu\rho\lambda}\epsilon^{\sigma\alpha\beta\gamma}=- 4! \delta_{[\mu}^{\sigma}\delta_{\nu}^{\alpha}\delta_{\rho}^{\beta}\delta_{\lambda]}^{\gamma} \label{eiden2}
	\end{equation}
	It is also worth noting that due to the non-tensorial nature of $\epsilon_{\mu\nu\rho\sigma}$\footnote{However, this can be converted to a tensor if we multiply by $\sqrt{-g}$, namely $\varepsilon_{\mu\nu\rho\sigma}:=\sqrt{-g}\epsilon_{\mu\nu\rho\sigma}$ does behave tensorial. Note that its contravariant tensor form is $\varepsilon^{\mu\nu\rho\sigma}:=\frac{\epsilon^{\mu\nu\rho\sigma}}{\sqrt{-g}}$} one has
	\begin{equation}
	\epsilon^{\mu\nu\rho\sigma}g_{\mu\alpha}g_{\nu\beta}g_{\rho\gamma}g_{\sigma\delta}=-g\epsilon_{\alpha\beta\gamma\delta} \label{c}
	\end{equation}
	and
	\begin{equation}
	\epsilon_{\mu\nu\rho\sigma}=-g\epsilon^{\mu\nu\rho\sigma}
	\end{equation}
	Now, multiplying (and contracting) equation $(\ref{c})$ by $g^{\kappa\delta}$ we arrive at
	\begin{equation}
	\epsilon^{\mu\nu\rho\kappa}g_{\mu\alpha}g_{\nu\beta}g_{\rho\gamma}=-g\epsilon_{\alpha\beta\gamma\delta}g^{\kappa\delta}
	\end{equation}
	which is going to be used in what follows. The determinant of the metric tensor is now given by
	\begin{equation}
	g:=\det{(g_{\mu\nu})}=\frac{1}{4!}\epsilon^{\mu\nu\rho\sigma}\epsilon^{\alpha\beta\gamma\delta}g_{\mu\alpha}g_{\nu\beta}g_{\rho\gamma}g_{\sigma\delta}
	\end{equation}
	Acting the covariant derivative on it, it follows that
	\begin{equation}
	\nabla_{\lambda}g=\frac{1}{4!}\epsilon^{\mu\nu\rho\sigma}\epsilon^{\alpha\beta\gamma\delta}\Big( (\nabla_{\lambda}g_{\mu\alpha})g_{\nu\beta}g_{\rho\gamma}g_{\sigma\delta}+...+g_{\mu\alpha}g_{\nu\beta}g_{\rho\gamma}(\nabla_{\lambda}g_{\sigma\delta}) \Big)
	\end{equation}
	and with a relabeling of the dummy indices the latter becomes
	\begin{gather}
	\nabla_{\lambda}g=\frac{4}{4!}\epsilon^{\mu\nu\rho\sigma}\epsilon^{\alpha\beta\gamma\delta}g_{\mu\alpha}g_{\nu\beta}g_{\rho\gamma}(\nabla_{\lambda}g_{\sigma\delta})= \nonumber \\
	=\frac{1}{3!}\epsilon^{\mu\nu\rho\sigma}\underbrace{\epsilon^{\alpha\beta\gamma\delta}g_{\mu\alpha}g_{\nu\beta}g_{\rho\gamma}}_{=-g\epsilon_{\mu\nu\rho\kappa}g^{\kappa\delta}} \nabla_{\lambda}g_{\sigma\delta}= \nonumber \\
	=-\frac{g}{3!}\underbrace{\epsilon^{\mu\nu\rho\sigma}\epsilon_{\mu\nu\rho\kappa}}_{=-3!\delta_{\kappa}^{\sigma}}g^{\kappa\delta}\nabla_{\lambda}g_{\sigma\delta}=+g g^{\sigma\delta}\nabla_{\lambda}g_{\sigma\delta} \Rightarrow  \nonumber
	\end{gather}
	\begin{equation}
	\nabla_{\lambda}g=+g g^{\mu\nu}\nabla_{\lambda}g_{\mu\nu}
	\end{equation}
	In addition, using 
	\begin{equation}
	0=\nabla_{\lambda}4=\nabla_{\lambda}(g_{\mu\nu}g^{\mu\nu})=g^{\mu\nu}\nabla_{\lambda} g_{\mu\nu}+g_{\mu\nu}\nabla_{\lambda}g^{\mu\nu}
	\end{equation}
	we may write the above as
	\begin{equation}
	\nabla_{\lambda}g=+g g^{\mu\nu}\nabla_{\lambda}g_{\mu\nu}=-g_{\mu\nu}\nabla_{\lambda}g^{\mu\nu}
	\end{equation}
	Recalling the definition of the non-metricity tensor
	\begin{equation}
	Q_{\lambda\mu\nu}=-\nabla_{\lambda}g_{\mu\nu}
	\end{equation}
	along with that of the Weyl vector
	\begin{equation}
	Q_{\lambda}:=g^{\mu\nu} Q_{\lambda\mu\nu}=Q_{\lambda\nu}^{\;\;\;\;\nu}=-g g^{\mu\nu}\nabla_{\lambda}g_{\mu\nu}
	\end{equation}
	we finally arrive at
	\begin{equation}
	\nabla_{\lambda}g=-g Q_{\lambda}
	\end{equation}
	Now, since $-g>0$ we have $g=\sqrt{-g}\sqrt{-g}$ and therefore
	\begin{equation}
	2\sqrt{-g}\nabla_{\lambda}\sqrt{-g}=g g^{\mu\nu}\nabla_{\lambda}g_{\mu\nu}\Rightarrow \frac{\nabla_{\lambda}\sqrt{-g}}{\sqrt{-g}}=\frac{1}{2} g^{\mu\nu}\nabla_{\lambda}g_{\mu\nu}
	\end{equation}
	or
	\begin{equation}
	\frac{\nabla_{\lambda}\sqrt{-g}}{\sqrt{-g}}=\nabla_{\lambda}\ln{\sqrt{-g}}=\frac{1}{2} g^{\mu\nu}\nabla_{\lambda}g_{\mu\nu}=-\frac{1}{2} g_{\mu\nu}\nabla_{\lambda}g^{\mu\nu}=-\frac{1}{2}Q_{\lambda} \label{d}
	\end{equation}
	Notice that the results derived so far not only hold true for $\nabla_{\lambda}$ but also for any linear operator $\hat{T}$. Indeed, our starting point was the following  expression of the determinant
	\begin{equation}
	g:=\det{(g_{\mu\nu})}=\frac{1}{4!}\epsilon^{\mu\nu\rho\sigma}\epsilon^{\alpha\beta\gamma\delta}g_{\mu\alpha}g_{\nu\beta}g_{\rho\gamma}g_{\sigma\delta}
	\end{equation}
	Acting on it with $\hat{T}$ and performing manipulations identical to the above ones, we would again arrive at
	\begin{equation}
	\frac{\hat{T}\sqrt{-g}}{\sqrt{-g}}=\hat{T}\ln{\sqrt{-g}}=\frac{1}{2} g^{\mu\nu}(\hat{T}g_{\mu\nu})=-\frac{1}{2} g_{\mu\nu}(\hat{T}g^{\mu\nu})
	\end{equation}
	thus, a similar expression holds for the ordinary derivative. Indeed, setting $\hat{T}\rightarrow \partial_{\mu}$ the latter takes the form
	\begin{equation}
	\frac{\partial_{\lambda}\sqrt{-g}}{\sqrt{-g}}=\partial_{\lambda}\ln{\sqrt{-g}}=\frac{1}{2} g^{\mu\nu} \partial_{\lambda}g_{\mu\nu}=-\frac{1}{2} g_{\mu\nu}\partial_{\lambda}g^{\mu\nu} \label{e}
	\end{equation}
	In addition, setting $\hat{T}\rightarrow \delta$ we also get the variation of the determinant
	\begin{equation}
	\frac{\delta \sqrt{-g}}{\sqrt{-g}}=\delta \ln{\sqrt{-g}}=\frac{1}{2} g^{\mu\nu}\delta g_{\mu\nu}=-\frac{1}{2} g_{\mu\nu}\delta g^{\mu\nu}
	\end{equation}
	Now, expanding (\ref{d}) and using (\ref{e}) it follows that
	\begin{gather}
	\frac{\nabla_{\lambda}\sqrt{-g}}{\sqrt{-g}}=\frac{1}{2} g^{\mu\nu}\nabla_{\lambda}g_{\mu\nu}=\underbrace{\frac{1}{2}g^{\mu\nu}\partial_{\lambda}g_{\mu\nu}}_{=\frac{\partial_{\lambda}\sqrt{-g}}{\sqrt{-g}}}-\frac{1}{2}\Big(\Gamma^{\alpha}_{\;\;\;\mu\lambda}g_{\nu\alpha}+\Gamma^{\alpha}_{\;\;\;\nu\lambda}g_{\mu\alpha}\Big)g^{\mu\nu}= \nonumber \\
	=\frac{\partial_{\lambda}\sqrt{-g}}{\sqrt{-g}}-\frac{1}{2}2\Gamma^{\mu}_{\;\;\;\mu\lambda} \Rightarrow \nonumber
	\end{gather}
	\begin{equation}
	\frac{\nabla_{\lambda}\sqrt{-g}}{\sqrt{-g}}=\frac{\partial_{\lambda}\sqrt{-g}}{\sqrt{-g}}-\Gamma^{\mu}_{\;\;\;\mu\lambda}  \nonumber
	\end{equation}
	\begin{equation}
	\nabla_{\lambda}\sqrt{-g}=\partial_{\lambda}\sqrt{-g}-\Gamma^{\mu}_{\;\;\;\mu\lambda}\sqrt{-g}  \nonumber
	\end{equation}
	and in terms of the Weyl vector
	\begin{equation}
	\Gamma^{\mu}_{\;\;\;\mu\lambda}=(\partial_{\lambda}-\nabla_{\lambda})\ln{\sqrt{-g}}=\partial_{\lambda}\ln{\sqrt{-g}}+\frac{1}{2}Q_{\lambda} \label{qv}
	\end{equation}
	Having this last relation we can bring the homothetic curvature tensor to the form
	\begin{gather}
	\hat{R}_{\mu\nu}:=\partial_{\mu}\Gamma^{\alpha}_{\;\;\;\alpha\nu}-\partial_{\nu}\Gamma^{\alpha}_{\;\;\;\alpha\mu}= \nonumber \\
	=\partial_{\mu}\partial_{\nu}(\ln{\sqrt{-g}})+\frac{1}{2}\partial_{\mu}Q_{\nu}-\partial_{\nu}\partial_{\mu}(\ln{\sqrt{-g}})-\frac{1}{2}\partial_{\nu}Q_{\mu}= \nonumber \\
	=\frac{1}{2}\Big(\partial_{\mu}Q_{\nu}-\partial_{\nu}Q_{\mu} \Big)=\partial_{[\mu}Q_{\nu]} \Rightarrow \nonumber
	\end{gather}
	\begin{equation}
	\hat{R}_{\mu\nu}=\partial_{[\mu}Q_{\nu]} 
	\end{equation}
	From this we see that the form of the homothetic curvature resembles that of the field strength of electromagnetism $F_{\mu\nu}=2\partial_{[\mu}A_{\nu]}$. Additionally, we conclude that non-metricity alone (the torsion does not enter at all) gives rise to the homothetic curvature. In particular the Weyl vector $Q_{\mu}$, constructed out of the non-metricity tensor, fully determines the homothetic curvature. Therefore, any attempt to brake the projective invariance of the Einstein-Hilbert action, by imposing $Q_{\mu}$ would automatically force the homothetic curvature to vanish as well. As a result, such a braking ($i.e.$ $Q_{\mu}=0$) should be avoided since it implies a vanishing homothetic curvature without any physical justification. Note that another place that the homothetic tensor arises naturally is when one takes the antisymmetrized covariant derivative of a scalar density (or more generally a tensor density). Indeed, considering the scalar density 
	\beq
	\Phi \equiv (\sqrt{-g})^{w}\phi
	\eeq
	of weight $w$, with $\phi$ being a scalar and acting the antisymmetrized covariant derivative on it we arrive at
	\beq
	[\nabla_{\mu},\nabla_{\nu}]\Phi=-w \hat{R}_{\mu\nu}\Phi +2 S_{\mu\nu}^{\;\;\;\;\lambda} \nabla_{\lambda}\Phi
	\eeq

	\subsection{The Levi-Civita Tensor}
	From the Levi-Civita symbol one can construct the Levi-Civita tensor. To see this we start by
	\begin{equation}
	\epsilon^{\mu\nu\rho\sigma}g_{\mu\alpha}g_{\nu\beta}g_{\rho\gamma}g_{\sigma\delta}=-g\epsilon_{\alpha\beta\gamma\delta} 
	\end{equation}
	and write it in the form
	\beq
	\frac{\epsilon^{\mu\nu\rho\sigma}}{\sqrt{-g}}g_{\mu\alpha}g_{\nu\beta}g_{\rho\gamma}g_{\sigma\delta}=\sqrt{-g}\epsilon_{\alpha\beta\gamma\delta} 
	\eeq
	From which we see that defining
	\beq
	\varepsilon_{\mu\nu\rho\sigma}:=\sqrt{-g}\epsilon_{\mu\nu\rho\sigma}
	\eeq
	and
	\beq
	\varepsilon^{\mu\nu\rho\sigma}:=\frac{\epsilon^{\mu\nu\rho\sigma}}{\sqrt{-g}}
	\eeq
	it follows that
	\beq
	\varepsilon^{\mu\nu\rho\sigma}g_{\mu\alpha}g_{\nu\beta}g_{\rho\gamma}g_{\sigma\delta}=\varepsilon_{\alpha\beta\gamma\delta} 
	\eeq
	which proves the tensorial nature of $\varepsilon_{\alpha\beta\gamma\delta}$. It is easy to check that this tensor also satisfies the set of identities ($\ref{eiden1}$) - ($\ref{eiden2}$) that hold for the Levi-Civita symbol, so for the Levi-Civita tensor one has\footnote{To see this just start with the identities for the Levi-Civita symbol, write $1=\frac{\sqrt{-g}}{\sqrt{-g}}$ and use the definition of the Levi-Civita tensor in its contravariant and covariant form.}

	\begin{equation}
	\varepsilon_{\mu\nu\rho\lambda}\varepsilon^{\mu\alpha\beta\gamma}=-1! 3! \delta_{[\nu}^{\alpha}\delta_{\rho}^{\beta}\delta_{\lambda]}^{\gamma} 
	\end{equation}
	\begin{equation}
	\varepsilon_{\mu\nu\rho\lambda}\varepsilon^{\mu\nu\kappa\sigma}=-2! 2! \delta_{[\rho}^{\kappa}\delta_{\lambda]}^{\sigma}
	\end{equation}
	\begin{equation}
	\varepsilon_{\mu\nu\rho\lambda}\varepsilon^{\mu\nu\rho\sigma}=-3!\delta_{\lambda}^{\sigma}
	\end{equation}
	\begin{equation}
	\varepsilon_{\mu\nu\rho\sigma}\varepsilon^{\mu\nu\rho\sigma}=-4!
	\end{equation}
	\begin{equation}
	\varepsilon_{\mu\nu\rho\lambda}\varepsilon^{\sigma\alpha\beta\gamma}=- 4! \delta_{[\mu}^{\sigma}\delta_{\nu}^{\alpha}\delta_{\rho}^{\beta}\delta_{\lambda]}^{\gamma} 
	\end{equation}

	\subsection{Covariant derivative of the Levi-Civita tensor}
	As it is well known an immediate implication of  $\nabla_{\alpha}g_{\mu\nu}=0$ is that the Levi-Civita tensor $\varepsilon_{\alpha\beta\gamma\delta}=\sqrt{-g}\epsilon_{\alpha\beta\gamma\delta}$,  (where $\epsilon_{\alpha\beta\gamma\delta}$ is the Levi-Civita symbol) is covariantly conserved. However, for a general non-vanishing non-metricity this statement is not true. In general it holds that
	\begin{equation}
	\nabla_{\mu}\varepsilon_{\alpha\beta\gamma\delta}=-\varepsilon_{\alpha\beta\gamma\delta}\frac{Q_{\mu}}{2}
	\end{equation}
	Let us prove this now. We have
	\begin{gather}
	\nabla_{\mu}\varepsilon_{\alpha\beta\gamma\delta}= \nonumber \\
	=\partial_{\mu}(\sqrt{-g}\epsilon_{\alpha\beta\gamma\delta})-\sqrt{-g}\Big(\Gamma^{\lambda}_{\;\;\;\alpha\mu}\epsilon_{\lambda\beta\gamma\delta}+\Gamma^{\lambda}_{\;\;\;\beta\mu}\epsilon_{\alpha\lambda\gamma\delta}+\Gamma^{\lambda}_{\;\;\;\gamma\mu}\epsilon_{\alpha\beta\lambda\delta}+\Gamma^{\lambda}_{\;\;\;\delta\mu}\epsilon_{\alpha\beta\gamma\lambda} \Big) \nonumber
	\end{gather}
	Now, if any of $\alpha,\beta,\gamma,\delta$ are equal this is zero, so we need only consider the case $\alpha=0,\beta=1,\gamma=2,\delta=3$ (any other possibility follows from circular permutations of it). Then
	\begin{gather}
	\nabla_{\mu}\varepsilon_{0123}=  \nonumber \\
	=(\partial_{\mu}\sqrt{-g})\epsilon_{0123}-\sqrt{-g}\Big( \Gamma^{\lambda}_{\;\;\;0\mu}\epsilon_{\lambda 123}+\Gamma^{\lambda}_{\;\;\;1\mu}\epsilon_{0\lambda 23}+\Gamma^{\lambda}_{\;\;\;2\mu}\epsilon_{01\lambda 3}+\Gamma^{\lambda}_{\;\;\;3\mu}\epsilon_{123\lambda} \Big) = \nonumber \\
	=(\partial_{\mu}\sqrt{-g})\epsilon_{0123}-\sqrt{-g}\epsilon_{0123}\Big( \Gamma^{0}_{\;\;\;0\mu}+ \Gamma^{1}_{\;\;\;1\mu} +\Gamma^{2}_{\;\;\;2\mu}+\Gamma^{3}_{\;\;\;3\mu} \Big)= \nonumber \\
	=(\partial_{\mu}\sqrt{-g})\epsilon_{0123}-\sqrt{-g}\epsilon_{0123}\Gamma^{\lambda}_{\;\;\;\lambda\mu} \Rightarrow  \nonumber
	\end{gather}
	\begin{equation}
	\nabla_{\mu}\varepsilon_{0123}=-\underbrace{\epsilon_{0123}\sqrt{-g}}_{\equiv \varepsilon_{0123}}\Big( \Gamma^{\lambda}_{\;\;\;\lambda\mu}-\frac{1}{\sqrt{-g}}\partial_{\mu}\sqrt{-g} \Big)
	\end{equation}
	In addition, using ($\ref{qv}$) the latter finally takes the form
	\begin{equation}
	\nabla_{\mu}\varepsilon_{0123}=-\varepsilon_{0123}\frac{Q_{\mu}}{2}
	\end{equation}
	which for general indices generalizes to
	\begin{equation}
	\nabla_{\mu}\varepsilon_{\alpha\beta\gamma\delta}=-\varepsilon_{\alpha\beta\gamma\delta}\frac{Q_{\mu}}{2}
	\end{equation}
	and indeed we see that when non-metricity is there ($Q_{\mu}\neq 0$) the Levi-Civita tensor is not covariantly conserved. However, when the theory is invariant under projective transformations of the connection, one can always use this freedom to define a volume-preserving connection, call it $^{\dagger }\Gamma^{\lambda}_{\;\;\;\mu\nu}$, for which 
	\begin{equation}
	^{\dagger }\nabla_{\mu}\varepsilon_{\alpha\beta\gamma\delta}=0
	\end{equation}
	Indeed, suppose that we have an affine connection $\Gamma^{\lambda}_{\;\;\;\mu\nu}$ for which $\nabla_{\mu}\varepsilon_{\alpha\beta\gamma\delta}\neq 0$. Then, consider the projective transformation
	\begin{equation}
	\Gamma^{\lambda}_{\;\;\;\mu\nu}\longrightarrow ^{\dagger }\Gamma^{\lambda}_{\;\;\;\mu\nu}=\Gamma^{\lambda}_{\;\;\;\mu\nu}+\delta_{\mu}^{\lambda}\xi_{\nu}
	\end{equation}
	For the daggered connection the covariant derivative on $\varepsilon_{\alpha\beta\gamma\delta}$ yields
	\begin{equation}
	^{\dagger }\nabla_{\mu}\varepsilon_{\alpha\beta\gamma\delta}=-\varepsilon_{\alpha\beta\gamma\delta}\Big( \Gamma^{\lambda}_{\;\;\;\lambda\mu}+4\xi_{\mu} -\frac{1}{\sqrt{-g}}\partial_{\mu}\sqrt{-g} \Big)= \nonumber \\
	=-\varepsilon_{\alpha\beta\gamma\delta}\Big( \frac{Q_{\mu}}{2} +4\xi_{\mu} \Big)
	\end{equation}
	and we see that if we choose  $\xi_{\mu}=-\frac{1}{8}Q_{\mu}$\footnote{For a general $n-dim$ spacetime the choice is $\xi_{\mu}=-Q_{\mu}/2n$} we have
	\begin{equation}
	^{\dagger }\nabla_{\mu}\varepsilon_{\alpha\beta\gamma\delta}=0
	\end{equation}
	Such a connection, namely one that goes like  
	\begin{equation}
	^{\dagger }\Gamma^{\lambda}_{\;\;\;\mu\nu}= \Gamma^{\lambda}_{\;\;\;\mu\nu}-\frac{1}{8}\delta_{\mu}^{\lambda}Q_{\nu}
	\end{equation}
	is called a volume preserving connection.

	\subsection{Variations of the Torsion tensor}
	Let us now derive the variations for the torsion tensor $(S_{\mu\nu}^{\;\;\;\;\alpha})$ and torsion vector $(S_{\mu}\equiv S_{\mu\alpha}^{\;\;\;\;\alpha})$ since we will be using them in the various theories we are going to study. Firstly, note that since the torsion does not depend on the metric, the $\delta g^{\mu\nu}$ variation is identically zero, namely
	\begin{equation}
	\delta_{g}S_{\mu\nu}^{\;\;\;\;\alpha}=\frac{\delta S_{\mu\nu}^{\;\;\;\;\alpha}}{\delta g^{\kappa\lambda}}\delta g^{\kappa\lambda}=0
	\end{equation}
	as well as\footnote{This is so because in order to form the torsion vector $S_{\mu}$ we need only contract an upper with a lower index without the use of any metric. Notice also that if we were to form another vector by contracting the first two indices of the torsion with the metric tensor, the result would yield zero due to the fact that the torsion is antisymmetric in its first two indices while the metric tensor is symmetric. In words, $\tilde{S}^{\mu}\equiv g^{\alpha\beta}S_{\alpha\beta}^{\;\;\;\;\mu}=0$. }
	\begin{equation}
	\delta_{g}S_{\mu}=0
	\end{equation}
	Now to proceed with the $\Gamma$-variation we recall that we want to have a common factor $\delta \Gamma^{\lambda}_{\;\;\;\mu\nu}$ appearing in the variation. Thus, we express the torsion tensor as
	\begin{gather}
	S_{\alpha\beta}^{\;\;\;\;\lambda}=\frac{1}{2}(\Gamma^{\lambda}_{\;\;\;\alpha\beta}-\Gamma^{\lambda}_{\;\;\;\beta\alpha})=\frac{1}{2}(\delta_{\alpha}^{\mu}\delta_{\beta}^{\nu}\Gamma^{\lambda}_{\;\;\;\mu\nu}-\delta_{\alpha}^{\nu}\delta_{\beta}^{\mu}\Gamma^{\lambda}_{\;\;\;\mu\nu})= \nonumber \\
	=\frac{1}{2}(\delta_{\alpha}^{\mu}\delta_{\beta}^{\nu}-\delta_{\alpha}^{\nu}\delta_{\beta}^{\mu})\Gamma^{\lambda}_{\;\;\;\mu\nu}=\delta_{\alpha}^{[\mu}\delta_{\beta}^{\nu]}\Gamma^{\lambda}_{\;\;\;\mu\nu} \Rightarrow  \nonumber
	\end{gather}
	\begin{equation}
	S_{\alpha\beta}^{\;\;\;\;\lambda}=\delta_{\alpha}^{[\mu}\delta_{\beta}^{\nu]}\Gamma^{\lambda}_{\;\;\;\mu\nu}
	\end{equation}
	such that
	\begin{equation}
	\delta_{\Gamma}S_{\alpha\beta}^{\;\;\;\;\lambda}=\delta_{\alpha}^{[\mu}\delta_{\beta}^{\nu]}\delta\Gamma^{\lambda}_{\;\;\;\mu\nu}
	\end{equation}
	So long as the torsion vector is concerned we contract the above in $\beta,\lambda$ to obtain
	\begin{equation}
	\delta_{\Gamma}S_{\alpha}=\delta_{\Gamma}S_{\alpha\lambda}^{\;\;\;\;\lambda}=\delta_{\alpha}^{[\mu}\delta_{\lambda}^{\nu]}\delta\Gamma^{\lambda}_{\;\;\;\mu\nu}
	\end{equation}
	and for the torsion pseudo-vector (in $4$-dim)
	\beq
	\delta_{\Gamma}\tilde{S}^{\alpha}=\epsilon^{\alpha\mu\nu}_{\;\;\;\;\;\; \lambda}\delta \Gamma^{\lambda}_{\;\;\;\mu\nu}
	\eeq
	Having performed the variations of the torsion, we know proceed to derive the variations of the non-metricity tensor with respect to both the metric tensor and the connection.

	\subsection{Variations of the Non-metricity tensor}
	Let us firstly obtain the variation of the non-metricity tensor with respect to the connection. To do so we single out a common $\Gamma^{\lambda}_{\;\;\;\mu\nu}$-factor in the expression of the non-metricity as we did with the torsion. We have
	\begin{gather}
	Q_{\rho\alpha\beta}=-\nabla_{\rho}g_{\alpha\beta}=-\partial_{\rho}g_{\alpha\beta}+\Gamma^{\lambda}_{\;\;\;\alpha\rho}g_{\lambda\beta}+\Gamma^{\lambda}_{\;\;\;\beta\rho}g_{\lambda\alpha}= \nonumber \\
	=-\partial_{\rho}g_{\alpha\beta}+\delta_{\alpha}^{\mu}\delta_{\rho}^{\nu}\Gamma^{\lambda}_{\;\;\;\mu\nu}g_{\lambda\beta}+\delta_{\beta}^{\mu}\delta_{\rho}^{\nu}\Gamma^{\lambda}_{\;\;\;\mu\nu}g_{\lambda\alpha}= \nonumber \\
	=-\partial_{\rho}g_{\alpha\beta}+\delta_{\rho}^{\nu}(\delta_{\alpha}^{\mu}g_{\lambda\beta}+\delta_{\beta}^{\mu}g_{\lambda\alpha})\Gamma^{\lambda}_{\;\;\;\mu\nu}\Rightarrow  \nonumber
	\end{gather}
	\begin{equation}
	Q_{\rho\alpha\beta}=-\partial_{\rho}g_{\alpha\beta}+\delta_{\rho}^{\nu}2\delta_{(\alpha}^{\mu} g_{\beta )\lambda}\Gamma^{\lambda}_{\;\;\;\mu\nu}
	\end{equation}
	Therefore, variation with respect to the connection, immediately gives
	\begin{equation}
	\delta_{\Gamma}Q_{\rho\alpha\beta}=\delta_{\rho}^{\nu}2\delta_{(\alpha}^{\mu} g_{\beta )\lambda}\delta\Gamma^{\lambda}_{\;\;\;\mu\nu} \label{nonqt}
	\end{equation}
	Let us now vary with respect to the metric tensor. Using the above definition of non-metricity along with the identity
	\begin{equation}
	\delta g_{\alpha\beta}=-g_{\mu\alpha}g_{\nu\beta}\delta g^{\mu\nu}
	\end{equation}
	it follows that
	\begin{gather}
	\delta_{g} Q_{\rho\alpha\beta}=-\partial_{\rho}\delta g_{\alpha\beta}+\Gamma^{\lambda}_{\;\;\;\alpha\rho}\delta g_{\lambda\beta}+\Gamma^{\lambda}_{\;\;\;\beta\rho}\delta g_{\lambda\alpha} = \nonumber \\
	=\partial_{\rho}(g_{\mu\alpha}g_{\nu\beta}\delta g^{\mu\nu})-\Gamma^{\lambda}_{\;\;\;\alpha\rho}g_{\lambda\mu}g_{\nu\beta}\delta g^{\mu\nu}-\Gamma^{\lambda}_{\;\;\;\beta\rho}g_{\lambda\mu}g_{\nu\alpha}\delta g^{\mu\nu}= \nonumber \\
	=\partial_{\rho}(g_{\mu\alpha}g_{\nu\beta}\delta g^{\mu\nu})-(\delta g^{\mu\nu})g_{\lambda\mu}2 g_{\nu (\alpha}\Gamma^{\lambda}_{\;\;\;\beta)\rho} \nonumber
	\end{gather}
	Thus, one has
	\begin{equation}
	\delta_{g} Q_{\rho\alpha\beta}=\partial_{\rho}(g_{\mu\alpha}g_{\nu\beta}\delta g^{\mu\nu})-(\delta g^{\mu\nu})2 g_{\lambda\mu} g_{\nu (\alpha}\Gamma^{\lambda}_{\;\;\;\beta)\rho} 
	\end{equation}
	We continue by varying the Weyl vector
	\begin{equation}
	Q_{\nu}\equiv -g^{\alpha\beta}\nabla_{\nu}g_{\alpha\beta}=-g^{\alpha\beta}\partial_{\nu}g_{\alpha\beta}+2 \Gamma^{\lambda}_{\;\;\;\lambda\nu}
	\end{equation}
	Variation with respect to the connection yields\footnote{This may also be obtained by contracting ($\ref{nonqt}$) with $g^{\alpha\beta}$. Of course, this can be done because the $\Gamma$-variation commutes with the metric tensor. However, this is not true for the $g$-variation.}
	\begin{equation}
	\delta_{\Gamma}Q_{\nu}=2\delta \Gamma^{\lambda}_{\;\;\;\lambda\nu}=\delta \Gamma^{\lambda}_{\;\;\;\mu\nu}2\delta_{\lambda}^{\mu} \Rightarrow\nonumber
	\end{equation}
	\beq
	\delta_{\Gamma}Q_{\rho}=2 \delta_{\rho}^{\nu}\delta_{\lambda}^{\mu}\delta \Gamma^{\lambda}_{\;\;\;\mu\nu}
	\eeq
	While variation with respect to the metric tensor gives
	\begin{equation}
	\delta_{g}Q_{\rho}=-(\delta g^{\mu\nu})\partial_{\rho}g_{\mu\nu}-g^{\alpha\beta}\partial_{\rho}\delta g_{\alpha\beta}
	\end{equation}
	Now, expanding the second term, we have
	\begin{gather}
	g^{\alpha\beta}\partial_{\rho}\delta g_{\alpha\beta}=-g^{\alpha\beta}\partial_{\rho}(g_{\mu\alpha}g_{\nu\beta}\delta g^{\mu\nu})= \nonumber \\
	=-g_{\mu\nu}\partial_{\rho}\delta g^{\mu\nu}-2(\delta g^{\mu\nu})\partial_{\rho}g_{\mu\nu}
	\end{gather}
	such that
	\begin{gather}
	\delta_{g}Q_{\rho}=-(\delta g^{\mu\nu})\partial_{\rho}g_{\mu\nu}+g_{\mu\nu}\partial_{\rho}\delta g^{\mu\nu}+2(\delta g^{\mu\nu})\partial_{\rho}g_{\mu\nu}= \nonumber \\
	=g_{\mu\nu}\partial_{\rho}\delta g^{\mu\nu}+(\delta g^{\mu\nu})\partial_{\rho}g_{\mu\nu}=\partial_{\rho}(g_{\mu\nu}\delta g^{\mu\nu})   \nonumber 
	\end{gather}
	Thus, the $g$-variation of the Weyl vector has the handy form
	\begin{equation}
	\delta_{g}Q_{\rho}=\partial_{\rho}(g_{\mu\nu}\delta g^{\mu\nu})  
	\end{equation}
	Let us now proceed by varying the second non-metricity vector $2nmv$. Recall that the latter is given by
	\begin{equation}
	\tilde{Q}_{\beta}=g^{\rho\alpha}Q_{\rho\alpha\beta=}=-g^{\rho\alpha}\partial_{\rho}g_{\alpha\beta}+(g^{\mu\nu}g_{\beta\lambda}+\delta_{\beta}^{\mu}\delta_{\lambda}^{\nu})\Gamma^{\lambda}_{\;\;\;\mu\nu}
	\end{equation}
	Variation with respect to the connection immediately gives
	\begin{equation}
	\delta_{\Gamma}\tilde{Q}_{\beta}=(g^{\mu\nu}g_{\beta\lambda}+\delta_{\beta}^{\mu}\delta_{\lambda}^{\nu})\delta \Gamma^{\lambda}_{\;\;\;\mu\nu}
	\end{equation}
	while variation with respect to the metric tensor reads
	\begin{gather}
	\delta_{g}\tilde{Q}_{\beta}=-(\delta g^{\mu\nu})\partial_{\mu}g_{\nu\beta}-g^{\rho\alpha}\partial_{\rho}\delta g_{\alpha\beta}+(\delta g^{\mu\nu})g_{\beta\lambda}\Gamma^{\lambda}_{\;\;\;\mu\nu}+g^{\mu\nu}\Gamma^{\lambda}_{\;\;\;\mu\nu}\delta g_{\beta\lambda}= \nonumber \\
	=\delta g^{\mu\nu}\Big[ -\partial_{\mu}g_{\nu\beta} +g_{\lambda\beta}\Gamma^{\lambda}_{\;\;\;\mu\nu}\Big] -g^{\rho\alpha}\partial_{\rho}\delta g_{\alpha\beta}+g^{\mu\nu}\Gamma^{\lambda}_{\;\;\;\mu\nu}\delta g_{\beta\lambda}
	\end{gather}
	Now using
	\begin{equation}
	\delta g_{\alpha\beta}=-g_{\alpha\mu}g_{\beta\nu}\delta g^{\mu\nu}
	\end{equation}
	it can easily be shown that
	\begin{equation}
	g^{\rho\alpha}\partial_{\rho}\delta g_{\alpha\beta}=-g_{\beta\nu}g^{\rho\alpha}(\partial_{\rho}g_{\mu\alpha})\delta g^{\mu\nu}-\partial_{\mu}(g_{\nu\beta}\delta g^{\mu\nu})
	\end{equation}
	as well as
	\begin{equation}
	g^{\mu\nu}\Gamma^{\lambda}_{\;\;\;\mu\nu}\delta g_{\beta\lambda}=-g^{\rho\sigma}\Gamma^{\alpha}_{\;\;\;\rho\sigma}g_{\mu\alpha}g_{\nu\beta}\delta g^{\mu\nu}
	\end{equation}
	and upon using these, the g-variation of $\tilde{Q}_{\beta}$ reads
	\begin{equation}
	\delta_{g} \tilde{Q}_{\beta}=\delta g^{\mu\nu}\Big[ g_{\nu\beta}g^{\rho\alpha}(\partial_{\rho}g_{\mu\alpha})+\Gamma^{\lambda}_{\;\;\;\mu\nu}g_{\lambda\beta}-g^{\rho\sigma}\Gamma^{\alpha}_{\;\;\;\rho\sigma}g_{\mu\alpha}g_{\nu\beta}\Big]+g_{\nu\beta}(\partial_{\mu}\delta g^{\mu\nu})
	\end{equation}
	Notice that there is a quicker and more elegant way to derive the $g-$variation of non-metricity. This comes about by first recalling that the general covariant derivative $\nabla_{\alpha}$ does not depend on the metric tensor. Then, using the definition of the variation, one has
	\beq
	\delta_{g}Q_{\alpha\mu\nu}=-\nabla_{\alpha}(g_{\mu\nu}+\delta g_{\mu\nu})+\nabla_{\alpha}g_{\mu\nu}=-\nabla_{\alpha}\delta g_{\mu\nu}
	\eeq
	and also
	\beq
	\delta_{g}Q_{\alpha}^{\;\;\mu\nu}=\nabla_{\alpha}(g^{\mu\nu}+\delta g^{\mu\nu})-\nabla_{\alpha}g_{\mu\nu}=+\nabla_{\alpha}\delta g^{\mu\nu}
	\eeq
	So, when coupled to a tensor filed (or a tensor density) $T^{\alpha}_{\;\;\mu\nu}$ we have
	\beq
	T^{\alpha}_{\;\;\mu\nu}\delta_{g}Q_{\alpha}^{\;\;\mu\nu}=\nabla_{\alpha}(T^{\alpha}_{\;\;\mu\nu}\delta g^{\mu\nu})-(\delta g^{\mu\nu})\nabla_{\alpha}T^{\alpha}_{\;\;\mu\nu}
	\eeq
	where we have employed Leibniz's rule for the covariant derivatives. Next we derive the variations of the Riemann tensor.

	\subsection{Variations of the Riemann tensor}
	For the sake of completeness we also give here the variations of the Riemann tensor (and its related contractions) with respect to the independent connection and the metric. First notice that the prototype of the Riemann tensor
	\begin{equation}
	R^{\mu}_{\;\;\;\nu\alpha\beta}:=2\partial_{[\alpha}\Gamma^{\mu}_{\;\;\;|\nu|\beta]}+2\Gamma^{\mu}_{\;\;\;\rho[\alpha}\Gamma^{\rho}_{\;\;\;|\nu|\beta]} \label{defriem}
	\end{equation}
	does not depend on the metric and therefore
	\beq
	\delta_{g} R^{\mu}_{\;\;\;\nu\alpha\beta}=0
	\eeq
	When the first index is brought down however we have a metric tensor dependence since
	\beq
	R_{\rho\nu\alpha\beta}=g_{\mu\rho}R^{\mu}_{\;\;\;\nu\alpha\beta}
	\eeq
	and thus
	\beq
	\delta_{g}R_{\rho\nu\alpha\beta}=(\delta g_{\mu\rho})R^{\mu}_{\;\;\;\nu\alpha\beta}=-(\delta g^{\kappa\lambda})g_{\mu\kappa}g_{\rho\lambda}R^{\mu}_{\;\;\;\nu\alpha\beta}=-(\delta g^{\kappa\lambda})g_{\rho\lambda}R_{\kappa\nu\alpha\beta}
	\eeq
	Now, to derive the variation with respect to the connection we start by $(\ref{defriem})$ and compute
	\beq
	\delta_{\Gamma}R^{\mu}_{\;\;\;\nu\alpha\beta}=R^{\mu}_{\;\;\;\nu\alpha\beta}[ \Gamma +\delta \Gamma]-R^{\mu}_{\;\;\;\nu\alpha\beta}[ \Gamma ]
	\eeq
	and expanding $R^{\mu}_{\;\;\;\nu\alpha\beta}[ \Gamma +\delta \Gamma]$ to linear order in $\delta \Gamma$ we finally arrive at
	\beq
	\delta_{\Gamma}R^{\mu}_{\;\;\;\nu\alpha\beta}=\nabla_{\alpha}(\delta \Gamma^{\mu}_{\;\;\;\nu\beta})-\nabla_{\beta}(\delta \Gamma^{\mu}_{\;\;\;\nu\alpha})-2 S_{\alpha\beta}^{\;\;\;\;\lambda}\delta\Gamma^{\mu}_{\;\;\;\nu\lambda}
	\eeq
	Having obtained all he necessary setup we are now in a position to study Metric-Affine Theories of Gravity. We do so in what follows.

	\chapter{Metric-Affine Theories of Gravity}
	
	This chapter deals with the general Model-building of MAG (in the coordinate formalism). After reviewing Einstein's theory in this formalism, we study  Palatini as well as Metric-Affine $f(R)$ theories and also present another way to brake the projective invariance in the aforementioned theories. We then go on and derive the field equations for more general theories.

	\section{Metric Affine f(R) Theories}
	Let us now study some characteristics of Metric Affine f(R) theories and spot any differences with their Metric counterpart. Firstly we consider the vacuum theories and then we add matter. Before considering the general $f(R)$ case we firstly consider the Einstein Hilbert action in the Metric-Affine Framework.

	\subsection{Einstein's Theory in the Metric-Affine Framework}
	We will show now that starting with the Einstein-Hilbert action and no matter fields, we end up with Einstein Gravity plus an additional unspecified vectorial degree of freedom that gives rise to both torsion and non-metricity but which can be eliminated by means of a projective transformation of the connection. This is possible because of the projective invariance of the Ricci tensor. However, this invariance is the very reason that renders the field equations problematic when one tries to add to the model a matter action that depends both on the metric and the connection. Then, one arrives at inconsistent field equations\footnote{This inconsistency arises due to the invariance of the Ricci scalar under projective transformations of the connection as we have already pointed out and is expressed as an unphysical constraint imposed on the matter fields. This is not an attribute only of the Einstein-Hilbert action, any action that is projective invariant will yield inconsistent field equations when matter is added.}. This inconsistency can be handled by fixing to zero the vector components of either the torsion or Weyl vectors but it seems that the situation suggests that in the MAG framework more general actions than the Einstein-Hilbert should be used. To this end we also present some actions that yield consistent field equations and in vacuum give Einstein Gravity with no additional fields.
	
	\subsubsection{Vacuum Einstein's Theory in MAG}
	Let us start with the Einstein-Hilbert action in $n$-dimensions
	\begin{equation}
	S_{EH}[g_{\mu\nu},\Gamma^{\lambda}_{\;\;\;\alpha\beta}]=\int d^{n}x\sqrt{-g}R=\int d^{n}x\sqrt{-g}g^{\mu\nu}R_{(\mu\nu)} \label{ei}
	\end{equation}
	and no matter fields. Here, no a priori relation between the metric tensor $g_{\mu\nu}$ and the connection $\Gamma^{\lambda}_{\;\;\;\alpha\beta}$ have been assumed and therefore we have not assumed any torsionless and metric compatibility of the connection to begin with. Varying ($\ref{ei}$) with respect to $g_{\mu\nu}$ and recalling that $R_{\mu\nu}$ is independent of the metric, we derive
	\begin{gather}
	\delta_{g}S_{EH}=0 \Rightarrow  \nonumber
	0=\int d^{n}x\sqrt{-g}\delta g^{\mu\nu}\Big[ R_{(\mu\nu)}-\frac{g_{\mu\nu}}{2}R\Big] \nonumber
	\end{gather}
	where we have used the identity
	\begin{equation}
	\delta_{g}\sqrt{-g}=-\frac{\sqrt{-g}}{2}g_{\mu\nu}\delta g^{\mu\nu}
	\end{equation}
	which is proved in the appendix. Now, since the latter must hold for any arbitrary variation $\delta g^{\mu\nu}$, we have
	\begin{equation}
	R_{(\mu\nu)}-\frac{g_{\mu\nu}}{2}R=0
	\end{equation}
	We should point out that at this point that we cannot identify the above as the Einstein equations yet since the torsionlessness and metric compatibility conditions have not been assumed. Now, using 
	\begin{equation}
	\delta_{\Gamma}R^{\mu}_{\;\;\;\nu\sigma\lambda}=\nabla_{\sigma}\delta \Gamma^{\mu}_{\;\;\;\nu\lambda}-\nabla_{\lambda}\delta \Gamma^{\mu}_{\;\;\;\nu\sigma}-2 S_{\sigma\lambda}^{\;\;\;\;\rho}\delta \Gamma^{\mu}_{\;\;\;\nu\rho}
	\end{equation}
	and varying  ($\ref{ei}$) with respect to the connection we get
	\begin{gather}
	\delta_{\Gamma}S_{EH}=0 \Rightarrow \\ \nonumber
	0=\int d^{n}x \delta\Gamma^{\lambda}_{\;\;\;\mu\nu}\Big[ -\nabla_{\lambda}(\sqrt{-g}g^{\mu\nu})+\nabla_{\sigma}(\sqrt{-g}g^{\mu\sigma})\delta^{\nu}_{\lambda} \\
	+2\sqrt{-g}(S_{\lambda}g^{\mu\nu}-S^{\mu}\delta_{\lambda}^{\nu}+g^{\mu\sigma}S_{\sigma\lambda}^{\;\;\;\;\nu})\Big]
	\end{gather}
	for this to hold true for any arbitrary variation $\delta\Gamma^{\lambda}_{\;\;\;\mu\nu}$ we must have
	\begin{equation}
	-\nabla_{\lambda}(\sqrt{-g}g^{\mu\nu})+\nabla_{\sigma}(\sqrt{-g}g^{\mu\sigma})\delta^{\nu}_{\lambda} \\
	+2\sqrt{-g}(S_{\lambda}g^{\mu\nu}-S^{\mu}\delta_{\lambda}^{\nu}+g^{\mu\sigma}S_{\sigma\lambda}^{\;\;\;\;\nu})=0 \label{nm}
	\end{equation}
	which is a relation that relates the metric tensor and the connection. It is common in the literature to denote the left hand side of the above equation (divided by $\sqrt{-g}$) as $P_{\lambda}^{\;\;\;\mu\nu}$ and call it the Palatini tensor. Namely,
	\beq
	P_{\lambda}^{\;\;\;\mu\nu}=-\frac{\nabla_{\lambda}(\sqrt{-g}g^{\mu\nu})}{\sqrt{-g}}+\frac{\nabla_{\sigma}(\sqrt{-g}g^{\mu\sigma})\delta^{\nu}_{\lambda}}{\sqrt{-g}} \\
	+2(S_{\lambda}g^{\mu\nu}-S^{\mu}\delta_{\lambda}^{\nu}+g^{\mu\sigma}S_{\sigma\lambda}^{\;\;\;\;\nu})
	\eeq
	Note that in the above case (Einstein-Hilbert action with no matter fields) the Palatini tensor vanishes identically. The Palatini tensor has only $n(n^{2}-1)$ instead of $n^{3}$ due to the fact that is traceless 
	\begin{equation}
	P_{\mu}^{\;\;\;\mu\nu}=0
	\end{equation}
	which is a general property and kills off $n$-equations\footnote{That is, in $4$-dim the Palatini tensor has $60$ components while the remaining $4$ components cannot be specified because of its traceless property.}. This implies that a vectorial degree of freedom is left unspecified and as a result the connection can only be determined up to a vector. More specifically, we state that equation ($\ref{nm}$) implies that the connection takes the following form
	\begin{equation}
	\Gamma^{\lambda}_{\;\;\;\mu\nu}= \tilde{\Gamma}^{\lambda}_{\;\;\;\mu\nu} -\frac{2}{(n-1)}S_{\nu}\delta_{\mu}^{\lambda}=\tilde{\Gamma}^{\lambda}_{\;\;\;\mu\nu}+\frac{1}{2 n}\delta_{\mu}^{\lambda}Q_{\nu}
	\end{equation}
	where $\tilde{\Gamma}^{\lambda}_{\;\;\;\mu\nu}$ is the Levi-Civita connection. To prove that, we start by contracting ($\ref{nm}$) in $\nu$ and $\lambda$ to get
	\begin{equation}
	(n-1)\nabla_{\sigma}(\sqrt{-g}g^{\mu\sigma})+2\sqrt{-g}(2-n)S^{\mu}=0 \Rightarrow \nonumber
	\end{equation}
	\begin{equation}
	S^{\mu}=\frac{(n-1)}{2(n-2)}\frac{\nabla_{\sigma}(\sqrt{-g}g^{\mu\sigma})}{\sqrt{-g}}
	\end{equation}
	or
	\begin{equation}
	\nabla_{\sigma}(\sqrt{-g}g^{\mu\sigma})=2\sqrt{-g}\left(\frac{n-2}{n-1}\right)S^{\mu} \label{expan}
	\end{equation}
	Substituting that very last equation back to ($\ref{nm}$) we obtain
	\begin{equation}
	-\nabla_{\lambda}(\sqrt{-g}g^{\mu\nu})
	+2\sqrt{-g}\left(S_{\lambda}g^{\mu\nu}+\frac{1}{1-n}S^{\mu}\delta_{\lambda}^{\nu}+g^{\mu\sigma}S_{\sigma\lambda}^{\;\;\;\;\nu}\right)=0 \label{nml}
	\end{equation}
	Playing a bit more, let us contract ($\ref{nml}$) by $g_{\mu\nu}$. We have
	\begin{gather}
	-g_{\mu\nu}\nabla_{\lambda}(\sqrt{-g}g^{\mu\nu})+2\sqrt{-g}\frac{n(n-2)}{(n-1)}S_{\lambda}=0 \Rightarrow  \nonumber \\
	-n\frac{\nabla_{\lambda}\sqrt{-g}}{\sqrt{-g}}-g_{\mu\nu}\nabla_{\lambda}g^{\mu\nu}+2\frac{n(n-2)}{(n-1)}S_{\lambda}=0
	\end{gather}
	Using the identity proved in the appendix
	\begin{equation}
	\frac{\nabla_{\lambda}\sqrt{-g}}{\sqrt{-g}}=\nabla_{\lambda}\ln{\sqrt{-g}}=\frac{1}{2} g^{\mu\nu}\nabla_{\lambda}g_{\mu\nu}=-\frac{1}{2} g_{\mu\nu}\nabla_{\lambda}g^{\mu\nu}=-\frac{1}{2}Q_{\lambda} \label{as}
	\end{equation}
	the latter recasts to
	\begin{equation}
	S_{\lambda}=-\frac{(n-1)}{4 n}Q_{\lambda} \label{sq}
	\end{equation}
	which relates the torsion and Weyl vectors. One can also relate the second non-metricity vector $\tilde{Q}^{\mu}=Q_{\sigma}^{\;\;\;\sigma\mu}=\nabla_{\sigma}g^{\sigma\mu}$ to $S^{\mu}$ and $Q^{\mu}$. To see this, we expand ($\ref{expan}$) and use ($\ref{as}$) to get
	\begin{gather}
	g^{\mu\sigma}\frac{\nabla_{\sigma}\sqrt{-g}}{\sqrt{-g}}+\nabla_{\sigma}g^{\sigma\mu}=2\frac{(n-2)}{(n-1)}S^{\mu} \Rightarrow \nonumber \\
	-\frac{1}{2}Q_{\sigma}g^{\mu\sigma}+ \tilde{Q}^{\mu}=2\frac{(n-2)}{(n-1)}S^{\mu}  \nonumber \\
	\end{gather}
	such that
	\begin{equation}
	\tilde{Q}^{\mu}=\frac{1}{2}Q^{\mu}+2\frac{(n-2)}{(n-1)}S^{\mu}
	\end{equation}
	Furthermore, using ($\ref{sq}$) we finally arrive at
	\begin{equation}
	\tilde{Q}^{\mu}=\frac{1}{n}Q^{\mu}=-\frac{4}{(n-1)}S^{\mu}
	\end{equation}
	Thus, all three vectors $S^{\mu},Q^{\mu}$ and $\tilde{Q}^{\mu}$ are related to one another. Going back to our proof now, we expand the first term in ($\ref{nml}$) and use equation ($\ref{as}$) along with the definition $Q_{\lambda}^{\;\;\;\mu\nu}\equiv +\nabla_{\lambda}g^{\mu\nu}$, to get
	\begin{equation}
	\frac{1}{2}g^{\mu\nu}Q_{\lambda}-Q_{\lambda}^{\;\;\;\mu\nu}+2 \left(S_{\lambda}g^{\mu\nu}+\frac{1}{1-n}S^{\mu}\delta_{\lambda}^{\nu}+g^{\mu\sigma}S_{\sigma\lambda}^{\;\;\;\;\nu}\right)=0
	\end{equation}
	Multiplying with $g^{\alpha\lambda}$ it follows that
	\begin{equation}
	\frac{1}{2}Q^{\alpha}g^{\mu\nu}-Q^{\alpha\mu\nu}+2\Big( g^{\mu\nu}S^{\alpha}+\frac{1}{1-n}S^{\mu}g^{\nu\alpha}\Big)+2 S^{\mu\alpha\nu}=0   \nonumber
	\end{equation}
	such that
	\begin{equation}
	Q^{\alpha\mu\nu}+2 S^{\alpha\mu\nu}=\frac{1}{2}g^{\mu\nu}Q^{\alpha}+2\Big( g^{\mu\nu}S^{\alpha}+\frac{1}{1-n}S^{\mu}g^{\nu\alpha}\Big) \label{qs}
	\end{equation}
	where the antisymmetry of $S^{\mu\alpha\nu}$ in $\mu,\alpha$ have been employed. Now we use the formula we had proved for the connection decomposition and try to pair the various terms in such a way as to be able to use the above equation. Recalling the decomposition,
	\begin{equation}
	\Gamma^{\lambda}_{\;\;\;\mu\nu}=\tilde{\Gamma}^{\lambda}_{\;\;\;\mu\nu}+\frac{1}{2}g^{\alpha\lambda}(Q_{\mu\nu\alpha}+Q_{\nu\alpha\mu}-Q_{\alpha\mu\nu}) -g^{\alpha\lambda}(S_{\alpha\mu\nu}+S_{\alpha\nu\mu}-S_{\mu\nu\alpha})
	\end{equation}
	we use the antisymmetry  $S_{\alpha\nu\mu}=-S_{\nu\alpha\mu}$ in order to re-express the latter as
	\begin{equation}
	\Gamma^{\lambda}_{\;\;\;\mu\nu}=\tilde{\Gamma}^{\lambda}_{\;\;\;\mu\nu}+\frac{1}{2}g^{\alpha\lambda}\Big[ -(Q_{\alpha\mu\nu}+2 S_{\alpha\mu\nu)}+(Q_{\mu\nu\alpha}+2 S_{\mu\nu\alpha})+(Q_{\nu\alpha\mu}+2 S_{\nu\alpha\mu})\Big]
	\end{equation}
	Now, multiplying ($\ref{qs}$) by $-1$ and adding the results obtained by successively permuting $\mu\rightarrow \nu$, $\nu\rightarrow \alpha$, $\alpha\rightarrow \mu$ we obtain
	\begin{gather}
	A_{\alpha\mu\nu}\equiv -(Q_{\alpha\mu\nu}+2 S_{\alpha\mu\nu)}+(Q_{\mu\nu\alpha}+2 S_{\mu\nu\alpha})+(Q_{\nu\alpha\mu}+2 S_{\nu\alpha\mu})= \nonumber \\
	=-\frac{1}{2}g_{\mu\nu}Q_{\alpha}-2\Big( g_{\mu\nu}S_{\alpha}+\frac{1}{1-n}S_{\mu}g_{\nu\alpha}\Big) \nonumber \\
	+\frac{1}{2}g_{\nu\alpha}Q_{\mu}-2\Big( g_{\nu\alpha}S_{\mu}+\frac{1}{1-n}S_{\nu}g_{\alpha\mu}\Big) \nonumber \\
	+\frac{1}{2}g_{\alpha\mu}Q_{\nu}-2\Big( g_{\alpha\mu}S_{\nu}+\frac{1}{1-n}S_{\alpha}g_{\mu\nu}\Big)
	\end{gather}
	Multiplying with $g^{\alpha\lambda}$ and grouping common terms we obtain 
	\begin{gather}
	g^{\alpha\lambda}A_{\alpha\mu\nu}=g^{\alpha\lambda}\Big[ -(Q_{\alpha\mu\nu}+2 S_{\alpha\mu\nu)}+(Q_{\mu\nu\alpha}+2 S_{\mu\nu\alpha})+(Q_{\nu\alpha\mu}+2 S_{\nu\alpha\mu})\Big]= \nonumber \\
	=-\frac{1}{2}g_{\mu\nu}\Big[ \underbrace{Q^{\lambda}+\frac{4 n}{n-1}S^{\lambda}}_{=0} \Big] + \delta_{(\mu}^{\lambda}Q_{\nu)}+\frac{2 n}{(n-1)}S_{\mu}\delta_{\nu}^{\lambda}+ \frac{2(n-2)}{(n-1)}S_{\nu}\delta_{\mu}^{\lambda}= \nonumber \\
	=\frac{1}{2}\delta_{\mu}^{\lambda}\Big[Q_{\nu}+\frac{4(n-2)}{(n-1)}S_{\nu} \Big]+\frac{1}{2}\delta_{\nu}^{\lambda}\Big[\underbrace{ Q_{\mu}+\frac{4 n}{(n-1)}S_{\mu}}_{=0} \Big]= \nonumber \\
	=\frac{1}{2}\delta_{\mu}^{\lambda}\Big[\underbrace{Q_{\nu}+\frac{4 n}{(n-1)}S_{\nu}}_{=0}-\frac{8}{(n-1)}S_{\nu} \Big] \Rightarrow \nonumber
	\end{gather}
	such that
	\begin{equation}
	g^{\alpha\lambda}A_{\alpha\mu\nu}=-\frac{4}{(n-1)}S_{\nu}\delta_{\mu}^{\lambda}=\frac{1}{n}Q_{\nu}\delta_{\mu}^{\lambda}
	\end{equation}
	where in all steps we have employed equation ($\ref{sq}$). It is worth noting that the coefficients in front of $g_{\mu\nu}$ and $\delta_{\nu}^{\lambda}$ are exactly equal to zero. Substituting this very last equation into the expression for the connection we complete the proof
	\begin{equation}
	\Gamma^{\lambda}_{\;\;\;\mu\nu}=\tilde{\Gamma}^{\lambda}_{\;\;\;\mu\nu}+\frac{1}{2}g^{\alpha\lambda}A_{\alpha\mu\nu} \Rightarrow  \nonumber
	\end{equation}
	
	\begin{equation}
	\Gamma^{\lambda}_{\;\;\;\mu\nu}= \tilde{\Gamma}^{\lambda}_{\;\;\;\mu\nu} -\frac{2}{(n-1)}S_{\nu}\delta_{\mu}^{\lambda}=\tilde{\Gamma}^{\lambda}_{\;\;\;\mu\nu}+\frac{1}{2 n}\delta_{\mu}^{\lambda}Q_{\nu} \label{g}
	\end{equation}
	Therefore, we conclude that indeed the connection is determined only up to an unspecified vectorial degree of freedom. This additional degree of freedom can be removed by means of a projective transformation of the connection
	\begin{equation}
	\Gamma^{\lambda}_{\;\;\;\mu\nu}\longrightarrow \Gamma^{\lambda}_{\;\;\;\mu\nu}+\delta_{\mu}^{\lambda}\xi_{\nu}
	\end{equation}
	if $\xi_{\nu}$ is chosen to be equal to -$Q_{\nu}/2n$. In addition, for connections of the form of ($\ref{g}$) only the Levi-Civita part contributes in both the Einstein-Hilbert action and Einstein's equations. Indeed, substituting ($\ref{g}$) in the definition of the Riemann tensor
	\begin{equation}
	R^{\mu}_{\;\;\;\nu\alpha\beta}:=2\partial_{[\alpha}\Gamma^{\mu}_{\;\;\;|\nu|\beta]}+2\Gamma^{\mu}_{\;\;\;\rho[\alpha}\Gamma^{\rho}_{\;\;\;|\nu|\beta]}
	\end{equation}
	It can easily be seen that
	\begin{equation}
	R^{\mu}_{\;\;\;\nu\alpha\beta}=\tilde{R}^{\mu}_{\;\;\;\nu\alpha\beta}+\frac{1}{n}\delta^{\mu}_{\nu}\partial_{[\alpha}Q_{\beta]}=\tilde{R}^{\mu}_{\;\;\;\nu\alpha\beta}+\frac{1}{n}\delta^{\mu}_{\nu}\hat{R}_{\alpha\beta}
	\end{equation}
	where $\tilde{R}^{\mu}_{\;\;\;\nu\alpha\beta}$ is the part of the Riemann tensor computed for the Levi-Civita connection, namely the Riemannian part while $\delta^{\mu}_{\nu}\partial_{[\alpha}Q_{\beta]}/n$ represents the non-Riemannian contribution. Subsequently, the Ricci tensor is given by
	\begin{equation}
	R_{\nu\beta}=\tilde{R}_{\nu\beta}+\frac{1}{n}\partial_{[\nu}Q_{\beta]}=\tilde{R}_{\nu\beta}+\frac{1}{n}\hat{R}_{\nu\beta}
	\end{equation}
	from which we conclude that its symmetric part (which is the one that contributes to Einstein equations\footnote{This is so because the Einstein Hilbert Lagrangian density is proportional to $R=g^{\mu\nu}R_{\mu\nu}=g^{\mu\nu}R_{(\mu\nu)}$ since the metric tensor is symmetric. As a result, the antisymmetric part of $R_{\mu\nu}$ gives no contribution to the equations of motion.}) is purely Riemannian
	\begin{equation}
	R_{(\nu\beta)}=\tilde{R}_{(\nu\beta)}=\tilde{R}_{\nu\beta}
	\end{equation}
	As a result
	\begin{equation}
	R=g^{\mu\nu}R_{\mu\nu}=g^{\mu\nu}\tilde{R}_{\mu\nu}
	\end{equation}
	and therefore the additional vectorial degree of freedom does not appear in the Einstein equations. Having solved exactly for the connection we can now compute the torsion and non-metricity tensors in closed form in terms of the unspecified torsion vector (or Weyl vector). Indeed, taking the antisymmetric part of ($\ref{g}$) we obtain for the torsion
	\begin{equation}
	S_{\mu\nu}^{\;\;\;\;\lambda}=\Gamma^{\lambda}_{\;\;\;[\mu\nu]}= \underbrace{\tilde{\Gamma}^{\lambda}_{\;\;\;[\mu\nu]}}_{=0} -\frac{2}{(n-1)}S_{[\nu}\delta_{\mu]}^{\lambda} \Rightarrow \nonumber
	\end{equation}
	\begin{equation}
	S_{\mu\nu}^{\;\;\;\;\lambda}=-\frac{2}{(n-1)}S_{[\nu}\delta_{\mu]}^{\lambda}=\frac{1}{n-1}\Big( S_{\mu}\delta_{\nu}^{\lambda}-S_{\nu}\delta_{\mu}^{\lambda} \Big)
	\end{equation}
	So long as the non-metricity tensor is concerned, by its definition we have
	\begin{gather}
	Q_{\alpha\mu\nu}=-\partial_{\alpha}g_{\mu\nu}+\Gamma^{\lambda}_{\;\;\;\mu\alpha}g_{\lambda\nu}+\Gamma^{\lambda}_{\;\;\;\nu\alpha}g_{\lambda\mu} = \nonumber \\
	=\underbrace{-\partial_{\alpha}g_{\mu\nu}+\tilde{\Gamma}^{\lambda}_{\;\;\;\mu\alpha}g_{\lambda\nu}+\tilde{\Gamma}^{\lambda}_{\;\;\;\nu\alpha}g_{\lambda\mu}}_{=0} +\frac{1}{2 n}(g_{\lambda\nu}\delta^{\lambda}_{\mu}Q_{\alpha}+g_{\lambda\mu}\delta^{\lambda}_{\nu}Q_{\alpha}) = \nonumber \\
	=\frac{1}{2 n}(g_{\mu\nu}Q_{\alpha}+g_{\nu\mu}Q_{\alpha})=\frac{1}{n}Q_{\alpha}g_{\mu\nu}  \nonumber
	\end{gather}
	where in the second line we used the fact that the non-metricity of the Levi-Civita connection is zero. Therefore,
	\begin{equation}
	Q_{\alpha\mu\nu}=\frac{1}{n}Q_{\alpha}g_{\mu\nu} 
	\end{equation}
	Thus we see that both the torsion and non-metricity are non vanishing and dependent on an unspecified vectorial degree of freedom. This is a consequence of the projective invariance of the Einstein-Hilbert action (which results in the tracelessness of the Palatini tensor $P_{\mu}^{\;\;\;\mu\nu}=0$). We conclude therefore that the Einstein-Hilbert action (without any matter fields) in the Metric-Affine framework does not reproduce Einstein's theory. What it gives is, Einstein field equations along with an additional vectorial degree of freedom that produces non-vanishing torsion and non-metricity. Then, one could ask which action, in Metric-Affine framework, does give Einstein equations in vacuum without any additional degree of freedom. We show in what follows that such an action is not unique and present a number of models consisting of such action. We will present this, in this chapter after studying Metric-Affine $f(R)$ Theories and projective invariance breaking by means of Lagrange multipliers. Let us therefore concentrate on $f(R)$ for the time being.

	\subsection{Vacuum f(R) Theories (Aka Palatini f(R))}
	Since we are in vacuum, our starting action will be
	\beq
	S=\frac{1}{2\kappa}\int d^{n}x \sqrt{-g}f(R) \label{ffrr}
	\eeq
	Varying with respect to the metric and using the principle of least action, we obtain
	\beq
	\delta_{g}S=\frac{1}{2\kappa}\int d^{n}x\sqrt{-g}\left[ f^{'}(R)R_{(\mu\nu)}-\frac{f(R)}{2}g_{\mu\nu} \right]=0 \Rightarrow  \nonumber
	\eeq
	\beq
	f^{'}(R)R_{(\mu\nu)}-\frac{f(R)}{2}g_{\mu\nu}=0 \label{efr}
	\eeq
	Now, using the fact that for a general tensor field (or tensor density) $B^{\mu\nu}$  it holds that\footnote{The proof can be found in the appendix.}
	\beq
	B^{\mu\nu}\delta_{\Gamma}R_{\mu\nu}=\delta \Gamma^{\lambda}_{\;\;\;\mu\nu}\Big( -\nabla_{\lambda}B^{\mu\nu}+\nabla_{\alpha}(B^{\mu\alpha}\delta_{\lambda}^{\nu})-2 B^{\mu\alpha}S_{\lambda\alpha}^{\;\;\;\;\nu}  \Big)+ A
	\eeq
	where 
	\beq
	A=\nabla_{\lambda}( B^{\mu\nu}\delta \Gamma^{\lambda}_{\;\;\;\mu\nu}-B^{\mu\lambda}\delta_{\alpha}^{\nu}\delta \Gamma^{\alpha}_{\;\;\;\mu\nu})
	\eeq
	we vary with respect to $\Gamma^{\alpha}_{\;\;\;\mu\nu}$, to get
	\beq
	-\nabla_{\lambda}(\sqrt{-g}f^{'}g^{\mu\nu})+\nabla_{\alpha}(\sqrt{-g}f^{'}g^{\mu\alpha}\delta_{\lambda}^{\nu})+\\ \nonumber
	2 \sqrt{-g}f^{'}(S_{\lambda}g^{\mu\nu}-S^{\mu}\delta_{\lambda}^{\nu}-  S_{\lambda}^{\;\;\;\mu\nu})=0 \label{eGf}
	\eeq
	Now we wish to solve the system of equations $(\ref{efr})$ and $(\ref{eGf})$. To do so, we first take the trace of $(\ref{efr})$ to arrive at
	\beq
	f^{'}(R)R-\frac{n}{2}f(R)=0 \label{rre}
	\eeq
	This is an algebraic equation on $R$ and it will have a number of solutions\footnote{When this equation has no solutions inconsistencies will arise as shown in $[]$.} $R=R_{\kappa}=c_{\kappa}=constant$, \; $\kappa=1,2,...,i$ where $i$ is the number of solutions. Notice that for the specific choice $f(R) \propto R^{n/2}$ the above is identically satisfied. We will study this case  separately and give its cosmological solutions (for our $n=4$ dim spacetime) in a next chapter. So, going back to our solutions, for $R=R_{\kappa}=c_{\kappa}=constant$ and using the latter equation, the field equations (\ref{efr}) take the form
	\beq
	R_{(\mu\nu)}-\frac{R_{\kappa}}{n}g_{\mu\nu}=0
	\eeq
	Also, since $f^{'}(R_{\kappa})$ is constant too, it can be pulled outside of the covariant derivative and ($\ref{eGf}$) becomes
	\beq
	f^{'}(R_{\kappa})\sqrt{-g} P_{\lambda}^{\;\;\;\mu\nu}=0
	\eeq
	where 
	\beq
	P_{\lambda}^{\;\;\;\mu\nu}=-\frac{\nabla_{\lambda}(\sqrt{-g}g^{\mu\nu})}{\sqrt{-g}}+\frac{\nabla_{\sigma}(\sqrt{-g}g^{\mu\sigma})\delta^{\nu}_{\lambda}}{\sqrt{-g}} \\
	+2(S_{\lambda}g^{\mu\nu}-S^{\mu}\delta_{\lambda}^{\nu}+g^{\mu\sigma}S_{\sigma\lambda}^{\;\;\;\;\nu})
	\eeq
	is the Palatini tensor which we had defined earlier. This last equation implies
	\beq
	P_{\lambda}^{\;\;\;\mu\nu}=0
	\eeq
	which in turn, as we have shown, says that the geometry is Riemannian but with an undetermined vectorial degree of freedom. More specifically, as we showed in the previous chapter, the vanishing of the Palatini tensor implies that
	\beq
	R_{(\mu\nu)}=\tilde{R}_{\mu\nu}\;,\; R=\tilde{R}
	\eeq
	and our field equations reduce to
	\beq
	\tilde{R}_{\mu\nu}-\frac{c_{\kappa}}{n}g_{\mu\nu}=0
	\eeq
	The last equation, is Einstein equation with a cosmological constant. In fact, this is GR with a whole set of Cosmological constants, for each solution $R=R_{\kappa}$ we pick we have a different theory with a Cosmological constant $\Lambda_{\kappa}=\frac{C_{\kappa}}{n}$. For a good discussion on this feature see also  
	\cite{ferraris1994universality}. So, this is an interesting result especially when compared with metric $f(R)$ theories of Gravity in vacuum. In  metric $f(R)$ theories in vacuum the field equations are of forth order and of course they are different from Einstein equations. On the other hand, Metric-Affine $f(R)$ theories in vacuum, are equivalent  to a class of Einstein Gravities, with different Cosmological constants which are solutions of ($\ref{rre}$) and each solution gives a different value for the Cosmological constant. In fact, we have $i$-different theories, where $i$ is the number of solutions of ($\ref{rre}$). One important point take home though, is that in each of these there is an undetermined vectorial degree of freedom which does not interfere with Einstein equations at this point but nevertheless it is there, and will cause inconsistence theories when matter is added as we will see later.

	\subsection{Metric Affine f(R) Theories With Matter}
	Let us now try to add a matter term to the gravity action  ($\ref{ffrr}$) and derive the field equations for Metric Affine theories with matter. Note that this matter action can depend both on the metric tensor and the connection $S_{M}=S_{M}[g_{\alpha\beta},\Gamma^{\lambda}_{\;\;\;\mu\nu}]$ and its variation with respect to the metric tensor defines as usual the energy-momentum tensor while the variation with respect to the connection gives the hypermomentum tensor. So, our full action will be
	\beq
	S=S_{G}+S_{M}=\frac{1}{2\kappa}\int d^{n}x \sqrt{-g}f(R) +\int d^{n}x \sqrt{-g} \mathcal{L}_{M}
	\eeq
	Varying the above with respect to the metric tensor, we obtain
	\beq
	f^{'}(R)R_{(\mu\nu)}-\frac{f(R)}{2}g_{\mu\nu}=\kappa T_{\mu\nu}
	\eeq
	where 
	\beq
	T_{\mu\nu} \equiv -\frac{2}{\sqrt{-g}}\frac{\delta S_{M}}{\delta g^{\mu\nu}}
	\eeq
	the usual energy-momentum (or stress-energy) tensor. Variation with respect to the independent connection gives
	\beq
	-\frac{\nabla_{\lambda}(\sqrt{-g}f^{'}g^{\mu\nu})}{\sqrt{-g}}+\frac{\nabla_{\alpha}(\sqrt{-g}f^{'}g^{\mu\alpha}\delta_{\lambda}^{\nu})}{\sqrt{-g}}+\\ \nonumber
	2 f^{'}(S_{\lambda}g^{\mu\nu}-S^{\mu}\delta_{\lambda}^{\nu}-  S_{\lambda}^{\;\;\;\mu\nu})=\kappa \Delta_{\lambda}^{\;\;\;\mu\nu} \label{frmt}
	\eeq
	where
	\beq
	\Delta_{\lambda}^{\;\;\;\mu\nu} \equiv -\frac{2}{\sqrt{-g}}\frac{\delta S_{M}}{\delta \Gamma^{\lambda}_{\;\;\;\mu\nu}}
	\eeq
	is the hypermomentum tensor which gives information of the spin, shear and dilation of matter. Notice now that the left hand side of ($\ref{frmt}$) is the Palatini tensor computed for the modified tensor\footnote{This is just a mathematical convenience, $h_{\mu\nu}$ has no physical significance.}
	\beq
	h_{\mu\nu}=f^{'}(R)g_{\mu\nu}
	\eeq
	With this observation, we may write
	\beq
	P_{\lambda}^{\;\;\;\mu\nu}(h)=\kappa\Delta_{\lambda}^{\;\;\;\mu\nu} 
	\eeq
	where 
	\begin{gather}
	P_{\lambda}^{\;\;\;\mu\nu}(h) \equiv -\frac{\nabla_{\lambda}(\sqrt{-g}f^{'}g^{\mu\nu})}{\sqrt{-g}}+\frac{\nabla_{\alpha}(\sqrt{-g}f^{'}g^{\mu\alpha}\delta_{\lambda}^{\nu})}{\sqrt{-g}}+ \\ \nonumber
	2 f^{'}(S_{\lambda}g^{\mu\nu}-S^{\mu}\delta_{\lambda}^{\nu}-  S_{\lambda}^{\;\;\;\mu\nu}) 
	\end{gather}
	and by applying the product rule for the covariant derivatives we find
	\beq
	P_{\lambda}^{\;\;\;\mu\nu}(h)=f^{'}P_{\lambda}^{\;\;\;\mu\nu}(g)+\delta_{\lambda}^{\nu}g^{\mu\alpha}\partial_{\alpha}f^{'}-g^{\mu\nu}\partial_{\lambda}f^{'}
	\eeq
	where $P_{\lambda}^{\;\;\;\mu\nu}(g)$ is the usual Palatini tensor computed with respect to the metric tensor $g_{\mu\nu}$.
	Now, as we have already  seen the Palatini tensor has zero trace when contracted in its two fist indices, that is\footnote{This is true irrespective of the metric used since $g_{\mu\nu}$ and $h_{\mu\nu}$ are conformally related.}
	\beq
	P_{\mu}^{\;\;\;\mu\nu}=0
	\eeq
	this is so because of the projective invariance of  the Ricci scalar $R$, and the above holds as an identity. This enforces
	\beq
	\Delta_{\mu}^{\;\;\;\mu\nu} =0
	\eeq
	and this, obviously, cannot be correct for any form of matter. We can find many examples of matter for which $\Delta_{\mu}^{\;\;\;\mu\nu} \neq 0$. For instance, suppose that we have a vector field $A_{\mu}$ whose matter action contains a term that goes like
	\beq
	S_{M}[g_{\alpha\beta},\Gamma^{\lambda}_{\;\;\;\mu\nu}]=-\frac{1}{4}\int d^{n}x \sqrt{-g} g^{\mu\alpha}g^{\nu\beta}(\nabla_{\mu}A_{\nu})(\nabla_{\alpha}A_{\beta})
	\eeq 
	The associated hypermomentum in this case, will be
	\beq
	\Delta_{\lambda}^{\;\;\;\mu\nu}=A_{\lambda}g^{\mu\alpha}g^{\nu\beta}(\nabla_{\beta}A_{\alpha})
	\eeq
	and therefore
	\beq
	\Delta_{\mu}^{\;\;\;\mu\nu}=A^{\alpha}(\nabla_{\beta}A_{\alpha})g^{\beta\nu} \neq 0
	\eeq
	So, we see that when one tries to add matter to Metric Affine $f(R)$ Gravities inconsistency\footnote{Inconsistency may be too strong a word here. As pointed out in \cite{jimenez2018teleparallel} these constraints on the matter fields, like eq.$(\ref{conseqw})$, are perfectly fine even desirable in some cases (see also \cite{jimenez2017born} for a similar discussion). In addition all standard matter fields, both bosonic and
		fermionic, respect the projective symmetry so no consistency
		problem arises. So, whether projective invariance should be broken or not is an interesting open subject. However, its discussion goes beyond the scope of this paper. In these notes we just present an another way to break the invariance given that one wants to break it.} arises due to the projective invariance of the Ricci scalar (and of course any function-$f(R)$ of it will respect this invariance too). To obtain a self-consistent theory one needs to somehow break this projective invariance by fixing a vectorial degree of freedom. This can be done by adding extra terms in the action that do not respect the projective invariance, but this is somewhat arbitrary. What seems more natural to do is to fix either the torsion  or non-metricity vectors to zero by means of a Lagrange multiplier added to the matter action. In \cite{1981GReGr..13.1037H,Lord2004MetricAftM} they fixed the Weyl vector $Q_{\mu}$ to zero\footnote{A similar way of breaking the projective invariance was also presented in \cite{smalley1979volume}.} but in \cite{sotiriou2007metric} it was shown that this is not a viable choice and works only for $f(R)=R$ that is, only for the Einstein Hilbert action, and the best way to proceed is to set $S_{\mu}=0$ by means of a Lagrange multiplier \cite{sotiriou2007metric}. We review both of them in the following chapter, and we also propose another possibility.

	\subsection{Braking the Projective Invariance}
	In order to break the projective invariance  one needs to fix a vectorial degree of freedom\footnote{In $4-dim$ for instance, we need to fix four degrees of freedom.}. So, what vectors do we have at our disposal? As we have seen, we can construct two vectors out of non-metricity by contracting with the metric. These are the Weyl
	\beq
	Q_{\alpha}=Q_{\alpha\mu\nu}g^{\mu\nu}
	\eeq
	and the second non-metricity vector 
	\beq
	\tilde{Q}_{\nu}=Q_{\alpha\mu\nu}g^{\alpha\mu}
	\eeq
	For torsion, because of its antisymmetry there is simply one vector to be constructed by contractions, and this is the torsion vector\footnote{The torsion vector can be defined without the use of the metric tensor!}
	\beq
	S_{\mu}=S_{\mu\lambda}^{\;\;\;\;\lambda}
	\eeq
	There is also another possibility, by contracting the torsion tensor with the Levi-Civita symbol we get the pseudo-vector
	\beq
	\tilde{S}^{\alpha} =-\epsilon^{\mu\nu\lambda\alpha}S_{\mu\nu\lambda}
	\eeq
	However, this quantity is itself invariant under projective transformations of the connection and therefore it cannot be used to break the projective invariance. As a result, the vectors that could potentially break the projective invariance and produce a self-consistent theory, are $\{ Q_{\alpha},\tilde{Q}_{\nu},S_{\mu}\}$ . We explore the possibility of fixing each of them to zero separately.

	\subsubsection{Fixing $S_{\mu}=0$}
    Let us now break the projective invariance and obtain a self-consistent theory by fixing the torsion vector to zero, as done in \cite{sotiriou2007metric}. To this end we add the part
	\beq
	S_{B}=\int d^{n}\sqrt{-g}B_{\mu}S^{\mu}
	\eeq
	where $B_{\mu}$ is a Lagrange multiplier that will fix $S_{\mu}$ to zero. Therefore, our total action will be
	\begin{gather}
	S[g_{\alpha\beta},\Gamma^{\lambda}_{\;\;\;\mu\nu},B_{\rho}]=S_{G}+S_{M}+S_{B}= \\ \nonumber
	=\int  d^{n}x\sqrt{-g} \left[ \frac{1}{2\kappa}f(R)+\mathcal{L}_{M}+B_{\mu}S^{\mu}\right]
	\end{gather}
	and the total variation will have three different parts to it
	\beq
	\delta S=\delta_{g}S+\delta_{\Gamma}S+\delta_{B}S
	\eeq
	so the least action principle will give
	\beq
	\delta S=0 \Rightarrow \;\delta_{g}S=0\;,\;\delta_{\Gamma}S=0\;,\;\delta_{B}S=0
	\eeq
	Now, the parts $S_{G}$ and $S_{M}$ we have already varied in the previous section, so we only need to focus on the variation of $S_{B}$, which contains the parts
	\beq
	\delta S_{B}=\delta_{g} S_{B}+\delta_{\Gamma} S_{B}+\delta_{B} S_{B}
	\eeq
	and an easy calculation reveals
	\beq
	\delta_{g} S_{B}= \int d^{n}x\sqrt{-g}(\delta g^{\mu\nu} )\left[ -\frac{1}{2}g_{\mu\nu}B_{\alpha}S^{\alpha}+B_{(\mu}S_{\nu)} \right]
	\eeq
	\beq
	\delta_{\Gamma} S_{B}= \int d^{n}x\sqrt{-g}(\delta \Gamma^{\lambda}_{\;\;\;\mu\nu}) \Big[ B^{[\mu}\delta^{\nu]}_{\lambda} \Big]
	\eeq
	and
	\beq
	\delta_{B} S_{B}= \int d^{n}x\sqrt{-g}(\delta B^{\mu})S_{\mu}
	\eeq
	respectively. So, varying the total action independently with respect to $g_{\alpha\beta},\;\Gamma^{\lambda}_{\;\;\;\mu\nu}$ and $B_{\rho}$ and applying the Least Action Principle, we obtain the set of field equations
	\beq
	f^{'}(R)R_{(\mu\nu)}-\frac{f(R)}{2}g_{\mu\nu}=\kappa \left( T_{\mu\nu}-\frac{1}{2}g_{\mu\nu}B_{\alpha}S^{\alpha}-B_{(\mu}S_{\nu)} \right)
	\eeq 
	\begin{gather}
	-\frac{\nabla_{\lambda}(\sqrt{-g}f^{'}g^{\mu\nu})}{\sqrt{-g}}+\frac{\nabla_{\alpha}(\sqrt{-g}f^{'}g^{\mu\alpha}\delta_{\lambda}^{\nu})}{\sqrt{-g}}+
	2 f^{'}(S_{\lambda}g^{\mu\nu}-S^{\mu}\delta_{\lambda}^{\nu}-  S_{\lambda}^{\;\;\;\mu\nu}) = \nonumber \\
	\kappa ( \Delta_{\lambda}^{\;\;\;\mu\nu}-  B^{[\mu}\delta^{\nu]}_{\lambda}  )
	\end{gather}
	\beq
	S_{\mu}=0
	\eeq
	Using the last equation ($S_{\mu}=0$)  the first two simplify and give
	\beq
	f^{'}(R)R_{(\mu\nu)}-\frac{f(R)}{2}g_{\mu\nu}=\kappa  T_{\mu\nu}
	\eeq
	\begin{gather}
	-\frac{\nabla_{\lambda}(\sqrt{-g}f^{'}g^{\mu\nu})}{\sqrt{-g}}+\frac{\nabla_{\alpha}(\sqrt{-g}f^{'}g^{\mu\alpha}\delta_{\lambda}^{\nu})}{\sqrt{-g}}-
	2 f^{'} S_{\lambda}^{\;\;\;\mu\nu} = \nonumber \\
	\kappa ( \Delta_{\lambda}^{\;\;\;\mu\nu}-  B^{[\mu}\delta^{\nu]}_{\lambda}  ) \label{Pam}
	\end{gather} 
	Now, taking the trace $\mu=\lambda$ in the last one, the left hand side is identically zero (since this is the contraction the modified Palatini tensor $P_{\mu}^{\;\;\;\mu\nu}(h)$) and we are left with
	\beq
	0=\Delta_{\mu}^{\;\;\;\mu\nu}-\frac{1}{2}(B^{\nu}-n B^{\nu}) \Rightarrow \nonumber
	\eeq
	\beq
	B^{\mu}=\frac{2}{1-n}\Delta_{\mu}^{\;\;\;\mu\nu}=\frac{2}{1-n}\tilde{\Delta}^{\nu}
	\eeq
	where we defined  $\Delta_{\mu}^{\;\;\;\mu\nu} \equiv \tilde{\Delta}^{\nu}$. Thus, this is the value we should pick for the Lagrange multiplier $B_{\mu}$ in order to obtain self-consistent field equations, which upon this last substitution, take their final form
	\beq
	f^{'}(R)R_{(\mu\nu)}-\frac{f(R)}{2}g_{\mu\nu}=\kappa  T_{\mu\nu} \label{emde}
	\eeq
	\begin{gather}
	-\frac{\nabla_{\lambda}(\sqrt{-g}f^{'}g^{\mu\nu})}{\sqrt{-g}}+\frac{\nabla_{\alpha}(\sqrt{-g}f^{'}g^{\mu\alpha}\delta_{\lambda}^{\nu})}{\sqrt{-g}}-
	2 f^{'} S_{\lambda}^{\;\;\;\mu\nu} = \nonumber \\
	\kappa \Big( \Delta_{\lambda}^{\;\;\;\mu\nu}+  \frac{2}{n-1}\tilde{\Delta}^{[\mu}\delta^{\nu]}_{\lambda} \Big) \label{Pam}
	\end{gather}  
	Along with the constraint $S_{\mu}=0$ this is a set of consistent field equations, whose dynamics have studied to some extend in \cite{sotiriou2007metric,vitagliano2011dynamics}. We will review it here and add some new calculations regarding the form of non-metricity when the matter action does not depend on the connection. More specifically, we claim that when the connection is decoupled from the matter action ($\Delta_{\lambda}^{\;\;\;\mu\nu}=0$) torsion vanishes and the non-metricity is not general but we have  the case of a  Weyl non-metricity. To prove this, setting the right hand side of ($\ref{Pam}$) equal to zero , we obtain
	\begin{gather}
	-\frac{\nabla_{\lambda}(\sqrt{-g}f^{'}g^{\mu\nu})}{\sqrt{-g}}+\frac{\nabla_{\alpha}(\sqrt{-g}f^{'}g^{\mu\alpha}\delta_{\lambda}^{\nu})}{\sqrt{-g}}-
	2 f^{'} S_{\lambda}^{\;\;\;\mu\nu} = 0 \label{Pamp}
	\end{gather} 
	and contracting in $\lambda=\nu$
	\beq
	\frac{(n-1)}{2}\frac{\nabla_{\alpha}(\sqrt{-g}f^{'}g^{\mu\alpha})}{\sqrt{-g}}-2 f^{'}S_{\lambda}^{\;\;\;\mu\lambda}=0
	\eeq
	but noticing that
	\beq
	S_{\lambda}^{\;\;\;\mu\lambda}=g_{\lambda\alpha}S^{\alpha\mu\lambda}=-g_{\lambda\alpha}S^{\mu\alpha\lambda}=-g^{\mu\kappa}S_{\kappa\lambda}^{\;\;\;\lambda} =-g^{\mu\kappa}S_{\kappa}=-S^{\mu}=0
	\eeq
	and substituting it above, we are left with
	\beq
	\frac{\nabla_{\alpha}(\sqrt{-g}f^{'}g^{\mu\alpha})}{\sqrt{-g}}=0
	\eeq
	which when itself is substituted back in ($\ref{Pamp}$) simplifies it to
	\beq
	\frac{\nabla_{\lambda}(\sqrt{-g}f^{'}g^{\mu\nu})}{\sqrt{-g}}+2 f^{'} S_{\lambda}^{\;\;\;\mu\nu} =0 \label{Plt}
	\eeq
	Taking the antisymmetric part in $\mu,\nu$ of the above we conclude that
	\beq
	S_{\lambda}^{\;\;\;[\mu\nu]}=0\Rightarrow S_{\lambda[\mu\nu]}=0\Rightarrow S_{\lambda\mu\nu}=S_{\lambda\nu\mu} 
	\eeq
	That is, torsion has to be symmetric on its second and third indices. But recall that torsion is antisymmetric when exchanging  first and second index. Any rank $3$ tensor that has both of these symmetries has to identically vanish . To see this, given that
	\beq
	S_{\mu\nu\lambda}=-S_{\nu\mu\lambda}\;,\; S_{\mu\nu\lambda}=S_{\mu\lambda\nu}
	\eeq
	exploiting these symmetries, we have
	\begin{gather}
	S_{\mu\nu\lambda}=S_{\mu\lambda\nu}=-S_{\lambda\mu\nu}=-S_{\lambda\nu\mu}=+S_{\nu\lambda\mu} = \nonumber \\
	=S_{\nu\mu\lambda}=-S_{\mu\nu\lambda} \Rightarrow \nonumber 
	\end{gather}
	\beq
	S_{\mu\nu\lambda}=0
	\eeq
	Thus, torsion vanishes and ($\ref{Plt}$) becomes
	\beq
	\nabla_{\lambda}(\sqrt{-g}f^{'}g^{\mu\nu})=0
	\eeq
	This very condition tells us that the non-metricity has to be of the Weyl type ( namely $Q_{\alpha\mu\nu}\propto Q_{\alpha}g_{\mu\nu}$ ). To see this, expand the covariant derivative
	\beq
	g^{\mu\nu}f^{'}\nabla_{\lambda}\sqrt{-g}+Q_{\lambda}^{\;\;\;\;\mu\nu}+g^{\mu\nu}\partial_{\lambda}f^{'}=0
	\eeq
	and use
	\beq
	\frac{\nabla_{\lambda}\sqrt{-g}}{\sqrt{-g}}=-\frac{1}{2}Q_{\lambda}
	\eeq
	to arrive at
	\beq
	-\frac{1}{2}Q_{\lambda} g^{\mu\nu}+Q_{\lambda}^{\;\;\;\;\mu\nu}+g^{\mu\nu}\frac{\partial_{\lambda}f^{'}}{f^{'}}=0
	\eeq
	Contracting this with the metric tensor $g_{\mu\nu}$ it follows that
	\beq
	Q_{\lambda}=\frac{2n}{n-2}\partial_{\lambda}\ln{f^{'}}
	\eeq
	Finally, substituting the latter in the former we get
	\beq
	Q_{\lambda\mu\nu}=\frac{Q_{\lambda}}{n}g_{\mu\nu}=\frac{2}{n-2}g_{\mu\nu}\partial_{\lambda}\ln{f^{'}}
	\eeq
	In addition, contraction of ($\ref{emde}$) with the metric tensor gives
	\beq
	f^{'}(R)R-\frac{n}{2}f(R)=\kappa T
	\eeq
	which defines the implicit function $R=R(T)$ and therefore  both $f(R)$ and $f^{'}(R)$  are functions of $T$  ($f(R)=f(R(T))=f(T)$ and $f^{'}(R)=f^{'}(R(T))=f^{'}(T)$). As a result, a given $T_{\mu\nu}$ will give rise to Weyl non-metricity
	\beq
	Q_{\lambda\mu\nu}=\frac{Q_{\lambda}}{n}g_{\mu\nu}=\frac{2}{n-2}g_{\mu\nu}\partial_{\lambda}\ln{f^{'}(T)}
	\eeq
	In fact, this is an Integrable Weyl Geometry (IWG) since the Weyl vector is exact ($Q_{\mu}\propto \partial_{\mu}\ln{f^{'}}$). So, to conclude, we have shown that a general $f(R)$ theory for which $S_{\mu}$ is fixed to zero and the matter fields do not couple to the connection ($\Delta_{\lambda}^{\;\;\;\mu\nu}=0$)  results in a theory with zero torsion and a Weyl Integrable Geometry. This result is of course too restricting since it does not allow for any torsion at all. To address this problem we will propose another way to break the projective invariance in what follows. Before doing so let us  explore first the possibility of fixing either of the  non-metricity vectors $Q_{\mu}$, $\tilde{Q}_{\mu}$ to zero.

	\subsubsection{Fixing $\tilde{Q}_{\mu}=0$}
	We now add the Lagrange multiplier $C_{\mu}$ and the new piece to our action is
	\beq
	S_{C}=\int d^{n}\sqrt{-g}C_{\mu}\tilde{Q}^{\mu}
	\eeq
	We could may as well have replaced $\tilde{Q}^{\mu}$ with $Q_{\mu}$ (this was the fixing proposed in \cite{1981GReGr..13.1037H}) in the above but identical results will follow as we show below.  Again, let us consider the vacuum case where the Lagrange multiplier itself vanishes.\footnote{Not a-priori but after taking the trace and expressing it in terms of the Hypermomentum as we saw before.} Varying with respect to the connection and the Lagrange multiplier respectively we derive
	\begin{gather}
	-\frac{\nabla_{\lambda}(\sqrt{-g}f^{'}g^{\mu\nu})}{\sqrt{-g}}+\frac{\nabla_{\alpha}(\sqrt{-g}f^{'}g^{\mu\alpha}\delta_{\lambda}^{\nu})}{\sqrt{-g}}+
	2 f^{'}(S_{\lambda}g^{\mu\nu}-S^{\mu}\delta_{\lambda}^{\nu}-  S_{\lambda}^{\;\;\;\mu\nu}) = 
	0 \label{Pampe}
	\end{gather}
	\beq
	\tilde{Q}_{\mu}=0
	\eeq
	Now, even though we have set $\tilde{Q}_{\mu}=0$ we will keep $\tilde{Q}_{\mu}$ in our calculations to see what causes the problem when one tries to fix either of the non-metricity vectors. To this end, contacting ($\ref{Pampe}$) in $\lambda=\nu$ we get 
	\beq
	(n-1)\frac{\nabla_{\alpha}(\sqrt{-g}f^{'}g^{\mu\alpha})}{\sqrt{-g}}+
	2 f^{'} (2-n)S^{\mu} = 0\Rightarrow  \nonumber
	\eeq
	\beq
	\frac{\nabla_{\alpha}(\sqrt{-g}f^{'}g^{\mu\alpha})}{\sqrt{-g}}=2\frac{(n-2)}{n-1}S^{\mu} \label{recal}
	\eeq
	which when substituted back above, gives
	\beq
	-\frac{\nabla_{\lambda}(\sqrt{-g}f^{'}g^{\mu\nu})}{\sqrt{-g}}+2 f^{'}(S_{\lambda}g^{\mu\nu}+\frac{1}{1-n}S^{\mu}\delta_{\lambda}^{\nu}-  S_{\lambda}^{\;\;\;\mu\nu})=0
	\eeq
	After expanding the term in the covariant derivative and using the definitions of non-metricity, the above recasts to
	\beq
	\frac{1}{2}Q_{\lambda}g^{\mu\nu} -Q_{\lambda}^{\;\;\;\;\mu\nu}-g^{\mu\nu}\frac{\partial_{\lambda}f^{'}}{f^{'}}          +2(S_{\lambda}g^{\mu\nu}+\frac{1}{1-n}S^{\mu}\delta_{\lambda}^{\nu}-  S_{\lambda}^{\;\;\;\mu\nu})=0 \label{kku}
	\eeq
	where we have also divided through by $f^{'}$. Contracting the latter with the metric tensor $g^{\mu\nu}$ it follows that
	\beq
	\frac{(n-2)}{2}Q_{\lambda}-n\frac{\partial_{\lambda}f^{'}}{f^{'}}+\frac{2 n(n-2)}{(n-1)}S_{\lambda}=0 \label{eqw1}
	\eeq
	Also, contracting ($\ref{kku}$) in $\lambda=\nu$ we obtain
	\beq
	-\frac{1}{2}Q^{\mu}+\tilde{Q}^{\mu}+\frac{\partial^{\mu} f^{'}}{f^{'}}-\frac{2(n-2)}{(n-1)}S^{\mu}=0
	\eeq
	Multiplying through by $n$ and bringing the index downstairs, we may write the last one as
	\beq
	-\frac{n}{2}Q_{\lambda}+n \tilde{Q}_{\lambda}+n\frac{\partial_{\lambda} f^{'}}{f^{'}}-\frac{2 n(n-2)}{(n-1)}S_{\lambda}=0 \label{eqw2}
	\eeq
	Therefore, adding up equations ($\ref{eqw1}$) and ($\ref{eqw2}$) it follows that
	\beq
	-Q_{\lambda}+n \tilde{Q}_{\lambda}=0
	\eeq
	From this we see that fixing either of $Q_{\lambda}$ or $\tilde{Q}_{\lambda}$ to zero, the other vector must vanish too. So, by adding either of the Lagrange multipliers the end result is the same $\tilde{Q}_{\mu}=Q_{\mu}=0$, and with this at hand, from $(\ref{eqw2})$ we conclude that
	\beq
	\frac{\partial_{\mu}f^{'}}{f^{'}}=2\frac{(n-2)}{(n-1)}S_{\mu}
	\eeq
	Substituting all of these back into $(\ref{kku})$ it follows that
	\beq
	Q_{\lambda}^{\;\;\;\;\mu\nu}+2 S_{\lambda}^{\;\;\;\mu\nu}=\frac{2}{n-1}\Big[ S_{\lambda}g^{\mu\nu}-S^{\mu}\delta^{\nu}_{\lambda} \Big]
	\eeq
	or bringing $\lambda$ upstairs
	\beq
	Q^{\alpha\mu\nu}+2 S^{\alpha\mu\nu}=\frac{2}{n-1}\Big[ S^{\alpha}g^{\mu\nu}-S^{\mu}g^{\alpha\nu} \Big] \label{qsw}
	\eeq
	Taking the symmetric part in $\alpha,\mu$ in the above we obtain
	\beq
	Q^{(\alpha\mu)\nu}=0
	\eeq
	where we have also used the fact that the torsion tensor is antisymmetric in its first two indices $(S_{(\alpha\mu)\nu}=0)$. The above equation implies that non-metricity has to be antisymmetric in its first two indices, but by definition it is symmetric in its last two. Any rank-$3$ tensor with such properties must identically vanish\footnote{We showed a similar result for torsion in the previous section}. Indeed, given that
	\beq
	Q_{\alpha\mu\nu}=-Q_{\mu\alpha\nu}\;\;and\;\; Q_{\alpha\mu\nu}=Q_{\alpha\nu\mu}
	\eeq
	we compute
	\beq
	Q_{\alpha\mu\nu}=-Q_{\mu\alpha\nu}=-Q_{\mu\nu\alpha}=Q_{\nu\mu\alpha}=Q_{\nu\alpha\mu}=-Q_{\alpha\nu\mu}=-Q_{\alpha\mu\nu}
	\eeq
	and therefore
	\beq
	Q_{\alpha\mu\nu}=0
	\eeq
	and we see that the whole non-metricity vanishes. In addition, taking the antisymmetric part of ($\ref{Pampe}$) and contracting in $\lambda=\mu$ we have 
	\beq
	\frac{\nabla_{\alpha}(\sqrt{-g}f^{'}g^{\mu\alpha})}{\sqrt{-g}}=-2 f^{'}S^{\mu}
	\eeq
	which when placed against $(\ref{recal})$ demands that
	\beq
	S^{\mu}=0
	\eeq
	and recalling that 
	\beq
	\frac{\partial_{\mu}f^{'}}{f^{'}}=2\frac{(n-2)}{(n-1)}S_{\mu}
	\eeq
	it follows that
	\beq
	\partial_{\mu}f^{'}=0\Rightarrow f^{'}=constant
	\eeq
	which is true only when $f(R)=R$ and therefore fixing either of $Q_{\mu}$ or $\tilde{Q_{\mu}}$ to zero leads to inconsistency since it forces the $f(R)$ to be linear in $R$. To recap, fixing either $Q_{\mu}=0$ or $\tilde{Q}_{\mu}=0$ in order to break the projective invariance works only for $f(R)=R$ and for general $f(R)$ leads to inconsistencies.\footnote{To be more specific, either of these constraints force the function $f(R)$ to be linear in $R$, which is unreasonable.}  Now, as we have seen fixing $S_{\mu}=0$ breaks the projective invariance and produces a consistent theory. Notice however, that this is not the most general case one can have, especially when one needs to study theories when both the torsion and non-metricity vectors are different from zero. To this end we propose another method that breaks the projective invariance that is more general and instead of setting a vector to zero, establishes a relation between the torsion and non-metricity vectors. We do so in what follows.

	\subsubsection{Our Proposal: Fixing $(\alpha S_{\mu}-\beta Q_{\mu}-n \gamma \tilde{Q}_{\mu})=0$}
	Instead of fixing any of the torsion and non-metricity vectors to zero, here we take a different route and impose a relation between them that can also break the projective invariance\footnote{This proposal we presented in \cite{iosifidis2018exactly}.} So, what we want to do is take a linear combination of the three vectors that we have and set it to zero, namely
	\beq
	\alpha S_{\mu}-\beta Q_{\mu}-n \gamma \tilde{Q}_{\mu}=0
	\eeq
	where $\alpha,\beta,\gamma \neq 0$ are numbers and the minus signs and the factor $n$ are put there just for convenience in the calculation. This constraint is imposed again by means of a Lagrange multiplier 
	\beq
	S_{A}=\int d^{n}x \sqrt{-g}A^{\mu}(\alpha S_{\mu}-\beta Q_{\mu}-n \gamma \tilde{Q}_{\mu})
	\eeq
	where $A^{\mu}$ is the Lagrange multiplier that establishes the relation between the three vectors. Our total action is
	\begin{gather}
	S[g_{\alpha\beta},\Gamma^{\lambda}_{\;\;\;\mu\nu},A_{\rho}]=S_{G}+S_{M}+S_{A}= \\ \nonumber
	=\int  d^{n}x\sqrt{-g} \left[ \frac{1}{2\kappa}f(R)+\mathcal{L}_{M}+A^{\mu}(\alpha S_{\mu}-\beta Q_{\mu}-n \gamma \tilde{Q}_{\mu})\right]
	\end{gather}
	Variation with respect to the Lagrange multiplier gives
	\beq
	\alpha S_{\mu}-\beta Q_{\mu}-n \gamma \tilde{Q}_{\mu}=0
	\eeq
	where the parameters $\alpha,\beta,\gamma $ are chosen such as not to preserve the projective invariance. Let us again consider the case where the matter decouples from the connection ($\Delta_{\lambda}^{\;\;\;\mu\nu}=0$) such that $A^{\mu}=0$ and the result after varying with respect to the connection is the same with the one we obtained in the previous subsections, namely
	\beq
	\frac{1}{2}Q_{\lambda}g^{\mu\nu} -Q_{\lambda}^{\;\;\;\;\mu\nu}-g^{\mu\nu}\frac{\partial_{\lambda}f^{'}}{f^{'}}          +2(S_{\lambda}g^{\mu\nu}+\frac{1}{1-n}S^{\mu}\delta_{\lambda}^{\nu}-  S_{\lambda}^{\;\;\;\mu\nu})=0 \label{kku}
	\eeq
	\beq
	\frac{(n-2)}{2}Q_{\lambda}-n\frac{\partial_{\lambda}f^{'}}{f^{'}}+\frac{2 n(n-2)}{(n-1)}S_{\lambda}=0 \label{eqw1}
	\eeq
	\beq
	-\frac{1}{2}Q^{\mu}+\tilde{Q}^{\mu}+\frac{\partial^{\mu} f^{'}}{f^{'}}-\frac{2(n-2)}{(n-1)}S^{\mu}=0
	\eeq
	and
	\beq
	Q_{\mu}-n\tilde{Q}_{\mu}=0
	\eeq
	Substituting this last equation into the constraint we get
	\beq
	S_{\mu}=\left( \frac{\beta +\gamma}{\alpha} \right)Q_{\mu}=\lambda Q_{\mu}
	\eeq
	where we have defined $\lambda=(\beta +\gamma)/\alpha$ and in order to brake the projective invariance it must hold that $\lambda \neq \frac{n-1}{4n}$.\footnote{For this value of the parameter $\lambda$ the combination $S_{\mu}-\lambda Q_{\mu}$ becomes projective invariant.} Now, after some straightforward manipulations of the above equations, one can show that
	\beq
	S_{\mu}=\lambda Q_{\mu}=\lambda n\tilde{Q}_{\mu}=a\frac{2 n \lambda}{(n-2)} \frac{\partial_{\mu}f^{'}}{f^{'}}
	\eeq
	where
	\beq
	a=\frac{1}{1+\frac{4 n}{n-1}}
	\eeq
	From which we see that all three vectors are related to each other and their source is the term $\frac{\partial_{\mu}f^{'}}{f^{'}}$. To gain more intuition on the above, let us vary the total action with respect to the metric tensor to obtain the field equations
	\beq
	f^{'}(R)R_{(\mu\nu)}-\frac{f(R)}{2}g_{\mu\nu}=\kappa  T_{\mu\nu} \label{emde}
	\eeq
	where we have also used the fact that $A_{\mu}=0$. Again, taking the trace of the above field equations it follows that
	\beq
	f^{'}(R)R-\frac{n}{2}f(R)=\kappa T
	\eeq
	which, as we have already discussed, defines the implicit function $R=R(T)$ and therefore  both $f(R)$ and $f^{'}(R)$  are functions of $T$  ($f(R)=f(R(T))=f(T)$ and $f^{'}(R)=f^{'}(R(T))=f^{'}(T)$). Therefore, a given $T_{\mu\nu}$ will give rise to torsion and non-metricity through its trace and the torsion and non-metricity vectors are related and are proportional to this source which is a function of $T$, that is
	\beq
	S_{\mu}=\lambda Q_{\mu}=\lambda n\tilde{Q}_{\mu}=a\frac{2 n \lambda}{(n-2)} \frac{\partial_{\mu}f^{'}(T)}{f^{'}(T)}
	\eeq
	We would now wish to solve explicitly for the torsion and non-metricity tensors and find their exact forms. To do so, we substitute the above relation into
	\beq
	\frac{1}{2}Q_{\lambda}g^{\mu\nu} -Q_{\lambda}^{\;\;\;\;\mu\nu}-g^{\mu\nu}\frac{\partial_{\lambda}f^{'}}{f^{'}}          +2(S_{\lambda}g^{\mu\nu}+\frac{1}{1-n}S^{\mu}\delta_{\lambda}^{\nu}-  S_{\lambda}^{\;\;\;\mu\nu})=0 \label{kku}
	\eeq
	to obtain
	\beq
	(Q_{\lambda}^{\;\;\;\;\mu\nu}+2 S_{\lambda}^{\;\;\;\mu\nu})=b g^{\mu\nu}Q_{\lambda}+\frac{2 \lambda }{1-n}Q^{\mu}\delta_{\lambda}^{\nu} 
	\eeq
	or
	\beq
	(Q_{\alpha\mu\nu}+2 S_{\alpha\mu\nu})=b Q_{\alpha} g_{\mu\nu}+\frac{2 \lambda }{1-n}Q_{\mu}g_{\nu\alpha} \label{qstn}
	\eeq
	where $b=\frac{1}{ n}+\frac{2\lambda}{n-1}$. Note now that this tensor combination along with some index permutations of it appears in the connection decomposition
	\begin{equation}
	\Gamma^{\lambda}_{\;\;\;\mu\nu}=\tilde{\Gamma}^{\lambda}_{\;\;\;\mu\nu}+\frac{1}{2}g^{\alpha\lambda}\Big( (Q_{\mu\nu\alpha}+2 S_{\mu\nu\alpha})+(Q_{\nu\alpha\mu}+2 S_{\nu\alpha\mu})-(Q_{\alpha\mu\nu}+2 S_{\alpha\mu\nu}) \Big) \label{solk}
	\end{equation}
	So, carrying out the calculations we finally arrive at
	\beq
	\Gamma^{\lambda}_{\;\;\;\mu\nu}=\tilde{\Gamma}^{\lambda}_{\;\;\;\mu\nu}+\frac{1}{2}g^{\alpha\lambda}\Big( A(Q_{\mu}g_{\alpha\nu}-Q_{\alpha}g_{\mu\nu})+B Q_{\nu}g_{\mu\alpha}  \Big)
	\eeq
	where $A=b-\frac{2 n}{n-1}\lambda$,\; $B=b+\frac{2 n}{n-1}\lambda$. Having this one can easily compute the torsion tensor
	\beq
	S_{\mu\nu}^{\;\;\;\;\lambda}=\Gamma^{\lambda}_{\;\;\;[\mu\nu]}=\frac{2}{n-1}\lambda Q_{[\mu}\delta_{\nu]}^{\lambda}
	\eeq
	and using $S_{\mu}=\lambda Q_{\mu}$ we also make the consistency check
	\beq
	S_{\mu\nu}^{\;\;\;\;\lambda}=\frac{2}{n-1} S_{[\mu}\delta_{\nu]}^{\lambda}
	\eeq
	So, we have the case of a vectorial torsion. As far as non-metricity is concerned, we substitute the last equation into $(\ref{qstn})$ and after some straightforward calculations we finally arrive at
	\beq
	Q_{\alpha\mu\nu}=\frac{Q_{\alpha}}{n}g_{\mu\nu}
	\eeq
	which is the case of a Weyl non-metricity. Note that the parameter $\lambda$ has canceled out in the expression for non-metricity.To conclude, what we have done here is to break the projective invariance and produce a viable metric affine $f(R)$ theory. Instead of  setting $S_{\mu}=0$ or $Q_{\mu}=0$ (or even $\tilde{Q}_{\mu}=0$) which singles out a vector out of  the three that are available, we took a different route  and imposed a constraint on the three vectors ($\alpha S_{\mu}-\beta Q_{\mu}-n \gamma \tilde{Q}_{\mu}=0$) that treats them on equal footing. Our result (when the connection decouples from the matter fields) is a fully consistent theory in which there exist both torsion and non-metricity, powered by a single vector that is sourced by the energy momentum tensor. More specifically, one has a vectorial torsion and a non-metricity of the Weyl type, with\footnote{Thus, all three vectors $S_{\mu},Q_{\mu},\tilde{Q}_{\mu}$ are proportional to one another. A similar relation was obtained in the $2-d$ MAG model of \cite{obukhov2004two}.}
	\beq
	S_{\mu\nu}^{\;\;\;\;\lambda}=\frac{2}{n-1} S_{[\mu}\delta_{\nu]}^{\lambda}
	\eeq
	\beq
	Q_{\alpha\mu\nu}=\frac{Q_{\alpha}}{n}g_{\mu\nu}
	\eeq
	\beq
	S_{\mu}=\lambda Q_{\mu}=\lambda n\tilde{Q}_{\mu}=a\frac{2 n \lambda}{(n-2)} \frac{\partial_{\mu}f^{'}(T)}{f^{'}(T)}
	\eeq
	
	Some comments are now in order. Firstly, notice that in vacuum ($T_{\mu\nu}=0$) both torsion and non-metricity vanish and therefore they are only introduced by matter fields. Secondly, the above expressions for the affine connection and subsequently for torsion and non-metricity, are algebraic ones since on the assumption that matter decouples from the connection ($\Delta_{\alpha\mu\nu}=0$) we have that $T_{\mu\nu}$ is independent of the connection as seen from $(\ref{emhpt})$. So, breaking the invariance this way we see that the simplest forms of torsion and non-metricity can be sourced by the energy momentum tensor alone, and for further degrees of freedom to be excited, a hypermomentum tensor is also needed. Therefore, when   $T_{\mu\nu}\neq 0$ and $\Delta_{\alpha\mu\nu}=0$ only the lowest excitations of torsion and non-metricity can be produced. To obtain more general forms one needs to have a non-zero hypermomentum.

	 Notice now that one can also break the projective invariance by adding scalars, into the original action, that do not respect this symmetry. The easiest way to do this is by adding a scalar term built from any of the torsion/non-metricity vectors, since non of them respects the projective symmetry. Even though this seems somewhat artificial we shall present three simple models illustrating this possibility and then generalize the results to more general actions. We will do this procedure for the Einstein-Hilbert action (i.e. Ricci scalar) but the results can also be generalized in $f(R)$.
	
	\subsubsection{Model $1$}
	Let us consider the model given by the action
	\begin{equation}
	S=\frac{1}{2\kappa}\int d^{n}x\Big[\sqrt{-g}R+\gamma\sqrt{-g}g^{\mu\nu}S_{\mu}S_{\nu}\Big]
	\end{equation}
	where $\gamma$ is a parameter and $S_{\alpha}\equiv S_{\alpha\beta}^{\;\;\;\;\beta}$ the torsion vector. We now state that the above action exactly yields Einstein equations in vacuum without any additional degree of freedom. To see this we first vary the latter with respect to $g_{\mu\nu}$ and apply the Principle of least action to arrive at
	\begin{equation}
	\delta_{g}S=0 \nonumber
	\end{equation}
	\begin{equation}
	R_{(\mu\nu)}-\frac{R}{2}g_{\mu\nu}=\gamma\left[\frac{1}{2}S_{\alpha}S^{\alpha}g_{\mu\nu}-S_{\mu}S_{\nu}\right] \label{ssa}
	\end{equation}
	which as they stand now seem to admit both torsion and non-metricity. However, varying with respect to the connection we obtain
	\begin{equation}
	\delta_{\Gamma}S=0 \nonumber
	\end{equation}
	\begin{equation}
	P_{\lambda}^{\;\;\;\mu\nu}+\gamma (S^{\mu}\delta_{\lambda}^{\nu}-S^{\nu}\delta_{\lambda}^{\mu})=0 \label{go}
	\end{equation}
	recall that the Palatini tensor is given by
	\begin{equation}
	P_{\lambda}^{\;\;\;\mu\nu}=-\frac{\nabla_{\lambda}(\sqrt{-g}g^{\mu\nu})}{\sqrt{-g}}+\frac{\nabla_{\sigma}(\sqrt{-g}g^{\mu\sigma})}{\sqrt{-g}}\delta^{\nu}_{\lambda}
	+2(S_{\lambda}g^{\mu\nu}-S^{\mu}\delta_{\lambda}^{\nu}+g^{\mu\sigma}S_{\sigma\lambda}^{\;\;\;\;\nu})
	\end{equation}
	and satisfies
	\begin{equation}
	P_{\mu}^{\;\;\;\mu\nu}=0
	\end{equation}
	As a result, contracting ($\ref{go}$) in $\mu, \lambda$ and using the very last equation we arrive at
	\begin{equation}
	\underbrace{P_{\mu}^{\;\;\;\mu\nu}}_{=0}+\gamma (1-n)S^{\nu}=0\Rightarrow \nonumber
	\end{equation}
	\begin{equation}
	S^{\nu}=0 \label{snm}
	\end{equation}
	which shows that the torsion vector vanishes. Substituting the latter back to ($\ref{go}$) it follows that
	\begin{equation}
	P_{\lambda}^{\;\;\;\mu\nu}=0
	\end{equation}
	and as we have already seen, this last condition implies that
	\begin{equation}
	S_{\lambda}=-\frac{(n-1)}{4 n}Q_{\lambda} 
	\end{equation}
	\begin{equation}
	S_{\mu\nu}^{\;\;\;\;\lambda}=-\frac{2}{(n-1)}S_{[\nu}\delta_{\mu]}^{\lambda}=\frac{1}{n-1}\Big( S_{\mu}\delta_{\nu}^{\lambda}-S_{\nu}\delta_{\mu}^{\lambda} \Big)
	\end{equation}
	\begin{equation}
	Q_{\alpha\mu\nu}=\frac{1}{n}Q_{\alpha}g_{\mu\nu} 
	\end{equation}
	which when combined with $(\ref{snm})$ yield
	\begin{equation}
	Q_{\lambda}=0
	\end{equation}
	\begin{equation}
	S_{\mu\nu}^{\;\;\;\;\lambda}=0
	\end{equation}
	\begin{equation}
	Q_{\alpha\mu\nu}=0
	\end{equation}
	Therefore, we see that both torsion and non-metricity vanish in the end. In addition, substituting $(\ref{snm})$ back to $(\ref{ssa})$ we recover Einstein's equations
	\begin{equation}
	R_{\mu\nu}-\frac{R}{2}g_{\mu\nu}=0
	\end{equation}
	where $R_{\mu\nu}$ is the symmetric Ricci tensor computed with respect to Levi-Civita connection. Thus, we have shown that the model considered here is equivalent to General Relativity in vacuum as claimed.

	\subsubsection{Model $2$}
	As a second model we consider
	\begin{equation}
	S=\frac{1}{2\kappa}\int d^{n}x\Big[\sqrt{-g}R+\lambda\sqrt{-g}g^{\mu\nu}Q_{\mu}Q_{\nu}\Big]
	\end{equation}
	where $\lambda$ is the model parameter and $Q_{\mu}=-g^{\alpha\beta}\nabla_{\mu}g_{\alpha\beta}$ the Weyl vector. This Lagrangian was also considered by \cite{hehl1976new} in order to brake the projective invariance of the Ricci scalar. Variation with respect to the metric gives
	\begin{equation}
	\delta_{g}S=0 \Rightarrow  \nonumber
	\end{equation}
	\begin{equation}
	R_{(\mu\nu)}-\frac{R}{2}g_{\mu\nu}=\lambda\left[\frac{1}{2}Q_{\alpha}Q^{\alpha}g_{\mu\nu}-Q_{\mu}Q_{\nu}+g_{\mu\nu}\frac{\partial_{\alpha}(2\sqrt{-g}Q^{\alpha})}{\sqrt{-g}}\right] \label{mjk}
	\end{equation}
	Meanwhile, variation with respect to the connection yields
	\begin{equation}
	\delta_{\Gamma}S=0 \Rightarrow  \nonumber
	\end{equation}
	\begin{equation}
	P_{\lambda}^{\;\;\;\mu\nu}+4\lambda Q^{\nu}\delta_{\lambda}^{\mu}=0 \label{zxc}
	\end{equation}
	Again, contracting in $\mu,\lambda$ and using the tracelessness of the Palatini tensor in the first two indices, we arrive at
	\begin{equation}
	4\lambda n Q^{\nu}=0\Rightarrow  Q^{\nu}=0 \label{qqqq}
	\end{equation}
	and substituting the latter back in $(\ref{zxc})$ we derive 
	\begin{equation}
	P_{\lambda}^{\;\;\;\mu\nu}=0
	\end{equation}
	which when combined with $Q^{\nu}=0$ gives
	\begin{equation}
	S_{\lambda}=0
	\end{equation}
	\begin{equation}
	S_{\mu\nu}^{\;\;\;\;\lambda}=0
	\end{equation}
	\begin{equation}
	Q_{\alpha\mu\nu}=0
	\end{equation}
	Thus, the torsion and non-metricity vanish in this model as well. Substituting ($\ref{qqqq}$) in ($\ref{mjk}$) we again end up with Einstein equations in vacuum
	\begin{equation}
	R_{\mu\nu}-\frac{R}{2}g_{\mu\nu}=0
	\end{equation}

	\subsubsection{Model $3$}
	In the previous models we added to the Einstein-Hilbert action, terms that looked like mass terms. Firstly a term of squared torsion vector and then the squared Weyl vector. We saw that in both models after some manipulations we end up with Einstein field equations with vanishing torsion and non-metricity. In this model we consider the other possibility left\footnote{Note that as we have already pointed out there are three independent vectors (before solving the field equation) we can construct out of torsion and non-metricity, these are  the torsion, Weyl, and second non-metricity vectors respectively denoted by $S_{\mu}$, $Q_{\mu}$ and $\tilde{Q}_{\mu}$. } namely adding a squared second non-metricity vector. In words,
	\begin{equation}
	S=\frac{1}{2\kappa}\int d^{n}x\Big[\sqrt{-g}R+\alpha \sqrt{-g}g^{\mu\nu}\tilde{Q}_{\mu}\tilde{Q}_{\nu}\Big]
	\end{equation}
	where $\alpha$ is the model parameter and $\tilde{Q}_{\mu}=-g^{\alpha\beta}\nabla_{\alpha}g_{\beta\mu}$ is the second non-metricity vector. Varying the above action with respect to the metric tensor and applying the least action principle we arrive at
	\begin{gather}
	R_{(\mu\nu)}-\frac{R}{2}g_{\mu\nu}=\alpha \Big[ \frac{1}{2}\tilde{Q}_{\alpha}\tilde{Q}^{\alpha}g_{\mu\nu}-\tilde{Q}_{\mu}\tilde{Q}_{\nu} \nonumber \\
	-2 g^{\rho\alpha}(\partial_{\rho}g_{\mu\alpha}g_{\nu\beta})\tilde{Q}^{\beta}-\Gamma^{\lambda}_{\;\;\;\mu\nu}\tilde{Q}_{\lambda}+ \tilde{Q}_{\nu}g_{\mu\alpha}g^{\rho\sigma}\Gamma^{\alpha}_{\;\;\;\rho\sigma}+2 g_{\nu\beta}\frac{\partial_{\mu}(\sqrt{-g}\tilde{Q}^{\beta})}{\sqrt{-g}} \Big] \label{zxcd}
	\end{gather}
	The variation with respect to the connection yields
	\begin{equation}
	P_{\lambda}^{\;\;\;\mu\nu}+g^{\mu\nu}2 \tilde{Q}_{\lambda}+\delta_{\lambda}^{\nu}2 \tilde{Q}^{\mu}=0
	\end{equation}
	Now, contracting the latter in $\mu,\lambda$ and using the fact that $P_{\mu}^{\;\;\;\mu\nu}=0$ it follows that
	\begin{equation}
	4 \tilde{Q}^{\nu}=0 \Rightarrow \tilde{Q}^{\nu}=0
	\end{equation}
	which implies that
	\begin{equation}
	P_{\lambda}^{\;\;\;\mu\nu}=0
	\end{equation}
	and the last two equations combined, give
	\begin{equation}
	Q_{\lambda}=0
	\end{equation}
	\begin{equation}
	S_{\lambda}=0
	\end{equation}
	\begin{equation}
	S_{\mu\nu}^{\;\;\;\;\lambda}=0
	\end{equation}
	\begin{equation}
	Q_{\alpha\mu\nu}=0
	\end{equation}
	namely, also in this model torsion and non-metricity vanish, and upon substituting $\tilde{Q}^{\nu}=0$ back in ($\ref{zxcd}$) we again end up with Einstein equations in vacuum
	\begin{equation}
	R_{(\mu\nu)}-\frac{R}{2}g_{\mu\nu}=0
	\end{equation}
	So we saw that by adding to the Einstein-Hilbert action a term that is quadratic in any of the torsion/non-metricity vectors, the resulting theory in vacuum is Einstein's Gravity. In fact this will hold true even if we were to add a coupling term between these vectors. For instance, adding the term $S_{\mu}Q^{\mu}$ would give the same result with the above. Interestingly the same result continuous to hold true if we add any function of the above combinations. We show this in what follows. First we start with torsion and then prove the generic result.
	
	\subsubsection{A class of of equivalent Theories}
	We will prove here (for the first time) that a generalized class of Theories in the Metric Affine Gravity (but with no matter) is equivallent to Einstein's Gravity in vacuum. To start with, first notice
	that the results of $Model-1$ we presented above continue to hold true even if we consider a general $f(S_{\mu}S^{\mu})$ added to the Einstein Hilbert action. Indeed, starting from
	\begin{equation}
	S=\frac{1}{2\kappa}\int d^{n}x\sqrt{-g} \Big[ R+f(\chi) \Big] \label{go1}
	\end{equation}
	where $\chi\equiv S_{\mu}S^{\mu} $,  variation with respect to the connection yields
	\beq
	P_{\lambda}^{\;\;\;\mu\nu}+f_{\chi} (S^{\mu}\delta_{\lambda}^{\nu}-S^{\nu}\delta_{\lambda}^{\mu})=0 
	\eeq
	where $f_{\chi}=\frac{\partial{f}}{\partial{\chi}}$.  This is identical to (\ref{go}) where the parameter $\gamma$ has now been replaced with the function $f_{\chi}$. Note  that this will  again give a vanishing $S_{\mu}$ when traced over $\mu=\lambda$, which when substituted back will give a zero Palatini tensor and therefore vanishing torsion and non-metricity as we saw earlier. Therefore, we conclude that theories of the form $(\ref{go1})$ will give Einstein Gravity ($S_{\alpha\mu\nu}=0$, $Q_{\alpha\mu\nu}=0$) in vacuum without a projective mode. In fact, this result holds true when one adds any quadratic term of the torsion or non-metricity vectors. For instance, the theories
	\begin{equation}
	S=\frac{1}{2\kappa}\int d^{n}x\sqrt{-g} \Big[ R+f(Q_{\alpha}Q^{\alpha}) \Big] \label{go2}
	\end{equation}
	and
	\begin{equation}
	S=\frac{1}{2\kappa}\int d^{n}x\sqrt{-g} \Big[ R+f(\tilde{Q}_{\alpha}\tilde{Q}^{\alpha}) \Big] \label{go3}
	\end{equation}
	will both give the same result as $(\ref{go1})$, that is, vacuum Einstein Gravity with vanishing torsion and non-metricity and no projective mode\footnote{This is so because the added term does not respect the projective invariance.}. This can be seen easily from the fact that when varying such quadratic terms with respect to the connection the end result is proportional to $Q^{\mu}$ (or $\tilde{Q}^{\mu}$). Then taking the trace in the first two indices (of the equation we get when we vary wrt the connection) forces this $Q_{\mu}=0$ (or $\tilde{Q}^{\mu}=0$) which again implies the vanishing of $P_{\lambda}^{\;\;\;\mu\nu}$ and as a result the geometry is Riemannian.
	
	The above considerations can also be generalized for gravitational actions given by
	\begin{equation}
	S=\frac{1}{2\kappa}\int d^{n}x\sqrt{-g} \Big[ R+f(\chi_{1},\chi_{2},\chi_{3}) \Big] \label{go4}
	\end{equation}
	where $\chi_{1}=S_{\mu}S^{\mu}$, \; $\chi_{2}=Q_{\mu}Q^{\mu}$, \;$\chi_{3}=  \tilde{Q}_{\mu}\tilde{Q}^{\mu}$. Indeed, variation of the above with respect to the connection, yields
	\begin{gather}
	P_{\lambda}^{\;\;\;\mu\nu}+ f_{\chi_{1}}4 Q^{\nu}\delta_{\lambda}^{\mu}+2 f_{\chi_{2}}\Big( \tilde{Q}_{\lambda}g^{\mu\nu}+\tilde{Q}^{\mu}\delta_{\lambda}^{\nu} \Big)+(1-n)f_{\chi_{3}}S^{[\mu}\delta_{\lambda}^{\nu]}=0 \label{conrr}
	\end{gather}
	where $f_{\chi_{i}}=\frac{\partial{f}}{\partial{\chi_{i}}}$,$i=1,2,3$. Taking the possible traces of the above, we arrive at
	\beq
	4n f_{\chi_{1}}Q^{\mu}+4 f_{\chi_{2}}\tilde{Q}^{\mu}+(1-n)f_{\chi_{3}}S^{\mu}=0
	\eeq
	\beq
	P^{\mu}+4 f_{\chi_{1}}Q^{\mu}+2 (n+1)f_{\chi_{2}}\tilde{Q}^{\mu}+(n-1)f_{\chi_{3}}S^{\mu}=0
	\eeq
	\beq
	\tilde{P}^{\mu}+4 f_{\chi_{1}}Q^{\mu}+2(n+1)f_{\chi_{2}}\tilde{Q}^{\mu}=0
	\eeq
	with
	\begin{equation}
	P^{\mu}=P_{\nu}^{\;\;\;\mu\nu}=(n-1)\left[ \tilde{Q}^{\mu}-\frac{1}{2}Q^{\mu}\right] +2(2-n)S^{\mu}
	\end{equation}
	and
	\begin{equation}
	\tilde{P}_{\lambda}= g_{\mu\nu}P_{\lambda}^{\;\;\;\mu\nu}=\frac{(n-3)}{2}Q_{\lambda}+\tilde{Q}_{\lambda}+2(n-2)S_{\lambda}
	\end{equation}
	Notice now, that the above is a homogeneous system of three equations with three unknowns (the vectors $S_{\mu}$,$Q_{\mu}$,$\tilde{Q}_{\mu}$) and this can only have a solution different from zero when the equations are linearly dependent or in other words, the determinant of the coefficients (in our case the functions $f_{\chi_{i}}$) of the unknowns is zero. This, however, will impose certain relations between the derivatives $f_{\chi_{i}}$ and  as a result restrict the  possible forms of the function $f$. So, we may assume that the determinant will be different from zero in general and therefore the solution of the above system will be $S_{\mu}=0=Q_{\mu}=\tilde{Q}_{\mu}$.  Then equation ($\ref{conrr}$) becomes
	\beq
	P_{\lambda}^{\;\;\;\mu\nu}=0
	\eeq
	which means that $Q_{\alpha\mu\nu}=0$, $S_{\alpha\mu\nu}=0$ and therefore the theory is equivalent to Einstein Gravity in vacuum. This result can also be generalized even further to include actions
	of the form
	\begin{equation}
	S=\frac{1}{2\kappa}\int d^{n}x\sqrt{-g} \Big[ R+f(\chi_{1},\chi_{2},\chi_{3},\chi_{4},\chi_{5},\chi_{6}) \Big]  \label{geo4e}
	\end{equation}
	where $\chi_{1}=S_{\mu}S^{\mu}$, $\chi_{2}=Q_{\mu}Q^{\mu}$, $\chi_{3}= \tilde{Q}_{\mu}\tilde{Q}^{\mu}$, $\chi_{4}=Q_{\mu}S^{\mu}$, $\chi_{5}=Q_{\mu}\tilde{Q}^{\mu}$, $\chi_{6}=S_{\mu}\tilde{Q}^{\mu}$ .
	So here we have proved  that  general classes of actions such as $(\ref{go4})$ and $(\ref{geo4e})$ are all equivalent to Einstein's Gravity in vacuum. To the best of our knowledge, this general result appears for the first time in the literature. It would also be interesting to generalize the above considerations even further and fine the general family of such Theories.
	
	\textbf{Comment:} Notice that even though the above actions are equivalent to Einstein's GR in vacuum, when matter is added (to these actions) the latter can differ greatly.

	\subsection{Special Case: The Palatini f(R) Gravity}
	The Palatini $f(R)$ Gravity with matter has been extensively studied in the literature (\cite{capozziello2007f},\cite{sotiriou2007metric},\cite{olmo2011palatini},\cite{roshan2008palatini}) and therefore we shall not examine it further here. Note that with the term Palatini here we mean that the hypermomentum tensor vanishes identically, that is the matter action is independent of the connection. With this simplification the Palatini $f(R)$ theory has been shown to be equivalent with a metric scalar tensor theory  \cite{sotiriou2007metric}. Also, constraints on Palatini f(R) have been studied in \cite{sotiriou2006constraining}. The situation changes radically however when the connection couples to matter. Then the connection becomes dynamical and propagates more degrees of freedom than GR \cite{vitagliano2011dynamics}. We shall now proceed with the discussion of more general families of Metric-Affine Theories.

	\section{General Metric-Affine Gravity Theories}

	\subsection{Meric-Affine $f(R,R_{\mu\nu}R^{\mu\nu})$ Gravity}
	
	Having studied the dynamics of metric affine $f(R)$ gravities let us now review a slight generalization that appears in the literature, the $f(R,R_{\mu\nu}R^{\mu\nu})$ theories. In fact the most common example that is studied in the literature  is the Palatini\footnote{Recall that in Palatini theories it is assumed that the connection does not couple to the matter fields, that is the hypermomentum is  zero.} $f(R,R_{(\mu\nu)}R^{(\mu\nu)})$ with zero torsion ( see \cite{olmo2011palatini,olmo2009dynamical} for instance). In these theories the symmetric part of the Ricci tensor appears in order to maintain the projective invariance\footnote{Even though the Ricci tensor $R_{\mu\nu}$ is not invariant under projective transformations, its symmetric part $R_{(\mu\nu)}$ is.}. However, this invariance will again cause problems when one wants to study the general affine theory (where the hypermomentum does not vanish). For this reason we will start our discussion as general as possible, taking the full Ricci scalar in the gravitation action, let the connection couple to the matter fields and also consider torsion. We will also denote some special cases and refer to the literature for more details.
	
	\subsubsection{Meric-Affine $f(R,R_{\mu\nu}R^{\mu\nu})$ with Matter}
	We start by the action
	\beq
	S=\frac{1}{2 \kappa}\int d^{n}x \sqrt{-g}f(R,\chi)+S_{M}[g_{\mu\nu},\Gamma^{\lambda}_{\;\;\;\alpha\beta},\psi]
	\eeq
	where we have abbreviated  $\chi\equiv R_{\mu\nu}R^{\mu\nu}$ and notice that we allow for both torsion and non-metricity. To proceed with the variations, let us first carefully vary $f(R,R_{\mu\nu}R^{\mu\nu})$. Variation with respect to the metric tensor yields
	\beq
	\delta_{g}f=\frac{\partial f}{\partial R}\delta_{g}R+\frac{\partial f}{\partial \chi}\delta_{g}\chi=f_{R}\delta_{g}R  +f_{\chi}\delta_{g}\chi
	\eeq
	Now
	\beq
	\delta_{g}R=\delta_{g}(R_{\mu\nu}g^{\mu\nu})=R_{\mu\nu}(\delta g^{\mu\nu})=R_{(\mu\nu)}(\delta g^{\mu\nu})
	\eeq
	where we have used the fact that $R_{\mu\nu}$ is independent of the connection. Regarding the other scalar, one has
	\begin{gather}
	\delta_{g}\chi=\delta_{g}(R_{\mu\nu}R^{\mu\nu})=\delta_{g}(R_{\mu\nu}R_{\nu\alpha}g^{\mu\alpha}g^{\nu\beta})=R_{\mu\nu}R_{\alpha\beta}\Big((\delta g^{\mu\alpha})g^{\nu\beta}+g^{\mu\alpha}(\delta g^{\nu\beta})\Big) \nonumber \\
	=(\delta g^{\mu\nu})\Big( R_{\mu\alpha}R_{\nu}^{\;\;\alpha}+R_{\alpha \mu}R^{\alpha}_{\;\;\nu} \Big)
	\end{gather}
	and note that the last combination is symmetric in $\mu,\nu$ as it should and that $R_{\mu\alpha}R_{\nu}^{\;\;\alpha}\neq R_{\alpha \mu}R^{\alpha}_{\;\;\nu}$ unless the Ricci tensor is symmetric. Now, let us vary $f$ with respect to the affine connection. We have
	\beq
	\delta_{\Gamma}f=\frac{\partial f}{\partial R}\delta_{\Gamma}R+\frac{\partial f}{\partial \chi}\delta_{\Gamma}\chi=f_{R}\delta_{\Gamma}R  +f_{\chi}\delta_{\Gamma}\chi \label{caivar}
	\eeq
	and 
	\beq
	\delta_{\Gamma}R=\delta_{\Gamma}(g^{\mu\nu}R_{\mu\nu})=g^{\mu\nu}(\delta_{\Gamma}R_{\mu\nu})
	\eeq
	\beq
	\delta_{\Gamma}\chi=\delta_{\Gamma}(R_{\mu\nu}R^{\mu\nu})= 2 R^{\mu\nu} (\delta_{\Gamma}R_{\mu\nu})
	\eeq
	Substituting the last two into $(\ref{caivar})$ it follows that
	\beq
	\delta_{\Gamma}f=\Big( f_{R}g^{\mu\nu}+2 f_{\chi}R^{\mu\nu} \Big)\delta_{\Gamma}R_{\mu\nu}
	\eeq
	and recalling that for any $M^{\mu\nu}$ (this can be either a tensor field or a tensor density)
	\beq
	M^{\mu\nu}\delta_{\Gamma}R_{\mu\nu}=\delta \Gamma^{\lambda}_{\;\;\;\mu\nu}\Big( -\nabla_{\lambda}M^{\mu\nu}+\nabla_{\alpha}(M^{\mu\alpha}\delta_{\lambda}^{\nu})-2 M^{\mu\alpha}S_{\lambda\alpha}^{\;\;\;\;\nu}  \Big)+ A
	\eeq
	where 
	\beq
	A=\nabla_{\lambda}( M^{\mu\nu}\delta \Gamma^{\lambda}_{\;\;\;\mu\nu}-M^{\mu\lambda}\delta_{\alpha}^{\nu}\delta \Gamma^{\alpha}_{\;\;\;\mu\nu})
	\eeq
	we have all the tools available to derive the variations. Using all the above we vary with respect to the metric and the connection, to arrive at
	\beq
	-\frac{f}{2}g_{\mu\nu}+f_{R}R_{(\mu\nu)}+f_{\chi}( R_{\mu\alpha}R_{\nu}^{\;\;\alpha}+R_{\alpha \mu}R^{\alpha}_{\;\;\nu}) =\kappa T_{\mu\nu}
	\eeq
	\begin{gather}
	-\nabla_{\lambda}(\sqrt{-g}B^{\mu\nu})+\nabla_{\alpha}(\sqrt{-g}B^{\mu\alpha})\delta^{\nu}_{\lambda} \nonumber \\
	+ 2\sqrt{-g}\Big[ -B^{\mu\alpha}(S_{\lambda\alpha}^{\;\;\;\;\nu}+S_{\alpha}\delta^{\nu}_{\lambda})+B^{\mu\nu}S_{\lambda} \Big]= \kappa \sqrt{-g} \Delta_{\lambda}^{\;\;\;\mu\nu}
	\end{gather}
	where
	\beq
	B^{\mu\nu} \equiv  f_{R}g^{\mu\nu}+2 f_{\chi}R^{\mu\nu}
	\eeq
	These are the field equations for a general Metric-Affine $f(R,R_{\mu\nu}R^{\mu\nu})$ . Notice that if we were to set torsion to zero ($S_{\alpha\mu\nu}=0$) from the onset, the set of field equations would read
	\beq
	-\frac{f}{2}g_{\mu\nu}+f_{R}R_{(\mu\nu)}+f_{\chi}( R_{\mu\alpha}R_{\nu}^{\;\;\alpha}+R_{\alpha \mu}R^{\alpha}_{\;\;\nu}) =\kappa T_{\mu\nu}
	\eeq
	\begin{gather}
	-\nabla_{\lambda}(\sqrt{-g}B^{\mu\nu})+\nabla_{\alpha}(\sqrt{-g}B^{(\mu|\alpha|})\delta^{\nu)}_{\lambda} 
	= \kappa \sqrt{-g} \Delta_{\lambda}^{\;\;\;\mu\nu}
	\end{gather}
	and observe that the last one is symmetrized in $\mu,\nu$ since the connection was symmetric to begin with. For a more detailed discussion on the theory that contains the general $R_{\mu\nu}$ term the reader is refereed to (\cite{vitagliano2011gravity} ,\cite{vitagliano2010dynamics}) and for the theories where only $R_{(\mu\nu)}$ enters see \cite{olmo2009dynamical}.

	\subsection{General $\mathcal{L}(g_{\mu\nu},R^{\alpha}_{\;\;\beta\gamma\rho})$ Meric-Affine  Gravity}
	Let us now generalize the considerations of the previous section and derive the field equations for a general action whose dependence on the connection (for the gravitational sector) comes entirely from the Riemann tensor\footnote{This means that the action will depend on scalars built from the Riemann tensor and its contractions entirely and not from torsion and non-metricity. This inclusion will be considered in the next section.} (and its contractions, of course). So the gravitational sector of our Lagrangian density will be $ \mathcal{L}_{G}(g_{\mu\nu},R^{\alpha}_{\;\;\beta\gamma\rho})$ and the total action of these theories, reads
	\beq
	S[g,\Gamma]=\frac{1}{2\kappa} \int d^{n}x\sqrt{-g}\mathcal{L}_{G}(g_{\mu\nu},R^{\alpha}_{\;\;\beta\gamma\rho})+ \int d^{n}x \sqrt{-g}\mathcal{L}_{M}(g_{\mu\nu},\Gamma^{\lambda}_{\;\;\;\alpha\beta},\psi)
	\eeq
	Note that both $f(R)$ and $f(R, R_{\mu\nu}R^{\mu\nu})$  theories are special cases of the above action. Variation of the above with respect to the metric tensor gives
	\begin{gather}
	\delta_{g}S=\frac{1}{2\kappa} \int d^{x}\Big[ \delta_{g}(\sqrt{-g}\mathcal{L}_{G})+\delta_{g}(\sqrt{-g}\mathcal{L}_{M})\Big]= \nonumber \\
	=\frac{1}{2 \kappa}\int d^{n}x \sqrt{-g}\Big[ -\frac{1}{2}g_{\mu\nu}\mathcal{L}_{G} \delta g^{\mu\nu}+\frac{\partial \mathcal{L}_{G} }{\partial g^{\mu\nu}}\delta g^{\mu\nu}-\kappa T_{\mu\nu}\delta g^{\mu\nu} \Big]= \nonumber \\
	=\frac{1}{2 \kappa}\int d^{n}x \sqrt{-g} (\delta g^{\mu\nu})\Big[ -\frac{1}{2}g_{\mu\nu}\mathcal{L}_{G}+\frac{\partial \mathcal{L}_{G} }{\partial g^{\mu\nu}}-\kappa T_{\mu\nu} \Big]=0
	\end{gather}
	So, the variation with respect to the metric tensor gives the filed equations
	\beq
	-\frac{1}{2}g_{\mu\nu}\mathcal{L}_{G}+\frac{\partial \mathcal{L}_{G} }{\partial g^{\mu\nu}}=\kappa T_{\mu\nu}
	\eeq
	Now, varying the action with respect  to the affine connection we get
	\beq
	\delta_{\Gamma}S= \frac{1}{2 \kappa} \int d^{n} x  \sqrt{-g}\Big[ \delta_{\Gamma}\mathcal{L}_{G}+2 \kappa \delta_{\Gamma}\mathcal{L}_{M} \Big]
	\eeq
	Now using the chain rule we may write
	\beq
	\delta_{\Gamma}\mathcal{L}_{G}=\frac{\partial \mathcal{L}_{G}}{\partial R^{\lambda}_{\;\;\mu\alpha\nu}} \delta_{\Gamma} R^{\lambda}_{\;\;\mu\alpha\nu}\equiv \Omega_{\lambda}^{\;\;\;\mu\alpha\nu} \delta_{\Gamma} R^{\lambda}_{\;\;\mu\alpha\nu}
	\eeq
	where we have defined $\Omega_{\lambda}^{\;\;\;\mu\alpha\nu}\equiv \frac{\partial \mathcal{L}_{G}}{\partial R^{\lambda}_{\;\;\mu\alpha\nu}}  $ and notice that this tensor is,  by construction,   antisymmetric in its last two indices ($\Omega_{\lambda}^{\;\;\;\mu\alpha\nu}=\Omega_{\lambda}^{\;\;\;\mu[\alpha\nu]}$). Now using 
	\beq
	\delta_{\Gamma}R^{\mu}_{\;\;\;\nu\alpha\beta}=\nabla_{\alpha}(\delta \Gamma^{\mu}_{\;\;\;\nu\beta})-\nabla_{\beta}(\delta \Gamma^{\mu}_{\;\;\;\nu\alpha})-2 S_{\alpha\beta}^{\;\;\;\;\lambda}\delta\Gamma^{\mu}_{\;\;\;\nu\lambda}
	\eeq
	we compute
	\begin{gather}
	\sqrt{-g}\delta_{\Gamma}\mathcal{L}_{G}=\Omega_{\alpha}^{\;\;\beta\gamma\delta} \Big[ \nabla_{\gamma}\delta \Gamma^{\alpha}_{\;\;\;\beta\delta}-\nabla_{\delta}\delta \Gamma^{\alpha}_{\;\;\;\beta\gamma}- 2 S_{\gamma\delta}^{\;\;\;\;\rho} \delta \Gamma^{\alpha}_{\;\;\;\beta\rho}\Big] =\nonumber \\
	=2\nabla_{\gamma}(\sqrt{-g}J^{\gamma}) -\delta \Gamma^{\lambda}_{\;\;\;\mu\nu} \Big[ \nabla_{\gamma}(\sqrt{-g}\Omega_{\lambda}^{\;\;\;\mu\gamma\nu}+\Omega_{\lambda}^{\;\;\;\mu\gamma\delta}S_{\gamma\delta}^{\;\;\;\;\nu}\Big]
	\end{gather}
	where we have set
	\beq
	J^{\gamma}=\Omega_{\alpha}^{\;\;\;\beta\gamma\delta}\delta \Gamma^{\alpha}_{\;\;\;\beta\delta}
	\eeq
	and now notice that 
	\beq
	\int d^{n}x \nabla_{\gamma}(\sqrt{-g}J^{\gamma})
	\eeq
	is not a surface term, but rather
	\begin{gather}
	\int d^{n}x \nabla_{\gamma}(\sqrt{-g}J^{\gamma})=\int d^{n}x \partial_{\gamma}(\sqrt{-g}J^{\gamma})+\int d^{n}x \sqrt{-g}2 S_{\mu}J^{\mu} =\nonumber \\
	=s.t.+\int d^{n}x \sqrt{-g}2 S_{\mu}J^{\mu} 
	\end{gather}
	Taking all the above into consideration along with the definition
	\beq
	\delta_{\Gamma}S_{M}\equiv \int d^{n}x\left( -\frac{\sqrt{-g}}{2}\Delta_{\lambda}^{\;\;\;\mu\nu}\right) \delta \Gamma^{\lambda}_{\;\;\;\mu\nu}
	\eeq
	we finally arrive at
	\begin{gather}
	\delta_{\Gamma}S=\frac{1}{2 \kappa} \int d^{n} x  \sqrt{-g}\Big[ \delta_{\Gamma}\mathcal{L}_{G}+2 \kappa \delta_{\Gamma}\mathcal{L}_{M} \Big]  \nonumber \\
	\frac{1}{2 \kappa}\int d^{n}x \sqrt{-g}  (\delta \Gamma^{\lambda}_{\;\;\;\mu\nu})\left( -2\frac{(\sqrt{-g}\Omega_{\lambda}^{\;\;\;\mu\alpha\nu})}{\sqrt{-g}}+4 \Omega_{\lambda}^{\;\;\;\mu\alpha\nu} S_{\alpha}- 2 \Omega_{\lambda}^{\;\;\;\mu\gamma\delta}S_{\gamma\delta}^{\;\;\;\;\nu} 
	-\kappa \Delta_{\lambda}^{\;\;\;\mu\nu} \right)
	\end{gather}
	Thus, the field equations coming from the variation of the connection read
	\beq
	-2\frac{(\sqrt{-g}\Omega_{\lambda}^{\;\;\;\mu\alpha\nu})}{\sqrt{-g}}+4 \Omega_{\lambda}^{\;\;\;\mu\alpha\nu} S_{\alpha}- 2 \Omega_{\lambda}^{\;\;\;\mu\gamma\delta}S_{\gamma\delta}^{\;\;\;\;\nu} 
	=\kappa \Delta_{\lambda}^{\;\;\;\mu\nu}
	\eeq
	or in the more compact form
	\beq
	(-\nabla_{\alpha}+ 2 S_{\alpha})(\sqrt{-g}\Omega_{\lambda}^{\;\;\;\mu\alpha\nu})-\sqrt{-g}\Omega_{\lambda}^{\;\;\;\mu\gamma\delta}S_{\gamma\delta}^{\;\;\;\;\nu}=\frac{\kappa}{2}\sqrt{-g}\Delta_{\lambda}^{\;\;\;\mu\nu}
	\eeq
	or better yet
	\beq
	(-\nabla_{\alpha}+ 2 S_{\alpha})(\mathcal{O}_{\lambda}^{\;\;\;\mu\alpha\nu})-\mathcal{O}_{\lambda}^{\;\;\;\mu\gamma\delta}S_{\gamma\delta}^{\;\;\;\;\nu}=\frac{\kappa}{2}\sqrt{-g}\Delta_{\lambda}^{\;\;\;\mu\nu}
	\eeq
	where we have defined the tensorial density $\mathcal{O}_{\lambda}^{\;\;\;\mu\alpha\nu} \equiv \sqrt{-g}\Omega_{\lambda}^{\;\;\;\mu\alpha\nu} $.
	So, to conclude, the field equations for a general  $\mathcal{L}(g_{\mu\nu},R^{\alpha}_{\;\;\beta\gamma\rho})$  metric-affine gravity, read  
	\beq
	-\frac{1}{2}g_{\mu\nu}\mathcal{L}_{G}+\frac{\partial \mathcal{L}_{G} }{\partial g^{\mu\nu}}=\kappa T_{\mu\nu}
	\eeq
	\beq
	(-\nabla_{\alpha}+ 2 S_{\alpha})(\sqrt{-g}\Omega_{\lambda}^{\;\;\;\mu\alpha\nu})-\sqrt{-g}\Omega_{\lambda}^{\;\;\;\mu\gamma\delta}S_{\gamma\delta}^{\;\;\;\;\nu}=\frac{\kappa}{2}\sqrt{-g}\Delta_{\lambda}^{\;\;\;\mu\nu}
	\eeq
	where
	\beq
	\Omega_{\lambda}^{\;\;\;\mu\alpha\nu}=\Omega_{\lambda}^{\;\;\;\mu[\alpha\nu]} \equiv \frac{\partial \mathcal{L}_{G}}{\partial R^{\lambda}_{\;\;\mu\alpha\nu}} 
	\eeq
	To make contact we our previous derived results, taking $\mathcal{L}_{G}=R$ (and let $\mathcal{L}_{M}=0$) we compute
	\beq
	\Omega_{\lambda}^{\;\;\;\mu\alpha\nu} =\frac{\partial R}{\partial  R^{\lambda}_{\;\;\mu\alpha\nu} }=\delta^{\beta}_{\gamma}g^{\kappa\rho}\frac{\partial R^{\gamma}_{\;\;\kappa\beta\rho}}{\partial  R^{\lambda}_{\;\;\mu\alpha\nu} }=g^{\mu[\nu} \delta^{\alpha]}_{\lambda}
	\eeq
	where we have used the  fact that
	\beq
	\frac{\partial R^{\gamma}_{\;\;\kappa\beta\rho}}{\partial  R^{\lambda}_{\;\;\mu\alpha\nu} }= \delta^{\gamma}_{\lambda}\delta^{\mu}_{\kappa} \delta^{[\alpha}_{\beta}\delta^{\nu]}_{\rho}
	\eeq
	Using this, along with the fact that
	\beq
	\frac{\partial R}{\partial g^{\mu\nu}}=R_{(\mu\nu)}
	\eeq
	the field equations take the form
	\begin{equation}
	R_{(\mu\nu)}-\frac{g_{\mu\nu}}{2}R=0
	\end{equation}
	\begin{equation}
	-\nabla_{\lambda}(\sqrt{-g}g^{\mu\nu})+\nabla_{\sigma}(\sqrt{-g}g^{\mu\sigma})\delta^{\nu}_{\lambda} \\
	+2\sqrt{-g}(S_{\lambda}g^{\mu\nu}-S^{\mu}\delta_{\lambda}^{\nu}+g^{\mu\sigma}S_{\sigma\lambda}^{\;\;\;\;\nu})=0 
	\end{equation}
	which are, of course, the ones we obtained when we studied the vacuum Einstein's gravity in the  Metric-Affine framework. In addition, taking $\mathcal{L}_{G}=f(R)$  we recover metric-affine $f(R)$ gravity and for $\mathcal{L}_{G}=f(R,R_{\mu\nu}R^{\mu\nu})$ one obtains the sub-class of theories we presented in the previous section. This is easily proved by using
	\beq
	\frac{\partial (R_{\kappa\rho}R^{\kappa\rho})}{\partial   R^{\lambda}_{\;\;\mu\alpha\nu} }=2R^{\mu[\nu}\delta^{\alpha]}_{\lambda}=R^{\mu\nu}\delta^{\alpha}_{\lambda}-R^{\mu\alpha}\delta^{\nu}_{\lambda}
	\eeq
	Notice also that the variations of the other contractions of the Riemman tensor read
	\beq
	\frac{\partial (\hat{R}_{\beta\gamma})}{\partial   R^{\lambda}_{\;\;\mu\alpha\nu} }=\delta^{\lambda}_{\mu}\delta^{\alpha}_{[\beta}\delta^{\nu}_{\gamma]} \;,  \;\;\;\;\; \frac{\partial (\hat{R}_{\beta\gamma}\hat{R}^{\beta\gamma})}{\partial   R^{\lambda}_{\;\;\mu\alpha\nu} }=2 \delta_{\lambda}^{\mu}\hat{R}^{\alpha\nu}
	\eeq
	\beq
	\frac{\partial (\check{R}_{\kappa\rho})}{\partial   R^{\lambda}_{\;\;\mu\alpha\nu} }=g_{\kappa\lambda}g^{\mu[\alpha} \delta^{\nu]}_{\rho} \;,  \;\;\;\;\; \frac{\partial (\check{R}_{\kappa\rho}\check{R}^{\kappa\rho})}{\partial   R^{\lambda}_{\;\;\mu\alpha\nu} }=2 g^{\mu[\alpha}\check{R}_{\lambda}^{\;\;\;\nu]}
	\eeq
	and recall that $\hat{R}_{\mu\nu}=R^{\lambda}_{\;\;\lambda\mu\nu}$ is the homothetic curvature and $\check{R}^{\mu}_{\;\;\beta} =g^{\nu\alpha}R^{\mu}_{\;\;\;\nu\alpha\beta}$ is the third independent contraction of the Riemann tensor (that can be formed once the space is endowed with a metric). Let us now generalize even further and derive the field equations when one also includes scalars built from torsion as well as non-metricity to the general action. That is we consider theories of the form $\mathcal{L}_{G}(g_{\mu\nu},R^{\alpha}_{\;\;\beta\gamma\rho}, S_{\mu\nu}^{\;\;\;\;\lambda},Q_{\alpha\mu\nu})$.

	\subsection{General $\mathcal{L}(g_{\mu\nu},R^{\alpha}_{\;\;\beta\gamma\rho}, S_{\mu\nu}^{\;\;\;\;\lambda},Q_{\alpha\mu\nu})$ Theories}
	To generalize the above considerations even further let us present the most general Gravity action that one can write down, whose dependence of the connection comes from scalars built from the Riemman, torsion and non-metricity tensors, and derive the field equations. Thus, the gravitation sector that we consider will be $\mathcal{L}(g_{\mu\nu},R^{\alpha}_{\;\;\beta\gamma\rho}, S_{\mu\nu}^{\;\;\;\;\lambda},Q_{\alpha\mu\nu})$. If we think about, this is indeed the most general Lagrangian one could write down (without including additional tensors constructed by the covariant  derivatives of these tensors), since the way that the connection enters the action is through the tensors $R^{\alpha}_{\;\;\beta\gamma\rho}(\Gamma, \partial \Gamma)$, $S_{\mu\nu}^{\;\;\;\;\lambda}(\Gamma)$  and $Q_{\alpha\mu\nu})(\Gamma)$ where torsion and non-metricity are linear in the connection while the Riemann tensor contains second order terms as well as derivatives of the connection. So, to begin with, we consider the theory given by
	\beq
	S[g,\Gamma]=\frac{1}{2\kappa} \int d^{n}x\sqrt{-g}\mathcal{L}_{G}(g_{\mu\nu},R^{\alpha}_{\;\;\beta\gamma\rho}, S_{\mu\nu}^{\;\;\;\;\lambda},Q_{\alpha\mu\nu})+ \int d^{n}x \sqrt{-g}\mathcal{L}_{M}(g_{\mu\nu},\Gamma^{\lambda}_{\;\;\;\alpha\beta},\psi)
	\eeq
	Now notice that as they stand, in their original forms, the tensors $R^{\alpha}_{\;\;\beta\gamma\rho}$ and $ S_{\mu\nu}^{\;\;\;\;\lambda}$ depend only on the connection and are independent of the metric\footnote{Of course, this is true when these tensors appear in their prototype forms and when one raises or lowers indices, multiplication with the metric is involved and as a result the final tensor does depend on the metric. For instance, even though $S_{\mu\nu}^{\;\;\;\;\lambda}$ is metric independent, the tensor $S_{\mu\nu\alpha}$  depends on the metric since $S_{\mu\nu\alpha}=g_{\lambda\alpha}S_{\mu\nu}^{\;\;\;\;\lambda}$. } while the non-metricity tensor depends on both the connection and the metric as can be seen by its very definition
	\begin{equation}
	Q_{\alpha\mu\nu}=-\nabla_{\alpha}g_{\mu\nu} =-\partial_{\alpha}g_{\mu\nu}+\Gamma^{\rho}_{\;\;\;\mu\alpha}g_{\rho\nu}+\Gamma^{\rho}_{\;\;\;\nu\alpha}g_{\mu\rho}
	\end{equation}
	Therefore, when we vary the general action with respect to the metric tensor the chain rule will be applied only for non-metricity 
	\beq
	\delta_{g}\mathcal{L}_{G}=\frac{\partial \mathcal{L}}{\partial g^{\mu\nu}}\delta g^{\mu\nu}+\frac{\partial \mathcal{L}}{\partial Q_{\rho\alpha\beta}}\delta_{g}Q_{\rho\alpha\beta}
	\eeq
	or we may write it as
	\beq
	\delta_{g}\mathcal{L}_{G}=\frac{\partial \mathcal{L}}{\partial g^{\mu\nu}}\delta g^{\mu\nu}+\frac{\partial \mathcal{L}}{\partial Q_{\rho}^{\;\;\;\alpha\beta}}\delta_{g}Q_{\rho}^{\;\;\;\alpha\beta}
	\eeq
	which will be more convenient for the calculations. Now, regarding the $\Gamma$-variation of $\mathcal{L}_{G}(g_{\mu\nu},R^{\alpha}_{\;\;\beta\gamma\rho}, S_{\mu\nu}^{\;\;\;\;\lambda},Q_{\alpha\mu\nu})$ , one has
	\beq
	\delta_{\Gamma}\mathcal{L}_{G}=\frac{\partial \mathcal{L}_{G}}{\partial R^{\lambda}_{\;\;\mu\alpha\nu}} \delta_{\Gamma} R^{\lambda}_{\;\;\mu\alpha\nu}+\frac{\partial \mathcal{L}_{G}}{\partial S_{\mu\nu}^{\;\;\;\;\lambda}} \delta_{\Gamma}S_{\mu\nu}^{\;\;\;\;\lambda}+\frac{\partial \mathcal{L}_{G}}{\partial Q_{\alpha\mu\nu}} \delta_{\Gamma}Q_{\alpha\mu\nu}
	\eeq
	or
	\beq
	\delta_{\Gamma}\mathcal{L}_{G}=\Omega_{\lambda}^{\;\;\;\mu\alpha\nu} \delta_{\Gamma}  R^{\lambda}_{\;\;\mu\alpha\nu}+V^{\mu\nu}_{\;\;\;\;\lambda}\delta_{\Gamma}S_{\mu\nu}^{\;\;\;\;\lambda}+W^{\alpha\mu\nu}\delta_{\Gamma}Q_{\alpha\mu\nu} \label{olavaria}
	\eeq
	where we have defined
	\beq
	\Omega_{\lambda}^{\;\;\;\mu\alpha\nu} \equiv \frac{\partial \mathcal{L}_{G}}{\partial R^{\lambda}_{\;\;\mu\alpha\nu}}\;,\;\; V^{\mu\nu}_{\;\;\;\;\lambda} \equiv \frac{\partial \mathcal{L}_{G}}{\partial S_{\mu\nu}^{\;\;\;\;\lambda}}\;,\; W^{\alpha\mu\nu}  \equiv \frac{\partial \mathcal{L}_{G}}{\partial Q_{\alpha\mu\nu}}
	\eeq
	and obviously,  they obey to the symmetries $\Omega_{\lambda}^{\;\;\;\mu\alpha\nu}=\Omega_{\lambda}^{\;\;\;\mu[\alpha\nu]}$, $V^{\mu\nu}_{\;\;\;\;\lambda}=V^{[\mu\nu]}_{\;\;\;\;\lambda}$, $W^{\alpha\mu\nu}=W^{\alpha(\mu\nu)}$ by construction. Now, notice that the first term on the right hand side of $(\ref{olavaria})$ we have already worked out in the previous section, so we only need to obtain the other two. Using
	\beq
	\delta_{\Gamma}S_{\alpha\beta}^{\;\;\;\;\lambda}=\delta_{[\alpha}^{\mu}\delta_{\beta]}^{\nu}\delta \Gamma^{\lambda}_{\;\;\;\mu\nu}
	\eeq
	\beq
	\delta_{\Gamma}Q_{\rho\alpha\beta}=2 \delta^{\nu}_{\rho}\delta^{\mu}_{(\alpha}g_{\beta)\lambda}\delta \Gamma^{\lambda}_{\;\;\;\mu\nu}
	\eeq
	the total variation of the gravitational sector reads
	\begin{gather}
	\delta_{\Gamma}\int d^{n}x \sqrt{-g}\frac{1}{2 \kappa}\mathcal{L}_{G}=\frac{1}{2 \kappa}\int d^{n}x \sqrt{-g}\delta_{\Gamma}\mathcal{L}_{G}=\nonumber \\
	=\frac{1}{2 \kappa}\int d^{n}x \sqrt{-g}\left( 2 W^{\mu\nu}_{\;\;\;\;\lambda} + V^{\mu\nu}_{\;\;\;\;\lambda}+A_{\lambda}^{\;\;\;\mu\nu} \right) +s.t.
	\end{gather}
	where
	\beq
	\sqrt{-g}A_{\lambda}^{\;\;\;\mu\nu}= (-\nabla_{\alpha}+ 2 S_{\alpha})(\sqrt{-g}\Omega_{\lambda}^{\;\;\;\mu\alpha\nu})-\sqrt{-g}\Omega_{\lambda}^{\;\;\;\mu\gamma\delta}S_{\gamma\delta}^{\;\;\;\;\nu}
	\eeq
	Also recalling that the variation of the matter action with respect to the connection gives
	\beq
	\delta_{\Gamma}S_{M}\equiv \int d^{n}x\left( -\frac{\sqrt{-g}}{2}\Delta_{\lambda}^{\;\;\;\mu\nu}\right) \delta \Gamma^{\lambda}_{\;\;\;\mu\nu}
	\eeq
	we may vary the total action with respect to the connection, to get
	\begin{gather}
	-2\frac{\nabla_{\alpha}(\sqrt{-g}\Omega_{\lambda}^{\;\;\;\mu\alpha\nu})}{\sqrt{-g}}+4 \Omega_{\lambda}^{\;\;\;\mu\alpha\nu}S_{\alpha}-\Omega_{\lambda}^{\;\;\;\mu\gamma\delta}S_{\gamma\delta}^{\;\;\;\;\nu} \nonumber \\
	+2 W^{\mu\nu}_{\;\;\;\;\lambda} + V^{\mu\nu}_{\;\;\;\;\lambda}= \kappa \Delta_{\lambda}^{\;\;\;\mu\nu}
	\end{gather}
	Going back to the metric tensor variation, using
	\beq
	\delta_{g}Q_{\alpha}^{\;\;\mu\nu}=\nabla_{\alpha}(g^{\mu\nu}+\delta g^{\mu\nu})-\nabla_{\alpha}g_{\mu\nu}=+\nabla_{\alpha}\delta g^{\mu\nu}
	\eeq
	it follows that
	\begin{gather}
	\delta_{g}S_{G} =\frac{1}{2 \kappa}\int d^{n}x \sqrt{-g} \delta g^{\mu\nu} \Big[ -\frac{1}{2}g_{\mu\nu}\mathcal{L}_{G}+\frac{\partial \mathcal{L}_{G} }{\partial g^{\mu\nu}} \nonumber \\
	+\frac{1}{\sqrt{-g}}( 2 S_{\alpha}-\nabla_{\alpha}) \sqrt{-g} \frac{\partial \mathcal{L}_{G}}{\partial Q_{\alpha}^{\;\;\;\mu\nu}}\Big]
	\end{gather}
	Therefore, by varying the total action with respect to the connection and applying the principle of least action , we finally get
	\beq
	-\frac{1}{2}g_{\mu\nu}\mathcal{L}_{G}+\frac{\partial \mathcal{L}_{G} }{\partial g^{\mu\nu}} 
	+\frac{1}{\sqrt{-g}}( 2 S_{\alpha}-\nabla_{\alpha})\sqrt{-g} \frac{\partial \mathcal{L}_{G}}{\partial Q_{\alpha}^{\;\;\;\mu\nu}}=\kappa T_{\mu\nu}
	\eeq
	Collecting everything, we conclude that the field equations for a general Metric-Affine $\mathcal{L}(g_{\mu\nu},R^{\alpha}_{\;\;\beta\gamma\rho}, S_{\mu\nu}^{\;\;\;\;\lambda},Q_{\alpha\mu\nu})$ Theory, are
	\begin{gather}
	-2\frac{\nabla_{\alpha}(\sqrt{-g}\Omega_{\lambda}^{\;\;\;\mu\alpha\nu})}{\sqrt{-g}}+4 \Omega_{\lambda}^{\;\;\;\mu\alpha\nu}S_{\alpha}-\Omega_{\lambda}^{\;\;\;\mu\gamma\delta}S_{\gamma\delta}^{\;\;\;\;\nu} \nonumber \\
	+2 W^{\mu\nu}_{\;\;\;\;\lambda} + V^{\mu\nu}_{\;\;\;\;\lambda}= \kappa \Delta_{\lambda}^{\;\;\;\mu\nu}
	\end{gather}
	\beq
	-\frac{1}{2}g_{\mu\nu}\mathcal{L}_{G}+\frac{\partial \mathcal{L}_{G} }{\partial g^{\mu\nu}} 
	+\frac{1}{\sqrt{-g}}( 2 S_{\alpha}-\nabla_{\alpha})\sqrt{-g} \frac{\partial \mathcal{L}_{G}}{\partial Q_{\alpha}^{\;\;\;\mu\nu}}=\kappa T_{\mu\nu}
	\eeq
	where 
	\beq
	\Omega_{\lambda}^{\;\;\;\mu\alpha\nu} \equiv \frac{\partial \mathcal{L}_{G}}{\partial R^{\lambda}_{\;\;\mu\alpha\nu}}\;,\;\; V^{\mu\nu}_{\;\;\;\;\lambda} \equiv \frac{\partial \mathcal{L}_{G}}{\partial S_{\mu\nu}^{\;\;\;\;\lambda}}\;,\; W^{\alpha\mu\nu}  \equiv \frac{\partial \mathcal{L}_{G}}{\partial Q_{\alpha\mu\nu}}
	\eeq
	These are the field equation of the most general Metric-Affine Gravity theory one could think of, since we have included a general dependence of the three basic objects of the underlying geometry namely, the curvature, torsion and non-metricity. Next we shall discuss a simple parity violating theory.

	\section{A Parity Violating Theory}
	Let us study now a parity violating theory of gravity. Our starting point will be the Einstein-Hilbert action
	\beq
	S_{EH}=\frac{1}{2 \kappa} \int d^{n}x \sqrt{-g}R
	\eeq
	plus the parity violating term\footnote{A similar theory but with torsion only was studied in\cite{mukhopadhyaya1999geometrical,sengupta1999parity}.}
	\beq
	\varepsilon^{\mu\nu\alpha\beta}R_{\mu\nu\alpha\beta}
	\eeq
	where $\varepsilon^{\mu\nu\alpha\beta}$ is the totally antisymmetric Levi-Civita tensor. Note that this term  vanishes identically in Metric theories of Gravity due to the identity $\tilde{R}^{\alpha}_{\;\;\;[\mu\nu\rho]}=0$ but when torsion and non-metricity are present this term is not zero\footnote{To be more precise, this term is not zero because of torsion only, as can be seen easily from the identity $R^{\alpha}_{\;\;\;[\beta\mu\nu]}=-2 \nabla_{[\beta}S_{\mu\nu]}^{\;\;\;\;\;\alpha}-4 S_{[\beta\mu}^{\;\;\;\;\;\lambda}S_{\nu]\lambda}^{\;\;\;\;\;\alpha}$.}. So, our total action reads
	\beq
	S=\frac{1}{2 \kappa} \int d^{n}x \sqrt{-g}\Big( R+\alpha \varepsilon^{\mu\nu\alpha\beta}R_{\mu\nu\alpha\beta}\Big) \label{geax}
	\eeq
	where $\alpha$ is a dimensionless parameter, the value of which we will discuss in what follows. Note that this term is not as arbitrary as it may seem at first sight. In fact, we may state that ($\ref{geax}$) is the most general action one can write down, that is linear in the Riemann tensor. Any scalar that is formed by contraction of the Riemann tensor with the metric tensor will give either the Ricci scalar (or a multiple of it) or zero. Then, the only other possibility to form a scalar is by contracting the Riemann tensor with the Levi-Civita tensor. Indeed, taking the various combinations that we mentioned, we have
	\begin{gather}
	R_{\mu\alpha\beta\gamma}( A g^{\mu\alpha}g^{\beta\gamma}+B g^{\mu\beta}g^{\alpha\gamma}+C  g^{\mu\gamma}g^{\alpha\beta}+D \varepsilon^{\mu\nu\alpha\beta})= \nonumber \\
	A\cdot 0+B R +C(-R)+D \varepsilon^{\mu\nu\alpha\beta}R_{\mu\alpha\beta\gamma}=(B-C) R+D \varepsilon^{\mu\nu\alpha\beta}R_{\mu\alpha\beta\gamma}
	\end{gather}
	where $A,B,C,D$ are parameters. Therefore, the most general gravity action that is linear in the Riemann tensor is the Einstein-Hilbert action plus the parity violating term $\varepsilon^{\mu\nu\alpha\beta}R_{\mu\alpha\beta\gamma}$.
	We should point out that the parity violating term is also (just like the Ricci scalar) invariant under projective transformations of the connection and therefore one expects the theory to possess an unspecified vectorial degree of freedom. Let us now examine the filed equations of the theory. In order to vary with respect to the metric tensor we first write
	\begin{gather}
	\delta_{g}S=\frac{1}{2 \kappa} \delta_{g}\int d^{n}x \sqrt{-g}\Big( R+\alpha \varepsilon^{\mu\nu\alpha\beta}R_{\mu\nu\alpha\beta}\Big)=
	\nonumber \\
	=\delta_{g}\int d^{n}x \Big( \sqrt{-g} R+\alpha g_{\mu\kappa}\sqrt{-g}\varepsilon^{\kappa\nu\alpha\beta}R^{\mu}_{\;\;\nu\alpha\beta}\Big) 
	\end{gather}
	and now notice that writing 
	\beq
	\delta_{g}(   \sqrt{-g}\varepsilon^{\kappa\nu\alpha\beta}g_{\mu\kappa})=g_{\mu\kappa}\delta_{g}(   \sqrt{-g}\varepsilon^{\kappa\nu\alpha\beta})+\sqrt{-g}\varepsilon^{\kappa\nu\alpha\beta}(\delta_{g}g_{\mu\kappa})
	\eeq
	and since 
	\beq
	\delta_{g}(   \sqrt{-g}\varepsilon^{\kappa\nu\alpha\beta})=\delta_{g}(   \sqrt{-g}\frac{\epsilon^{\kappa\nu\alpha\beta}}{\sqrt{-g}})=\delta_{g}(   \epsilon^{\kappa\nu\alpha\beta})=0
	\eeq
	we are left with
	\beq
	\delta_{g}(   \sqrt{-g}\varepsilon^{\kappa\nu\alpha\beta}g_{\mu\kappa})=\sqrt{-g}\varepsilon^{\kappa\nu\alpha\beta} (\delta_{g}g_{\mu\kappa})
	\eeq
	Also, recalling that the Riemann tensor $R^{\mu}_{\;\;\nu\alpha\beta}$ is independent of the metric, it follows that
	\beq
	\delta_{g}(\sqrt{-g}\varepsilon^{\mu\nu\alpha\beta}R_{\mu\nu\alpha\beta})=-\sqrt{-g}\varepsilon_{(\nu}^{\;\;\;\lambda\rho\sigma}R_{\mu)\lambda\rho\sigma}\delta g^{\mu\nu}
	\eeq
	Therefore, the total variation with respect to the metric tensor reads\footnote{Notice that now the Levi-Civita tensor $\varepsilon^{\mu\nu\alpha\beta}$ appears and not the symbol. These are related through $\varepsilon^{\mu\nu\alpha\beta}=\frac{1}{\sqrt{-g}}\epsilon^{\mu\nu\alpha\beta}$} 
	\begin{gather}
	\delta_{g}S=\frac{1}{2\kappa}\int d^{n}x\sqrt{-g}(\delta g^{\mu\nu})\left[ R_{(\mu\nu)}-\frac{1}{2}g_{\mu\nu}R-\alpha \varepsilon_{(\nu}^{\;\;\;\alpha\beta\gamma}R_{\mu)\alpha\beta\gamma} \right] =0
	\end{gather}
	and the first set of the field equations, reads
	\beq
	R_{(\mu\nu)}-\frac{1}{2}g_{\mu\nu}R-\alpha \varepsilon_{(\nu}^{\;\;\;\alpha\beta\gamma}R_{\mu)\alpha\beta\gamma}=0
	\eeq
	by contracting this with $g_{\mu\nu}$ we get the relation
	\beq
	\alpha \varepsilon^{\mu\nu\alpha\beta}R_{\mu\nu\alpha\beta}=\Big( 1-\frac{n}{2}\Big) R
	\eeq
	Variation with respect to the connection reads
	\beq
	P_{\lambda}^{\;\;\;\mu\nu}+2 \alpha \left( -\frac{\nabla_{\alpha}(\sqrt{-g}\varepsilon^{\mu\nu\alpha\kappa}g_{\lambda\kappa})}{\sqrt{-g}}  +\varepsilon^{\kappa\mu\beta\gamma}S_{\beta\gamma}^{\;\;\;\;\nu} g_{\lambda\kappa} +4 S_{\alpha}\varepsilon^{\mu\nu\alpha\kappa}g_{\lambda\kappa} \right)=0
	\eeq
	Of course, contracting the latter in $\mu=\lambda$ gives no new identity because of the projective invariance of the action. Contracting the above one time in $\nu=\lambda$ and one time by $g^{\mu\nu}$ we get
	\beq
	P^{\mu}-2 \alpha \tilde{S}^{\mu}=0 \label{pal1}
	\eeq
	and
	\beq
	\tilde{P}_{\lambda}+2 \alpha \tilde{S}_{\lambda}=0 \label{pal2}
	\eeq
	where
	\begin{equation}
	P^{\mu}\equiv P_{\nu}^{\;\;\;\mu\nu}=(n-1)\left[ \tilde{Q}^{\mu}-\frac{1}{2}Q^{\mu}\right] +2(2-n)S^{\mu}
	\end{equation}
	and
	\begin{equation}
	\tilde{P}_{\lambda} \equiv g_{\mu\nu}P_{\lambda}^{\;\;\;\mu\nu}=\frac{(n-3)}{2}Q_{\lambda}+\tilde{Q}_{\lambda}+2(n-2)S_{\lambda}
	\end{equation}
	Furthermore, using
	\beq
	P^{\mu}+\tilde{P}^{\mu}=n \tilde{Q}^{\mu}-Q^{\mu}
	\eeq
	and
	\beq
	P^{\mu}-\tilde{P}^{\mu}=(n-2) (\tilde{Q}^{\mu}-Q^{\mu}-4 S^{\mu})
	\eeq
	we add and subtract equations $(\ref{pal1})$ and $(\ref{pal2})$ to arrive at
	\beq
	n \tilde{Q}^{\mu}-Q^{\mu}=0
	\eeq
	and
	\beq
	(n-2) (\tilde{Q}^{\mu}-Q^{\mu}-4 S^{\mu})= 4 \alpha \tilde{S}^{\mu}
	\eeq
	In addition, multiplying $()$ by $g^{\lambda\alpha}$ we get
	\beq
	P^{\alpha\mu\nu}+ 2 \alpha \left( \frac{Q_{\rho}}{2}\varepsilon^{\rho\alpha\mu\nu}- \nabla_{\rho}\varepsilon^{\rho\alpha\mu\nu}+Q_{\beta\gamma}^{\;\;\;\;\alpha}\varepsilon^{\mu\nu\beta\gamma}+\varepsilon^{\alpha\mu\beta\gamma}S_{\beta\gamma}^{\;\;\;\;\nu}+4 S_{\rho}\varepsilon^{\rho\alpha\mu\nu} \right)=0
	\eeq
	To get another identity for the torsion and non-metricity vectors we contract this by $\varepsilon_{\lambda\alpha\mu\nu}$ and use the identities
	\begin{gather}
	\varepsilon^{\rho\alpha\mu\nu}\varepsilon_{\lambda\alpha\mu\nu}=-3! \delta^{\rho}_{\lambda}  \;,\;\; \varepsilon^{\mu\nu\beta\gamma}\varepsilon_{\lambda\alpha\mu\nu}=-2!2! \delta^{[\beta}_{\lambda}\delta^{\gamma]}_{\alpha} \nonumber \\
	\nabla_{\rho}\varepsilon_{\lambda\alpha\mu\nu} =-\frac{Q_{\rho}}{2} \varepsilon_{\lambda\alpha\mu\nu} \; ,\;\; \varepsilon_{\lambda\alpha\mu\nu}\nabla_{\rho}\varepsilon^{\rho\alpha\mu\nu}=-3! \frac{Q_{\lambda}}{2} \nonumber \\
	\frac{\nabla_{\lambda}\sqrt{-g}}{\sqrt{-g}}=-\frac{Q_{\lambda}}{2}\;,\;\; \varepsilon_{\lambda\alpha\mu\nu}P^{[\alpha\mu\nu]}=- 2 S^{[\alpha\mu\nu]}\varepsilon_{\lambda\alpha\mu\nu}=-2 \tilde{S}_{\lambda}
	\end{gather}
	to obtain
	\beq
	-2 \tilde{S}_{\lambda}+ 2 \alpha \Big( -3! \frac{Q_{\lambda}}{2}+3! \frac{Q_{\lambda}}{2}-4\delta^{[\beta}_{\lambda}\delta^{\gamma]}_{\alpha}Q_{\beta\gamma}^{\;\;\;\;\alpha}-4 \delta^{[\beta}_{\lambda}\delta^{\gamma]}_{\nu}S_{\beta\gamma}^{\;\;\;\;\nu}-3! 4 S_{\lambda} \Big) =0 
	\eeq
	or
	\beq
	\tilde{S}_{\mu}+\alpha \Big[ 2( Q_{\mu}-\tilde{Q}_{\mu}) + 28 S_{\mu}\Big]=0
	\eeq
	Collecting everything, our set of equations for the torsion and non-metricity vectors, is
	\begin{gather}
	Q^{\mu}=n \tilde{Q}^{\mu}  \\
	\frac{(n-1)}{2n}Q_{\mu}+2 S_{\mu}=\frac{2 \alpha}{2-n}\tilde{S}_{\mu}  \\
	\tilde{S}_{\mu}+\alpha \Big[ 2( Q_{\mu}-\tilde{Q}_{\mu}) + 28 S_{\mu}\Big]=0
	\end{gather}
	Note now, that this is a system of $4$ unknowns with $3$ equations and therefore there is no unique solution. Of course, this we already knew since there is an unspecified vectorial degree of freedom due to the projective invariance of the action. What we would like to do is to solve everything in terms of one of these vectors (say $\tilde{S}_{\mu}$). We have some interesting cases depending on the value of  the dimensionless parameter $\alpha$. Let us examine them. Substituting the first two of the above equations into the third one we get
	\beq
	\left( \frac{1}{4}-\frac{2 \alpha^{2}}{n-2} \right)\tilde{S}_{\mu}+5\alpha S_{\mu}=0
	\eeq
	From this we see that for
	\beq
	\alpha= \pm \sqrt{\frac{n-2}{8}}
	\eeq
	we have that
	\beq
	S_{\mu}=0
	\eeq
	and subsequently
	\beq
	\tilde{Q}_{\mu} =\frac{1}{n}Q_{\mu}=\pm \frac{\sqrt{2}}{\sqrt{n-2}{(1-n)}}\tilde{S}_{\mu} \label{stilq}
	\eeq
	So, for this particular value of $\alpha$ we have vanishing torsion vector and the non-metricity vectors are related to the torsion pseudo-vector by equation ($\ref{stilq}$). For $\alpha \neq \pm \sqrt{\frac{n-2}{8}}$ we have that
	\beq
	S_{\mu}=-\frac{1}{5 \alpha}\left( \frac{1}{4}-\frac{2 \alpha^{2}}{n-2} \right)\tilde{S}_{\mu}
	\eeq
	and now the non-metricity vectors are given by
	\beq
	Q_{\mu}=n\tilde{Q}_{\mu}=\frac{n}{5\alpha (n-1)((n-2))}( n-2-28 \alpha^{2}) \tilde{S}_{\mu}
	\eeq
	From which we see that when
	\beq
	n-2-28 \alpha^{2}=0 \Longrightarrow \alpha=\pm \sqrt{\frac{n-2}{28}}
	\eeq
	both the non-metricity vectors vanish
	\beq
	Q_{\mu}=0 \;,\;\; \tilde{Q}_{\mu}=0
	\eeq
	and the torsion vector is given by
	\beq
	S_{\mu}=\mp \frac{1}{2 \sqrt{7(n-2)}} \tilde{S}_{\mu}
	\eeq
	Now, if $\alpha \neq \pm \sqrt{\frac{n-2}{8}}$ and $\alpha=\pm \sqrt{\frac{n-2}{28}}$ then, non of the vectors vanishes and they are related through
	\beq
	S_{\mu}=-\frac{1}{5 \alpha}\left( \frac{1}{4}-\frac{2 \alpha^{2}}{n-2} \right)\tilde{S}_{\mu}
	\eeq
	\beq
	Q_{\mu}=n\tilde{Q}_{\mu}=\frac{n}{5\alpha (n-1)((n-2))}( n-2-28 \alpha^{2}) \tilde{S}_{\mu}
	\eeq
	We should note that it would be interesting to find solutions for the above parity violating Theory. For the time being however we will focus our attention on another subject and see how one can generate torsional and non-metric degrees of freedom by coupling total derivatives to scalars.

	\chapter{Exciting Torsional/Non-mmetric degrees of freedom}
	
	In this short chapter we will study in some detail a general procedure that may be used in order to generate torsional and non-metric degrees of freedom. The recipe here is to couple total derivative terms (that otherwise would be surface terms) to some spacetime function and add them to the Einstein-Hilbert action. We will start with a known model that generates torsion and then present some ways to excite also non-metric degrees of freedom. Then we also present ways that can generate both torsional and non-metric degrees of freedom.

	\section{A way to excite Torsional/Non-Metric d.o.f.}	
	Let us firstly review a model that has been studied in \cite{d1982gravity,leigh2009torsion,petkou2010torsional} but now in the coordinate formalism. It is easy to show that the Nieh-Yan term considered there, translates to
	\begin{equation}
	\epsilon^{\mu\nu\rho\sigma}\partial_{\mu}S_{\nu\rho\sigma}
	\end{equation}
	in the coordinate formalism. Therefore, in our formalism the total action reads
	\begin{gather}
	S=\frac{1}{2\kappa}\int d^{4}x \sqrt{-g}R_{(\mu\nu)}g^{\mu\nu}+\frac{1}{2\kappa}\int d^{4}x F(x)\epsilon^{\mu\nu\rho\sigma}\partial_{\mu}S_{\nu\rho\sigma}= \nonumber \\
	=\frac{1}{2\kappa}\int d^{4}x \sqrt{-g}R_{(\mu\nu)}g^{\mu\nu}-\frac{1}{2\kappa}\int d^{4}x \epsilon^{\mu\nu\rho\sigma}(\partial_{\mu}F)S_{\nu\rho\sigma}+s.t.
	\end{gather}
	where $s.t.$ stands for surface term. Variation with respect to the connection yields
	\begin{gather}
	P_{\lambda}^{\;\;\;\mu\nu}-\frac{1}{\sqrt{-g}}\delta_{\beta}^{[\mu}\delta^{\nu]}_{\gamma}\epsilon^{\alpha\beta\gamma\delta}g_{\lambda\delta}(\partial_{\alpha}F)=0 \Rightarrow \nonumber \\
	\sqrt{-g}P_{\lambda}^{\;\;\;\mu\nu}-\epsilon^{\alpha\mu\nu\delta}g_{\lambda\delta}(\partial_{\alpha}F)=0 \label{aab}
	\end{gather}
	Now, since in this model the non-metricity is zero, the Palatini tensor reads
	\begin{equation}
	P_{\lambda}^{\;\;\;\mu\nu}=2\Big( g^{\mu\nu}S_{\lambda}-S^{\mu}\delta^{\nu}_{\lambda}+g^{\mu\sigma}S_{\sigma\lambda}^{\;\;\;\;\nu} \Big)
	\end{equation}
	and thus
	\begin{equation}
	P^{\nu\rho\sigma}=2\Big( g^{\rho\sigma}S^{\nu}-g^{\sigma\nu}S^{\rho}+ S^{\rho\nu\sigma} \Big) \label{ddf}
	\end{equation}
	Now, equation ($\ref{aab}$) can also be written as\footnote{After a contraction with the metric tensor and some relabeling of the indices.}
	\begin{equation}
	\sqrt{-g}P^{\nu\rho\sigma}-\epsilon^{\alpha\rho\sigma\nu}(\partial_{\alpha}F)=0
	\end{equation}
	and contracting the above with $\epsilon^{\mu\rho\sigma\nu}$ and using ($\ref{ddf}$) we obtain
	\begin{gather}
	2\sqrt{-g} \epsilon_{\mu\rho\sigma\nu}S^{\rho\nu\sigma}-3! \delta^{\alpha}_{\mu}(\partial_{\alpha}F)=0\Rightarrow  \nonumber \\
	\varepsilon_{\mu\nu\rho\sigma}S^{\rho\sigma\nu}=3(\partial_{\mu}F)
	\end{gather}
	where $\varepsilon_{\mu\nu\rho\sigma}\equiv \sqrt{-g} \epsilon_{\mu\rho\sigma\nu}$ is the Levi-Civita tensor. So, we may also write
	\begin{gather}
	\varepsilon_{\mu}^{\;\;\;\rho\sigma\nu}S_{\rho\sigma\nu}=3 (\partial_{\mu}F)
	\end{gather}
	or
	\begin{equation}
	\epsilon^{\mu\nu\rho\sigma}S_{\nu\rho\sigma}=3\sqrt{-g}g^{\mu\nu}(\partial_{\nu}F)
	\end{equation}
	Substituting the latter in our action we arrive at
	\begin{equation}
	S=\frac{1}{2\kappa}\int d^{4}x\Big[ \sqrt{-g}R-\sqrt{-g}3 g^{\mu\nu}(\partial_{\mu}F)(\partial_{\nu}F) \Big]
	\end{equation}
	We can also immediately see that
	\begin{equation}
	S_{\mu}=0
	\end{equation}
	\begin{equation}
	P^{\mu\nu\alpha}=-\varepsilon^{\rho\mu\nu\alpha}\partial_{\rho}F
	\end{equation}
	and
	\begin{equation}
	S_{\mu\nu\alpha}=-\frac{1}{2}\varepsilon_{\mu\nu\alpha\lambda}\partial^{\lambda}F
	\end{equation}
	which when plugged into the connection decomposition yield
	\begin{equation}
	\Gamma^{\lambda}_{\;\;\;\;\mu\nu}=\tilde{\Gamma}^{\lambda}_{\;\;\;\;\mu\nu}+\frac{1}{2}\varepsilon_{\mu\nu}^{\;\;\;\;\;\rho\lambda}\partial_{\rho}F
	\end{equation}
	From which we conclude that this kind of torsion (being totally antisymmetric) has no effect on the autoparallels and the latter coincide with the geodesics. Now, we can fully decompose our original action to a Riemannian part plus an axion field. Indeed, to see this first recall the Ricci scalar decomposition
	\begin{equation}
	R=\tilde{R}+ \tilde{\nabla}_{\mu}( A^{\mu}-B^{\mu})+ B_{\mu}A^{\mu}-N_{\alpha\mu\nu}N^{\mu\nu\alpha}
	\end{equation}
	Note now that the second term is a surface term and can therefore be dropped when taken into the action integral. Regarding the other quantities appearing, we compute for our case
	\begin{equation}
	N_{\mu\nu\alpha}=-\frac{1}{2}\varepsilon_{\mu\nu\alpha\rho}\partial^{\rho}F
	\end{equation}
	\begin{equation}
	A^{\mu}= N^{\mu}_{\;\;\;\nu\beta}g^{\nu\beta}=0, \;\;\; B^{\mu}=N^{\alpha\mu}_{\;\;\;\;\alpha}=0
	\end{equation}
	\begin{equation}
	N_{\alpha\mu\nu}N^{\mu\nu\alpha}=-\frac{3}{2}\partial_{\mu}F\partial^{\mu}F
	\end{equation}
	so that, when substituted back to our action give
	\begin{equation}
	S=\frac{1}{2\kappa}\int d^{4}x\sqrt{-g}\Big[ \tilde{R}-\frac{3}{2} g^{\mu\nu}(\partial_{\mu}F)(\partial_{\nu}F) \Big]
	\end{equation}
	which is the action of Einstein gravity plus an axionic massless field. For this model, the affine connection takes the form
	\begin{equation}
	\Gamma^{\lambda}_{\;\;\;\;\mu\nu}=\tilde{\Gamma}^{\lambda}_{\;\;\;\;\mu\nu}+ N^{\lambda}_{\;\;\;\;\mu\nu}=\tilde{\Gamma}^{\lambda}_{\;\;\;\;\mu\nu}+\frac{1}{2}\varepsilon_{\mu\nu}^{\;\;\;\;\rho\lambda}\partial_{\rho}F
	\end{equation}
	Note now that this type of torsion (totally) antisymmetric has no effect on the autoparallels and the latter coincide with the geodesics. However, for general torsion (even with vanishing non-metricity) the two are not the same.

	\section{Trying to excite non-metricity}
	We consider the model
	\begin{equation}
	S=\frac{1}{2\kappa}\int d^{4}x \sqrt{-g}R_{(\mu\nu)}g^{\mu\nu}+\frac{1}{2\kappa}\int d^{4}x F(x)\partial_{\mu}\epsilon^{\mu\nu\rho\sigma} Q_{\nu}\partial_{\rho}Q_{\sigma} \label{1}
	\end{equation}
	where $Q_{\mu}=-g^{\alpha\beta}\nabla_{\mu}g_{\alpha\beta}$ is the Weyl vector. Notice that  $\partial_{\mu}\epsilon^{\mu\nu\rho\sigma}Q_{\nu}\partial_{\rho}Q_{\sigma}$ alone is a surface term, but when coupled to some function $F(x)$ (as above) cannot be disregarded. In what follows we shall also use the homothetic curvature $\hat{R}_{\mu\nu}$ defined by
	\begin{equation}
	\hat{R}_{\mu\nu}:=\partial_{[\mu}Q_{\nu]}
	\end{equation} 
	Varying $(\ref{1})$ with respect with the metric tensor we obtain
	\begin{equation}
	R_{(\mu\nu)}-\frac{R}{2}g_{\mu\nu}=0
	\end{equation}
	which are the modified Einstein equations. Now, the variation with respect to the connection yields
	\begin{gather}
	-\nabla_{\lambda}(\sqrt{-g}g^{\mu\nu})+\nabla_{\sigma}(\sqrt{-g}g^{\mu\sigma})\delta_{\lambda}^{\nu}+2\sqrt{-g}(g^{\mu\nu}S_{\lambda}-S^{\mu}\delta^{\nu}_{\lambda}+g^{\mu\sigma}S_{\sigma\lambda}^{\;\;\;\nu}) \nonumber \\
	-\delta^{\mu}_{\lambda}\partial_{\alpha}(4F\epsilon^{\alpha\nu\rho\sigma}\hat{R}_{\rho\sigma})=0
	\end{gather}
	where $S_{\mu\nu}^{\;\;\;\;\alpha}:=\Gamma^{\alpha}_{\;\;\;[\mu\nu]}$ is the torsion tensor and $S_{\mu}:=S_{\mu\nu}^{\;\;\;\;\nu}$ the torsion vector. Contracting the latter in $\mu,\lambda$ we arrive at
	\begin{equation}
	\partial_{\mu}(F\epsilon^{\mu\nu\rho\sigma}\hat{R}_{\rho\sigma})=0 \Rightarrow  \nonumber
	\end{equation}
	\begin{equation}
	\partial_{\mu}(\epsilon^{\mu\nu\rho\sigma}F\partial_{\rho}Q_{\sigma})=0 \label{2}
	\end{equation}
	Since we have included an additional field $F(x)$ we must vary with respect to it as well. The variation reads
	\begin{equation}
	\partial_{\mu}(\epsilon^{\mu\nu\rho\sigma} Q_{\nu}\partial_{\rho}Q_{\sigma})=0 \label{3}
	\end{equation} 
	Upon some examination of the field equations, one can show that $Q_{\mu}=0$ and $P_{\lambda}^{\;\;\;\mu\nu}=0$ and thus the model is trivial, leading to Einstein equations in vacuum. In what follows we try a couple of different things to excite non-metric degrees of freedom.

	\section{A Simple Non-metric Model}
	We consider the model given by the action
	\begin{equation}
	I=I_{R}+I_{\Lambda}+2 I_{Q}
	\end{equation}
	where 
	\begin{equation}
	I_{R}=\int _{\mathcal{M}}\epsilon_{abcd}e^{a}\wedge e^{b}\wedge R^{cd}
	\end{equation}
	is the Einstein-Hilbert action(up to numerical factors), $I_{\Lambda}$ is the cosmological constant term
	\begin{equation}
	I_{\Lambda}=\Lambda \int _{\mathcal{M}}\epsilon_{abcd}e^{a}\wedge e^{b}\wedge e^{c}\wedge e^{d}
	\end{equation}
	and we have also considered, in the total action, the presence of the term
	\begin{equation}
	I_{Q}=\int _{\mathcal{M}}F(x)dC_{W}=\frac{1}{2}\int _{\mathcal{M}}F(x)d(Q\wedge R^{a}_{\;\;a})
	\end{equation}
	where $F(x)$ is a scalar and $Q=Q_{ab}\eta^{ab}$ , with $Q_{ab}$ being the non-metricity $1$-form. Notice that the term $dC_{w}$ alone, being a total derivative term, would not affect the field equations. However, if we couple it to a scalar $F(x)$ we a get a non-vanishing contribution. Indeed, one has
	\begin{gather}
	I_{Q}=\frac{1}{2}\int _{\mathcal{M}}F(x)d(Q\wedge R^{a}_{\;\;a})=  \nonumber \\
	=\frac{1}{2}\int _{\mathcal{M}}d\Big[F(x)(Q\wedge R^{a}_{\;\;a})\Big]-\frac{1}{2}\int _{\mathcal{M}} dF(x)\wedge (Q\wedge R^{a}_{\;\;a})= \nonumber \\
	=-\frac{1}{2}\int _{\mathcal{M}} dF(x)\wedge (Q\wedge R^{a}_{\;\;a})+st
	\end{gather} 
	where $st$ stands for surface terms. Having the total action we perform independent variations with respect to $e^{a}$, $\omega^{a}_{\;b}$ and $F(x)$ respectively. We should mention that in this model no a priori assumptions about the torsionlessness and metricity of spacetime have been made.\footnote{That is, the spacetime will possess both torsion and non-metricity in general.} Writing down the total action, we have
	\begin{equation}
	I=\int _{\mathcal{M}}\left[\epsilon_{abcd}e^{a}\wedge e^{b}\wedge R^{cd}+\Lambda \epsilon_{abcd}e^{a}\wedge e^{b}\wedge e^{c}\wedge e^{d}-dF(x)\wedge (Q\wedge R^{a}_{\;\;a}) \right] \nonumber
	\end{equation}
	Variation with respect to the vierbeins ($e^{a}$) yields
	\begin{equation}
	\epsilon_{abcd} e^{b}\wedge \left[ R^{cd}-\frac{\Lambda}{3}e^{c}\wedge e^{d} \right] =0
	\end{equation}
	which we recognize as the modified Einstein equations with torsion and non-metricity.\footnote{Note that here the spin connection $\omega^{a}_{\;\;b}$ is not related to $e^{a}$ with the usual way. A relation between them may be found after solving the field equations.} Varying with respect to $\omega^{a}_{\;\; b}$ we obtain
	\begin{equation}
	\eta_{bc}\mathcal{D}\left[ \epsilon_{lmae}e^{l}\wedge e^{m}\eta^{ce}-\delta_{a}^{c}dF\wedge Q\right]+2n_{ab}dF\wedge R^{c}_{\;\; c}=0
	\end{equation}
	where $\mathcal{D}$ represents covariant differentiation and $Q_{ab}=-\mathcal{D}\eta_{ab}=2\omega_{(ab)}$ is the non-metricity tensor. Finally, the $F$-variation gives the constraint 
	\begin{equation}
	d(Q\wedge R^{c}_{\;\; c})=0
	\end{equation}

	\section{Model in the coordinate formalism}
	Let us consider the model of the previous section in the coordinate formulation\footnote{That is, the action is expressed in terms of the metric instead of the vielbeins.}
	\begin{equation}
	S=\frac{1}{2\kappa}\int d^{4}x \sqrt{-g}R_{(\mu\nu)}g^{\mu\nu}+\frac{1}{2\kappa}\int d^{4}x F(x)\partial_{\mu}\epsilon^{\mu\nu\rho\sigma} Q_{\nu}\partial_{\rho}Q_{\sigma} \label{1}
	\end{equation}
	where $Q_{\mu}=-g^{\alpha\beta}\nabla_{\mu}g_{\alpha\beta}$ is the Weyl vector. Notice that  $\partial_{\mu}\epsilon^{\mu\nu\rho\sigma}Q_{\nu}\partial_{\rho}Q_{\sigma}$ alone is a surface term, but when coupled to some function $F(x)$ (as above) cannot be disregarded. In what follows we shall also use the homothetic curvature $\hat{R}_{\mu\nu}$ defined by
	\begin{equation}
	\hat{R}_{\mu\nu}:=\partial_{[\mu}Q_{\nu]}
	\end{equation} 
	Varying $(\ref{1})$ with respect with the metric tensor we obtain
	\begin{equation}
	R_{(\mu\nu)}-\frac{R}{2}g_{\mu\nu}=0
	\end{equation}
	which are the modified Einstein equations. Now, the variation with respect to the connection yields
	\begin{gather}
	-\nabla_{\lambda}(\sqrt{-g}g^{\mu\nu})+\nabla_{\sigma}(\sqrt{-g}g^{\mu\sigma})\delta_{\lambda}^{\nu}+2\sqrt{-g}(g^{\mu\nu}S_{\lambda}-S^{\mu}\delta^{\nu}_{\lambda}+g^{\mu\sigma}S_{\sigma\lambda}^{\;\;\;\nu}) \nonumber \\
	-\delta^{\mu}_{\lambda}\partial_{\alpha}(4F\epsilon^{\alpha\nu\rho\sigma}\hat{R}_{\rho\sigma})=0 \label{f}
	\end{gather}
	where $S_{\mu\nu}^{\;\;\;\;\alpha}:=\Gamma^{\alpha}_{\;\;\;[\mu\nu]}$ is the torsion tensor and $S_{\mu}:=S_{\mu\nu}^{\;\;\;\;\nu}$ the torsion vector. Contracting the latter in $\mu,\lambda$ we arrive at
	\begin{equation}
	\partial_{\mu}(F\epsilon^{\mu\nu\rho\sigma}\hat{R}_{\rho\sigma})=0 \Rightarrow  \nonumber
	\end{equation}
	\begin{equation}
	\partial_{\mu}(\epsilon^{\mu\nu\rho\sigma}F\partial_{\rho}Q_{\sigma})=0 \label{2}
	\end{equation}
	Since we have included an additional field $F(x)$ we must vary with respect to it as well. The variation reads
	\begin{equation}
	\partial_{\mu}(\epsilon^{\mu\nu\rho\sigma} Q_{\nu}\partial_{\rho}Q_{\sigma})=0 \label{3}
	\end{equation} 
	Now, substituting ($\ref{2}$) back in $(\ref{f})$ we arrive at
	\begin{equation}
	-\nabla_{\lambda}(\sqrt{-g}g^{\mu\nu})+\nabla_{\sigma}(\sqrt{-g}g^{\mu\sigma})\delta_{\lambda}^{\nu}+2\sqrt{-g}(g^{\mu\nu}S_{\lambda}-S^{\mu}\delta^{\nu}_{\lambda}+g^{\mu\sigma}S_{\sigma\lambda}^{\;\;\;\nu})=0
	\end{equation}
	which, as we have already seen, after some manipulations implies that
	\begin{equation}
	S_{\mu\nu}^{\;\;\;\;\lambda}=\Gamma^{\lambda}_{\;\;\;[\mu\nu]}=0
	\end{equation}
	\begin{equation}
	Q_{\alpha}^{\;\;\;\mu\nu}=\nabla_{\alpha}g^{\mu\nu}=0
	\end{equation}
	Thus, the theory considered here is equivalent to General Relativity.

	\subsection{Side Note: Connecting the two formalisms}
	Having introduced both the coordinate (here the fields are $g_{\mu\nu}$, $\Gamma^{\lambda}_{\;\;\;\mu\nu}$) and the vierbein ($e^{a}$, $\omega_{ab}$) formalisms let us now give some connecting identities that help one to switch from one formalism to another. The starting point of everything that follows is the vielbein postulate. This states that the vielbeins are covariantly conserved, namely\footnote{Note that our definition of the index placing in both the vielbein and the spin connection is the following:  the coordinate index goes first (left side) and then the Lorentz indices follow.}
	\begin{equation}
	\nabla_{\nu}e_{\mu}^{\;\; a}=0
	\end{equation}
	Expanding the latter we derive
	\begin{equation}
	\partial_{\nu}e_{\mu}^{\;\; a}-\Gamma^{\rho}_{\;\;\;\mu\nu}e_{\rho}^{\;\;a}+\omega_{\nu\;\;b}^{\;\;a}e_{\mu}^{\;\;b}=0
	\end{equation}
	Now, to solve for $\Gamma^{\lambda}_{\;\;\;\mu\nu}$ we simply multiply (and contract) with $e^{\lambda}_{\;\;a}$ to get
	\begin{equation}
	\Gamma^{\lambda}_{\;\;\;\mu\nu}=e^{\lambda}_{\;\; a}\partial_{\nu}e_{\mu}^{\;\; a}+\omega_{\nu \;\; b}^{\;\;a}e^{\lambda}_{\;\; a}e_{\mu}^{\;\;b}
	\end{equation}
	while, multiplication with $e^{\mu}_{\;\; c}$ solves for $\omega_{\mu a b}$,
	\begin{equation}
	\omega_{\nu \;\; c}^{\;\; a}=\Gamma^{\lambda}_{\;\;\;\mu\nu}e_{\lambda}^{\;\; a}e^{\mu}_{\;\; c}-e^{\mu}_{\;\; c}\partial_{\nu}e_{\mu}^{\;\; a}
	\end{equation}
	or
	\begin{gather}
	\omega_{\nu ab}=\Gamma^{\lambda}_{\;\;\;\mu\nu}e_{\lambda a}e^{\mu}_{\;\; b}-e^{\mu}_{\;\; b}\partial_{\nu}e_{\mu a}= \nonumber \\
	=\Gamma^{\lambda}_{\;\;\;\mu\nu}e_{\lambda a}e^{\mu}_{\;\; b}+e_{\mu a}\partial_{\nu}e^{\mu}_{\;\; b} \Rightarrow \nonumber
	\end{gather}
	\beq
		\omega_{\nu ab}=\Big( \Gamma^{\lambda}_{\;\;\;\mu\nu}e^{\mu}_{\;\; b} +\partial_{\nu}e^{\lambda}_{\;\; b}\Big) e_{\lambda a}
	\eeq
	where we have also made some relabeling of the indices and used the fact that $e^{\mu}_{\;\; b}e_{\mu a}=\eta_{ba}$. We should point out that one would like to have an expression of $\omega_{\mu a b}$ in terms of the affine connection and the metric tensor only\footnote{in the above the spin connection depends on the vielbeins as well.}. However, such a possibility does not exist since the vielbeins carry more degrees of freedom than  the metric tensor does and therefore the former cannot be solved in terms of the latter. Notice now that if we define the 'Lorentz blind'covariant derivative 
	\begin{equation}
	\hat{\nabla}_{\mu}A^{\lambda}_{\;\;b}:=\partial_{\mu}A^{\lambda}_{\;\;b}+\Gamma^{\lambda}_{\;\;\;\nu\mu}A^{\nu}_{\;\;b}
	\end{equation}
	that is the covariant derivative that 'sees' only the coordinate indices, than the above can be written as
	\begin{equation}
	\omega_{\mu a b}=e_{\lambda a}\hat{\nabla}_{\mu}e^{\lambda}_{\;\;b}
	\end{equation}
	In a similar manner we define the 'coordinate blind' covariant derivative or else the gauge covariant derivative to be
	\begin{equation}
	D_{\nu}A_{\mu}^{\;\;b}:= \partial_{\nu}A_{\mu}^{\;\;b}+\omega_{\nu\;\;a}^{\;\;b}A_{\mu}^{\;\;a}
	\end{equation}
	which when applied on the vielbeins, yields
	\begin{equation}
	D_{\nu}e_{\mu}^{\;\;b}:= \partial_{\nu}e_{\mu}^{\;\;b}+\omega_{\nu\;\;a}^{\;\;b}e_{\mu}^{\;\;a}
	\end{equation}
	and as a result, ($\ref{gamma}$) can be written in the handy form
	\begin{equation}
	\Gamma^{\lambda}_{\;\;\;\mu\nu}=e^{\lambda}_{\;\;b}D_{\nu}e_{\mu}^{\;\;b} \label{plk}
	\end{equation}
	We would now like to show that the definitions of both the torsion and non-metricity coincide (up to sign factors due to definitions) for the two formalisms. Let us start with torsion. In the coordinate formalism, the latter is defined by
	\begin{equation}
	S_{\mu\nu}^{\;\;\;\;\lambda}:=\Gamma^{\lambda}_{\;\;\;[\mu\nu]}
	\end{equation}
	and upon using ($\ref{plk}$) it can also be expressed as
	\begin{equation}
	S_{\mu\nu}^{\;\;\;\;\lambda}=e^{\lambda}_{\;\;b}D_{[\nu}e_{\mu]}^{\;\;b}
	\end{equation}
	In the first order formalism, torsion is defined by the $2$-form
	\begin{equation}
	T^{b}:=De^{b}=d e^{b}+\omega^{b}_{\;\;a}\wedge e^{a}
	\end{equation}
	which when expanded in the $\{ dx^{\mu}\}$ basis, gives
	\begin{equation}
	T^{b}=T_{\nu\mu}^{\;\;\;\;b}dx^{\nu}\wedge dx^{\mu}=(\partial_{[\nu}e_{\mu]}^{\;\;b}+\omega_{[\nu\;\;a}^{\;\;\;b}e_{\mu]}^{\;\;a})dx^{\nu}\wedge dx^{\mu}= D_{[\nu}e_{\mu]}^{\;\;b}dx^{\nu}\wedge dx^{\mu}
	\end{equation}
	that is
	\begin{equation}
	T_{\nu\mu}^{\;\;\;\;b}= D_{[\nu}e_{\mu]}^{\;\;b}
	\end{equation}
	We therefore, conclude that
	\begin{equation}
	T_{\nu\mu}^{\;\;\;\;\lambda}=e^{\lambda}_{\;\;b}T_{\nu\mu}^{\;\;\;\;b}=e^{\lambda}_{\;\;b}D_{[\nu}e_{\mu]}^{\;\;b}=S_{\mu\nu}^{\;\;\;\;\lambda} \Rightarrow \nonumber
	\end{equation}
	\begin{equation}
	T_{\mu\nu}^{\;\;\;\;\lambda}=-S_{\mu\nu}^{\;\;\;\;\lambda}
	\end{equation}
	which is what we wanted to show. Notice the appearance of the minus sign which is purely conventional and a remnant of the index placing definition in the affine connection. 
	Keeping this in mind and sticking strictly to our definitions, this sign difference is not going to cause any problems. Another thing we can comment on is that even though torsion depends only on the affine connection (no metric dependence) in the coordinate formalism, when working on the vielbein formalism, the torsion $2$-form depends both on $e^{a}$ and $\omega^{a}_{\;\;b}$.
	So long as non-metricity is concerned, in the coordinate formalism the definition reads
	\begin{equation}
	Q_{\lambda\mu\nu}:= -\nabla_{\lambda}g_{\mu\nu} \label{nmt}
	\end{equation}
	while in the vielbein formalism is given by the one form
	\begin{equation}
	Q_{ab}:=-D\eta_{ab}=\omega_{a b}+\omega_{b a}=2\omega_{(a b)}\Rightarrow  \nonumber
	\end{equation}
	\begin{equation}
	Q_{ab}=Q_{\lambda a b}dx^{\lambda}=(\omega_{\mu a b}+\omega_{\lambda b a})dx^{\lambda}
	\end{equation}
	What we are to show now is that given one from the above we can compute the other one by some multiplication with the vielbeins. More specifically, we show that
	\begin{equation}
	Q_{\lambda \mu\nu}=e_{\mu}^{\;\;a}e_{\nu}^{\;\;b}Q_{\lambda a b}=2e_{\mu}^{\;\;a}e_{\nu}^{\;\;b}\omega_{\mu(a b)}
	\end{equation}
	To start with, first recall the relation relating the metric tensor with the vielbeins
	\begin{equation}
	g_{\mu\nu}=e_{\mu}^{\;\;a}e_{\nu}^{\;\;b}\eta_{a b}
	\end{equation}
	which when substituted in ($\ref{nmt}$) yields
	\begin{equation}
	Q_{\lambda\mu\nu}= -\nabla_{\lambda}g_{\mu\nu} =-e_{\mu}^{\;\;a}e_{\nu}^{\;\;b}\nabla_{\lambda}\eta_{a b}=+2 e_{\mu}^{\;\;a}e_{\nu}^{\;\;b}\omega_{\lambda (a b)}=e_{\mu}^{\;\;a}e_{\nu}^{\;\;b}Q_{\lambda a b}
	\end{equation}
	as desired. In addition, taking the interior product of 
	\begin{equation}
	Q_{ab}=Q_{\lambda a b }dx^{\lambda}=Q_{d a b}e^{d}
	\end{equation} 
	with $e_{c}$, it follows that
	\begin{equation}
	Q_{c a b}=e_{c}\rfloor Q_{a b}=2 e_{c}\rfloor \omega_{(a b)}
	\end{equation}
	such that
	\begin{equation}
	Q_{\lambda\mu\nu}=e_{\lambda}^{\;\;c}e_{\mu}^{\;\;a}e_{\nu}^{\;\;b} 2(e_{c}\rfloor \omega_{(ab)})
	\end{equation}
	Note now that while in the first order formalism the non-metricity $1$-form depends only on the spin connection (vielbein independent) when switching to coordinate formalism, the non-metricity tensor depends both on $g_{\mu\nu}$ and $\Gamma^{\lambda}_{\;\;\;\mu\nu}$. For more on the frame and exterior form formalism of gravity one may consult \cite{aldrovandi2010introduction,hehl1978metric}. We may now proceed  searching for ways to excite non-metric degrees of freedom.

	\section{A way to excite non-metric d.o.f.}
	Let us start again with the Einstein-Hilbert action and couple the surface term
	\begin{equation}
	\partial_{\mu}( \sqrt{-g} Q^{\mu})
	\end{equation}
	where $Q_{\mu}$ is the Weyl vector, to a scalar $F$. In words, the action reads
	\begin{equation}
	S=\frac{1}{2\kappa}\int d^{n}x \sqrt{-g}R_{(\mu\nu)}g^{\mu\nu}+\frac{1}{2\kappa}\int d^{n}x F(x)\partial_{\mu}( \sqrt{-g} Q^{\mu})
	\end{equation}
	Variation of the above with respect to $g_{\mu\nu}$ yields the modified Einstein equations, while the variation with respect to the connection gives
	\begin{equation}
	P_{\lambda}^{\;\;\;\mu\nu}-2\delta_{\lambda}^{\mu}g^{\alpha\nu}\partial_{\alpha}F=0 \label{2}
	\end{equation}
	Now, contracting in $\mu=\lambda$ and using the tracelessness of the Palatini tensor in its first two indices ($P_{\mu}^{\;\;\;\mu\nu}=0$), we arrive at
	\begin{equation}
	g^{\alpha\nu}\partial_{\alpha}F=0 \label{3}
	\end{equation}
	which when substituted back in ($\ref{2}$) implies that $P_{\lambda}^{\;\;\;\mu\nu}=0$ and therefore the theory considered here is again trivial. Same goes when one tries to add a term that includes the second non-metricity vector $\tilde{Q}_{\mu}$ instead of the Weyl vector. Indeed, if we consider
	\begin{equation}
	S=\frac{1}{2\kappa}\int d^{n}x \sqrt{-g}R_{(\mu\nu)}g^{\mu\nu}+\frac{1}{2\kappa}\int d^{n}x F(x)\partial_{\mu}( \sqrt{-g} \tilde{Q}^{\mu})
	\end{equation}
	then upon varying with respect to the connection we arrive at
	\begin{equation}
	P_{\lambda}^{\;\;\;\mu\nu}- (\delta_{\lambda}^{\alpha}g^{\mu\nu}+g^{\alpha\mu}\delta_{\lambda}^{\nu})\partial_{\alpha}F=0
	\end{equation}
	and again it follows that $P_{\lambda}^{\;\;\;\mu\nu}=0$ and $\partial_{\mu}F=0$ as before. So, we now want to find a way to avoid this triviality. This triviality comes about due to the fact that while the Ricci scalar is invariant under projective transformations of the connection
	\begin{equation}
	\Gamma^{\lambda}_{\;\;\mu\nu}\longrightarrow \Gamma^{\lambda}_{\;\;\mu\nu}+ \delta_{\mu}^{\lambda}\xi_{\nu} \label {4}
	\end{equation}
	the additional term in the action (in both cases) does not respect this symmetry and thus we have a vanishing vectorial degree of freedom. So, what we want, is to add a term that its variation with respect to the connection will yield a tensor that is also traceless in its first two indices. To this end, we add to the action a term that goes like
	\begin{equation}
	I \propto \int d^{n}x F(x)\partial_{\mu}\Big[ \sqrt{-g} (\alpha Q^{\mu}+\beta\tilde{Q}^{\mu})\Big]
	\end{equation}
	and choose the parameters $\alpha$, $\beta$ in such a way so that the tensor obtained after varying with respect to the connection is identically traceless without imposing any field equation.  We have
	\begin{gather}
	A_{\lambda}^{\;\;\;\mu\nu} \equiv \frac{\delta}{\delta \Gamma^{\lambda}_{\;\;\;\mu\nu}}\left(\sqrt{-g}(\partial_{\alpha}F)(\alpha Q^{\alpha}+\beta\tilde{Q}^{\alpha})  \right) = \\ \nonumber
	=\sqrt{-g}(\partial_{\alpha}F) \Big[ 2\alpha \delta_{\lambda}^{\mu}g^{\alpha\nu}+ \beta (\delta_{\lambda}^{\alpha}g^{\mu\nu}+g^{\mu\alpha}\delta_{\lambda}^{\nu}) \Big]
	\end{gather}
	and 
	\begin{equation}
	A_{\mu}^{\;\;\;\mu\nu}=\sqrt{-g}(\partial_{\alpha}F)g^{\alpha\nu}2(\alpha n+ \beta)
	\end{equation}
	So, we see that if we choose $\alpha=-\beta/n$,  $A_{\mu}^{\;\;\;\mu\nu}$ is identically zero. In fact, there is a deeper reason why one should choose this relation between the parameters. The reason being the projective invariance of the action that we discussed before. To see this, we first note that under a projective transformation of the form of ($\ref{4}$), the non-metricity tensor changes as follows
	\begin{equation}
	Q_{\alpha\mu\nu}\longrightarrow Q_{\alpha\mu\nu} +2\xi_{\alpha}g_{\mu\nu}
	\end{equation}
	and therefore the Weyl and second non-metricity vectors change correspondingly as
	\begin{equation}
	Q_{\mu}\longrightarrow Q_{\mu} +2 n \xi_{\mu}
	\end{equation}
	\begin{equation}
	\tilde{Q}_{\mu}\longrightarrow \tilde{Q}_{\mu} +2 \xi_{\mu}
	\end{equation}
	Therefore, $Q_{\mu}$  and $\tilde{Q}_{\mu}$ are not projective invariant individually, but the combination
	\begin{equation}
	Q_{\mu}-n \tilde{Q}_{\mu}
	\end{equation}
	remains invariant under ($\ref{4}$), which of course amounts to the same choice for the parameters $\alpha$, $\beta$ as above. Having said this let us now go back to our model. The action reads
	\begin{gather}
	S=\frac{1}{2\kappa}\int d^{n}x \sqrt{-g}R_{(\mu\nu)}g^{\mu\nu}+\frac{\beta}{2 \kappa}\int d^{n}x F(x)\partial_{\mu}\left( \sqrt{-g} (-\frac{Q^{\mu}}{4}+\tilde{Q}^{\mu}) \right) \\ \nonumber
	=\frac{1}{2\kappa}\int d^{n}x \sqrt{-g}R_{(\mu\nu)}g^{\mu\nu}-\frac{\beta}{2 \kappa}\int d^{n}x \sqrt{-g}(\partial_{\mu}F)(-\frac{Q^{\mu}}{4}+\tilde{Q}^{\mu}) + s.t. \label{6}
	\end{gather} 
	From now on we shall focus in four dimensions ($n=4$) and set $\beta=1$ because it can always be absorbed in the definition of $F$. Variation of the above with respect to the metric tensor gives us the modified Einstein equations with non-metricity, while the variation with respect to the connection yields
	\begin{equation}
	P_{\lambda}^{\;\;\;\mu\nu}- (\partial_{\alpha}F) \Big[ -\frac{1}{2} g^{\alpha\nu}\delta_{\lambda}^{\mu}+\delta_{\lambda}^{\alpha}g^{\mu\nu}+\delta_{\lambda}^{\nu}g^{\mu\alpha} \Big]=0 \label{5}
	\end{equation}
	Note now that if we contract the in $\mu=\lambda$ we do not get any constraint since the action we consider is projective invariant. It is easy to show that the affine connection for this model is
	\begin{equation}
	\Gamma^{\lambda}_{\;\;\;\;\mu\nu}=\tilde{\Gamma}^{\lambda}_{\;\;\;\;\mu\nu}+\frac{1}{4}g_{\mu\nu}\partial^{\lambda}F+\frac{1}{4}\delta^{\lambda}_{\nu}\partial_{\mu}F-\frac{5}{4}\delta^{\lambda}_{\mu}\partial_{\nu}F
	\end{equation}
	Having this we compute for the torsion and non-metricity tensors
	\begin{equation}
	S_{\mu\nu}^{\;\;\;\;\lambda}=\frac{3}{2}\delta_{[\nu}^{\lambda}\partial_{\mu]}F
	\end{equation}
	and
	\begin{equation}
	Q_{\alpha\mu\nu}=-2 g_{\alpha(\mu}\partial_{\nu)}F+\frac{1}{2}g_{\mu\nu}\partial_{\alpha}F
	\end{equation}
	respectively. We also compute
	\begin{equation}
	-\frac{1}{4}Q_{\mu}+\tilde{Q}_{\mu}=\frac{9}{4}\partial_{\mu}F
	\end{equation}

	Upon substitution of the latter two into our action we arrive at
	\begin{equation}
	S= \frac{1}{2\kappa}\int d^{4}x \sqrt{-g}R_{(\mu\nu)}g^{\mu\nu}-\frac{1}{2 \kappa}\int d^{4}x \sqrt{-g}\frac{9}{4}g^{\mu\nu}\partial_{\mu}F\partial_{\nu}F
	\end{equation}
	
	\section{Exciting both torsional and non-metric d.o.f.}
	
	\subsection{Model $1$}
	Let us now try to excite torsional degrees of freedom along with the non-metric ones. To this end, we must add a surface term (coupled to a scalar) that also includes the torsion vector and the total combination has again to be invariant under projective transformations. As we have already seen, the Weyl and second non-metricity vectors, transform as
	\begin{equation}
	Q_{\mu}\longrightarrow Q_{\mu} +2 n \xi_{\mu}
	\end{equation}
	\begin{equation}
	\tilde{Q}_{\mu}\longrightarrow \tilde{Q}_{\mu} +2 \xi_{\mu}
	\end{equation}
	under a projective transformation of the connection. From the definition of torsion it is obvious that under (\ref{4}) the latter transforms as
	\begin{equation}
	S_{\mu\nu}^{\;\;\;\;\lambda}\longrightarrow S_{\mu\nu}^{\;\;\;\;\lambda}+\delta_{[\mu}^{\lambda}\xi_{\nu]}
	\end{equation}
	from which we deduce the transformation law for the torsion vector
	\begin{equation}
	S_{\mu} \longrightarrow S_{\mu} + \frac{(1-n)}{2}\xi_{\mu}
	\end{equation}
	From the above we see that an obvious combination (including all three) that remains invariant, under a projective transformation, is the following
	\begin{equation}
	Q_{\mu}-\tilde{Q}_{\mu}+4 S_{\mu}
	\end{equation}
	So, our model consists of the action
	\begin{gather}
	S=\frac{1}{2\kappa}\int d^{n}x \sqrt{-g}R_{(\mu\nu)}g^{\mu\nu}+\frac{1}{2 \kappa}\int d^{n}x F(x)\partial_{\mu}\Big( \sqrt{-g} (Q^{\mu}-\tilde{Q}^{\mu} + 4 S^{\mu}) \Big) \\ \nonumber
	=\frac{1}{2\kappa}\int d^{n}x \sqrt{-g}R_{(\mu\nu)}g^{\mu\nu}-\frac{1}{2 \kappa}\int d^{n}x \sqrt{-g}(\partial_{\mu}F)(Q^{\mu}-\tilde{Q}^{\mu} + 4 S^{\mu})   + s.t. \label{9}
	\end{gather} 
	Let us concentrate on the variation with respect to the connection. This yields
	\begin{gather}
	P_{\lambda}^{\;\;\;\mu\nu}=(\partial_{\alpha}F)\Big[ 2 \delta_{\lambda}^{\mu}g^{\alpha\nu}-(\delta_{\lambda}^{\alpha}g^{\mu\nu}+\delta_{\lambda}^{\nu}g^{\alpha\mu})+2(\delta_{\lambda}^{\nu}g^{\alpha\mu}-\delta_{\lambda}^{\mu}g^{\alpha\nu}) \Big]= \nonumber \\
	=(\partial_{\alpha}F)(\delta_{\lambda}^{\nu}g^{\mu\alpha}-\delta_{\lambda}^{\alpha}g^{\mu\nu})=2(\partial_{\alpha}F)g^{\mu[\alpha}\delta_{\lambda}^{\nu]} \label{10}
	\end{gather}
	Contracting in $\nu=\lambda$ we arrive at
	\begin{equation}
	(n-1)\left( \tilde{Q}^{\mu}-\frac{Q^{\mu}}{2} \right) + 2(2-n)S^{\mu}=(n-1)g^{\mu\alpha}(\partial_{\alpha}F)
	\end{equation}
	or 
	\begin{equation}
	(n-1)\left( \tilde{Q}_{\mu}-\frac{Q_{\mu}}{2} \right) + 2(2-n)S_{\mu}=(n-1)(\partial_{\mu}F)
	\end{equation}
	In addition, contracting (\ref{10}) with $g_{\mu\nu}$ it follows that
	\begin{equation}
	\frac{(n-3)}{2}Q_{\mu}+\tilde{Q}_{\mu}+2(n-2)S_{\mu}= (1-n)(\partial_{\mu}F)
	\end{equation}
	Adding the latter two we obtain
	\begin{equation}
	Q_{\mu}= n \tilde{Q}_{\mu}
	\end{equation}
	which when substituted to either one of the above, yields
	\begin{equation}
	(n-1)\tilde{Q}_{\mu}+4 S_{\mu}= 2\left(\frac{n-1}{2-n}\right) \partial_{\mu}F
	\end{equation}
	Now, observe that this very quantity is the same appearing in the action. Indeed, using the latter two equations we may write 
	\begin{gather}
	Q_{\mu}-\tilde{Q}_{\mu}+4 S_{\mu}=(n-1)\tilde{Q}_{\mu}+4 S_{\mu}= 2\left(\frac{n-1}{2-n}\right) \partial_{\mu}F
	\end{gather}
	Therefore, substituting the above algebraic into our action, we arrive at
	\begin{equation}
	\frac{1}{2\kappa}\int d^{n}x \sqrt{-g}R_{(\mu\nu)}g^{\mu\nu}+\frac{1}{2 \kappa}\int d^{n}x \sqrt{-g}2\left(\frac{n-1}{n-2}\right)g^{\mu\nu} (\partial_{\mu}F)(\partial_{\nu}F)  + s.t. 
	\end{equation}
	
	\subsection{Model $2$}
	
	Let us now consider the action
	\begin{gather}
	S=\frac{1}{2\kappa}\int d^{4}x \sqrt{-g}R_{(\mu\nu)}g^{\mu\nu}+ \nonumber \\
	\frac{1}{2 \kappa}\int d^{4}x \left[ \phi(x)\partial_{\mu}\left( \sqrt{-g} (-\frac{Q^{\mu}}{4}+\tilde{Q}^{\mu}) \right) + \chi (x)\epsilon^{\mu\nu\rho\sigma}\partial_{\mu}S_{\nu\rho\sigma} \right] \\ \nonumber
	=\frac{1}{2\kappa}\int d^{4}x \sqrt{-g}R_{(\mu\nu)}g^{\mu\nu} \nonumber \\ -\frac{1}{2 \kappa}\int d^{4}x \sqrt{-g}\left[ (\partial_{\mu}\phi)(-\frac{Q^{\mu}}{4}+\tilde{Q}^{\mu})+ \epsilon^{\mu\nu\rho\sigma}(\partial_{\mu}\chi)S_{\nu\rho\sigma} \right] + s.t. 
	\end{gather} 
	where we now consider two scalar fields, $\phi(x)$ and $\chi(x)$. Varying with respect to the connection we obtain
	\begin{equation}
	P_{\lambda}^{\;\;\;\mu\nu}-\frac{1}{\sqrt{-g}}\epsilon^{\alpha\mu\nu\delta}g_{\lambda\delta}(\partial_{\alpha}\phi)-(\partial_{\alpha}\chi)\Big[ -\frac{1}{2}g^{\alpha\nu}\delta_{\lambda}^{\mu}+ g^{\mu\nu}\delta_{\lambda}^{\alpha}+g^{\mu\alpha}\delta_{\lambda}^{\nu} \Big] =0
	\end{equation}
	The latter can also be written as
	\begin{equation}
	P^{\nu\rho\sigma}-\frac{1}{\sqrt{-g}}\epsilon^{\alpha\rho\sigma\nu}(\partial_{\alpha}\phi)-(\partial_{\alpha}\chi)\Big[ -\frac{1}{2}g^{\alpha\sigma}g^{\rho\nu}+g^{\alpha\nu}g^{\rho\sigma}+g^{\sigma\nu}g^{\rho\alpha} \Big] =0 \label{mmn}
	\end{equation}
	and upon contracting with $\epsilon_{\mu\nu\rho\sigma}$ we arrive at
	\begin{gather}
	3 g^{\mu\alpha}\partial_{\mu}\phi =\varepsilon^{\alpha\nu\rho\sigma}S_{\nu\rho\sigma} \Rightarrow  \nonumber \\
	\epsilon^{\mu\nu\rho\sigma}S_{\nu\rho\sigma}=3 \sqrt{-g}g^{\mu\nu} (\partial_{\nu}\phi) \label{dc}
	\end{gather}
	Now, contracting ($\ref{mmn}$) in $\lambda=\nu$ we obtain
	\begin{equation}
	3 \Big[ \tilde{Q}^{\mu}-\frac{1}{2}Q^{\mu} \Big] -4 S^{\mu} =\frac{9}{2}g^{\mu\nu}(\partial_{\nu}\chi) \label{bbf}
	\end{equation}
	In addition, a contraction with $g_{\mu\nu}$ yields
	\begin{equation}
	\frac{1}{2}Q_{\lambda}+\tilde{Q}_{\lambda}+ 4 S_{\lambda}-\frac{9}{2}(\partial_{\lambda}\chi) =0 
	\end{equation}
	or in its contravariant form
	\begin{equation}
	\frac{1}{2}Q^{\mu}+\tilde{Q}^{\mu}+ 4 S^{\mu} =\frac{9}{2}g^{\mu\nu}(\partial_{\nu}\chi) 
	\end{equation}
	Adding the latter with $(\ref{bbf})$ we obtain
	\begin{gather}
	4\tilde{Q}^{\mu}-Q^{\mu}=9 g^{\mu\nu}(\partial_{\nu}\chi) \Rightarrow \nonumber
	-\frac{1}{4}Q^{\mu}+\tilde{Q}^{\mu}=\frac{9}{4} g^{\mu\nu}(\partial_{\nu}\chi)
	\end{gather}
	Note now that, remarkably, the torsion vector has been dropped out and the combination on the left hand side is the very combination appearing in the action. Therefore, substituting the latter equation along with ($\ref{dc}$) into our original action, we finally arrive at
	\begin{gather}
	S=\frac{1}{2\kappa}\int d^{4}x \sqrt{-g}R_{(\mu\nu)}g^{\mu\nu}-\frac{1}{2 \kappa}\int d^{4}x \sqrt{-g}\left[ 3 g^{\mu\nu}(\partial_{\mu}\phi)(\partial_{\nu}\phi)+ \frac{9}{4} g^{\mu\nu}(\partial_{\mu}\chi)(\partial_{\nu}\chi) \right] + s.t. 
	\end{gather}
	We note that extensions of the above are possible and one can also generate tensorial parts of torsion and non-metricity by coupling surface terms to tensor fields.

	\chapter{Solving for the Connection in Metric-Affine Gravity}

	\section{Exactly Solvable Models/Solving for the Affine Connection}
	Let us  now give a systematic way to solve for the affine connection in Metric-Affine Theories. We state and prove our results as three subsequent Theorems. The Theorems we state and prove appear for the first time in the literature of MAG.  First we start by allowing actions that are linear in the connection to be added to the Einstein Hilbert. The expression for the connection is then given by Theorem-1. Then, in Theorem-$2$ we generalize for $f(R)$ and in the last case we assume no restriction on the additional part of the action (Theorem-3). We then see some applications of our results with three simple examples and discuss the conditions for obtaining dynamical/non-dynamical connections.
	\subsection{Expression for an Exactly Solvable Connection}
	Let us start with our first Theorem\footnote{Here we will follow a step by step proof, in order to make the procedure of solving with respect to affine connection completely clear, since we think that such a  systematic procedure is absent from the literature.}
	\newline \\
	\textbf{Theorem 1:} Consider the action
	\begin{equation}
	S[ g_{\mu\nu},\Gamma^{\lambda}_{\;\;\;\alpha\beta},\phi]=\frac{1}{2\kappa}\int d^{n}x\sqrt{-g}R+S_{1}[ g_{\mu\nu},\Gamma^{\lambda}_{\;\;\;\alpha\beta},\phi] \label{sena}
	\end{equation}
	where $\phi$ denotes any other additional fields that may be present in the space and 
	\begin{equation}
	S_{1}[ g_{\mu\nu},\Gamma^{\lambda}_{\;\;\;\alpha\beta},\phi]= \int d^{n}x\sqrt{-g} \mathcal{L}_{1}(g,\Gamma,\phi)
	\end{equation}
	Now given any general action $S_{1}[g,\Gamma,\phi]$ that is\footnote{Notice that we made no assumption about the origin of the action. It may include both matter and gravitational parts so long as it satisfies the requirements that we impose! However, a gravitational sector that is linear in the connection is difficult to come up with, we just include it for generality. In the second Theorem we will assume that $S_{1}$ contains only a matter sector.} 
	\begin{itemize}
		\item At most linear in $\Gamma^{\lambda}_{\;\;\;\mu\nu}$ and its partial derivatives
		\item Projective invariant
	\end{itemize}
	we state that the affine connection can solely be expressed in terms of variations of $\mathcal{L}_{1}$\footnote{Of course the result also contains the metric tensor and its derivatives as they appear for instance in the Levi-Civita part, but since this is too obvious we will omit mentioning it.} and its form is the following
	\begin{equation}
	\Gamma^{\lambda}_{\;\;\;\mu\nu}=\tilde{\Gamma}^{\lambda}_{\;\;\;\mu\nu}-\frac{g^{\lambda\alpha}}{2}(B_{\alpha\mu\nu}-B_{\nu\alpha\mu}-B_{\mu\nu\alpha})-\frac{g^{\alpha\lambda}}{(n-2)}g_{\nu[\mu}(B_{\alpha]}-\tilde{B}_{\alpha]})
	\end{equation}
	where
	\begin{equation}
	B_{\lambda}^{\;\;\;\mu\nu}\equiv  \frac{2 \kappa}{\sqrt{-g}}\frac{\delta  S_{1}}{\delta \Gamma^{\lambda}_{\;\;\;\mu\nu}}=\frac{2 \kappa}{\sqrt{-g}}\frac{\partial (\sqrt{-g}\mathcal{L}_{1})}{\partial \Gamma^{\lambda}_{\;\;\;\mu\nu}}
	\end{equation}
	and $B^{\mu}\equiv B_{\lambda}^{\;\;\;\mu\lambda}$, $\tilde{B}^{\mu}\equiv g_{\alpha\beta}B^{\mu\alpha\beta}$.
	\newline \\
	\textbf{Proof:} Varying $(\ref{sena})$ with respect to the affine connection, we derive
	\begin{equation}
	P_{\lambda}^{\;\;\;\mu\nu}+B_{\lambda}^{\;\;\;\mu\nu}=0 \label{pala}
	\end{equation}
	where
	\begin{equation}
	B_{\lambda}^{\;\;\;\mu\nu}\equiv  \frac{2 \kappa}{\sqrt{-g}}\frac{\delta  S_{1}}{\delta \Gamma^{\lambda}_{\;\;\;\mu\nu}}=\frac{2 \kappa}{\sqrt{-g}}\frac{\partial (\sqrt{-g}\mathcal{L}_{1})}{\partial \Gamma^{\lambda}_{\;\;\;\mu\nu}}
	\end{equation}
	and $P_{\lambda}^{\;\;\;\mu\nu}$ is the Palatini tensor which is defined by
	\begin{gather}
	P_{\lambda}^{\;\;\;\mu\nu}\equiv  \frac{1}{\sqrt{-g}} \frac{\delta  S_{EH}}{\delta \Gamma^{\lambda}_{\;\;\;\mu\nu}} =\frac{1}{\sqrt{-g}} \frac{\partial( \sqrt{-g} R)}{\partial \Gamma^{\lambda}_{\;\;\;\mu\nu}}= \nonumber \\
	=-\frac{\nabla_{\lambda}(\sqrt{-g}g^{\mu\nu})}{\sqrt{-g}}+\frac{\nabla_{\sigma}(\sqrt{-g}g^{\mu\sigma})}{\sqrt{-g}}\delta_{\lambda}^{\nu}+2(g^{\mu\nu}S_{\lambda}-S^{\mu}\delta^{\nu}_{\lambda}+g^{\mu\sigma}S_{\sigma\lambda}^{\;\;\;\nu})
	\end{gather}
	as we have already seen. Now, as we show in the appendix, the latter can also be written in the form
	\begin{equation}
	P^{\alpha\mu\nu}=\left( \frac{Q^{\alpha}}{2}+2 S^{\alpha}\right) g^{\mu\nu}-(Q^{\alpha\mu\nu}+2 S^{\alpha\mu\nu})+\left( \tilde{Q}^{\mu}-\frac{Q^{\mu}}{2}-2 S^{\mu} \right)g^{\nu\alpha}
	\end{equation}
	With this at hand and recalling the connection decomposition in terms of the Riemannian part, non-metricity and torsion
	\begin{equation}
	\Gamma^{\lambda}_{\;\;\;\mu\nu}=\tilde{\Gamma}^{\lambda}_{\;\;\;\mu\nu}+\frac{1}{2}g^{\alpha\lambda}(Q_{\mu\nu\alpha}+Q_{\nu\alpha\mu}-Q_{\alpha\mu\nu}) -g^{\alpha\lambda}(S_{\alpha\mu\nu}+S_{\alpha\nu\mu}-S_{\mu\nu\alpha}) \label{affconection}
	\end{equation}
	we observe that the combination $(Q^{\alpha\mu\nu}+2 S^{\alpha\mu\nu})$ appears in both and can, therefore, be eliminated\footnote{Fun Fact: This observation came to me as an insight during my visit in a CERN Winter School in $2015$.  }. Indeed, pairing up a bit the terms of the last equation, we may re-write it as
	\begin{equation}
	\Gamma^{\lambda}_{\;\;\;\mu\nu}=\tilde{\Gamma}^{\lambda}_{\;\;\;\mu\nu}+\frac{1}{2}g^{\alpha\lambda}\Big( (Q_{\mu\nu\alpha}+2 S_{\mu\nu\alpha})+(Q_{\nu\alpha\mu}+2 S_{\nu\alpha\mu})-(Q_{\alpha\mu\nu}+2 S_{\alpha\mu\nu}) \Big) \label{solk}
	\end{equation}
	where we have used the fact that $S_{\alpha\mu\nu}=-S_{\mu\alpha\nu}$. In addition, we observe that 
	\begin{equation}
	P_{\alpha\mu\nu}-P_{\nu\alpha\mu}-P_{\mu\nu\alpha}=A_{\alpha\mu\nu}-g_{\alpha\mu}\tilde{Q}_{\nu}+2 g_{\nu[\alpha}(\tilde{Q}_{\mu]}-Q_{\mu]}-4 S_{\mu]})
	\end{equation}
	where
	\begin{equation}
	A_{\mu\nu\alpha}=(Q_{\mu\nu\alpha}+2 S_{\mu\nu\alpha})+(Q_{\nu\alpha\mu}+2 S_{\nu\alpha\mu})-(Q_{\alpha\mu\nu}+ 2 S_{\alpha\mu\nu})
	\end{equation}
	Thus, substituting the above combination into $(\ref{solk})$ we obtain
	\begin{equation}
	\Gamma^{\lambda}_{\;\;\;\mu\nu}=\tilde{\Gamma}^{\lambda}_{\;\;\;\mu\nu}+\frac{g^{\lambda\alpha}}{2}(P_{\alpha\mu\nu}-P_{\nu\alpha\mu}-P_{\mu\nu\alpha})+g^{\alpha\lambda}g_{\nu[\mu}(\tilde{Q}_{\alpha]}-Q_{\alpha]}-4 S_{\alpha]})+\frac{1}{2}\delta_{\mu}^{\lambda}\tilde{Q}_{\mu}
	\end{equation}
	Now, as we also prove in the appendix, it holds that
	\begin{equation}
	P^{\mu}\equiv P_{\lambda}^{\;\;\;\mu\lambda}=(n-1)\left( \tilde{Q}^{\mu}-\frac{1}{2}Q^{\mu}\right) +2(2-n)S^{\mu}
	\end{equation}
	\begin{equation}
	\tilde{P}^{\mu}\equiv g_{\alpha\beta}P^{\mu\alpha\beta}=\frac{(n-3)}{2}Q^{\mu}+\tilde{Q}^{\mu}+2(n-2)S^{\mu}
	\end{equation}
	such that
	\begin{equation}
	P^{\mu}-\tilde{P}^{\mu}=(n-2)(\tilde{Q}^{\mu}-Q^{\mu}-4 S^{\mu})
	\end{equation}
	Using this fact, the connection recasts to
	\begin{equation}
	\Gamma^{\lambda}_{\;\;\;\mu\nu}=\tilde{\Gamma}^{\lambda}_{\;\;\;\mu\nu}+\frac{g^{\lambda\alpha}}{2}(P_{\alpha\mu\nu}-P_{\nu\alpha\mu}-P_{\mu\nu\alpha})+\frac{g^{\alpha\lambda}}{(n-2)}g_{\nu[\mu}(P_{\alpha]}-\tilde{P}_{\alpha]})+\frac{1}{2}\delta_{\mu}^{\lambda}\tilde{Q}_{\nu}  \label{gg}
	\end{equation}
	Notice now that our total action is projective invariant by assumption. This means, as we have already seen, that the theory is invariant under
	\begin{equation}
	\Gamma^{\lambda}_{\;\;\;\mu\nu} \rightarrow \Gamma^{\lambda}_{\;\;\;\mu\nu} +\delta^{\lambda}_{\mu}\xi_{\nu}
	\end{equation}
	for any vector $\xi_{\nu}$. That is, there exists an unspecified vectorial degree of freedom. Using this very fact we can always make any gauge choice that we may like. As it is apparent from ($\ref{gg}$) in order to get rid of the last term (which is unspecified) we make the gauge choice
	\begin{equation}
	\xi_{\nu}=-\frac{1}{2}\delta_{\mu}^{\lambda}\tilde{Q}_{\nu} 
	\end{equation}
	Then, the connection assumes the form
	\begin{equation}
	\Gamma^{\lambda}_{\;\;\;\mu\nu}=\tilde{\Gamma}^{\lambda}_{\;\;\;\mu\nu}+\frac{g^{\lambda\alpha}}{2}(P_{\alpha\mu\nu}-P_{\nu\alpha\mu}-P_{\mu\nu\alpha})+\frac{g^{\alpha\lambda}}{(n-2)}g_{\nu[\mu}(P_{\alpha]}-\tilde{P}_{\alpha]})
	\end{equation}
	Upon using ($\ref{pala}$) and defining $B^{\mu}\equiv B_{\lambda}^{\;\;\;\mu\lambda}$ along with
	$\tilde{B}^{\mu}\equiv g_{\alpha\beta}B^{\mu\alpha\beta}$, we finally arrive at
	\begin{equation}
	\Gamma^{\lambda}_{\;\;\;\mu\nu}=\tilde{\Gamma}^{\lambda}_{\;\;\;\mu\nu}-\frac{g^{\lambda\alpha}}{2}(B_{\alpha\mu\nu}-B_{\nu\alpha\mu}-B_{\mu\nu\alpha})-\frac{g^{\alpha\lambda}}{(n-2)}g_{\nu[\mu}(B_{\alpha]}-\tilde{B}_{\alpha]}) \label{theo1}
	\end{equation}
	as stated.
	
	\textbf{Comment 1:} The projective invariance of $S_{1}$ is only necessary in order to remove the term 
	$\frac{1}{2}\delta_{\mu}^{\lambda}\tilde{Q}_{\nu}$ from $(\ref{gg})$. If  $S_{1}$  does not respect projective invariance one has to add the aforementioned term in the general result $(\ref{theo1})$.
	
	\textbf{Comment 2:} If there is no gravitational sector to $S_{1}$, i.e the latter is a purely matter action $S_{1}=S_{M}$ then $B_{\lambda}^{\;\;\;\mu\nu}=-\kappa \Delta_{\lambda}^{\;\;\;\mu\nu}$ and the connection is found to be
	\beq
	\Gamma^{\lambda}_{\;\;\;\mu\nu}=\tilde{\Gamma}^{\lambda}_{\;\;\;\mu\nu}+\kappa\frac{g^{\lambda\alpha}}{2}(\Delta_{\alpha\mu\nu}-\Delta_{\nu\alpha\mu}-\Delta_{\mu\nu\alpha})+\frac{g^{\alpha\lambda}}{(n-2)}g_{\nu[\mu}(\Delta_{\alpha]}-\tilde{\Delta}_{\alpha]})
	\eeq
	where  $\Delta^{\mu}\equiv \Delta_{\lambda}^{\;\;\;\mu\lambda}$, $\tilde{\Delta}^{\mu}\equiv g_{\alpha\beta}\Delta^{\mu\alpha\beta}$ which, as it stands, is an algebraic equation for the connection given the fact that for a matter sector linear in $\Gamma$ the hypermomentum is independent of the connection.

	\subsubsection{Expressions for torsion and non-metricity}
	Having the above decomposition we can easily derive the expressions for torsion and non-metricity by their very definitions. Starting with torsion, we have
	\begin{equation}
	S_{\mu\nu}^{\;\;\;\;\lambda}\equiv \Gamma^{\lambda}_{\;\;\;[\mu\nu]}=\frac{1}{2}\Big( B_{[\mu\nu]}^{\;\;\;\;\;\lambda}+B_{[\nu \;\;\;\;\mu]}^{\;\;\;\lambda}-B^{\lambda}_{\;\;\;[\mu\nu]}\Big)-\frac{1}{2(n-2)}\delta^{\lambda}_{\nu}(B_{\mu}-\tilde{B}_{\mu})
	\end{equation}
	As long as non-metricity is concerned, from its definition it follows that
	\begin{equation}
	Q_{\alpha\mu\nu}\equiv -\nabla_{\alpha}g_{\mu\nu}=-\tilde{\nabla}_{\alpha}g_{\mu\nu}+\frac{1}{2}\Big( B_{\alpha\mu\nu}+B_{\nu\alpha\mu}+B_{\alpha\nu\mu}+B_{\mu\alpha\nu}-B_{\mu\nu\alpha}-B_{\nu\mu\alpha} \Big)
	\end{equation}
	Now, using the fact that the Levi-Civita connection is metric compatible ($\tilde{\nabla}_{\alpha}g_{\mu\nu}=0$) we obtain for the non-metricity
	\begin{equation}
	Q_{\alpha\mu\nu}=B_{(\mu\nu)\alpha}+B_{(\mu|\alpha|\nu)}-B_{\alpha(\mu\nu)}
	\end{equation}

	\textbf{Comment:} Since $S_{1}[g,\Gamma]$ is linear in the connection, its variation $B_{\alpha\mu\nu}$ is independent of the connection. Then, expression ($\ref{theo1}$) is an algebraic equation for the connection. So in this case, not surprisingly, the connection caries no dynamics.

	\subsection{Generalizing the Theorem}
	Now, our above result may be readily generalized for actions of the form
	\begin{equation}
	S[ g_{\mu\nu},\Gamma^{\lambda}_{\;\;\;\alpha\beta},\phi]=\frac{1}{2\kappa}\int d^{n}x\sqrt{-g}f(R)+S_{1}[ g_{\mu\nu},\Gamma^{\lambda}_{\;\;\;\alpha\beta},\phi] \label{sena}
	\end{equation}
	where we have replaced $R$ with a general $f(R)$ function. In addition, we will now consider the additional part $S_{1}$ to be a purely matter part, that is  $S_{1}[g,\Gamma,\phi]=S_{M}[g,\Gamma,\phi]$. We do so in order to see how the energy tensors (energy momentum and hyper-momentum) enter the picture, especially with regards to the dynamical content of the connection\footnote{Similar results hold if we consider also a gravitational sector to $S_{1}$ but then there is no direct contact with the energy tensors.}. So, we may now state and prove a second theorem.
	\newline \\
	\textbf{Theorem 2:} Consider the action
	\begin{equation}
	S[ g_{\mu\nu},\Gamma^{\lambda}_{\;\;\;\alpha\beta},\phi]=\frac{1}{2\kappa}\int d^{n}x\sqrt{-g}f(R)+S_{1}[ g_{\mu\nu},\Gamma^{\lambda}_{\;\;\;\alpha\beta},\phi] \label{sena2}
	\end{equation}
	where $\phi$ denotes any other additional fields that may be present in the spacetime and 
	\begin{equation}
	S_{1}[ g_{\mu\nu},\Gamma^{\lambda}_{\;\;\;\alpha\beta},\phi]=S_{M}[ g_{\mu\nu},\Gamma^{\lambda}_{\;\;\;\alpha\beta},\phi]=\frac{1}{2\kappa} \int d^{n}x\sqrt{-g} \mathcal{L}_{M}(g,\Gamma,\phi)
	\end{equation}
	Now given any general matter action $S_{1}[g,\Gamma,\phi]=S_{M}[g,\Gamma,\phi]$ that is
	\begin{itemize}
		\item At most linear in $\Gamma^{\lambda}_{\;\;\;\mu\nu}$ and its partial derivatives
		\item Projective invariant
	\end{itemize}
	we state that the affine connection can solely be expressed in terms of $f^{'}(T)$ (where T is the trace of the energy momentum tensor) and of $\Gamma$-variations of $\mathcal{L}_{M}$ (i.e. the hypermomentum $\Delta_{\lambda}^{\;\;\;\mu\nu}$)  and its form is the following
	\begin{equation}
	\Gamma^{\lambda}_{\;\;\;\mu\nu}=\tilde{\Gamma}^{\lambda}_{\;\;\;\mu\nu}+\frac{g^{\lambda\alpha}}{2}(H_{\alpha\mu\nu}-H_{\nu\alpha\mu}-H_{\mu\nu\alpha})+\frac{g^{\alpha\lambda}}{(n-2)}g_{\nu[\mu}(H_{\alpha]}-\tilde{H}_{\alpha]})
	\end{equation}
	where
	\begin{equation}
	H_{\lambda}^{\;\;\;\mu\nu}\equiv - \frac{ 2 \kappa}{f^{'}\sqrt{-g}}\frac{\delta  \mathcal{L}_{M}}{\delta \Gamma^{\lambda}_{\;\;\;\mu\nu}}+\frac{1}{f^{'}}( g^{\mu\nu} \partial_{\lambda}f^{'}-\delta_{\lambda}^{\nu}\partial^{\mu} f^{'} )=\frac{\kappa}{f^{'}}\Delta_{\lambda}^{\;\;\;\mu\nu}+\frac{1}{f^{'}}( g^{\mu\nu} \partial_{\lambda}f^{'}-\delta_{\lambda}^{\nu}\partial^{\mu} f^{'})
	\end{equation}
	and $H^{\mu}\equiv H_{\lambda}^{\;\;\;\mu\lambda}$, \;$\tilde{H}^{\mu}\equiv g_{\alpha\beta}H^{\mu\alpha\beta}$,\; $f'=f'(T)$,\; $T\equiv g^{\mu\nu}T_{\mu\nu}=g^{\mu\nu}\frac{2}{\sqrt{-g}}\frac{\partial(\sqrt{-g}\mathcal{L}_{M})}{\partial g^{\mu\nu}}$  and the prime denotes differentiation with respect to the Ricci scalar.
	\newline \\
	\textbf{Proof:} Varying ($\ref{sena2}$) with respect to the connection we obtain
	\beq
	P_{\lambda}^{\;\;\;\mu\nu}(h)=\kappa \Delta_{\lambda}^{\;\;\;\mu\nu} \label{pafr}
	\eeq
	where
	\begin{gather}
	P_{\lambda}^{\;\;\;\mu\nu}(h) \equiv -\frac{\nabla_{\lambda}(\sqrt{-g}f^{'}g^{\mu\nu})}{\sqrt{-g}}+\frac{\nabla_{\alpha}(\sqrt{-g}f^{'}g^{\mu\alpha}\delta_{\lambda}^{\nu})}{\sqrt{-g}}+ \\ \nonumber
	2 f^{'}(S_{\lambda}g^{\mu\nu}-S^{\mu}\delta_{\lambda}^{\nu}-  S_{\lambda}^{\;\;\;\mu\nu}) \label{pakah1}
	\end{gather}
	is the Palatini tensor of the metric $h_{\mu\nu}=f'(R)g_{\mu\nu}$, which is conformally related to $g_{\mu\nu}$ and prime here denotes differentiation with respect to the Ricci scalar. $ \Delta_{\lambda}^{\;\;\;\mu\nu}$ is the usual hypermomentum tensor we have defined earlier. Now, expanding the covariant derivatives in the above we see that
	\beq
	P_{\lambda}^{\;\;\;\mu\nu}(h)=f^{'}P_{\lambda}^{\;\;\;\mu\nu}(g)+\delta_{\lambda}^{\nu}g^{\mu\alpha}\partial_{\alpha}f^{'}-g^{\mu\nu}\partial_{\lambda}f^{'}
	\eeq
	where $P_{\lambda}^{\;\;\;\mu\nu}(g)$ is the usual Palatini tensor of $g_{\mu\nu}$. Then for $f'(R)\neq 0$ we may solve for the latter
	\beq
	P_{\lambda}^{\;\;\;\mu\nu}(g)=\frac{1}{f'}\left( P_{\lambda}^{\;\;\;\mu\nu}(h)-\delta_{\lambda}^{\nu}g^{\mu\alpha}\partial_{\alpha}f^{'}+g^{\mu\nu}\partial_{\lambda}f^{'}\right)
	\eeq
	or by virtue of ($\ref{pafr}$)
	\beq
	P_{\lambda}^{\;\;\;\mu\nu}(g)=\frac{1}{f'}\left( \kappa\Delta_{\lambda}^{\;\;\;\mu\nu}-\delta_{\lambda}^{\nu}g^{\mu\alpha}\partial_{\alpha}f^{'}+g^{\mu\nu}\partial_{\lambda}f^{'}\right)
	\eeq
	Then recalling eq. ($\ref{gg}$) that we obtained in the first Theorem,
	\begin{equation}
	\Gamma^{\lambda}_{\;\;\;\mu\nu}=\tilde{\Gamma}^{\lambda}_{\;\;\;\mu\nu}+\frac{g^{\lambda\alpha}}{2}\Big(P_{\alpha\mu\nu}(g)-P_{\nu\alpha\mu}(g)-P_{\mu\nu\alpha}(g)\Big)+\frac{g^{\alpha\lambda}}{(n-2)}g_{\nu[\mu}\Big(P_{\alpha]}(g)-\tilde{P}_{\alpha]}(g)\Big)+\frac{1}{2}\delta_{\mu}^{\lambda}\tilde{Q}_{\nu}  
	\end{equation}  
	and using the above, we find
	\begin{gather}
	\Gamma^{\lambda}_{\;\;\;\mu\nu}=\tilde{\Gamma}^{\lambda}_{\;\;\;\mu\nu}+\frac{\kappa}{f'}\frac{g^{\lambda\alpha}}{2}(\Delta_{\alpha\mu\nu}-\Delta_{\nu\alpha\mu}-\Delta_{\mu\nu\alpha})+\frac{\kappa}{f'}\frac{g^{\alpha\lambda}}{(n-2)}g_{\nu[\mu}(\Delta_{\alpha]}-\tilde{\Delta}_{\alpha]}) \nonumber \\
	+\frac{1}{(n-2)f'}\Big( \delta^{\lambda}_{\nu}\partial_{\mu}f' -g_{\mu\nu} \partial^{\lambda}f' \Big)
	+\frac{1}{2}\delta_{\mu}^{\lambda}\tilde{Q}_{\nu}  
	\end{gather}  
	where at this point $f'=f'(R)$. Now, variation of our total action with respect to the metric, yields
	\beq
	f^{'}(R)R_{(\mu\nu)}-\frac{f(R)}{2}g_{\mu\nu}=\kappa T_{\mu\nu}
	\eeq
	where 
	\beq
	T_{\mu\nu} \equiv -\frac{2}{\sqrt{-g}}\frac{\delta S_{M}}{\delta g^{\mu\nu}}
	\eeq
	which we may contract with the metric tensor to obtain
	\beq
	f^{'}(R)R-\frac{n}{2}f(R)=\kappa T \label{tmn}
	\eeq
	The latter defines the implicit function $R=R(T)$\footnote{Except in the case $f(R)\propto R^{2}$ for which the left send hide of ($\ref{tmn}$) is identically zero and the model allows only for conformally invariant matter ($T=0$). This exception have been studied in \cite{iosifidis2018torsion} where also the  cosmological solutions were given for this case.} and therefore  both $f(R)$ and $f^{'}(R)$  are all functions of $T$  ($f(R)=f(R(T))=f(T)$ and $f^{'}(R)=f^{'}(R(T))=f^{'}(T)$). With this at hand, and using the fact that our total action is projective invariant we may remove the term $\frac{1}{2}\delta_{\mu}^{\lambda}\tilde{Q}_{\nu} $ and write
	\begin{gather}
	\Gamma^{\lambda}_{\;\;\;\mu\nu}=\tilde{\Gamma}^{\lambda}_{\;\;\;\mu\nu}+\frac{\kappa}{f'}\frac{g^{\lambda\alpha}}{2}(\Delta_{\alpha\mu\nu}-\Delta_{\nu\alpha\mu}-\Delta_{\mu\nu\alpha})+\frac{\kappa}{f'}\frac{g^{\alpha\lambda}}{(n-2)}g_{\nu[\mu}(\Delta_{\alpha]}-\tilde{\Delta}_{\alpha]}) \nonumber \\
	+\frac{1}{(n-2)f'}\Big( \delta^{\lambda}_{\nu}\partial_{\mu}f' -g_{\mu\nu} \partial^{\lambda}f' \Big) 
	\end{gather} 
	where $f'$ is a function of $T$ now. Finally, defining
	\beq
	H_{\alpha\mu\nu}\equiv \frac{1}{f'}\left( \kappa \Delta_{\alpha\mu\nu}+ g_{\mu\nu}\partial_{\alpha}f' -g_{\nu\alpha}\partial_{\mu}f' \right)
	\eeq
	we complete the proof
	\begin{equation}
	\Gamma^{\lambda}_{\;\;\;\mu\nu}=\tilde{\Gamma}^{\lambda}_{\;\;\;\mu\nu}+\frac{g^{\lambda\alpha}}{2}(H_{\alpha\mu\nu}-H_{\nu\alpha\mu}-H_{\mu\nu\alpha})+\frac{g^{\alpha\lambda}}{(n-2)}g_{\nu[\mu}(H_{\alpha]}-\tilde{H}_{\alpha]})
	\end{equation}

	\subsection{Generalized Theorem}
	We may now relax our assumptions and let $S_{1}[ g_{\mu\nu},\Gamma^{\lambda}_{\;\;\;\alpha\beta},\phi]$ have an arbitrary dependence on the connection and its derivatives and may not respect the projective symmetry in general. This leads us to the third Theorem.
	\newline \\
	\textbf{Theorem 3:} Consider the action
	\begin{equation}
	S[ g_{\mu \nu},\Gamma^{\lambda}_{\;\;\;\alpha\beta},\phi]=\frac{1}{2\kappa}\int d^{n}x\sqrt{-g}R+S_{1}[ g_{\mu\nu},\Gamma^{\lambda}_{\;\;\;\alpha\beta},\phi] \label{sena}
	\end{equation}
	where
	\begin{equation}
	S_{1}[ g_{\mu\nu},\Gamma^{\lambda}_{\;\;\;\alpha\beta},\phi]=\frac{1}{2\kappa} \int d^{n}x\sqrt{-g} \mathcal{L}_{1}(g,\Gamma,\phi)
	\end{equation}
	has an arbitrary dependence on the affine connection and its derivatives. Then, the connection is given by the solution of
	\begin{equation}
	\Gamma^{\lambda}_{\;\;\;\mu\nu}=\tilde{\Gamma}^{\lambda}_{\;\;\;\mu\nu}-\frac{g^{\lambda\alpha}}{2}(B_{\alpha\mu\nu}-B_{\nu\alpha\mu}-B_{\mu\nu\alpha})-\frac{g^{\alpha\lambda}}{(n-2)}g_{\nu[\mu}(B_{\alpha]}-\tilde{B}_{\alpha]}) 
	\end{equation}
	where
	\begin{equation}
	B_{\lambda}^{\;\;\;\mu\nu}\equiv  \frac{2 \kappa}{\sqrt{-g}}\frac{\delta  S_{1}}{\delta \Gamma^{\lambda}_{\;\;\;\mu\nu}}=\frac{2 \kappa}{\sqrt{-g}}\frac{\partial (\sqrt{-g}\mathcal{L}_{1})}{\partial \Gamma^{\lambda}_{\;\;\;\mu\nu}}
	\end{equation}
	$B^{\mu}\equiv B_{\lambda}^{\;\;\;\mu\lambda}$, $\tilde{B}^{\mu}\equiv g_{\alpha\beta}B^{\mu\alpha\beta}$
	and the above will be a differential equation for the connection in general since B has an arbitrary dependence  on the connection and its derivatives. \newline \\
	\textbf{Proof:} Following identical steps with  Theorem-1 but now keeping in mind that $B_{\lambda}^{\;\;\;\mu\nu}(\Gamma, \partial \Gamma)$ is a general function of the connection and its derivatives, we get
	\begin{equation}
	\Gamma^{\lambda}_{\;\;\;\mu\nu}=\tilde{\Gamma}^{\lambda}_{\;\;\;\mu\nu}-\frac{g^{\lambda\alpha}}{2}(B_{\alpha\mu\nu}-B_{\nu\alpha\mu}-B_{\mu\nu\alpha})-\frac{g^{\alpha\lambda}}{(n-2)}g_{\nu[\mu}(B_{\alpha]}-\tilde{B}_{\alpha]}) \label{theo3}
	\end{equation}
	where $B^{\mu}\equiv B_{\lambda}^{\;\;\;\mu\lambda}$,
	$\;\tilde{B}^{\mu}\equiv g_{\alpha\beta}B^{\mu\alpha\beta}$ and since $B_{\lambda}^{\;\;\;\mu\nu}(\Gamma, \partial \Gamma)$ has an arbitrary dependence of the connection and its derivatives, the above is a dynamical equation for the connection in contrast to equation ($\ref{theo1}$) which is an algebraic one. Having presented and proved the three Theorems we may now see some examples where the latter can by applied.

	\subsection{Example 1: Exciting Torsional d.o.f.}
	Let us now use the results we obtained for the connection decomposition (the $3$ Theorems) in order to review the model studied in \cite{d1982gravity,leigh2009torsion,petkou2010torsional} but now in the coordinate formalism. Same way we did in the previous chapter we start with the action\footnote{We studied exactly this model in the previous chapter so we will skip most of the calculations now.}
	\begin{gather}
	S=\frac{1}{2\kappa}\int d^{4}x \sqrt{-g}R_{(\mu\nu)}g^{\mu\nu}+\frac{1}{2\kappa}\int d^{4}x F(x)\epsilon^{\mu\nu\rho\sigma}\partial_{\mu}S_{\nu\rho\sigma}= \nonumber \\
	=\frac{1}{2\kappa}\int d^{4}x \sqrt{-g}R_{(\mu\nu)}g^{\mu\nu}-\frac{1}{2\kappa}\int d^{4}x \epsilon^{\mu\nu\rho\sigma}(\partial_{\mu}F)S_{\nu\rho\sigma}+s.t.
	\end{gather}
	where $F(x)$ is a scalar and $s.t.$ stands for surface term. Notice that the additional piece here is linear in the connection and therefore falls in the category of our Theorem-1. So, we may proceed and use the result we obtained for the connection.
	Variation with respect to the connection yields\footnote{Where we have used the properties of the Levi-Civita symbol and also raised an index with the metric.}
	\begin{equation}
	\sqrt{-g}P^{\nu\rho\sigma}-\epsilon^{\alpha\rho\sigma\nu}(\partial_{\alpha}F)=0 \label{epsif}
	\end{equation}
	Also, since in this model the non-metricity is zero, the Palatini tensor reads
	\begin{equation}
	P^{\nu\rho\sigma}=2\Big( g^{\rho\sigma}S^{\nu}-g^{\sigma\nu}S^{\rho}+ S^{\rho\nu\sigma} \Big) \label{ddf}
	\end{equation}
	Now, contracting ($\ref{epsif}$) with $\epsilon^{\mu\rho\sigma\nu}$ and also using the above, we obtain
	\begin{gather}
	\varepsilon_{\mu\nu\rho\sigma}S^{\rho\sigma\nu}=3(\partial_{\mu}F)
	\end{gather}
	where $\varepsilon_{\mu\nu\rho\sigma}\equiv \sqrt{-g} \epsilon_{\mu\rho\sigma\nu}$ is the Levi-Civita tensor. So, we may also write
	\begin{equation}
	\epsilon^{\mu\nu\rho\sigma}S_{\nu\rho\sigma}=3\sqrt{-g}g^{\mu\nu}(\partial_{\nu}F)
	\end{equation}
	Substituting the latter in our action we arrive at
	\begin{equation}
	S=\frac{1}{2\kappa}\int d^{4}x\Big[ \sqrt{-g}R-\sqrt{-g}3 g^{\mu\nu}(\partial_{\mu}F)(\partial_{\nu}F) \Big]
	\end{equation}
	We can also immediately see that
	\begin{equation}
	S_{\mu}=0\;,\;\;P^{\mu\nu\alpha}=-\varepsilon^{\rho\mu\nu\alpha}\partial_{\rho}F \;,\;\;S_{\mu\nu\alpha}=-\frac{1}{2}\varepsilon_{\mu\nu\alpha\lambda}\partial^{\lambda}F
	\end{equation}
	which when plugged into the connection decomposition ($\ref{theo1}$) of Theorem-1 yield
	\begin{equation}
	\Gamma^{\lambda}_{\;\;\;\;\mu\nu}=\tilde{\Gamma}^{\lambda}_{\;\;\;\;\mu\nu}+\frac{1}{2}\varepsilon_{\mu\nu}^{\;\;\;\;\;\rho\lambda}\partial_{\rho}F
	\end{equation}
	From which we conclude that this kind of torsion (being totally antisymmetric) has no effect on the autoparallels and the latter coincide with the geodesics. Now, we can fully decompose our original action to a Riemannian part plus an axion field. Indeed, to see this first recall the Ricci scalar decomposition
	\begin{equation}
	R=\tilde{R}+ \tilde{\nabla}_{\mu}( A^{\mu}-B^{\mu})+ B_{\mu}A^{\mu}-N_{\alpha\mu\nu}N^{\mu\nu\alpha}
	\end{equation}
	Note now that the second term is a surface term and can therefore be dropped when taken into the action integral. Regarding the other quantities appearing, we compute for our case
	\begin{equation}
	N_{\mu\nu\alpha}=-\frac{1}{2}\varepsilon_{\mu\nu\alpha\rho}\partial^{\rho}F
	\end{equation}
	\begin{equation}
	A^{\mu}= N^{\mu}_{\;\;\;\nu\beta}g^{\nu\beta}=0, \;\;\; B^{\mu}=N^{\alpha\mu}_{\;\;\;\;\alpha}=0
	\end{equation}
	\begin{equation}
	N_{\alpha\mu\nu}N^{\mu\nu\alpha}=-\frac{3}{2}\partial_{\mu}F\partial^{\mu}F
	\end{equation}
	so that, when substituted back to our action give
	\begin{equation}
	S=\frac{1}{2\kappa}\int d^{4}x\sqrt{-g}\Big[ \tilde{R}-\frac{3}{2} g^{\mu\nu}(\partial_{\mu}F)(\partial_{\nu}F) \Big]
	\end{equation}
	which is the action of Einstein gravity plus an axionic massless field. To recap, for this model, the affine connection takes the form
	\begin{equation}
	\Gamma^{\lambda}_{\;\;\;\;\mu\nu}=\tilde{\Gamma}^{\lambda}_{\;\;\;\;\mu\nu}+ N^{\lambda}_{\;\;\;\;\mu\nu}=\tilde{\Gamma}^{\lambda}_{\;\;\;\;\mu\nu}+\frac{1}{2}\varepsilon_{\mu\nu}^{\;\;\;\;\rho\lambda}\partial_{\rho}F
	\end{equation}
	Note now that this type of torsion (totally) antisymmetric has no effect on the autoparallels and the latter coincide with the geodesics. However, for general torsion (even with vanishing non-metricity) the two are not the same.

	\subsection{Example 2: Metric-Affine $f(R)$ with projective invariant matter}
	Let us now apply the results of our connection decomposition and study some characteristics of Metric Affine f(R) theories (\cite{sotiriou2007metric,olmo2011palatini,vitagliano2010dynamics,olmo2009dynamical}). In the case where the matter is not projective invariant one may break the invariance by one of the ways we saw in the previous chapter.
	Interestingly, if matter fields that respect the projective invariant are added to $f(R)$ we have exactly the case we presented in Theorem-2. Then, applying the results of our second Theorem we immediately get for the affine connection
	\begin{gather}
	\Gamma^{\lambda}_{\;\;\;\mu\nu}=\tilde{\Gamma}^{\lambda}_{\;\;\;\mu\nu}+\frac{\kappa}{f'}\frac{g^{\lambda\alpha}}{2}(\Delta_{\alpha\mu\nu}-\Delta_{\nu\alpha\mu}-\Delta_{\mu\nu\alpha})+\frac{\kappa}{f'}\frac{g^{\alpha\lambda}}{(n-2)}g_{\nu[\mu}(\Delta_{\alpha]}-\tilde{\Delta}_{\alpha]}) \nonumber \\
	+\frac{1}{(n-2)f'}\Big( \delta^{\lambda}_{\nu}\partial_{\mu}f' -g_{\mu\nu} \partial^{\lambda}f' \Big)\;,\;\;\;\;where\;\;\;\;\;f'=f'(T)
	\end{gather}  
	With the above connection being dynamical when $T_{\mu\nu}$ depends on the connection, and lacking dynamics when the latter is independent of the connection.

	\subsection{Example 3: A Theory with a dynamical connection}
	As an application of our third Theorem let us consider the theory
	\beq
	S[g,\Gamma]=\int d^{4} x \sqrt{-g}\left( \frac{1}{2 \kappa}R+ \frac{\lambda}{2 \kappa} R_{\mu\nu}R^{\mu\nu} \right)
	\eeq
	where $\lambda$ is a parameter. Notice that there is no motivation behind the choice of this action, we consider it here as a simple example in order to apply our Theorem-3. It is known in the literature (see \cite{vitagliano2010dynamics} for instance) that Theories of the family $f(R,R_{\mu\nu}R^{\mu\nu})$ admit a dynamical connection in general. Therefore, we expect that in the above Theory the connection is dynamical. This can be easily verified by using our third Theorem. To see this, let us vary the above action with respect to the connection, to get
	\beq
	P_{\lambda}^{\;\;\;\mu\nu}(g)=-2\lambda P_{\lambda}^{\;\;\;\mu\nu}(R)
	\eeq
	where $P_{\lambda}^{\;\;\;\mu\nu}(g)$ is the usual Palatini tensor computed with respect to the metric and
	\beq
	P_{\lambda}^{\;\;\;\mu\nu}(R)\equiv -\frac{\nabla_{\lambda}(\sqrt{-g}R^{\mu\nu})}{\sqrt{-g}}+\frac{\nabla_{\sigma}(\sqrt{-g}R^{\mu\sigma})}{\sqrt{-g}}\delta_{\lambda}^{\nu}+2(R^{\mu\nu}S_{\lambda}-S_{\alpha}R^{\mu\alpha}\delta^{\nu}_{\lambda}+R^{\mu\sigma}S_{\sigma\lambda}^{\;\;\;\nu})
	\eeq
	Then using the result $(\ref{theo3})$ of our third Theorem, we have
	\begin{equation}
	\Gamma^{\lambda}_{\;\;\;\mu\nu}=\tilde{\Gamma}^{\lambda}_{\;\;\;\mu\nu}-g^{\lambda\alpha}\lambda\Big(P_{\alpha\mu\nu}(R)-P_{\nu\alpha\mu}(R)-P_{\mu\nu\alpha}(R)\Big)-2 \lambda\frac{g^{\alpha\lambda}}{(n-2)}g_{\nu[\mu} \Big(P_{\alpha]}(R)-\tilde{P}_{\alpha]}(R)\Big)+\frac{1}{2}\delta_{\mu}^{\lambda}\tilde{Q}_{\nu}  
	\end{equation}
	and the above is a dynamical equation for the connection. This is easily understood by the appearance of the terms such as $\nabla_{\lambda}R^{\mu\nu}$ which contain higher order terms and derivatives of the connection. So, with this simple example we see an immediate application of our third Theorem. It goes beyond the purposes of this letter to investigate the above theory any further but we mention that a similar theory\footnote{The additional piece they added to the Einstein Hilbert part there was $c_{1}R^{(\mu\nu)}R_{(\mu\nu)}+ c_{2}R^{[\mu\nu]}R_{[\mu\nu]}   $.    } was studied in \cite{vitagliano2010dynamics} . In particular it was shown there that for vanishing torsion, the Theory is equivalent to Einstein's Gravity plus a Proca field \cite{vitagliano2010dynamics}. Similar results (again for vanishing torsion) for an action containing the anti-symmetric part of the Ricci tensor and a quadratic non-metricity term were also found in \cite{allemandi2004accelerated}. However, for projective actions of the form $f(R,R_{(\mu\nu)}R^{(\mu\nu)})$  the connection lacks dynamics \cite{vitagliano2010dynamics}. It would therefore be interesting to classify other actions that give similar results and the conditions upon which the connection lacks/gains. These subjects certainly worth further investigation.

		\section{Auxiliary Relation used for the Theorems}
	In this mini appendix we are going to express the Palatini tensor in terms of torsion and non-metricity (and their related vectors), an equation we used in order to prove our 3-Theorems. We start by writing down the definition of the Palatini tensor and expand the various terms to arrive at
	\begin{gather}
	P_{\lambda}^{\;\;\;\mu\nu}=-g^{\mu\nu}\frac{\nabla_{\lambda}\sqrt{-g}}{\sqrt{-g}}-\nabla_{\lambda}g^{\mu\nu}+g^{\mu\sigma}\frac{\nabla_{\sigma}\sqrt{-g}}{\sqrt{-g}}\delta_{\lambda}^{\nu}+\delta^{\nu}_{\lambda}\nabla_{\sigma}g^{\mu\sigma} \\
	+2(S_{\lambda}g^{\mu\nu}-S^{\mu}\delta_{\lambda}^{\nu}+g^{\mu\sigma}S_{\sigma\lambda}^{\;\;\;\;\nu})
	\end{gather}
	and by using 
	\begin{equation}
	Q_{\lambda}^{\;\;\;\mu\nu}=+\nabla_{\lambda}g^{\mu\nu} \nonumber
	\end{equation}
	\begin{equation}
	\frac{\nabla_{\lambda}\sqrt{-g}}{\sqrt{-g}}=-\frac{1}{2}Q_{\lambda} \nonumber
	\end{equation}
	\begin{equation}
	\tilde{Q}^{\mu}=\nabla_{\sigma}g^{\sigma\mu} \nonumber
	\end{equation}
	it follows that
	\begin{gather}
	P_{\lambda}^{\;\;\;\mu\nu}=-g^{\mu\nu}\frac{Q_{\lambda}}{2}-Q_{\lambda}^{\;\;\;\mu\nu}+\delta_{\lambda}^{\nu}\left( \tilde{Q}^{\mu}- \frac{Q^{\mu}}{2}\right) \\
	+2(S_{\lambda}g^{\mu\nu}-S^{\mu}\delta_{\lambda}^{\nu}+g^{\mu\sigma}S_{\sigma\lambda}^{\;\;\;\;\nu})
	\end{gather}
	and upon multiplying (and contracting) with $g^{\alpha\lambda}$ we finally obtain
	\begin{equation}
	P^{\alpha\mu\nu}=g^{\mu\nu}\left( \frac{Q^{\alpha}}{2}+2 S^{\alpha} \right)-(Q^{\alpha\mu\nu}+2 S^{\alpha\mu\nu})+g^{\nu\alpha}\left( \tilde{Q}^{\mu}- \frac{Q^{\mu}}{2}-2 S^{\mu} \right)
	\end{equation}
	Note now that the second combination $(Q^{\alpha\mu\nu}+2 S^{\alpha\mu\nu})$ plus circular permutations is the exact one appearing on the decomposition of the connection. Then following the steps we outlined previously (Theorems) we can solve the affine connection in terms of the Palatini tensor as we showed.

	\chapter{A Peculiar f(R) Case: The model $f(R)\propto R^{2}$}

	In this Chapter we will focus on the peculiar $f(R)\propto R^{2}$ case which as we have mentioned in previous Chapter does not give $R=constant$ in vacuum. We will first focus on the theory with vanishing torsion and compare our results with the work of \cite{capozziello2008f}\footnote{There, they considered the same theory but with vanishing non-metricity and non-vanishing torsion.} and then consider the general case with both torsion and non-metricity. We will see how the effects of torsion and non-metricity are indistinguishable when only vectorial degrees of freedom of the latter are excited.

	\section{Propagating Non-Metricity and its Cosmological implications}
	
	As it is well known \cite{sotiriou2010f,sotiriou2009f}, generic Palatini $f(R)$ theories of gravity with matter (where we have both torsion and non-metricity but the matter fields do not couple to the connection) ,  are equivalent to  Brans-Dicke gravity with parameter $\omega_{0}=-3/2$. The same holds when one turns on only torsion (with vanishing non-metricity) or turns on only non-metricity (with zero torsion). So, Palatini $f(R)$ with torsion only is in fact equivalent to Palatini $f(R)$ with only non-metricity, since both of them are equivalent to the same Brans-Dicke theory. In vacuum $f(R)$ theories are equivalent to General Relativity with a cosmological constant                \footnote{Actually many models with different cosmological constant corresponding to the roots of $f'(R)R-2f(R)=0$.} except in the case were $f(R)=\alpha R^{2}$. This is the case under investigation here, that is our starting action is
	\beq
	S=\frac{1}{2\kappa}\int d^{4}x\sqrt{-g}\alpha R^{2}
	\eeq 
	where our connection is torsionless but non-metricity is present. Varying with respect to the metric we get
	\beq
	2R \left( R_{(\mu\nu)}-\frac{R}{4}g_{\mu\nu}  \right)=0
	\eeq
	which gives us two possibilities, either
	\beq
	R=0
	\eeq
	or
	\beq
	R_{(\mu\nu)}-\frac{R}{4}g_{\mu\nu} =0
	\eeq
	Here we consider that $R\neq 0$ since this is the trivial case, so the field equations are
	\beq
	R_{(\mu\nu)}-\frac{R}{4}g_{\mu\nu} =0
	\eeq
	Note that these look like Einstein equations in vacuum, but there is an extra factor of $1/2$ in front of the Ricci scalar. This factor does make a huge difference because if we were to take the trace of the above we would get no additional equation since the left hand side is identically zero in $4$-dimensions.\footnote{Of course, this contraction gives $R=0$ for Einstein's equations.}  Now, upon varying with respect to the (symmetric) connection it follows that
	\beq
	\nabla_{\alpha}\Big(R \sqrt{-g}g^{\mu\nu}\Big)-\nabla_{\beta}\Big(R\sqrt{-g}g^{\beta(\mu} \Big) \delta^{\nu)}_{\alpha}=0
	\eeq
	taking the trace in $\alpha=\mu$ we get
	\beq
	\nabla_{\beta}\Big(R\sqrt{-g}g^{\beta\mu}\Big)=0
	\eeq
	which when substituted back on the first one gives
	\beq
	\nabla_{\alpha}\Big(R \sqrt{-g}g^{\mu\nu}\Big)=0
	\eeq
	Expanding the latter, we arrive at
	\beq
	g^{\mu\nu}\partial_{\lambda}R-\frac{R}{2}g^{\mu\nu}Q_{\lambda}+R Q_{\lambda}^{\;\;\;\mu\nu}=0
	\eeq
	which when contracted by $g_{\mu\nu}$, gives
	\beq
	\frac{\partial_{\mu}R}{R}=\frac{1}{4}Q_{\lambda}  \Rightarrow  \partial_{\lambda}(\ln{R})=\frac{1}{4}Q_{\lambda} \label{nonm}
	\eeq
	The latter is the equation that gives dynamics to non-metricity, even though we are in vacuum. Therefore in this model we see an example of propagating non-metricity. To see this more clearly, we can decompose the Ricci scalar in its Riemannian and non-metric parts
	\beq
	R=\tilde{R}-\frac{3}{4}\tilde{\nabla}_{\mu}Q^{\mu}-\frac{3}{32}Q_{\mu}Q^{\mu}
	\eeq
	and by substituting the latter into ($\ref{nonm}$) it follows that
	\beq
	\partial_{\lambda}\Big( \tilde{R}-\frac{3}{4}\tilde{\nabla}_{\mu}Q^{\mu}-\frac{3}{32}Q_{\mu}Q^{\mu} \Big) =\frac{1}{4}\Big( \tilde{R}-\frac{3}{4}\tilde{\nabla}_{\mu}Q^{\mu}-\frac{3}{32}Q_{\mu}Q^{\mu} \Big) Q_{\lambda}
	\eeq
	which is an equation containing only the non-metricity and the metric. Furthermore, it is easy to show that
	\beq
	Q_{\lambda\mu\nu}=\frac{1}{4}g_{\mu\nu}Q_{\lambda}
	\eeq
	which is the case of a Weyl non-metricity. With this at hand, the affine connection is easily found to be
	\beq
	\Gamma^{\lambda}_{\;\;\;\;\mu\nu}=\tilde{\Gamma}^{\lambda}_{\;\;\;\;\mu\nu}+\frac{1}{2 n}\left( 2 \delta^{\lambda}_{(\nu}Q_{\mu)} - Q^{\lambda} g_{\mu\nu} \right) 
	\eeq
	where $\tilde{\Gamma}^{\lambda}_{\;\;\;\;\mu\nu}$ is the Levi-Civita connection.

	From this point on let us be more specific and study the cosmological implication of this model. To this end we consider a spatially flat FLRW universe equipped with the metric
	\beq
	ds^{2}=-dt^{2}+a^{2}(t)\Big(dx^{2}+dy^{2}+dz^{2}\Big)
	\eeq
	for which the non-vanishing Christoffel (Levi-Civita) symbols are
	\beq
	\tilde{\Gamma}^{0}_{\;\;ij}=a\dot{a}\delta_{ij}
	\eeq
	\beq
	\tilde{\Gamma}^{i}_{\;\;j0}=\frac{\dot{a}}{a}\delta^{i}_{j}
	\eeq
	and the Riemannian parts of the Ricci tensor and scalar can be easily computed
	\beq
	\tilde{R}_{00}=-3\frac{\ddot{a}}{a}
	\eeq
	\beq
	\tilde{R}_{ij}=6\left[ \frac{\ddot{a}}{a}+\left(\frac{\dot{a}}{a}\right)^{2} \right]g_{ij}
	\eeq
	Note now that the only non-vanishing component of non-metricity in such a universe is $Q_{0}=Q(t)$ (since $Q_{i}$ defines a direction and therefore must identically vanish). Then, ($\ref{nonm}$) can be directly integrated to give
	\beq
	R=Ce^{\frac{1}{4}\int Q dt}
	\eeq
	which may be expanded to give
	\beq
	6(\dot{H}+2H^{2})+\frac{3}{4}\dot{Q}+\frac{9}{4}HQ +\frac{3}{32}Q^{2}=Ce^{\frac{1}{4}\int Qdt} \label{QH1}
	\eeq
	Furthermore, taking the $00$-component of the field equations we have
	\beq
	R_{00}-\frac{1}{4}g_{00}R=0 \Rightarrow  \nonumber
	\eeq
	\beq
	\frac{\ddot{a}}{a}=\frac{C}{12}e^{\frac{1}{4}\int Qdt}-\frac{\dot{Q}}{8}-\frac{1}{8}HQ
	\eeq
	or
	\beq
	(\dot{H}+H^{2})+\frac{\dot{Q}}{8}+\frac{1}{8}HQ      =\frac{C}{12}e^{\frac{1}{4}\int Qdt} \label{QH2}
	\eeq
	where we have used the fact that for Weyl non-metricity, the full Ricci tensor reads\footnote{This is easily seen by decomposing the connection into the Levi-Civita plus the contribution for torsion and then substituting the result into the decomposition of the full Ricci tensor in terms of its Riemannian and non-metric part. }
	\begin{gather}
	R_{\mu\nu}=\tilde{R}_{\mu\nu}+\frac{1}{2 n}\left( \tilde{\nabla}_{\mu}Q_{\nu}+\tilde{\nabla}_{\nu}Q_{\mu}-(\tilde{\nabla}_{\alpha}Q^{\alpha}) g_{\mu\nu} \right)-\frac{1}{2}\tilde{\nabla}_{\nu}Q_{\mu} \nonumber \\
	+\frac{(n-2)}{ (2 n)^{2}}\Big( Q_{\mu}Q_{\nu}-(Q_{\alpha}Q^{\alpha})g_{\mu\nu} \Big)
	\end{gather}
	From which, we get the component\footnote{We also set $n=4$ for the spacetime dimension.}
	\beq
	R_{00}=\tilde{R}_{00}-\frac{3}{8}( \dot{Q}+HQ)
	\eeq
	Now, upon combining $(\ref{QH1})$ and $(\ref{QH2})$, we arrive at
	\beq
	H^{2}+\frac{HQ}{4}+\frac{Q^{2}}{64}=\frac{C}{12}e^{\frac{1}{4}\int Qdt}
	\eeq
	Observing now that the terms at the left hand side form  a complete square, the latter can be written as
	\beq
	\left( H+\frac{Q}{8} \right)^{2}=\frac{C}{12}e^{\frac{1}{4}\int Qdt}
	\eeq
	which gives
	\beq
	H=H_{0}e^{\frac{1}{8}\int Q dt}-\frac{Q}{8} \label{H}
	\eeq
	where $H_{0}$ is a constant that can be both positive or negative. Note now that this is exactly the same expression, for the Hubble parameter, with the one found in \cite{capozziello2008f} upon the exchange $T \leftrightarrow  \frac{3}{8}Q$. We comment more on this duality and make it more clear in what follows. Let us now find solutions for the scale factor.

	\subsubsection{ Solutions}
	Integrating ($\ref{H}$) for a general function $Q(t)$ we get for the scale factor
	\beq
	a(t)=a_{0}e^{\Lambda(t)}
	\eeq
	where $a_{0}$ is another integration constant and
	\beq
	\Lambda(t)=\int \Big[ H_{0}e^{\frac{1}{8}\int Q dt} -\frac{Q}{8}  \Big]  dt
	\eeq
	which is the most general solution. Let us now assume that the non-metric component is constant, that is $Q(t)=Q_{0}$, then the above reads
	\beq
	a(t)= a_{0}e^{\frac{8 H_{0}}{Q_{0}}e^{\frac{Q_{0}}{8}t}-\frac{Q_{0}}{8}t}
	\eeq
	which is a non-metric cosmological expansion! It is interesting to note that one gets accelerated expansion even when there is on non-metricity $Q_{\mu}=0$. Indeed, looking at ($\ref{nonm}$) we see that for $Q_{\mu}=0$ one gets $\partial_{\mu}R=0\Rightarrow R=constant=\tilde{R}$. Then, we effectively have a cosmological constant sourced solely by the curvature scalar $\tilde{R}$ as can be easily seen by the field equations
	\beq
	\tilde{R}_{\mu\nu}=\frac{\tilde{R}}{4}g_{\mu\nu}=\Lambda g_{\mu\nu}
	\eeq
	where we have set $\Lambda =\frac{\tilde{R}}{4}$. Then,  assuming that $\tilde{R}>0$, the solutions for the Hubble parameter and the scale factor read
	\beq
	H=H_{0}=\sqrt{\frac{\tilde{R}}{12}}=\sqrt{\frac{\Lambda}{3}}
	\eeq
	\beq
	a(t)=a_{0}e^{H_{0}(t-t_{0})}
	\eeq
	Note, of course, that the above cannot be regarded as a realistic cosmological model since we have a free function ($Q(t)$) in our Theory due to the conformal nature of the $R^{2}$ term. We do however consider it as a useful toy model to illustrate the connection between torsion and non-metricity, for simple models on a cosmological context.

	\subsubsection{Impact on vector's lengths}
	As we know the presence of non-metricity changes the lengths of vectors (when the vector is parallely transported along a given curve) according to
	\beq
	\frac{d(w^{\mu}w_{\mu})}{d\lambda}=-Q_{\alpha\mu\nu}\frac{dx^{\alpha}}{d\lambda}w^{\mu}w^{\nu}
	\eeq
	As we showed, in our case the non metricity takes the simple Weyl form
	\beq
	Q_{\alpha\mu\nu}=\frac{1}{4}Q_{\alpha}g_{\mu\nu}
	\eeq
	for which the above is written as
	\beq
	dl=-\frac{1}{4}l Q_{\alpha}dx^{\alpha}
	\eeq
	where $l=w^{\mu}w_{\mu}$. The later is easily integrated to give
	\beq
	l \propto e^{-\frac{1}{4}\int Q_{\mu}dx^{\mu}}
	\eeq
	and in the case of an FLRW universe
	\beq
	l \propto e^{-\frac{1}{4}\int Q_{0}(t)dt}
	\eeq
	and for the particular case where $Q_{0}(t)=Q_{0}=cnst.$ takes the form
	\beq
	l=l_{0}e^{-\frac{Q_{0}}{4}t}
	\eeq

	\section{Torsion-Non metricity duality}
	Note that in our previous model, the Ricci scalar decomposition in terms of its Riemannian and non metric parts is given by
	\beq
	R=\tilde{R}-\frac{3}{4}\tilde{\nabla}_{\mu}Q^{\mu}-\frac{3}{32}Q_{\mu}Q^{\mu} 
	\eeq
	or
	\beq
	R=\tilde{R}+\frac{3}{4}\dot{Q}+\frac{9}{4}HQ+\frac{3}{32}Q^{2} 
	\eeq
	Now we state that this expression is dual to the one (which had only torsion) appearing in [] which was found there to be
	\beq
	R=\tilde{R}+2\dot{T}+6 HT+\frac{2}{3}T^{2} 
	\eeq
	Indeed, it can be easily seen that one maps to another by making the exchange
	\beq
	T \leftrightarrow  \frac{3}{8}Q
	\eeq
	This duality not only holds for an FLRW geometry but for any other geometry as can be easily seen from the general Ricci scalars
	\beq
	R=\tilde{R}-2\tilde{\nabla}_{\mu}T^{\mu}-\frac{2}{3}T_{\mu}T^{\mu} \label{ricciS}
	\eeq
	\beq
	R=\tilde{R}-\frac{3}{4}\tilde{\nabla}_{\mu}Q^{\mu}-\frac{3}{32}Q_{\mu}Q^{\mu} \label{ricciQ}
	\eeq
	with the exchange
	\beq
	T^{\mu}\leftrightarrow  \frac{3}{8}Q^{\mu}
	\eeq
	for a general dimension $n$ the latter is generalized to
	\beq
	T^{\mu}\leftrightarrow  \frac{n-1}{2n}Q^{\mu}
	\eeq
	We should point out that this $T^{\mu}$ vector that appears in \cite{capozziello2008f} is related to the torsion vector through $T^{\mu}=2S^{\mu}$, so the duality really looks like
	\beq
	S^{a} \leftrightarrow    \frac{n-1}{4 n}Q^{a} \label{sqduality}
	\eeq
	Note now that this duality\footnote{Similar dualities of torsion and non-metricity have reported previously in the literature \cite{berthias1993torsion,sotiriou2009f}. Note however that such dualities arise only when the theories at hand are projective invariant.} can also be seen when one is looking at the autoparallel equation. In the case of torsion the connection is found to be
	\beq
	\Gamma^{\lambda}_{\;\;\mu\nu}=\tilde{\Gamma}^{\lambda}_{\;\;\mu\nu}-\frac{2}{n-1}\Big( S^{\lambda}g_{\mu\nu}-S_{\mu}\delta^{\lambda}_{\nu} \Big)
	\eeq
	so that the autoparallel equation is
	\beq
	\ddot{x}^{\alpha}+\tilde{\Gamma}^{\alpha}_{\;\;\mu\nu}\dot{x}^{\mu}\dot{x}^{\nu}=-\frac{2}{n-1}(S_{\mu}\dot{x}^{\mu})\dot{x}^{a}+\frac{2}{n-1}S^{a}\dot{x}^{2}
	\eeq
	where $\dot{x}^{2}=g_{\mu\nu}\dot{x}^{\mu}\dot{x}^{\nu}$ and the dot represents differentiation with respect to the affine parameter $\lambda$. In the non-metric case the connection is
	\beq
	\Gamma^{\lambda}_{\;\;\mu\nu}=\tilde{\Gamma}^{\lambda}_{\;\;\mu\nu}-\frac{1}{2n}\Big( Q^{\lambda}g_{\mu\nu}-2 Q_{(\mu}\delta_{\nu)}^{\lambda} \Big)
	\eeq
	with autoparallels 
	\beq
	\ddot{x}^{\alpha}+\tilde{\Gamma}^{\alpha}_{\;\;\mu\nu}\dot{x}^{\mu}\dot{x}^{\nu}=-\frac{1}{n}(Q_{\mu}\dot{x}^{\mu})\dot{x}^{a}+\frac{1}{2n}Q^{a}\dot{x}^{2}
	\eeq
	Note now that in both cases the first term on the right hand side of the autoparallel equation is of the form $f(\lambda)\dot{x}^{a}$ which as we know can be dropped by a re-parametrization  of the curve. So we see that the duality is also apparent in the autoparallel equation when one exchanges
	\beq
	\frac{2}{n-1}S^{a} \leftrightarrow    \frac{1}{2n}Q^{a} \label{sqduality}
	\eeq
	As far as the Ricci tensors are concerned, when only torsion is present one has\footnote{For now on we will do the calculations for general dimension $n$ and only set $n=4$ when we study the cosmological implications.}
	\beq
	R_{\mu\nu}=\tilde{R}_{\mu\nu}-\frac{2(n-2)}{(n-1)}\tilde{\nabla}_{\nu}S_{\mu}-\frac{2}{(n-1)}(\tilde{\nabla}_{\alpha}S^{\alpha})g_{\mu\nu}+4\frac{(n-2)}{(n-1)^{2}}\Big[ S_{\mu}S_{\nu}-(S_{\alpha}S^{\alpha})g_{\mu\nu} \Big] \label{rics}
	\eeq
	while, when we have only non-metricity
	\begin{gather}
	R_{\mu\nu}=\tilde{R}_{\mu\nu}+\frac{1}{2 n}\left( \tilde{\nabla}_{\mu}Q_{\nu}+\tilde{\nabla}_{\nu}Q_{\mu}-(\tilde{\nabla}_{\alpha}Q^{\alpha}) g_{\mu\nu} \right)-\frac{1}{2}\tilde{\nabla}_{\nu}Q_{\mu} \nonumber \\
	+\frac{(n-2)}{ (2 n)^{2}}\Big( Q_{\mu}Q_{\nu}-(Q_{\alpha}Q^{\alpha})g_{\mu\nu} \Big) \label{ricq}
	\end{gather}
	Next we ask the question, what happens when both torsion and non-metricity are present and find  generalizations to the solutions we found so far.

	\section{Mixed Torsion and Non-metricity}
	As we have already pointed out, when one allows only torsion to be present (for $f(R)=R^{\frac{n}{2}}$) one gets accelerated expansion due to the torsion vector $S_{\mu}$ and it can be seen that the full Ricci scalar reads\footnote{In general dimension $n$.}
	\beq
	R=C e^{\frac{4}{n-1}\int S_{\mu}dx^{\mu}}
	\eeq
	while when only non-metricity is present
	\beq
	R=C e^{\frac{1}{n}\int Q_{\mu}dx^{\mu}}
	\eeq
	Now let us see what happens for a general $f(R)=R^{\frac{n}{2}}$ theory in vacuum when both torsion and non-metricity are different from zero. Variation with respect to the metric gives
	\beq
	n R^{n/2-1} \Big( R_{(\mu\nu)}-\frac{R}{n}g_{\mu\nu}\Big) =0
	\eeq
	and disregarding the trivial solution $R=0$, the field equations follow
	\beq
	R_{(\mu\nu)}-\frac{R}{n}g_{\mu\nu}=0
	\eeq
	Varying with respect to the affine connection we get
	\begin{gather}
	-\nabla_{\lambda}(\sqrt{-g}g^{\mu\nu}R^{n/2-1})+\nabla_{\sigma}(\sqrt{-g}g^{\mu\sigma}R^{n/2-1})\delta^{\nu}_{\lambda} \\
	+2\sqrt{-g}R^{n/2-1}(S_{\lambda}g^{\mu\nu}-S^{\mu}\delta_{\lambda}^{\nu}+g^{\mu\sigma}S_{\sigma\lambda}^{\;\;\;\;\nu})=0 
	\end{gather}
	and after some contractions it follows that
	\beq
	\frac{\partial_{\mu}R}{R}=\frac{1}{n}Q_{\mu}+\frac{4}{n-1}S_{\mu}
	\eeq
	which can be directly integrated to give
	\beq
	R=C e^{\int (\frac{1}{n}Q_{\mu}+ \frac{4}{n-1}S_{\mu} )  dx^{\mu}}
	\eeq
	After some lengthy calculations, we can also solve for the affine connection
	\beq
	\Gamma^{\lambda}_{\;\;\;\;\mu\nu}=\tilde{\Gamma}^{\lambda}_{\;\;\;\;\mu\nu}+\frac{1}{2 n}\left( \delta^{\lambda}_{\nu}\Big( Q_{\mu}+\frac{4 n}{n-1}S_{\mu}\Big) -g_{\mu\nu}\Big( Q^{\lambda} +\frac{4 n}{n-1}S_{\mu}\Big)  \right) +\frac{1}{2}\delta^{\lambda}_{\mu}\tilde{Q}_{\nu}
	\eeq
	Note now that the last term containing $\tilde{Q}_{\nu}$ can be ignored due to the invariance of our starting action under projective transformations. Defining 
	\beq
	w_{\mu} :=  \frac{1}{n}Q_{\mu}+ \frac{4}{n-1}S_{\mu} = \frac{\partial_{\mu}R}{R}
	\eeq
	and dropping the last term, the latter may also be written as
	\beq
	\Gamma^{\lambda}_{\;\;\;\;\mu\nu}=\tilde{\Gamma}^{\lambda}_{\;\;\;\;\mu\nu}+\frac{1}{2}\left( \delta^{\lambda}_{\nu}w_{\mu} -g_{\mu\nu}w^{\lambda} \right) 
	\eeq
	We should stress out from this final result, that there exists a possibility where we can have both torsion and non-metricity but their contributions in $w_{\mu}$ be such that they cancel out. This happens when $\frac{1}{n}Q_{\mu}+ \frac{4}{n-1}S_{\mu}=0$. Then, one has effectively a Riemannian space with the usual Levi-Civita connection. Also, using the above definitions, the Ricci tensor and scalar are decomposed according to
	\begin{gather}
	R_{\mu\nu}=\tilde{R}_{\mu\nu}-\frac{(n-2)}{2}\tilde{\nabla}_{\nu}w_{\mu}-\frac{1}{2}(\tilde{\nabla}_{\alpha}w^{\alpha})g_{\mu\nu} +\frac{(n-2)}{4}\Big[ w_{\mu} w_{\nu}-(w_{\alpha}w^{\alpha})g_{\mu\nu} \Big]
	\end{gather}
	\beq
	R=\tilde{R}+(1-n)\tilde{\nabla}_{\mu}w^{\mu}-\frac{(n-2)(n-1)}{4}w_{\mu}w^{\mu}
	\eeq
	Notice that for vanishing torsion ($S_{\mu}=0$) the above Ricci scalar reduces to ($\ref{ricciQ}$) while when non-metricity is zero ($Q_{\mu}=0$) it reduces to $(\ref{ricciS})$. As far as the Ricci tensor is concerned, for vanishing torsion one gets $(\ref{ricq})$\footnote{Note that in order to prove this one also needs to use the fact that because $Q_{\mu}$ is exact $(Q_{\mu}\propto \partial_{\mu}\ln|R|)$, it holds that $\tilde{\nabla}_{[\nu}w_{\mu]}=0$ and therefore in this case the Ricci tensor is symmetric.} and for vanishing non-metricity $(\ref{rics})$. Having these, we can again find the evolution of the Hubble parameter
	\beq
	H=H_{0}e^{\frac{1}{2}\int w dt}-\frac{w}{2}
	\eeq
	where $w=w_{0}$.
	
	\section{Torsion/Non-Metricity interrelation}
	
	As we saw above, a geometry with vectorial torsion and zero non-metricity seems to have the same effects with a geometry that has Weyl non-metricity (i.e. vectorial form) and zero torsion. In fact this is always true if the theory at hand has projective invariance. This has been mentioned sometimes in the literature but no mathematical proof of the equivalence was ever given. We will now state and prove this equivalence for the first time in the literature (at least to our knowledge).  So, we state and prove the following proposition.
	\newline
	\textbf{ Proposition.} 
	A projective invariant theory of gravity with vectorial torsion and zero non-metricity can be switched with a torsionless theory with Weyl non-metricity, by means of a projective transformation of the affine connection, and vice versa: 
	\beq
	S_{\mu\nu}^{\;\;\;\;\lambda}=\frac{2}{n-1}S_{[\mu}\delta_{\nu]}^{\lambda}\;,\; Q_{\alpha\mu\nu}=0 \Longleftrightarrow S_{\mu\nu}^{\;\;\;\;\lambda}=0\;,\; Q_{\alpha\mu\nu}=\frac{Q_{\alpha}}{n}g_{\mu\nu}
	\eeq
	\newline
	\textbf{Proof:}
	Let us first prove the $''$ $\Rightarrow$ $''$ part. Starting with
	\beq
	S_{\mu\nu}^{\;\;\;\;\lambda}=\frac{2}{n-1}S_{[\mu}\delta_{\nu]}^{\lambda}\;,\; Q_{\alpha\mu\nu}=0
	\eeq
	let us perform the projective transformation (since our theory is respects projective invariance by assumption)
	\begin{equation}
	\Gamma^{\lambda}_{\;\;\mu\nu}\longrightarrow\hat{\Gamma}^{\lambda}_{\;\;\mu\nu} =\Gamma^{\lambda}_{\;\;\mu\nu}+ \delta_{\mu}^{\lambda}\xi_{\nu} 
	\end{equation}
	Then, the torsion tensor transforms as
	\beq
	\hat{S}_{\mu\nu}^{\;\;\;\;\lambda}=\frac{1}{n-1}\Big( S_{\mu}-\frac{n-1}{2}\xi_{\mu}\Big)\delta_{\nu}^{\lambda}-\frac{1}{n-1}\Big( S_{\nu}-\frac{n-1}{2}\xi_{\nu}\Big)\delta_{\mu}^{\lambda}
	\eeq
	From which we see that for the gauge choice 
	\beq
	\xi_{\mu}=\frac{2}{n-1}S_{\mu}
	\eeq
	we have 
	\beq
	\hat{S}_{\mu\nu}^{\;\;\;\;\lambda}=0
	\eeq
	and 
	\beq
	\hat{Q}_{\nu\alpha\mu}=2 N_{(\alpha\mu)\nu}=2g_{\alpha\mu}\xi_{\nu}\Rightarrow \hat{Q}_{\alpha\mu\nu}=\frac{4}{n-1}S_{\alpha} g_{\mu\nu}
	\eeq
	which is the case of a Weyl non-metricity with Weyl vector $Q_{\mu}=\frac{4 n}{n-1}S_{\mu}$ and vanishing torsion. Conversely ( $''$ $\Leftarrow$ $''$), starting with 
	\beq
	S_{\mu\nu}^{\;\;\;\;\lambda}=0\;,\; Q_{\alpha\mu\nu}=\frac{Q_{\alpha}}{n}g_{\mu\nu}
	\eeq
	and performing a projective transformation but now for the gauge choice
	\beq
	\xi_{\mu}=-\frac{1}{2 n}Q_{\mu}
	\eeq
	leaves us with
	\beq
	\hat{Q}_{\alpha\mu\nu}=0
	\eeq
	\beq
	\hat{S}_{\mu\nu}^{\;\;\;\;\lambda}=\delta^{\lambda}_{[\mu}\xi_{\nu]}=\frac{1}{2 n}Q_{[\mu}\delta_{\nu]}^{\lambda}
	\eeq
	which is the case of a metric theory with vectorial torsion! So, to conclude when we have very restricted forms of torsion and non-metricity, torsion and non-metricity can be exchanged with one another and therefore the effect of torsion is the same with that of non-metricity. Of course this is not true for general forms of torsion and non-metricity since the non-metricity tensor has more degrees of freedom from the torsion tensor. However, as we have seen for restricted forms the two are related. To state it one more time, a projective invariant theory with vectorial torsion and zero non-metricity can  be traded by a (projective invariant) theory with Weyl non-metricity and zero torsion. Note that projective invariance is key here.

	\chapter{Cosmology with Torsion and Non-metricity}

	In this Chapter we study to some degree the kinematics of torsion and non-metricity firstly for general spacetimes and then for FLRW Cosmologies.  We then derive the allowed forms of torsion and non-metricity that can live in such highly symmetric Cosmological spacetimes. In addition we obtain, for the first time, the form of fixed length vector non-metricity that is allowed in FLRW spacetimes. We  find cosmological solutions of universes with torsion and also derive, for the first time in the literature, the modified Friedmann equations in the presence of non-metricity.

	\section{Kinematics with Torsion and Non-Metricity}
	In this section we wish to study how does  the  time projected continuity equation
	\beq
	u_{\nu}\nabla_{\mu}T^{\mu\nu}=0
	\eeq
	modify in the presence of torsion and non-metricity and then apply our results in Cosmology. In Einstein's Gravity this is a direct consequence of the contracted Bianchi identities $\nabla_{\mu}G^{\mu\nu}=0$where $G_{\mu\nu}$ is the Einstein tensor\footnote{In Riemannian Geometry $R_{\mu\nu}R_{(\mu\nu)}$.}
	\beq
	G_{\mu\nu}=R_{\mu\nu}-\frac{R}{2}g_{\mu\nu}
	\eeq
	and any quantity that appears is purely Riemannian. Now, when torsion and non-metricity are present, the Bianchi identities have a more complicated form\footnote{Of course their Riemannian parts themselves satisfy all the identities of Einstein's Gravity but the quantities appearing in the field equations are the total ones with torsion and non-metric contributions too. These satisfy sets of identities that are much more complicated than their Riemannian counterparts. }and one expects to have an equation that goes like
	\begin{gather}
	u_{\nu}\nabla_{\mu}T^{\mu\nu}= \nabla_{\mu}G^{\mu\nu}=\nabla_{\mu} \left(R^{(\mu\nu)}-\frac{R}{2}g^{\mu\nu} \right)= \nonumber \\ 
	Torsion+NonMetricity \;\;Terms \neq 0
	\end{gather}
	In what follows we find exactly what the right hand terms look like firstly when one allows torsion (but with vanishing non-metricity) and later when one allows non-metricity (but with vanishing torsion). The entire analysis is done for the case when the Field equations are formalistically the same with Einstein Equations
	\beq
	G_{\mu\nu}\equiv R_{(\mu\nu)}-\frac{R}{2}g_{\mu\nu}=\kappa T_{\mu\nu}
	\eeq
	but with the quantities appearing in it having both the Riemannian and non-Riemannian contributions. So, lets start with torsion.

	\subsection{Kinematics with Torsion}
	Let us consider the Einstein-Cartan model of Gravity, that is a torsion-full and metric theory of gravity given by
	\beq
	S=\frac{1}{2\kappa}\int d^{4}x \sqrt{-g}R+ S_{M}[g_{\mu\nu}, \Gamma^{\lambda}_{\;\;\;\alpha\beta}]
	\eeq
	Variation with respect to the metric tensor gives the modified torsion-full Einstein equations
	\beq
	G_{\mu\nu}\equiv R_{(\mu\nu)}-\frac{R}{2}g_{\mu\nu}=\kappa T_{\mu\nu}
	\eeq
	Note now that in Einstein equations only the symmetric part of the Ricci tensor contributes. The covariant derivative of the above reads
	\beq
	\kappa \nabla^{\mu}T_{\mu\nu}=\nabla^{\mu}G_{\mu\nu} \label{Ein}
	\eeq
	where $\nabla^{\mu}G_{\mu\nu}\neq 0$, and depends on torsion. Let us find the kinematics of the latter in the case of a perfect fluid
	\beq
	T_{\mu\nu}=\rho u_{\mu}u_{\nu}+p h_{\mu\nu}
	\eeq
	where $u_{\mu}$ is an observers'$4$-velocity and $h_{\mu\nu}=g_{\mu\nu}+u_{\mu}u_{\nu}$ the projective tensor. For such a fluid, we readily derive
	\beq
	\nabla^{\mu}T_{\mu\nu}=\Big[ \dot{\rho}+\Theta(\rho+p)\Big]u_{\nu}+(\rho+p)A_{\nu}+D_{\nu}p \label{T}
	\eeq
	Now, regarding the part involving the Einstein tensor we proceed as follows. By contracting the Bianchi identities we obtain
	\beq
	\nabla^{\mu}\left(R_{\mu\nu}-\frac{1}{2}R g_{\mu\nu} \right)=-( 2R^{\alpha\beta}S_{\nu\alpha\beta}+R_{\alpha\beta\lambda\nu}S^{\alpha\beta\lambda})
	\eeq
	where $R_{\mu\nu}$ is now the full Ricci tensor (both symmetric and antisymmetric parts). Decomposing 
	\beq
	R_{\mu\nu}=R_{(\mu\nu)}+R_{[\mu\nu]}
	\eeq
	and using the fact that
	\beq
	R_{[\mu\nu]}=\nabla^{\alpha}S_{\mu\nu\alpha}+2\nabla_{[\mu}S_{\nu]}-2S_{\mu\nu\alpha}S^{\alpha}
	\eeq
	where $S_{\mu}\equiv S_{\mu\lambda}^{\;\;\;\;\lambda}$, it follows that
	\beq
	\nabla^{\mu}G_{\mu\nu}+\nabla^{\mu}R_{[\mu\nu]}=-( 2R^{\alpha\beta}S_{\nu\alpha\beta}+R_{\alpha\beta\lambda\nu}S^{\alpha\beta\lambda}) \Rightarrow  \nonumber
	\eeq
	\begin{gather}
	\nabla^{\mu}G_{\mu\nu}=-\Big[ \nabla^{\mu}\nabla^{\alpha}S_{\mu\nu\alpha}+\Box S_{\nu}-\nabla^{\mu}\nabla_{\nu}S_{\mu}-2\nabla^{\mu}(S_{\mu\nu\alpha}S^{\alpha}) \nonumber \\
	+ 2R^{\alpha\beta}S_{\nu\alpha\beta}+R_{\alpha\beta\lambda\nu}S^{\alpha\beta\lambda} \Big]
	\end{gather}
	Combining this with $(\ref{Ein})$ and $(\ref{T})$ it follows that
	\begin{gather}
	\kappa\Big[ \dot{\rho}+\Theta(\rho+p)\Big] u_{\nu}+\kappa(\rho+p)A_{\nu}+\kappa D_{\nu}p =\nonumber \\
	=-\Big[ \nabla^{\mu}\nabla^{\alpha}S_{\mu\nu\alpha}+\Box S_{\nu}-\nabla^{\mu}\nabla_{\nu}S_{\mu}-2\nabla^{\mu}(S_{\mu\nu\alpha}S^{\alpha}) 
	+ 2R^{\alpha\beta}S_{\nu\alpha\beta}+R_{\alpha\beta\lambda\nu}S^{\alpha\beta\lambda} \Big]
	\end{gather}
	Projecting this along $u^{\nu}$ we have
	\begin{gather}
	\kappa\Big[ \dot{\rho}+\Theta(\rho+p)\Big]
	=\Big[ \nabla^{\mu}\nabla^{\alpha}S_{\mu\nu\alpha}+\Box S_{\nu}-\nabla^{\mu}\nabla_{\nu}S_{\mu}-2\nabla^{\mu}(S_{\mu\nu\alpha}S^{\alpha}) \Big]u^{\nu} \nonumber \\
	+ 2R^{\alpha\beta}S_{\nu\alpha\beta}u^{\nu}+R_{\alpha\beta\lambda\nu}S^{\alpha\beta\lambda} u^{\nu} \label{kinematics}
	\end{gather}
	Note now that one would like to fully eliminate the last two terms (Riemann tensor depended) appearing on the RHS of the above and solely express everything in terms of torsion. Let us see how we can deal with the first one. We start by decomposing
	\beq
	2R^{\alpha\beta}S_{\nu\alpha\beta}u^{\nu}=2R^{[\alpha\beta]}S_{\nu\alpha\beta}u^{\nu}+2R^{(\alpha\beta)}S_{\nu\alpha\beta}u^{\nu}
	\eeq
	Now, since the form of $R_{[\alpha\beta]}$ is known, the first term on the RHS of the latter is written as
	\beq
	2R^{[\alpha\beta]}S_{\nu\alpha\beta}u^{\nu}=2\Big[ \nabla_{\alpha}S_{\mu\nu}^{\;\;\;\;\;\alpha}+2\nabla_{[\mu}S_{\nu]}-2 S_{\mu\nu}^{\;\;\;\;\;\alpha}S_{\alpha}\Big] S^{\lambda\mu\nu}u_{\lambda}
	\eeq
	So long as the second one is concerned, we start by writing down the field equations
	\beq
	R^{(\alpha\beta)}-\frac{1}{2}R g^{\alpha\beta}=\kappa T^{\alpha\beta}
	\eeq
	which when contracted with $S_{\nu\alpha\beta}$  yield
	\beq
	R^{(\alpha\beta)}S_{\nu\alpha\beta}-\frac{1}{2}R S_{\nu}=\kappa T^{\alpha\beta}S_{\nu\alpha\beta}
	\eeq
	and upon using
	\beq
	R=\kappa (\rho- 3 p)
	\eeq
	along with
	\beq
	T^{\alpha\beta}S_{\nu\alpha\beta}=(\rho+p)S_{\nu\alpha\beta}u^{\alpha}u^{\beta}+p S_{\nu}
	\eeq
	we obtain
	\beq
	2R^{(\alpha\beta)}S_{\nu\alpha\beta}=\kappa \Big[ 2(\rho+ p)S_{\nu\alpha\beta}u^{\alpha}u^{\beta}+(\rho -p)S_{\nu}  \Big]
	\eeq
	Now, contract with $u^{\nu}$ to finally arrive at
	\beq
	2R^{(\alpha\beta)}S_{\nu\alpha\beta}u^{\nu}=\kappa (\rho -p)S_{\nu} u^{\nu}
	\eeq
	where we have used the fact that $S_{\nu\alpha\beta}u^{\nu}u^{\alpha}u^{\beta}=0$ since $S_{\nu\alpha\beta}$ is antisymmetric in $\nu,\alpha$ and $u^{\nu}u^{\alpha}u^{\beta}$ fully symmetric in all its indices. Now we want to deal with the term $R_{\alpha\beta\lambda\nu}S^{\alpha\beta\lambda} u^{\nu}$. Note that the fact that the index contracted with $u^{\nu}$ is the last one\footnote{If it were the second one (or the first one) we could have readily eliminate it by the very definition of the Riemann tensor (antisymmetrized covariant derivative acting on $u^{\nu}$).} makes this term more elaborate to work with. A way to proceed goes as follows. Take the antisymmetrized Riemann tensor and contract it with the torsion tensor to obtain
	\beq
	R_{\alpha[\beta\mu\nu]}S^{\alpha\beta\mu}=\frac{1}{3}\Big( R_{\alpha\beta\mu\nu}S^{\alpha\beta\mu} +R_{\alpha\nu\beta\mu}S^{\alpha\beta\mu}+R_{\alpha\mu\nu\beta}S^{\alpha\beta\mu}     \Big)
	\eeq
	Circularly permuting $\alpha\rightarrow \mu\rightarrow \beta\rightarrow \alpha$ in the torsion tensor, we also have
	\beq
	R_{\alpha[\beta\mu\nu]}S^{\mu\alpha\beta}=\frac{1}{3}\Big( R_{\alpha\beta\mu\nu}S^{\mu\alpha\beta} +R_{\alpha\nu\beta\mu}S^{\mu\alpha\beta}+R_{\alpha\mu\nu\beta}S^{\mu\alpha\beta}     \Big)
	\eeq
	and permuting once more
	\beq
	R_{\alpha[\beta\mu\nu]}S^{\beta\mu\alpha}=\frac{1}{3}\Big( R_{\alpha\beta\mu\nu}S^{\beta\mu\alpha} +R_{\alpha\nu\beta\mu}S^{\beta\mu\alpha}+R_{\alpha\mu\nu\beta}S^{\beta\mu\alpha}     \Big)
	\eeq
	Now, using the symmetries of both the Riemann and torsion tensor and some relabeling of the dummy indices, the last two equations may be written as
	\beq
	R_{\alpha[\beta\mu\nu]}S^{\mu\alpha\beta}=\frac{1}{3}\Big( R_{\alpha\mu\nu\beta}S^{\alpha\beta\mu}+R_{\alpha\nu\beta\mu}S^{\mu\alpha\beta}+R_{\alpha\beta\mu\nu}S^{\alpha\beta\mu}   \Big)
	\eeq 
	and
	\beq
	R_{\alpha[\beta\mu\nu]}S^{\beta\mu\alpha}=\frac{1}{3}\Big( R_{\alpha\mu\nu\beta}S^{\alpha\beta\mu}+R_{\alpha\nu\beta\mu}S^{\beta\mu\alpha}  + R_{\alpha\mu\nu\beta}S^{\alpha\beta\mu}   \Big)
	\eeq
	Combining all three\footnote{That is subtracting the last one from the former two.}, we obtain
	\beq
	R_{\alpha\beta\mu\nu}S^{\alpha\beta\mu}u^{\nu}=\frac{3}{2}(S^{\alpha\beta\mu}+S^{\mu\alpha\beta}-S^{\beta\mu\alpha})\Big( R_{\alpha[\beta\mu\nu]}u^{\nu}-\frac{1}{3}R_{\alpha\nu\beta\mu}u^{\nu} \Big)
	\eeq
	Notice now, that the above combination of the torsion tensor is (up to a minus sign) exactly the contorsion tensor. More precisely, it holds that
	\beq
	S^{\alpha\beta\mu}+S^{\mu\alpha\beta}-S^{\beta\mu\alpha}=-K^{\mu\alpha\beta}
	\eeq
	so that
	\beq
	R_{\alpha\beta\mu\nu}S^{\alpha\beta\mu}u^{\nu}=K^{\mu\alpha\beta} \Big(\frac{1}{2}R_{\alpha\nu\beta\mu}u^{\nu} - \frac{3}{2}R_{\alpha[\beta\mu\nu]}u^{\nu} \Big)
	\eeq
	From this point on is simply a matter of application of identities to express everything in terms of torsion. Indeed, using 
	\beq
	R^{\alpha}_{\;\;\;[\beta\mu\nu]}=-2 \nabla_{[\beta}S_{\mu\nu]}^{\;\;\;\;\;\alpha}-4 S_{[\beta\mu}^{\;\;\;\;\;\lambda}S_{\nu]\lambda}^{\;\;\;\;\;\alpha}
	\eeq
	along with
	\begin{equation}
	2\nabla_{[\alpha} \nabla_{\beta]}u^{\mu}=R^{\mu}_{\;\;\;\nu\alpha\beta} u^{\nu}+2 S_{\alpha\beta}^{\;\;\;\;\;\nu}\nabla_{\nu}u^{\mu} \Rightarrow  \nonumber
	\end{equation}
	\beq
	R^{\mu}_{\;\;\;\nu\alpha\beta} u^{\nu}=2\Big(\nabla_{[\alpha} \nabla_{\beta]}-S_{\alpha\beta}^{\;\;\;\;\;\nu}\nabla_{\nu} \Big)u^{\mu}
	\eeq
	it follows that
	\beq
	R_{\alpha\beta\mu\nu}S^{\alpha\beta\mu}u^{\nu}=K^{\mu\alpha\beta}\Big[ (\nabla_{[\beta}\nabla_{\mu]}-S_{\mu\beta}^{\;\;\;\;\lambda}\nabla_{\lambda})u_{\alpha}+3 u^{\nu}( \nabla_{[\beta}S_{\mu\nu]\alpha}+2S_{[\beta\mu}^{\;\;\;\;\lambda}S_{\nu]\lambda\alpha} )\Big] \nonumber
	\eeq
	Substituting all the above in $(\ref{kinematics})$ we finally arrive at
	\begin{gather}
	\kappa \big[\dot{\rho}+\Theta (\rho +p)\Big] = (u^{\nu}\nabla^{\mu}+2 u_{\lambda}S^{\lambda\mu\nu})(\nabla^{\alpha}S_{\mu\nu\alpha}+2\nabla_{[\mu}S_{\nu]}-2S_{\mu\nu\alpha}S^{\alpha}) \nonumber \\
	+K^{\mu\alpha\beta}\Big[ (\nabla_{[\beta}\nabla_{\mu]}-S_{\mu\beta}^{\;\;\;\;\lambda}\nabla_{\lambda})u_{\alpha}+3 u^{\nu}( \nabla_{[\beta}S_{\mu\nu]\alpha}+2S_{[\beta\mu}^{\;\;\;\;\lambda}S_{\nu]\lambda\alpha})\Big] \nonumber \\
	+\kappa (\rho -p)S_{\nu}u^{\nu} \label{contin}
	\end{gather}
	Note now that the latter can be generalized for any kind of matter that might be present in spacetime. Indeed, going back to our derivation one can easily check that the only term that was $T_{\mu\nu}$-dependent was $2R^{(\alpha\beta)}S_{\nu\alpha\beta}u^{\nu}$, which in the case of a general fluid takes the form
	\beq
	2R^{(\alpha\beta)}S_{\nu\alpha\beta}u^{\nu}=\kappa \Big( -T S_{\nu}u^{\nu}+T^{\alpha\beta}u^{\nu}S_{\nu\alpha\beta} \Big)
	\eeq
	as can be easily seen by contracting the field equations 
	\beq
	R^{(\alpha\beta)}-\frac{1}{2}R g^{\alpha\beta}=\kappa T^{\alpha\beta}
	\eeq
	with $u^{\nu}S_{\nu\alpha\beta}$. So, when no assumption about the matter filling the spacetime is made, the kinematic equation generalizes to
	\begin{gather}
	\kappa u^{\nu}\nabla^{\mu}T_{\mu\nu}= -(u^{\nu}\nabla^{\mu}+2 u_{\lambda}S^{\lambda\mu\nu})(\nabla^{\alpha}S_{\mu\nu\alpha}+2\nabla_{[\mu}S_{\nu]}-2S_{\mu\nu\alpha}S^{\alpha}) \nonumber \\
	-K^{\mu\alpha\beta}\Big[ (\nabla_{[\beta}\nabla_{\mu]}-S_{\mu\beta}^{\;\;\;\;\lambda}\nabla_{\lambda})u_{\alpha}+3 u^{\nu}( \nabla_{[\beta}S_{\mu\nu]\alpha}+2S_{[\beta\mu}^{\;\;\;\;\lambda}S_{\nu]\lambda\alpha})\Big] \nonumber \\
	-\kappa ( -T S_{\nu}u^{\nu}+T^{\alpha\beta}u^{\nu}S_{\nu\alpha\beta})
	\end{gather}
	This is the kinematic's equation generalization in the presence of torsion but with vanishing non-metricity. Next we do the same for non-vanishing non-metricity but zero torsion.

	\subsection{Kinematics of Non-metricity}
	Same way we did with torsion, we now consider a theory with non-metricity but vanishing torsion, that is given by 
	\beq
	S=\frac{1}{2\kappa}\int d^{4}x \sqrt{-g}R+ S_{M}[g_{\mu\nu}, \Gamma^{\lambda}_{\;\;\;\alpha\beta}]
	\eeq
	Variation with respect to the metric tensor gives the modified  Einstein equations
	\beq
	G_{\mu\nu}\equiv R_{(\mu\nu)}-\frac{R}{2}g_{\mu\nu}=\kappa T_{\mu\nu}
	\eeq
	which now contain non-metricity but vanishing torsion. Again, using the generalized  Bianchi identities, we find 
	\begin{gather}
	\nabla_{\mu}G^{\mu\nu}=C^{\nu}+(\tilde{Q}_{\mu}-\nabla_{\mu})R^{[\mu\nu]}-Q_{\rho\mu\beta}R^{\mu\beta}g^{\rho\nu} \nonumber \\
	+( R^{(\mu\nu)}-Rg^{\mu\nu})\tilde{Q}_{\mu}-Q_{\rho\beta\mu}R^{\rho\beta\mu\nu}
	\end{gather}
	where
	\beq
	C^{\nu}=\nabla_{\mu}\Big( g^{\mu\lambda}g^{\rho\nu}g^{\kappa\alpha}(\nabla_{[\alpha}Q_{\rho]\kappa\lambda} ) \Big) +Q^{\alpha\rho\nu}g^{\mu\lambda}\nabla_{[\lambda}Q_{\rho]\mu\alpha}
	\eeq
	We now wish to express everything that appears at the right hand side, in terms of non-metricity. So, what we want to do again, is to express the Riemann tensor and its contractions in terms of the non-metricity and  maybe matter ($T_{\mu\nu}$).In pretty much the same way as we did with torsion, using the generalized curvature identities and the field equations, after some heavy calculations we arrive at
	\begin{gather}
	u_{\nu}\nabla_{\mu}G^{\mu\nu}=\kappa \left[ T^{\mu\nu}(\tilde{Q}_{\mu}u_{\nu}-u^{\alpha}Q_{\alpha\mu\nu})+\frac{T}{2}( \tilde{Q}_{\mu}+Q_{\mu})u^{\mu}  \right] \nonumber \\
	-Q^{\mu\alpha\beta}\left[  u^{\nu}\Big( 2\nabla_{[\beta}Q_{\nu]\alpha\mu}-\nabla_{[\mu}Q_{\nu]\alpha\beta}   \Big)-2 g_{\alpha\lambda}\nabla_{[\beta}\nabla_{\mu]}u^{\lambda}                      \right] \nonumber \\
	+u_{\nu} \left[  \nabla_{\mu}\Big( g^{\mu\lambda}g^{\rho\nu}g^{\kappa\alpha}(\nabla_{[\alpha}Q_{\rho]\kappa\lambda} ) \Big) +Q^{\alpha\rho\nu}g^{\mu\lambda}\nabla_{[\lambda}Q_{\rho]\mu\alpha} \right] \nonumber \\
	+\frac{1}{2}u_{\nu}(\tilde{Q}_{\mu}-\nabla_{\mu})\partial^{[\mu}Q^{\nu]}
	\end{gather}
	So, in this case the kinematic equation is
	\begin{gather}
	\kappa u_{\nu}\nabla_{\mu}T^{\mu\nu}=\kappa \left[ T^{\mu\nu}(\tilde{Q}_{\mu}u_{\nu}-u^{\alpha}Q_{\alpha\mu\nu})+\frac{T}{2}( \tilde{Q}_{\mu}+Q_{\mu})u^{\mu}  \right] \nonumber \\
	-Q^{\mu\alpha\beta}\left[  u^{\nu}\Big( 2\nabla_{[\beta}Q_{\nu]\alpha\mu}-\nabla_{[\mu}Q_{\nu]\alpha\beta}   \Big)-2 g_{\alpha\lambda}\nabla_{[\beta}\nabla_{\mu]}u^{\lambda}                      \right] \nonumber \\
	+u_{\nu} \left[  \nabla_{\mu}\Big( g^{\mu\lambda}g^{\rho\nu}g^{\kappa\alpha}(\nabla_{[\alpha}Q_{\rho]\kappa\lambda} ) \Big) +Q^{\alpha\rho\nu}g^{\mu\lambda}\nabla_{[\lambda}Q_{\rho]\mu\alpha} \right] \nonumber \\
	+\frac{1}{2}u_{\nu}(\tilde{Q}_{\mu}-\nabla_{\mu})\partial^{[\mu}Q^{\nu]}
	\end{gather}
	and this is true for generic matter fields contained in $T_{\mu\nu}$. Next we see some applications of the above considerations in Cosmology.

	\subsubsection{FLRW Cosmology with Torsion}
	Let us consider a flat universe filled with a fluid that generates torsion. The isotropy and homogeneity of such a space allow only one degree of freedom for torsion (call it $\phi(t)$)
	\beq
	S_{0i}^{\;\;\;\;j}=\delta_{i}^{j}\phi(t)
	\eeq
	or
	\beq
	S_{0ij}=g_{ij}\phi(t) \label{sij}
	\eeq
	The modified Friedmann equations, in the presence of torsion, are then
	\beq
	\frac{\ddot{a}}{a}=-\frac{4\pi G}{3}(\rho + 3p)-2\dot{\phi}
	\eeq
	\beq
	\frac{\ddot{a}}{a}+\left(\frac{\dot{a}}{a}\right)^{2}=-\frac{4\pi G}{3}(-\rho + 3p)-2(\dot{\phi}+4\phi^{2})-6\frac{\dot{a}}{a} \phi
	\eeq
	and by subtracting the two, we get
	\beq
	\left(\frac{\dot{a}}{a}\right)^{2}=\frac{8\pi G}{3}\rho -2\left( 4\phi^{2}+\frac{\dot{a}}{a}\phi \right)
	\eeq
	Another combination, gives
	\beq
	\dot{H}=-4\pi G(\rho+p)+2 (4\phi^{2}-\dot{\phi}-H\phi)
	\eeq
	Noticing now that
	\beq
	h_{0i}=0=h_{00}
	\eeq
	for a comoving observer, we can suggest for $(\ref{sij})$ the covariant form\footnote{A similar ansatz was given in \cite{tsamparlis1981methods}.}
	\beq
	S_{\mu\nu\alpha}=2u_{[\mu}h_{\nu]\alpha}\phi
	\eeq
	Taking the above considerations into account, the continuity equation becomes
	\beq
	\dot{\rho}+\Theta (\rho +p)=3 \phi(\rho -p)+\frac{24}{\kappa}H\phi^{2}
	\eeq
	where
	\beq
	\Theta \equiv \nabla_{\mu}u^{\mu}=3H- 6\phi
	\eeq
	Now, one might ask what is the physical significance of the scalar $\phi$ torsional degree of freedom. If you assume that torsion is related to spin (which seems to be true) we can relate the torsion vector $S_{\alpha}$ to a spin vector. Now, we can also observe that for the above ansatz for torsion, the torsion vector is written as
	\beq
	S_{\alpha}=3\phi u_{\alpha}
	\eeq
	Notice now, that if the latter is regarded as a spin current (just as a regular current $\vec{J}= \rho \vec{\upsilon } $ ) we can see that $3\phi$ is its spin density! As a result, spin conservation is ensured as long as
	\beq
	\nabla^{\alpha}S_{\alpha}=0
	\eeq
	From which it follows the continuity equation for $\phi$,
	\beq
	\dot{\phi}+\Theta \phi =0 
	\eeq
	The latter closes the system of equations we need to solve to obtain solutions. To find a solution let us assume that $\phi/H<<1$,(and just focus on dust $p=0$) so that the system of equations becomes
	\beq
	\dot{\phi}+\Theta \phi=0 \Rightarrow 
	\eeq
	\beq
	\dot{\phi}+3H\phi \approx 0 \label{phiH}
	\eeq
	\beq
	\dot{\rho}+\Theta \rho \approx 3\phi \rho \label{rhop}
	\eeq
	\beq
	(H+\phi)^{2} \approx\frac{\kappa}{3}\rho \label{Ha}
	\eeq
	Now, $(\ref{phiH})$ can be directly integrated to give
	\beq
	\phi=\frac{C_{0}}{a^{3}}
	\eeq
	Next we proceed by integrating $(\ref{rhop})$ in two ways. First by eliminating $\Theta$ (using the equation for $\phi$)
	\beq
	\dot{\rho}-\frac{\dot{\phi}}{\phi}\rho \approx 3\phi \rho
	\eeq
	which upon integration results in
	\beq
	\frac{\rho}{\phi}=C_{1}e^{3 \int \phi dt}
	\eeq
	and by expanding $\Theta$,
	\beq
	\dot{\rho}-3 H\rho \approx 9\phi \rho
	\eeq
	which leads to
	\beq
	\rho a^{3}=C_{2}e^{9\int \phi dt}
	\eeq
	Combining the latter two equations, it follows that
	\beq
	\rho= \frac{A}{a^{3}}
	\eeq
	where $A=(C_{0}C_{1})^{3/2}/\sqrt{C_{2}}$. Now, let us find the evolution of the scale factor. Upon using $(\ref{Ha})$ we obtain
	\beq
	H=\pm \sqrt{\frac{\kappa}{3}}\frac{\sqrt{A}}{\alpha^{3/2}}-\frac{C_{0}}{a^{3}}
	\eeq
	and by separating variables and integrating we finally arrive at
	\beq
	a^{3/2}+\frac{C_{0}}{\lambda}\ln{\left( \lambda a^{3/2}-C_{0} \right)}=\frac{3\lambda}{2}t+C
	\eeq
	where $\lambda=\pm  \sqrt{\frac{\kappa}{3}}\sqrt{A} $. Now, for a general barotropic fluid ($p=w\rho$) the solution of the system is
	\beq
	\phi=\frac{C_{0}}{a^{3}}
	\eeq
	\beq
	\rho= \frac{B}{a^{3(1+w)}}
	\eeq
	which shows that when torsion is not strong enough ($\phi H^{-1}<<1$) the matter decouples and evolves as if torsion was not there.
	\subsubsection{Radiation Solution}
	When radiation dominates the universe ($w=1/3$) the solution for the scale factor reads
	\beq
	C_{1}^{2}t+C_{2}=\frac{C_{1}}{2}a^{2}+C_{0}a+\frac{C_{0}}{C_{1}}\ln{(c_{1}a-C_{0})}
	\eeq
	where $C_{1},C_{2}$ are integration constants. Notice that in this case the scale factor cannot become zero and its minimum value is $a_{min}=C_{0}/C_{1}$.

	\subsubsection{Stiff matter solution}
	For a stiff matter model $(w=1)$ the solution is
	\beq
	a(t)=\Big[ 3(C_{0}-C_{1})t+3 C_{2} \Big]^{1/3}
	\eeq
	and we see that in this case also the scale factor is nonzero for $t=0$ in general(expect if $C_{2}=0$).

	\subsubsection{Inflation}
	When torsion is present, during inflation era,the universe expands as
	\beq
	a(t)=\left[ \frac{C_{0}}{C_{1}}+\frac{C_{2}}{C_{1}}e^{3 C_{1}t} \right]^{1/3}
	\eeq

	\subsubsection{Static Universe Solutions}
	
	Let us now study the case of a static universe scale factor-wise($\dot{a}=0$). The conservation equation for $\phi$ then becomes 
	\beq
	\dot{\phi}=6\phi^{2}
	\eeq
	which when integrated, gives
	\beq
	\phi(t)=\frac{1}{c_{1}-6t}
	\eeq
	Now, from Friedmann equation, we have
	\beq
	\frac{\kappa}{3}\rho=8\phi^{2}\Rightarrow  \nonumber
	\eeq
	\beq
	\rho(t)=\frac{24}{\kappa}\frac{1}{(c_{1}-6t)^{2}}
	\eeq
	and from the continuity equation we find that
	\beq
	p(t)=\frac{\dot{\rho}}{3\phi}-3\rho=\frac{24}{\kappa}\frac{1}{(c_{1}-6t)^{2}}
	\eeq
	and we wee that for such a universe the only possible matter form is stiff matter ($p=\rho$)

	\section{Cosmology with non-metricity}
	
	Let us consider a flat $FLRW$ cosmology with zero torsion but a non-vanishing non-metricity. As shown in \cite{minkevich1998isotropic} in such a universe the non-metricity tensor has three independent (a priori) degrees of freedom, call them $A(t)$, $B(t)$, $C(t)$ and it holds that
	\beq
	Q_{000}=A(t)
	\eeq
	\beq
	Q_{ij0}=\delta_{ij}\tilde{B}(t)=g_{ij}B(t)
	\eeq
	\beq
	Q_{0ij}=\delta_{ij}\tilde{C}(t)=g_{ij}C(t)
	\eeq
	Considering an Einstein-Hilbert action and the presence of a perfect fluid, the Friedmann equations with non-metricity are
	\beq
	\frac{\ddot{a}}{a}+\left(\frac{\dot{a}}{a}\right)^{2}=-\frac{\kappa}{6}(-\rho +3 p)+\frac{1}{8}( C^{2}+6 B^{2}+AC+BC-3AB )+\frac{3}{4}(\dot{B}-\dot{C})
	\eeq
	\beq
	\frac{\ddot{a}}{a}=-\frac{\kappa}{6}(\rho +3p)-\frac{1}{2}\left[ \frac{\dot{a}}{a}A+\dot{B}+\frac{1}{2}(AB+C^{2}) \right]
	\eeq
	
	\subsection{General form of the Non-metricity tensor in FLRW Universes}
	Let us now extend our previous considerations and derive the most general form that the non-metricity tensor can have in a general FLRW background (that is, for any value of the curvature). To do so, consider an observer with $4$-velocity $u^{\mu}$, then our building blocks for constructing $Q_{\alpha\mu\nu}$ can only be $\{ u^{\mu}, g_{\mu\nu}\}$ and demanding symmetry in the last two indices of non-metricity, the only possible combinations will be
	\beq
	u_{\alpha}g_{\mu\nu}\;, \;\; g_{\alpha(\mu}u_{\nu)}\;,\;\; u_{\alpha}u_{\mu}u_{\nu}
	\eeq
	and any of these combinations can have a factor in front of it that depends mostly on time (the $x^{0}=t$ coordinate of the observer). These requirements  leave no other choice than
	\beq
	Q_{\alpha\mu\nu}=F_{1}(t)u_{\alpha}g_{\mu\nu}+F_{2}(t)g_{\alpha(\mu}u_{\nu)}+F_{3}(t)u_{\alpha}u_{\mu}u_{\nu} \label{Qnmcos}
	\eeq
	This is the most general form of the non-metricity tensor in FLRW Universes and is presented here for the first time in the literature. Note that in such a spacetimes, non-metricity is specified by three functions of time.
	\subsubsection{Fixed Length Vectors in FLRW Universes}
	As we have seen many times by now, one effect of the non-metricity is that it changes the length of any vector that lives in spacetime. However, a certain form  of non-metricity can be found for which the space possesses what is known as fixed length vectors\footnote{Note that the length of any vector is fixed but the angle between two vectors will change due to non-metricity even for this special case.}. This specific kind of non-metricity, as we have already seen, obeys
	\beq
	Q_{\alpha\mu\nu}v^{\alpha}v^{\mu}v^{\nu}=0 \label{fix1}
	\eeq 
	for any vector $v^{\mu}$, that is
	\beq
	Q_{(\alpha\mu\nu)}=0 \label{fix}
	\eeq 
	namely, the completely symmetric part of non-metricity is zero. In a given spacetime, any non-metric configuration that respects ($\ref{fix}$) possesses fixed length vectors. Let us now find the general for for such a tensor in an FLRW spacetime. Firstly note that since the totally symmetric part of $Q_{\alpha\mu\nu}$ has to vanish, the term $u_{\alpha}u_{\mu}u_{\nu}$  has to be absent from $(\ref{Qnmcos})$ and therefore $F_{3}=0$ for this kind of non-metricity. Furthermore, expanding equation $(\ref{fix1})$ we have
	\beq
	F_{1}u^{4}+F_{2}u^{4}=0
	\eeq
	and for this to hold true for any $u_{\mu}$ we must have $F_{2}=-F_{1}$. Then, substituting this relation back in $(\ref{Qnmcos})$ we conclude that in an FLRW universe, with a fixed length non-metricity, the non-metricity tensor reads\footnote{Again, as far as we know, this equation is presented  for the first time in the literature here.}
	\beq
	Q_{\alpha\mu\nu}=A(t)\Big( u_{\alpha}g_{\mu\nu}-g_{\alpha(\mu}u_{\nu)} \Big)
	\eeq
	where $A(t)=F_{1}(t)$. Therefore, in this universe the fixed length vector non-metricity evolution is determined by one time function.

	\chapter{The Raychaudhuri Equation in Spaces with Torsion and Non-metricity}

	In this chapter we introduce and carefully develop the $1+3$ spacetime splitting for general non-Riemannian spaces. We therefore let the space possess generic torsion and non-metricity along with curvature and also allow the dimension to be arbitrary.
	Focusing on timelike observers, we identify and discuss the main differences between their kinematics and those of their counterparts living in standard Riemannian spacetimes. At the centre of our analysis lies the Raychaudhuri equation, which is the fundamental formula monitoring the convergence/divergence, namely the collapse/expansion, of timelike worldline congruences. To the best of our knowledge, we provide the most general\footnote{Forms of the Raychaudhuri equation with torsion have been developed in literature previously(see refs in what follows) however  its form with non-metricity was not known until now. We derive here the most general expression with both torsion and non-metricity.} expression so far of the Raychaudhuri equation, with applications to an extensive range of non-standard astrophysical and cosmological studies. Assuming that metricity holds, but allowing for nonzero torsion, we recover the results of analogous previous treatments. Focusing on non-metricity alone, we identify a host of effects that depend on the nature of the timelike congruence and on the type of the adopted non-metricity. We also demonstrate that in spaces of high symmetry one can recover the pure-torsion results from their pure non-metricity analogues, and vice-versa, via a simple ansatz between torsion and non-metricity. We then proceed to derive the most generic equation for the evolution of vorticity. Some of the results of this chapter we  have published in \cite{iosifidis2018raychaudhuri}.

	\section{Spacetime Splitting with Torsion and Non-Metricity}
	Let us generalize the $1+3$ formulation in the presence of both torsion and non-metricity\footnote{The Raychaudhuri equation in spaces with torsion has been presented in some previous works (see \cite{pasmatsiou2017kinematics},\cite{luz2017raychaudhuri},\cite{kar2007raychaudhuri} for instance and for a spin fluid in \cite{fennelly1991including} ) but for generic non-metricity no formula was ever given.}. The crucial thing now is that since non-metricity does not vanish, the length of every vector changes as one moves in spacetime. This has several important implications some of which include, the non-uniform passing of time for a co-moving observer, the fact that the velocity of the observer is no longer perpendicular to the acceleration (because the velocity is no longer of unit length) and the existence of two $'accelerations'$ as we shall see. In the following we define and use the   needed  set up for obtaining the expansion equation.

	\subsection{Velocity, acceleration and spatial projections}
	Let $u^{\mu}$ be the tangent vector of a curve and accordingly the $4-$velocity of an observer, then the scalar $u_{\mu}u^{\mu}$ cannot be normalized to $-1$ (or $1$) because the length of every vector changes in spacetime due to non-metricity. Therefore the normalization now reads
	\beq
	u_{\mu}u^{\mu}=g_{\mu\nu}u^{\mu}u^{\nu}=-l^{2}(x)\equiv -\phi(x) 
	\eeq
	with $u^{\mu}=\frac{dx^{\mu}}{d\lambda}$. As we will see later this very equation tells us that now the $4$-velocity and $4$-acceleration of an observer are no longer perpendicular to each other. Let us define below the whole set up to be used in our analysis. First we define the observer's spatial metric. The naive generalization
	\beq
	h_{\mu\nu}=g_{\mu\nu}+u_{\mu}u_{\nu}
	\eeq 
	does not seem to work here since the basic properties $h_{\mu\nu}u^{\mu}=0=h_{\mu\nu}u^{\nu}$ and $h_{\mu\nu}h^{\mu\nu}=n-1$ are not met. To fix this, we simply normalize the velocity term and define
	\beq
	h_{\mu\nu}=g_{\mu\nu}+\frac{u_{\mu}u_{\nu}}{l^{2}}
	\eeq
	which now satisfies $h_{\mu\nu}u^{\mu}=0=h_{\mu\nu}u^{\nu}$ and $h_{\mu\nu}h^{\mu\nu}=n-1$ as can be easily checked. In addition, it also satisfies 
	\beq
	h_{\mu\alpha}h^{\nu\alpha}=h_{\mu}^{\;\nu}=\delta_{\mu}^{\nu}+\frac{u_{\mu}u^{\nu}}{l^{2}}
	\eeq
	We now define the projections along time and spatial space in the usual manner
	\beq
	\dot{T}_{\alpha_{1}...\alpha_{n}}^{\;\;\;\;\;\;\beta_{1}...\beta_{m}}=u^{\mu}\nabla_{\mu}T_{\alpha_{1}...\alpha_{n}}^{\;\;\;\;\;\;\beta_{1}...\beta_{m}}
	\eeq
	\beq
	D_{\mu}T_{\alpha_{1}...\alpha_{n}}^{\;\;\;\;\;\;\beta_{1}...\beta_{m}}=h_{\mu}^{\;\lambda}h_{\alpha_{1}}^{\;\gamma_{1}}...h_{\alpha_{n}}^{\;\gamma_{n}}h_{\delta_{1}}^{\;\beta_{1}}...h_{\delta_{m}}^{\;\beta_{m}}\nabla_{\lambda}T_{\gamma_{1}...\gamma_{n}}^{\;\;\;\;\;\;\delta_{1}...\delta_{m}}
	\eeq
	respectively. With this at hand we can define the observer's acceleration, but we have to be careful with the indices since the covariant derivative does not commute with the metric due to the presence of non-metricity. For this reason we devote a full subsection on the actual definition. Before doing so, let us first discuss some subtle points regarding the definition of the proper time and affine parametrization.
	
	\subsubsection{Proper $'time'$ and Affine Parametrization}
	When we defined the projection along time we defined it as (on a scalar for instance)
	\beq
	\dot{F}=u^{\mu}\nabla_{\mu}F
	\eeq
	where
	\beq
	u^{\mu}\equiv \frac{dx^{\mu}}{d\lambda}
	\eeq
	is the tangent vector along the curve the observer follows, parametrized by the affine parameter $\lambda$. We should stress out that for generic non-metricity this affine parameter cannot be the proper time, and the two coincide only when the theory possesses fixed length vectors.\footnote{Fixed length vectors can also exist in non-metric spaces given that the non-metricity tensor satisfies $Q_{(\lambda\mu\nu)}=0$. This is certainly not true for Weyl non-metricity where $Q_{\alpha\mu\nu}=1/n Q_{\alpha}g_{\mu\nu}$.} To me more specific, given that
	\beq
	g_{\mu\nu}u^{\mu}u^{\nu}=g_{\mu\nu}\frac{dx^{\mu}}{d\lambda}\frac{dx^{\nu}}{d\lambda}=-l^{2} \label{lc}
	\eeq
	and the definition of proper time
	\beq
	d \tau^{2}=-ds^{2}=-g_{\mu\nu}dx^{\mu}dx^{\nu} \Rightarrow  \nonumber
	\eeq
	\beq
	1=-g_{\mu\nu}\frac{dx^{\mu}}{d\tau}\frac{dx^{\nu}}{d\tau}
	\eeq
	By using the chain rule, it follows that
	\beq
	-l^{2}=g_{\mu\nu}\frac{dx^{\mu}}{d\lambda}\frac{dx^{\nu}}{d\lambda}= \underbrace{g_{\mu\nu}\frac{dx^{\mu}}{d\tau}\frac{dx^{\nu}}{d\tau}}_{=-1}\left(\frac{d\tau}{d\lambda}\right)^{2}=-\left(\frac{d\tau}{d\lambda}\right)^{2} \Rightarrow \nonumber
	\eeq
	\beq
	\left(\frac{d\tau}{d\lambda}\right)^{2}=l^{2}
	\eeq
	or
	\beq
	\left(\frac{d\tau}{d\lambda}\right)=l \label{tl}
	\eeq
	and since $l=l(x)$ is not constant in a generic non-metric space, the parameter $\lambda$ cannot be identified with the proper time-$\tau$ and their relation is given by ($\ref{tl}$). Given that $l(x)$ can be specified when the non-metricity tensor is known, the latter can be integrated to give
	\beq
	\tau =\int l(x) d\lambda + C
	\eeq
	Notice however that the time $x^{0}$ that a co-moving observer ($u^{i}=0$) measures identifies with the proper time when $g_{00}=-1$ as in the case of an FLRW universe. To see this, using $u^{0}=\frac{d x^{0}}{d \lambda}$ and $u^{i}=0$, equation ($\ref{lc}$)  becomes
	\beq
	g_{00}(u^{0})^{2}=-l^{2} \Rightarrow \nonumber
	\eeq
	\beq
	\frac{d x^{0}}{d \lambda} =\frac{l}{\sqrt{-g_{00}}}
	\eeq
	Comparing the last one with ($\ref{tl}$) we see that indeed when $g_{00}=-1$ we have that $dx^{0}=d\tau$ and the two $'$times$'$ are the same. To conclude, when taking the derivative one should be careful and the proper time is not a good parametrization for the curve. So, derivatives with respect to the affine parameter $\lambda$ we will denote with a dot
	\beq
	\dot{F}\equiv \frac{dF}{d\lambda}=u^{\mu}\nabla_{\mu}F
	\eeq
	and derivatives with respect to the proper time (and also with respect to $x^{0}$ when $g_{00}=-1$) shall be denoted by a prime
	\beq
	F^{'}\equiv \frac{d F}{d \tau}
	\eeq
	and the two are related by
	\beq
	\dot{F}\equiv \frac{d F}{d \lambda}= \frac{d F}{d \tau}\frac{d \tau}{d \lambda}=l F^{'}
	\eeq
	or in operator form
	\beq
	\frac{d}{d \lambda}=l \frac{d}{d\tau}
	\eeq
	Having clarified this point, we can move on and define the acceleration when non-metricity is present.

	\subsubsection{Path and Hyper Acceleration}
	The fact that the metric tensor is not covariantly conserved (non-metricity) means that we cannot freely raise and lower indices with the metric tensor inside the covariant derivative. This allows one to define two kinds of acceleration, one is the usual one
	\beq
	A^{\mu}\equiv \dot{u}^{\mu}\equiv u^{\lambda}\nabla_{\lambda}u^{\mu}
	\eeq
	which we shall call the \textbf{path acceleration} since the vanishing of it implies that we have autoparallel motion.\footnote{A particle follows an autoparallel trajectory when its path acceleration is zero, that is $A^{\mu}\equiv \dot{u}^{\mu}\equiv u^{\lambda}\nabla_{\lambda}u^{\mu}=0$ in contrast to the geodesic motion in which $ \tilde{u}^{\lambda}\tilde{\nabla}_{\lambda}\tilde{u}^{\mu}=0$, where $\tilde{\nabla}_{\lambda}$ is the covariant derivative computed with respect to the Levi-Civita connection.} Note now that one may define another acceleration through
	\beq
	a_{\mu}\equiv \dot{u}_{\mu}\equiv  u^{\lambda}\nabla_{\lambda}u_{\mu}
	\eeq
	and notice that
	\beq
	A^{\mu} \neq a^{\mu}
	\eeq
	but rather
	\beq
	A^{\mu}=u^{\lambda}\nabla_{\lambda}u^{\mu}=u^{\lambda}\nabla_{\lambda}(u_{\nu}g^{\mu\nu}) \Rightarrow \nonumber
	\eeq
	\beq
	A^{\mu}=a^{\mu}+Q^{\lambda\mu\nu}u_{\lambda}u_{\nu}
	\eeq
	Also, it is worth pointing out that for autoparallel motion, the fact that $A^{\mu}=0$ does not force $a^{\mu}$ to vanish but rather fixes it to $a^{\mu}=-Q^{\lambda\mu\nu}u_{\lambda}u_{\nu}$. From this we see that $a^{\mu}$ arises due to non-metricity and we shall call it \textbf{hyper acceleration} since it does not vanish even for autoparallel motion.
	\subsubsection{Identities for the accelerations}
	As we have already mentioned the fact the the length of the $4-velocity$ changes due to non-metricity, implies that the acceleration (both the path and the hyper one) is no longer perpendicular to the velocity. In fact, this allows one two obtain some identities among them regarding their inner products.  Starting with the relation
	\beq
	u_{\mu}u^{\mu}=g_{\mu\nu}u^{\mu}u^{\nu}=-l^{2}(x)\equiv -\phi(x) 
	\eeq
	and taking the covariant derivative of it, one time writing $u_{\mu}u^{\mu}=g_{\mu\nu}u^{\mu}u^{\nu}$ and the other writing it as $u_{\mu}u^{\mu}=g^{\mu\nu}u_{\mu}u_{\nu}$ we obtain
	\beq
	-Q_{\lambda\mu\nu}u^{\mu}u^{\nu}+2 u_{\mu}\nabla_{\lambda}u^{\mu}=-\nabla_{\lambda}\phi
	\eeq
	and
	\beq
	Q_{\lambda}^{\;\;\;\mu\nu}u_{\mu}u_{\nu}+2 u^{\mu}\nabla_{\lambda}u_{\mu}=-\nabla_{\lambda}\phi
	\eeq
	Now, contracting both with $u^{\lambda}$ it follows that
	\beq
	-Q_{\lambda\mu\nu}u^{\lambda}u^{\mu}u^{\nu}+2 u_{\mu}A^{\mu}=-\dot{\phi}
	\eeq
	as well as
	\beq
	Q_{\lambda\mu\nu}u^{\lambda}u^{\mu}u^{\nu}+2 u_{\mu}a^{\mu}=-\dot{\phi}
	\eeq
	which when added and subtracted lead to
	\beq
	u_{\mu}(A^{\mu}+a^{\mu})=-\dot{\phi} \label{i1}
	\eeq
	and
	\beq
	u_{\mu}(A^{\mu}-a^{\mu})=Q_{\lambda\mu\nu}u^{\lambda}u^{\mu}u^{\nu} \label{i2}
	\eeq
	respectively. From the above two we see that neither of the accelerations is perpendicular to the velocity when non-metricity is present. We should stress out that these identities are general and for any curve one considers, next we study how do these modify when one assumes autoparallel motion.

	\subsubsection{Hyper-Acceleration For Autoparallel Motion}\label{ssPHn-A}
	Let us now specialize and see what happens when our curve\footnote{That is the observer we are considering is moving with zero path acceleration.} is an autoparallel. In this case we have
	\beq
	A^{\mu}=\dot{u}^{\mu}\equiv u^{\lambda}\nabla_{\lambda}u^{\mu}=0
	\eeq
	which, recalling the relation 
	\beq
	A^{\mu}=a^{\mu}+Q^{\lambda\mu\nu}u_{\lambda}u_{\nu}
	\eeq
	implies that
	\beq
	a^{\mu}=-Q^{\lambda\mu\nu}u_{\lambda}u_{\nu}
	\eeq
	and the identities $(\ref{i1})$ and $(\ref{i2})$  become
	\beq
	\dot{\phi}=-u_{\mu}a^{\mu}
	\eeq
	\beq
	-u_{\mu}a^{\mu}=Q_{\lambda\mu\nu}u^{\lambda}u^{\mu}u^{\nu}
	\eeq
	or upon combining them
	\beq
	\dot{\phi}=Q_{\lambda\mu\nu}u^{\lambda}u^{\mu}u^{\nu}=Q_{(\lambda\mu\nu)}u^{\lambda}u^{\mu}u^{\nu}
	\eeq
	from this we see that when $Q_{(\lambda\mu\nu)}=0$ we have that $\dot{\phi}=0\Rightarrow \phi=const.=l^{2}$ and the spacetime possesses fixed length vectors. This, of course, is not true for generic non-metricity. So, the length change will in general be (upon integrating the last one)
	\beq
	\phi =l^{2}=\int Q_{\lambda\mu\nu}u^{\lambda}u^{\mu}u^{\nu} d\lambda +C \label{phidot}
	\eeq
	Next we see how all these simplify when the theory possesses fixed length vectors.

	\subsubsection{Fixed Length Vectors}
	As we stated in earlier chapter, for a theory to possess fixed length vectors we must have $Q_{\lambda\mu\nu}u^{\lambda}u^{\mu}u^{\nu}=0$ for any vector or equivalently $Q_{(\lambda\mu\nu)}=0$. Looking now at ($\ref{phidot}$) we see that indeed this is the exact condition we must impose so as to achieve
	\beq
	u_{\mu} u^{\mu}=-l^{2}=-\phi=-C=constant
	\eeq
	we might as well normalize this constant to be $C=1$ so as to have the standard normalization
	\beq
	u_{\mu} u^{\mu}=-1
	\eeq
	If we further assume an autoparallel motion
	\beq
	u^{\alpha}\nabla_{\alpha}u^{\mu}=0
	\eeq
	a whole set of interesting identities follows which we will give later on, after we have defined all the appropriate quantities. So, we devote the next section in defining all these quantities.

	\subsection{Expansion, Shear, Vorticity and the rest}
	Now, in order to derive the Raychaudhuri equation we have to carefully define the various quantities appearing in it. Firstly, for convenience we define
	\beq
	\xi_{\mu} \equiv u^{\alpha}\nabla_{\mu}u_{\alpha}
	\eeq
	Next we define the expansion, rotation and shear\footnote{Each of these quantities can be split in its Riemannian and non-Riemannian pieces but this decomposition does not simplify things in any way. For instance the expansion can be written as $\Theta=\tilde{\Theta}+\Big( -\tilde{Q}_{\mu}+1/2 Q_{\mu}+2S_{\mu}\Big)u^{\mu}$.}
	\beq
	\Theta \equiv g^{\mu\nu}\nabla_{\mu}u_{\nu}
	\eeq
	\beq
	\omega_{\nu\mu} \equiv D_{[\mu}u_{\nu]}
	\eeq
	\beq
	\sigma_{\nu\mu} \equiv D_{<\mu}u_{\nu>} \equiv D_{(\mu}u_{\nu)}-\frac{(h^{\alpha\beta}D_{\alpha}u_{\beta})}{n-1}h_{\mu\nu}
	\eeq
	respectively. Notice now that $\Theta \neq \nabla_{\mu}u^{\mu}$ but rather
	\beq
	\nabla_{\mu}u^{\mu} = \Theta +u^{\mu}\tilde{Q}_{\mu }
	\eeq 
	It also holds that
	\beq
	\nabla_{\mu}u_{\nu}=D_{\mu}u_{\nu}-\frac{\xi_{\mu}u_{\nu}+u_{\mu}a_{\nu}}{l^{2}}-\frac{u_{\mu}u_{\nu}(a\cdot u)}{l^{4}}
	\eeq
	and
	\beq
	D_{\mu}u_{\nu}=\omega_{\nu\mu}+\sigma_{\nu\mu}+\Big( \Theta +\frac{(a\cdot u)}{l^{2}} \Big)\frac{h_{\mu\nu}}{n-1}
	\eeq
	and upon combining them
	\beq
	\nabla_{\mu}u_{\nu}=\omega_{\nu\mu}+\sigma_{\nu\mu}+\Big( \Theta +\frac{(a\cdot u)}{l^{2}} \Big)\frac{h_{\mu\nu}}{n-1}-\frac{\xi_{\mu}u_{\nu}+u_{\mu}a_{\nu}}{l^{2}}-\frac{u_{\mu}u_{\nu}(a\cdot u)}{l^{4}} \label{nabldec}
	\eeq
	where $(a\cdot u)=a_{\mu}u^{\mu}=g^{\mu\nu}a_{\mu}u_{\nu}$. In addition with some basic calculations it can be seen that the whole set of equations
	\beq
	0=\sigma_{\mu\nu}u^{\mu}=g^{\mu\nu}\sigma_{\mu\nu}=h^{\mu\nu}\sigma_{\mu\nu}=\omega_{\mu\nu}u^{\mu}=g^{\mu\nu}\omega_{\mu\nu}=h^{\mu\nu}\omega_{\mu\nu}= \nonumber
	\eeq
	\beq
	=\omega_{\mu\nu}\sigma^{\mu\nu}=u^{\mu}D_{\mu}u_{\nu}=u^{\nu}D_{\mu}u_{\nu}
	\eeq
	is satisfied.
	
	\subsection{The Raychaudhuri equation}

	With this equipment at hand, we are now in a position to derive the evolution equations. As far as the expansion equation is concerned, we start from Ricci identity for $u_{\mu}$
	\beq
	2\nabla_{[\alpha}\nabla_{\beta]}u_{\mu}=-R^{\lambda}_{\;\;\mu\alpha\beta}u_{\lambda}+2 S_{\alpha\beta}^{\;\;\;\;\nu}\nabla_{\nu}u_{\mu}
	\eeq
	and contract it by $g^{\mu\beta}u^{\alpha}$ to obtain
	\beq
	g^{\mu\beta}u^{\alpha}( \nabla_{\alpha}\nabla_{\beta}u_{\mu}-\nabla_{\beta}\nabla_{\alpha}u_{\mu})=-R_{\lambda\mu\alpha\beta}u^{\lambda}u^{\alpha}g^{\mu\beta}+2 S_{\alpha}^{\;\;\mu\nu}u^{\alpha}\nabla_{\nu}u_{\mu} \label{rrq}
	\eeq
	We first carry out the calculations for the left hand side. After some partial integrations and using the definitions, a rather lengthy calculation reveals 
	\begin{gather}
	g^{\mu\beta}u^{\alpha}( \nabla_{\alpha}\nabla_{\beta}u_{\mu}-\nabla_{\beta}\nabla_{\alpha}u_{\mu})=\dot{\Theta}+\frac{\left(\Theta +\frac{a\cdot u}{l^{2}} \right)^{2}}{n-1}+\sigma^{2}-\omega^{2} \nonumber \\
	-\frac{(a\cdot u)^{2}}{l^{4}}-2\frac{(a\cdot \xi)}{l^{2}}-u_{\alpha}Q^{\alpha\beta\mu}\nabla_{\beta}u_{\mu}+u_{\alpha}Q^{\mu\nu\alpha}\nabla_{\nu}u_{\mu}-g^{\mu\nu}\nabla_{\mu}a_{\nu}
	\end{gather}
	where $\sigma^{2} \equiv \sigma_{\mu\nu}\sigma^{\mu\nu}$, $\;\omega^{2}\equiv \omega_{\mu\nu}\omega^{\mu\nu}$ and notice that $g^{\mu\nu}\nabla_{\mu}a_{\nu} \neq \nabla_{\mu}a^{\mu}$ but rather
	\beq
	\nabla_{\mu}a^{\mu}=g^{\mu\nu}\nabla_{\mu}a_{\nu}+\tilde{Q}^{\mu}a_{\mu}
	\eeq
	Now as far as the right hand side is concerned the only tricky term is the one involving the Riemann tensor. Because of the limited symmetries now, the contraction does not give the Ricci tensor directly, but one has
	\beq
	-R_{\lambda\mu\alpha\beta}g^{\mu\beta}=+R_{\lambda\mu\beta\alpha}g^{\mu\beta}=\check{R}_{\lambda\alpha}
	\eeq
	which is the third independent contraction of the Riemann tensor. To express this in terms of the Ricci tensor, which can latter on be eliminated in terms of the energy momentum tensor upon using the field equations, we use an identity we have proven
	\beq
	R_{(\mu\nu)\alpha\beta}=\nabla_{[\alpha}Q_{\beta]\mu\nu}-S_{\alpha\beta}^{\;\;\;\;\lambda}Q_{\lambda\mu\nu}
	\eeq
	So by adding a zero we have
	\begin{gather}
	\check{R}_{\lambda\alpha}=R_{\lambda\mu\beta\alpha}g^{\mu\beta}=g^{\mu\beta}\Big( R_{\lambda\mu\beta\alpha}+R_{\mu\lambda\beta\alpha}-R_{\mu\lambda\beta\alpha}\Big)= \nonumber \\
	=g^{\mu\beta}2R_{(\lambda\mu)\beta\alpha}-R_{\lambda\alpha}= \nonumber \\
	=2 g^{\mu\beta}\Big( \nabla_{[\beta}Q_{\alpha]\lambda\mu}-S_{\beta\alpha}^{\;\;\;\;\rho}Q_{\rho\lambda\mu} \Big)-R_{\lambda\alpha}
	\end{gather}
	Thus, the relation
	\beq
	\check{R}_{\lambda\alpha}=2g^{\mu\beta}\Big( \nabla_{[\beta}Q_{\alpha]\lambda\mu}-S_{\beta\alpha}^{\;\;\;\;\rho}Q_{\rho\lambda\mu} \Big)-R_{\lambda\alpha}
	\eeq
	holds as an identity! Using all the above eq. $(\ref{rrq})$ takes the form
	\begin{gather}
	\dot{\Theta}+\frac{\left(\Theta +\frac{a\cdot u}{l^{2}} \right)^{2}}{n-1}+\sigma^{2}-\omega^{2} -g^{\mu\nu}\nabla_{\mu}a_{\nu} \nonumber \\
	-\frac{(a\cdot u)^{2}}{l^{4}}-2\frac{(a\cdot \xi)}{l^{2}}-u_{\alpha}Q^{\alpha\beta\mu}\nabla_{\beta}u_{\mu}+u_{\alpha}Q^{\mu\nu\alpha}\nabla_{\nu}u_{\mu}= \nonumber \\
	=-R_{\mu\nu}u^{\mu}u^{\nu}+ 2 u^{\mu}u^{\beta}  \Big( g^{\nu\alpha} \nabla_{[\alpha}Q_{\beta]\mu\nu}-S_{\alpha\beta}^{\;\;\;\;\lambda}Q_{\lambda\mu}^{\;\;\;\;\alpha} \Big)+2 u_{\alpha}S^{\alpha\mu\nu}\nabla_{\nu}u_{\mu} \label{13gen}
	\end{gather}
	Note that if we define the expansion rate
	\beq
	\Theta_{D}\equiv g^{\mu\nu}D_{\mu}u_{\nu}=\Theta +\frac{a\cdot u}{l^{2}}
	\eeq
	the above can be written as
	\begin{gather}
	\dot{\Theta}_{D} +\frac{\Theta^{2}_{D}}{n-1}-\frac{d}{d\lambda}\Big(\frac{(a\cdot u)}{l^{2}}\Big)+\sigma^{2}-\omega^{2} -g^{\mu\nu}\nabla_{\mu}a_{\nu} \nonumber \\
	-\frac{(a\cdot u)^{2}}{l^{4}}-2\frac{(a\cdot \xi)}{l^{2}}-u_{\alpha}Q^{\alpha\beta\mu}\nabla_{\beta}u_{\mu}+u_{\alpha}Q^{\mu\nu\alpha}\nabla_{\nu}u_{\mu}= \nonumber \\
	=-R_{\mu\nu}u^{\mu}u^{\nu}+2 u^{\mu}u^{\beta}  \Big( g^{\nu\alpha} \nabla_{[\alpha}Q_{\beta]\mu\nu}-S_{\alpha\beta}^{\;\;\;\;\lambda}Q_{\lambda\mu}^{\;\;\;\;\alpha} \Big)+2 u_{\alpha}S^{\alpha\mu\nu}\nabla_{\nu}u_{\mu} \label{13gen}
	\end{gather}
	The last equation is the generalization of the Raychaudhuri equation in spaces where apart from curvature and torsion, there is also non-metricity and to our knowledge appears for the first time in literature. Notice that in this form the latter holds as a geometric identity and only becomes an equation when an energy momentum tensor $T_{\mu\nu}$ and a hyper-momentum tensor $\Delta_{\lambda}^{\;\;\;\;\mu\nu}$ are given. The former giving rise to curvature through the field equations and the latter giving rise to both torsion and non-metricity through the Palatini equations (equations obtained after varying with respect to the general affine connection). We should also point out that in deriving the equation above no specific choice of the curve (that the observer follows) was made, that is if one considers autoparallel motion ( $A^{\mu}=0$) the equation changes accordingly as we discuss in what follows. Before discussing some special cases of the above derived generalized Raychaudhuri equation let us point out that the above form is not in its irreducible form. Further decomposition of the terms $\nabla_{\mu}u_{\nu}$,  $\nabla_{\mu}a_{\nu}$
	leads to the fully irreducible form of the generalized Raychaudhuri equation in $n-dim$ spacetime with curvature, torsion and non-metricity, which reads
	\begin{eqnarray}
	\dot{\Theta}&=& -{1\over n-1}\,\Theta^2- R_{\mu\nu}u^{\mu}u^{\nu}- 2\left(\sigma^2-\omega^2\right)+ {\rm D}^{\mu}a_{\mu}+ {1\over\ell^2}\,a_{\mu}A^{\mu} \nonumber\\ &&+{2\over n-1} \left(\Theta+{1\over\ell^2}\,a_{\nu}u^{\nu}\right)S_{\mu}u^{\mu}+ 2S_{\mu\nu\lambda}u^{\mu}(\sigma^{\nu\lambda}+\omega^{\nu\lambda})+ {2\over\ell^2}\,S_{\mu\nu\lambda}a^{\mu}u^{\nu}u^{\lambda} \nonumber\\ &&-{1\over\ell^2}(a_{\mu}u^{\mu})^{\cdot}- {2\Theta\over\ell^2(n-1)}\,a_{\mu}u^{\mu}+ {n-2\over\ell^4(n-1)}(a_{\mu}u^{\mu})^2+ {2\over\ell^2}\,a_{\mu}\xi^{\mu}- \dot{\tilde{Q}}_{\mu}u^{\mu} \nonumber\\ &&+{1\over n-1} \left(\Theta+{1\over\ell^2}\,a_{\nu}u^{\nu}\right) (Q_{\mu}-\tilde{Q}_{\mu})u^{\mu}-Q_{\mu\nu\lambda} (\sigma^{\mu\nu}+\omega^{\mu\nu}) u^{\lambda}- {1\over\ell^2}\,Q_{\mu\nu\lambda}u^{\mu}u^{\nu} (a^{\lambda}+\xi^{\lambda}) \nonumber\\ &&+Q_{\mu\nu\lambda}u^{\mu}\sigma^{\nu\lambda}+ {1\over\ell^2}\,Q_{\mu\nu\lambda}(u^{\mu}\xi^{\nu}+a^{\mu}u^{\nu}) u^{\lambda}+ u^{\mu}u^{\nu}\nabla^{\lambda}Q_{\mu\nu\lambda}+ Q_{\mu}{}^{\lambda\beta}Q_{\beta\lambda\nu}u^{\mu}u^{\nu} \nonumber\\  &&+2S_{\mu}{}^{\lambda\beta} Q_{\beta\lambda\nu}u^{\mu}u^{\nu}\,.  \label{nmtRay}
	\end{eqnarray}
	Note that only the terms in the first line on the right-hand side of the above have Riemannian analogues. More specifically, in the absence of torsion and in the presence of metricity (i.e.~when $S_{\mu\nu\lambda}\equiv 0\equiv Q_{\mu\nu\lambda}$), the rest of the terms on the right-hand side of (\ref{nmtRay}) vanish identically. Then, setting $n=4$, we recover the standard form of the Raychaudhuri equation (e.g.~see~ \cite{pasmatsiou2017kinematics,luz2017raychaudhuri,capozziello2001geometric,kar2007raychaudhuri}  and also keep in mind that $a_{\mu}\equiv A_{\mu}$, with $a_{\mu}u^{\mu}=0= A_{\mu}u^{\mu}$, and that $\xi_{\mu}\equiv0$ when metricity holds). Let us now see how the above generalized Raychaudhuri equation is simplified in the cases of pure torsion and pure non-metricity respectively.

	\subsection{The case of pure torsion}\label{ssCPT}
	%%%%%%%%%%%%%%%%%%%%%%%%%%%%%%%%%%%%%%%%%%%%%%%%%%
	The terms in the second line on the right-hand side of Eq.~(\ref{nmtRay}) are purely torsional in nature, with the exception of the first which has a additional contribution from the non-metricity of the space (through the inner product $a_{\mu}u^{\mu}$, which vanishes when metricity holds). Then, when dealing with a $n$-dimensional spacetime that has nonzero torsion but satisfies the metricity condition, expression (\ref{nmtRay}) reduces to
	\begin{eqnarray}
	{\Theta}^{\prime}&=& -{1\over n-1}\,\Theta^2- R_{\mu\nu}u^{\mu}u^{\nu}- 2\left(\sigma^2-\omega^2\right)+ {\rm D}^{\mu}A_{\mu}+ A^{\mu}A_{\mu} \nonumber\\ &&+{2\over n-1}\,\Theta S_{\mu}u^{\mu}+ 2S_{\mu\nu\lambda}u^{\mu} \left(\sigma^{\nu\lambda}+\omega^{\nu\lambda}\right)+ 2S_{\mu\nu\lambda}A^{\mu}u^{\nu}u^{\lambda}\,,  \label{tRay}
	\end{eqnarray}
	with the prime indicating differentiation with respect to proper time. Applying the above to a 4-dimensional spacetime, one recovers the Raychaudhuri equation of the Riemann-Cartan geometry derived in \cite{pasmatsiou2017kinematics}(Alternative derivations of the Raychaudhuri equation with torsion have also been given in \cite{luz2017raychaudhuri,capozziello2001geometric,kar2007raychaudhuri}). Note that, when doing the aforementioned identification, one should also take into account the differences in the definitions of the torsion tensor and of the torsion vector between the two studies.
	
	Following (\ref{tRay}), torsion affects the convergence/divergence of a timelike congruence in a variety of ways, which depend on whether these worldlines are geodesics or not, as well as on whether they have nonzero shear or vorticity. The most straightforward effect of torsion propagates via the first term in the second line on the right-hand side of the above. More specifically, torsion enhances/inhibits the expansion/contraction of the worldline congruence depending on the sign of the inner product ($S_{\mu}u^{\mu}$) between the torsion vector and the $n$-velocity (i.e.~on the relative orientation of the two vector fields (see also  \cite{pasmatsiou2017kinematics}  for further discussion).
	
	As we mentioned in the previous section, Eq.~(\ref{tRay}) is of purely geometrical nature, since no matter fields have been introduced yet. In order to investigate the effects of gravity, we need to relate both the Ricci tensor and the torsion tensor to the material component of the spacetime. This can be done by means of, say, the Einstein-Cartan and the Cartan field equations.

	\subsection{The case of pure non-metricity}\label{ssCPN-M}
	%%%%%%%%%%%%%%%%%%%%%%%%%%%%%%%%%%%%%%%%%%%%%%%%%%%%%%%%%%
	Finally, the terms seen in lines three to six on the right-hand side of (\ref{nmtRay}) are due to the non-metricity of the space, with the last of them carrying a torsional contribution as well. Therefore, in the presence of non-metricity but in the absence of torsion, we may write
	\begin{eqnarray}
	\dot{\Theta}&=& -{1\over n-1}\,\Theta^2- R_{\mu\nu}u^{\mu}u^{\nu}- 2\left(\sigma^2-\omega^2\right)+ {\rm D}^{\mu}a_{\mu}+ {1\over\ell^2}\,A^{\mu}a_{\mu} \nonumber\\ &&+{1\over n-1} \left(\Theta+{1\over\ell^2}\,a_{\nu}u^{\nu}\right) \left(Q_{\mu}-\tilde{Q}_{\mu}\right)u^{\mu}+ Q_{\mu\nu\lambda}u^{\mu}\sigma^{\nu\lambda}- {1\over\ell^2}\,Q_{\mu\nu\lambda}u^{\mu}u^{\nu} \left(a^{\lambda}+\xi^{\lambda}\right) \nonumber\\ &&+{1\over\ell^2(n-1)} \left(\Theta-{n-2\over\ell^2}\,a_{\beta}u^{\beta}\right) Q_{\mu\nu\lambda}u^{\mu}u^{\nu}u^{\lambda}+ u^{\mu}u^{\nu}\nabla^{\lambda}Q_{\mu\nu\lambda}+ Q_{\mu}{}^{\lambda\beta}Q_{\beta\lambda\nu}u^{\mu}u^{\nu}- \dot{\tilde{Q}}_{\mu}u^{\mu} \nonumber\\ &&-{1\over\ell^2}\left(a_{\mu}u^{\mu}\right)^{\cdot}- {1\over\ell^2(n-1)}\,\Theta\left(a_{\mu}+A_{\mu}\right)u^{\mu}+ {n-2\over\ell^2(n-1)}\,A_{\mu}a_{\nu}u^{\mu}u^{\nu}\nonumber\\  &&+{1\over\ell^2}\left(a_{\mu}\zeta^{\mu}+A_{\mu}\xi^{\mu}\right)\,.  \label{nmRay}
	\end{eqnarray}
	Here, in contrast to Eq.~(\ref{tRay}), the overdot implies differentiation in terms of the affine parameter (i.e.~relative to $\lambda$ -- see definitions in the beginning of the chapter.). According to the above, the implications of non-metricity for the convergence/divergence of a timelike congruence are multiple and not straightforward to decode. Similarly to the case of pure torsion seen before, the most transparent effects are those depending on the orientation of the non-metricity vectors and their derivatives ($Q_{\mu}$, $\tilde{Q}_{\mu}$ and $\dot{\tilde{Q}}_{\mu}$) relative to the $u_{\mu}$-field.
	
	Before  closing this subsection, we should point out that the Raychaudhuri formulae given in expressions (\ref{nmtRay})-(\ref{nmRay}), are purely geometrical relations, which acquire physical relevance after the energy-momentum and the hyper-momentum tensors are introduced. The former gives rise to spacetime curvature and the latter to both torsion and non-metricity, through the field equations and the Palatini equations respectively. Also note that the nature of the observers' worldlines, namely of the curves tangent to the $n$-velocity vector $u_{\mu}$, has so far been left unspecified. Assuming, for example, motion along autoparallel curves the path-acceleration vanishes (i.e.~$A_{\mu}=0$ -- see \S~\ref{ssPHn-A} earlier), in which case the Raychaudhuri equation simplifies considerably. Let us now focus on specific forms of torsion and non-metricity, see how the Raychaudhuri equation simplifies and seek cosmological solutions.

	\subsubsection{Vectorial Torsion}
	For a vectorial torsion of the form 
	\beq
	S_{\mu\nu}^{\;\;\;\;\lambda}=\frac{2}{n-1}S_{[\mu}\delta_{\nu]}^{\lambda}
	\eeq
	and vanishing non-metricity, the expansion equation takes the form
	\beq
	\dot{\Theta}+\frac{\Theta^{2}}{n-1} +\sigma^{2}-\omega^{2}-\nabla_{\mu}a^{\mu}=-R_{\mu\nu}u^{\mu}u^{\nu}+\frac{2}{n-1}\Big( u^{\mu}S_{\mu}\Theta -a^{\mu}S_{\mu} \Big)
	\eeq
	which for autoparallel motion ($a_{\mu}=0$) simplifies to
	\beq
	\dot{\Theta}+\frac{\Theta^{2}}{n-1} +\sigma^{2}-\omega^{2}=-R_{\mu\nu}u^{\mu}u^{\nu}+\frac{2}{n-1} (u^{\mu}S_{\mu})\Theta 
	\eeq
	\subsubsection{Exact Cosmological Solution For Generic Torsion Vector}
	If we consider an empty  and flat FLRW universe, the above equation can be solved exactly for random torsion vector $S_{\mu}$. Indeed, in this case we have
	\beq
	\dot{\Theta}+\frac{\Theta^{2}}{3} =\frac{2}{3} (u^{\mu}S_{\mu})\Theta \label{T}
	\eeq
	Notice now that the expansion can be expressed as
	\beq
	\Theta = \tilde{\Theta} +2S_{\mu}u^{\mu}
	\eeq
	where
	\beq
	\tilde{\Theta}=3\frac{\dot{a}}{a}
	\eeq
	is the Riemannian part and $2S_{\mu}u^{\mu}$ the contribution from torsion. Solving the latter equation for $2S_{\mu}u^{\mu}$ and substituting back into $(\ref{T})$ we obtain
	\beq
	\dot{\Theta}+\frac{\Theta^{2}}{3}=\frac{\Theta^{2}}{3}-\frac{\tilde{\Theta}\Theta}{3} \Rightarrow \nonumber
	\eeq
	\beq
	\dot{\Theta}+\frac{\tilde{\Theta}\Theta}{3}=0 \Rightarrow \nonumber
	\eeq
	\beq
	\frac{\dot{\Theta}}{\Theta}+\frac{\dot{a}}{a}=0 \Rightarrow \nonumber
	\eeq
	\beq
	\frac{d}{dt}\Big( \ln{\Theta\cdot a}\Big)=0 \Rightarrow \nonumber 
	\eeq
	\beq
	\Theta\cdot a=const.=c_{0}
	\eeq
	Expanding $\Theta$ in the last one it follows that
	\beq
	\dot{a}+\frac{2}{3}(S_{\mu}u^{\mu})a=C_{0}
	\eeq
	and multiplying through by $e^{\frac{2}{3}\int S_{\mu}dx^{\mu}}$ we get
	\beq
	\dot{a}e^{\frac{2}{3}\int S_{\mu}dx^{\mu}}+\frac{2}{3}(S_{\mu}u^{\mu})e^{\frac{2}{3}\int S_{\mu}dx^{\mu}}a=C_{0}e^{\frac{2}{3}\int S_{\mu}dx^{\mu}}
	\eeq
	Observe now that the left hand side can be written as the product derivative
	\beq
	\frac{d}{dt}\Big( a \cdot e^{\frac{2}{3}\int S_{\mu}dx^{\mu}} \Big)=C_{0}e^{\frac{2}{3}\int S_{\mu}dx^{\mu}}
	\eeq
	and by integrating the last one we can solve for the scale factor
	\beq
	a(t)=e^{-\frac{2}{3}\int S_{\mu}dx^{\mu}}\left[ C_{1}+C_{0}\int e^{\frac{2}{3}\int S_{\mu}dx^{\mu}} dt \right]
	\eeq
	Thus, we have find the scale factor for any given torsion vector $S_{\mu}$. Notice though that since we have considered a flat FLRW the only non-zero component of $S_{\mu}$ is $S_{0}(t)$ and depends only on time, so we can write the above as
	\beq
	a(t)=e^{-\frac{2}{3}\int S_{0}(t)dt}\left[ C_{1}+C_{0}\int e^{\frac{2}{3}\int S_{0}(t)dt} dt \right] \label{asl}
	\eeq

	\subsection{Form for Weyl Non-metricity}
	Let us see apply now our generalized Raychaudhuri equation for a Weyl non-metricity. Recall that for Weyl non-metricity we have a single vector $Q_{\mu}$ defining non metricity, that is
	\beq
	Q_{\alpha\mu\nu}=\frac{1}{n}Q_{\alpha}g_{\mu\nu}
	\eeq
	and torsion is zero. For such an arrangement along with the demand that our curve is an autoparallel ($A^{\mu}=u^{\alpha}\nabla_{\alpha}u^{\mu}=0$)
	the following hold true
	\beq
	\tilde{Q}_{\mu}=\frac{1}{n}Q_{\mu}\;,\;  a_{\mu}=-\frac{(Q_{\lambda}u^{\lambda})}{n}u_{\mu}
	\eeq
	\beq
	Q_{\mu}u^{\mu}=-2 n\frac{\dot{l}}{l}\;,\; a_{\mu}u^{\mu}=-2 l \dot{l}
	\eeq
	and equation $(\ref{13gen})$ takes the form
	\beq
	\dot{\Theta}+\frac{(\Theta -  L)^{2}}{n-1}+\sigma^{2}-\omega^{2}-\dot{L}=-R_{\mu\nu}u^{\mu}u^{\nu}
	\eeq
	where we have set
	\beq
	L =2\frac{\dot{l}}{l}
	\eeq
	for convenience. In addition, splitting $\Theta$ into its Riemannian and non-Riemannian parts we have
	\beq
	\Theta =\tilde{\nabla}_{\mu}u^{\mu}+\Big( -\tilde{Q}_{\mu}-\frac{Q_{\mu}}{2}\Big)u^{\mu} \Rightarrow \nonumber
	\eeq
	\beq
	\Theta =\partial_{\mu}u^{\mu}+\tilde{\Gamma}^{\mu}_{\;\;\;\;\lambda\mu}u^{\lambda}+\frac{n-2}{2 n}Q_{\mu}u^{\mu}
	\eeq
	Notice now that since $u^{\mu}=\delta^{\mu}_{0}l$ the partial derivative appearing on the right hand side is not zero but rather 
	\beq
	\partial_{\mu}u^{\mu}= \partial_{0}u^{0} =     \frac{d l}{d \tau}=\frac{\dot{l}}{l}
	\eeq
	Taking all the above into consideration\footnote{And also using the Christoffel symbols for a flat FRW spacetime. } and setting $n=4$ we finally arrive at
	\beq
	\Theta =-\frac{\dot{l}}{l}+3\frac{\dot{a}}{a}
	\eeq
	We should stress out again that here the dot denotes differentiation with respect to the affine parameter $\lambda$ which is not the proper time.

	\subsubsection{Solution for a $\sigma=0=\omega$ empty universe in $4$-dim}
	Let us now seek a solution when both shear and rotation are zero and in the absence of matter for a $n=4$ dimensional universe. We then, have to solve
	\beq
	\dot{\Theta}-\dot{L}+\frac{(\Theta -L)^{2}}{3}=0
	\eeq
	This can be immediately integrated to give
	\beq
	\Theta= \frac{1}{\frac{\lambda}{3}+C}+L
	\eeq
	and as we noted before, the decomposition of $\Theta$ in Riemannian and non-Riemannian parts reads
	\beq
	\Theta=3 \frac{\dot{a}}{a}-\frac{\dot{l}}{l}
	\eeq
	such that
	\beq
	\Theta -L=3 \frac{\dot{a}}{a}-3 \frac{\dot{l}}{l}=\frac{d}{d\lambda}\left[ \ln{\Big(\frac{a^{3}}{l^{3}}\Big)} \right]
	\eeq
	upon a second integration we can solve for the scale factor
	\beq
	a=C_{0}(C_{1}+\lambda )l
	\eeq
	where
	\beq
	l=  l_{0}e^{-\frac{1}{8}\int Q_{\mu}u^{\mu}d \lambda }= l_{0}e^{-\frac{1}{8}\int Q_{0}(\tau)d\tau}
	\eeq
	and
	\beq
	\lambda =\int \frac{d \tau}{l}=\int \frac{d\tau}{l_{0}}e^{\frac{1}{8}\int Q_{0}(\tau)d\tau}
	\eeq
	So, we may write
	\beq
	a(\lambda)=C_{0}l_{0}(C_{1}+\lambda) e^{-\frac{1}{8}\int Q_{\mu}u^{\mu}d \lambda }
	\eeq
	or
	\beq
	a(\tau)= C_{0}l_{0}e^{-\frac{1}{8}\int Q_{0}(\tau)d\tau} \left[ C_{1}+\int \frac{1}{l_{0}}e^{\frac{1}{8}\int Q_{0}(\tau)d\tau} d\tau \right]
	\eeq
	Now, from the last one we see an astonishing result. This solution looks similar to the solution ($\ref{asl}$) we found for vectorial torsion and zero non-metricity. In fact the two solutions are identical upon the exchange\footnote{For general dimension-$n$ this duality reads $S_{\mu}\leftrightarrow  \frac{n-1}{4n}  Q_{\mu}$.}
	\beq
	S_{\mu} \longleftrightarrow \frac{3}{16}Q_{\mu}
	\eeq
	and notice that this duality is exactly the same with the one that appears in $[our \;other\; paper\; with \; R^{2}]$. This of course is due to the fact that vectorial torsion can be traded with Weyl non-metricity as we saw in a previous chapter.

	\subsection{Form for Fixed Length Vectors Non-metricity}
	As it can be easily checked for a fixed length vector theory and assuming autoparallel motion, the set of identities\footnote{Notice that no issue of affine parametrization not identified with the proper time arises since the magnitude of every vector remains constant $(u_{\mu}u^{\mu}=-1)$. That is, $\lambda=\tau=t$.}
	\begin{gather}
	u^{\mu}a_{\mu}=0 \;,\; Q_{\alpha\mu\nu}u^{\mu}u^{\nu}=0 \;,\;  a^{\nu}=-Q_{\alpha\mu\nu}u^{\alpha}u^{\mu}=0 \;,\;\xi_{\alpha}=u^{\mu}\nabla_{\alpha}u^{\mu}=0 \\ \nonumber
	u^{\alpha}Q_{\alpha\mu\nu}=-A(t)h_{\mu\nu} \;,\; Q_{\alpha\mu\nu}u^{\nu}=\frac{1}{2}A(t)h_{\alpha\mu} \\ \nonumber
	u^{\alpha}Q_{\alpha\mu\nu}\nabla_{\mu}u_{\nu}=-A(t)\Theta \;,\; u_{\alpha}Q^{\mu\nu\alpha}\nabla_{\nu}u_{\mu}=\frac{1}{2}A(t)\Theta \\ \nonumber
	Q_{\mu}=A(t)(n-1)u_{\mu}  \;,\; \tilde{Q}_{\mu}=-\frac{1}{2}A(t)(n-1)u_{\mu}
	\end{gather}
	is satisfied. Note that all these hold true for general dimension $n$. Considering the $4-dim$ spacetime of our world, only the last two fix to
	\beq
	Q_{\mu}=3 A(t)u_{\mu}  \;,\; \tilde{Q}_{\mu}=-\frac{3}{2}A(t)u_{\mu}
	\eeq
	In this case we also have
	\beq
	u^{\mu}u^{\beta}g^{\nu\alpha}\nabla_{\beta}Q_{\alpha\mu\nu}=\frac{3}{2}(\dot{A}+A^{2})
	\eeq
	and 
	\beq
	u^{\mu}u^{\beta}g^{\nu\alpha}\nabla_{\alpha}Q_{\beta\mu\nu}=\frac{A}{2}\nabla_{\mu}u^{\mu}
	\eeq
	where $\nabla_{\mu}u^{\mu}=\Theta+\tilde{Q}_{\mu}u^{\mu}$, which combine to give
	\beq
	u^{\mu}u^{\beta}g^{\nu\alpha}\nabla_{[\alpha}Q_{\beta ]\mu\nu}=-\frac{3}{4}\left( \dot{A}+\frac{1}{2}A^{2}-\frac{1}{3}A \Theta  \right)
	\eeq
	Using these results, the expansion equation takes the form
	\beq
	\dot{\Theta}+\frac{\Theta^{2}}{3}=-\frac{1}{4}\left( 5 A \Theta +3 \dot{A} +\frac{3}{2}A^{2} \right)
	\eeq
	and this is Raychaudhuri's equation for a  fixed length theory in an FLRW spacetime.

	\subsubsection{Solution for $A=A_{0}=const.$}
	Let us now seek solutions for $A=A_{0}=constant$. Then, the expansion equation becomes
	\beq
	\dot{\Theta}+\frac{\Theta^{2}}{3}=-\frac{1}{4}\left( 5 A_{0} \Theta  +\frac{3}{2}A_{0}^{2} \right)
	\eeq
	which after completing the square, can be brought to
	\beq
	\frac{d}{dt}\left( \Theta+ \frac{15}{8}A_{0} \right)+\frac{1}{3}\left( \Theta+ \frac{15}{8}A_{0} \right)^{2}=\frac{C^{2}}{3}
	\eeq
	where we have set $C^{2}/3=51 A^{2}_{0}/64 $. Considering the change of variable
	\beq
	\Theta+ \frac{15}{8}A_{0} =y
	\eeq
	the later is written as
	\beq
	\dot{y}=\frac{1}{3}( C^{2}-y^{2})
	\eeq
	which can be easily integrated to give
	\beq
	y=C \left(  \frac{C_{1}e^{\frac{2C}{3}t}-1}{C_{1}e^{\frac{2C}{3}t}+1}\right)
	\eeq
	or
	\beq
	\Theta (t)=C \left( \frac{C_{1}e^{\frac{2C}{3}t}-1}{C_{1}e^{\frac{2C}{3}t}+1}\right) -\frac{15}{8}A_{0}  \label{sing}
	\eeq
	where $C_{1}>0$ is an integration constant. Now we would like to study if our solution exhibits singularity as $t\rightarrow 0$ as in the case of Einstein's General Relativity. Notice that one avoids an initial singularity as long as 
	\beq
	\dot{\Theta}+\frac{1}{3}\Theta^{2}>0
	\eeq
	From the above solution, one trivially shows that
	\beq
	\dot{\Theta}=\frac{4 C^{2}}{3}\frac{C_{1}e^{\frac{2C}{3}t}}{(1+C_{1}e^{\frac{2C}{3}t})^{2}} >0
	\eeq
	which is strictly positive for any $C_{1}>0$ and for every $t$, that is it retains its initial sign for all times. Therefore, by enhancing the above inequality it follows that
	\beq
	\dot{\Theta}+\frac{1}{3}\Theta^{2}>0
	\eeq
	for every $t$ and as a result the minimum and maximum values of $\Theta$ are obtained for $t=0$ and $t\rightarrow  \infty$ respectively.  Notice that  our solution can also be written as
	\beq
	\Theta(t)= C-\frac{15}{8}A_{0}-\frac{2 C}{C_{1}e^{\frac{2C}{3}t}+1}
	\eeq
	or, using the fact that $C \approx 3A_{0}/2$
	\beq
	\Theta(t)=-\frac{3 A_{0}}{8}\left( 1+ \frac{8}{C_{1}e^{A_{0}t}+1} \right)
	\eeq
	Furthermore, decomposing $\Theta$ inti its Riemannian and non-metric parts
	\beq
	\Theta =\tilde{\Theta}-3 A_{0}
	\eeq
	we can integrate the above and solve for the scale factor
	\beq
	a(t)=C_{2}( C_{1}+e^{-A_{0}t} ) e^{\frac{7}{8}A_{0}t}
	\eeq
	where $C_{2}>0$ is another integration constant. Assuming the initial conditions $a(t=0)=a_{0}$ and $H(t=0)=H_{0}$ for the scale factor and Hubble parameter respectively, the above can also be written as
	\beq
	a(t)=a_{0}+a_{0}\Big( \frac{7}{8}A_{0}-H_{0}\Big) ( e^{-\frac{1}{8}A_{0}t} - e^{\frac{7}{8}A_{0}t}  )
	\eeq
	where $a_{0}>0$. Since both $C_{1}, C_{2}>0$ we see that we have accelerated (inflation-like) expansion irrespective of the sign of $A_{0}$.

	\subsubsection{Pseudo-vectorial Torsion}
	For a pseudo-vectorial form of torsion, one has
	\beq
	S_{\mu\nu\lambda}=\frac{1}{3!}\epsilon_{\mu\nu\lambda\rho}\tilde{S}^{\rho}
	\eeq
	such that
	\beq
	\dot{\Theta}+\frac{\Theta^{2}}{n-1} +\sigma^{2}-\omega^{2}-\nabla_{\mu}a^{\mu}=-R_{\mu\nu}u^{\mu}u^{\nu}+\frac{1}{3}\epsilon^{\alpha\beta\mu\nu}u_{\alpha}\tilde{S}^{\beta}\omega_{\mu\nu}
	\eeq
	which, for autoparallel motion reads
	\beq
	\dot{\Theta}+\frac{\Theta^{2}}{n-1} +\sigma^{2}-\omega^{2}=-R_{\mu\nu}u^{\mu}u^{\nu}+\frac{1}{3}\epsilon^{\alpha\beta\mu\nu}u_{\alpha}\tilde{S}^{\beta}\omega_{\mu\nu}
	\eeq
	Notice now that for an observer with $u_{\mu}=\delta_{\mu}^{0}$ only the spatial part $\omega_{ij}$ of rotation arises due to torsion.

	\section{Vorticity Evolution}
	Let us derive now, the evolution equation for the vorticity tensor in general non-Riemannian spaces, just for the sake of completeness. We prove the most general expression, for the first time in the literature. The starting point is again the Ricci identity
	\beq
	\nabla_{\alpha}\nabla_{\beta}u_{\mu}-\nabla_{\beta}\nabla_{\alpha}u_{\mu}=-R^{\lambda}_{\;\;\mu\alpha\beta}u_{\lambda}+2 S_{\alpha\beta}^{\;\;\;\;\nu}\nabla_{\nu}u_{\mu}
	\eeq
	which we may contract now by $u^{\alpha}$ and antisymmetrize in  $\beta,\mu$ to arrive at
	\beq
	u^{\alpha}\nabla_{\alpha}\nabla_{[\beta}u_{\mu]}-u^{\alpha}\nabla_{[\beta}\nabla_{|\alpha|}u_{\mu]}=\frac{1}{2}(R_{\lambda\beta\alpha\mu}-R_{\alpha\mu\lambda\beta})+2 u^{\alpha} S_{\alpha[\beta}^{\;\;\;\;\nu}\nabla_{|\nu |}u_{\mu]}
	\eeq
	Now, using decomposition ($\ref{nabldec}$) we may take the antisymmetric part of it
	\beq
	\nabla_{[\beta}u_{\mu]}=\omega_{\mu\beta}-\frac{1}{l^{2}}(\xi_{[\beta}u_{\mu]}+u_{[\beta}a_{\mu]})
	\eeq
	which will help us compute the first term of the left hand side of the above
	\beq
	u^{\alpha}\nabla_{\alpha}\nabla_{[\beta}u_{\mu]}=\dot{\omega}_{\mu\beta}+\frac{d}{d\lambda}\left(\frac{1}{l^{2}} t_{[\mu}u_{\beta]} \right)
	\eeq
	where we have defined $t_{\mu} \equiv \xi_{\mu}-a_{\mu}$. Regarding the second term on the left hand side of the contracted Ricci identity, using partial integration we obtain
	\beq
	u^{\alpha}\nabla_{[\beta}\nabla_{|\alpha|}u_{\mu]}=\nabla_{[\mu}a_{\beta]}-\nabla_{\alpha}u_{[\beta}\nabla_{\mu]}u^{\alpha}
	\eeq
	To compute the last term of the above we first note that\footnote{To derive this we simply use $u^{\alpha}=g^{\alpha\nu}u_{\nu}$ and the definition of non-metricity.}
	\beq
	(\nabla_{\alpha}u_{\beta})\nabla_{\mu}u^{\alpha}=(\nabla_{\alpha}u_{\beta})u_{\nu}Q_{\mu}^{\;\;\nu\alpha}+(\nabla_{\alpha}u_{\beta})g^{\nu\alpha}\nabla_{\mu}u_{\nu}
	\eeq
	and we only need to take care of the last term of the latter. After some lengthy calculations, we finally arrive at
	\begin{gather}
	(\nabla_{\alpha}u_{\beta})g^{\nu\alpha}(\nabla_{\mu}u_{\nu})=\omega_{\beta\alpha}\omega^{\alpha}_{\;\;\mu}+\sigma_{\beta\alpha}\sigma^{\alpha}_{\;\;\mu}+\omega_{\beta\alpha}\sigma^{\alpha}_{\;\;\mu}+\sigma_{\beta\alpha}\omega^{\alpha}_{\;\;\mu} \nonumber \\
	+\frac{2}{n-1}\Theta_{D}(\omega_{\beta\mu}+\sigma_{\beta\mu})-\frac{1}{l^{2}}u_{\mu}a^{\alpha}(\omega_{\beta\alpha}+\sigma_{\beta\alpha})-\frac{1}{l^{2}}u_{\beta}\xi_{\alpha}(\omega^{\alpha}_{\;\;\mu} +\sigma^{\alpha}_{\;\;\mu}) \nonumber \\
	+\frac{\Theta_{D}^{2}}{(n-1)^{2}}h_{\mu\beta}+\frac{u_{\mu}u_{\beta}}{l^{4}}\left[ -\frac{2}{n-1}\Theta_{D}(a\cdot u)+\frac{1}{l^{2}}(a\cdot u)(\xi \cdot u)+(\xi \cdot a) \right] \nonumber \\
	-\frac{1}{l^{2}}\frac{\Theta_{D}}{n-1}(a_{\beta}u_{\mu}+\xi_{\mu}u_{\beta}+u_{\mu}a_{\beta})-\frac{1}{l^{2}}a_{\beta}\xi_{\mu}
	\end{gather}
	and taking its antisymmetric part in $\beta,\mu$ it follows that
	\begin{gather}
	(\nabla_{\alpha}u_{[\beta})g^{\nu\alpha}\nabla_{\mu]}u_{\nu}=2 \sigma_{\alpha[\beta}\omega^{\alpha}_{\;\;\mu]}+\frac{2\Theta_{D}}{n-1}\omega_{\beta\mu}-\frac{\Theta_{D}}{(n-1)l^{2}}u_{[\beta}t_{\mu]} \nonumber \\
	+\frac{1}{l^{2}}(p^{\alpha}\omega_{\alpha[\beta}+t^{\alpha}\sigma_{\alpha[\beta})u_{\mu]}-\frac{1}{l^{2}}a_{[\beta}\xi_{\mu]}
	\end{gather}
	where we have defined $p_{\mu}\equiv \xi_{\mu}+a_{\mu}$. Using, all the above, the contracted Ricci identity finally becomes
	\begin{gather}
	\dot{\omega}_{\lambda\nu}=\nabla_{[\nu}a_{\lambda]}+2 \sigma_{\alpha[\nu}\omega^{\alpha}_{\;\;\lambda]}-\frac{2 \Theta_{D}}{n-1}\omega_{\lambda\nu} \nonumber \\
	+(\nabla_{\alpha}u_{[\nu})\Big( Q_{\lambda]\kappa}^{\;\;\;\;\;\alpha}+2  S_{\lambda]\kappa}^{\;\;\;\;\alpha}  \Big)u^{\kappa} \nonumber \\
	+\frac{d}{d\lambda}\left( \frac{1}{l^{2}}u_{[\nu}t_{\lambda]} \right) -\frac{\Theta_{D}}{(n-1)l^{2}}u_{[\nu}t_{\lambda]}-\frac{1}{l^{2}}a_{[\nu}\xi_{\lambda]} \nonumber \\
	+\frac{1}{l^{2}}\Big(p^{\alpha}\omega_{\alpha[\nu}+t^{\alpha}\sigma_{\alpha[\nu}\Big)u_{\lambda]}+\frac{1}{2}\Big( R_{\mu\nu\kappa\lambda}-R_{\kappa\lambda\mu\nu}\Big)u^{\mu}u^{\kappa} \label{oomegad}
	\end{gather}
	where $t_{\mu} \equiv \xi_{\mu}-a_{\mu}$,\; $p_{\mu}\equiv \xi_{\mu}+a_{\mu}$ and recall that $\Theta_{D}=\Theta +(a\cdot u)/l^{2}$. Now, notice that the last term of the above involving the Riemman tensor, namely
	\beq
	\frac{1}{2}\Big( R_{\mu\nu\kappa\lambda}-R_{\kappa\lambda\mu\nu}\Big)u^{\mu}u^{\kappa}
	\eeq
	does not vanish as in Riemannian case\footnote{In the Riemannian case, where both torsion and non-metricity vanish, one has $ R_{\mu\nu\kappa\lambda}=R_{\kappa\lambda\mu\nu} $ and the aforementioned term is zero.}  as one cannot freely interchange the first two with the last two indices. To deal with this term let us recall identity ($\ref{xamoulhs}$)
	\begin{gather}
	R_{\mu\nu\kappa\lambda}-R_{\kappa\lambda\mu\nu}=3\Big( g_{\mu\alpha}\nabla_{[\nu}S_{\lambda\kappa]}^{\;\;\;\;\;\alpha} + g_{\nu\alpha}\nabla_{[\mu}S_{\lambda\kappa]}^{\;\;\;\;\;\alpha}+ g_{\kappa\alpha}\nabla_{[\lambda}S_{\mu\nu]}^{\;\;\;\;\;\alpha}+g_{\lambda\alpha}\nabla_{[\kappa}S_{\mu\nu]}^{\;\;\;\;\;\alpha}
	\Big) \nonumber \\
	+6\Big( g_{\mu\alpha}S_{[\nu\lambda}^{\;\;\;\;\;\beta}S_{\kappa]\beta}^{\;\;\;\;\;\alpha}+g_{\nu\alpha}S_{[\mu\lambda}^{\;\;\;\;\;\beta}S_{\kappa]\beta}^{\;\;\;\;\;\alpha} +g_{\kappa\alpha}S_{[\lambda\mu}^{\;\;\;\;\;\beta}S_{\nu]\beta}^{\;\;\;\;\;\alpha}+g_{\lambda\alpha}S_{[\kappa\mu}^{\;\;\;\;\;\beta}S_{\nu]\beta}^{\;\;\;\;\;\alpha}   \Big) \nonumber \\
	+\nabla_{[\lambda}Q_{\kappa]\nu\mu}+\nabla_{[\nu}Q_{\lambda]\kappa\mu}+\nabla_{[\kappa}Q_{\mu]\lambda\nu}+\nabla_{[\mu}Q_{\nu]\lambda\kappa}+\nabla_{[\mu}Q_{\lambda]\nu\kappa}+\nabla_{[\nu}Q_{\kappa]\lambda\mu} \nonumber \\
	-\Big( S_{\lambda\kappa}^{\;\;\;\;\;\alpha}Q_{\alpha\nu\mu}+ S_{\nu\lambda}^{\;\;\;\;\;\alpha}Q_{\alpha\kappa\mu}+ S_{\kappa\mu}^{\;\;\;\;\;\alpha}Q_{\alpha\lambda\nu} \nonumber \\
	+ S_{\mu\nu}^{\;\;\;\;\;\alpha}Q_{\alpha\lambda\kappa}+ S_{\mu\lambda}^{\;\;\;\;\;\alpha}Q_{\alpha\nu\kappa}+ S_{\nu\kappa}^{\;\;\;\;\;\alpha}Q_{\alpha\lambda\mu}\Big)
	\end{gather}
	which we may contract by $u^{\mu}u^{\kappa}$ to arrive at 
	\begin{gather}
	\Big(R_{\mu\nu\kappa\lambda}-R_{\kappa\lambda\mu\nu}\Big)u^{\mu}u^{\kappa}= \nonumber \\
	3\Big( g_{\mu\alpha}\nabla_{[\nu}S_{\lambda\kappa]}^{\;\;\;\;\;\alpha} + g_{\nu\alpha}\nabla_{[\mu}S_{\lambda\kappa]}^{\;\;\;\;\;\alpha}+ g_{\kappa\alpha}\nabla_{[\lambda}S_{\mu\nu]}^{\;\;\;\;\;\alpha}+g_{\lambda\alpha}\nabla_{[\kappa}S_{\mu\nu]}^{\;\;\;\;\;\alpha}
	\Big)u^{\mu}u^{\kappa} \nonumber \\
	+6\Big( g_{\mu\alpha}S_{[\nu\lambda}^{\;\;\;\;\;\beta}S_{\kappa]\beta}^{\;\;\;\;\;\alpha}+g_{\nu\alpha}S_{[\mu\lambda}^{\;\;\;\;\;\beta}S_{\kappa]\beta}^{\;\;\;\;\;\alpha} +g_{\kappa\alpha}S_{[\lambda\mu}^{\;\;\;\;\;\beta}S_{\nu]\beta}^{\;\;\;\;\;\alpha}+g_{\lambda\alpha}S_{[\kappa\mu}^{\;\;\;\;\;\beta}S_{\nu]\beta}^{\;\;\;\;\;\alpha}   \Big)u^{\mu}u^{\kappa} \nonumber \\
	+\Big( \nabla_{[\lambda}Q_{\kappa]\nu\mu}+\nabla_{[\nu}Q_{\lambda]\kappa\mu}+\nabla_{[\kappa}Q_{\mu]\lambda\nu}+\nabla_{[\mu}Q_{\nu]\lambda\kappa}+\nabla_{[\mu}Q_{\lambda]\nu\kappa}+\nabla_{[\nu}Q_{\kappa]\lambda\mu}\Big) u^{\mu}u^{\kappa} \nonumber \\
	-\Big( S_{\lambda\kappa}^{\;\;\;\;\;\alpha}Q_{\alpha\nu\mu}+ S_{\nu\lambda}^{\;\;\;\;\;\alpha}Q_{\alpha\kappa\mu}+ S_{\kappa\mu}^{\;\;\;\;\;\alpha}Q_{\alpha\lambda\nu} \nonumber \\
	+ S_{\mu\nu}^{\;\;\;\;\;\alpha}Q_{\alpha\lambda\kappa}+ S_{\mu\lambda}^{\;\;\;\;\;\alpha}Q_{\alpha\nu\kappa}+ S_{\nu\kappa}^{\;\;\;\;\;\alpha}Q_{\alpha\lambda\mu}\Big)u^{\mu}u^{\kappa}
	\end{gather}
	Substituting this back to ($\ref{oomegad}$) we obtain
	\begin{gather}
	\dot{\omega}_{\lambda\nu}=\nabla_{[\nu}a_{\lambda]}+2 \sigma_{\alpha[\nu}\omega^{\alpha}_{\;\;\lambda]}-\frac{2 \Theta_{D}}{n-1}\omega_{\lambda\nu} \nonumber \\
	+(\nabla_{\alpha}u_{[\nu})\Big( Q_{\lambda]\kappa}^{\;\;\;\;\;\alpha}+2  S_{\lambda]\kappa}^{\;\;\;\;\alpha}  \Big)u^{\kappa}+\frac{1}{l^{2}}\Big(p^{\alpha}\omega_{\alpha[\nu}+t^{\alpha}\sigma_{\alpha[\nu}\Big)u_{\lambda]} \nonumber \\
	+\frac{d}{d\lambda}\left( \frac{1}{l^{2}}u_{[\nu}t_{\lambda]} \right) -\frac{\Theta_{D}}{(n-1)l^{2}}u_{[\nu}t_{\lambda]}-\frac{1}{l^{2}}a_{[\nu}\xi_{\lambda]} \nonumber \\
	\frac{3}{2}\Big( g_{\mu\alpha}\nabla_{[\nu}S_{\lambda\kappa]}^{\;\;\;\;\;\alpha} + g_{\nu\alpha}\nabla_{[\mu}S_{\lambda\kappa]}^{\;\;\;\;\;\alpha}+ g_{\kappa\alpha}\nabla_{[\lambda}S_{\mu\nu]}^{\;\;\;\;\;\alpha}+g_{\lambda\alpha}\nabla_{[\kappa}S_{\mu\nu]}^{\;\;\;\;\;\alpha}
	\Big)u^{\mu}u^{\kappa} \nonumber \\
	+3\Big( g_{\mu\alpha}S_{[\nu\lambda}^{\;\;\;\;\;\beta}S_{\kappa]\beta}^{\;\;\;\;\;\alpha}+g_{\nu\alpha}S_{[\mu\lambda}^{\;\;\;\;\;\beta}S_{\kappa]\beta}^{\;\;\;\;\;\alpha} +g_{\kappa\alpha}S_{[\lambda\mu}^{\;\;\;\;\;\beta}S_{\nu]\beta}^{\;\;\;\;\;\alpha}+g_{\lambda\alpha}S_{[\kappa\mu}^{\;\;\;\;\;\beta}S_{\nu]\beta}^{\;\;\;\;\;\alpha}   \Big)u^{\mu}u^{\kappa} \nonumber \\
	+\frac{1}{2}\Big( \nabla_{[\lambda}Q_{\kappa]\nu\mu}+\nabla_{[\nu}Q_{\lambda]\kappa\mu}+\nabla_{[\kappa}Q_{\mu]\lambda\nu}+\nabla_{[\mu}Q_{\nu]\lambda\kappa}+\nabla_{[\mu}Q_{\lambda]\nu\kappa}+\nabla_{[\nu}Q_{\kappa]\lambda\mu}\Big) u^{\mu}u^{\kappa} \nonumber \\
	-\frac{1}{2}\Big( S_{\lambda\kappa}^{\;\;\;\;\;\alpha}Q_{\alpha\nu\mu}+ S_{\nu\lambda}^{\;\;\;\;\;\alpha}Q_{\alpha\kappa\mu}+ S_{\kappa\mu}^{\;\;\;\;\;\alpha}Q_{\alpha\lambda\nu} \nonumber \\
	+ S_{\mu\nu}^{\;\;\;\;\;\alpha}Q_{\alpha\lambda\kappa}+ S_{\mu\lambda}^{\;\;\;\;\;\alpha}Q_{\alpha\nu\kappa}+ S_{\nu\kappa}^{\;\;\;\;\;\alpha}Q_{\alpha\lambda\mu}\Big)u^{\mu}u^{\kappa}
	\end{gather}
	which is the evolution equation for the vorticity! To make the decomposition irreducible (as if in this form it is not already horrifying enough!) we may also substitute $\nabla_{\alpha}u_{\nu}$ (appearing in the second line) by its decomposition to finally arrive at
	\begin{gather}
	\dot{\omega}_{\lambda\nu}=\nabla_{[\nu}a_{\lambda]}+2 \sigma_{\alpha[\nu}\omega^{\alpha}_{\;\;\lambda]}-\frac{2 \Theta_{D}}{n-1}\omega_{\lambda\nu} \nonumber \\
	+(\sigma_{\alpha[\nu}-\omega_{\alpha[\nu})\Big( Q_{\lambda]\kappa}^{\;\;\;\;\;\alpha}+2Q_{\lambda]\kappa}^{\;\;\;\;\;\alpha}\Big)u^{\kappa}
	+\frac{\Theta_{D}}{n-1}\Big(Q_{[\lambda|\kappa|\nu]} +2S_{[\lambda|\kappa|\nu]} \Big)u^{\kappa} \nonumber \\
	+\frac{\Theta_{D}}{(n-1)l^{2}}u_{[\nu}\Big( Q_{\lambda]\kappa\alpha} +2S_{\lambda]\kappa\alpha} \Big)u^{\kappa}u^{\alpha}-\frac{1}{l^{2}}(\xi^{\alpha}u_{[\nu}+u^{\alpha}a_{[\nu})\Big(Q_{\lambda]\kappa\alpha} +2S_{\lambda]\kappa\alpha} \Big)u^{\kappa} \nonumber \\
	-\frac{1}{l^{4}}(a\cdot u)u_{[\nu}\Big(Q_{\lambda]\kappa\alpha} +2S_{\lambda]\kappa\alpha} \Big)u^{\kappa}u^{\alpha} 
	+\frac{1}{l^{2}}\Big(p^{\alpha}\omega_{\alpha[\nu}+t^{\alpha}\sigma_{\alpha[\nu}\Big)u_{\lambda]} \nonumber \\
	+\frac{d}{d\lambda}\left( \frac{1}{l^{2}}u_{[\nu}t_{\lambda]} \right) -\frac{\Theta_{D}}{(n-1)l^{2}}u_{[\nu}t_{\lambda]}-\frac{1}{l^{2}}a_{[\nu}\xi_{\lambda]} \nonumber \\
	\frac{3}{2}\Big( g_{\mu\alpha}\nabla_{[\nu}S_{\lambda\kappa]}^{\;\;\;\;\;\alpha} + g_{\nu\alpha}\nabla_{[\mu}S_{\lambda\kappa]}^{\;\;\;\;\;\alpha}+ g_{\kappa\alpha}\nabla_{[\lambda}S_{\mu\nu]}^{\;\;\;\;\;\alpha}+g_{\lambda\alpha}\nabla_{[\kappa}S_{\mu\nu]}^{\;\;\;\;\;\alpha}
	\Big)u^{\mu}u^{\kappa} \nonumber \\
	+3\Big( g_{\mu\alpha}S_{[\nu\lambda}^{\;\;\;\;\;\beta}S_{\kappa]\beta}^{\;\;\;\;\;\alpha}+g_{\nu\alpha}S_{[\mu\lambda}^{\;\;\;\;\;\beta}S_{\kappa]\beta}^{\;\;\;\;\;\alpha} +g_{\kappa\alpha}S_{[\lambda\mu}^{\;\;\;\;\;\beta}S_{\nu]\beta}^{\;\;\;\;\;\alpha}+g_{\lambda\alpha}S_{[\kappa\mu}^{\;\;\;\;\;\beta}S_{\nu]\beta}^{\;\;\;\;\;\alpha}   \Big)u^{\mu}u^{\kappa} \nonumber \\
	+\frac{1}{2}\Big( \nabla_{[\lambda}Q_{\kappa]\nu\mu}+\nabla_{[\nu}Q_{\lambda]\kappa\mu}+\nabla_{[\kappa}Q_{\mu]\lambda\nu}+\nabla_{[\mu}Q_{\nu]\lambda\kappa}+\nabla_{[\mu}Q_{\lambda]\nu\kappa}+\nabla_{[\nu}Q_{\kappa]\lambda\mu}\Big) u^{\mu}u^{\kappa} \nonumber \\
	-\frac{1}{2}\Big( S_{\lambda\kappa}^{\;\;\;\;\;\alpha}Q_{\alpha\nu\mu}+ S_{\nu\lambda}^{\;\;\;\;\;\alpha}Q_{\alpha\kappa\mu}+ S_{\kappa\mu}^{\;\;\;\;\;\alpha}Q_{\alpha\lambda\nu} \nonumber \\
	+ S_{\mu\nu}^{\;\;\;\;\;\alpha}Q_{\alpha\lambda\kappa}+ S_{\mu\lambda}^{\;\;\;\;\;\alpha}Q_{\alpha\nu\kappa}+ S_{\nu\kappa}^{\;\;\;\;\;\alpha}Q_{\alpha\lambda\mu}\Big)u^{\mu}u^{\kappa}
	\end{gather}
	This is the evolution equation for vorticity in spaces with both torsion and non-metricity and is the most general one can have. This is the most  general form of the vorticity evolution equation and (to the best of our knowledge) it is presented here for the first time in the literature. Again, at this point the above holds as an identity an only becomes an equation once an energy momentum and a hyper-momentum tensors are given. 
	
	To complete the analysis we should also give the evolution equation for shear but since this is a lot more complicated even from the above, we will refrain from writing it down for obvious reasons. We may now change subject and study scale transformations in Metric-Affine Geometry.

	\chapter{Scale Transformations in Metric-Affine Geometry}
	
	In this chapter we define some transformations that can be performed in a non-Riemannian space. In particular, we define the projective, conformal and frame rescaling transformations, and see the effect of those transformations on the basic geometrical quantities. We also state and prove some useful identities that follow when a scalar quantity is invariant under any of the above separately. We then compute how quadratic torsion and non-metricity scalars transform under those transformations and construct invariant actions with respect to each transformation. The results of this chapter we have published in \cite{iosifidis2018raychaudhuri}.
	
	\section{Transformations in Metric Affine Manifolds}
	 To start with, it will be helpful for our discussion later to define the scalars \newline
	\textbf{Pure Non-Metricity Scalars}
	\begin{gather}
	A_{1}=Q_{\alpha\mu\nu}Q^{\alpha\mu\nu} \\
	A_{2}=Q_{\alpha\mu\nu}Q^{\mu\nu\alpha} \\
	A_{3}=Q_{\mu}Q^{\mu}
	\\
	A_{4}=q_{\mu}q^{\mu}
	\\
	A_{5}=Q_{\mu}q^{\mu}
	\\
	A_{6}=\epsilon^{\alpha\beta\gamma\delta}Q_{\alpha\beta\mu}Q_{\gamma\delta}^{\;\;\;\;\mu}
	\end{gather}
	where $Q_{\alpha}\equiv Q_{\alpha\mu\nu}g^{\mu\nu}$ and $\tilde{Q}_{\mu}=Q_{\lambda\nu\mu}g^{\lambda\nu}$.\newline
	\textbf{Pure Torsion Scalars}
	\begin{gather}
	B_{1}=S_{\alpha\mu\nu}S^{\alpha\mu\nu} \\
	B_{2}=S_{\alpha\mu\nu}S^{\mu\nu\alpha} \\
	B_{3}=S_{\mu}S^{\mu}
	\\
	B_{4}=t_{\mu}t^{\mu}
	\\
	B_{5}=S_{\mu}t^{\mu}
	\\
	B_{6}=\epsilon^{\alpha\beta\gamma\delta}S_{\alpha\beta\mu}S_{\gamma\delta}^{\;\;\;\;\mu}
	\\
	B_{7}=\epsilon^{\alpha\beta\gamma\delta}S_{\lambda\alpha\beta}S^{\lambda}_{\;\;\gamma\delta}
	\\
	B_{8}=\epsilon^{\alpha\beta\gamma\delta}S_{\mu\alpha\beta}S_{\gamma\delta}^{\;\;\;\;\mu}
	\end{gather}
	where $S_{\mu}\equiv S_{\mu\lambda}^{\;\;\;\;\lambda}$ and $t^{\alpha}\equiv \epsilon^{\alpha\beta\gamma\delta}S_{\beta\gamma\delta}$.\newline
	\textbf{Mixed}
	\begin{gather}
	C_{1}=Q_{\alpha\mu\nu}S^{\alpha\mu\nu}
	\\
	C_{2}=Q_{\mu}S^{\mu} 
	\\
	C_{3}=q_{\mu}S^{\mu}
	\\
	C_{4}=Q^{\mu}t_{\mu}
	\\
	C_{5}=q^{\mu}t_{\mu}
	\\
	C_{6}=\epsilon^{\alpha\beta\gamma\delta}Q_{\alpha\beta\mu}S_{\gamma\delta}^{\;\;\;\;\mu}
	\\
	C_{7}=\epsilon^{\alpha\beta\gamma\delta}Q_{\alpha\beta\mu}S^{\mu}_{\;\;\;\gamma\delta}
	\end{gather}
	We should remark that the parity-even scalars $A_{1},...,A_{5}$ $\; B_{1},B_{2},B_{3}$ $C_{1},C_{2},C_{3}$ exist for any dimension-$n$ while the remaining parity-odd only for $n=4$. Let us now proceed to discuss the transformations that can be formed in a Metric-Affine Geometry.

	\subsection{Projective Transformations}
	In this section we define and  discuss more thoroughly projective transformations of the affine connection and see how some important tensors transform under such transformations. In addition, we present some projective invariant tensors and prove the connection of tracelessness and projective invariance. So, to start our discussion, a projective transformation of the affine connection is defined as
	\begin{equation}
	\Gamma^{\lambda}_{\;\;\mu\nu}\longrightarrow \Gamma^{\lambda}_{\;\;\mu\nu}+ \delta_{\mu}^{\lambda}\xi_{\nu} \label{ptf}
	\end{equation}
	where $\xi_{\nu}(x)$ is a random vector field. When this vector field is exact, namely when
	\beq
	\xi_{\nu}=\partial_{\nu}\lambda
	\eeq
	for some scalar function $\lambda(x)$ we have what is known as  a $\bf{special}$  $\bf{projective}$   $\bf{transformation}$
	\begin{equation}
	\Gamma^{\lambda}_{\;\;\mu\nu}\longrightarrow \Gamma^{\lambda}_{\;\;\mu\nu}+ \delta_{\mu}^{\lambda} \partial_{\nu}\lambda
	\end{equation}
	Under transformation $(\ref{ptf})$, the Riemann tensor transforms as
	\beq
	R^{\mu}_{\;\;\nu\alpha\beta}\longrightarrow R^{\mu}_{\;\;\nu\alpha\beta}-2\delta_{\nu}^{\mu}\partial_{[\alpha}\xi_{\beta]}
	\eeq
	and therefore the transformation rule for the Ricci tensor is
	\beq
	R_{\alpha\beta}\longrightarrow R_{\alpha\beta}-2 \partial_{[\alpha}\xi_{\beta]}
	\eeq
	and for homothetic curvature
	\beq
	\hat{R}_{\alpha\beta}\longrightarrow \hat{R}_{\alpha\beta}-2 n \partial_{[\alpha}\xi_{\beta]}
	\eeq
	Notice that all of the above tensors are invariant under special projective transformations ( since $\partial_{[\alpha}\partial_{\beta]}\lambda =0$ ) but not general ones. In addition, if the space is endowed with a metric we also have the transformation rule
	\beq
	R_{\mu\nu\alpha\beta}\longrightarrow R_{\mu\nu\alpha\beta}-2 g_{\mu\nu}\partial_{[\alpha}\xi_{\beta]}
	\eeq
	and the third contraction of the Riemann tensor $\check{R}_{\mu\beta}=R_{\mu\nu\alpha\beta} g^{\nu\alpha}$ transforms according to
	\beq
	\check{R}_{\mu\beta}\longrightarrow \check{R}_{\mu\beta} -2\partial_{[\mu}\xi_{\beta]}
	\eeq
	At this point we should stress out that even though the Ricci tensor $R_{\mu\nu}$ is not projective invariant, its symmetric part $R_{(\mu\nu)}$ is unaltered under projective transformations
	\beq
	R_{(\mu\nu)}\longrightarrow R_{(\mu\nu)}
	\eeq
	and as a result the Ricci scalar is invariant under projective transformations
	\beq
	R\longrightarrow R
	\eeq
	Thus, the Einstein-Hilbert action (and of course any $f(R)$ theory) has a gravitation sector that is invariant under projective transformations and this invariance is the reason why the Palatini tensor is traceless in its first two indices. Now regarding the transformations of torsion and non-metricity tensors, one can easily check that under projective transformations
	\beq
	S_{\mu\nu}^{\;\;\;\;\lambda}\longrightarrow S_{\mu\nu}^{\;\;\;\;\lambda}+\delta^{\lambda}_{[\mu}\xi_{\nu]}
	\eeq
	or
	\beq
	S_{\mu\nu\alpha}\longrightarrow S_{\mu\nu\alpha} +g_{\alpha[\mu}\xi_{\nu]} \label{tortr}
	\eeq
	and
	\beq
	Q_{\alpha\mu\nu}\longrightarrow Q_{\alpha\mu\nu}+2 \xi_{\alpha}g_{\mu\nu}
	\eeq
	For their associate vectors we have
	\begin{equation}
	S_{\mu} \longrightarrow S_{\mu} + \frac{(1-n)}{2}\xi_{\mu}
	\end{equation}
	\begin{equation}
	Q_{\mu}\longrightarrow Q_{\mu} +2 n \xi_{\mu}
	\end{equation}
	\begin{equation}
	\tilde{Q}_{\mu}\longrightarrow \tilde{Q}_{\mu} +2 \xi_{\mu}
	\end{equation}
	Also, since
	\beq
	S_{[\mu\nu\alpha]}\longrightarrow S_{[\mu\nu\alpha]} 
	\eeq
	as is easily seen from ($\ref{tortr}$) the torsion pseudo-vector remains invariant
	\beq
	\tilde{S}_{\mu} \longrightarrow \tilde{S}_{\mu} 
	\eeq

	With these at hand, we can find how the quadratic scalars transform under projective transformations. A straightforward calculation yields
	\begin{gather}
	\hat{A}_{1}=A_{1}+4 Q_{\mu}\xi^{\mu}+4 n \xi_{\mu}\xi^{\mu}
	\\
	\hat{A}_{2}=A_{2}+4 q_{\mu}\xi^{\mu}+4  \xi_{\mu}\xi^{\mu}
	\\
	\hat{A}_{3}=A_{3}+4 n Q_{\mu}\xi^{\mu}+4 n^{2} \xi_{\mu}\xi^{\mu}
	\\
	\hat{A}_{4}=A_{4}+4 q_{\mu}\xi^{\mu}+4  \xi_{\mu}\xi^{\mu}
	\\
	\hat{A}_{5}=A_{5}+2 ( Q_{\mu}+n q_{\mu})\xi^{\mu}+4 n \xi_{\mu}\xi^{\mu}
	\\
	\hat{A}_{6}=A_{6}
	\end{gather}
	for the pure non-metricity quadratic scalars. For the pure torsion scalars we fine
	\begin{gather}
	\hat{B}_{1}=B_{1}-2 S_{\mu}\xi^{\mu}+\frac{(n-1)}{2} \xi_{\mu}\xi^{\mu}
	\\
	\hat{B}_{2}=B_{2}+ S_{\mu}\xi^{\mu}+\frac{(1-n)}{4} \xi_{\mu}\xi^{\mu}
	\\
	\hat{B}_{3}=B_{3}+(1-n) S_{\mu}\xi^{\mu}+\frac{(n-1)^{2}}{4} \xi_{\mu}\xi^{\mu}
	\\
	\hat{B}_{4}=B_{4}
	\\
	\hat{B}_{5}=B_{5}+\frac{(1-n)}{2}t_{\mu}\xi^{\mu}
	\\
	\hat{B}_{6}=B_{6}-2 t_{\mu}\xi^{\mu}
	\end{gather}
	and for the mixed terms
	\begin{gather}
	\hat{C}_{1}=C_{1}+\frac{1}{2}(q_{\mu}-Q_{\mu}+4 S_{\mu})\xi^{\mu}+(1-n)\xi_{\mu}\xi^{\mu}
	\\
		\hat{C}_{2}=C_{2}+\frac{(1-n)}{2}\Big( Q_{\mu}+\frac{4}{1-n} S_{\mu})\xi^{\mu}+(1-n)\xi_{\mu}\xi^{\mu}
		\\
			\hat{C}_{3}=C_{3}+\frac{(1-n)}{2}\Big( q_{\mu}+\frac{4n}{1-n} S_{\mu})\xi^{\mu}+n(1-n)\xi_{\mu}\xi^{\mu}
	\\
	\hat{C}_{4}=C_{4}+2n t_{\mu} \xi^{\mu}
	\\
	\hat{C}_{5}=C_{5}+2 t_{\mu} \xi^{\mu}
	\\
	\hat{C}_{6}=C_{6}+2 t_{\mu} \xi^{\mu}
	\end{gather}

	Next we digress a bit and discuss more combinations that give projective invariant quantities.
	
	\subsubsection{Projective Invariant Combinations}
	As we have already seen, some quantities remain invariant under projective transformation, for instance the Ricci scalar
	\beq
	R\longrightarrow R
	\eeq
	is unchanged. Let us  enumerate some (of the many) projective invariant combinations that can be formed. These include
	
	\begin{itemize}
		\item $R\longrightarrow R$
		\item $R_{(\mu\nu)}\longrightarrow R_{(\mu\nu)} $
		\item $R_{[\mu\nu]\alpha\beta}\longrightarrow R_{[\mu\nu]\alpha\beta} $
		\item $(R_{\mu\nu\alpha\beta}-g_{\mu\nu}R_{\alpha\beta})\longrightarrow ( R_{\mu\nu\alpha\beta}-g_{\mu\nu}R_{\alpha\beta})$
		\item $ \epsilon^{\mu\nu\alpha\beta}R_{\mu\nu\alpha\beta} \longrightarrow \epsilon^{\mu\nu\alpha\beta}R_{\mu\nu\alpha\beta} $
		\item $(n R_{\mu\nu\alpha\beta}-g_{\mu\nu}\hat{R}_{\alpha\beta})\longrightarrow ( n R_{\mu\nu\alpha\beta}-g_{\mu\nu}\hat{R}_{\alpha\beta})$
		\item $(\hat{R}_{\mu\nu}-n R_{\mu\nu} ) \longrightarrow  (\hat{R}_{\mu\nu}-n R_{\mu\nu} ) $
		\item $S_{[\alpha\mu\nu]} \longrightarrow  S_{[\alpha\mu\nu]}  $
		\item $\tilde{S}_{\mu} \longrightarrow \tilde{S}_{\mu}   $
		\item $( n\tilde{Q}_{\mu}-Q_{\mu})\longrightarrow    (n\tilde{Q}_{\mu}-Q_{\mu} )$
		\item $ (\tilde{Q}_{\mu}-Q_{\mu}-4 S_{\mu}) \longrightarrow ( \tilde{Q}_{\mu}-Q_{\mu}-4 S_{\mu} ) $
		\item $(Q_{\mu}-\frac{4n}{1-n}S_{\mu}) \longrightarrow (Q_{\mu}-\frac{4n}{1-n}S_{\mu})$
		\item $(\alpha Q_{\mu}+\beta  \tilde{Q}_{\mu}+\gamma S_{\mu})\longrightarrow (\alpha Q_{\mu}+\beta  \tilde{Q}_{\mu}+\gamma S_{\mu})\;,\;\;with\;\;\; n\alpha+\beta+\frac{(1-n)}{4}\gamma=0$
		\item $(Q_{[\mu\nu]\alpha}+2S_{\mu\nu\alpha})\longrightarrow (Q_{[\mu\nu]\alpha}+2S_{\mu\nu\alpha})$
		\item $ (Q_{\alpha\mu\nu}-\frac{Q_{\alpha}}{n}g_{\mu\nu} )\longrightarrow (Q_{\alpha\mu\nu}-\frac{Q_{\alpha}}{n}g_{\mu\nu} )$
		\item $ (Q_{\alpha\mu\nu}-\tilde{Q}_{\alpha}g_{\mu\nu})\longrightarrow  (Q_{\alpha\mu\nu}-\tilde{Q}_{\alpha}g_{\mu\nu})$
		\item $ ( S_{\mu\nu\alpha}-\frac{2}{1-n}\delta_{[\mu}^{\lambda}S_{\nu]}) \longrightarrow   ( S_{\mu\nu\alpha}-\frac{2}{1-n}\delta_{[\mu}^{\lambda}S_{\nu]})       $
		\item $  ( R_{\mu\nu}+\frac{4}{1-n}\partial_{[\mu}S_{\nu]}) \longrightarrow   ( R_{\mu\nu}+\frac{4}{1-n}\partial_{[\mu}S_{\nu]})   $
		\item $      ( \hat{R}_{\mu\nu}+\frac{4n}{1-n}\partial_{[\mu}S_{\nu]}) \longrightarrow   ( \hat{R}_{\mu\nu}+\frac{4n}{1-n}\partial_{[\mu}S_{\nu]})                $
	\end{itemize}
	These are just a few projective invariant combinations one can form, and by no means all the possible ones. Notice that any scalar that it's build up out of these combinations will also be projective invariant and as a result its variation with respect to the connection will yield a tensor that is traceless in its first two indices. We prove this (along with other two identities) after defining the conformal and frame rescaling transformations. Now let us continue with the conformal transformations.
	
	\subsection{Conformal Transformations}
	A conformal transformation of the metric is defined as
	\beq
	g_{\mu\nu}\rightarrow \bar{g}_{\mu\nu}=e^{2 \phi}g_{\mu\nu} \label{qtr}
	\eeq
	\beq
	\Gamma^{\lambda}_{\;\;\;\mu\nu}\rightarrow \bar{\Gamma}^{\lambda}_{\;\;\;\mu\nu}=\Gamma^{\lambda}_{\;\;\;\mu\nu}
	\eeq
	that is under a conformal transformation the metric tensor picks up a conformal factor $e^{2 \phi} $ while the affine connection is left unchanged. Note that the contravariant form of the metric tensor transforms as
	\beq
	\bar{g}^{\mu\nu}=e^{-2\phi}g^{\mu\nu} 
	\eeq
	as can be easily seen from the relation $g_{\mu\nu}g^{\nu\lambda}=\delta_{\mu}^{\lambda}$. In addition, the square root of the determinant of the metric obeys the transformation rule
	\beq
	\sqrt{-\bar{g}}=e^{n \phi} \sqrt{-g}
	\eeq 
	and for $n=4$ 
	\beq
	\sqrt{-\bar{g}}=e^{4 \phi} \sqrt{-g}
	\eeq 
	which is obtained directly by first taking the determinant of ($\ref{qtr}$) and then taking the square root of the result. From this last relation we infer the transformation rule for the Levi-Civita tensor
	\beq
	\bar{\epsilon}_{\mu\nu\rho\sigma}=e^{4 \phi}\epsilon_{\mu\nu\rho\sigma}
	\eeq
	\beq
	\bar{\epsilon}^{\mu\nu\rho\sigma}=e^{-4 \phi}\epsilon^{\mu\nu\rho\sigma}
	\eeq
	and recall that $\epsilon_{\mu\nu\rho\sigma}=\sqrt{-g}\eta_{\mu\nu\rho\sigma}$ where $\eta_{\mu\nu\rho\sigma}$ is the Levi-Civita symbol. Using the above we see that torsion and non-metricity transform as
	\beq
	\bar{S}_{\mu\nu}^{\;\;\;\;\lambda}=S_{\mu\nu}^{\;\;\;\;\lambda}
	\eeq
	\beq
	\bar{Q}_{\alpha\mu\nu}=e^{2 \phi}\Big( Q_{\alpha\mu\nu}-2(\partial_{\alpha} \phi)g_{\mu\nu} \Big)
	\eeq
	and the related vectors
	\begin{gather}
	\bar{Q}_{\mu}=Q_{\mu}-2 n \partial_{\mu} \phi
	\\
	\bar{q}_{\mu}=q_{\mu}-2 \partial_{\mu}\phi
	\\
	\bar{S}_{\mu}=S_{\mu}
	\\
	\bar{t}_{\mu}=t_{\mu}
	\end{gather}
	Then, it follows that all pure torsion scalars
	\begin{gather}
	B_{1}=S_{\alpha\mu\nu}S^{\alpha\mu\nu} \\
	B_{2}=S_{\alpha\mu\nu}S^{\mu\nu\alpha} \\
	B_{3}=S_{\mu}S^{\mu}
	\\
	B_{4}=t_{\mu}t^{\mu}
	\\
	B_{5}=S_{\mu}t^{\mu}
	\\
	B_{6}=\epsilon^{\alpha\beta\gamma\delta}S_{\alpha\beta\mu}S_{\gamma\delta}^{\;\;\;\;\mu}
	\\
	B_{7}=\epsilon^{\alpha\beta\gamma\delta}S_{\lambda\alpha\beta}S^{\lambda}_{\;\;\gamma\delta}
	\\
	B_{8}=\epsilon^{\alpha\beta\gamma\delta}S_{\mu\alpha\beta}S_{\gamma\delta}^{\;\;\;\;\mu}
	\end{gather}
	transform conformaly, that is
	\beq
	\bar{B}_{i}=e^{-2 \phi}B_{i}
	\eeq
	for any $i=1,2,...,8$. This means that any combination $\sqrt{-g}B_{i}B_{j}$ is conformally invariant\footnote{Weyl-Invariant extensions of the Metric-Affine Gravity were studied in \cite{vazirian2015weyl} in $4$-dim. The quadratic conformally invariant action was also given there.}.
	Regarding the pure non-metricity scalars, one can verify the transformation laws
	\begin{gather}
	\bar{A}_{1}=\bar{Q}_{\alpha\mu\nu}\bar{Q}^{\alpha\mu\nu}=e^{-2 \phi}\Big[ A_{1}-4 Q^{\mu}\partial_{\mu} \phi +4 n (\partial \phi)^{2}\Big]\\
	\bar{A}_{2}=\bar{Q}_{\alpha\mu\nu}\bar{Q}^{\mu\nu\alpha}=e^{-2 \phi}\Big[ A_{2}-4 q^{\mu}\partial_{\mu} \phi  +4  (\partial \phi)^{2}\Big]
	\\
	\bar{A}_{3}=\bar{Q}_{\mu}\bar{Q}^{\mu}=e^{- 2\phi}\Big[ A_{3}-4 n Q^{\mu}\partial_{\mu}\phi +4 n^{2} (\partial \phi)^{2}\Big] 
	\\
	\bar{A}_{4}=\bar{q}_{\mu}\bar{q}^{\mu}=e^{- 2\phi}\Big[ A_{4}-4  q^{\mu}\partial_{\mu}\phi +4  (\partial \phi)^{2}\Big] 
	\\
	\bar{A}_{5}=\bar{Q}_{\mu}\bar{q}^{\mu}=e^{- 2\phi}\Big[ A_{5}-2(Q^{\mu} +n  q^{\mu})\partial_{\mu}\phi +4 n (\partial \phi)^{2}\Big] 
	\\
	\bar{A}_{6}=\bar{\epsilon}^{\alpha\beta\gamma\delta}\bar{Q}_{\alpha\beta\mu}\bar{Q}_{\gamma\delta}^{\;\;\;\;\mu}=e^{-2 \phi}A_{6}
	\end{gather}
	and for the mixed terms 
	\begin{gather}
	\bar{C}_{1}=\bar{Q}_{\alpha\mu\nu}\bar{S}^{\alpha\mu\nu}=e^{-2 \phi}\Big[C_{1} - 2  S^{\mu}\partial_{\mu}\phi \Big]
	\\
	\bar{C}_{2}=\bar{Q}_{\mu}\bar{S}^{\mu} =e^{-2 \phi}\Big[C_{2} - 2 n S^{\mu}\partial_{\mu}\phi \Big]
	\\
	\bar{C}_{3}=\bar{q}_{\mu}\bar{S}^{\mu}=e^{-2 \phi}\Big[C_{3} - 2  S^{\mu}\partial_{\mu}\phi \Big]
	\\
	\bar{C}_{4}=\bar{Q}^{\mu}\bar{t}_{\mu}=e^{-2 \phi}\Big[C_{4} - 2 n  t^{\mu}\partial_{\mu}\phi \Big]
	\\
	\bar{C}_{5}=\bar{q}^{\mu}\bar{t}_{\mu}=e^{-2 \phi}\Big[C_{5} - 2  t^{\mu}\partial_{\mu}\phi \Big]
	\\
	\tilde{C}_{6}=\bar{\epsilon}^{\alpha\beta\gamma\delta}\bar{Q}_{\alpha\beta\mu}\bar{S}_{\gamma\delta}^{\;\;\;\;\mu}=e^{-2 \phi}\Big[C_{6} - 2  t^{\mu}\partial_{\mu}\phi \Big]
	\\
	\bar{C}_{7}=\bar{\epsilon}^{\alpha\beta\gamma\delta}\bar{Q}_{\alpha\beta\mu}\bar{S}^{\mu}_{\;\;\;\gamma\delta}=e^{-2 \phi}\Big[C_{7} - 2  t^{\mu}\partial_{\mu}\phi \Big]
	\end{gather}

	\subsection{Frame Resclaling}
	A frame rescaling transformation results in a combination of  a conformal metric transformation$+$ a special projective transformation. More specifically, we have
	\beq
	\Gamma^{\lambda}_{\;\;\;\mu\nu}\rightarrow \tilde{\Gamma}^{\lambda}_{\;\;\;\mu\nu}=\Gamma^{\lambda}_{\;\;\;\mu\nu}+\delta^{\lambda}_{\mu}\partial_{\nu}\phi
	\eeq
	\beq
	g_{\mu\nu}\rightarrow \tilde{g}_{\mu\nu}=e^{2 \phi}g_{\mu\nu}
	\eeq
	with the same scalar field $\phi(x)$ appearing in both above. Interestingly, under the above transformations, the non-metricity tensor does not change and it just picks-up a conformal factor.  In words
	\beq
	\tilde{Q}_{\alpha\mu\nu}=e^{2\phi}Q_{\alpha\mu\nu}
	\eeq
	as can be easily seen by applying both transformations on the non-metricity tensor. This makes the procedure  of computing quadratic non-metricity scalars, extremely simple. Indeed, let us consider the scalars
	\begin{gather}
	A_{1}=Q_{\alpha\mu\nu}Q^{\alpha\mu\nu} \\
	A_{2}=Q_{\alpha\mu\nu}Q^{\mu\nu\alpha} \\
	A_{3}=Q_{\mu}Q^{\mu}
	\\
	A_{4}=q_{\mu}q^{\mu}
	\\
	A_{5}=Q_{\mu}q^{\mu}
	\\
	A_{6}=\epsilon^{\alpha\beta\gamma\delta}Q_{\alpha\beta\mu}Q_{\gamma\delta}^{\;\;\;\;\mu}
	\end{gather}
	It is an easy matter to show that under frame rescalings
	\beq
	\tilde{A}_{i}=e^{-2\phi}A_{i}
	\eeq
	for any $i=1,2,...,6$. Therefore any combination $\sqrt{-g}A_{i}A_{j}$ is invariant under frame rescalings.  As far as torsion is concerned, we have the transformation law
	\beq
	\tilde{S}_{\mu\nu}^{\;\;\;\;\lambda}=S_{\mu\nu}^{\;\;\;\;\lambda}+\delta^{\lambda}_{[\mu}\partial_{\nu]}\phi
	\eeq
	and for the torsion vector
	\beq
	\tilde{S}_{\mu}=S_{\mu}+\frac{(1-n)}{2}\partial_{\mu}\phi
	\eeq
	Then, considering the scalars
	\begin{align}
	B_{1}=S_{\mu\nu\alpha}S^{\mu\nu\alpha} \\
	B_{2}=S_{\alpha\mu\nu}S^{\mu\nu\alpha} \\
	B_{3}=S_{\mu}S^{\mu}
	\end{align}
	we see that under a frame rescaling, the above transform as
	\begin{align}
	\tilde{B}_{1}=e^{-2\phi} \Big[ B_{1}-2S^{\mu}\partial_{\mu}\phi
	+\frac{(n-1)}{2} (\partial \phi)^{2}\Big] \\
	\tilde{B}_{2}=e^{-2\phi}\Big[ B_{2}+S^{\mu}\partial_{\mu}\phi +\frac{(1-n)}{4}(\partial \phi)^{2} \Big] \\
	\tilde{B}_{3}=e^{-2\phi}\Big[B_{3}+(1-n)S^{\mu} \partial_{\mu}\phi +\frac{(1-n)^{2}}{4}(\partial \phi)^{2} \Big]
	\end{align}
	Notice that the combinations $B_{1}+2B_{2}$,\;$(n-1)B_{1}-2B_{3}$ and $(n-1)B_{2}+B_{3}$ transform conformally. For the rest of the quadratic torsion scalars one finds
	\begin{gather}
	\tilde{B}_{4}=e^{-2\phi} B_{4}
	\\
	\tilde{B}_{5}=e^{-2\phi} \Big[ B_{5}+\frac{(1-n)}{2}t^{\mu}\partial_{\mu}\phi \Big]
	\\
	\tilde{B}_{6}=e^{-2\phi} \Big[ B_{6}-2 t^{\mu}\partial_{\mu}\phi \Big]
	\\
	\tilde{B}_{7}=e^{-2\phi} \Big[ B_{7}+ t^{\mu}\partial_{\mu}\phi \Big]
	\\
	\tilde{B}_{8}=e^{-2\phi} \Big[ B_{8}-\frac{1}{2} t^{\mu}\partial_{\mu}\phi \Big]
	\end{gather}

	For the mixed terms one finds
	\begin{gather}
	\tilde{C}_{1}=\tilde{Q}_{\alpha\mu\nu}\tilde{S}^{\alpha\mu\nu}=e^{-2\phi} \Big[ C_{1}+\frac{1}{2}(q^{\mu}-Q^{\mu})\partial_{\mu}\phi \Big]
	\\
	\tilde{C}_{2}=\tilde{Q}_{\mu}\tilde{S}^{\mu}=e^{-2\phi} \Big[ C_{2}+\frac{1}{2}(1-n)Q^{\mu}\partial_{\mu}\phi \Big]
	\\
	\tilde{C}_{3}=\tilde{q}_{\mu}\tilde{S}^{\mu}=e^{-2\phi} \Big[ C_{3}+\frac{1}{2}(1-n)q^{\mu}\partial_{\mu}\phi \Big]
	\\
	\tilde{C}_{4}=\tilde{Q}_{\mu}\tilde{t}^{\mu}=e^{-2\phi}C_{4}
	\\
	\tilde{C}_{5}=\tilde{q}_{\mu}\tilde{t}^{\mu}=e^{-2\phi}C_{5}
	\\
	\tilde{C}_{6}=\tilde{\epsilon}^{\alpha\beta\gamma\delta}\tilde{Q}_{\alpha\beta\mu}\tilde{S}_{\gamma\delta}^{\;\;\;\;\mu}=e^{-2 \phi}C_{6}
	\\
	\tilde{C}_{7}=\tilde{\epsilon}^{\alpha\beta\gamma\delta}\tilde{Q}_{\alpha\beta\mu}\tilde{S}^{\mu}_{\;\;\;\gamma\delta}=e^{-2 \phi}C_{6}
	\end{gather}
	Having obtained the needed setup we are now in a position to study theories invariant under the three above transformations we defined. Before doing so, let us make an aside and see what identities do these invariances impose on the theories that are invariant under those.

	\subsection{Invariances and Identities}
	We will show now that the variations of actions that are invariant  under a transformation produce tensors that satisfy certain identities. More specifically we show that \newline 
	$\textbf{1)}$ If an action is invariant under projective transformations then its variation with respect to the connection produces a tensor that is traceless in its first two indices. \newline $\textbf{2)}$ An action invariant under conformal transformations, produces a traceless tensor when varied with respect to the metric. \newline $\textbf{3)}$ If an action is invariant under frame rescalings then the trace of its variation with respect to the metric tensor is related to the divergence of the trace in the first two indices of its variation with respect to the connection. We prove each of the above statements below.
	
	\subsubsection{Projective Invariance and tracelessness}
	As we have already pointed out the Palatini Tensor
	\begin{equation}
	P_{\lambda}^{\;\;\;\mu\nu}\equiv \frac{\delta R}{\delta \Gamma^{\lambda}_{\;\;\;\mu\nu}}
	\end{equation}
	has zero trace when contracted in its first two indices, that is
	\begin{equation}
	P_{\mu}^{\;\;\;\mu\nu}=0
	\end{equation}
	In fact as we have argued before, any tensor constructed out of a projective invariant quantity has this property. Let us prove this here. Consider the scalar quantity
	$\Psi$ that is invariant under projective transformations. Then define
	\begin{equation}
	\Xi_{\lambda}^{\;\;\;\mu\nu}\equiv \frac{\delta \Psi}{\delta \Gamma^{\lambda}_{\;\;\;\mu\nu}}
	\end{equation}
	Now consider the projective transformation
	\begin{equation}
	\Gamma^{\lambda}_{\;\;\mu\nu}\longrightarrow \hat{\Gamma}^{\lambda}_{\;\;\mu\nu} =\Gamma^{\lambda}_{\;\;\mu\nu}+ \delta_{\mu}^{\lambda}\xi_{\nu} 
	\end{equation}
	such that\footnote{{$\delta_{p}$} denotes a projective variation of the connection.}
	\begin{equation}
	\delta_{p} \Gamma^{\lambda}_{\;\;\;\mu\nu}=\hat{\Gamma}^{\lambda}_{\;\;\mu\nu} -\Gamma^{\lambda}_{\;\;\mu\nu}= \delta_{\mu}^{\lambda}\xi_{\nu}
	\end{equation}
	Applying the latter transformation to $\Psi$, we have
	\begin{equation}
	\delta_{p}\Psi =\frac{\delta \Psi}{\delta_{p} \Gamma^{\lambda}_{\;\;\;\mu\nu}} \delta_{p}\Gamma^{\lambda}_{\;\;\;\mu\nu}=\Xi_{\lambda}^{\;\;\;\mu\nu}\delta_{\mu}^{\lambda}\xi_{\nu}=\Xi_{\mu}^{\;\;\;\mu\nu}\xi_{\nu}
	\end{equation}
	Now since $\Psi$ is invariant, we have that $\delta_{p}\Psi=0$. Thus, using this, along with the fact that the vector $\xi_{\nu}$ is arbitrary, from the above we conclude that 
	\begin{equation}
	\Xi_{\mu}^{\;\;\;\mu\nu}=0
	\end{equation}
	as we stated.

	\subsubsection{Conformal Invariance and Tracelessness}
	As we have proved earlier, if a scalar quantity is invariant under projective transformations then its variation with respect to the connection yields a tensor (or tensor density if we do not divide the result by $\sqrt{-g}$) that is traceless in its first two indices. Similarly, if a scalar density (which we may integrate to construct an action of course) is invariant under conformal transformations then its variation with respect to the metric tensor yields a tensor that is traceless. Let us prove this here.\newline
	\textbf{Proof:} Consider the scalar density
	\beq
	\sqrt{-g} \Psi \label{psig}
	\eeq 
	where $\Psi$ is a scalar. Then define the variation
	\beq
	M_{\mu\nu} \equiv \frac{1}{\sqrt{-g}}\frac{\delta (\sqrt{-g} \Psi)}{\delta g^{\mu\nu}}
	\eeq
	and denote its trace by $M\equiv M_{\mu\nu}g^{\mu\nu}$.
	Consider now a conformal transformation of the metric
	\beq
	\bar{g}_{\mu\nu}=e^{2 \phi}g_{\mu\nu}
	\eeq
	or in its contravariant form
	\beq
	\bar{g}^{\mu\nu}=e^{-2 \phi}g^{\mu\nu}
	\eeq
	Expanding the latter for infinitesimal transformations, it follows that
	\beq
	\bar{g}^{\mu\nu} \approx (1-2\phi)  g^{\mu\nu} \Rightarrow \delta_{c}g^{\mu\nu}=-2 \phi g^{\mu\nu}
	\eeq
	where $\delta_{c}g^{\mu\nu}\equiv \bar{g}^{\mu\nu}-g^{\mu\nu} $ denotes the infinitesimal change the metric undergoes under the conformal transformation. Given that ($\ref{psig}$) is invariant under conformal transformations we have 
	\begin{gather}
	\delta_{c}(\sqrt{-g} \Psi)=0\Rightarrow M_{\mu\nu}\delta_{c}g^{\mu\nu}=0\Rightarrow \nonumber
	-2\phi M_{\mu\nu}g^{\mu\nu}=0
	\end{gather}
	and since the last one must hold true for arbitrary $\phi$ we conclude that
	\beq
	M=M_{\mu\nu}g^{\mu\nu}=0
	\eeq
	as stated.\newline
	\textbf{Examples:} Let us confirm the strength of the above statement with two examples. First consider the scalar density (in $4$dimensions)\footnote{This of course generalizes to any dimension and takes the form $\sqrt{-g}R^{\frac{n}{2}}$ where $n$ is the dimension of the space.}
	\beq
	\sqrt{-g}R^{2}
	\eeq
	which is conformally invariant as can be easily seen. Its metric variation is found to be
	\beq
	M_{\mu\nu}=\frac{1}{\sqrt{-g}}\frac{\delta (\sqrt{-g}R^{2})
	}{\delta g^{\mu\nu}}=2 R\left( R_{\mu\nu}-\frac{1}{4}g_{\mu\nu}R\right)
	\eeq
	and therefore
	\beq
	M=M_{\mu\nu}g^{\mu\nu}=2R\left( R-R\right)=0
	\eeq
	as expected. As a second example consider
	\beq
	\sqrt{-g}R_{\mu\nu}R^{\mu\nu}
	\eeq
	which is also a conformally invariant quantity in $4-dim$. Variation with respect to the metric yields 
	\begin{gather}
	M_{\mu\nu}=\frac{1}{\sqrt{-g}}\frac{\delta (\sqrt{-g}R_{\mu\nu}R^{\mu\nu})
	}{\delta g^{\mu\nu}}= \nonumber \\
	=-\frac{1}{2}R_{\alpha\beta}R^{\alpha\beta}g_{\mu\nu}+R_{\mu}^{\;\;\beta}R_{\nu\beta}+R^{\beta}_{\;\;\nu}R_{\beta\mu}
	\end{gather}
	which again gives a vanishing trace since
	\beq
	M=M_{\mu\nu}g^{\mu\nu}=-2R_{\mu\nu}R^{\mu\nu}+R_{\mu\nu}R^{\mu\nu}+R_{\mu\nu}R^{\mu\nu}=0
	\eeq

	\subsubsection{Invariance under frame rescalings}
	As we have seen a frame rescaling results in a conformal transformation $+$ a special projective transformation both powered by a single scalar field $\phi(x)$.\footnote{This is most important because one can also have projective and conformal transformations that are powered by different fields. Then invariance means that both metric and connection conjugates have zero traces and they are not related. As an example consider $\sqrt{-g}R^{2}$ which is independently invariant under $ \Gamma^{\lambda}_{\;\;\mu\nu} \rightarrow\Gamma^{\lambda}_{\;\;\mu\nu}+ \delta_{\mu}^{\lambda}\xi_{\nu} $  and $g_{\mu\nu} \rightarrow e^{2 \phi}g_{\mu\nu}$ (where $\xi_{\nu}$ and $\phi$ are not related to any way) and as a result $M=0$ and $\Xi^{\mu}=0$.} We will now prove that if a scalar density is invariant under frame rescalings then, the trace of its metric conjugate and the divergence of the trace of its connection conjugate are related to one another.\newline
	\textbf{Proof:} Consider the action
	\beq
	S=\int d^{n}x \sqrt{-g}\mathcal{L} \label{Theor}
	\eeq
	and recall the definitions of the metric and connection conjugates
	\beq
	M_{\mu\nu} \equiv \frac{1}{\sqrt{-g}}\frac{\delta (\sqrt{-g} \mathcal{L})}{\delta g^{\mu\nu}}
	\eeq
	\begin{equation}
	\Xi_{\lambda}^{\;\;\;\mu\nu}\equiv \frac{1}{\sqrt{-g}}\frac{\delta (\sqrt{-g} \mathcal{L})}{\delta \Gamma^{\lambda}_{\;\;\;\mu\nu}}=\frac{\delta \mathcal{L}}{\delta \Gamma^{\lambda}_{\;\;\;\mu\nu}}
	\end{equation}
	and define the traces
	\beq
	M=g^{\mu\nu}M_{\mu\nu}\;,\;\;\; \Xi^{\mu} =\Xi_{\lambda}^{\;\;\;\lambda\mu}
	\eeq
	We now state that if ($\ref{Theor}$) is invariant under frame rescalings then
	\beq
	2M+\frac{\partial_{\mu}(\sqrt{-g}\Xi^{\mu})}{\sqrt{-g}}=0
	\eeq
	To prove this let us compute the change in ($\ref{Theor}$) under frame rescalings. Using
	\beq
	\delta_{fr}g^{\mu\nu}=\tilde{g}^{\mu\nu}-g^{\mu\nu}=-2\phi g^{\mu\nu}
	\eeq
	\beq
	\delta_{fr}\Gamma^{\lambda}_{\;\;\;\;\mu\nu}=\tilde{\Gamma}^{\lambda}_{\;\;\;\;\mu\nu}-\Gamma^{\lambda}_{\;\;\;\;\mu\nu}=\delta_{\mu}^{\lambda}\partial_{\nu}\phi
	\eeq
	where $\delta_{fr}$ denotes the change under frame rescalings, we compute
	\begin{gather}
	\delta_{fr}S=\int d^{n}x \Big[ \sqrt{-g}M_{\mu\nu}\delta_{fr}g^{\mu\nu}+\sqrt{-g}\Xi_{\lambda}^{\;\;\;\mu\nu}\delta_{fr}\Gamma^{\lambda}_{\;\;\;\;\mu\nu}\Big]=\nonumber \\
	=\int d^{n}x \Big[ \sqrt{-g}(-2\phi g^{\mu\nu}M_{\mu\nu})+\sqrt{-g}\Xi_{\lambda}^{\;\;\;\mu\nu}\delta_{\mu}^{\lambda}\partial_{\nu}\phi \Big]=\nonumber \\
	=\int d^{n}x \Big[ -\sqrt{-g}2 \phi M +\sqrt{-g}\Xi^{\mu}\partial_{\mu}\phi \Big]= \nonumber \\
	=-\int d^{n}x \Big[ \sqrt{-g}2 \phi M +\phi(\partial_{\mu}\sqrt{-g}\Xi^{\mu}) \Big]+\int d^{n}x \partial_{\mu}( \sqrt{-g}\Xi^{\mu}\phi)= \nonumber \\
	=-\int d^{n}x  \phi \Big[ \sqrt{-g}2 M +(\partial_{\mu}\sqrt{-g}\Xi^{\mu}) \Big]+s.t.
	\end{gather}
	Then, ignoring surface terms, since $S$ is invariant it follows that
	\beq
	\delta_{fr}S\Rightarrow \phi \Big[ \sqrt{-g}2 M +(\partial_{\mu}\sqrt{-g}\Xi^{\mu}) \Big]=0
	\eeq
	and since the last must be true for arbitrary $\phi$ we conclude that
	\beq
	2M+\frac{\partial_{\mu}(\sqrt{-g}\Xi^{\mu})}{\sqrt{-g}}=0
	\eeq
	as stated. \newline
	\textbf{Example:}As an example consider the scalar density
	\beq
	\sqrt{-g}A^{2}=\sqrt{-g}(Q_{\mu}Q^{\mu})^{2}
	\eeq
	which is invariant under frame rescalings in $4-dim$ as can be easily seen. Its metric conjugate reads (where we have dropped a total derivative that is assumed to vanish on the boundary)
	\beq
	M_{\mu\nu}=-\frac{1}{2}g_{\mu\nu}A^{2}+2A Q_{\mu}Q_{\nu}-4 g_{\mu\nu}\frac{\partial_{\alpha}{(\sqrt{-g}Q^{\alpha}A)}}{\sqrt{-g}}
	\eeq
	with trace
	\beq
	M=-16\frac{\partial_{\alpha}{(\sqrt{-g}Q^{\alpha}A)}}{\sqrt{-g}}
	\eeq
	The associated connection conjugate is found to be
	\beq
	\Xi_{\lambda}^{\;\;\;\mu\nu}=8AQ^{\nu}\delta^{\mu}_{\lambda}
	\eeq
	with trace (in the first two indices)
	\beq
	\Xi^{\nu}=32AQ^{\nu} 
	\eeq
	So we observe that
	\begin{gather}
	\frac{\partial_{\mu}(\sqrt{-g}\Xi^{\mu})}{\sqrt{-g}}=32\frac{\partial_{\alpha}{(\sqrt{-g}Q^{\alpha}A)}}{\sqrt{-g}}=-2M\Rightarrow \nonumber
	\end{gather}
	\beq
	2M+\frac{\partial_{\mu}(\sqrt{-g}\Xi^{\mu})}{\sqrt{-g}}=0
	\eeq
	as expected. Let us see now what do these invariances imply for the matter part of the action.
	
	\subsubsection{Identities for the Matter Sector}
	When the above invariances are also respected  by the matter sector of the theory, they impose certain identities for the energy momentum and hypermomentum tensors. In particular, from the above discussion we immediately conclude that, if the matter action is invariant under projective transformations it holds that
	\beq
	\Delta ^{\mu} \equiv \Delta_{\lambda}^{\;\;\;\lambda\mu}=0
	\eeq
	Namely projective invariance means that the hypermomentum tensor is traceless in its first two indices. A weaker condition on $\Delta_{\lambda}^{\;\;\;\mu\nu}$ is generated when the action is invariant only under special projective transformations but not general ones. Then, it holds that
	\beq
	\partial_{\mu}(\sqrt{-g}\Delta_{\lambda}^{\;\;\;\lambda\mu})=0
	\eeq
	for matter that is invariant under special projective transformations.

	On the other hand, if the matter action is conformally invariant, then the associated theory calls only for a traceless energy momentum tensor, viz.
	\beq
	T=T_{\mu\nu}g^{\mu\nu}=0
	\eeq
	Notice that if the theory is independently invariant under both of the above transformations, then it will satisfy both of the above identities. Lastly, for matter that is invariant under frame rescalings the traces of the energy momentum and hypermomentum tensors are related through
	\beq
	2 T +\frac{\partial_{\mu}(\sqrt{-g}\Delta_{\lambda}^{\;\;\;\lambda\mu})}{\sqrt{-g}} =0
	\eeq

	\subsubsection{Variations}
	Let us gather here the various variations that we will use in what follows. We start with torsion and compute variations with respect to the metric first. We have
	\beq
	T_{\mu\nu\lambda}(\delta_{g}S^{\mu\nu\lambda})=\delta g^{\mu\nu}\Big( T_{\mu\alpha\beta}S_{\nu}^{\;\;\;\alpha\beta}-T_{\alpha\nu\beta}S_{\mu}^{\;\;\;\alpha\beta}\Big)=\delta g^{\mu\nu}(2T_{[\nu\alpha]\beta}S_{\mu}^{\;\;\;\alpha\beta})
	\eeq
	and also
	\beq
	T^{\mu\nu\lambda}(\delta_{g}S_{\mu\nu\lambda})=-\delta g^{\mu\nu} \Big( T^{\alpha\beta}_{\;\;\;\;\nu}S_{\alpha\beta\mu} \Big)
	\eeq
	where $T_{\mu\nu\lambda}$ is an arbitrary tensor field (or tensor density). Then setting $T_{\mu\nu\lambda}=S_{\mu\nu\lambda}$ one has
	\beq
	S_{\mu\nu\lambda}(\delta_{g}S^{\mu\nu\lambda})=\delta g^{\mu\nu}(2S_{\nu\alpha\beta}S_{\mu}^{\;\;\;\alpha\beta})
	\eeq
	as well as
	\beq
	S^{\mu\nu\lambda}(\delta_{g}S_{\mu\nu\lambda})=-\delta g^{\mu\nu}(S_{\alpha\beta\mu}S^{\alpha\beta}_{\;\;\;\;\nu})
	\eeq
	such that
	\beq
	\delta_{g}(S_{\mu\nu\lambda}S^{\mu\nu\lambda})=\delta g^{\mu\nu}\Big( 2S_{\nu\alpha\beta}S_{\mu}^{\;\;\;\alpha\beta}-S_{\alpha\beta\mu}S^{\alpha\beta}_{\;\;\;\;\nu} \Big)
	\eeq
	In addition, setting $T_{\mu\nu\lambda}=S_{\lambda\mu\nu}$ we conclude that
	\beq
	\delta_{g}(S_{\mu\nu\lambda}S^{\lambda\mu\nu})=-S_{\nu\alpha\beta}S_{\mu}^{\;\;\;\alpha\beta}(\delta g^{\mu\nu})
	\eeq
	Now, using
	\beq
	\delta_{g}\epsilon_{\alpha\beta\gamma\delta}=\delta_{g}(\sqrt{-g}\eta_{\alpha\beta\gamma\delta})=-\frac{1}{2}\epsilon_{\alpha\beta\gamma\delta} g_{\mu\nu}\delta  g^{\mu\nu}
	\eeq
	we compute
	\beq
	A^{\alpha}\delta_{g}t_{\alpha}=\delta g^{\mu\nu} \left[ -\frac{1}{2}g_{\mu\nu}A_{\alpha}t^{\alpha}+2 A^{\lambda}\epsilon_{\lambda\nu\alpha\beta}S_{\mu}^{\;\;\;\alpha\beta} \right]
	\eeq
	where $A^{\mu}$ is an arbitrary vector. Then, also using that $\delta_{g}S_{\mu}=0$ we find
	\beq
	\delta_{g}(t_{\alpha}S^{\alpha})=\delta g^{\mu\nu} \left[ -\frac{1}{2}g_{\mu\nu}S_{\alpha}t^{\alpha}+2 S^{\lambda}\epsilon_{\lambda\nu\alpha\beta}S_{\mu}^{\;\;\;\alpha\beta} +t_{\mu}S_{\nu}\right]
	\eeq
	and also 
	\beq
	\delta_{g}(S_{\alpha}S^{\alpha})=\delta g^{\mu\nu}( S_{\mu}S_{\nu})
	\eeq
	Following the same procedure for the rest of the quadratic torsion scalars, we finally derive the $g-$variations
	\begin{gather}
	\delta_{g}B_{1}=\delta g^{\mu\nu}\Big( 2S_{\nu\alpha\beta}S_{\mu}^{\;\;\;\alpha\beta}-S_{\alpha\beta\mu}S^{\alpha\beta}_{\;\;\;\;\nu} \Big) \nonumber \\
	\delta_{g}B_{2}=\delta g^{\mu\nu}(-S_{\nu\alpha\beta}S_{\mu}^{\;\;\;\alpha\beta}) \nonumber \\
	\delta_{g}B_{3}=\delta g^{\mu\nu}( S_{\mu}S_{\nu}) \nonumber \\
		\delta_{g}B_{4}=\delta g^{\mu\nu} \left[ t_{\mu} t_{\nu}-g_{\mu\nu}t_{\alpha}t^{\alpha}+4 t^{\lambda}\epsilon_{\lambda\nu\alpha\beta}S_{\mu}^{\;\;\;\alpha\beta} \right] \nonumber \\
	\delta_{g}B_{5}=\delta g^{\mu\nu} \left[ -\frac{1}{2}g_{\mu\nu}S_{\alpha}t^{\alpha}+2 S^{\lambda}\epsilon_{\lambda\nu\alpha\beta}S_{\mu}^{\;\;\;\alpha\beta} +t_{\mu}S_{\nu}\right] \nonumber \\
	\delta_{g}B_{6}=\delta g^{\mu\nu}\left( \frac{1}{2}g_{\mu\nu}B_{6}-\epsilon^{\alpha\beta\gamma\delta}S_{\alpha\beta\mu}S_{\gamma\delta\nu} \right)\nonumber \\
	\delta_{g}B_{7}=\delta g^{\mu\nu}\left(\frac{1}{2}g_{\mu\nu}B_{7}+2 S^{\alpha\beta}_{\;\;\;\;\mu}S_{\alpha}^{\;\;\;\gamma\delta}\epsilon_{\nu\beta\gamma\delta}+\epsilon^{\alpha\beta\gamma\delta}S_{\mu\alpha\beta}S_{\nu\gamma\delta} \right) \nonumber \\
	\delta_{g}B_{8}=\delta g^{\mu\nu}\left( \frac{1}{2}g_{\mu\nu}B_{8}-\epsilon^{\beta}_{\;\;\nu\gamma\delta}S^{\gamma\delta\alpha}S_{\mu\alpha\beta} \right)
	\end{gather}

	\subsection{$\Gamma$-Variations}

	For the $\Gamma$-variations of non-metricity scalars we find
	\begin{gather}
	\delta_{\Gamma}A_{1}=\delta_{\Gamma}(Q_{\alpha\mu\nu}Q^{\alpha\mu\nu})=(4 Q^{\nu\mu}_{\;\;\;\;\lambda} ) \delta \Gamma^{\lambda}_{\;\;\;\mu\nu} \nonumber \\
	\delta_{\Gamma}A_{2}=\delta_{\Gamma}(Q_{\alpha\mu\nu}Q^{\mu\nu\alpha})=2 (Q^{\mu\nu}_{\;\;\;\;\lambda} +Q_{\lambda}^{\;\;\mu\nu})\delta \Gamma^{\lambda}_{\;\;\;\mu\nu} \nonumber \\
	\delta_{\Gamma}A_{3}=\delta_{\Gamma}(Q_{\mu}Q^{\mu})=(4 Q^{\nu} \delta^{\mu}_{\;\;\;\;\lambda} ) \delta \Gamma^{\lambda}_{\;\;\;\mu\nu} \nonumber \\
	\delta_{\Gamma}A_{4}=\delta_{\Gamma}(\tilde{Q}_{\mu} \tilde{Q}^{\mu})=2(\tilde{Q}_{\lambda}g^{\mu\nu}+\tilde{Q}^{\mu}\delta_{\lambda}^{\nu}) \delta \Gamma^{\lambda}_{\;\;\;\mu\nu} \nonumber \\
	\delta_{\Gamma}A_{5}=\delta_{\Gamma}( Q_{\mu} \tilde{Q}^{\mu})=( 2 \tilde{Q}^{\nu}\delta_{\lambda}^{\mu}+ Q_{\lambda}g^{\mu\nu}+Q^{\mu}\delta^{\nu}_{\lambda}) \delta \Gamma^{\lambda}_{\;\;\;\mu\nu} \nonumber \\
	\end{gather}
	and for the pure torsion and mixed scalars
	\begin{gather}
\delta_{\Gamma}B_{1}=\delta_{\Gamma}(S_{\mu\nu\lambda}S^{\mu\nu\lambda})=2 S^{\mu\nu}_{\;\;\;\;\lambda}\delta \Gamma^{\lambda}_{\;\;\;\mu\nu} \nonumber \\
\delta_{\Gamma}B_{2}=\delta_{\Gamma}(S_{\mu\nu\lambda}S^{\lambda\mu\nu})=2 S_{\lambda}^{\;\;\;[\mu\nu]} \delta \Gamma^{\lambda}_{\;\;\;\mu\nu} \nonumber \\
\delta_{\Gamma}B_{3}=\delta_{\Gamma}(S_{\mu} S^{\mu})=2 S^{[\mu}\delta^{\nu]}_{\lambda}\delta \Gamma^{\lambda}_{\;\;\;\mu\nu} \nonumber \\
\delta_{\Gamma}C_{1}=\delta_{\Gamma}(Q_{\alpha\mu\nu}S^{\alpha\mu\nu}) =(S^{\nu\mu}_{\;\;\;\;\lambda}-S_{\lambda}^{\;\;\nu\mu}+Q^{[\mu\nu]}_{\;\;\;\;\;\lambda})\delta \Gamma^{\lambda}_{\;\;\;\mu\nu} \nonumber \\
\delta_{\Gamma}C_{2}=\delta_{\Gamma}(Q_{\mu}S^{\mu}) =(2 S^{\nu}\delta_{\lambda}^{\mu}+Q^{[\mu}\delta^{\nu]}_{\lambda})\delta \Gamma^{\lambda}_{\;\;\;\mu\nu} \nonumber \\
\delta_{\Gamma}C_{3}=\delta_{\Gamma}(q_{\mu}S^{\mu}) =(S_{\lambda}g^{\mu\nu}+S^{\mu}\delta^{\nu}_{\lambda}+q^{[\mu}\delta^{\nu]}_{\lambda})\delta \Gamma^{\lambda}_{\;\;\;\mu\nu} \nonumber
	\end{gather}

	\subsection{ A Simple Conformally Invariant Theory}
	As a warm up, let us study now a conformally invariant theory by coupling the Ricci scalar, to a scalar field $\psi$, in the metric-affine framework\footnote{Conformally invariant theories in the context of teleparallel gravity have been studied in \cite{maluf2012conformally}. Notice however the difference between the transformation law for torsion tensor in their formalism compared to ours.}. The nice thing now is that one does not need the existence of an additional gauge field $A_{\mu}$ in order to define the gauge covariant derivative on $\psi$ since torsion and non-metricity offer enough room to accommodate it into them. To be more specific, consider the action
	\beq
	S=\frac{1}{2\kappa}\int d^{n}x \Big[\sqrt{-g}\psi^{2}R+\lambda \sqrt{-g} g^{\mu\nu}D_{\mu}\psi D_{\nu}\psi \Big] \label{confthe}
	\eeq
	where $\lambda$ is a parameter and $D_{\mu}\psi$ the gauge covariant derivative on the field, to be defined in a moment.
	Notice now that the first term in the above action is invariant under conformal transformations of the metric
	\beq
	g_{\mu\nu} \rightarrow \bar{g}_{\mu\nu}=e^{2 \theta}g_{\mu\nu}
	\eeq
	provided that we simultaneously transform the scalar field as
	\beq
	\psi \rightarrow \bar{\psi}=e^{\frac{(2-n)}{2}\theta}\psi
	\eeq
	In order to keep this invariance on the kinetic term too, one needs to replace the partial derivative $\partial_{\mu}$ with a covariant one $D_{\mu}=\partial_{\mu}+A_{\mu}$ and  also impose a gauge transformation on the field $A_{\mu}$ ($A_{\mu}\rightarrow A_{\mu}+\partial_{\mu}\chi$) so as to have the transformation
	\beq
	\bar{D}_{\mu}\bar{\psi}=e^{\frac{(2-n)}{2}\theta}D_{\mu}\psi
	\eeq
	and subsequently
	\beq
	\sqrt{-\bar{g}}\bar{g}^{\mu\nu}\bar{D}_{\mu}\bar{\psi}\bar{D}_{\nu}\bar{\psi}=g^{\mu\nu}D_{\mu}\psi D_{\nu}\psi \label{Deq}
	\eeq
	which will ensure the conformal invariance of the total action. Now, what's interesting is that we do not have to add this gauge field $A_{\mu}$ by hand, we have a generalized geometry offering torsion and non-metricity vectors that can do the job. Notice now that since the torsion vector $S_{\mu}$ does not change under conformal transformations, it cannot be regarded as our desired gauge field. The non-metricity (Weyl) vector however, transforms as
	\beq
	\bar{Q}_{\mu}=Q_{\mu}-2 n \partial_{\mu} \theta
	\eeq
	under a conformal transformation. Therefore, defining the covariant derivative on the scalar field as
	\beq
	D_{\mu}\equiv \partial_{\mu}+\frac{2-n}{4n}Q_{\mu}
	\eeq
	ensures that (\ref{Deq}) is satisfied. So, building the action this way, let us derive the field equations of ($\ref{confthe}$). Variation with respect to the metric tensor yields
	\begin{gather}
	-\frac{1}{2}g_{\mu\nu}\Big( \psi^{2}R +\lambda (D\psi)^{2}\Big)+\psi^{2}R_{(\mu\nu)}+\lambda D_{\mu}\psi D_{\nu}\psi  \nonumber \\ 
	+ \lambda \frac{(n-2)}{2 n}g_{\mu\nu} \frac{\partial_{\alpha}(\sqrt{-g}\psi D^{\alpha} \psi)}{\sqrt{-g}}=0
	\end{gather}
	where we have abbreviated $(D\psi)^{2}=g^{\mu\nu}D_{\mu}\psi D_{\nu}\psi$. Now, since our initial action is conformally invariant one would expect that the trace of the above equation identically vanishes. In fact, the trace of the above equation gives the same equation that one gets when varying with respect to the scalar field $\psi$. Therefore, when the equation of motion for $\psi$ is on shell, the above trace vanishes identically. To see this first note that the trace of the above field equations is
	\beq
	\psi^{2}R +\lambda (D\psi)^{2}-\lambda \frac{\partial_{\alpha}(\sqrt{-g}\psi D^{\alpha} \psi)}{\sqrt{-g}} =0 \label{Rpsi}
	\eeq
	On the other hand, varying the action with respect to $\psi$, we obtain
	\beq
	R \psi -\lambda \frac{\partial_{\alpha}(\sqrt{-g} D^{\alpha} \psi)}{\sqrt{-g}}-\lambda \frac{(n-2)}{4 n}Q^{\mu}(D_{\mu}\psi)=0
	\eeq
	Multiplying this by $\psi$ (given that $\psi \neq 0$) and doing a partial integration it follows that
	\beq
	R \psi^{2}-\lambda \frac{\partial_{\alpha}(\sqrt{-g}\psi D^{\alpha} \psi)}{\sqrt{-g}}+\lambda \Big(\partial_{\mu}+\frac{2-n}{4n}Q_{\mu}\Big)D^{\mu}\psi =0
	\eeq
	or equivalently 
	\beq
	\psi^{2}R +\lambda (D\psi)^{2}-\lambda \frac{\partial_{\alpha}(\sqrt{-g}\psi D^{\alpha} \psi)}{\sqrt{-g}} =0
	\eeq 
	which is indeed the same equation with ($\ref{Rpsi}$). Lastly, variation of the action with respect to the connection yields
	\beq
	P_{\lambda}^{\;\;\;\mu\nu}(h)+\lambda\frac{(2-n)}{n}\delta_{\lambda}^{\mu}(D^{\nu}\psi)=0 \label{Palahx}
	\eeq
	where
	\begin{gather}
	P_{\lambda}^{\;\;\;\mu\nu}(h) \equiv -\frac{\nabla_{\lambda}(\sqrt{-g}\psi^{2}g^{\mu\nu})}{\sqrt{-g}}+\frac{\nabla_{\alpha}(\sqrt{-g}\psi^{2}g^{\mu\alpha}\delta_{\lambda}^{\nu})}{\sqrt{-g}}+ \\ \nonumber
	2 \psi^{2}(S_{\lambda}g^{\mu\nu}-S^{\mu}\delta_{\lambda}^{\nu}-  S_{\lambda}^{\;\;\;\mu\nu}) 
	\end{gather}
	is the Palatini tensor computed with respect to the metric $h_{\mu\nu}=\psi^{2} g_{\mu\nu}$. This tensor can also be written as
	\beq
	P_{\lambda}^{\;\;\;\mu\nu}(h)=\psi^{2}P_{\lambda}^{\;\;\;\mu\nu}(g)+\delta_{\lambda}^{\nu}g^{\mu\alpha}\partial_{\alpha}\psi^{2}-g^{\mu\nu}\partial_{\lambda}\psi^{2}
	\eeq
	where $P_{\lambda}^{\;\;\;\mu\nu}(g)$ is the usual Palatini tensor computed with respect to the metric tensor $g_{\mu\nu}$. Looking back at ($\ref{Palahx}$), contracting in $\mu=\lambda$ and using the fact that the Palatini tensor is traceless in its first two indices,\footnote{Note that both $P_{\mu}^{\;\;\;\mu\nu}(g)=0$ and $P_{\mu}^{\;\;\;\mu\nu}(h)=0$, that is any Palatini tensor that is built from a metric conformally related to $g_{\mu\nu}$ is also traceless in its first two indices. }it follows that
	\beq
	D^{\nu}\psi=0 \label{psieq}
	\eeq
	which when substituted back at ($\ref{Palahx}$) gives
	\begin{gather}
	P_{\lambda}^{\;\;\;\mu\nu}(h) =0 \Rightarrow \nonumber \\
	\psi^{2}P_{\lambda}^{\;\;\;\mu\nu}(g)=-\delta_{\lambda}^{\nu}g^{\mu\alpha}\partial_{\alpha}\psi^{2}+g^{\mu\nu}\partial_{\lambda}\psi^{2}
	\end{gather}
	with this at hand we can use the connection decomposition (Theorem-$1$, Chapter $4$) and easily find the affine connection
	\beq
	\Gamma^{\lambda}_{\;\;\;\;\mu\nu}=\tilde{\Gamma}_{\;\;\;\;\mu\nu}+\frac{2}{n-2}g_{\mu\nu}\frac{\partial^{\lambda}\psi}{\psi}-\frac{2}{n-2}\delta^{\lambda}_{\nu}\frac{\partial_{\mu}\psi}{\psi}+\frac{1}{2}\delta^{\lambda}_{\mu}\tilde{Q}_{\nu}
	\eeq
	Before finding the expressions for torsion and non-metricity that follow from the above, let us expand ($\ref{psieq}$) to get
	\beq
	\partial_{\mu}\psi -\frac{(n-2)}{4n}Q_{\mu} \psi =0
	\eeq
	from which we conclude that
	\beq
	Q_{\mu}=\frac{4 n}{n-2}\frac{\partial_{\mu}\psi}{\psi}
	\eeq
	that is, the Weyl vector is exact and powered by the scalar field-$\psi$. Now, using the above connection decomposition and the fact that
	\beq
	S_{\mu\nu}^{\;\;\;\;\lambda}=N^{\lambda}_{\;\;\;\;[\mu\nu]}
	\eeq
	and
	\beq
	Q_{\alpha\mu\nu}=2 N_{(\alpha\mu)\nu}
	\eeq
	where $N^{\lambda}_{\;\;\;\;\mu\nu} \equiv \Gamma^{\lambda}_{\;\;\;\;\mu\nu}-\tilde{\Gamma}_{\;\;\;\;\mu\nu} $, it follows that
	\beq
	S_{\mu\nu}^{\;\;\;\;\lambda}=-2\frac{\partial_{[\mu}\psi \delta^{\lambda}_{\nu]}}{\psi}+\frac{1}{2}\delta^{\lambda}_{\mu} \tilde{Q}_{\nu]}
	\eeq
	and 
	\beq
	Q_{\alpha\mu\nu}=\tilde{Q}_{\alpha}g_{\mu\nu}
	\eeq
	Contracting the last equation with $g^{\mu\nu}$ we conclude that $Q_{\alpha}=n\tilde{Q}_{\alpha}$
	\beq
	Q_{\alpha\mu\nu}=\frac{1}{n} Q_{\alpha}g_{\mu\nu}
	\eeq
	also recalling that $Q_{\mu}=\frac{4 n}{n-2}\frac{\partial_{\mu}\psi}{\psi}$ we have
	\beq
	Q_{\alpha\mu\nu}=\frac{4}{n-2}g_{\mu\nu}\frac{\partial_{\mu}\psi}{\psi}
	\eeq
	which is the case of a Weyl integrable non-metricity. Also, using the above, the torsion tensor may be expressed as
	\beq
	S_{\mu\nu}^{\;\;\;\;\lambda}=\frac{4}{n-2}\frac{\delta_{[\mu}^{\lambda}\partial_{\nu]}\psi}{\psi}
	\eeq
	with torsion vector
	\beq
	S_{\mu}=-\frac{2(n-1)}{(n-2)}\frac{\partial_{\mu}\psi}{\psi}
	\eeq
	and the above s a case of vectorial torsion with an exact torsion vector. Finally, using the above results, the field equations for the scalar field and the metric imply
	\beq
	R=0 
	\eeq
	and 
	\beq
	R_{\mu\nu}=0
	\eeq
	Note however that these do not mean that we have Einstein's Gravity in vacuum since the curvature has more degrees of freedom coming from torsion and non-metricity. So, in this simple conformally invariant model we have a Weyl non-metricity and vectorial torsion both sourced by the scalar field $\psi$.

	\subsection{Generalized Quadratic Theory}
	The most general Theory of Gravity that is quadratic in torsion and non-metricity is given by
	\beq
	S=\frac{1}{2 \kappa}\int d^{n}x \sqrt{-g} \Big[ \mathcal{L}_{Q}+ \mathcal{L}_{T}+ \mathcal{L}_{QT} \Big] +S_{Matter}
	\eeq
	where
	\beq
	\mathcal{L}_{Q}=\sum_{i=1}^{6} a_{i}A_{i}
	\eeq
	\beq
	\mathcal{L}_{T}=\sum_{i=1}^{8} b_{i}B_{i}
	\eeq
	\beq
	\mathcal{L}_{QT}=\sum_{i=1}^{7} c_{i}C_{i}
	\eeq
	and $a_{i},b_{i},c_{i}$ are constant parameters. Furthermore, demanding a parity preserving theory, one is left with
	\begin{gather}
	S=\frac{1}{2 \kappa}\int d^{n}x \sqrt{-g} \Big[a_{4}A_{4}+a_{5}A_{5} +\sum_{i=1}^{3} (a_{i}A_{i}+b_{i}B_{i}+c_{i}C_{i})  \Big] +S_{Matter}= \nonumber \\
	=\frac{1}{2 \kappa}\int d^{n}x \sqrt{-g} \Big[   
	b_{1}S_{\alpha\mu\nu}S^{\alpha\mu\nu} +
	b_{2}S_{\alpha\mu\nu}S^{\mu\nu\alpha} +
	b_{3}S_{\mu}S^{\mu} \nonumber \\
	a_{1}Q_{\alpha\mu\nu}Q^{\alpha\mu\nu} +
	a_{2}Q_{\alpha\mu\nu}Q^{\mu\nu\alpha} +
	a_{3}Q_{\mu}Q^{\mu}+
	a_{4}q_{\mu}q^{\mu}+
	a_{5}Q_{\mu}q^{\mu} \nonumber \\
	+c_{1}Q_{\alpha\mu\nu}S^{\alpha\mu\nu}+
	c_{2}Q_{\mu}S^{\mu} +
	c_{3}q_{\mu}S^{\mu} \Big] +S_{Matter} \label{genact}
	\end{gather}
	Notice now that for the parameter choice $b_{1}=1$, $b_{2}=-2$, $b_{3}=-4$, $a_{i}=0=c_{i}$ and imposing a vanishing curvatute and non-metricity, one recovers the teleparallel equivalent of GR.  In addition, demanding vanishing curvature and torsion and taking $a_{1}=-a_{3}=1/4$, $a_{2}=-a_{5}=-1/2$, $a_{4}=0$, $b_{i}=0=c_{i}$ one obtains the symmetric teleparallel equivalent of GR. Furthermore if we pick $b_{1}=1$, $b_{2}=-1$, $b_{3}=-4$ , $a_{1}=-a_{3}=1/4$, $a_{2}=-a_{5}=-1/2$, $a_{4}=0$,  $c_{1}=-c_{2}=c_{3}=2$ and impose only vanishing curvature we reproduce a generalized equivalent to GR that admits both torsion and non-metricity. Now, in order to obtain a conformally invariant theory we should first restrict the above parameters and find a specific combination for which the total Lagrangian density transforms conformally, namely it only picks up a factor $e^{- 2 \theta}$. To do so, we use the transformation laws for the quadratic scalars that we obtained earlier. Then under a conformal transformation, we have
	\beq
	\bar{\mathcal{L}}_{T}=e^{-2\theta}\mathcal{L}_{T}
	\eeq
	\begin{gather}
	\bar{\mathcal{L}}_{Q}=e^{-2\theta}\mathcal{L}_{Q}-e^{-2\theta}Q^{\mu}\partial_{\mu}\theta (4a_{1}+4 n a_{3}+2 a_{5}) \nonumber \\
	-e^{-2\theta}q^{\mu}\partial_{\mu}\theta ( 4 a_{2}+4 a_{4}+2 n a_{5}) \nonumber \\
	+e^{-2\theta}(\partial \theta)^{2}4( n a_{1}+a_{2}+ n^{2} a_{3}+a_{4}+n a_{5})
	\end{gather}
	\begin{gather}
	\bar{\mathcal{L}}_{QT}=e^{-2\theta}\mathcal{L}_{QT}-e^{-2\theta}2 S^{\mu}\partial_{\mu}\theta (  c_{1} +n c_{2}+c_{3})
	\end{gather}
	From these we conclude that the parameter choice
	\begin{gather}
	4a_{1}+4 n a_{3}+2 a_{5}=0  \nonumber \\ 4 a_{2}+4 a_{4}+2 n a_{5}=0 \nonumber \\ n a_{1}+a_{2}+ n^{2} a_{3}+a_{4}+n a_{5}=0 \nonumber \\
	c_{1}+n c_{2}+c_{3}=0 \label{confinv}
	\end{gather}
	and whatever $b_{i}'s$ guarantee that
	\beq
	\bar{\mathcal{L}}_{Q}+\bar{\mathcal{L}}_{T}+\bar{\mathcal{L}}_{QT}=e^{-2\theta}\Big(\mathcal{L}_{Q}+ \mathcal{L}_{T}+\mathcal{L}_{QT} \Big)
	\eeq
	as we desired. The above parameter choice ensures that the total action is conformally invariant! Now, let us consider frame rescalings, then it can be easily seen that
	\beq
	\tilde{\mathcal{L}}_{Q}= e^{-2 \theta}\mathcal{L}_{Q}
	\eeq
	\begin{gather}
	\tilde{\mathcal{L}}_{T}= e^{-2 \theta}\mathcal{L}_{T}+e^{-2 \theta}S^{\mu}\partial_{\mu}\theta \Big( -2 b_{1}+b_{2}+(1-n)b_{3} \Big) \nonumber \\
	+\frac{(n-1)}{4} e^{-2 \theta}(\partial \theta)^{2}\Big( 2 b_{1} -b_{2}+(n-1)b_{3} \Big)
	\end{gather}
	\begin{gather}
	\tilde{\mathcal{L}}_{QT}= e^{-2 \theta}\mathcal{L}_{QT}+\frac{1}{2}e^{-2 \theta}Q^{\mu}\partial_{\mu}\theta \Big( -c_{1}+(1-n)c_{2} \Big) \nonumber \\
	+\frac{1}{2} e^{-2 \theta}q^{\mu}\partial_{\mu}\theta  \Big(c_{1}+(1-n)c_{3} \Big)
	\end{gather}
	Then, frame rescaling invariance 
	\beq
	\tilde{\mathcal{L}}_{Q}+\tilde{\mathcal{L}}_{T}+\tilde{\mathcal{L}}_{QT}=e^{-2\theta}\Big(\mathcal{L}_{Q}+ \mathcal{L}_{T}+\mathcal{L}_{QT} \Big)
	\eeq
	is ensured so long as
	\begin{gather}
	-2 b_{1}+b_{2}+(1-n)b_{3}=0   \nonumber \\
	- c_{1}+(1-n) c_{2}=0 \nonumber \\
	c_{1}+(1-n) c_{3}=0 \label{frinv}
	\end{gather}
	and whatever $a_{i}'s$. Now, let us see how our action changes under projective transformations of the connection 
	\begin{gather}
	\Gamma^{\lambda}_{\;\;\mu\nu}\longrightarrow \hat{\Gamma}^{\lambda}_{\;\;\mu\nu} =\Gamma^{\lambda}_{\;\;\mu\nu}+ \delta_{\mu}^{\lambda}\xi_{\nu}  \nonumber \\
	g_{\mu\nu}\longrightarrow \hat{g}_{\mu\nu}=g_{\mu\nu}
	\end{gather}
	We compute
	\begin{gather}
	\hat{\mathcal{L}}_{Q}= \mathcal{L}_{Q}+(4 a_{1}+4 n a_{3}+2 a_{5})Q_{\mu}\xi^{\mu}+(4 a_{2}+4 a_{4}+2 n a_{5})q_{\mu}\xi^{\mu} \nonumber \\
	+(4 n a_{1}+4 a_{2}+4 n^{2} a_{3}+4 a_{4}+4 n a_{5})\xi_{\mu}\xi^{\mu}
	\end{gather}
	\begin{gather}
	\hat{\mathcal{L}}_{T}= \mathcal{L}_{T}+\Big[-2 b_{1}+b_{2}+(1-n) b_{3}  \Big] S_{\mu}\xi^{\mu} \nonumber \\
	-\frac{(n-1)}{4}\Big[-2 b_{1}+b_{2}+(1-n) b_{3} \Big] \xi_{\mu}\xi^{\mu}
	\end{gather}
	\begin{gather}
	\hat{\mathcal{L}}_{QT}= \mathcal{L}_{QT}+\frac{1}{2}\Big[ -c_{1}+(1-n)c_{2} \Big]Q_{\mu}\xi^{\mu}+\frac{1}{2}\Big[ c_{1}+(1-n)c_{3} \Big]q_{\mu}\xi^{\mu} \nonumber \\
	+2 (c_{1}+n c_{2}+ c_{3})S_{\mu}\xi^{\mu}+(1-n)(c_{1}+n c_{2}+ c_{3})\xi_{\mu}\xi^{\mu}
	\end{gather}
	Therefore, the total action changes according to
	\begin{gather}
	\hat{\mathcal{L}}_{Q}+\hat{\mathcal{L}}_{T}+\hat{\mathcal{L}}_{QT}=\mathcal{L}_{Q}+\mathcal{L}_{T}+ \mathcal{L}_{QT} \nonumber \\
	+\left[ 2(2 a_{1}+2 n a_{3}+a_{5})+\frac{1}{2}\Big( -c_{1}+(1-n)c_{2} \Big) \right]Q_{\mu}\xi^{\mu} \nonumber \\
	+\left[ 2(2 a_{2}+2  a_{4}+n a_{5})+\frac{1}{2}\Big( c_{1}+(1-n)c_{3} \Big) \right]q_{\mu}\xi^{\mu} \nonumber \\
	+\Big[ -2 b_{1}+b_{2}+(1-n) b_{3}+2(c_{1} +n c_{2}+c_{3}) \Big]S_{\mu}\xi^{\mu} \nonumber \\
	\left[ 4(n a_{1}+ a_{2}+ n^{2} a_{3}+ a_{4}+ n a_{5}) +\frac{(n-1)}{4}\Big( 2 b_{1}-b_{2}+(n-1) b_{3}\Big) 
	-(n-1)(c_{1}+n c_{2}+c_{3})  \right]\xi_{\mu}\xi^{\mu} 
	\end{gather}
	Then, projective invariance is ensured if the parameters satisfy
	\beq
	4 (2 a_{1}+2 n a_{3}+a_{5})-c_{1}+(1-n)c_{2}=0
	\eeq
	\beq
	4 (2 a_{2}+2  a_{4}+n a_{5})+c_{1}+(1-n)c_{3}=0
	\eeq
	\beq
	-2 b_{1}+b_{2}+(1-n) b_{3}+2(c_{1} +n c_{2}+c_{3})=0
	\eeq
	\begin{gather}
	16 (n a_{1}+ a_{2}+ n^{2} a_{3}+ a_{4}+ n a_{5}) \nonumber \\
	+(n-1)\Big( 2 b_{1}-b_{2}+(n-1) b_{3}-4 (c_{1}+n c_{2}+c_{3})\Big) =0
	\end{gather}
	The important thing to note here is that the parameters $a_{i},b_{i},c_{i}$ mix when one demands projective invariance.  This means that $\mathcal{L}_{Q},\;\mathcal{L}_{T}$ and $ \mathcal{L}_{QT}$ are not independently projective invariant but their sum is. This was not the case when we considered conformal and frame rescaling transformations where the parameters did not mix and  $\mathcal{L}_{Q},\;\mathcal{L}_{T}$ and $ \mathcal{L}_{QT}$ where all independently invariant under the associated transformations.
	
	Having restricted the parameter space in each of the transformations we can now obtain an invariant theory by coupling the above to $\psi^{2}$. We first combine the case of conformal and frame rescaling transformations in a single action given by
	\begin{gather}
	S=\frac{1}{2 \kappa}\int d^{n}x \sqrt{-g} \psi^{2} \Big[a_{4}A_{4}+a_{5}A_{5} +\sum_{i=1}^{3} (a_{i}A_{i}+b_{i}B_{i}+c_{i}C_{i})  \Big] +S_{\psi}= \nonumber \\
	=\frac{1}{2 \kappa}\int d^{n}x \sqrt{-g} \Big[ \psi^{2}\Big(  
	b_{1}S_{\alpha\mu\nu}S^{\alpha\mu\nu} +
	b_{2}S_{\alpha\mu\nu}S^{\mu\nu\alpha} +
	b_{3}S_{\mu}S^{\mu} \nonumber \\
	a_{1}Q_{\alpha\mu\nu}Q^{\alpha\mu\nu} +
	a_{2}Q_{\alpha\mu\nu}Q^{\mu\nu\alpha} +
	a_{3}Q_{\mu}Q^{\mu}+
	a_{4}q_{\mu}q^{\mu}+
	a_{5}Q_{\mu}q^{\mu} \nonumber \\
	+c_{1}Q_{\alpha\mu\nu}S^{\alpha\mu\nu}+
	c_{2}Q_{\mu}S^{\mu} +
	c_{3}q_{\mu}S^{\mu}\Big) +\lambda g^{\mu\nu}D_{\mu}\psi D_{\nu}\psi\Big] =\nonumber \\
	=\int d^{n}x \sqrt{-g}\Big[ \psi^{2}\mathcal{L} +\lambda g^{\mu\nu}D_{\mu}\psi D_{\nu}\psi\Big] \label{genactionS}
	\end{gather}
	where again $\lambda$ is a parameter, $D_{\mu}$ is the gauge covariant derivative to be defined later, and $\mathcal{L}=\mathcal{L}_{Q}+\mathcal{L}_{T}+ \mathcal{L}_{QT}$ with
	\beq
	\mathcal{L}_{Q} =a_{1}Q_{\alpha\mu\nu}Q^{\alpha\mu\nu} +
	a_{2}Q_{\alpha\mu\nu}Q^{\mu\nu\alpha} +
	a_{3}Q_{\mu}Q^{\mu}+
	a_{4}q_{\mu}q^{\mu}+
	a_{5}Q_{\mu}q^{\mu}
	\eeq
	\beq
	\mathcal{L}_{T}=b_{1}S_{\alpha\mu\nu}S^{\alpha\mu\nu} +
	b_{2}S_{\alpha\mu\nu}S^{\mu\nu\alpha} +
	b_{3}S_{\mu}S^{\mu} 
	\eeq
	\beq
	\mathcal{L}_{QT}=c_{1}Q_{\alpha\mu\nu}S^{\alpha\mu\nu}+
	c_{2}Q_{\mu}S^{\mu} +
	c_{3}q_{\mu}S^{\mu}
	\eeq
	Now, it will be convenient for the calculations to define the 'superpotentials'
	\beq
	\Omega^{\alpha\mu\nu} \equiv a_{1}Q^{\alpha\mu\nu}+a_{2} Q^{\mu\nu\alpha}+a_{3} g^{\mu\nu}Q^{\alpha}+a_{4}g^{\alpha\mu}q^{\nu}+a_{5}g^{\alpha\mu}Q^{\nu}
	\eeq
	\beq
	\Sigma^{\alpha\mu\nu} \equiv b_{1}S^{\alpha\mu\nu}+b_{2}S^{\mu\nu\alpha}+b_{3}g^{\mu\nu}S^{\alpha}
	\eeq 
	\beq
	\Pi^{\alpha\mu\nu} \equiv c_{1}S^{\alpha\mu\nu}+c_{2}g^{\mu\nu}S^{\alpha}+c_{3}g^{\alpha\mu}S^{\nu}
	\eeq
	for non-metricity, torsion and their mixing, respectively. With these, the above are written as
	\beq
	\mathcal{L}_{Q}=Q_{\alpha\mu\nu}\Omega^{\alpha\mu\nu}
	\eeq
	\beq
	\mathcal{L}_{T}=S_{\alpha\mu\nu}\Sigma^{\alpha\mu\nu}
	\eeq
	\beq
	\mathcal{L}_{QT}=Q_{\alpha\mu\nu}\Pi^{\alpha\mu\nu}
	\eeq
	We are now in a position to derive the variations of the above. Let us first compute variations with respect to the metric. We have
	\begin{gather}
	\sqrt{-g}\psi^{2}\delta_{g}\mathcal{L}_{Q}=(\delta g^{\mu\nu}) \Big[ \sqrt{-g}\psi^{2} L_{(\mu\nu)}+ (2S_{\lambda}-\nabla_{\lambda})J^{\lambda}_{\;\;\;(\mu\nu)}+g_{\mu\nu}(2S_{\lambda}-\nabla_{\lambda})\zeta^{\lambda} \nonumber \\
	+\alpha_{4}(2S_{(\mu}-\nabla_{(\mu})( \sqrt{-g}\psi^{2}q_{\nu)}) \Big]
	\end{gather}
	where
	\begin{gather}
	L_{\mu\nu}=(a_{1}Q_{\mu\alpha\beta}+a_{2}Q_{\alpha\beta\mu})Q_{\nu}^{\;\;\;\alpha\beta}+(a_{3}Q_{\mu}+a_{5}q_{\mu})Q_{\nu}+a_{3}Q_{\alpha\mu\nu}Q^{\alpha} \nonumber \\
	+Q_{\mu\nu\alpha}(a_{4}q^{\alpha}+a_{5}Q^{\alpha})-\Omega^{\alpha\beta}_{\;\;\;\;\nu}Q_{\alpha\beta\mu}-\Omega_{\alpha\mu\beta}Q^{\alpha\beta}_{\;\;\;\;\nu}
	\end{gather}
	and we have also defined the tensor densities
	\beq
	J^{\lambda}_{\;\;\;(\mu\nu)} \equiv \sqrt{-g}\psi^{2}( \alpha_{1}Q^{\lambda}_{\;\;\;\mu\nu}+a_{2}Q_{\mu\nu}^{\;\;\;\;\lambda}+\Omega^{\lambda}_{\;\;\;\mu\nu})
	\eeq
	\beq
	\zeta^{\lambda}=\sqrt{-g}\psi^{2}(a_{3}Q^{\lambda}+a_{5}q^{\lambda})
	\eeq
	Continuing with the  pure torsion and mixed part, we obtain
	\beq
	\sqrt{-g}\psi^{2}\delta_{g}\mathcal{L}_{T}=(\delta g^{\mu\nu}) \sqrt{-g}\psi^{2}\Big[ b_{1}(2S_{\nu\alpha\beta}S_{\mu}^{\;\;\;\alpha\beta}-S_{\alpha\beta\mu}S^{\alpha\beta}_{\;\;\;\;\nu})-b_{2}S_{\nu\alpha\beta}S_{\mu}^{\;\;\;\alpha\beta}+b_{3}S_{\mu}S_{\nu} \Big]
	\eeq
	\begin{gather}
	\sqrt{-g}\psi^{2}\delta_{g}\mathcal{L}_{QT}=(\delta g^{\mu\nu}) \sqrt{-g}\psi^{2}\Big[\Pi_{\mu\alpha\beta}Q_{\nu}^{\;\;\;\alpha\beta}\nonumber \\
	-( c_{1}S_{\alpha\beta\nu}Q^{\alpha\beta}_{\;\;\;\;\mu}+c_{2}S^{\alpha}Q_{\alpha\mu\nu}+c_{3}S^{\alpha}Q_{\mu\nu\alpha}) 
	+\frac{1}{\sqrt{-g}\psi^{2}}(2S_{\lambda}-\nabla_{\lambda})(\sqrt{-g}\psi^{2}\Pi^{\lambda}_{\;\;\;\mu\nu}) \Big]
	\end{gather}
	Using all the above we can now derive the field equations for the conformally  and frame rescaling invariant theories. To obtain a conformally invariant theory, the parameters must satisfy ($\ref{confinv}$) and the gauge covariant derivative on the scalar field has to be defined as
	\beq
	D_{\mu}\equiv \partial_{\mu}-\left(\frac{n-2}{4n}\right)Q_{\mu} \label{gcovQ}
	\eeq 
	On the other hand, in order to obtain a frame rescaling invariant theory, the parameter space is restricted to ($\ref{frinv}$) and the gauge derivative is defined as
	\beq
	D_{\mu}\equiv \partial_{\mu}-\left(\frac{n-2}{n-1}\right)S_{\mu} \label{gcovS}
	\eeq 
	Having clarified this, the field equations after varying with respect to the metric tensor are
	\beq
	\psi^{2}\Big( Z_{(\mu\nu)}-\frac{1}{2}g_{\mu\nu}\mathcal{L}\Big)-\frac{1}{2}g_{\mu\nu}\lambda (D\psi)^{2}+ \lambda \Big( D_{\mu}\psi D_{\nu}\psi +K_{\mu\nu}\Big)=0
	\eeq
	where 
	\begin{gather}
	Z_{\mu\nu}\equiv L_{\mu\nu}+\xi_{\mu\nu}+b_{1}(2S_{\nu\alpha\beta}S_{\mu}^{\;\;\;\alpha\beta}-S_{\alpha\beta\mu}S^{\alpha\beta}_{\;\;\;\;\nu})-b_{2}S_{\nu\alpha\beta}S_{\mu}^{\;\;\;\alpha\beta}+b_{3}S_{\mu}S_{\nu} \nonumber \\
	+\Pi_{\mu\alpha\beta}Q_{\nu}^{\;\;\;\alpha\beta}-( c_{1}S_{\alpha\beta\nu}Q^{\alpha\beta}_{\;\;\;\;\mu}+c_{2}S^{\alpha}Q_{\alpha\mu\nu}+c_{3}S^{\alpha}Q_{\mu\nu\alpha}) \nonumber \\
	+\frac{1}{\sqrt{-g}\psi^{2}}(2S_{\lambda}-\nabla_{\lambda})(\sqrt{-g}\psi^{2}\Pi^{\lambda}_{\;\;\;\mu\nu})
	\end{gather}
	\begin{gather}
	\xi_{\mu\nu} \equiv \frac{1}{\sqrt{-g}\psi^{2}}\Big[ (2S_{\lambda}-\nabla_{\lambda})J^{\lambda}_{\;\;\;(\mu\nu)}+g_{\mu\nu}(2S_{\lambda}-\nabla_{\lambda})\zeta^{\lambda} \nonumber \\
	+\alpha_{4}(2S_{(\mu}-\nabla_{(\mu})( \sqrt{-g}\psi^{2}q_{\nu)}) \Big]
	\end{gather}
	\beq
	\int d^{n}x\sqrt{-g} K_{\mu\nu}\equiv \int d^{n}x \sqrt{-g}(D^{\alpha}\psi)\frac{\delta(D_{\alpha}\psi)}{\delta g^{\mu\nu}}
	\eeq
	and therefore
	\beq
	K_{\mu\nu}= \frac{(n-2)}{2 n}g_{\mu\nu} \frac{\partial_{\alpha}(\sqrt{-g}\psi D^{\alpha} \psi)}{\sqrt{-g}}
	\eeq
	for the conformally invariant theory and
	\beq
	K_{\mu\nu}=0
	\eeq
	for the frame rescaling invariant theory\footnote{This is so because in this case the gauge covariant derivative is constructed in terms of $S_{\mu}$ and the latter is independent of the metric tensor.}. Let us continue with the rest of the field equations. Variation with respect to the connection gives
	\begin{gather}
	\psi^{2} \Big( H^{\mu\nu}_{\;\;\;\;\lambda}+\delta^{\mu}_{\lambda}k^{\nu}+\delta^{\nu}_{\lambda}h^{\mu}+g^{\mu\nu}h_{\lambda}+f^{[\mu}\delta^{\nu ]}_{\lambda} \Big)+ \Theta^{\mu\nu}_{\;\;\;\;\lambda}=0
	\end{gather}
	where 
	\begin{gather}
	H^{\mu\nu}_{\;\;\;\;\lambda} \equiv a_{1}Q^{\nu\mu}_{\;\;\;\;\lambda}+2 a_{2}(Q^{\mu\nu}_{\;\;\;\;\lambda}+Q_{\lambda}^{\;\;\;\mu\nu})+2 b_{1}S^{\mu\nu}_{\;\;\;\;\lambda}+2 b_{2}S_{\lambda}^{\;\;\;[\mu\nu]} \nonumber \\
	+c_{1}( S^{\nu\mu}_{\;\;\;\;\lambda}-S_{\lambda}^{\;\;\;\nu\mu}+Q^{[\mu\nu]}_{\;\;\;\;\;\lambda}
	\end{gather}
	\beq
	k_{\mu} \equiv 4 a_{3}Q_{\mu}+2 a_{5}q_{\mu}+2 c_{2}S_{\mu}
	\eeq
	\beq
	h_{\mu} \equiv a_{5} Q_{\mu}+2 a_{4}q_{\mu}+c_{3}S_{\mu}
	\eeq
	\beq
	f_{\mu} \equiv c_{2} Q_{\mu}+ c_{3}q_{\mu}+2 b_{3}S_{\mu}
	\eeq
	and 
	\beq
	\Theta^{\mu\nu}_{\;\;\;\;\lambda} \equiv \frac{\partial}{\partial \Gamma^{\lambda}_{\;\;\;\mu\nu}}\Big( \lambda g^{\alpha\beta}D_{\alpha}\psi D_{\beta}\psi \Big)
	\eeq
	which for the conformally invariant case takes the form
	\beq
	\Theta^{\mu\nu}_{\;\;\;\;\lambda} =-\lambda \left( \frac{n-2}{n} \right)\psi (D^{\nu}\psi)\delta^{\mu}_{\lambda}
	\eeq
	and for the frame rescaling invariant theory
	\beq
	\Theta^{\mu\nu}_{\;\;\;\;\lambda} =-2 \lambda \left( \frac{n-2}{n-1} \right)\psi (D^{[\mu}\psi)\delta^{\nu]}_{\lambda}
	\eeq
	with the gauge covariant derivative given by ($\ref{gcovQ}$) for the former and ($\ref{gcovS}$) for the latter respectively.
	To conclude, for the conformally invariant case the $\Gamma$-field equations read
	\begin{gather}
	\psi^{2} \Big( H^{\mu\nu}_{\;\;\;\;\lambda}+\delta^{\mu}_{\lambda}k^{\nu}+\delta^{\nu}_{\lambda}h^{\mu}+g^{\mu\nu}h_{\lambda}+f^{[\mu}\delta^{\nu ]}_{\lambda} \Big)=\lambda \left( \frac{n-2}{n} \right)\psi (D^{\nu}\psi)\delta^{\mu}_{\lambda} \label{gammaeqn}
	\end{gather}
	and for the frame rescaling invariant case
	\begin{gather}
	\psi^{2} \Big( H^{\mu\nu}_{\;\;\;\;\lambda}+\delta^{\mu}_{\lambda}k^{\nu}+\delta^{\nu}_{\lambda}h^{\mu}+g^{\mu\nu}h_{\lambda}+f^{[\mu}\delta^{\nu ]}_{\lambda} \Big)=2 \lambda \left( \frac{n-2}{n-1} \right)\psi (D^{[\mu}\psi)\delta^{\nu]}_{\lambda} \label{gammaeqn2}
	\end{gather}
	Now, to close the system of the field equations it remains to vary with respect to the scalar $\psi$. For the conformally invariant case we find
	\beq
	\psi \mathcal{L}=\lambda \left( \frac{n-2}{4 n}Q_{\mu}D^{\mu}\psi+\frac{\partial_{\mu}(\sqrt{-g} D^{\mu}\psi)}{\sqrt{-g}}   \right)
	\eeq
	while for the frame rescaling invariant theory, one obtains
	\beq
	\psi \mathcal{L}=\lambda \left( \frac{n-2}{n-1}S_{\mu}D^{\mu}\psi+\frac{\partial_{\mu}(\sqrt{-g} D^{\mu}\psi)}{\sqrt{-g}}   \right)
	\eeq
	Before gathering our results let us examine (\ref{gammaeqn}) and (\ref{gammaeqn2}) a little further. To do so, notice that we can consider three operations on (\ref{gammaeqn}) and (\ref{gammaeqn2}). We can contract in $\mu=\lambda$, contact in $\nu=\lambda$ and multiply (and contact) by $g^{\mu\nu}$. Then we get three vector equations that we may formally write as
	\begin{gather}
	\alpha_{1}Q_{\mu}+\alpha_{2}q_{\mu} +\alpha_{3}S_{\mu}=\frac{\partial_{\mu}\psi}{\psi} \nonumber \\
	\beta_{1}Q_{\mu}+\beta_{2}q_{\mu} +\beta_{3}S_{\mu}=\frac{\partial_{\mu}\psi}{\psi} \nonumber \\
	\gamma_{1}Q_{\mu}+\gamma_{2}q_{\mu} +\gamma_{3}S_{\mu}=\frac{\partial_{\mu}\psi}{\psi}
	\end{gather}
	where the $\alpha_{i},\beta_{i},\gamma_{i}$ are all combinations of $a_{i},b_{i},c_{i}$ and $\lambda$. Then the above system of equations can be formally solved\footnote{Assuming that the determinant of the matrix corresponding to the system does not vanish.} to give
	\beq
	Q_{\mu}=\lambda_{1} \frac{\partial_{\mu}\psi}{\psi}\;,\; q_{\mu}=\lambda_{2} \frac{\partial_{\mu}\psi}{\psi}\;,\;S_{\mu}=\lambda_{3} \frac{\partial_{\mu}\psi}{\psi}
	\eeq
	where  the $\lambda_{i}'s$ depend on $\alpha_{i},\beta_{i},\gamma_{i}$. This result when substituted back at (\ref{gammaeqn})and (\ref{gammaeqn2}) yield
	\beq
	H^{\mu\nu}_{\;\;\;\;\lambda}=\sigma_{1}\delta^{\mu}_{\lambda}\frac{\partial_{\nu}\psi}{\psi}+\sigma_{2}\delta^{\nu}_{\lambda}\frac{\partial_{\mu}\psi}{\psi}+\sigma_{3}g^{\mu\nu}\frac{\partial^{\lambda}\psi}{\psi}
	\eeq
	where again $\sigma_{i}'s$ depend on $a_{i},b_{i},c_{i}$ and $\lambda$. We are now in a position to recap our results. So, starting with
	\begin{gather}
	S=\int d^{n}x \sqrt{-g}\Big[ \psi^{2}\mathcal{L} +\lambda g^{\mu\nu}D_{\mu}\psi D_{\nu}\psi\Big] \nonumber \\
	=\frac{1}{2 \kappa}\int d^{n}x \sqrt{-g} \Big[ \psi^{2}\Big(  
	b_{1}S_{\alpha\mu\nu}S^{\alpha\mu\nu} +
	b_{2}S_{\alpha\mu\nu}S^{\mu\nu\alpha} +
	b_{3}S_{\mu}S^{\mu} \nonumber \\
	a_{1}Q_{\alpha\mu\nu}Q^{\alpha\mu\nu} +
	a_{2}Q_{\alpha\mu\nu}Q^{\mu\nu\alpha} +
	a_{3}Q_{\mu}Q^{\mu}+
	a_{4}q_{\mu}q^{\mu}+
	a_{5}Q_{\mu}q^{\mu} \nonumber \\
	+c_{1}Q_{\alpha\mu\nu}S^{\alpha\mu\nu}+
	c_{2}Q_{\mu}S^{\mu} +
	c_{3}q_{\mu}S^{\mu}\Big) +\lambda g^{\mu\nu}D_{\mu}\psi D_{\nu}\psi\Big] =\nonumber \\  \label{genact2}
	\end{gather}
	A \textbf{conformally} invariant theory is produced when the parameters satisfy
	\begin{gather}
	4a_{1}+4 n a_{3}+2 a_{5}=0  \nonumber \\ 4 a_{2}+4 a_{4}+2 n a_{5}=0 \nonumber \\ n a_{1}+a_{2}+ n^{2} a_{3}+a_{4}+n a_{5}=0 \nonumber \\
	c_{1}+n c_{2}+c_{3}=0  \nonumber \\
	b_{i}'s=no\;constraint
	\end{gather}
	and the gauge covariant derivative is defined as
	\beq
	D_{\mu}\equiv \partial_{\mu}-\left(\frac{n-2}{4n}\right)Q_{\mu} 
	\eeq 
	Then the field equations following from the above, read
	\newline 
	\textbf{g-Variation:}
	\beq
	\psi^{2}\Big( Z_{(\mu\nu)}-\frac{1}{2}g_{\mu\nu}\mathcal{L}\Big)-\frac{1}{2}g_{\mu\nu}\lambda (D\psi)^{2}+ \lambda \Big( D_{\mu}\psi D_{\nu}\psi +K_{\mu\nu}\Big)=0
	\eeq
	where 
	\begin{gather}
	Z_{\mu\nu}\equiv L_{\mu\nu}+\xi_{\mu\nu}+b_{1}(2S_{\nu\alpha\beta}S_{\mu}^{\;\;\;\alpha\beta}-S_{\alpha\beta\mu}S^{\alpha\beta}_{\;\;\;\;\nu})-b_{2}S_{\nu\alpha\beta}S_{\mu}^{\;\;\;\alpha\beta}+b_{3}S_{\mu}S_{\nu} \nonumber \\
	+\Pi_{\mu\alpha\beta}Q_{\nu}^{\;\;\;\alpha\beta}-( c_{1}S_{\alpha\beta\nu}Q^{\alpha\beta}_{\;\;\;\;\mu}+c_{2}S^{\alpha}Q_{\alpha\mu\nu}+c_{3}S^{\alpha}Q_{\mu\nu\alpha}) \nonumber \\
	+\frac{1}{\sqrt{-g}\psi^{2}}(2S_{\lambda}-\nabla_{\lambda})(\sqrt{-g}\psi^{2}\Pi^{\lambda}_{\;\;\;\mu\nu})
	\end{gather}
	\begin{gather}
	\xi_{\mu\nu} \equiv \frac{1}{\sqrt{-g}\psi^{2}}\Big[ (2S_{\lambda}-\nabla_{\lambda})J^{\lambda}_{\;\;\;(\mu\nu)}+g_{\mu\nu}(2S_{\lambda}-\nabla_{\lambda})\zeta^{\lambda} \nonumber \\
	+\alpha_{4}(2S_{(\mu}-\nabla_{(\mu})( \sqrt{-g}\psi^{2}q_{\nu)}) \Big]
	\end{gather}
	\begin{gather}
	L_{\mu\nu}=(a_{1}Q_{\mu\alpha\beta}+a_{2}Q_{\alpha\beta\mu})Q_{\nu}^{\;\;\;\alpha\beta}+(a_{3}Q_{\mu}+a_{5}q_{\mu})Q_{\nu}+a_{3}Q_{\alpha\mu\nu}Q^{\alpha} \nonumber \\
	+Q_{\mu\nu\alpha}(a_{4}q^{\alpha}+a_{5}Q^{\alpha})-\Omega^{\alpha\beta}_{\;\;\;\;\nu}Q_{\alpha\beta\mu}-\Omega_{\alpha\mu\beta}Q^{\alpha\beta}_{\;\;\;\;\nu}
	\end{gather}
	\beq
	J^{\lambda}_{\;\;\;(\mu\nu)} \equiv \sqrt{-g}\psi^{2}( \alpha_{1}Q^{\lambda}_{\;\;\;\mu\nu}+a_{2}Q_{\mu\nu}^{\;\;\;\;\lambda}+\Omega^{\lambda}_{\;\;\;\mu\nu})
	\eeq
	\beq
	\zeta^{\lambda}=\sqrt{-g}\psi^{2}(a_{3}Q^{\lambda}+a_{5}q^{\lambda})
	\eeq
	\beq
	K_{\mu\nu}= \frac{(n-2)}{2 n}g_{\mu\nu} \frac{\partial_{\alpha}(\sqrt{-g}\psi D^{\alpha} \psi)}{\sqrt{-g}}
	\eeq
	\textbf{$\Gamma$-Variation:}
	\begin{gather}
	\psi^{2} \Big( H^{\mu\nu}_{\;\;\;\;\lambda}+\delta^{\mu}_{\lambda}k^{\nu}+\delta^{\nu}_{\lambda}h^{\mu}+g^{\mu\nu}h_{\lambda}+f^{[\mu}\delta^{\nu ]}_{\lambda} \Big)=\lambda \left( \frac{n-2}{n} \right)\psi (D^{\nu}\psi)\delta^{\mu}_{\lambda} 
	\end{gather}
	where
	\beq
	k_{\mu} \equiv 4 a_{3}Q_{\mu}+2 a_{5}q_{\mu}+2 c_{2}S_{\mu}
	\eeq
	\beq
	h_{\mu} \equiv a_{5} Q_{\mu}+2 a_{4}q_{\mu}+c_{3}S_{\mu}
	\eeq
	\beq
	f_{\mu} \equiv c_{2} Q_{\mu}+ c_{3}q_{\mu}+2 b_{3}S_{\mu}
	\eeq
	\textbf{$\psi$-Variation:}
	\beq
	\psi \mathcal{L}=\lambda \left( \frac{n-2}{4 n}Q_{\mu}D^{\mu}\psi+\frac{\partial_{\mu}(\sqrt{-g} D^{\mu}\psi)}{\sqrt{-g}}   \right)
	\eeq
	For the \textbf{frame\;rescaling} invariant theory the action is again ($\ref{genact2}$) but now the parameters have to satisfy
	\begin{gather}
	a_{i}=no\;constraint \nonumber \\
	-2 b_{1}+b_{2}+(1-n)b_{3}=0   \nonumber \\
	- c_{1}+(1-n) c_{2}=0 \nonumber \\
	c_{1}+(1-n) c_{3}=0 \label{frinv}
	\end{gather}
	and the gauge covariant derivative must be defined as
	\beq
	D_{\mu}\equiv \partial_{\mu}-\left(\frac{n-2}{n-1}\right)S_{\mu} \label{gcovS}
	\eeq 
	For this case the field equations are
	\textbf{g-Variation:}
	\beq
	\psi^{2}\Big( Z_{(\mu\nu)}-\frac{1}{2}g_{\mu\nu}\mathcal{L}\Big)-\frac{1}{2}g_{\mu\nu}\lambda (D\psi)^{2}+ \lambda \Big( D_{\mu}\psi D_{\nu}\psi \Big)=0
	\eeq
	where 
	\begin{gather}
	Z_{\mu\nu}\equiv L_{\mu\nu}+\xi_{\mu\nu}+b_{1}(2S_{\nu\alpha\beta}S_{\mu}^{\;\;\;\alpha\beta}-S_{\alpha\beta\mu}S^{\alpha\beta}_{\;\;\;\;\nu})-b_{2}S_{\nu\alpha\beta}S_{\mu}^{\;\;\;\alpha\beta}+b_{3}S_{\mu}S_{\nu} \nonumber \\
	+\Pi_{\mu\alpha\beta}Q_{\nu}^{\;\;\;\alpha\beta}-( c_{1}S_{\alpha\beta\nu}Q^{\alpha\beta}_{\;\;\;\;\mu}+c_{2}S^{\alpha}Q_{\alpha\mu\nu}+c_{3}S^{\alpha}Q_{\mu\nu\alpha}) \nonumber \\
	+\frac{1}{\sqrt{-g}\psi^{2}}(2S_{\lambda}-\nabla_{\lambda})(\sqrt{-g}\psi^{2}\Pi^{\lambda}_{\;\;\;\mu\nu})
	\end{gather}
	\begin{gather}
	\xi_{\mu\nu} \equiv \frac{1}{\sqrt{-g}\psi^{2}}\Big[ (2S_{\lambda}-\nabla_{\lambda})J^{\lambda}_{\;\;\;(\mu\nu)}+g_{\mu\nu}(2S_{\lambda}-\nabla_{\lambda})\zeta^{\lambda} \nonumber \\
	+\alpha_{4}(2S_{(\mu}-\nabla_{(\mu})( \sqrt{-g}\psi^{2}q_{\nu)}) \Big]
	\end{gather}
	\begin{gather}
	L_{\mu\nu}=(a_{1}Q_{\mu\alpha\beta}+a_{2}Q_{\alpha\beta\mu})Q_{\nu}^{\;\;\;\alpha\beta}+(a_{3}Q_{\mu}+a_{5}q_{\mu})Q_{\nu}+a_{3}Q_{\alpha\mu\nu}Q^{\alpha} \nonumber \\
	+Q_{\mu\nu\alpha}(a_{4}q^{\alpha}+a_{5}Q^{\alpha})-\Omega^{\alpha\beta}_{\;\;\;\;\nu}Q_{\alpha\beta\mu}-\Omega_{\alpha\mu\beta}Q^{\alpha\beta}_{\;\;\;\;\nu}
	\end{gather}
	\beq
	J^{\lambda}_{\;\;\;(\mu\nu)} \equiv \sqrt{-g}\psi^{2}( \alpha_{1}Q^{\lambda}_{\;\;\;\mu\nu}+a_{2}Q_{\mu\nu}^{\;\;\;\;\lambda}+\Omega^{\lambda}_{\;\;\;\mu\nu})
	\eeq
	\beq
	\zeta^{\lambda}=\sqrt{-g}\psi^{2}(a_{3}Q^{\lambda}+a_{5}q^{\lambda})
	\eeq
	\textbf{$\Gamma$-Variation:}
	\begin{gather}
	\psi^{2} \Big( H^{\mu\nu}_{\;\;\;\;\lambda}+\delta^{\mu}_{\lambda}k^{\nu}+\delta^{\nu}_{\lambda}h^{\mu}+g^{\mu\nu}h_{\lambda}+f^{[\mu}\delta^{\nu ]}_{\lambda} \Big)=2 \lambda \left( \frac{n-2}{n-1} \right)\psi (D^{[\mu}\psi)\delta^{\nu]}_{\lambda}
	\end{gather}
	\textbf{$\psi$-Variation:}
	\beq
	\psi \mathcal{L}=\lambda \left( \frac{n-2}{n-1}S_{\mu}D^{\mu}\psi+\frac{\partial_{\mu}(\sqrt{-g} D^{\mu}\psi)}{\sqrt{-g}}   \right)
	\eeq
	Apart from the two above invariant theories, another case is of interest and this is the case of projective invariance. Note that in order to obtain a projective invariant theory no scalar field is required, so we may set $\psi=1$ and $\lambda=0$ to our starting action to arrive at
	\begin{gather}
	S=\int d^{n}x \sqrt{-g}\mathcal{L}  \nonumber \\
	=\frac{1}{2 \kappa}\int d^{n}x \sqrt{-g} \Big[   
	b_{1}S_{\alpha\mu\nu}S^{\alpha\mu\nu} +
	b_{2}S_{\alpha\mu\nu}S^{\mu\nu\alpha} +
	b_{3}S_{\mu}S^{\mu} \nonumber \\
+	a_{1}Q_{\alpha\mu\nu}Q^{\alpha\mu\nu} +
	a_{2}Q_{\alpha\mu\nu}Q^{\mu\nu\alpha} +
	a_{3}Q_{\mu}Q^{\mu}+
	a_{4}q_{\mu}q^{\mu}+
	a_{5}Q_{\mu}q^{\mu} \nonumber \\
	+c_{1}Q_{\alpha\mu\nu}S^{\alpha\mu\nu}+
	c_{2}Q_{\mu}S^{\mu} +
	c_{3}q_{\mu}S^{\mu} \Big] 
	\end{gather}

	The above defines  $projective$ invariant theories so long as the parameters satisfy
	\beq
	4 (2 a_{1}+2 n a_{3}+a_{5})-c_{1}+(1-n)c_{2}=0
	\eeq
	\beq
	4 (2 a_{2}+2  a_{4}+n a_{5})+c_{1}+(1-n)c_{3}=0
	\eeq
	\beq
	-2 b_{1}+b_{2}+(1-n) b_{3}+2(c_{1} +n c_{2}+c_{3})=0
	\eeq
	\begin{gather}
	16 (n a_{1}+ a_{2}+ n^{2} a_{3}+ a_{4}+ n a_{5}) \nonumber \\
	+(n-1)\Big( 2 b_{1}-b_{2}+(n-1) b_{3}-4 (c_{1}+n c_{2}+c_{3})\Big) =0
	\end{gather}

	\subsection{Including the Parity-Odd terms}
	Now, let us also add the parity violating terms into our general quadratic action. One thing we should clarify however is the redundancy among these terms. In particular, in $4$-dimensions there are only two independent parity-odd quadratic pure torsion terms. That is, from the four parity-odd torsion scalars we considered $B_{5},B_{6},B_{7},B_{8}$ only two are independent. A way to see this is starting by
	\beq
	t^{\rho}=\epsilon^{\rho\kappa\lambda\sigma}S_{\kappa\lambda\sigma}
	\eeq
	which when  contracted by $\epsilon_{\rho\alpha\beta\mu}$ and using $\epsilon^{\rho\kappa\lambda\sigma}\epsilon_{\rho\alpha\beta\mu}=-3! \delta^{[\kappa}_{\alpha}\delta^{\lambda}_{\beta}\delta^{\sigma]}_{\mu}$ gives
	\beq
	\epsilon_{\rho\alpha\beta\mu}t^{\rho}=-3! S_{[\alpha\beta\mu]}
	\eeq
	Exploiting the antisymmetry of the torsion tensor in its first two indices the above may be expressed as
	\beq
	\epsilon_{\rho\alpha\beta\mu}t^{\rho}=-2(  S_{\alpha\beta\mu}+S_{\mu\alpha\beta} +S_{\beta\mu\alpha}    )
	\eeq
	Furthermore, contracting the above with $\epsilon^{\alpha\beta\gamma\delta}$ and using $\epsilon_{\rho\alpha\beta\mu}\epsilon^{\alpha\beta\gamma\delta}=-4 \delta^{[\gamma}_{\rho}\delta^{\delta]}_{\mu}$ we finally arrive at
	\beq
	2 t^{[\gamma}\delta^{\delta]}_{\mu}=\epsilon^{\alpha\beta\gamma\delta}S_{\alpha\beta\mu}+2 \epsilon^{\alpha\beta\gamma\delta} S_{\mu\alpha\beta} \label{tepseq}
	\eeq
	The latter is the key equation that gives the relations among the parity-odd terms. To obtain these, we first contract ($\ref{tepseq}$) by $S_{\gamma\delta}^{\;\;\;\;\mu}$ and use the definitions of $B_{i}'s$ to obtain
	\beq
	2 B_{5}=B_{6}+2B_{8}
	\eeq
	In addition, contracting with $S^{\mu}_{\;\;\gamma\delta}$ this time, gives
	\beq
	-B_{5}=B_{8}+2 B_{7}
	\eeq
	Therefore, we have two equations relating the $B_{5},...,B_{8}$ and so only two of the four are independent. We may choose the $B_{5}$ and $B_{6}$. Regarding the mixed\footnote{For the pure non-metricity parity-odd scalars we just have one term-$A_{6}$ so we may not worry about redundancy here.} parity-odd terms, we note that out of the four combinations $C_{4}, C_{5},C_{6},C_{7}$ only the three are independent. This is easily seen by contracting ($\ref{tepseq}$) with $Q_{\gamma\delta}^{\;\;\;\;\mu}$ to arrive at
	\beq
	C_{4}-C_{5}=C_{6}+2 C_{7}
	\eeq
	therefore one scalar is redundant and we choose to disregard $C_{7}$. Before writing down the general quadratic action including the above parity-odd terms to the original action, let us point out another redundant term, this time though a parity-even one, and that is $B_{4}=t_{\mu}t^{\mu}$.By a direct calculation, this is found to be
	\begin{gather}
	B_{4}=t_{\mu}t^{\mu}=\epsilon_{\mu\alpha\beta\gamma}\epsilon^{\mu\kappa\lambda\rho}S^{\alpha\beta\gamma}S_{\kappa\lambda\rho}=-3! \delta^{[\kappa}_{\alpha}\delta^{\lambda}_{\beta}\delta^{\rho]}_{\gamma}S^{\alpha\beta\gamma}S_{\kappa\lambda\rho}= \nonumber \\
	=-3! S^{\alpha\beta\gamma}S_{[\alpha\beta\gamma]}=-2 S^{\alpha\beta\gamma}( S_{\alpha\beta\gamma}+S_{\gamma\beta\alpha}+S_{\beta\gamma\alpha}    ) =\nonumber \\
	=-2(B_{1}+2B_{2})
	\end{gather}
	In conclusion, for the parity-odd case we have $2$ independent pure torsion scalars, $3$ independent mixed scalars  and $1$ pure non-metric scalar. Note that our results are in perfect agreement with the number of parity-odd scalars that were considered in \cite{pagani2015quantum}.  Thus, in $4-dim$ there will be $6$ parity-odd quadratic scalars added to our general action. With the points we presented above on the redundant scalars we may set
	\beq
	b_{7}=b_{8}=c_{7}=b_{4}=0
	\eeq
	since the scalars with those coefficients all depend on the other basic scalars as we showed earlier. So, our total action including the parity-odd terms reads
	\begin{gather}
	S
	=\frac{1}{2 \kappa}\int d^{4}x \sqrt{-g} \psi^{2}\Big[   
	b_{1}S_{\alpha\mu\nu}S^{\alpha\mu\nu} +
	b_{2}S_{\alpha\mu\nu}S^{\mu\nu\alpha} +
	b_{3}S_{\mu}S^{\mu} \nonumber \\
	a_{1}Q_{\alpha\mu\nu}Q^{\alpha\mu\nu} +
	a_{2}Q_{\alpha\mu\nu}Q^{\mu\nu\alpha} +
	a_{3}Q_{\mu}Q^{\mu}+
	a_{4}q_{\mu}q^{\mu}+
	a_{5}Q_{\mu}q^{\mu} \nonumber \\
	+c_{1}Q_{\alpha\mu\nu}S^{\alpha\mu\nu}+
	c_{2}Q_{\mu}S^{\mu} +
	c_{3}q_{\mu}S^{\mu} \nonumber \\
	+a_{6}\epsilon^{\alpha\beta\gamma\delta}Q_{\alpha\beta\mu}Q_{\gamma\delta\;\;\;\;}^{\mu}+b_{5}S_{\mu}t^{\mu}+b_{6}\epsilon^{\alpha\beta\gamma\delta}S_{\alpha\beta\mu}S_{\gamma\delta}^{\;\;\;\;\mu} \nonumber \\
	c_{4}Q_{\mu}t^{\mu}+c_{5}q^{\mu}t_{\mu}+c_{6}\epsilon^{\alpha\beta\gamma\delta}Q_{\alpha\beta\mu}S_{\gamma\delta}^{\;\;\;\;\mu}
	\Big] \nonumber \\
	+\frac{1}{2 \kappa}\int d^{4}x \sqrt{-g}\lambda g^{\mu\nu}D_{\mu}\psi D_{\nu}\psi 
	\end{gather}
	Note that the first three lines of the above represent the quadratic parity-even terms (as appear in $\ref{genactionS}$), and in the forth and fifth line we have included the $6$ parity-odd terms $A_{6},B_{5},B_{6},C_{4},C_{5},C_{6}$. Let us now find the parameter space for the above action to be invariant under each of the three transformations. Let us start with the conformal transformations. The newly added parity-odd terms transform as
	\begin{gather}
	\bar{A}_{6}=e^{-2\theta}A_{6}\;,\; \bar{B}_{5}=e^{-2\theta}B_{5}\;,\; \bar{B}_{6}=e^{-2\theta}B_{6}\;,\; \nonumber \\
	\bar{C}_{4}=e^{-2\theta}( C_{4}-2n t^{\mu}\partial_{\mu}\theta ) \nonumber \\
	\bar{C}_{5}=e^{-2\theta}( C_{5}-2 t^{\mu}\partial_{\mu}\theta ) \nonumber \\
	\bar{C}_{6}=e^{-2\theta}( C_{6}-2 t^{\mu}\partial_{\mu}\theta ) \nonumber
	\end{gather}
	under a conformal metric transformation. Defining then the parity-odd Lagrangian densities
	\beq
	\mathcal{L}_{Q}^{p-odd} \equiv a_{6}\epsilon^{\alpha\beta\gamma\delta}Q_{\alpha\beta\mu}Q_{\gamma\delta\;\;\;\;}^{\mu}=a_{6}A_{6}
	\eeq
	\beq
	\mathcal{L}_{T}^{p-odd} \equiv b_{5}S_{\mu}t^{\mu}+b_{6}\epsilon^{\alpha\beta\gamma\delta}S_{\alpha\beta\mu}S_{\gamma\delta}^{\;\;\;\;\mu} =b_{5}B_{5}+b_{6}B_{6}
	\eeq
	\beq
	\mathcal{L}_{QT}^{p-odd} \equiv c_{4}Q_{\mu}t^{\mu}+c_{5}q^{\mu}t_{\mu}+c_{6}\epsilon^{\alpha\beta\gamma\delta}Q_{\alpha\beta\mu}S_{\gamma\delta}^{\;\;\;\;\mu}=c_{4}C_{4}+c_{5}C_{5}+c_{6}C_{6}
	\eeq
	\beq
	\mathcal{L}^{p-odd} \equiv \mathcal{L}_{Q}^{p-odd}+\mathcal{L}_{T}^{p-odd}+\mathcal{L}_{QT}^{p-odd}
	\eeq
	we see that the latter transform as
	\beq
	\bar{\mathcal{L}}_{Q}^{p-odd}=e^{-2\theta}\mathcal{L}_{Q}^{p-odd}\;, \;\; \bar{\mathcal{L}}_{T}^{p-odd}=e^{-2\theta}\mathcal{L}_{T}^{p-odd}
	\eeq
	\beq
	\bar{\mathcal{L}}_{QT}^{p-odd}=e^{-2\theta}\mathcal{L}_{Q}^{p-odd}-e^{-2\theta}2t^{\mu}(\partial_{\mu}\theta )(n c_{4}+c_{5}+c_{6})
	\eeq
	The transformation for the parity-even part of the Lagrangian we have already computed in the previous section. So, for the total action to be invariant under \textbf{conformal\; transformations} we must have\footnote{Of course in what follows $n=4$.}
	\begin{gather}
	4a_{1}+4 n a_{3}+2 a_{5}=0  \nonumber \\ 4 a_{2}+4 a_{4}+2 n a_{5}=0 \nonumber \\ n a_{1}+a_{2}+ n^{2} a_{3}+a_{4}+n a_{5}=0 \nonumber \\
	c_{1}+n c_{2}+c_{3}=0\;,\;\;b_{i} 's=no\;constraint  \nonumber \\
	nc_{4}+c_{5}+c_{6}=0\;,\; a_{6}=whatever
	\end{gather}
	Note that the first four constraints in the above are the ones we had derived previously for the pure parity-even Lagrangian and the last constraint is imposed on the parity-odd part. We should mention that the additional constraint establishes a relation only between the coefficients of the parity-odd terms and does not mix them with the parameters of the parity-even scalars! Now, under a frame rescaling the parity-odd parts transform as
	\beq
	\tilde{\mathcal{L}}_{Q}^{p-odd}=e^{-2\theta}\mathcal{L}_{Q}^{p-odd}\;, \;\; \tilde{\mathcal{L}}_{T}^{p-odd}=e^{-2\theta}\mathcal{L}_{T}^{p-odd}+e^{-2 \theta}t^{\mu}\partial_{\mu}\theta \Big( \frac{1-n}{2}b_{5}-2 b_{6} \Big)
	\eeq
	\beq
	\tilde{\mathcal{L}}_{QT}^{p-odd}=e^{-2\theta}\mathcal{L}_{Q}^{p-odd}
	\eeq
	And for the total action to be invariant under \textbf{frame\; rescalings}, the parameters must satisfy
	\begin{gather}
	a_{i}=no\;constraint \nonumber \\
	-2 b_{1}+b_{2}+(1-n)b_{3}=0   \nonumber \\
	- c_{1}+(1-n) c_{2}=0 \nonumber \\
	c_{1}+(1-n) c_{3}=0 \nonumber \\
	\frac{1-n}{2}b_{5}-2 b_{6}=0 \;,\; b_{5},b_{6}=no\;constraint
	\end{gather}
	Again, the first four constraints above are the same with the pure parity-even theory and the last one is imposed among the parameters of the parity-odd terms. Now, in order to study the parameter space for the projective invariant case, again there is no need for a scalar field $\psi$ to compensate for the invariance, and our action is therefore 
	\begin{gather}
	S
	=\frac{1}{2 \kappa}\int d^{4}x \sqrt{-g} \Big[   
	b_{1}S_{\alpha\mu\nu}S^{\alpha\mu\nu} +
	b_{2}S_{\alpha\mu\nu}S^{\mu\nu\alpha} +
	b_{3}S_{\mu}S^{\mu} \nonumber \\
	a_{1}Q_{\alpha\mu\nu}Q^{\alpha\mu\nu} +
	a_{2}Q_{\alpha\mu\nu}Q^{\mu\nu\alpha} +
	a_{3}Q_{\mu}Q^{\mu}+
	a_{4}q_{\mu}q^{\mu}+
	a_{5}Q_{\mu}q^{\mu} \nonumber \\
	+c_{1}Q_{\alpha\mu\nu}S^{\alpha\mu\nu}+
	c_{2}Q_{\mu}S^{\mu} +
	c_{3}q_{\mu}S^{\mu} \nonumber \\
	+a_{6}\epsilon^{\alpha\beta\gamma\delta}Q_{\alpha\beta\mu}Q_{\gamma\delta\;\;\;\;}^{\mu}+b_{5}S_{\mu}t^{\mu}+b_{6}\epsilon^{\alpha\beta\gamma\delta}S_{\alpha\beta\mu}S_{\gamma\delta}^{\;\;\;\;\mu} \nonumber \\
	c_{4}Q_{\mu}t^{\mu}+c_{5}q^{\mu}t_{\mu}+c_{6}\epsilon^{\alpha\beta\gamma\delta}Q_{\alpha\beta\mu}S_{\gamma\delta}^{\;\;\;\;\mu}
	\Big] \nonumber 
	\end{gather}
	for the most general case (including the parity-odd scalars too). As we have already seen, the parity-even part transforms as
	\begin{gather}
	\hat{\mathcal{L}}_{Q}^{p-even}+\hat{\mathcal{L}}_{T}^{p-even}+\hat{\mathcal{L}}_{QT}^{p-even}=\mathcal{L}_{Q}^{p-even}+\mathcal{L}_{T}^{p-even}+ \mathcal{L}_{QT}^{p-even} \nonumber \\
	+\left[ 2(2 a_{1}+2 n a_{3}+a_{5})+\frac{1}{2}\Big( -c_{1}+(1-n)c_{2} \Big) \right]Q_{\mu}\xi^{\mu} \nonumber \\
	+\left[ 2(2 a_{2}+2  a_{4}+n a_{5})+\frac{1}{2}\Big( c_{1}+(1-n)c_{3} \Big) \right]q_{\mu}\xi^{\mu} \nonumber \\
	+\Big[ -2 b_{1}+b_{2}+(1-n) b_{3}+2(c_{1} +n c_{2}+c_{3}) \Big]S_{\mu}\xi^{\mu} \nonumber \\
	\left[ 4(n a_{1}+ a_{2}+ n^{2} a_{3}+ a_{4}+ n a_{5}) +\frac{(n-1)}{4}\Big( 2 b_{1}-b_{2}+(n-1) b_{3}\Big) 
	-(n-1)(c_{1}+n c_{2}+c_{3})  \right]\xi_{\mu}\xi^{\mu} 
	\end{gather}
	where
	\begin{gather}
	\mathcal{L}_{Q}^{p-even}=a_{1}Q_{\alpha\mu\nu}Q^{\alpha\mu\nu} +
	a_{2}Q_{\alpha\mu\nu}Q^{\mu\nu\alpha} +
	a_{3}Q_{\mu}Q^{\mu}+
	a_{4}q_{\mu}q^{\mu}+
	a_{5}Q_{\mu}q^{\mu} \nonumber \\
	\mathcal{L}_{T}^{p-even}=b_{1}S_{\alpha\mu\nu}S^{\alpha\mu\nu} +
	b_{2}S_{\alpha\mu\nu}S^{\mu\nu\alpha} +
	b_{3}S_{\mu}S^{\mu} \nonumber \\
	\mathcal{L}_{QT}^{p-even}=c_{1}Q_{\alpha\mu\nu}S^{\alpha\mu\nu}+
	c_{2}Q_{\mu}S^{\mu} +
	c_{3}q_{\mu}S^{\mu} 
	\end{gather}
	Now, the parity-odd part transforms according to
	\begin{gather}
	\hat{\mathcal{L}}_{Q}^{p-odd}=\mathcal{L}_{Q}^{p-odd} \nonumber \\
	\hat{\mathcal{L}}_{T}^{p-odd}=\mathcal{L}_{T}^{p-odd}+b_{5}\frac{1-n}{2}t_{\mu}\xi^{\mu}-2b_{6}t_{\mu}\xi^{\mu} \nonumber \\
	\hat{\mathcal{L}}_{QT}^{p-odd}=\mathcal{L}_{QT}^{p-odd}+2t_{\mu}\xi^{\mu}(nc_{4}+c_{5}+c_{6}) \nonumber
	\end{gather}
	as can be easily checked. As a result, the total Lagrangian density 
	$\mathcal{L}=\mathcal{L}^{p-even}+\mathcal{L}^{p-odd}$ undergoes the transformation
	\begin{gather}
	\hat{\mathcal{L}}=\hat{\mathcal{L}}^{p-even}+\hat{\mathcal{L}}^{p-odd}=\mathcal{L} \nonumber \\
	+\left[ 2(2 a_{1}+2 n a_{3}+a_{5})+\frac{1}{2}\Big( -c_{1}+(1-n)c_{2} \Big) \right]Q_{\mu}\xi^{\mu} \nonumber \\
	+\left[ 2(2 a_{2}+2  a_{4}+n a_{5})+\frac{1}{2}\Big( c_{1}+(1-n)c_{3} \Big) \right]q_{\mu}\xi^{\mu} \nonumber \\
	+\Big[ -2 b_{1}+b_{2}+(1-n) b_{3}+2(c_{1} +n c_{2}+c_{3}) \Big]S_{\mu}\xi^{\mu} \nonumber \\
	\left[ 4(n a_{1}+ a_{2}+ n^{2} a_{3}+ a_{4}+ n a_{5}) +\frac{(n-1)}{4}\Big( 2 b_{1}-b_{2}+(n-1) b_{3}\Big) 
	-(n-1)(c_{1}+n c_{2}+c_{3})  \right]\xi_{\mu}\xi^{\mu} \nonumber \\
	+2\Big( \frac{1-n}{4}b_{5}-b_{6}+nc_{4}+c_{5}+c_{6}\Big)t_{\mu}\xi^{\mu}
	\end{gather}
	So, \textbf{projective\; invariance} is ensured if the parameters satisfy
	\beq
	4 (2 a_{1}+2 n a_{3}+a_{5})-c_{1}+(1-n)c_{2}=0
	\eeq
	\beq
	4 (2 a_{2}+2  a_{4}+n a_{5})+c_{1}+(1-n)c_{3}=0
	\eeq
	\beq
	-2 b_{1}+b_{2}+(1-n) b_{3}+2(c_{1} +n c_{2}+c_{3}) =0
	\eeq
	\begin{gather}
	16 (n a_{1}+ a_{2}+ n^{2} a_{3}+ a_{4}+ n a_{5}) \nonumber \\
	+(n-1)\Big( 2 b_{1}-b_{2}+(n-1) b_{3}-4 (c_{1}+n c_{2}+c_{3})\Big) =0
	\end{gather}
	\beq
	\frac{1-n}{4}b_{5}-b_{6}+nc_{4}+c_{5}+c_{6}=0
	\eeq
	Note that in comparison with the pure parity-even case, the first four constraints remain the same, and a fifth additional constraint is imposed only among the parameters of the parity-odd scalars. The important thing is that the constraints again do not mix the parameters of the parity-even with the parameters of the parity-odd scalars.

	\subsection{Conformally Invariant Quartic Actions}
	Having established the transformation laws for the quadratic torsion and non-metricity scalars let us now find some (of the many!) quartic combinations that remain invariant under conformal metric transformations. To start with, let us first note that
	\begin{gather}
	(n \bar{A}_{1}-\bar{A}_{3})=e^{-2 \theta}(n A_{1}-A_{3}) \nonumber \\
	( \bar{A}_{2}-\bar{A}_{4})=e^{-2 \theta}( A_{2}-A_{4}) \nonumber \\
	\Big( \bar{A}_{5}-\frac{n}{2}\bar{A}_{4} -\frac{1}{2n}\bar{A}_{3}\Big)=e^{-2 \theta}\Big( A_{5}-\frac{n}{2}A_{4} -\frac{1}{2n} A_{3}\Big) \nonumber \\
	\bar{B}_{i}=e^{-2 \theta}B_{i}\;,\;\;\; \forall \; i \nonumber \\
	( \bar{C}_{1}-\bar{C}_{3})=e^{-2 \theta}(C_{1}-C_{3}) \nonumber \\
	(n \bar{C}_{1}-\bar{C}_{2})=e^{-2 \theta}( n C_{1}-C_{2}) \nonumber \\
	(n \bar{C}_{3}-\bar{C}_{2})=e^{-2 \theta}( n C_{3}-C_{2}) \nonumber \\
	(2 \bar{C}_{2}-n \bar{C}_{1}-n \bar{C}_{3})=e^{-2 \theta}( 2 C_{2}- n C_{1}-n C_{3}) \nonumber 
	\end{gather} 
	under $\bar{g}_{\mu\nu}=e^{2\theta}g_{\mu\nu}$. This in turn means that any of the above combinations when squared or multiplied by another combination of the list, yields a conformally invariant scalar. For instance
	\beq
	\sqrt{-g}(n A_{1}-A_{3})^{2}
	\eeq
	\beq
	\sqrt{-g}( A_{2}-A_{4})B_{2}
	\eeq
	are both conformally invariant. Following the above procedure one can find more conformally  invariant quartic scalars.

	\chapter{Discussion/Conclusions}
	
	Metric-Affine Theories of Gravity have certainly paved the way towards a better understanding of gravitation. The enriched (non-Riemannian) geometry in such theories is well understood (in terms of torsion and non-metricity) and the modifications compared to GR have also a nice geometrical meaning. In addition, one of the main advantages of MAG is that it has direct link with the microscopic properties of matter\cite{lobo2015crystal}.  Therefore, it makes it easier to unify gravitation with other forces and as a possible result, may allow for the quantization of gravity\cite{hehl1995metric}. As pointed out in \cite{puetzfeld2008probing} the non-Riemannian spacetime geometry can only be detected by probing  matter with
	microstructure\footnote{See for instance \cite{roychowdhury2017non}.}. 
	
	In this thesis we tackled  and answered but few of the many questions that arise in such interesting geometries. Let us review what we have done here. After introducing the basic geometric elements that constitute a non-Riemannian geometry, we gave many of examples in order to  illustrate the role of torsion and non-metricity on geometrical grounds. Then, we went on and derived the field equations for specific Theories and later on we generalized for general Metric-Affine Theories. We paid special attention to $f(R)$ theories and presented another way to break the projective invariance in these theories. The peculiar case $f(R)=\alpha R^{2}$ was also separately studied in a next chapter and the cosmological solutions were given. We then extended a known method to generate torsion (by coupling surface terms to scalars) to include non-metricity. Some models where both torsion and non-metricity can be excited were presented.
	
	Having stressed out how important it is to have a tool to solve for the affine connection we proved, for the first time in the literature, a step by step way to solve for the affine connection. We started with certain assumptions about the additional action\footnote{More specifically we considered actions that are linear in the connection.} that may be added to the Einstein-Hilbert, generalized our result to $f(R)$ and then to arbitrary actions. We presented and proved our results as $3$ subsequent Theorems and applied each Theorem to a specific example to illustrate the procedure. We also discussed the cases of dynamical/non-dynamical connections and how our Theorems may applied to classify the dynamical content of the connection of a given theory.
	
	Continuing, we focused our attention on the effects of torsion and non-metricity in Cosmology. More specifically, after discussing the kinematics of FLRW universes with torsion and non-metricity, we presented the most general form that torsion can have in such highly symmetric spacetimes (the result was already known in the literature). In addition, using symmetry arguments we derived the most general form of non-metricity in such spacetimes. We also showed how the results are simplified in the case of Weyl and fixed length vector non-metricity. The modified Friedmann equations, in the presence of non-metricity, were also given in this Chapter. 
	
	Then, in Chapter $7$ we derived for the first time  in the literature the Raychaudhuri equation with both torsion and non-metricity. We should point out that in the presence of torsion only (along with curvature) the Raychaudhuri equation was known in the literature and has been derived independently from many groups (see for instance \cite{luz2017raychaudhuri,pasmatsiou2017kinematics,capozziello2001geometric,kar2007raychaudhuri} ). However, in the presence of non-metricity such expression was not known till now. We therefore derived the most general form of the Raychaudhuri equation by allowing the presence of both torsion and non-metricity (and also considered general dimension $n$). We then applied the results to cosmology and found the cosmological solutions of theories that have torsion of vectorial form and vanishing non-metricity. Then, we switched on Weyl non-metricity and considered vanishing torsion. It is worth noting that the solutions for the scale factor look identical upon exchanging the torsion and Weyl vectors. This is a consequence of the interrelation between vectorial torsion and Weyl non-metricity for projective invariant theories. We also found a solution for fixed length vector non-metricity and derived the evolution of vorticity for general non-Riemannian spaces.
	
	Finally, we defined  three possible scale transformations that one can consider in a Metric-Affine Geometry. These are, conformal transformations of the metric (with fixed connection), projective transformations of the connection (with fixed metric) and frame rescalings of the orthonormal frame that result in a combination of a conformal transformation of the metric along with a special projective transformation of the connection. We carefully obtained all the independent quadratic scalars for  pure torsion, pure non-metricity and mixed terms. We considered general quadratic theories and obtained the parameter space of the theories respecting each of the aforementioned transformations. We then, extended the discussion and also included parity violating terms. For this case we also found the parameter space for theories invariant under each transformation. The general field equations for all cases were also derived.
	
	Let us now discuss some future extensions of the above study. First of all let us note that the three Theorems for the affine connection may be used in order to study wide classes of MAG Theories and possibly classify theories with regards to the dynamical content of their connections. For instance, which subclasses of the general Horndeski's theory admit a non-dynamical connection? Upon what assumptions the connection becomes dynamical? These are but few cases where the connection Theorems can be applied. For instance we may just as well use the results of the Theorems to study teleparallel and symmetric teleparallel theories of gravity in the coordinate formalism. Turning our attention to the generalized Raychaudhuri equation, the applications are endless. For instance we could study how the singularity Theorems are modified, find solutions of astronomical and cosmological interest,  examine further what effects the non-metric terms produce an so on. Also it would be interesting to find solutions for the modified Friedmann equations with non-metricity that we presented in Chapter $6$ and compare them with the ones of pure torsion dominated universes.

	\chapter*{Appendix $A$}

	\subsection{Flat Space}
	As a first example let us consider the theory (in the Metric-Affine framework) given by the Einstein-Hilbert action alone. As we had shown, this theory admits an additional vectorial unspecified degree of freedom. In addition, we saw that this degree of freedom does not affect Einstein equations but it does affect the autoparallels as we will show now. As we proved, the affine connection for this theory takes the form
	\begin{equation}
	\Gamma^{\lambda}_{\;\;\;\mu\nu}= \tilde{\Gamma}^{\lambda}_{\;\;\;\mu\nu} -\frac{2}{(n-1)}S_{\nu}\delta_{\mu}^{\lambda}=\tilde{\Gamma}^{\lambda}_{\;\;\;\mu\nu}+\frac{1}{2 n}\delta_{\mu}^{\lambda}Q_{\nu}
	\end{equation}
	Now, to investigate whether there is a difference or not between geodesics and autoparallels let us consider a flat space, namely one that there is always a coordinate system in which we have $g_{\mu\nu}=\eta_{\mu\nu}=diag(-1,1,1,...,1)$\footnote{This holds for Lorentzian spaces. For a Riemannian space (only positive inner products) one would have $g_{\mu\nu}=\delta_{\mu\nu}$=diag(1,1,1,...,1).} and $\tilde{\Gamma}^{\lambda}_{\;\;\;\mu\nu}=0$. Then, the geodesic equation becomes
	\begin{equation}
	\frac{d^{2} x^{\mu}}{d\lambda^{2}}+\tilde{\Gamma}^{\mu}_{\;\;\;\alpha\beta}\frac{dx^{\alpha}}{d\lambda}\frac{dx^{\beta}}{d\lambda}=0\Rightarrow  \nonumber
	\end{equation}
	\begin{equation}
	\ddot{x}^{\mu}=0
	\end{equation}
	and integrating twice, it follows that
	\begin{equation}
	x^{\mu}(\lambda)=c^{\mu}\lambda+b^{\mu}, \;\;\;\;\; c^{\mu},b^{\mu}=const.
	\end{equation}
	which is the equation for a straight line as expected. Now let us find the autoparallel curves. Before doing so, let us slightly generalize and compute how a general vector field $u^{\mu}$ changes under parallel transport along a given curve. To this end we use the parallel transport equation that we have already given previously, for the connection at hand. We have
	\begin{gather}
	\dot{u}^{\mu}+\Gamma^{\mu}_{\;\;\;\alpha\beta}u^{\alpha}\dot{x}^{\beta}=0
	\end{gather}
	but for the given model and since we are considering flat space, it holds that
	\begin{equation}
	\Gamma^{\lambda}_{\;\;\;\mu\nu}= \underbrace{\tilde{\Gamma}^{\lambda}_{\;\;\;\mu\nu}}_{=0}+\frac{1}{2 n}\delta_{\mu}^{\lambda}Q_{\nu}=\frac{1}{2 n}\delta_{\mu}^{\lambda}Q_{\nu}
	\end{equation}
	so that
	\begin{gather}
	\dot{u}^{\mu}+\frac{1}{2 n}Q_{\nu}\frac{d x^{\nu}}{d\lambda}u^{\mu}=0 \Rightarrow \nonumber
	du^{\mu}+\frac{1}{2 n}Q_{\nu}d x^{\nu}u^{\mu}=0
	\end{gather}
	Now, multiplying through by $e^{\frac{1}{2 n}\int Q_{\nu}d x^{\nu}}$ it follows that
	\begin{gather}
	e^{\frac{1}{2 n}\int Q_{\nu}d x^{\nu}}du^{\mu}+\frac{1}{2 n}Q_{\nu}d x^{\nu}e^{\frac{1}{2 n}\int Q_{\nu}d x^{\nu}}u^{\mu}\Rightarrow  \nonumber \\
	d\Big( u^{\mu} e^{\frac{1}{2 n}\int Q_{\nu}d x^{\nu}}  \Big)= 0 \Rightarrow \nonumber \\
	u^{\mu} e^{\frac{1}{2 n}\int Q_{\nu}d x^{\nu}}=const.=u^{\mu}(0)
	\end{gather}
	Therefore
	\begin{equation}
	u^{\mu}=u^{\mu}(0)e^{-\frac{1}{2 n}\int Q_{\nu}d x^{\nu}}
	\end{equation}
	where $u^{\mu}(0)$ is the initial value of the vector field. If the loop is closed the above becomes
	\begin{equation}
	u^{\mu}=u^{\mu}(0)e^{-\frac{1}{2 n}\oint_{C} Q_{\nu}d x^{\nu}}
	\end{equation}
	Thus, the magnitude changes according to
	\begin{equation}
	\| u\|^{2}=u^{\mu}u_{\mu}= \| u(0)\|^{2}e^{-\frac{1}{ n}\oint_{C} Q_{\nu}d x^{\nu}}
	\end{equation}
	From the last two equations we conclude that when we parallel transport a vector in flat space (but in the presence of torsion and non-metricity!\footnote{To be more specific, in the presence of a vectorial degree of freedom that gives torsion and non-metricity as given in this model.}) its direction remains the same but its magnitude changes. If we take now $u^{\mu}$ to be the tangent vector on the curve, that is $u^{\mu}=\dot{x}^{\mu}=\frac{d x^{\mu}}{d\lambda}$, the latter becomes
	\begin{equation}
	\dot{x}^{\mu}=\dot{x}^{\mu}(0)e^{-\frac{1}{2 n}\oint_{C} Q_{\nu}d x^{\nu}}
	\end{equation}
	Integrating the above once more, we derive the autoparallel curves
	\begin{equation}
	x^{\mu}(\lambda)=x^{\mu}(0)+\dot{x}^{\mu}(0)\int e^{-\frac{1}{2 n}\oint_{C} Q_{\nu}d x^{\nu}} d\lambda
	\end{equation}
	As we can see now, it happens that autoparallels are also straight lines in this space. Indeed, using the above, one can write
	\begin{equation}
	\frac{x^{0}(\lambda)-x^{0}(0)}{\dot{x}^{0}(0)}=\frac{x^{1}(\lambda)-x^{1}(0)}{\dot{x}^{1}(0)}=...=\frac{x^{n-1}(\lambda)-x^{n-1}(0)}{\dot{x}^{n-1}(0)}
	\end{equation}
	which are the equations of straight lines in n-dim space.

	\subsection{Poincare half-plane}
	Next we consider the Poincare half-plane which is given by the metric
	\begin{equation}
	ds^{2}=\frac{1}{y^{2}}\Big( dx^{2}+dy^{2} \Big) \;, \;\;\;y>0
	\end{equation}
	that is, a space with metric tensor $g_{ij}=\frac{1}{y^{2}} diag(1,1)=\frac{1}{y^{2}} \delta_{ij}$, $i,j=1,2$. As  it can be easily seen, the non-vanishing Christoffel symbols for the given metric are
	\begin{equation}
	\tilde{\Gamma}^{2}_{\;\;\;22}=\tilde{\Gamma}^{1}_{\;\;\;12}=\tilde{\Gamma}^{1}_{\;\;\;21}=-\tilde{\Gamma}^{2}_{\;\;\;11}=-\frac{1}{y}
	\end{equation} 
	Therefore, the geodesic equations
	\begin{equation}
	\ddot{x}^{k}+\tilde{\Gamma}^{k}_{\;\;\; ij}\dot{x}^{i}\dot{x}^{j}=0
	\end{equation}
	read
	\begin{equation}
	\ddot{x}-2\frac{\dot{x}\dot{y}}{y}=0 \label{poin1}
	\end{equation}
	\begin{equation}
	\ddot{y}+\frac{\dot{x}^{2}}{y}-\frac{\dot{y}^{2}}{y}=0
	\end{equation}
	To solve this system, first assume that $\dot{x}\neq 0$, then dividing ($\ref{poin1}$) by it we obtain
	\begin{gather}
	\frac{\ddot{x}}{\dot{x}}-2\frac{\dot{y}}{y}=0\Rightarrow  \nonumber \\
	\frac{d}{d\lambda}\Big( \ln{\dot{x}}-\ln{y^{2}} \Big)=0\Rightarrow \frac{d}{d\lambda}\left(\ln{\frac{\dot{x}}{y^{2}}}\right)=0 \Rightarrow  \nonumber \\
	\frac{\dot{x}}{y^{2}}=const.=c_{1} \label{firstint}
	\end{gather}
	In addition, multiplying the second equation by $1/y$ we have
	\begin{equation}
	\frac{y\ddot{y}-\dot{y}^{2}}{y^{2}}+\frac{\dot{x}}{y^{2}}\dot{x}=0
	\end{equation}
	Now, noting that
	\begin{equation}
	\frac{d}{d\lambda}\left( \frac{\dot{y}}{y}\right)=\frac{y\ddot{y}-\dot{y}^{2}}{y^{2}}
	\end{equation}
	and substituting $\frac{\dot{x}}{y^{2}}=c_{1}$ from ($\ref{firstint}$), it follows that
	\begin{gather}
	\frac{d}{d\lambda}\left( \frac{\dot{y}}{y}\right)+c_{1}\dot{x}=0 \Rightarrow  \nonumber
	\frac{d}{d\lambda}\left( \frac{\dot{y}}{y}+c_{1}x\right)=0 
	\end{gather}
	that is
	\begin{equation}
	\frac{\dot{y}}{y}+c_{1}x= const.=c_{2}
	\end{equation}
	and multiplying through by $y^{2}$ we finally arrive at
	\begin{gather}
	y\dot{y}+\underbrace{c_{1}y^{2}}_{=\dot{x}}x=\underbrace{y^{2}}_{=\dot{x}/c_{1}}c_{2} \Rightarrow \nonumber \\
	y\dot{y}+x\dot{x}-\frac{c_{2}}{c_{1}}\dot{x}=0 \Rightarrow \nonumber \\
	\frac{d}{d\lambda}\left(\frac{y^{2}}{2}+\frac{x^{2}}{2}-\frac{c_{2}}{c_{1}}x \right)=0 \Rightarrow \nonumber \\
	y^{2}+x^{2}-2\frac{c_{2}}{c_{1}}=c_{3}
	\end{gather}
	completing the square in $x$ we then find that
	\begin{gather}
	y^{2}+x^{2}-2\frac{c_{2}}{c_{1}}+\left(\frac{c_{2}}{c_{1}}\right)^{2}=c_{3}+\left(\frac{c_{2}}{c_{1}}\right)^{2}\equiv a^{2}\Rightarrow  \nonumber \\
	\left(x-\frac{c_{2}}{c_{1}}\right)^{2}+y^{2}=a^{2}
	\end{gather}
	which represent half-circles in the upper half-plane ($y>0$). These are the one type of geodesics in the Poincare half-plane. Notice that in arriving to this result we have assumed that $\dot{x}\neq 0$. So, we should also solve the geodesic equations in the case $\dot{x}= 0$ $\Rightarrow x(\lambda)=const.=c_{4}$. Then, the first geodesic equation is trivially satisfied and from the second one we have
	\begin{gather}
	\frac{y\ddot{y}-\dot{y}^{2}}{y^{2}}=0\Rightarrow 
	\frac{d}{d\lambda}\left( \frac{\dot{y}}{y}\right)=0 \Rightarrow 
	\frac{\dot{y}}{y}=const.=c_{5}
	\end{gather}
	which, once integrated gives
	\begin{equation}
	y(\lambda) \propto e^{c_{5}\lambda}
	\end{equation}
	The latter one along with $x(\lambda)=const.=c_{4}$ represent half-lines in the upper half-plane. Therefore we conclude that in the Poincare half-plane there exist two kinds of geodesics\footnote{The initial conditions, of course, specify in which one among these one is on.}, half-circles and half-lines. It is interesting to look now for solutions of the autoparallel equations in the Poincare half-plane and see whether they are the same or not with geodesics. We again consider the model where the connection is given by
	\begin{equation}
	\Gamma^{k}_{\;\;\;ij}=\tilde{\Gamma}^{k}_{\;\;\;ij}+\frac{1}{2 n}\delta_{i}^{k}Q_{j}
	\end{equation}  
	Then, the autoparallel equations are
	\begin{equation}
	\ddot{x}^{k}+\tilde{\Gamma}^{k}_{\;\;\;ij}\dot{x}^{i}\dot{x}^{j}=-\frac{1}{2 n}Q_{j}\dot{x}^{j}\dot{x}^{k}
	\end{equation}
	Thus, setting $k=1$ and $k=2$ respectively, we get
	\begin{equation}
	\ddot{x}-2\frac{\dot{x}\dot{y}}{y}=-\frac{1}{2 n}(Q_{j}\dot{x}^{j})\dot{x}
	\end{equation}
	and
	\begin{equation}
	\ddot{y}+\frac{\dot{x}^{2}}{y}-\frac{\dot{y}^{2}}{y}=-\frac{1}{2 n}(Q_{j}\dot{x}^{j})\dot{y}
	\end{equation} 
	For $\dot{x}\neq0$ the first one becomes
	\begin{gather}
	\frac{\ddot{x}}{\dot{x}}-2\frac{\dot{y}}{y}=-\frac{1}{2 n}(Q_{j}\dot{x}^{j}) \Rightarrow \nonumber \\
	\frac{d}{d\lambda}\Big( \ln{\frac{\dot{x}}{y^{2}}} \Big)=-\frac{1}{2 n}(Q_{j}\dot{x}^{j})
	\end{gather}
	such that
	\begin{gather}
	\ln{\frac{\dot{x}}{y^{2}}} =-\frac{1}{2 n}\int Q_{j}dx^{j}+C_{1} \Rightarrow \nonumber \\
	\frac{\dot{x}}{y^{2}}=A e^{-\frac{1}{2 n}\int Q_{j}dx^{j}} \label{aqwe}
	\end{gather}
	where $A=e^{C_{1}}$. This is a first integral of the system. Taking the second equation now, and dividing by $y$ it follows that
	\begin{gather}
	\frac{\ddot{y}y-\dot{y}^{2}}{y^{2}}+\frac{\dot{x}}{y^{2}}\dot{x}=-\frac{1}{2 n}(Q_{j}\dot{x}^{j})\frac{\dot{y}}{y} \Rightarrow \nonumber \\
	\frac{d}{d\lambda}\left(\frac{\dot{y}}{y} \right)+ A e^{-\frac{1}{2 n}\int Q_{j}dx^{j}} \dot{x}=-\frac{1}{2 n}(Q_{j}\dot{x}^{j})\frac{\dot{y}}{y}
	\end{gather}
	where we have employed ($\ref{aqwe}$). Multiplying the latter by $e^{+\frac{1}{2 n}\int Q_{j}dx^{j}}$ we obtain 
	\begin{equation}
	\underbrace{e^{\frac{1}{2 n}\int Q_{j}dx^{j}}d\left(\frac{\dot{y}}{y} \right)+\frac{1}{2 n}(Q_{i}dx^{i})e^{\frac{1}{2 n}\int Q_{j}dx^{j}}\frac{\dot{y}}{y}}_{\equiv \frac{d}{d\lambda}\left( \frac{\dot{y}}{y} e^{\frac{1}{2 n}\int Q_{j}dx^{j}} \right)}+A\dot{x}=0 \Rightarrow \nonumber
	\end{equation}
	\begin{equation}
	\frac{d}{d\lambda}\left( \frac{\dot{y}}{y} e^{\frac{1}{2 n}\int Q_{j}dx^{j}} +A x \right)=0
	\end{equation}
	such that 
	\begin{equation}
	\frac{\dot{y}}{y} e^{\frac{1}{2 n}\int Q_{j}dx^{j}} +A x =const.=B
	\end{equation}
	Now, in order to eliminate the exponential factor from the latter we use equation ($\ref{aqwe}$), namely
	\begin{equation}
	e^{\frac{1}{2 n}\int Q_{j}dx^{j}}=\frac{A y^{2}}{\dot{x}}
	\end{equation}
	and the above recasts to
	\begin{equation}
	A\frac{y \dot{y}}{\dot{x}}+A x-B=0
	\end{equation}
	or
	\begin{gather}
	y\dot{y}+x\dot{x}-\frac{B}{A}\dot{x}=0\Rightarrow \nonumber \\
	\frac{d}{d\lambda}\left(\frac{y^{2}}{2}+\frac{x^{2}}{2}-\frac{B}{A}x \right)=0 \Rightarrow \nonumber
	y^{2}+x^{2}-2\frac{B}{A}x=C
	\end{gather}
	Completing the square again, we arrive at
	\begin{equation}
	\left( x-\frac{B}{A}\right)^{2}+y^{2}=r_{0}^{2}, \;\;\; r_{0}^{2}=C+\left(\frac{B}{A}\right)^{2}
	\end{equation}
	Thus, one type of geodesics is again half-circles. Now, for $\dot{x}=0$ we have
	\begin{gather}
	\ddot{y}-\frac{\dot{y}^{2}}{y}=-\frac{1}{2 n}Q_{2}\dot{y}^{2}\Rightarrow  \nonumber \\
	\frac{d}{d\lambda}\left( \frac{\dot{y}}{y}\right)=-\frac{1}{2 n}Q_{2}\dot{y} \frac{\dot{y}}{y}
	\end{gather}
	Setting $u \frac{\dot{y}}{y}=$ it follows that
	\begin{gather}
	\frac{du}{u}=-\frac{1}{2 n}Q_{2}dy\Rightarrow u=C_{0}e^{-\frac{1}{2 n}\int Q_{2}dy} \Rightarrow  \nonumber \\
	\dot{y}=y C_{0}e^{-\frac{1}{2 n}\int Q_{2}dy}\Rightarrow \int \frac{1}{y}e^{-\frac{1}{2 n}\int Q_{2}dy} dy =C_{0}\lambda +const.
	\end{gather}
	When the latter is reversed to define $y=y(\lambda)$ together with $x(\lambda)=const.$ will again represent straight half-lines in the upper half-plane. Therefore we conclude that the autoparallels are exactly the same with the geodesics in the Poincare half-plane and for the given metric. The reason for that is due to the fact that the connection
	\begin{equation}
	\Gamma^{k}_{\;\;\;ij}=\tilde{\Gamma}^{k}_{\;\;\;ij}+\frac{1}{2 n}\delta_{i}^{k}Q_{j}
	\end{equation}
	is projectively equivalent to $\tilde{\Gamma}^{k}_{\;\;\;ij}$. And, as it is well known from theory, two connections $\Gamma^{k}_{\;\;\;ij}$, $C^{k}_{\;\;\;ij}$ that are projectively equivalent, i.e. there exists a vector $a_{i}$ such that
	\begin{equation}
	\Gamma^{k}_{\;\;\;(ij)}=C^{k}_{\;\;\;(ij)}+\delta_{i}^{k}a_{j}+\delta_{j}^{k}a_{i}
	\end{equation}
	share the same autoparallel curves! In our case 
	\begin{equation}
	\Gamma^{k}_{\;\;\;(ij)}=\tilde{\Gamma}^{k}_{\;\;\;(ij)}+\frac{1}{ 4 n}\delta_{i}^{k}Q_{j}+\frac{1}{ 4 n}\delta_{j}^{k}Q_{i}
	\end{equation}
	and by comparison with the above we conclude that $C^{k}_{\;\;\;(ij)}=\tilde{\Gamma}^{k}_{\;\;\;(ij)}$ and $a_{i}=\frac{1}{4 n}Q_{i}$. That is, the connections $\Gamma^{k}_{\;\;\;ij}$ and $\tilde{\Gamma}^{k}_{\;\;\;ij}$ are projectively equivalent and therefore share the same autoparallels which are the geodesics of $\tilde{\Gamma}^{k}_{\;\;\;ij}$ that we have already found.
	\subsection{Illustrative examples}
	Let us now consider two simple cases in which the effect of torsion and non-metricity is apparent and produces deviations from the geodesic motion. Rewriting the equation for autoparallels 
	\begin{gather}
	\frac{d^{2} x^{\lambda}}{d\lambda^{2}}+\tilde{\Gamma}^{\lambda}_{\;\;\;\mu\nu}\frac{dx^{\mu}}{d\lambda}\frac{dx^{\nu}}{d\lambda}= \nonumber \\
	=-g^{\alpha\lambda}\frac{dx^{\mu}}{d\lambda}\frac{dx^{\nu}}{d\lambda}\left[ \frac{1}{2}Q_{\mu\nu\alpha}+\frac{1}{2}Q_{\mu\alpha\nu}-\frac{1}{2}Q_{\alpha\mu\nu}-2S_{\alpha\mu\nu}\right]
	\end{gather}
	let us suppose that a configuration of torsion and non-metricity exists such that
	\begin{equation}
	\left[\frac{1}{2}Q_{(\mu\nu)\alpha}+\frac{1}{2}Q_{(\mu\mid \alpha\mid \nu)}-\frac{1}{2}Q_{\alpha(\mu\nu)}-2S_{\alpha(\mu\nu)}\right]= a_{\alpha}g_{\mu\nu}
	\end{equation}
	where $a_{\mu}$ represents a vectorial degree of freedom. Then, the above equation becomes
	\begin{gather}
	\frac{d^{2} x^{\lambda}}{d\lambda^{2}}+\tilde{\Gamma}^{\lambda}_{\;\;\;\mu\nu}\frac{dx^{\mu}}{d\lambda}\frac{dx^{\nu}}{d\lambda}= \nonumber \\
	=-g^{\alpha\lambda} a_{\alpha}g_{\mu\nu}=-a^{\lambda}g_{\mu\nu}\frac{dx^{\mu}}{d\lambda}\frac{dx^{\nu}}{d\lambda}\Rightarrow  \nonumber
	\end{gather}
	\begin{equation}
	\ddot{x}^{\lambda}+\tilde{\Gamma}^{\lambda}_{\;\;\;\mu\nu}\dot{x}^{\mu}\dot{x}^{\nu}=-a^{\lambda}\dot{x}^{\mu}\dot{x}_{\mu}
	\end{equation}
	We now proceed by solving the latter in the case of $2-dim$ Euclidean flat space  as well as for the Poincare half-plane, for specific choices of $a_{\mu}$.
	\subsubsection{$2-dim$ Euclidean flat space}
	For this space the geodesics are of course straight lines. However, as we will show, the autoparallels (for this configuration) are not. We have
	\begin{equation}
	g_{ij}=\delta_{ij}
	\end{equation}
	and
	\begin{equation}
	\tilde{\Gamma}^{i}_{\;\;\;jk}=0
	\end{equation}
	Furthermore, taking the vector $a_{i}$ to be
	\begin{equation}
	a_{i}=(1,1)
	\end{equation}
	the differential equations giving the autoparallel curves, become
	\begin{gather}
	\ddot{x}=-(\dot{x}^{2}+\dot{y}^{2}) \label{sdx} \\
	\ddot{y}=-(\dot{x}^{2}+\dot{y}^{2})
	\end{gather}
	Therefore
	\begin{equation}
	\ddot{x}=\ddot{y} \Rightarrow 
	\end{equation}
	\begin{equation}
	\dot{x}=\dot{y}+c_{1} \Rightarrow \label{iok}
	\end{equation}
	\begin{equation}
	x=y+c_{1}\lambda+c_{2} 
	\end{equation}
	Now, inserting ($\ref{iok}$) into ($\ref{sdx}$) in order to to eliminate $\dot{y}$, it follows that
	\begin{equation}
	\ddot{x}=-\Big[ \dot{x}^{2}+(\dot{x}-c_{1})^{2}\Big]
	\end{equation}
	To solve this, consider the transformation 
	\begin{equation}
	z=\dot{x}-\frac{c_{1}}{2} \Rightarrow 
	\end{equation}
	\begin{equation}
	\dot{z}=\ddot{x}
	\end{equation}
	such that
	\begin{equation}
	\dot{z}=-\Big[ \left( z+\frac{c_{1}}{2}\right)^{2}+\left( z-\frac{c_{1}}{2}\right)^{2}\Big]= -2\left[ z^{2}+\left(\frac{c_{1}}{2}\right)^{2} \right]\Rightarrow 
	\end{equation}
	\begin{equation}
	\frac{dz}{\left[ z^{2}+\left(\frac{c_{1}}{2}\right)^{2} \right]}=-2d\lambda
	\end{equation}
	which, upon integration, gives
	\begin{equation}
	\frac{1}{\left(\frac{c_{1}}{2}\right)}\arctan{\frac{z}{\left(\frac{c_{1}}{2}\right)}}=-2\lambda +\tilde{c}_{3} \Rightarrow \nonumber
	\end{equation}
	\begin{equation}
	z=\frac{c_{1}}{2}\tan{\Big(c_{3}-c_{1}\lambda \Big)}
	\end{equation}
	where $c_{3}=\tilde{c}_{3}c_{1}/2$. Therefore
	\begin{equation}
	\dot{x}=\frac{c_{1}}{2}\left[ 1+\tan{\Big(c_{3}-c_{1}\lambda \Big)} \right] 
	\end{equation}
	which with a final integration results in
	\begin{equation}
	x(\lambda)=\frac{c_{1}}{2}\left[ \lambda+\frac{1}{c_{1}}\ln{\mid \cos{(c_{3}-c_{1}\lambda)} \mid} \right]+c_{4}
	\end{equation}
	Furthermore, using the fact that
	\begin{equation}
	x=y+c_{1}\lambda +c_{2} \label{ddf}
	\end{equation}
	we also find $y$ in terms of $\lambda$,
	\begin{equation}
	y(\lambda )=\frac{c_{1}}{2}\left[ -\lambda+\frac{1}{c_{1}}\ln{\mid \cos{(c_{3}-c_{1}\lambda)} \mid} \right] +(c_{4}-c_{2})
	\end{equation}
	and we have find the parametric solution for autoparallels. Going one step further we can solve ($\ref{ddf}$) for $\lambda$ and eliminate it to express the solution as
	\begin{equation}
	x+y=\ln{\mid \cos{(c_{3}+c_{2}+y-x)} \mid} +(2 c_{4}-c_{2})
	\end{equation}
	from which it is now apparent that the solutions are not, in general ( that is for $c_{1}\neq 0$ ), straight lines.
	\subsubsection{Poincare half-plane}
	As another example we consider again the Poincare half-plane. For this space, as we proved before, there exist two kind of geodesics, half-circles and half-lines on the upper half-plane. It is interesting to see know how the autoparallels would look like considering again a torsion and non-metricity configuration which gives a connection of the form 
	\begin{equation}
	\Gamma^{i}_{\;\;\;j k}=\tilde{\Gamma}^{i}_{\;\;\;j k}+a^{i}g_{j k}
	\end{equation}
	Thus, the autoparallel curves satisfy
	\begin{equation}
	\ddot{x}^{i}+\tilde{\Gamma}^{i}_{\;\;\;j k}\dot{x}^{j}\dot{x}^{k}=-\dot{x}^{j}\dot{x}^{k}a^{i}g_{j k}
	\end{equation}
	namely (setting $i=1,2$ respectively)
	\begin{equation}
	\ddot{x}-2\frac{\dot{x}\dot{y}}{y}=-\frac{1}{y^{2}}(\dot{x}^{2}+\dot{y}^{2}) a^{1}
	\end{equation}
	\begin{equation}
	\ddot{y}+\frac{\dot{x}^{2}}{y}-\frac{\dot{y}^{2}}{y}=-\frac{1}{y^{2}}(\dot{x}^{2}+\dot{y}^{2}) a^{2}
	\end{equation}
	Now, this can be solved most easily\footnote{And at the same time avoiding the triviality $a^{i}=(0,0)$.} by taking a vector $a^{i}$ that goes like 
	\begin{equation}
	a^{i}=(0,-y)
	\end{equation}
	Then, the above equations read
	\begin{equation}
	\ddot{x}=2\frac{\dot{x}\dot{y}}{y}
	\end{equation}
	\begin{equation}
	\ddot{y}=\frac{2\dot{y}^{2}}{y}
	\end{equation}
	For $\dot{x}\neq 0$ we divide the two to arrive at
	\begin{equation}
	\frac{\ddot{x}}{\dot{x}}=\frac{\ddot{y}}{\dot{y}}
	\end{equation}
	That is
	\begin{equation}
	\frac{d}{d\lambda}\left( \ln{\frac{\dot{y}}{\dot{x}}} \right)= 0 \Rightarrow  \nonumber 
	\end{equation}
	\begin{equation}
	\dot{y}=c_{1}\dot{x}\Rightarrow \nonumber
	\end{equation}
	\begin{equation}
	y=c_{1}x+c_{2}
	\end{equation}
	which represent straight lines on the upper half-plane. For $\dot{x}=0$ the solutions are
	\begin{gather}
	x=c_{1} \nonumber \\
	y=\frac{1}{c_{3}-c_{2}\lambda}
	\end{gather}
	which are again straight lines that are parallel to the $y-axis$. Therefore, we conclude that for the given torsion- non-metricity configuration, on the Poincare plane, the autoparallels are only straight lines.

	\section{Projective equivalent connections}
	Let us state and prove here a well-known Theorem about equivalent connections. The Theorem states that given two symmetric connections $\Gamma^{\lambda}_{\;\;\;\;\mu\nu}$ and
	$\tilde{\Gamma}^{\lambda}_{\;\;\;\;\mu\nu}$ related by
	\beq
	\Gamma^{\lambda}_{\;\;\;\;\mu\nu}=\tilde{\Gamma}^{\lambda}_{\;\;\;\;\mu\nu}+\delta^{\lambda}_{\mu}A_{\nu}+\delta^{\lambda}_{\nu}A_{\mu} \label{auto}
	\eeq
	where $A_{\mu}$ is an arbitrary vector field, the two connections define the same autoparallel curves, just with a different parametrization. Indeed, suppose $\tilde{C}:\tilde{x}^{\mu}=\tilde{x}^{\mu}{(\lambda)}$( $\lambda$ being the curve parameter) is the autoparallel curve derived from $\tilde{\Gamma}^{\lambda}_{\;\;\;\;\mu\nu}$ and therefore satisfies
	\beq
	\frac{d^{2}\tilde{x}^{\alpha}}{d\lambda^{2}}+\tilde{\Gamma}^{\alpha}_{\;\;\;\;\mu\nu}\frac{d\tilde{x}^{\mu}}{d\lambda}\frac{d\tilde{x}^{\nu}}{d\lambda}=0
	\eeq
	Now let $C:x^{\mu}=x^{\mu}{(\lambda)}$ be the autoparallel curve satisfied by $\Gamma^{\lambda}_{\;\;\;\;\mu\nu}$  and so
	\beq
	\frac{d^{2}x^{\alpha}}{d\lambda^{2}}+\Gamma^{\alpha}_{\;\;\;\;\mu\nu}\frac{d x^{\mu}}{d\lambda}\frac{d x^{\nu}}{d\lambda}=0
	\eeq
	we state that the connection $\Gamma^{\lambda}_{\;\;\;\;\mu\nu}$  in the latter equation can be replaced by $\tilde{\Gamma}^{\lambda}_{\;\;\;\;\mu\nu}$ if the parametrization of the curved is changed. To see this let us expand $\Gamma^{\lambda}_{\;\;\;\;\mu\nu}$  in the above using equation $(\ref{auto})$ to obtain
	\beq
	\frac{d^{2}x^{\alpha}}{d\lambda^{2}}+\tilde{\Gamma}^{\alpha}_{\;\;\;\;\mu\nu}\frac{d x^{\mu}}{d\lambda}\frac{d x^{\nu}}{d\lambda}=-(\delta^{\alpha}_{\mu}A_{\nu}+\delta^{\alpha}_{\nu}A_{\mu})\frac{d x^{\mu}}{d\lambda}\frac{d x^{\nu}}{d\lambda} \nonumber \Rightarrow 
	\eeq
	\beq
	\frac{d^{2}x^{\alpha}}{d\lambda^{2}}+\tilde{\Gamma}^{\alpha}_{\;\;\;\;\mu\nu}\frac{d x^{\mu}}{d\lambda}\frac{d x^{\nu}}{d\lambda}=-\left( 2 A_{\mu}\frac{d x^{\mu}}{d \lambda} \right) \frac{d x^{\alpha}}{d \lambda} \nonumber \Rightarrow 
	\eeq
	\beq
	\frac{d^{2}x^{\alpha}}{d\lambda^{2}}+\tilde{\Gamma}^{\alpha}_{\;\;\;\;\mu\nu}\frac{d x^{\mu}}{d\lambda}\frac{d x^{\nu}}{d\lambda}=f(\lambda) \frac{d x^{\alpha}}{d \lambda}  
	\eeq
	where we have set 
	\beq
	f(\lambda)= - 2 A_{\mu}\frac{d x^{\mu}}{d \lambda} 
	\eeq
	Now consider the change of variables $s=s(\lambda)$. Using the chain rule it follows that
	\beq
	\frac{d x^{\alpha}}{d \lambda}=\frac{d x^{\alpha}}{d s}\frac{d s}{d \lambda}=\frac{d x^{\alpha}}{d \lambda} \dot{s}
	\eeq 
	as well as
	\beq
	\frac{d^{2}x^{\alpha}}{d\lambda^{2}}=\frac{d^{2}x^{\alpha}}{d s^{2}}\dot{s}^{2}+\frac{d x^{\alpha}}{d \lambda} \ddot{s}
	\eeq
	where the dot denotes differentiation with respect to $\lambda$. Plugging these into our autoparallel equation, we obtain
	\beq
	\frac{d^{2}x^{\alpha}}{d s^{2}}+\tilde{\Gamma}^{\alpha}_{\;\;\;\;\mu\nu}\frac{d x^{\mu}}{d s}\frac{d x^{\nu}}{d s}=\frac{1}{\dot{s}^{2}}\Big(f(\lambda)\dot{s}-\ddot{s}\Big) \frac{d x^{\alpha}}{d s}  
	\eeq
	from which we see that if we choose $s(\lambda)$  such that 
	\beq
	f(\lambda)\dot{s}-\ddot{s}=0
	\eeq
	the right hand side vanishes and the autoparallel equation is identical to the one satisfied by $\tilde{\Gamma}^{\alpha}_{\;\;\;\;\mu\nu}$. In addition, integrating twice the latter differential equation we find the exact re-parametrization that we need to perform
	\beq
	s(\lambda)=\int g(\lambda)d\lambda
	\eeq
	where
	\beq
	g(\lambda)= e^{\int f(\lambda) d \lambda}=e^{-\int 2 A_{\mu}dx^{\mu}}
	\eeq
	Therefore we conclude that two symmetric connections $\Gamma^{\lambda}_{\;\;\;\;\mu\nu}$ and
	$\tilde{\Gamma}^{\lambda}_{\;\;\;\;\mu\nu}$ related by
	\beq
	\Gamma^{\lambda}_{\;\;\;\;\mu\nu}=\tilde{\Gamma}^{\lambda}_{\;\;\;\;\mu\nu}+\delta^{\lambda}_{\mu}A_{\nu}+\delta^{\lambda}_{\nu}A_{\mu} \label{auto}
	\eeq
	share the same autoparallel curves but with a different  parametrization in general. As we do throughout  the thesis we use this result where we take $\Gamma^{\lambda}_{\;\;\;\;\mu\nu}$ to be our general affine connection and identify $\tilde{\Gamma}^{\lambda}_{\;\;\;\;\mu\nu}$ with the Levi-Civita connection.

	\section{Projective equivalent connections-Extension to non-symmetric connections}
	Let us generalize now the previous Theorem in the case where the connection is not necessarily symmetric, that is we allow torsion in our space. We claim that two connections $\Gamma^{\lambda}_{\;\;\;\;\mu\nu}$ and $\tilde{\Gamma}^{\lambda}_{\;\;\;\;\mu\nu}$ related by
	\beq
	\Gamma^{\lambda}_{\;\;\;\;\mu\nu}=\tilde{\Gamma}^{\lambda}_{\;\;\;\;\mu\nu}+\delta^{\lambda}_{\mu}A_{\nu}+\delta^{\lambda}_{\nu}B_{\mu} +K_{\mu\nu}^{\;\;\;\lambda}
	\eeq
	where $A_{\mu}$, $B_{\mu}$ are arbitrary vector fields and  $K_{\mu\nu}^{\;\;\;\lambda}$ is a tensor that is antisymmetric in $\mu,\nu$. As before suppose $\tilde{C}:\tilde{x}^{\mu}=\tilde{x}^{\mu}{(\lambda)}$ is the autoparallel curve for $\tilde{\Gamma}^{\lambda}_{\;\;\;\;\mu\nu}$, that is
	\beq
	\frac{d^{2}\tilde{x}^{\alpha}}{d\lambda^{2}}+\tilde{\Gamma}^{\alpha}_{\;\;\;\;\mu\nu}\frac{d\tilde{x}^{\mu}}{d\lambda}\frac{d\tilde{x}^{\nu}}{d\lambda}=0
	\eeq
	Now let $C:x^{\mu}=x^{\mu}{(\lambda)}$ be the autoparallel curve of $\Gamma^{\lambda}_{\;\;\;\;\mu\nu}$  and so
	\beq
	\frac{d^{2}x^{\alpha}}{d\lambda^{2}}+\Gamma^{\alpha}_{\;\;\;\;\mu\nu}\frac{d x^{\mu}}{d\lambda}\frac{d x^{\nu}}{d\lambda}=0 \nonumber \Rightarrow 
	\eeq
	\beq
	\frac{d^{2}x^{\alpha}}{d\lambda^{2}}+\tilde{\Gamma}^{\alpha}_{\;\;\;\;\mu\nu}\frac{d x^{\mu}}{d\lambda}\frac{d x^{\nu}}{d\lambda}=-(\delta^{\alpha}_{\mu}A_{\nu}+\delta^{\alpha}_{\nu}B_{\mu})\frac{d x^{\mu}}{d\lambda}\frac{d x^{\nu}}{d\lambda}+K_{\mu\nu}^{\;\;\;\lambda}\frac{d x^{\mu}}{d\lambda}\frac{d x^{\nu}}{d\lambda}  \nonumber \Rightarrow 
	\eeq
	\beq
	\frac{d^{2}x^{\alpha}}{d\lambda^{2}}+\tilde{\Gamma}^{\alpha}_{\;\;\;\;\mu\nu}\frac{d x^{\mu}}{d\lambda}\frac{d x^{\nu}}{d\lambda}=-\left( ( A_{\mu}+B_{\mu})\frac{d x^{\mu}}{d \lambda} \right) \frac{d x^{\alpha}}{d \lambda} \nonumber \Rightarrow 
	\eeq
	\beq
	\frac{d^{2}x^{\alpha}}{d\lambda^{2}}+\tilde{\Gamma}^{\alpha}_{\;\;\;\;\mu\nu}\frac{d x^{\mu}}{d\lambda}\frac{d x^{\nu}}{d\lambda}=f(\lambda) \frac{d x^{\alpha}}{d \lambda}  
	\eeq
	where now we have set
	\beq
	f(\lambda)= - (A_{\mu}+B_{\mu})\frac{d x^{\mu}}{d \lambda} 
	\eeq
	and on going from the second to the third line we have used the fact that $K_{\mu\nu}^{\;\;\;\lambda}\frac{d x^{\mu}}{d\lambda}\frac{d x^{\nu}}{d\lambda}=0$ since $K_{\mu\nu}^{\;\;\;\lambda}$ is antisymmetric in $\mu,\nu$.
	In the exact same way we did for the symmetric connection, we again consider the re-parametrization $s=s(\lambda)$ and the above equation takes the form
	\beq
	\frac{d^{2}x^{\alpha}}{d s^{2}}+\tilde{\Gamma}^{\alpha}_{\;\;\;\;\mu\nu}\frac{d x^{\mu}}{d s}\frac{d x^{\nu}}{d s}=\frac{1}{\dot{s}^{2}}\Big(f(\lambda)\dot{s}-\ddot{s}\Big) \frac{d x^{\alpha}}{d s}  
	\eeq
	and as before we choose $s(\lambda)$  such that 
	\beq
	f(\lambda)\dot{s}-\ddot{s}=0
	\eeq
	Therefore we end up with
	\beq
	\frac{d^{2}x^{\alpha}}{d s^{2}}+\tilde{\Gamma}^{\alpha}_{\;\;\;\;\mu\nu}\frac{d x^{\mu}}{d s}\frac{d x^{\nu}}{d s}=0
	\eeq
	where
	\beq
	s(\lambda)=\int g(\lambda)d\lambda
	\eeq
	and
	\beq
	g(\lambda)= e^{\int f(\lambda) d \lambda}=e^{-\int ( A_{\mu}+B_{\mu})dx^{\mu}}
	\eeq
	So we conclude that if we have an affine connection of the form
	\beq
	\Gamma^{\lambda}_{\;\;\;\;\mu\nu}=\tilde{\Gamma}^{\lambda}_{\;\;\;\;\mu\nu}+\delta^{\lambda}_{\mu}A_{\nu}+\delta^{\lambda}_{\nu}B_{\mu} +K_{\mu\nu}^{\;\;\;\lambda}
	\eeq
	where $\tilde{\Gamma}^{\lambda}_{\;\;\;\;\mu\nu}$ is the Levi-Civita connection and the rest is terms coming from torsion and non-metricity, this connection has the same autoparallel curves with $\tilde{\Gamma}^{\lambda}_{\;\;\;\;\mu\nu}$, that is, for such a connection autoparallels and geodesics coincide.

	\subsubsection{Autoparallel-Geodesic deviation for small torsion and non-metricity}
	Now, one may ask, how much do autoparallels differ from geodesics when the non-Riemannian effects (torsion and non-metricity) are small? To this end let us consider a geodesic curve $x^{\mu}=x^{\mu}(\lambda)$, which of course satisfies the equation
	\beq
	\ddot{x}^{\alpha}+\tilde{\Gamma}^{\alpha}_{\;\;\;\mu\nu}\dot{x}^{\mu}\dot{x}^{\nu}=0
	\eeq
	In addition, let us consider an autoparallel curve $y^{\mu}=y^{\mu}(\lambda)$, for which as we know it holds that
	\beq
	\ddot{y}^{\alpha}+\Gamma^{\alpha}_{\;\;\;\mu\nu}\dot{y}^{\mu}\dot{y}^{\nu}=0
	\eeq
	Now let us suppose that the non-Riemannian effect (deviation from the Levi-Civita connection) is small enough. Then, the affine connection can be written as
	\beq
	\Gamma^{\alpha}_{\;\;\;\mu\nu}\approx \tilde{\Gamma}^{\alpha}_{\;\;\;\mu\nu}+\delta \Gamma^{\alpha}_{\;\;\;\mu\nu}
	\eeq
	where $\delta \Gamma^{\alpha}_{\;\;\;\mu\nu}$ represents the small deviation from the Levi-Civita connection. Accordingly, the two curves also differ by a small amount $\delta x^{\mu}$ and one has
	\beq
	y^{\mu}\approx x^{\mu}+\delta x^{\mu}
	\eeq
	Taking all the above into consideration and neglecting higher order terms in $\delta \Gamma^{\alpha}_{\;\;\;\mu\nu}$,\; $\delta x^{\mu}$, \; it follows that
	\beq
	\ddot{(\delta x^{\alpha})}+2\tilde{\Gamma}^{\alpha}_{\;\;\;\mu\nu}\dot{x}^{\mu}(\dot{\delta x^{\nu}})+\delta \Gamma^{\alpha}_{\;\;\;\mu\nu}\dot{x}^{\mu}\dot{x}^{\nu}=0
	\eeq

	\section{Geodesic/Autoparallel deviation}
	Let us study here a concept that is of great importance in General Relativity and that is the geodesic deviation equation. Physically this describes how a congruence of free falling particles that initially rest on nearby geodesics (in GR geodesics and autoparallels coincide) deviate or converge as they move under the presence of curvature. Mathematically it expresses the deviation of nearby geodesics (as we move along them) from being parallel due to the curvature of space. We will study here how the equations modify when also torsion and non-metricity are present in the space along with curvature. A crucial point here is that, as we have mentioned many times before, geodesics and autoparallels are different curves in general. Therefore, we will study how nearby geodesics/autoparallels deviate from on another as we move along them. Let us start the discussion by considering a curve $\mathcal{C}:x^{\mu}=x^{\mu}(t)$ which for the most part will remain general and only assume it to be a geodesic or autoparallel after the calculations are performed. In the usual manner (as done in most textbooks) let us consider the tangent vector to our reference curve
	\beq
	T^{\mu} \equiv \frac{\partial x^{\mu}}{\partial t}
	\eeq
	and the deviation vector pointing at nearby geodesics/autoparallels
	\beq
	X^{\mu} \equiv \frac{\partial x^{\mu}}{\partial s}
	\eeq
	where $t$ is the affine parameter along the fixed curve (geodesic/autoparallel) and $s$ the parameter pointing at nearby geodesics/autoparallels. Then, $x^{\mu}=x^{\mu}(t,s)$ defines a surface, the vectors $T^{\mu}, X^{\mu}$ form a coordinate basis and it holds that
	\beq
	[X,T]^{\alpha}=X^{\beta}\partial_{\beta}T^{\alpha}-T^{\beta}\partial_{\beta}X^{\alpha}=0
	\eeq
	Now, in expressing the partial derivatives, appearing in the commutator above, with the covariant ones we note that there is an extra term appearing due to torsion. More precisely, expanding the above one has
	\beq
	X^{\beta}\nabla_{\beta}T^{\alpha}-T^{\beta}\nabla_{\beta}X^{\alpha}+2 S_{\beta\gamma}^{\;\;\;\;\alpha}X^{\beta}T^{\gamma}=0   \; \Rightarrow  \nonumber
	\eeq 
	\beq
	T^{\beta}\nabla_{\beta}X^{\alpha}=X^{\beta}\nabla_{\beta}T^{\alpha}+2 S_{\beta\gamma}^{\;\;\;\;\alpha}X^{\beta}T^{\gamma} \label{tx}
	\eeq 
	Now, define the 'relative velocity'of geodesics/autoparallels via
	\beq
	V^{\alpha}\equiv T^{\beta}\nabla_{\beta}X^{\alpha}
	\eeq
	and subsequently, the 'relevant acceleration' 
	\beq
	a^{\alpha}\equiv T^{\gamma}\nabla_{\gamma}(T^{\beta}\nabla_{\beta}X^{\alpha})
	\eeq
	Expanding the latter, we obtain
	\begin{gather}
	a^{\alpha}= T^{\gamma}\nabla_{\gamma}(T^{\beta}\nabla_{\beta}X^{\alpha})= \nonumber \\
	=T^{\gamma}\nabla_{\gamma}(X^{\beta}\nabla_{\beta}T^{\alpha}+2 S_{\beta\gamma}^{\;\;\;\;\alpha}X^{\beta}T^{\gamma} ) = \nonumber \\
	=(T^{\gamma}\nabla_{\gamma}X^{\beta})(\nabla_{\beta}T^{\alpha})+T^{\gamma}X^{\beta}\nabla_{\gamma}\nabla_{\beta}T^{\alpha}+T^{\gamma}\nabla_{\gamma}(2 S_{\beta\gamma}^{\;\;\;\;\alpha}X^{\beta}T^{\gamma} )
	\end{gather}
	and upon using $(\ref{tx})$ and the definition of the anti-symmetrized covariant derivative acting on a vector, a straightforward but rather lengthy calculation yields
	\beq
	a^{\mu}=R^{\mu}_{\;\;\;\nu\rho\sigma}T^{\nu}T^{\rho}X^{\sigma}+T^{\lambda}\nabla_{\lambda}(2 S_{\alpha\beta}^{\;\;\;\;\mu}X^{\alpha}T^{\beta})+X^{\lambda}\nabla_{\lambda}(T^{\beta}\nabla_{\beta}T^{\mu})
	\eeq
	Note now that the third term on the RHS of the above is zero only for autoparallels and not for geodesics. So, for autoparallels we have that
	\beq
	T^{\beta}\nabla_{\beta}T^{\mu}=0
	\eeq
	and as a result the autoparallel deviation equation looks like
	\beq
	a^{\mu}=R^{\mu}_{\;\;\;\nu\rho\sigma}T^{\nu}T^{\rho}X^{\sigma}+T^{\lambda}\nabla_{\lambda}(2 S_{\alpha\beta}^{\;\;\;\;\mu}X^{\alpha}T^{\beta})
	\eeq
	As for the geodesic, it holds that
	\beq
	T^{\beta}\tilde{\nabla}_{\beta}T^{\mu}=0
	\eeq
	where $\tilde{\nabla}_{\beta}$ is the covariant derivative computed with respect to the Levi-Civita connection. Therefore, on a geodesic $T^{\beta}\nabla_{\beta}T^{\mu}$ is not zero but rather
	\beq
	T^{\beta}\nabla_{\beta}T^{\mu}=T^{\beta}\tilde{\nabla}_{\beta}T^{\mu}+N^{\mu}_{\;\;\;\nu\rho}T^{\nu}=N^{\mu}_{\;\;\;\nu\rho}T^{\nu}
	\eeq
	and so, the geodesic deviation equation is given by
	\beq
	a^{\mu}=R^{\mu}_{\;\;\;\nu\rho\sigma}T^{\nu}T^{\rho}X^{\sigma}+T^{\lambda}\nabla_{\lambda}(2 S_{\alpha\beta}^{\;\;\;\;\mu}X^{\alpha}T^{\beta})+X^{\lambda}\nabla_{\lambda}(N^{\mu}_{\;\;\;\nu\rho}T^{\nu}T^{\rho})
	\eeq

	\subsection{Expressing the connection in terms of the Palatini tensor}
	In this section we are going to express the general affine connection in terms of the Palatini tensor plus vectorial torsion and non-metricity contributions\footnote{Of course there is also the Levi-Civita part to it.} We start by writing down the definition of the Palatini tensor and expand the various terms to arrive at
	\begin{gather}
	P_{\lambda}^{\;\;\;\mu\nu}=-g^{\mu\nu}\frac{\nabla_{\lambda}\sqrt{-g}}{\sqrt{-g}}-\nabla_{\lambda}g^{\mu\nu}+g^{\mu\sigma}\frac{\nabla_{\sigma}\sqrt{-g}}{\sqrt{-g}}\delta_{\lambda}^{\nu}+\delta^{\nu}_{\lambda}\nabla_{\sigma}g^{\mu\sigma} \\
	+2(S_{\lambda}g^{\mu\nu}-S^{\mu}\delta_{\lambda}^{\nu}+g^{\mu\sigma}S_{\sigma\lambda}^{\;\;\;\;\nu})
	\end{gather}
	and by using 
	\begin{equation}
	Q_{\lambda}^{\;\;\;\mu\nu}=+\nabla_{\lambda}g^{\mu\nu} \nonumber
	\end{equation}
	\begin{equation}
	\frac{\nabla_{\lambda}\sqrt{-g}}{\sqrt{-g}}=-\frac{1}{2}Q_{\lambda} \nonumber
	\end{equation}
	\begin{equation}
	\tilde{Q}^{\mu}=\nabla_{\sigma}g^{\sigma\mu} \nonumber
	\end{equation}
	it follows that
	\begin{gather}
	P_{\lambda}^{\;\;\;\mu\nu}=-g^{\mu\nu}\frac{Q_{\lambda}}{2}-Q_{\lambda}^{\;\;\;\mu\nu}+\delta_{\lambda}^{\nu}\left( \tilde{Q}^{\mu}- \frac{Q^{\mu}}{2}\right) \\
	+2(S_{\lambda}g^{\mu\nu}-S^{\mu}\delta_{\lambda}^{\nu}+g^{\mu\sigma}S_{\sigma\lambda}^{\;\;\;\;\nu})
	\end{gather}
	and upon multiplying (and contracting) with $g^{\alpha\lambda}$ we finally obtain
	\begin{equation}
	P^{\alpha\mu\nu}=g^{\mu\nu}\left( \frac{Q^{\alpha}}{2}+2 S^{\alpha} \right)-(Q^{\alpha\mu\nu}+2 S^{\alpha\mu\nu})+g^{\nu\alpha}\left( \tilde{Q}^{\mu}- \frac{Q^{\mu}}{2}-2 S^{\mu} \right)
	\end{equation}
	Note now that the second combination $(Q^{\alpha\mu\nu}+2 S^{\alpha\mu\nu})$ plus circular permutations is the exact one appearing on the decomposition of the connection.

	\chapter{Appendix $B$}

	\section{Properties of the Palatini tensor}
	We prove here some basic properties of the Palatini tensor that we have been using throughout the thesis. Recalling its definition
	\begin{equation}
	P_{\lambda}^{\;\;\;\mu\nu}=-\frac{\nabla_{\lambda}(\sqrt{-g}g^{\mu\nu})}{\sqrt{-g}}+\frac{\nabla_{\sigma}(\sqrt{-g}g^{\mu\sigma})\delta^{\nu}_{\lambda}}{\sqrt{-g}} \\
	+2(S_{\lambda}g^{\mu\nu}-S^{\mu}\delta_{\lambda}^{\nu}+g^{\mu\sigma}S_{\sigma\lambda}^{\;\;\;\;\nu})  \nonumber
	\end{equation}
	and contracting in $\mu,\lambda$, immediately follows that
	\begin{equation}
	P_{\mu}^{\;\;\;\mu\nu}=-\frac{\nabla_{\mu}(\sqrt{-g}g^{\mu\nu})}{\sqrt{-g}}+\frac{\nabla_{\sigma}(\sqrt{-g}g^{\nu\sigma})}{\sqrt{-g}} \\
	+2(S^{\nu}-S^{\nu}+0)=0 \Rightarrow \nonumber
	\end{equation}
	\begin{equation}
	P_{\mu}^{\;\;\;\mu\nu}=0
	\end{equation}
	thus, the Palatini tensor is traceless in first and second index. Contracting now in $\nu,\lambda$ we have
	\begin{equation}
	P_{\nu}^{\;\;\;\mu\nu}=(n-1)\frac{\nabla_{\sigma}(\sqrt{-g}g^{\mu\sigma})}{\sqrt{-g}}+2(2-n)S^{\mu}
	\end{equation}
	and upon using 
	\begin{equation}
	\nabla_{\sigma}g^{\mu\sigma}=\tilde{Q}^{\mu}
	\end{equation}
	along with
	\begin{equation}
	\frac{\nabla_{\sigma}\sqrt{-g}}{\sqrt{-g}}=-\frac{1}{2}Q_{\sigma}
	\end{equation}
	the latter recasts to
	\begin{equation}
	P_{\nu}^{\;\;\;\mu\nu}=(n-1)\left[ \tilde{Q}^{\mu}-\frac{1}{2}Q^{\mu}\right] +2(2-n)S^{\mu}
	\end{equation}
	To obtain a third identity, we multiply (and contract) with $g_{\mu\nu}$ and use the above relations for the Weyl and second non-metricity vector, to arrive at
	\begin{equation}
	g_{\mu\nu}P_{\lambda}^{\;\;\;\mu\nu}=\frac{(n-3)}{2}Q_{\lambda}+\tilde{Q}_{\lambda}+2(n-2)S_{\lambda}
	\end{equation}
	Now, defining $P^{\mu}\equiv P_{\nu}^{\;\;\;\mu\nu}$ and $\tilde{P}^{\mu}\equiv g_{\alpha\beta}P^{\mu\alpha\beta}$  adding and subtracting the above two, we get
	\beq
	P^{\mu}+\tilde{P}^{\mu}=n \tilde{Q}^{\mu}-Q^{\mu}
	\eeq
	and
	\beq
	P^{\mu}-\tilde{P}^{\mu}=(n-2) (\tilde{Q}^{\mu}-Q^{\mu}-4 S^{\mu})
	\eeq
	respectively, and notice that both of the above combinations are projective invariant! Another useful relation comes about by taking the antisymmetric part of the Palatini tensor, which is equal to
	\begin{equation}
	P_{\lambda}^{\;\;\;[\mu\nu]}=2 A^{[\mu}\delta^{\nu]}_{\lambda}+2 g^{\sigma[\mu}S_{\sigma\lambda}^{\;\;\;\;\;\nu]}
	\end{equation}
	where
	\begin{equation}
	A^{\mu}=\frac{1}{2}\tilde{Q}^{\mu}-\frac{1}{4}Q^{\mu}-S^{\mu}
	\end{equation}
	Using the definitions of non-metricity tensor and vectors we can easily express the Palatini tensor in the form
	\begin{equation}
	P_{\lambda}^{\;\;\;\mu\nu}=\delta^{\nu}_{\lambda}\left( \tilde{Q}^{\mu}-\frac{1}{2}Q^{\mu}-2 S^{\mu} \right) + g^{\mu\nu}\left( \frac{1}{2}Q_{\lambda}+2 S_{\lambda} \right)-( Q_{\lambda}^{\;\;\;\mu\nu}+2 S_{\lambda}^{\;\;\;\;\mu\nu})
	\end{equation}
	such that
	\begin{equation}
	P^{\alpha\mu\nu}= g^{\alpha\nu}\left( \tilde{Q}^{\mu}-\frac{1}{2}Q^{\mu}-2 S^{\mu}\right)+ g^{\mu\nu}\left( \frac{1}{2}Q^{\alpha}+2 S^{\alpha} \right)-( Q^{\alpha\mu\nu}+2 S^{\alpha\mu\nu})
	\end{equation}
	Note now, that the fully antisymmetric part of the Palatini tensor is determined only by the torsion tensor (the non-metricity part drops out)
	\begin{equation}
	P^{[\alpha\mu\nu]}=-2  S^{[\alpha\mu\nu]}
	\end{equation}
	In addition, the completely symmetric part of it is solely determined by non-metricity. Indeed, the above can also be written as
	\begin{equation}
	P^{\alpha\mu\nu}= g^{\alpha\nu} \tilde{Q}^{\mu}+ 2 g^{\nu[\mu}\left( \frac{1}{2}Q^{\alpha]}+2 S^{\alpha]} \right)-( Q^{\alpha\mu\nu}+2 S^{\alpha\mu\nu})
	\end{equation}
	and by taking the fully symmetric part it follows that
	\begin{equation}
	P^{(\alpha\mu\nu)}= g^{(\alpha\nu} \tilde{Q}^{\mu)}-Q^{(\alpha\mu\nu)}
	\end{equation}

	\subsection{The Hodge Star Operator}
	Given a $p$-form $\Psi$ expanded in the coordinate $1$-form basis $\{dx^{a}\}$,
	\begin{equation}
	\Psi=\frac{1}{p!}\Psi_{a_{1}a_{2}...a_{p}}dx^{a_{1}}\wedge dx^{a_{2}}\wedge ...\wedge dx^{a_{p}}
	\end{equation}
	the operation of the Hodge star $\ast $ maps it into the $(n-p)$
	-form
	\begin{equation}
	\ast \Psi :=\frac{1}{(n-p)!p!}\sqrt{|g|}g^{a_{1}c_{1}}...g^{a_{p}c_{p}}\epsilon_{a_{1}...a_{p}b_{1}...b_{n-p}} \Psi_{c_{1}...c_{p}}\vartheta^{b_{1}}\wedge ...\wedge \vartheta^{b_{n-p}}
	\end{equation}

	\subsection{Calculus of Variations}
	Let us see how the Calculus of Variations arises out of pure mathematical curiosity. Consider the integral\footnote{We call $I[y(x)]$ a functional. A functional is a map $\mathcal{C}^{n} \to \mathbb{R}$ , where $\mathcal{C}^{n}$ is the set of $n-times$ differentiable continuous functions. It should be differentiated from a function which is a map $\mathbb{R} \to \mathbb{R}$.}
	\begin{equation}
	I[y(x)]=\int_{x_{1}}^{x_{2}}dx f(y(x),y'(x),x)
	\end{equation}
	where $y(x)$ is a function of $x$, $y'(x)=\frac{dy}{dx}$ and $f$ an arbitrary function of the given arguments. We ask now, for which curve $y(x)$ does the above integral get its extreme value (maximum or minimum)? Note that if we had a function $I(\epsilon)$ we would know how to proceed, the extremum (or extrema if there are more than one) occurs exactly there where
	\begin{equation}
	\frac{dI}{d\epsilon}=0
	\end{equation}
	However, in our case, after the integration is performed we are left with a number (which of course cannot be varied). Here is where the magic of mathematics comes in. Suppose we have found that function which extremizes the integral, call it $y(x)$. We then consider a family of curves parametrized by $\epsilon$ , deviating by this solution. We denote them by
	\begin{equation}
	Y(x,\epsilon)
	\end{equation}
	and we demand that they continuously depend on $\epsilon$ and for $\epsilon=0$ we recover our extreme curve $y(x)$, namely
	\begin{equation}
	Y(x,0)=y(x)
	\end{equation}
	In addition all the curves should end up to the same points on the plane, namely
	\begin{equation}
	Y(x_{1},\epsilon)=y(x_{1})=y_{1} , \;\;\;  Y(x_{2},\epsilon)=y(x_{2})=y_{2} \label{endpoints}
	\end{equation}
	Now, we have constructed a function-$I(\epsilon)$ which we are allowed to vary with respect to $\epsilon$. One then has
	\begin{gather}
	\frac{dI}{d\epsilon} =\frac{d}{d\epsilon}\int_{x_{1}}^{x_{2}}dx f(Y(x,\epsilon),Y'(x,\epsilon),x)= \nonumber \\
	\int_{x_{1}}^{x_{2}}dx \frac{\partial }{\partial \epsilon}f(Y(x,\epsilon),Y'(x,\epsilon),x)= \nonumber
	\end{gather}
	and using the chain rule
	\begin{gather}
	\frac{dI}{d\epsilon} =\int_{x_{1}}^{x_{2}}dx \Big[ \frac{\partial f}{\partial Y}\frac{\partial Y}{\partial \epsilon}+\frac{\partial f}{\partial Y'}\frac{\partial Y'}{\partial \epsilon}\Big]
	\end{gather}
	but since  partial derivatives commute, we may write
	\begin{equation}
	\frac{\partial Y'}{\partial \epsilon}=\frac{\partial}{\partial x}\Big( \frac{\partial Y}{\partial \epsilon}\Big) 
	\end{equation}
	and therefore
	\begin{gather}
	\frac{dI}{d\epsilon} =\int_{x_{1}}^{x_{2}}dx \Big[ \frac{\partial f}{\partial Y}\frac{\partial Y}{\partial \epsilon}+\frac{\partial f}{\partial Y'}\frac{\partial}{\partial x}\Big( \frac{\partial Y}{\partial \epsilon}\Big) \Big]= \nonumber \\
	=\int_{x_{1}}^{x_{2}}dx \Big[ \frac{\partial f}{\partial Y}\frac{\partial Y}{\partial \epsilon}+\frac{\partial}{\partial x}\Big(\frac{\partial f}{\partial Y'}\frac{\partial Y}{\partial \epsilon}\Big)- \frac{\partial Y}{\partial \epsilon}\frac{\partial}{\partial x}\Big(\frac{\partial f}{\partial Y'}\Big)\Big]
	\end{gather}
	Now we remove the parameter $\epsilon$ by setting $\epsilon=0$ since the introduction of it was made for auxiliary reasons. Then, partial derivatives with respect to $x$ are reduced to total ones
	\begin{equation}
	\frac{\partial}{\partial x} \rightarrow \frac{d}{dx}
	\end{equation}
	and by definition
	\begin{equation}
	Y(x,\epsilon)\Big|_{\epsilon=0}=Y(x,0)=y(x)
	\end{equation}
	Note also that since the parametrization is regular, one has
	\begin{equation}
	\frac{\partial Y}{\partial \epsilon} \neq 0, \;\;\; \forall \;\; \epsilon
	\end{equation}
	The extremum then occurs exactly there where
	\begin{equation}
	\frac{dI}{d\epsilon}\Big|_{\epsilon=0} =0
	\end{equation}
	Taking all the above into account, one  arrives at
	\begin{gather}
	0=\int_{x_{1}}^{x_{2}}dx \Big[ \frac{\partial f}{\partial Y}\Big|_{\epsilon=0}\frac{\partial Y}{\partial \epsilon}\Big|_{\epsilon=0}+\frac{d}{d x}\Big(\frac{\partial f}{\partial Y'}\frac{\partial Y}{\partial \epsilon}\Big)\Big|_{\epsilon=0}- \frac{\partial Y}{\partial \epsilon}\Big|_{\epsilon=0}\frac{d}{d x}\Big(\frac{\partial f}{\partial Y'}\Big)\Big|_{\epsilon=0}\Big]= \nonumber \\
	=\Big[\Big(\frac{\partial f}{\partial Y'}\frac{\partial Y}{\partial \epsilon}\Big)\Big|_{\epsilon=0}\Big]\Big|_{x_{1}}^{x_{2}}+\int_{x_{1}}^{x_{2}}dx \Big(\frac{\partial Y}{\partial \epsilon}\Big)\Big|_{\epsilon=0}\Big[ \frac{\partial f}{\partial y}- \frac{d}{d x}\Big(\frac{\partial f}{\partial y'}\Big)\Big] \label{big}
	\end{gather}
	Now as an immediate consequence of (\ref{endpoints}) we have that
	\begin{equation}
	\frac{\partial Y}{\partial \epsilon}\Big|_{\epsilon=0, \; x=x_{i}}=0, \;\;\; i=1,2
	\end{equation}
	at the end points $x_{1}, x_{2}$. Indeed, Taylor expanding $Y(x,\epsilon)$ in $\epsilon$ around the solution $y(x)$ it follows that
	\begin{equation}
	Y(x,\epsilon) \approx y(x)+\frac{\partial Y}{\partial \epsilon}\Big|_{\epsilon=0} \epsilon +O(\epsilon^{2})
	\end{equation}
	and by evaluating the latter at $x_{i}$ ($i=1,2$) we obtain
	\begin{gather}
	y_{i} \approx y_{i}+\frac{\partial Y}{\partial \epsilon}\Big|_{\epsilon=0, x=x_{i}} \epsilon \Rightarrow \nonumber \\
	\frac{\partial Y}{\partial \epsilon}\Big|_{\epsilon=0, x=x_{i}} =0
	\end{gather}
	to first order in $\epsilon$. Using this we see that the first term in the second line of (\ref{big}) vanishes and we are left with
	\begin{equation}
	\int_{x_{1}}^{x_{2}}dx \Big(\frac{\partial Y}{\partial \epsilon}\Big)\Big|_{\epsilon=0}\Big[ \frac{\partial f}{\partial y}- \frac{d}{d x}\Big(\frac{\partial f}{\partial y'}\Big)\Big]
	\end{equation}
	and since this must be true for any $\Big(\frac{\partial Y}{\partial \epsilon}\Big)\Big|_{\epsilon=0}$ we conclude that
	\begin{equation}
	\frac{\partial f}{\partial y}- \frac{d}{d x}\Big(\frac{\partial f}{\partial y'}\Big) \label{y}=0
	\end{equation}
	We are now in a position to answer the question we raised at the beginning of this section. The function $y(x)$ that extremizes the integral
	\begin{equation}
	I[y(x)]=\int_{x_{1}}^{x_{2}}dx f(y(x),y'(x),x)
	\end{equation}
	can be found by solving the differential equation (\ref{y}). It is, in fact, an astonishing result. However, the whole derivation was somewhat involved. There exists an equivalent method that can be used in a more straightforward manner and appears to be more practical in applications. Let us unfold it here. To first order in $\epsilon$ we can write
	\begin{equation}
	Y(x,\epsilon) \approx y(x)+\epsilon g(x)
	\end{equation}
	where $g(x)=(\partial_{\epsilon}Y)|_{\epsilon=0}$. We then define  the deviation  from the extreme path $y(x)$ via
	\begin{equation}
	\delta y := Y(x,\epsilon)-y(x)=\epsilon g(x)
	\end{equation}
	so that $Y(x,\epsilon)= y(x)+\delta y$.
	We also define the variation of the functional through
	\begin{equation}
	\delta I:=I[Y]-I[y]=I[y+\delta y]-I[y]
	\end{equation}
	Then,  for small deviations,  expanding in $\epsilon$ one has
	\begin{gather}
	\delta I =I[y+\delta y]-I[y]=I[y+\epsilon g(x)]-I[y] \approx \nonumber \\
	\approx I[y]+\frac{dI}{d \epsilon}\Big|_{\epsilon=0}\underbrace{\epsilon g(x)}_{=\delta y}-I[y] 
	=\frac{dI}{d \epsilon}\Big|_{\epsilon=0} \delta y \nonumber
	\end{gather}
	and therefore
	\begin{equation}
	\delta I =I[y+\delta y]-I[y] \approx \frac{dI}{d \epsilon}\Big|_{\epsilon=0} \delta y
	\end{equation}
	to first order in $\epsilon$. The latter implies that the condition-$\frac{dI}{d \epsilon}\Big|_{\epsilon=0}=0$ for the extreme curve $y(x)$ is translated into the equivalent restriction
	\begin{equation}
	\delta I =I[y+\delta y]-I[y] =0
	\end{equation}
	and due to the fact that the derivation was made for small $\epsilon$ to first order, the above equation is called the first variation. This method is indeed more straightforward to use as the following examples demonstrate.

	\subsection{Principle of Least Action}
	The principle of least action seems to be the most profound and useful tool in theoretical physics. Simply put, it is the application of Calculus of Variations for physical problems. There are several reasons contributing to this privileged position of this principle. First of all, its apparent simplicity and elegance lead to straightforward examinations of the systems under consideration . In addition,  it provides a new, powerful way to  look at known problems, gives additional intuition, and the symmetries of the system appear in an apparent way. Another reason is its Universality. Namely, it is used in many different areas of physics with great success in all cases. For instance, it is used in Classical mechanics, Electrodynamics, Particle Physics,  and in Gravity as we have solely used it throughout this thesis. These are but few reasons telling us why the Principle of Least Action is considered to be the most profound concept of modern theoretical physics. In the following subsections we present some of its many applications and examine the physical significance in each case.

	\subsection{With dependence on higher derivatives}
	Let us now allow for a Lagrangian which depends on higher order (greater than the first) derivatives. The form of such a Lagrangian is
	\begin{equation}
	L=L\Big(y(x),y'(x),...,y^{(n)}(x),x\Big) 
	\end{equation} 
	where, with $y^{(n)}(x)$ we denote the $n-th$ derivative of $y(x)$ with respect to $x$, namely $y^{(n)}(x):= \frac{d^{n}}{dx^{n}}y(x)$. Now we also assume that all the derivatives up to $(n-1)$-order of the variation vanish at the boundaries, along with $\delta y $. In words
	\begin{equation}
	\delta y |_{t_{i}}=0=\delta y'|_{t_{i}}=...=\delta y^{(n-1)}|_{t_{i}}
	\end{equation}
	where $i=1,2$. By definition we have
	\begin{equation}
	\delta L=L(y+\delta y ,y'+\delta y',...,y^{(n)}+\delta y^{(n)},x )-L(y,y',...,y^{(n)},x)
	\end{equation}
	and Taylor expanding $L(y+\delta y ,y'+\delta y',...,y^{(n)},x)$ we arrive at (keeping only linear terms\footnote{Since the variations $\delta y$,... etc are small we can drop quadratic and higher order terms.})
	\begin{gather}
	L(y+\delta y ,y'+\delta y',...,y^{(n)},x)  \approx L(y,y',...,y^{(n)},x)+ \\ \nonumber
	\frac{\partial L}{\partial y}\delta y+\frac{\partial L}{\partial y'}\delta y'+...+ \frac{\partial L}{\partial y^{(n)}}\delta y^{(n)}
	\end{gather}
	and therefore
	\begin{equation}
	\delta L= \frac{\partial L}{\partial y}\delta y+\frac{\partial L}{\partial y'}\delta y'+...+ \frac{\partial L}{\partial y^{(n)}}\delta y^{(n)}
	\end{equation}
	The variation of the action will then be
	\begin{gather}
	\delta S=\delta \int dx L(y,y'',...,y^{(n)},x)=\int dx \delta L= \nonumber \\
	=\int_{x_{1}}^{x_{2}} dx \Big(\frac{\partial L}{\partial y}\delta y+\frac{\partial L}{\partial y'}\delta y'+...+ \frac{\partial L}{\partial y^{(n)}}\delta y^{(n)} \Big)
	\end{gather}
	Now, to manipulate the terms involving derivatives of $\delta y$ we first compute
	\begin{gather}
	\int_{x_{1}}^{x_{2}} dx \frac{\partial L}{\partial y'}\delta y'= \int_{x_{1}}^{x_{2}} dx \Big[ \frac{d}{dx}\Big(\frac{\partial L}{\partial y'}\delta y\Big)-\delta y\frac{d}{dx}\Big(\frac{\partial L}{\partial y'}\Big) \Big] =\nonumber \\
	=\underbrace{\Big(\frac{\partial L}{\partial y'}\delta y\Big)\Big|_{x_{1}}^{x_{2}}}_{=0}-\int_{x_{1}}^{x_{2}} dx \delta y\Big[\frac{d}{dx}\Big(\frac{\partial L}{\partial y'}\Big)\Big]=-\int_{x_{1}}^{x_{2}} dx \delta y\Big[\frac{d}{dx}\Big(\frac{\partial L}{\partial y'}\Big)\Big]\Rightarrow  \nonumber
	\end{gather}
	\begin{equation}
	\int_{x_{1}}^{x_{2}} dx \frac{\partial L}{\partial y'}\delta y'=-\int_{x_{1}}^{x_{2}} dx \delta y\Big[\frac{d}{dx}\Big(\frac{\partial L}{\partial y'}\Big)\Big]
	\end{equation}
	continue with
	\begin{gather}
	\int_{x_{1}}^{x_{2}} dx \frac{\partial L}{\partial y''}\delta y''=\int_{x_{1}}^{x_{2}} dx \Big[ \frac{d}{dx}\Big( \frac{\partial L}{\partial y''}\delta y'\Big)-\delta y'\frac{d}{dx}\Big(\frac{\partial L}{\partial y''}\Big) \Big]= \nonumber \\
	=\underbrace{\Big( \frac{\partial L}{\partial y''}\delta y'\Big)\Big|_{x_{1}}^{x_{2}}}_{=0}-\int_{x_{1}}^{x_{2}} dx\Big[ \frac{d}{dx}\Big(\delta y\frac{d}{dx}\frac{\partial L}{\partial y''}\Big)-\delta y \frac{d^{2}}{dx^{2}}\Big( \frac{\partial L}{\partial y ''}\Big)\Big]= \nonumber \\
	=\underbrace{\Big[\delta y\frac{d}{dx}\Big(\frac{\partial L}{\partial y''}\Big)\Big]\Big|_{x_{1}}^{x_{2}}}_{=0}+\int_{x_{1}}^{x_{2}} dx\delta y\Big[ \frac{d^{2}}{dx^{2}}\Big(\frac{\partial L}{\partial y''}\Big)\Big]=\nonumber \\
	=\int_{x_{1}}^{x_{2}} dx\delta y\Big[ \frac{d^{2}}{dx^{2}}\Big(\frac{\partial L}{\partial y''}\Big)\Big] \Rightarrow  \nonumber
	\end{gather}
	\begin{equation}
	\int_{x_{1}}^{x_{2}} dx \frac{\partial L}{\partial y''}\delta y''=\int_{x_{1}}^{x_{2}} dx\delta y\Big[ \frac{d^{2}}{dx^{2}}\Big(\frac{\partial L}{\partial y''}\Big)\Big]
	\end{equation}
	now we see the pattern, all the derivatives on $\delta y$ are now acting on the terms that contain partial derivatives of $L$ with respect to $y$, and for an odd number of partial integrations we pick up a factor $-1$ whilst for even number a factor of $+1$. Indeed, for any term $\frac{\partial L}{\partial y^{(k)}}$ , with $k=0,1,2,...,n$ the $k-th$ derivative it will hold
	\begin{gather}
	\int _{x_{1}}^{x_{2}}dx \frac{\partial L}{\partial y^{(k)}} \delta y^{(k)}=\int _{x_{1}}^{x_{2}}dx\Big[ \frac{d}{dx}\Big( \delta y^{(k-1)}\frac{\partial L}{\partial y^{(k)}}\Big)-\delta y^{(k-1)}\frac{d}{dx}\Big(\frac{\partial L}{\partial y^{(k)}}\Big) \Big]= \nonumber \\
	=\underbrace{\Big( \delta y^{(k-1)}\frac{\partial L}{\partial y^{(k)}}\Big)\Big|_{x_{1}}^{x_{2}}}_{=0}- \int _{x_{1}}^{x_{2}}dx\delta y^{(k-1)}\Big[ \frac{d}{dx}\Big(\frac{\partial L}{\partial y^{(k)}}\Big) \Big]=\underbrace{...}_{k-times}= \nonumber \\
	=(-1)^{k}\int_{x_{1}}^{x_{2}}dx\delta y \Big[ \frac{d^{k}}{dx^{k}}\Big( \frac{\partial L}{\partial y^{(k)}}\Big) \Big] \Rightarrow \nonumber
	\end{gather}
	and so, indeed 
	\begin{equation}
	\int _{x_{1}}^{x_{2}}dx \frac{\partial L}{\partial y^{(k)}} \delta y^{(k)}=(-1)^{k}\int_{x_{1}}^{x_{2}}dx\delta y \Big[ \frac{d^{k}}{dx^{k}}\Big( \frac{\partial L}{\partial y^{(k)}}\Big) \Big] 
	\end{equation}
	Therefore, the variation yields
	\begin{gather}
	\delta S=\delta \int_{x_{1}}^{x_{2}} dx L(y,y'',...,y^{(n)},x)=\int_{x_{1}}^{x_{2}} dx \delta L= \nonumber \\
	=\int_{x_{1}}^{x_{2}} dx \Big(\frac{\partial L}{\partial y}\delta y+\frac{\partial L}{\partial y'}\delta y'+...+ \frac{\partial L}{\partial y^{(n)}}\delta y^{(n)} \Big)= \nonumber \\
	=\int_{x_{1}}^{x_{2}}dx \delta y \Big[ \frac{\partial L}{\partial y}-\frac{d}{dx}\Big(\frac{\partial L}{\partial y'}\Big)+...+(-1)^{n}\frac{d^{n}}{dx^{n}}\Big( \frac{\partial L}{\partial y^{(n)}}\Big)\Big]= \nonumber \\
	=\int_{x_{1}}^{x_{2}}dx \delta y \Big[ \sum_{k=0}^{n}(-1)^{k}\frac{d^{k}}{dx^{k}}\Big(\frac{\partial L}{\partial y^{(k)}}\Big) \Big]=0
	\end{gather}
	and for the last one to hold true we must have
	\begin{equation}
	\sum_{k=0}^{n}(-1)^{k}\frac{d^{k}}{dx^{k}}\Big(\frac{\partial L}{\partial y^{(k)}}\Big)=0
	\end{equation}

	\subsection{Classical Particle Mechanics}
	Consider now a classical particle moving in an $1-dim$ potential $V(q)$, where q is a canonical coordinate. The Lagrangian for this system is simply
	\begin{equation}
	L(q,\dot{q})=T-V= \frac{1}{2}m \dot{q}^{2}-V(q)
	\end{equation}
	The Principle of Least Action gives
	\begin{equation}
	\delta \int_{t_{1}}^{t_{2}}dt L(q,\dot{q})=0 \Rightarrow
	\end{equation}
	\begin{equation}
	\int_{t_{1}}^{t_{2}}dt \; \delta L(q,\dot{q})=0  \label{La}
	\end{equation}
	and as has been proven before for the general case
	\begin{equation}
	\delta L=\frac{\partial L}{\partial q} \delta q+\frac{\partial L}{\partial \dot{q}} \delta \dot{q}
	\end{equation}
	with $\delta q=0$ at the endpoints $t_{1},t_{2}$. Partially integrating the second term, it follows that
	\begin{equation}
	\delta L=\frac{\partial L}{\partial q} \delta q+\frac{d}{dt}\Big(\frac{\partial L}{\partial \dot{q}} \delta q \Big)-\delta q \frac{d}{dt}\Big(\frac{\partial L}{\partial \dot{q}} \Big)
	\end{equation}
	so that, equation (\ref{La}) gives
	\begin{gather}
	0= \int_{t_{1}}^{t_{2}}dt \Big[ \frac{\partial L}{\partial q} \delta q+\frac{d}{dt}\Big(\frac{\partial L}{\partial \dot{q}} \delta q \Big)-\delta q \frac{d}{dt}\Big(\frac{\partial L}{\partial \dot{q}} \Big)\Big]= \nonumber \\
	=\Big(\frac{\partial L}{\partial \dot{q}}\delta q \Big)\Big|_{t_{1}}^{t_{2}}+ \int _{t_{1}}^{t_{2}}dt \delta q  \Big[ \frac{\partial L}{\partial q} -\frac{d}{dt}\Big(\frac{\partial L}{\partial \dot{q}} \Big)\Big]= \nonumber \\
	=\int _{t_{1}}^{t_{2}}dt \delta q  \Big[ \frac{\partial L}{\partial q} -\frac{d}{dt}\Big(\frac{\partial L}{\partial \dot{q}} \Big)\Big]
	\end{gather}
	Now since $t_{1}$ and $t_{2}$ are arbitrary and the latter should vanish for any variation $\delta g$ we conclude that
	\begin{equation}
	\frac{\partial L}{\partial q} -\frac{d}{dt}\Big(\frac{\partial L}{\partial \dot{q}}\Big)=0
	\end{equation}
	this is  the famous Euler-Lagrange equation.

	\subsection{Some basic Variations}
	We present and prove here some of the basic variations that have ,extensively,been used  throughout the derivations. Let us start by the definition of the inverse metric tensor
	\begin{equation}
	g_{\mu\nu}g^{\nu\alpha}=\delta_{\mu}^{\alpha}
	\end{equation}
	and vary the last by noting that $\delta_{\mu}^{\alpha}$ is constant, to get
	\begin{equation}
	0=(\delta g_{\mu\nu})g^{\nu\alpha}+(\delta g^{\nu\alpha})g_{\mu\nu}
	\end{equation}
	Now, contracting with $g_{\alpha\beta}$ we arrive at
	\begin{equation}
	0=(\delta g_{\mu\nu})\delta^{\nu}_{\beta}+(\delta g^{\nu\alpha})g_{\mu\nu}g_{\alpha\beta} \Rightarrow \nonumber
	\end{equation}
	\begin{equation}
	\delta g_{\mu\beta}=-g_{\mu\nu}g_{\alpha\beta} (\delta g^{\nu\alpha})
	\end{equation}
	An alternative derivation of the same relation goes as follows,
	\begin{equation}
	\delta g_{\mu\nu}= \delta (g_{\mu\alpha}g_{\nu\beta}g^{\alpha\beta})= \nonumber
	\end{equation}
	\begin{equation}
	= (\delta g_{\mu\alpha})g_{\nu\beta}g^{\alpha\beta}+ g_{\mu\alpha}  (\delta g_{\nu\beta})g^{\alpha\beta}+  g_{\mu\alpha})g_{\nu\beta} (\delta g^{\alpha\beta})= \nonumber
	\end{equation}
	\begin{equation}
	= (\delta g_{\mu\alpha})g_{\nu\beta}g^{\alpha\beta}+ \delta g_{\mu\nu}+ \delta g_{\nu\mu} \Rightarrow \nonumber
	\end{equation}
	\begin{equation}
	\delta g_{\mu\nu}= (\delta g_{\mu\alpha})g_{\nu\beta}g^{\alpha\beta}+ \delta g_{\mu\nu}+ \delta g_{\nu\mu} \Rightarrow \nonumber
	\end{equation}
	\begin{equation}
	\delta g_{\mu\nu}=-g_{\mu\alpha}g_{\nu\beta}\delta g^{\alpha\beta}
	\end{equation}
	Now varying
	\begin{equation}
	g_{\mu\nu}g^{\mu\nu}=4
	\end{equation}
	we get
	\begin{equation}
	g^{\mu\nu}\delta{g_{\mu\nu}}=-g_{\mu\nu}\delta{g^{\mu\nu}}
	\end{equation}
	Of course, the above results hold true for an general $n-dimensional$ Riemannian (or pseudo-Remannian) space namely
	\begin{equation}
	g_{ab}g^{bc}=\delta_{a}^{c}
	\end{equation}
	and 
	\begin{equation}
	g_{ab}g^{ab}=n
	\end{equation}
	so that
	\begin{equation}
	\delta g_{ab}=-g_{ac}g_{bd}\delta g^{cd}
	\end{equation}
	as well as
	\begin{equation}
	g^{ab}\delta g_{ab}=-g_{ab}\delta g^{ab}
	\end{equation}
	where the indices $a,b,c,d,...$ run over the dimensionality of the space (namely from $1$ to $n$). Let us now compute the variation  of the square root of the determinant of the metric tensor. We do the calculation fully general  for $n-dim$ Riemannian spaces  and then apply it for the pseudo-Riemannian $4-dim$ space of General relativity . Firstly, we write the determinant simply as
	\begin{equation}
	g \equiv det(g_{ab})
	\end{equation}
	We have that
	\begin{equation}
	\delta(\sqrt{g})=\frac{1}{2\sqrt{g}}\delta g
	\end{equation}
	Now, for any square $n \times n$- matrix $\mathcal{A}$ it holds that
	\begin{equation}
	det(\mathcal{A})=e^{Tr(\mathcal{A})}
	\end{equation}
	Setting $\mathcal{A} \rightarrow g_{ab}$ in the above, we arrive at
	\begin{equation}
	g=det(g_{ab})=e^{Tr(g_{ab})}
	\end{equation}
	and under the variation $g_{ab}\rightarrow g_{ab}+\delta g_{ab}$ it follows that
	\begin{equation}
	det(g_{ab}+\delta g_{ab})=e^{Tr(g_{ab}+\delta g_{ab})}=e^{Tr(g_{ab})+Tr(\delta g_{ab})}=\underbrace{e^{Tr(g_{ab})}}_{\equiv g}e^{Tr(\delta g_{ab})}
	\end{equation}
	where on going from the second to the third equality we employed the linearity of the trace. Since the variations $\delta g$ are small, in the expansion of  $e^{Tr(\delta g_{ab})}$ we can neglect second and higher order terms $\Big ((\delta g)^{2} \approx 0 \Big)$ and we shall have
	\begin{equation}
	e^{Tr(\delta g_{ab})} \approx 1+ Tr(\delta g_{ab})
	\end{equation}
	and therefore
	\begin{equation}
	det(g_{ab}+\delta g_{ab}) \approx g\Big(1+ Tr(\delta g_{ab})\Big)
	\end{equation}
	but, by the definition of the trace
	\begin{equation}
	Tr(\delta g_{ab})=g^{ab}\delta g_{ab}
	\end{equation}
	so that
	\begin{equation}
	det(g_{ab}+\delta g_{ab}) \approx g\Big(1+ g^{ab}\delta g_{ab}\Big)
	\end{equation}
	Using the latter to the definition of the variation, we arrive at
	\begin{gather}
	\delta g=\delta \Big( det (g_{ab}) \Big)= det(g_{ab}+\delta g_{ab}) -det (g_{ab}) \approx \nonumber \\
	\approx g\Big(1+ g^{ab}\delta g_{ab}\Big)-g = g g^{ab}\delta g_{ab}
	\end{gather}
	It has been proven before that
	\begin{equation}
	g^{ab}\delta g_{ab}=-g_{ab}\delta g^{ab}
	\end{equation}
	thus we finally obtain
	\begin{equation}
	\delta g= -g g_{ab}\delta g^{ab} \label{nice}
	\end{equation}
	It also follows that
	\begin{equation}
	\delta (\sqrt{g})=\frac{1}{2\sqrt{g}}\delta g=-\frac{1}{2}\frac{g}{\sqrt{g}}g_{ab}\delta g^{ab}=-\frac{1}{2}\sqrt{g} g_{ab}\delta g^{ab} \Rightarrow \nonumber
	\end{equation}
	\begin{equation}
	\delta (\sqrt{g})=-\frac{1}{2}\sqrt{g} g_{ab}\delta g^{ab} 
	\end{equation}
	Now, in order to get the expression for the $4-dim$ pseudo-Riemannian space of General relativity, we simply replace $g \rightarrow -g$ and let the indices run from $0$ to $3$ (The usual Greek ones). We then have
	\begin{equation}
	\delta (\sqrt{-g})=-\frac{1}{2}\sqrt{-g} g_{\mu\nu}\delta g^{\mu\nu} 
	\end{equation}
	where $\mu,\nu=0,1,2,3$. In addition eq. (\ref{nice}) provides a nice formula that allows one to compute in compact form, variations of any power of the determinant. Indeed, for a general Riemannian space, the variation of the $n-th$\footnote{This n here has nothing to do with the dimension of spacetime, it is merely an arbitrary real number} power of g will be
	\begin{equation}
	\delta (g^{n})=ng^{n-1}\delta g= -ng^{n-1}g g_{ab}\delta g^{ab}=-ng^{n}g_{ab}\delta g^{ab} \Rightarrow  \nonumber
	\end{equation}
	\begin{equation}
	\delta (g^{n})=-ng^{n}g_{ab}\delta g^{ab} 
	\end{equation}
	If we now sum over the body of all natural numbers $(n \in )$ we obtain
	\begin{equation}
	\sum_{n=0}^{\infty}\delta(g^{n})=-\Big( \sum_{n=0}^{\infty}ng^{n}\Big) g_{ab}\delta g^{ab} \label{athroisma}
	\end{equation}
	Assuming now that $g<1$, in order to calculate the sum  appearing on the right hand side we start by
	\begin{equation}
	\frac{1}{1-x}=\sum_{n=0}^{\infty}x^{n} \; ,\;\; |x|<1
	\end{equation}
	and differentiate with respect to x,  to get
	\begin{equation}
	\frac{1}{(1-x)^{2}}=\sum_{n=0}^{\infty}nx^{n-1} \; ,\;\; |x|<1
	\end{equation}
	multiplying through by x
	\begin{equation}
	\frac{x}{(1-x)^{2}}=\sum_{n=0}^{\infty}nx^{n} \; ,\;\; |x|<1
	\end{equation}
	Thus, setting $x=g$ to the last one, we obtain
	\begin{equation}
	\frac{g}{(1-g)^{2}}=\sum_{n=0}^{\infty}ng^{n} \; ,\;\; |g|<1
	\end{equation}
	so that equation (\ref{athroisma}) assumes the nice form
	\begin{equation}
	\sum_{n=0}^{\infty}\delta(g^{n})=-  \frac{g}{(1-g)^{2}}g_{ab}\delta g^{ab} \; , \;\; |g|<1
	\end{equation}
	We should mention here that we arrived to this compact form because of our assumption that $|g|<1$. This is a necessary condition we must impose in order for the sum to converge. When  $|g|\geq 1$ the sum diverges and such a compact formula does not exist.
	
	\section{The Affine Group}
	\subsection{The rigid Affine Group $A(n,R)$}
	Consider a flat $n-dim$ affine space $R^{n}$. We define the rigid affine group to be the semi-direct product group
	\begin{equation}
	\mathcal{A}(n,R) :=R^{n}\odot  GL(n,R)
	\end{equation}
	where $R^{n}$ represents the group of $n-dim$ translations\footnote{Usually, one also uses the symbol $T^{n}$ for $n-dim$ translations.}, and $GL(n,R)$ is the group of  $n-dim$ general linear transformations. Note that the symbol $\odot$, in the above, denotes the semi-direct product of two groups. The affine group can be seen as the generalization of the Poincare group (in $n-dim$)
	\begin{equation}
	P := R^{n} \odot SO(1,n-1)
	\end{equation}
	with the group $SO(1,n-1)$  being generalized to $GL(n,R)$. The action of the affine group on an affine n-vector $x=\{x^{a} \}, a=0,1,...,n-1$, goes as follows
	\begin{equation}
	x \longrightarrow x^{\prime}=\Lambda x +\tau
	\end{equation}
	where $\Lambda=\{\Lambda^{a}_{\;\;b} \}$ is an element of $GL(n,R)$ and $\tau=\{\tau^{a} \}$ $\in$ $R^{n}$. Considering two successive actions of the affine group on $x$, it follows that
	\begin{equation}
	x^{\prime\prime}=\Lambda^{\prime}x^{\prime}+\tau^{\prime}=\Lambda^{\prime}(\Lambda x +\tau)+\tau^{\prime} \Rightarrow \nonumber
	\end{equation}
	\begin{equation}
	x^{\prime\prime}=(\Lambda^{\prime}\Lambda )x+(\Lambda^{\prime}\tau+\tau^{\prime})
	\end{equation}
	Thus, on the group level, we have the composition law
	\begin{equation}
	(\Lambda^{\prime},\tau^{\prime}) \circ (\Lambda,\tau) =(\Lambda^{\prime}\Lambda, \Lambda^{\prime} \tau+\tau^{\prime})
	\end{equation}
	Note that due to the mixing  of the elements of the two groups, as seen in the second argument of the right-hand side we do not have a direct product group.\footnote{Recall that the direct product group $G$, of two groups $h$ and $g$ with composition laws $\circ$  and $\bullet$ respectively, is defined by $G=h\otimes g$ and the composition law reads $G_{1}$ $\ast$ $G_{2}$ =$(h_{1}\circ h_{2}$ , $g_{1}\bullet g_{2})$, with $h_{1},h_{2}$ $\in$ $h$ , $g_{1},g_{2}$ $\in$ $g$. That is the subgroups do not mix under the composition. } We therefore have a semi-direct product group.
	It is of convenience now, to use a $M\ddot{o}bius$ type representation. The latter consists of exactly that subgroup of $GL(n+1,R)$ which leaves the $n-dim$ hyperplane $\bar{R}^{n} :=$ \Big\{$\bar{x}=$
	$
	\begin{pmatrix}
	x  \\
	1 
	\end{pmatrix}
	$
	$\in$ $R^{n+1}$ \Big\} invariant. In words \newline
	
	\begin{equation}
	A(n,R)= \Big\{ 
	\begin{pmatrix}
	\Lambda  & \tau  \\
	0 & 1
	\end{pmatrix}
	\in GL(n+1,R) \Big| \Lambda \in GL(n,R), \; \tau \in R^{n}\Big\}
	\end{equation}
	As a result, an affine transformation on $\bar{x}$ will yield
	\begin{equation}
	\bar{x}^{\prime}=A\bar{x}= \begin{pmatrix}
	\Lambda  & \tau  \\
	0 & 1
	\end{pmatrix} 
	\begin{pmatrix}
	x  \\
	1 
	\end{pmatrix}
	=\begin{pmatrix}
	\Lambda x+\tau  \\
	1
	\end{pmatrix}
	\end{equation}
	which of course reproduces the transformation law $x\rightarrow x^{\prime}=\Lambda x+\tau$,  for the $R^{n}$ subpart, as required. We now proceed by giving the algebra $a(n,R)$ of the rigid affine group. For the translational part we have the $n-dim$ translation operators $P_{c}$. In addition, the $gl(n,R)$ algebra is spanned by the generators $L^{a}_{\;\; b}$. Thus, the full algebra of the rigid affine group is given by
	\begin{align}
	& [P_{a},P_{b}]=0 \\
	& [L^{a}_{\;\; b}, P_{c}]= \delta^{a}_{c}P_{b}  \\
	& [L^{a}_{\;\; b},L^{c}_{\;\; d}]=\delta^{a}_{d}L^{c}_{\;\; b}-\delta^{c}_{b}L^{a}_{\;\; d}
	\end{align}

	\subsection{The Gauged Affine Group $\mathcal{A}(n,R)$ }
	We now want to find an expression for the affine gauge group. In order to do so, we let the transformation parameters (that is $\Lambda,\tau$) go local. Namely 
	\begin{equation}
	\Lambda\rightarrow \Lambda(x) \nonumber
	\end{equation}
	\begin{equation}
	\tau\rightarrow \tau(x) \nonumber
	\end{equation}
	We then define the affine gauge group to be
	\begin{equation}
	\mathcal{A}(n,R)= \Big\{ 
	\begin{pmatrix}
	\Lambda (x)  & \tau (x)  \\
	0 & 1
	\end{pmatrix}
	\Big| \Lambda (x) \in \mathcal{G}\mathcal{L}(n,R), \; \tau (x) \in \mathcal{T}(n,R)\Big\}
	\end{equation}
	In a Yang-Mills like manner, we go on introducing the generalized affine connection
	\begin{equation}
	\bar{\Gamma}=
	\begin{pmatrix}
	\Gamma^{(L)}  & \Gamma^{(T)}  \\
	0 & 0
	\end{pmatrix}=
	\begin{pmatrix}
	\Gamma^{(L)\;b}_{a}L^{a}_{\;\; b}  & \Gamma^{(T)\;a}P_{a}  \\
	0 & 0
	\end{pmatrix}
	\end{equation}
	where $ \Gamma^{(L)}$ is an $n \times n$ matrix corresponding to the linear transformations part, and $\Gamma^{(T)}$ an n-dim row vector related to the translational part. The above connection is also an $1$-form and can be expanded as
	\begin{equation}
	\bar{\Gamma}=\bar{\Gamma}_{\mu}dx^{\mu}
	\end{equation}
	In addition, it transforms inhomogeneously under an affine gauge transformation, according to
	\begin{equation}
	\bar{\Gamma}\stackrel{A^{-1}(x)}{\longrightarrow}  \; \bar{\Gamma}^{\prime}=A^{-1}(x)\bar{\Gamma}A(x)+A^{-1}(x)dA(x) \label{a}
	\end{equation}
	with $A(x), A^{-1}(x)$ $\;$ $\in $ $\mathcal{A}(n,R)$. Our definition of an $active$ transformation is that which is formed by the action of the inverse group element $A(x)^{-1}$. The components of the latter can be found by using
	\begin{equation}
	A(x)A^{-1}(x)= \mathbb{1}_{(n+1)\times (n+1)} 
	\end{equation}
	which readily gives
	\begin{equation}
	A^{-1}(x)=
	\begin{pmatrix}
	\Lambda^{-1} (x)  & - \Lambda^{-1} (x)\tau (x)  \\
	0 & 1
	\end{pmatrix}	
	\end{equation}
	Now, the curvature $2$-form $\bar{R}$, associated with the above connection, will be\footnote{The exterior product of Lie algebra-valued forms is evaluated with respect to the adjoint group representation}
	\begin{equation}
	\bar{R}:= d\bar{\Gamma}+\bar{\Gamma}\wedge \bar{\Gamma}=
	\begin{pmatrix}
	d\Gamma^{(L)}  & d\Gamma^{(T)}  \\
	0 & 0
	\end{pmatrix}+
	\begin{pmatrix}
	\Gamma^{(L)}\wedge   & \Gamma^{(T)} \wedge  \\
	0 & 0
	\end{pmatrix}
	\begin{pmatrix}
	\Gamma^{(L)}  & \Gamma^{(T)}  \\
	0 & 0
	\end{pmatrix} \Rightarrow \nonumber
	\end{equation}
	\begin{equation}
	\bar{R}=\begin{pmatrix}
	d\Gamma^{(L)}+\Gamma^{(L)}\wedge \Gamma^{(L)}   & d\Gamma^{(T)}+ \Gamma^{(L)} \wedge \Gamma^{(T)} \\
	0 & 0
	\end{pmatrix}=\begin{pmatrix}
	R^{(L)}  & R^{(T)}   \\
	0 & 0
	\end{pmatrix}
	\end{equation}
	Note that the curvature does transform covariantly under the action of the affine group. Indeed, we have the transformation rule
	\begin{equation}
	\bar{R}\stackrel{A^{-1}(x)}{\longrightarrow}  \; \bar{R}^{\prime}=A^{-1}(x)\bar{R}A(x)
	\end{equation}
	under the group action. Consider now an affine $p$-form 
	\begin{equation}
	\bar{\Psi}=\begin{pmatrix}
	\Psi   \\
	1
	\end{pmatrix}
	\end{equation}
	Then, the exterior covariant derivative ($\bar{D}:=d+\bar{\Gamma}\wedge $) will act on it, according to
	\begin{equation}
	\bar{D}\bar{\Psi}=\begin{pmatrix}
	d\Psi+\Gamma^{(L)}\wedge \Psi+ \Gamma^{(T)}   \\
	0
	\end{pmatrix}=
	\begin{pmatrix}
	D\Psi++ \Gamma^{(T)}   \\
	0
	\end{pmatrix}
	\end{equation}
	Therefore, in order to recover the covariant exterior derivative $D:= d+\Gamma^{(L)}$, with $\Gamma^{(L)}=\Gamma_{a}^{\;\;b}\rho(L^{a}_{\;\;b})$,\footnote{The quantity $\rho(L^{a}_{\;\;b})$ denotes representation type and depends on the field on which the covariant derivative acts upon. } one must impose $\Gamma^{(T)}=0$. Acting once more with the covariant exterior derivative operator on the above, it follows that
	\begin{equation}
	\bar{D}\bar{D}\bar{\Psi}=
	\begin{pmatrix}
	DD\Psi++ D\Gamma^{(T)}   \\
	0
	\end{pmatrix}=\bar{R}\bar{\Psi}
	\end{equation}
	Now, having the form of $A^{-1}(x)$ and the transformation law ($\ref{a}$) for the generalized affine connection-$\bar{\Gamma}$, we find the transformation laws for 
	the linear and translation parts, to be
	\begin{equation}
	\bar{\Gamma}^{(L)}\stackrel{A^{-1}(x)}{\longrightarrow}  \; \bar{\Gamma}^{(L)\prime}=\Lambda^{-1}(x)\bar{\Gamma}^{(L)}\Lambda (x)+\Lambda^{-1}(x)d\Lambda (x)
	\end{equation}
	\begin{equation}
	\bar{\Gamma}^{(T)}\stackrel{A^{-1}(x)}{\longrightarrow}  \; \bar{\Gamma}^{(T)\prime}=\Lambda^{-1}(x)\bar{\Gamma}^{(T)}+\Lambda^{-1}(x)D\tau (x)
	\end{equation}
	respectively. From the above we conclude that the  transformation rule for the $\Gamma^{(L)}$ part is that of a Yang-Mills type connection (closes to itself without including any $\tau (x)'s$ from the group $\mathcal{T}(n,R)$) for $\mathcal{G}\mathcal{L}(n,R)$ and so we make the identification
	\begin{equation}
	\Gamma^{(L)}=\Gamma=\Gamma_{a}^{\;\;b}L^{a}_{\;\;b}
	\end{equation}
	Namely, we identify $\Gamma^{(L)}$ with the linear connection $\Gamma$. Now, as long as the $\Gamma^{(T)}$ part is concerned, we see that the latter does not transform as a covector (due to the additional term $D\tau (x)$) and therefore acquires no identification with Lie algebra-valued connection of $R^{n}$.\footnote{Recall that the generators of the $R^{n}$ algebra are the translation operators $P_{a}$.}

	\nocite{poplawski2007massive}
	\nocite{koivisto2007viable}
	\nocite{kalmykov1994projective}
	\nocite{cacciatori2006chern}
	\nocite{zanelli2005lecture}
	\nocite{poplawski2009spacetime}
	\nocite{jimenez2017born}
	\nocite{jimenez2016cosmology}
	\nocite{carollc2004}
	\nocite{carroll1997lecture}
	\nocite{blau2011lecture}
	\nocite{hehl1976general}
	\nocite{maluf2016teleparallel}
	\nocite{heisenberg2018systematic}
	\nocite{cai2016f}
	\nocite{harko2018coupling}
	\nocite{poplawski2006acceleration}
	\nocite{garcia2000plane}
	\nocite{karahan2013scalars}
	\nocite{gronwald1997metric}
	\nocite{hehl1989progress}
	\nocite{hehl1977hadron}
	\nocite{maluf2003dirac}
	\nocite{shapiro2002physical}
	\nocite{obukhov1997effective}
	\nocite{aoki2018galileon}
	\nocite{gotay1992stress}
	\nocite{koivisto2018integrable}
	\nocite{gotay1992stress}
	\nocite{conroy2018spectrum}
	\nocite{jimenez2016cosmology}
	\nocite{heisenberg2018scalar}
	\nocite{jimenez2018born}
	\nocite{jimenez2016cosmology}
	\nocite{berthias1993torsion}
	\nocite{afonso2017role}
	\nocite{wald1984general}
	\nocite{trautman2006einstein}
	\nocite{obukhov1987weyssenhoff}
	\nocite{gasperini1986spin}
	\nocite{tsagas2008relativistic}
	\nocite{julia1998currents}
	\nocite{weinberg1972gravitation}
	\nocite{weinberg1989cosmological}
	\nocite{winberg1972gravitation}
	\nocite{nakahara2003geometry}
	\nocite{berthias1993torsion}
	\nocite{kolb1990early}
	\nocite{ortin2004gravity}
	\nocite{percacci1991higgs}
	\nocite{percacci2009gravity}
	\nocite{obukhov2003metric}
	\nocite{kobayashi1963foundations}
	\nocite{palatini1919deduzione}
	\nocite{jimenez2016cosmology}
	\nocite{jimenez2016cosmology}
	\nocite{jimenez2014extended}
	\nocite{capozziello2015hybrid}
	\nocite{krssak2018teleparallel}
	\nocite{stelmach1991nonmetricity}
	\nocite{cartan1926onriemannian}
	\nocite{einstein1946generalization}
	\nocite{hehl2007elie}
    \nocite{hartley1995normal}
	\nocite{afonso2018correspondence}
	\nocite{kofinas2017use}
	\nocite{kofinas2014cosmological}
	\nocite{cacciatori2006chern}
	\nocite{sobreiro2010aspects}
	\nocite{barragan2010isotropic}
	\nocite{puetzfeld2005prospects}

	\chapter*{Acknowledgments}

	%%%%%%%%%%%%%%%%%%%%%%%%%%%%%%
	Regarding the construction of the Thesis and the publications that came along with it I would like to thank my supervisor Anastasios Petkou for guidance and useful discussions. In addition I would like to thank our main collaborator (and also my second supervisor) Christos Tsagas for guidance suggestions and useful discussions as well as Jaehoon Joeng for fruitful conversations. I would also like to thank Tomi Koivisto for extremely valuable email discussions, comments and correspondence. In addition, I would like to thank Christos Charmousis for useful conversations and collaboration and  Lavinia Heisenberg  for email discussions. Of course there are also other people outside academia like family (especially mother), friends, girlfriend etc. that I would like to thank but since this is a scientific work I would not digress by naming everyone here and right unnecessary facts outside of this thesis. However, I would like to say a big thank you to all of them.

	\bibliographystyle{unsrt}
	\bibliography{ref}

\begin{thebibliography}{100}

\bibitem{charmousis2015self}
Christos Charmousis and Damianos Iosifidis.
\newblock Self tuning scalar tensor black holes.
\newblock In {\em Journal of Physics: Conference Series}, volume 600, page
  012003. IOP Publishing, 2015.

\bibitem{hehl1995metric}
Friedrich~W Hehl, J~Dermott McCrea, Eckehard~W Mielke, and Yuval Ne'eman.
\newblock Metric-affine gauge theory of gravity: field equations, noether
  identities, world spinors, and breaking of dilation invariance.
\newblock {\em Physics Reports}, 258(1-2):1--171, 1995.

\bibitem{iosifidis2018exactly}
Damianos Iosifidis.
\newblock Exactly solvable connections in metric-affine gravity.
\newblock {\em arXiv preprint arXiv:1812.04031}, 2018.

\bibitem{iosifidis2018scale}
Damianos Iosifidis and Tomi Koivisto.
\newblock Scale transformations in metric-affine geometry.
\newblock {\em arXiv preprint arXiv:1810.12276}, 2018.

\bibitem{iosifidis2018torsion}
Damianos Iosifidis, Anastasios~C Petkou, and Christos~G Tsagas.
\newblock Torsion/non-metricity duality in f (r) gravity.
\newblock {\em arXiv preprint arXiv:1810.06602}, 2018.

\bibitem{kranas2018friedmann}
D~Kranas, CG~Tsagas, JD~Barrow, and D~Iosifidis.
\newblock Friedmann-like universes with torsion.
\newblock {\em arXiv preprint arXiv:1809.10064}, 2018.

\bibitem{iosifidis2018raychaudhuri}
Damianos Iosifidis, Christos~G Tsagas, and Anastasios~C Petkou.
\newblock Raychaudhuri equation in spacetimes with torsion and nonmetricity.
\newblock {\em Physical Review D}, 98(10):104037, 2018.

\bibitem{Weyl:1918ib}
H.~Weyl.
\newblock {Gravitation and electricity}.
\newblock {\em Sitzungsber. Preuss. Akad. Wiss. Berlin (Math. Phys.)},
  1918:465, 1918.
\newblock [,24(1918)].

\bibitem{cartan1922equations}
Elie Cartan.
\newblock Sur les {\'e}quations de la gravitation d'einstein.
\newblock {\em Journal de Math{\'e}matiques pures et appliqu{\'e}es},
  1:141--204, 1922.

\bibitem{vitagliano2011dynamics}
Vincenzo Vitagliano, Thomas~P Sotiriou, and Stefano Liberati.
\newblock The dynamics of metric-affine gravity.
\newblock {\em Annals of Physics}, 326(5):1259--1273, 2011.

\bibitem{olmo2011palatini}
Gonzalo~J Olmo.
\newblock Palatini approach to modified gravity: f (r) theories and beyond.
\newblock {\em International Journal of Modern Physics D}, 20(04):413--462,
  2011.

\bibitem{sotiriou2007metric}
Thomas~P Sotiriou and Stefano Liberati.
\newblock Metric-affine f (r) theories of gravity.
\newblock {\em Annals of Physics}, 322(4):935--966, 2007.

\bibitem{vitagliano2010dynamics}
V~Vitagliano, TP~Sotiriou, and S~Liberati.
\newblock The dynamics of generalised palatini theories of gravity (2010).
\newblock {\em arXiv preprint arXiv:1007.3937}.

\bibitem{olmo2009dynamical}
Gonzalo~J Olmo, Helios Sanchis-Alepuz, and Swapnil Tripathi.
\newblock Dynamical aspects of generalized palatini theories of gravity.
\newblock {\em Physical Review D}, 80(2):024013, 2009.

\bibitem{sotiriou2010f}
Thomas~P Sotiriou and Valerio Faraoni.
\newblock f (r) theories of gravity.
\newblock {\em Reviews of Modern Physics}, 82(1):451, 2010.

\bibitem{sotiriou2009f}
Thomas~P Sotiriou.
\newblock f (r) gravity, torsion and non-metricity.
\newblock {\em Classical and Quantum Gravity}, 26(15):152001, 2009.

\bibitem{allemandi2004accelerated}
Gianluca Allemandi, Andrzej Borowiec, and Mauro Francaviglia.
\newblock Accelerated cosmological models in ricci squared gravity.
\newblock {\em Physical Review D}, 70(10):103503, 2004.

\bibitem{pagani2015quantum}
Carlo Pagani and Roberto Percacci.
\newblock Quantum gravity with torsion and non-metricity.
\newblock {\em Classical and Quantum Gravity}, 32(19):195019, 2015.

\bibitem{vitagliano2014role}
Vincenzo Vitagliano.
\newblock The role of nonmetricity in metric-affine theories of gravity.
\newblock {\em Classical and Quantum Gravity}, 31(4):045006, 2014.

\bibitem{1981GReGr..13.1037H}
F.~W. {Hehl}, E.~A. {Lord}, and L.~L. {Smalley}.
\newblock {Metric-affine variational principles in general relativity II.
  Relaxation of the Riemannian constraint}.
\newblock {\em General Relativity and Gravitation}, 13:1037--1056, November
  1981.

\bibitem{capozziello2008f}
Salvatore Capozziello, R~Cianci, C~Stornaiolo, and S~Vignolo.
\newblock f (r) cosmology with torsion.
\newblock {\em Physica Scripta}, 78(6):065010, 2008.

\bibitem{aldrovandi2010introduction}
R~Aldrovandi and JG~Pereira.
\newblock An introduction to teleparallel gravity.
\newblock {\em Instituto de Fisica Teorica, UNSEP, Sao Paulo}, 2010.

\bibitem{poplawski2006nonsymmetric}
Nikodem~J Poplawski.
\newblock On the nonsymmetric purely affine gravity.
\newblock {\em arXiv preprint gr-qc/0610132}, 2006.

\bibitem{koivisto2006note}
Tomi Koivisto.
\newblock A note on covariant conservation of energy--momentum in modified
  gravities.
\newblock {\em Classical and Quantum Gravity}, 23(12):4289, 2006.

\bibitem{hehl1976hypermomentum}
Friedrich~W Hehl, G~David Kerlick, and Paul von~der Heyde.
\newblock On hypermomentum in general relativity iii. coupling hypermomentum to
  geometry.
\newblock {\em Zeitschrift fuer Naturforschung A}, 31(7):823--827, 1976.

\bibitem{hehl1977hadron}
FW~Hehl, EA~Lord, and Y~Ne'Eman.
\newblock Hadron dilation, shear and spin as components of the intrinsic
  hypermomentum current and metric-affine theory of gravitation.
\newblock {\em Physics Letters B}, 71(2):432--434, 1977.

\bibitem{hehl1989progress}
Friedrich~W Hehl, J~Dermott McCrea, Eckehard~W Mielke, and Yuval Ne'Eman.
\newblock Progress in metric-affine gauge theories of gravity with local scale
  invariance.
\newblock {\em Foundations of Physics}, 19(9):1075--1100, 1989.

\bibitem{hehl1999metric}
Friedrich~W Hehl and Alfredo Macias.
\newblock Metric--affine gauge theory of gravity ii: Exact solutions.
\newblock {\em International Journal of Modern Physics D}, 8(04):399--416,
  1999.

\bibitem{obukhov1996exact}
Yu~N Obukhov, EJ~Vlachynsky, W~Esser, R~Tresguerres, and FW~Hehl.
\newblock An exact solution of the metric-affine gauge theory with dilation,
  shear, and spin charges.
\newblock {\em arXiv preprint gr-qc/9604027}, 1996.

\bibitem{tresguerres1995exact}
Romualdo Tresguerres.
\newblock Exact static vacuum solution of four-dimensional metric-affine
  gravity with nontrivial torsion.
\newblock {\em Physics Letters A}, 200(6):405--410, 1995.

\bibitem{hehl1998gauge}
Friedrich~W Hehl and Jos{\'e} Socorro.
\newblock Gauge theory of gravity: Electrically charged solutions within the
  metric--affine framework.
\newblock {\em arXiv preprint gr-qc/9803037}, 1998.

\bibitem{puetzfeld2001cosmological}
Dirk Puetzfeld and Romualdo Tresguerres.
\newblock A cosmological model in weyl-cartan spacetime.
\newblock {\em Classical and Quantum Gravity}, 18(4):677, 2001.

\bibitem{Shimada:2018lnm}
Keigo Shimada, Katsuki Aoki, and Kei-ichi Maeda.
\newblock {Metric-affine Gravity and Inflation}.
\newblock 2018.

\bibitem{cai2016f}
Yi-Fu Cai, Salvatore Capozziello, Mariafelicia De~Laurentis, and Emmanuel~N
  Saridakis.
\newblock f (t) teleparallel gravity and cosmology.
\newblock {\em Reports on Progress in Physics}, 79(10):106901, 2016.

\bibitem{nester1999symmetric}
James~M Nester and Hwei-Jang Yo.
\newblock Symmetric teleparallel general relativity.
\newblock {\em Chinese Journal of Physics}, 37(2):113--117, 1999.

\bibitem{jimenez2018teleparallel}
Jose~Beltr{\'a}n Jim{\'e}nez, Lavinia Heisenberg, and Tomi Koivisto.
\newblock Teleparallel palatini theories.
\newblock {\em arXiv preprint arXiv:1803.10185}, 2018.

\bibitem{mccrea1992irreducible}
J~Dermott McCrea.
\newblock Irreducible decompositions of nonmetricity, torsion, curvature and
  bianchi identities in metric-affine spacetimes.
\newblock {\em Classical and Quantum Gravity}, 9(2):553, 1992.

\bibitem{obukhov1997irreducible}
Yu~N Obukhov, EJ~Vlachynsky, W~Esser, and FW~Hehl.
\newblock Irreducible decompositions in metric-affine gravity models.
\newblock {\em arXiv preprint gr-qc/9705039}, 1997.

\bibitem{schrodinger1985space}
Erwin Schr{\"o}dinger.
\newblock {\em Space-time structure}.
\newblock Cambridge University Press, 1985.

\bibitem{jimenez2018coincident}
Jose~Beltran Jimenez, Lavinia Heisenberg, and Tomi Koivisto.
\newblock Coincident general relativity.
\newblock {\em Physical Review D}, 98(4):044048, 2018.

\bibitem{carroll1997lecture}
Sean~M Carroll.
\newblock Lecture notes on general relativity.
\newblock {\em arXiv preprint gr-qc/9712019}, 1997.

\bibitem{ferraris1994universality}
Marco Ferraris, Mauro Francaviglia, and Igor Volovich.
\newblock The universality of vacuum einstein equations with cosmological
  constant.
\newblock {\em Classical and Quantum Gravity}, 11(6):1505, 1994.

\bibitem{jimenez2017born}
Jose~Beltr{\'a}n Jim{\'e}nez, Lavinia Heisenberg, Gonzalo~J Olmo, and Diego
  Rubiera-Garcia.
\newblock Born--infeld inspired modifications of gravity.
\newblock {\em Physics Reports}, 2017.

\bibitem{Lord2004MetricAftM}
Eric~A. Lord.
\newblock Metric-aft ' me variational principles in general relativity ii .
  relaxation of the riemannian constraint.
\newblock 2004.

\bibitem{smalley1979volume}
LL~Smalley.
\newblock Volume preserving and conformal transformations in the metric-affine
  gravitational theory.
\newblock {\em Lettere al Nuovo Cimento (1971-1985)}, 24(11):406--410, 1979.

\bibitem{obukhov2004two}
Yuri~N Obukhov.
\newblock Two-dimensional metric-affine gravity.
\newblock {\em Physical Review D}, 69(6):064009, 2004.

\bibitem{hehl1976new}
FW~Hehl, GD~Kerlick, and P~Von Der~Heyde.
\newblock On a new metric affine theory of gravitation.
\newblock {\em Physics Letters B}, 63(4):446--448, 1976.

\bibitem{capozziello2007f}
Salvatore Capozziello, R~Cianci, C~Stornaiolo, and S~Vignolo.
\newblock f (r) gravity with torsion: the metric-affine approach.
\newblock {\em Classical and Quantum Gravity}, 24(24):6417, 2007.

\bibitem{roshan2008palatini}
Mahmood Roshan and Fatimah Shojai.
\newblock Palatini f (r) cosmology and noether symmetry.
\newblock {\em Physics Letters B}, 668(3):238--240, 2008.

\bibitem{sotiriou2006constraining}
Thomas~P Sotiriou.
\newblock Constraining f (r) gravity in the palatini formalism.
\newblock {\em Classical and Quantum Gravity}, 23(4):1253, 2006.

\bibitem{vitagliano2011gravity}
Vincenzo Vitagliano.
\newblock Gravity beyond general relativity: theory and phenomenology.
\newblock 2011.

\bibitem{mukhopadhyaya1999geometrical}
Biswarup Mukhopadhyaya and Soumitra Sengupta.
\newblock A geometrical interpretation of parity violation in gravity with
  torsion.
\newblock {\em Physics Letters B}, 458(1):8--12, 1999.

\bibitem{sengupta1999parity}
Soumitra Sengupta.
\newblock Parity violation in a gravitational theory with torsion: A
  geometrical interpretation.
\newblock {\em Pramana}, 53(6):1115--1119, 1999.

\bibitem{d1982gravity}
R~d'Auria and T~Regge.
\newblock Gravity theories with asymptotically flat instantons.
\newblock {\em Nuclear Physics B}, 195(2):308--324, 1982.

\bibitem{leigh2009torsion}
Robert~G Leigh, Nam~Nguyen Hoang, and Anastasios~C Petkou.
\newblock Torsion and the gravity dual of parity breaking in ads4/cft3
  holography.
\newblock {\em Journal of High Energy Physics}, 2009(03):033, 2009.

\bibitem{petkou2010torsional}
Anastasios~C Petkou.
\newblock Torsional degrees of freedom in ads4/cft3.
\newblock {\em arXiv preprint arXiv:1004.1640}, 2010.

\bibitem{hehl1978metric}
Friedrich~W Hehl and G~David Kerlick.
\newblock Metric-affine variational principles in general relativity. i.
  riemannian space-time.
\newblock {\em General Relativity and Gravitation}, 9(8):691--710, 1978.

\bibitem{berthias1993torsion}
J-P Berthias and Bahman Shahid-Saless.
\newblock Torsion and nonmetricity in scalar-tensor theories of gravity.
\newblock {\em Classical and Quantum Gravity}, 10(5):1039, 1993.

\bibitem{tsamparlis1981methods}
Michael Tsamparlis.
\newblock Methods for deriving solutions in generalized theories of
  gravitation: The einstein-cartan theory.
\newblock {\em Physical Review D}, 24(6):1451, 1981.

\bibitem{minkevich1998isotropic}
AV~Minkevich and AS~Garkun.
\newblock Isotropic cosmology in metric-affine gauge theory of gravity.
\newblock {\em arXiv preprint gr-qc/9805007}, 1998.

\bibitem{pasmatsiou2017kinematics}
Klaountia Pasmatsiou, Christos~G Tsagas, and John~D Barrow.
\newblock Kinematics of einstein-cartan universes.
\newblock {\em Physical Review D}, 95(10):104007, 2017.

\bibitem{luz2017raychaudhuri}
Paulo Luz and Vincenzo Vitagliano.
\newblock Raychaudhuri equation in spacetimes with torsion.
\newblock {\em Physical Review D}, 96(2):024021, 2017.

\bibitem{kar2007raychaudhuri}
Sayan Kar and Soumitra Sengupta.
\newblock The raychaudhuri equations: A brief review.
\newblock {\em Pramana}, 69(1):49--76, 2007.

\bibitem{fennelly1991including}
AJ~Fennelly, Jean~P Krisch, John~R Ray, and Larry~L Smalley.
\newblock Including spin in the raychaudhuri equation.
\newblock {\em Journal of mathematical physics}, 32(2):485--487, 1991.

\bibitem{capozziello2001geometric}
S~Capozziello, G~Lambiase, and C~Stornaioloi.
\newblock Geometric classification of the torsion tensor of space-time.
\newblock {\em Annalen der Physik}, 10(8):713--727, 2001.

\bibitem{vazirian2015weyl}
R~Vazirian, MR~Tanhayi, and ZA~Motahar.
\newblock Weyl-invariant extension of the metric-affine gravity.
\newblock {\em Advances in High Energy Physics}, 2015, 2015.

\bibitem{maluf2012conformally}
JW~Maluf and FF~Faria.
\newblock Conformally invariant teleparallel theories of gravity.
\newblock {\em Physical Review D}, 85(2):027502, 2012.

\bibitem{lobo2015crystal}
Francisco~SN Lobo, Gonzalo~J Olmo, and D~Rubiera-Garcia.
\newblock Crystal clear lessons on the microstructure of spacetime and modified
  gravity.
\newblock {\em Physical Review D}, 91(12):124001, 2015.

\bibitem{puetzfeld2008probing}
Dirk Puetzfeld and Yuri~N Obukhov.
\newblock Probing non-riemannian spacetime geometry.
\newblock {\em Physics Letters A}, 372(45):6711--6716, 2008.

\bibitem{roychowdhury2017non}
Ayan Roychowdhury and Anurag Gupta.
\newblock Non-metric connection and metric anomalies in materially uniform
  elastic solids.
\newblock {\em Journal of Elasticity}, 126(1):1--26, 2017.

\bibitem{poplawski2007massive}
Nikodem~J Poplawski.
\newblock Massive vectors from projective-invariance breaking.
\newblock {\em arXiv preprint arXiv:0709.3652}, 2007.

\bibitem{koivisto2007viable}
Tomi Koivisto.
\newblock Viable palatini-f (r) cosmologies with generalized dark matter.
\newblock {\em Physical Review D}, 76(4):043527, 2007.

\bibitem{kalmykov1994projective}
M~Yu Kalmykov, PI~Pronin, and KV~Stepanyantz.
\newblock Projective invariance and one-loop effective action in affine metric
  gravity interacting with a scalar field.
\newblock {\em Classical and Quantum Gravity}, 11(11):2645, 1994.

\bibitem{cacciatori2006chern}
Sergio~L Cacciatori, Marco~M Caldarelli, Alex Giacomini, Dietmar Klemm, and
  Diego~S Mansi.
\newblock Chern--simons formulation of three-dimensional gravity with torsion
  and nonmetricity.
\newblock {\em Journal of Geometry and Physics}, 56(12):2523--2543, 2006.

\bibitem{zanelli2005lecture}
Jorge Zanelli.
\newblock Lecture notes on chern-simons (super-) gravities. (february 2008).
\newblock {\em arXiv preprint hep-th/0502193}, 2005.

\bibitem{poplawski2009spacetime}
Nikodem~J Poplawski.
\newblock Spacetime and fields.
\newblock {\em arXiv preprint arXiv:0911.0334}, 2009.

\bibitem{jimenez2016cosmology}
Jose~Beltr{\'a}n Jim{\'e}nez, Lavinia Heisenberg, and Tomi~S Koivisto.
\newblock Cosmology for quadratic gravity in generalized weyl geometry.
\newblock {\em Journal of Cosmology and Astroparticle Physics}, 2016(04):046,
  2016.

\bibitem{carollc2004}
Sean~M Caroll.
\newblock c2004: Spacetime and geometry: An introduction to general relativity.
\newblock {\em P.--, SanFrancisco: Addition Wesley}.

\bibitem{blau2011lecture}
Matthias Blau.
\newblock {\em Lecture notes on general relativity}.
\newblock Albert Einstein Center for Fundamental Physics Bern Germany, 2011.

\bibitem{hehl1976general}
Friedrich~W Hehl, Paul Von~der Heyde, G~David Kerlick, and James~M Nester.
\newblock General relativity with spin and torsion: Foundations and prospects.
\newblock {\em Reviews of Modern Physics}, 48(3):393, 1976.

\bibitem{maluf2016teleparallel}
Jos{\'e}~Wadih Maluf.
\newblock The teleparallel equivalent of general relativity and the
  gravitational centre of mass.
\newblock {\em Universe}, 2(3):19, 2016.

\bibitem{heisenberg2018systematic}
Lavinia Heisenberg.
\newblock A systematic approach to generalisations of general relativity and
  their cosmological implications.
\newblock {\em arXiv preprint arXiv:1807.01725}, 2018.

\bibitem{harko2018coupling}
Tiberiu Harko, Tomi~S Koivisto, Francisco~SN Lobo, Gonzalo~J Olmo, and Diego
  Rubiera-Garcia.
\newblock Coupling matter in modified q gravity.
\newblock {\em Physical Review D}, 98(8):084043, 2018.

\bibitem{poplawski2006acceleration}
Nikodem~J Pop{\l}awski.
\newblock Acceleration of the universe in the einstein frame of a metric-affine
  f (r) gravity.
\newblock {\em Classical and Quantum Gravity}, 23(6):2011, 2006.

\bibitem{garcia2000plane}
Alberto Garcia, Alfredo Macias, Dirk Puetzfeld, and Jose Socorro.
\newblock Plane-fronted waves in metric-affine gravity.
\newblock {\em Physical Review D}, 62(4):044021, 2000.

\bibitem{karahan2013scalars}
Canan~N Karahan, Asl{\i} Alta{\c{s}}, and Durmu{\c{s}}~A Demir.
\newblock Scalars, vectors and tensors from metric-affine gravity.
\newblock {\em General Relativity and Gravitation}, 45(2):319--343, 2013.

\bibitem{gronwald1997metric}
Frank Gronwald.
\newblock Metric-affine gauge theory of gravity: I. fundamental structure and
  field equations.
\newblock {\em International Journal of Modern Physics D}, 6(03):263--303,
  1997.

\bibitem{maluf2003dirac}
JW~Maluf.
\newblock Dirac spinor fields in the teleparallel gravity: comment on
  “metric-affine approach to teleparallel gravity”.
\newblock {\em Physical Review D}, 67(10):108501, 2003.

\bibitem{shapiro2002physical}
Ilya~Lvovitch Shapiro.
\newblock Physical aspects of the space--time torsion.
\newblock {\em Physics Reports}, 357(2):113--213, 2002.

\bibitem{obukhov1997effective}
Yu~N Obukhov, EJ~Vlachynsky, W~Esser, and FW~Hehl.
\newblock Effective einstein theory from metric-affine gravity models via
  irreducible decompositions.
\newblock {\em Physical Review D}, 56(12):7769, 1997.

\bibitem{aoki2018galileon}
Katsuki Aoki and Keigo Shimada.
\newblock Galileon and generalized galileon with projective invariance in
  metric-affine formalism.
\newblock {\em arXiv preprint arXiv:1806.02589}, 2018.

\bibitem{gotay1992stress}
Mark~J Gotay and Jerrold~E Marsden.
\newblock Stress-energy-momentum tensors and the belinfante-rosenfeld formula.
\newblock {\em Contemporary Mathematics}, (132):367--392, 1992.

\bibitem{koivisto2018integrable}
Tomi Koivisto.
\newblock On an integrable geometrical foundation of gravity.
\newblock {\em arXiv preprint arXiv:1802.00650}, 2018.

\bibitem{conroy2018spectrum}
Aindri{\'u} Conroy and Tomi Koivisto.
\newblock The spectrum of symmetric teleparallel gravity.
\newblock {\em The European Physical Journal C}, 78(11):923, 2018.

\bibitem{heisenberg2018scalar}
Lavinia Heisenberg.
\newblock Scalar-vector-tensor gravity theories.
\newblock {\em arXiv preprint arXiv:1801.01523}, 2018.

\bibitem{jimenez2018born}
Jose~Beltr{\'a}n Jim{\'e}nez, Lavinia Heisenberg, Gonzalo~J Olmo, and Diego
  Rubiera-Garcia.
\newblock Born--infeld inspired modifications of gravity.
\newblock {\em Physics Reports}, 727:1--129, 2018.

\bibitem{afonso2017role}
Victor~I Afonso, Cecilia Bejarano, Jose~Beltran Jimenez, Gonzalo~J Olmo, and
  Emanuele Orazi.
\newblock The role of torsion in projective invariant theories of gravity with
  non-minimally coupled matter fields.
\newblock {\em arXiv preprint arXiv:1705.03806}, 2017.

\bibitem{wald1984general}
Robert~M Wald.
\newblock General relativity, chicago, usa: Univ.
\newblock {\em Pr. 491p}, 1984.

\bibitem{trautman2006einstein}
Andrzej Trautman.
\newblock Einstein-cartan theory.
\newblock {\em arXiv preprint gr-qc/0606062}, 2006.

\bibitem{obukhov1987weyssenhoff}
Yu~N Obukhov and VA~Korotky.
\newblock The weyssenhoff fluid in einstein-cartan theory.
\newblock {\em Classical and Quantum Gravity}, 4(6):1633, 1987.

\bibitem{gasperini1986spin}
M~Gasperini.
\newblock Spin-dominated inflation in the einstein-cartan theory.
\newblock {\em Physical review letters}, 56(26):2873, 1986.

\bibitem{tsagas2008relativistic}
Christos~G Tsagas, Anthony Challinor, and Roy Maartens.
\newblock Relativistic cosmology and large-scale structure.
\newblock {\em Physics Reports}, 465(2-3):61--147, 2008.

\bibitem{julia1998currents}
B~Julia and S~Silva.
\newblock Currents and superpotentials in classical gauge-invariant theories:
  I. local results with applications to perfect fluids and general relativity.
\newblock {\em Classical and Quantum Gravity}, 15(8):2173, 1998.

\bibitem{weinberg1972gravitation}
Steven Weinberg.
\newblock {\em Gravitation and cosmology: principles and applications of the
  general theory of relativity}, volume~1.
\newblock Wiley New York, 1972.

\bibitem{weinberg1989cosmological}
Steven Weinberg.
\newblock The cosmological constant problem.
\newblock {\em Reviews of modern physics}, 61(1):1, 1989.

\bibitem{winberg1972gravitation}
S~Winberg.
\newblock Gravitation and cosmology.
\newblock {\em ed. John Wiley and Sons, New York}, 1972.

\bibitem{nakahara2003geometry}
Mikio Nakahara.
\newblock {\em Geometry, topology and physics}.
\newblock CRC Press, 2003.

\bibitem{kolb1990early}
Edward~W Kolb and Michael~S Turner.
\newblock The early universe (redwood city, 1990.

\bibitem{ortin2004gravity}
Tom{\'a}s Ort{\'\i}n.
\newblock {\em Gravity and strings}.
\newblock Cambridge University Press, 2004.

\bibitem{percacci1991higgs}
R~Percacci.
\newblock The higgs phenomenon in quantum gravity.
\newblock {\em Nuclear Physics B}, 353(1):271--290, 1991.

\bibitem{percacci2009gravity}
Roberto Percacci.
\newblock Gravity from a particle physicists' perspective.
\newblock {\em arXiv preprint arXiv:0910.5167}, 2009.

\bibitem{obukhov2003metric}
Yu~N Obukhov and Jos{\'e}~G Pereira.
\newblock Metric-affine approach to teleparallel gravity.
\newblock {\em Physical Review D}, 67(4):044016, 2003.

\bibitem{kobayashi1963foundations}
Shoshichi Kobayashi and Katsumi Nomizu.
\newblock {\em Foundations of differential geometry}, volume~1.
\newblock Interscience publishers New York, 1963.

\bibitem{palatini1919deduzione}
Attilio Palatini.
\newblock Deduzione invariantiva delle equazioni gravitazionali dal principio
  di hamilton.
\newblock {\em Rendiconti del Circolo Matematico di Palermo (1884-1940)},
  43(1):203--212, 1919.

\bibitem{jimenez2014extended}
Jose~Beltran Jimenez and Tomi~S Koivisto.
\newblock Extended gauss--bonnet gravities in weyl geometry.
\newblock {\em Classical and quantum gravity}, 31(13):135002, 2014.

\bibitem{capozziello2015hybrid}
Salvatore Capozziello, Tiberiu Harko, Tomi~S Koivisto, Francisco~SN Lobo, and
  Gonzalo~J Olmo.
\newblock Hybrid metric-palatini gravity.
\newblock {\em Universe}, 1(2):199--238, 2015.

\bibitem{krssak2018teleparallel}
M~Krssak, RJ~Van Den~Hoogen, JG~Pereira, CG~Boehmer, and AA~Coley.
\newblock Teleparallel theories of gravity: Illuminating a fully invariant
  approach.
\newblock {\em arXiv preprint arXiv:1810.12932}, 2018.

\bibitem{stelmach1991nonmetricity}
J~Stelmach.
\newblock Nonmetricity driven inflation.
\newblock {\em Classical and Quantum Gravity}, 8(5):897, 1991.

\bibitem{cartan1926onriemannian}
{\'E}lie Cartan and Jan~Arnoldus Schouten.
\newblock Onriemannian geometries admitting an absolute parallelism.
\newblock Koninklijke Akademie van Wetenschappen te Amsterdam, 1926.

\bibitem{einstein1946generalization}
Albert Einstein and Ernst~G Straus.
\newblock A generalization of the relativistic theory of gravitation, ii.
\newblock {\em Annals of Mathematics}, pages 731--741, 1946.

\bibitem{hehl2007elie}
Friedrich~W Hehl and Yuri~N Obukhov.
\newblock {\'E}lie cartan's torsion in geometry and in field theory, an essay.
\newblock {\em arXiv preprint arXiv:0711.1535}, 2007.

\bibitem{hartley1995normal}
David Hartley.
\newblock Normal frames for non-riemannian connections.
\newblock {\em Classical and Quantum Gravity}, 12(11):L103, 1995.

\bibitem{afonso2018correspondence}
Victor~I Afonso, Gonzalo~J Olmo, Emanuele Orazi, and Diego Rubiera-Garcia.
\newblock A correspondence between modified gravity and general relativity with
  scalar fields.
\newblock {\em arXiv preprint arXiv:1810.04239}, 2018.

\bibitem{kofinas2017use}
Georgios Kofinas.
\newblock Is the use of christoffel connection in gravity theories conceptually
  correct?
\newblock {\em arXiv preprint arXiv:1712.02215}, 2017.

\bibitem{kofinas2014cosmological}
Georgios Kofinas and Emmanuel~N Saridakis.
\newblock Cosmological applications of f (t, t g) gravity.
\newblock {\em Physical Review D}, 90(8):084045, 2014.

\bibitem{sobreiro2010aspects}
Rodrigo~F Sobreiro and Victor J~Vasquez Otoya.
\newblock Aspects of nonmetricity in gravity theories.
\newblock {\em Brazilian Journal of Physics}, 40(4):370--374, 2010.

\bibitem{barragan2010isotropic}
Carlos Barrag{\'a}n and Gonzalo~J Olmo.
\newblock Isotropic and anisotropic bouncing cosmologies in palatini gravity.
\newblock {\em Physical Review D}, 82(8):084015, 2010.

\bibitem{puetzfeld2005prospects}
Dirk Puetzfeld.
\newblock Prospects of non-riemannian cosmology.
\newblock {\em arXiv preprint astro-ph/0501231}, 2005.

\end{thebibliography}

\end{document}